\documentclass[12pt,twoside,notitlepage]{book}

\RequirePackage{lineno}

\usepackage{mathpple}
\usepackage{float,graphicx}
\usepackage[figuresright]{rotating}
\usepackage[usenames,dvipsnames]{color}
\usepackage{microtype}
\usepackage[small,bf]{caption}
\usepackage{tabulary}
\usepackage{booktabs,multirow,dcolumn,bigdelim}

\usepackage[pdftex,plainpages=false,pdfpagelabels,backref,pdfborder={0 0 0}]{hyperref}

\usepackage{url}
\usepackage[toc,page,titletoc]{appendix}

\usepackage{caption}
\usepackage{moresize}
\usepackage[scaled]{helvet}
\usepackage{sectsty}
\allsectionsfont{\normalfont\sffamily}
\usepackage{titletoc}
\titlecontents{chapter}
  [1.5em]
  {\linespread{0.9}\normalfont\sffamily\bfseries}
  {\contentslabel{1em}} 
  {\hspace*{-2.3em}} 
  {\mdseries\titlerule*[1pc]{.}\contentspage} 
\titlecontents{section}
  [3.5em]
  {\linespread{0.9}\normalfont\sffamily}
  {\contentslabel{2.3em}} 
  {\hspace*{-2.3em}} 
  {\titlerule*[1pc]{.}\contentspage} 
\titlecontents{subsection}
  [4.5em]
  {\linespread{0.9}\normalfont\sffamily}
  {\contentslabel{2.3em}} 
  {\hspace*{-2.3em}} 
  {\titlerule*[1pc]{.}\contentspage} 
\titlecontents{subsubsection}
  [5.5em]
  {\linespread{0.9}\normalfont\sffamily}
  {\contentslabel{2.3em}} 
  {\hspace*{-2.3em}} 
  {\titlerule*[1pc]{.}\contentspage} 
\usepackage[text={6.5in,8.75in},headheight=15pt,centering]{geometry}

\usepackage{fancyhdr}
\fancypagestyle{plain}{%
  \fancyhf{}
  \fancyfoot[RO,LE]{\sffamily \thepage}
}

\newcommand {\pp}       {\mbox{$p$$+$$p$}}

\newcommand{\be}{\begin{equation}}
\newcommand{\ee}{\end{equation}}

\usepackage{amsmath}
\usepackage{amssymb}

\usepackage{subfig}
\usepackage{subfloat}

\begin{document}  

\frontmatter

\pagestyle{empty}

\renewcommand*\familydefault{\sfdefault}
{\sffamily
\vspace{2cm}

\vspace{1.5cm}

\centerline{\huge \emph{A Proposal for the Muon Piston}}
\centerline{\huge \emph{Calorimeter Extension (MPC-EX)}}
\centerline{\huge \emph{to the PHENIX Experiment at RHIC}}

\vfill

\centerline{\Large Brookhaven National Laboratory}

\vspace{0.5cm}

\centerline{\Large Relativistic Heavy Ion Collider}

\vspace{0.5cm}


\vspace{0.25cm}
\centerline{\today}

\vfill
}
\begin{figure}[H]
  \begin{center}
  \includegraphics[width=0.8\linewidth]{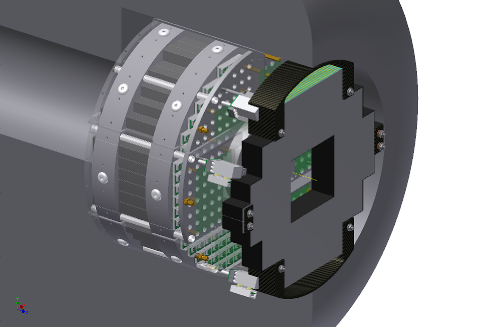}
  \end{center}
\end{figure}

\vfill
\begin{table}[hbt]
\begin{flushleft}
\begin{tabbing}
\end{tabbing}
\end{flushleft}
\end{table}
\renewcommand*\familydefault{\rmdefault}


\cleardoublepage

\pagestyle{fancy}

\chapter*{Introduction and Executive Summary}
\label{exec_summ}
\setcounter{page}{1}

The Muon Piston Calorimeter (MPC) Extension, or MPC-EX, is a Si-W preshower detector that will 
be installed in front of the existing PHENIX MPC's. This detector consists of eight layers
of Si ``minipad'' sensors interleaved with tungsten absorber and enables the 
identification and reconstruction of $\pi^{0}$ mesons at energies up to $\sim$80~GeV. 

The MPC and MPC-EX sit at forward rapidities ($3.1<\eta<3.8$) and are 
uniquely positioned to measure phenomena related to either low-x partons (in the target 
hadron or nucleus) or high-x partons (in the projectile nucleon or nucleus).  We propose 
to use the power and capabilities of the MPC-EX to make critical new measurements that will
elucidate the gluon distribution at low-x in nuclei as well as the origin of large transverse
single spin asymmetries in polarized p+p collisions. 

The collision of deuterons and Au nuclei at RHIC offers an exciting window into the initial 
state of HI collisions as well a probe of partonic phenomena in nuclei that are interesting in
their own right. Measurements of the production of $\pi^{0}$ mesons at forward rapidities 
(in the deuteron direction) at RHIC have already shown a suppression that could be interpreted 
in terms of partonic shadowing or the formation of a condensate of gluons below a saturation scale 
(the Color Glass Condensate, or CGC). The MPC-EX will be able to extend these measurements to a 
new kinematic regime, and through correlations, down to a partonic x of $10^{-3}$. Such 
measurements will provide high statistics data that can be further used to constrain models 
of the gluon saturation at low-x in nuclei. However, measurements of hadrons will be limited 
by uncertainties in the $\pi^{0}$ fragmentation functions and contamination and dilution from partonic
processes other than those of interest.  

With the capability of the MPC-EX to reconstruct and reject $\pi^{0}$ mesons (as well
as other hadronic sources of photons) at very high energies comes the capability to separate 
prompt (direct and fragmentation) photons from other sources of photons. Direct photons are extremely interesting as a 
complimentary observable to hadronic measurements. At leading order the direct photon kinematics 
are much more easily related to the parton kinematics because there is no smearing due to 
fragmentation. However, a measurement of direct photons is more difficult experimentally and will 
involve different systematic errors when compared to measurements of hadrons. 

We propose to investigate gluon saturation in nuclei at low-x through the measurement of $R_{dAu}$ for
$\pi^{0}$ mesons and direct photons. These measurements will provide strong constraints on
existing models of the gluon PDF in nuclei, such as the EPS09 PDF sets. 
These measurements will be timely and competitive with measurements from the LHC. 
The timing of a future d+Au run at the LHC is not
known, although it is certainly under discussion. Both ATLAS and CMS
have electromagnetic calorimeters in the forward region. However, a
crucial element of the direct photon measurement is the ability of the MPC-EX at RHIC to
measure relatively low p$_T$ direct photons to measure R$_{G}$ at low
Q$^2$ where the suppression is strongest, which is not accessible at the LHC. 

The large transverse single-spin asymmetries observed in polarized p+p collisions 
at RHIC are believed to be related to either initial or final state effects that originate
primarily in the valence region of the projectile nucleon (the Sivers or transversity distributions in the 
TMD approach, or parton correlations in a collinear factorized framework). While data in semi-inclusive
deep-inelastic scattering has been used to constrain these effects, the situation is more complicated
in p+p collisions due to the presence of both strong initial- and final-state corrections arising from the 
soft exchange of gluons. 

A key issue in making progress in the theoretical understanding of
transverse spin asymmetries in p+p collisions are measurements that can elucidate the origin 
of the single hadron asymmetries. One approach is to measure the single spin asymmetry $A_{N}$ for 
prompt photons, which is dominated by initial state correlations between partonic motion and 
proton spin. Because of the ability of the MPC-EX to reject high momentum 
$\pi^{0}$ mesons as well as measure the asymmmetry of background contributions from
$\pi^{0}$ and $\eta$ mesons, the SSA of prompt photons can be measured with good
precision in 200~GeV transversely polarized p+p collsions. 

Another approach is to directly measure the asymmetry in 
the fragmentation of spin-polarized quarks that arise from the hard scattering of partons in a polarized 
p+p collision. In addition to providing fine-grained information on the development of electromagnetic showers, 
the MPC-EX is also capable of tracking minimum ionizing particles (charged hadrons) that do not 
shower in the detector. While we do not have an energy or momentum measurement for these hadrons, 
this capability can be exploited to reconstruct a proxy for the jet axis for a fragmenting parton. Because 
$\pi^{+}$ and $\pi^{-}$ hadrons, the dominant charged particle species in the jet, exhibit a roughly equal and 
opposite transverse spin asymmetry, the effect of the asymmetry on the determination of the jet axis 
is minimized. This jet axis can then be used to correlate the azimuthal angle of $\pi^{0}$ mesons around the jet axis, 
with respect to the spin direction. An asymmetry measured in this way would arise from the combination
of quark transversity and the Collins spin-dependent fragmentation function (in the TMD framework). 
Measurements made with the MPC-EX would be sensitive to this source of the single particle transverse 
spin asymmetry if it made up as little as $\sim$27\% of the single-particle transverse spin asymmetry.

The structure of this proposal is organized as follows. In the first chapter we highlight the 
MPC-EX physics case for cold nuclear matter and nucleon spin. The second chapter describes the 
hardware design of the MPC-EX and its integration into the existing PHENIX detector. In the third 
chapter we detail the simulations completed to characterize the performance of the MPC-EX detector for
the reconstruction of electromagnetic showers and the separation of direct photons from other sources. 
In the last two sections of this chapter we detail a full simulation of two key physics observables 
in the MPC-EX, the direct photon and the measurement of azimuthal asymmetries in fragmentation. Finally, we 
conclude with chapters on the budget and management of the MPC-EX project. Appendix A contains additional 
information on events rates, cross sections, and triggering schemes that were used to make the projections
in the third chapter. 

\begin{figure}
  \hspace*{-0.12in}
  \centering
  \includegraphics[width=0.7\linewidth]{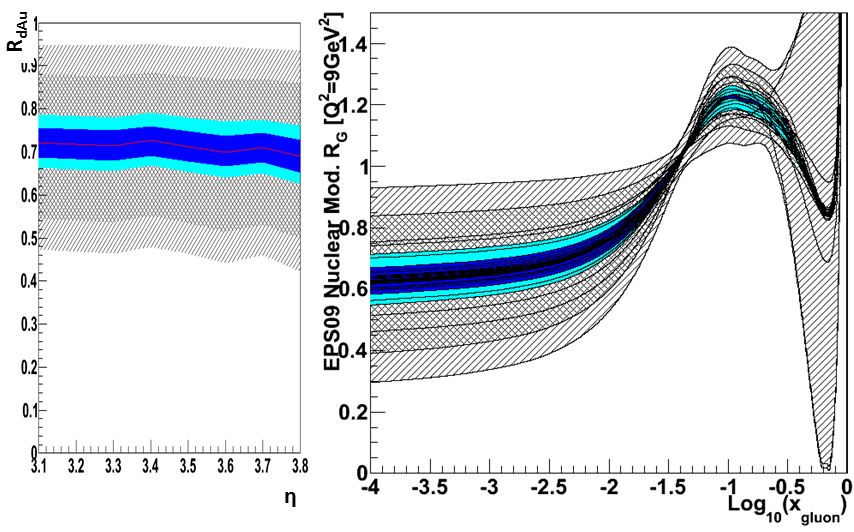}
  \vspace*{-0.12in}
  \caption{\label{Fig:rf_limits2} EPS09 exclusion plots in R$_{dAu}$
    (left) and R$_G$ (right). The outer hatched lines are the 90\%
    confidence level envelope of all the EPS09 curves. The light blue
    areas represent the 90\% confidence level limits of the simulated
    measurement, while the dark blue represent the 1$\sigma$
    limits.  The nominal value is taken as the central EPS09 curve. See Section~\ref{sim:dphot} for details.}
\end{figure}

\begin{figure}[htbp]
\centerline{
\includegraphics[width=0.6\linewidth]{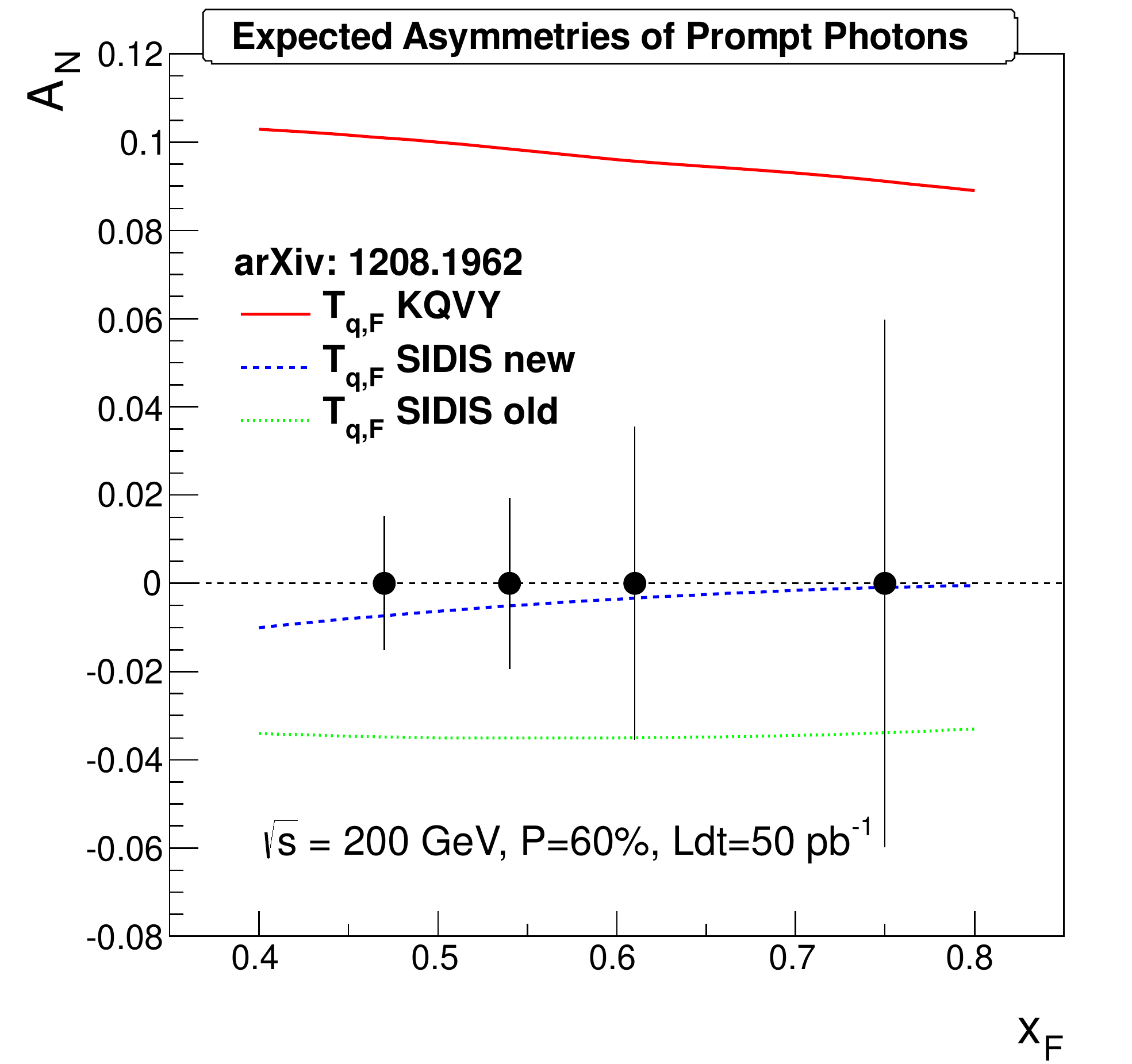}
}
\caption{Projected sensitivity for the prompt photon single spin asymmetry with the MPC-EX assuming an integrated luminosity of 
50 pb$^{-1}$ and 60\% beam polarization at $\sqrt{s}=200$ GeV. The sensitivities are shown compared to calculations in the 
collinear factorized approach ~\protect\cite{spin:KANG2011_1,Gamberg:2012iq} using a direct extraction of the 
quark-gluon correlation function from polarized $p+p$ data (upper solid curve), compared to the correlation
function derived from SIDIS extractions (lower dotted and dashed curves). See Section~\ref{sim:photon_AN} for details. 
} 
\end{figure}

\begin{figure}
\centering
\includegraphics[width=0.6\linewidth]{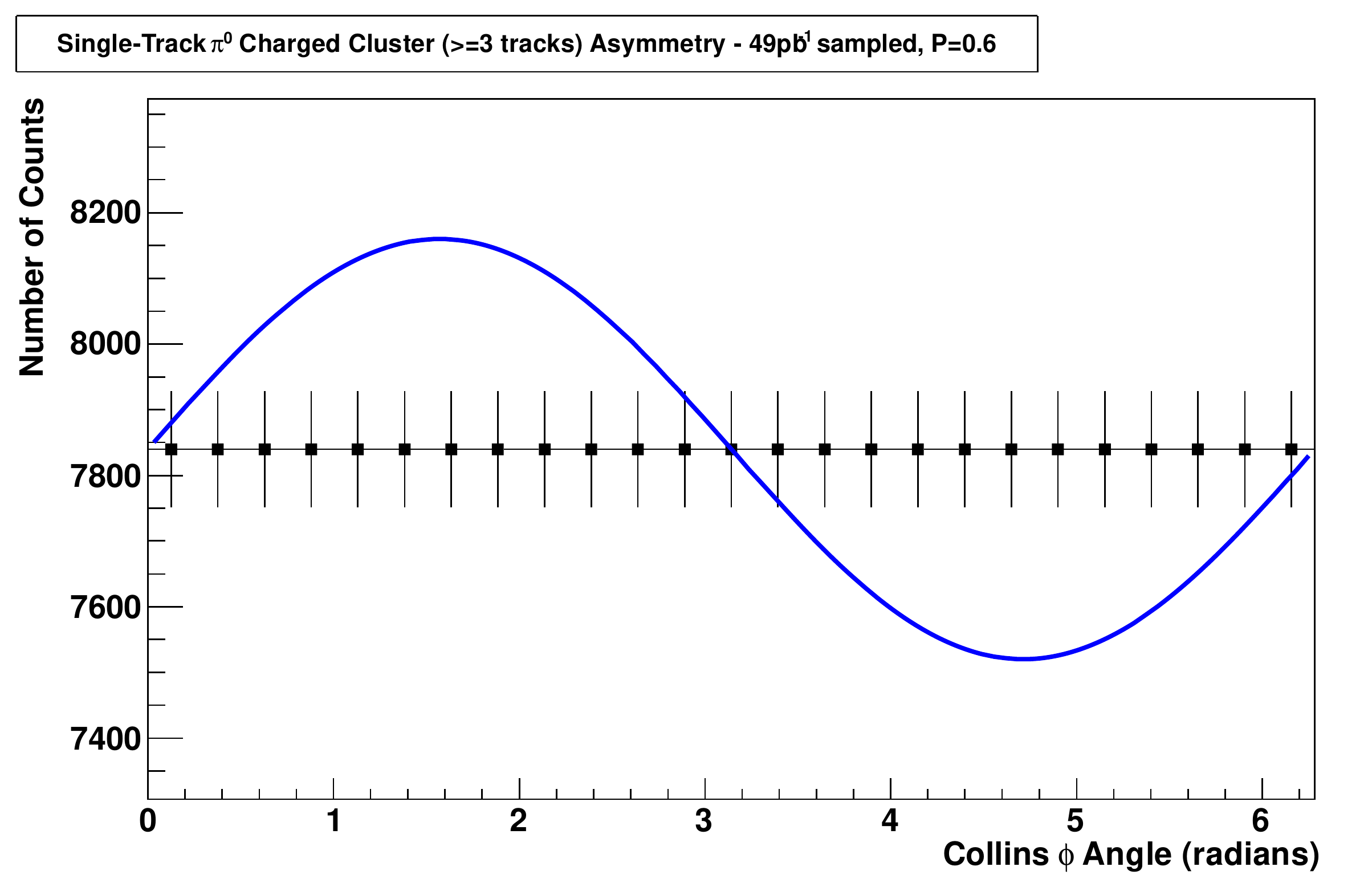}
\caption{\label{fig:st_proj2} Anticipated statistics as a function of Collins angle for 49~$pb^{-1}$ sampled luminosity and 
average polarization of 60\% using single-track $\pi^0$'s correlated with a jet axis determined by three or more charged particles. 
The blue curve is the anticipated asymmetry for the data sample from the Monte Carlo, corrected for the beam polarization 
of 60\%. See Section~\ref{sim:collins} for details.}
\end{figure}

\cleardoublepage

\resetlinenumber

\cleardoublepage

\resetlinenumber

\tableofcontents

\cleardoublepage

\resetlinenumber

\cleardoublepage

\mainmatter

\resetlinenumber

  \chapter{Physics Overview}
  \label{physics}

In this proposal we focus on two key questions in QCD - the suppression of partons at small-x in nuclei, and
how the spin of the nucleon is carried by its constituent partons. Both of these questions address 
fundamental issues in our understanding of QCD, and measurements with the MPC-EX hold the potential 
to greatly expand our understanding of the strong nuclear force. 

\clearpage
  \label{cnm}

\section[Cold Nuclear Matter]{Cold Nuclear Matter, the Initial State of 
the sQGP and  low-$x$ Physics}
\label{sec:cnm}

\subsection{Introduction}
\begin{figure}[hbt]
  \begin{center}
    \hspace*{-0.12in}
    \includegraphics[width=0.8\linewidth]{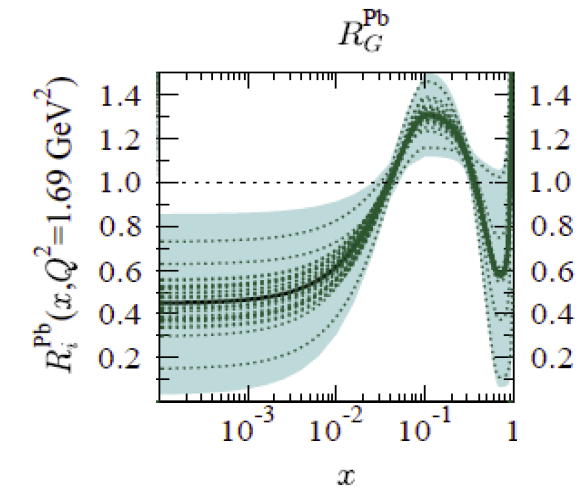}
  \end{center}\vspace*{-0.12in}
  \caption{\label{fig:mpcexcnm1} EPS09 gluon nuclear modification
    ratio, i.e. the ratio between the gluon PDF in a heavy nucleus
    (Pb) and in a proton. The lines correspond to the various
    possibilities which are consistent with world data. At low x,
    there is virtually no constraint. }
\end{figure}

The behavior of parton distributions in a heavy nucleus such as Au is
of interest since they are not simply a
superposition of nucleon parton distributions, but display effects
related to the nuclear environment.  These phenomena vary as a function
of patonic longitudinal momentum fraction x. Of particular importance is the gluon distribution at low-x
where a variety of models predict strong suppression.  Very little is
known about the gluon distribution function at x$_{gluon}$$<10^{-2}$ (for
the rest of this section x$_{gluon}$ in the heavy nucleus will be referred to as x$_2$).
Figure~\ref{fig:mpcexcnm1} shows a variety of fits to the data of the
gluon nuclear modification factor
$$R_g^A(x,Q^2)=\frac{f_g^A(x,Q^2)}{f_g^{proton}(x,Q^2)} $$ the ratio
of the gluon distribution function in a nucleus as compared to the
proton.  A strong suppression could explain the reduction in p+A
collisions relative to p+p collisions of pions and pion pairs at
forward rapidity \cite{Arsene:2004fa,Adare:2011sc} as well as the
stronger suppression of $J/\psi$ at forward rapidity as compared to
mid-rapidity \cite{Adare:2010fn}.

The need to understand such effects has taken on a new urgency because
of the discovery of the sQGP at RHIC.  The measurement of the low-x
gluon distribution of the nucleus is the first step in understanding
the formation of the sQGP at RHIC. To make a first order estimate, the bulk
of the particles at p$_T \sim$ few times the initial temperature
($\sim$ 1 GeV, assuming an initial temperature of 300-600 MeV), are
formed from gluons within a nucleus with x$_2<10^{-2}$, precisely where 
there is little constraint. In addition,
with the observation that the matter seen in heavy ion collisions at
the LHC is very similar to RHIC, the study of these effects is very
timely and important since at forward rapidity we probe the same low-x
which is relevant for bulk dynamics at the LHC.

A careful measurement of the gluons in a
nucleus would set the initial conditions of the initial entropy and
entropy fluctuations which lead to the creation of the sQGP. This in
turn would allow for the interpretation of jet and flow measurements
in terms of interesting physical quantities, e.g. the sheer and bulk
viscosity, diffusion coefficients, the
speed of sound, and the jet quenching parameter $\hat{q}$. 
For creation of the bulk hot-dense matter in A+A collisions, the
relevant x is below $10^{-2}$. For x$_{gluon}$ less than $10^{-2}$,
the uncertainty is large, hence the region
most necessary for setting the initial state of the sQGP is not well
known.

\subsection{Models including the Color Glass Condensate and EPS09}
\begin{figure}[hbt]
  \begin{center}
    \hspace*{-0.12in}
    \includegraphics[width=0.8\linewidth]{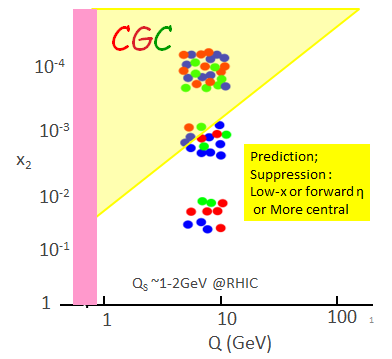}
  \end{center}\vspace*{-0.12in}
  \caption{\label{fig:cgc} A schematic of the CGC region of validity.
    The plot shows the region of validity (shaded in yellow) as a
    function of x$_2$ of the gluon in the Au nucleus in p(d)+Au
    collisions and Q. The model is valid at high density - which
    occurs at low-x. For RHIC collisions in Au nuclei, Q$_{sat}\sim$ 1-2
    GeV.}
\end{figure}

A variety of physical pictures have been used to model gluons
at low-x, or forward rapidity. These fall into several
classes. The first class of these models extend pQCD calculations into
the non-pertubative regime, via the addition of multiple scattering, coherence
 or higher twist effects\cite{Armesto:2006ph}.
A second class of models is referred to as the Color Class Condensate
(CGC)\cite{McLerran:1993ni,Kharzeev:2002ei} and assume that the
density of gluons is high enough that to first order, they can be
treated classically. Quantum corrections are added as a second order
effect.  In its region of applicability (see Figure~\ref{fig:cgc}) the CGC is a rigorous QCD
calculation with essentially one free parameter - the saturation scale
Q$_{sat}$, although in practice other parameters or assumptions are
invoked in order to make comparisons with experimental data.  The two
contrasting sorts of models could be two equivalent descriptions of
the same phenomena, with one being more appropriate than the other
depending on the kinematic range in question. An example of this ``duality''
is mentioned below in the discussion on transverse momentum dependent
gluon distributions and the CGC.

The CGC is valid for very high density systems and is a
non-pertubative model.  However it requires that the system be weakly
interacting and is appropriate only in a regime in which the density
is high enough that $\alpha_S(Q_{sat})$ is small.  Hence, one must
establish whether such calculations are applicable at RHIC.  The
partons which produce the bulk of particles constituting the hot-dense
matter at RHIC have an x$_{2}\sim10^{-2}$ with the saturation
parameter Q$_{sat}$ in the CGC model $\sim$3T$_{init}$. Assuming a
value of T$_{init} \sim$ 300-600 MeV, coming from the PHENIX thermal
photon measurement, gives Q$_{sat} \sim$ 1-2
GeV/c\cite{Kharzeev:2000ph}. Pion suppression and correlation data
from RHIC\cite{Arsene:2004fa,Adare:2011sc} at forward rapidities seem
to be consistent with the CGC hypothesis, however alternate
explanations also may explain the data.  Mid-rapidity d+Au pion data
at RHIC showed no suppression\cite{Adler:2003ii}, while it is almost
certain that similar data from the LHC will show suppression if the
CGC model is correct. If the CGC model is a good description at RHIC,
the MPC-EX should be able to measure the parameter Q$_{sat}$.

For the purposes of this proposal, a third class of models is used,
which are parametrizations of the modification of the gluon
distribution function in nuclei, $R_g^A(x,Q^2)$. They are obtained by
fitting deep-inelastic scattering events, Drell Yan pairs, and RHIC
mid-rapidity $\pi^0$s\cite{Eskola:2009uj} and are shown in
Figure~\ref{fig:mpcexcnm1}. The various lines represent different sets
of parametrizations consistent with the data, where the colored region
corresponds to the 90\% confidence level band. 
We have added to the
EPS09 distributions, a centrality dependence coming from a Glauber
model. This class of models does not invoke a physical picture save that
the gluons can be legitimately described via x$_2$, the fraction of
the nucleon momentum carried by an individual gluon.
 It must be stressed that this is simply one model and may not be a good
representation of reality; for instance it does not consider the k$_T$
of the parton with respect to in its hadron; it also may be that 
gluons should not be considered as individual entities, but rather
as a collective state.


\subsection{Direct Photons}

Low-x phenomena can be studied using direct photon production at
forward rapidities with the MPC-EX. Direct photons can either be used
on their own, or they can be correlated
with either a pion or a jet opposite in azimuth, to determine
x$_{gluon}$ to leading order with reasonable accuracy.  In the CGC model these
opposite side correlated particles are suppressed since the recoil is
absorbed by the CGC (like the Mossbauer effect). In fact the gluon PDF
which gives the distribution of gluons with a fraction x of the
nucleon's momentum, assumes a pQCD like picture. One can use three handles
to constrain the theory: the rapidity dependence, centrality
dependence (i.e.  dependence on Q$_{sat}$), and the p$_T$ balance of
the recoiling particles.  This would yield a centrality and
x-dependent set of measurements, allowing a differentiation between
various models. The x in question here would be the effective x as
measured in the experiment since the variable x$_{2}$ is not well
defined in the CGC model. The centrality dependence of most pQCD
inspired models follows a Glauber distribution, since they are
proportional to the thickness function of the nucleus, while for the
CGC it is given by the relationship between the saturation parameter
Q$_{sat}$ and the assumed gluon density. Other models, which involve
radiative energy loss of quarks traversing cold nuclear matter or
absorption, in the case of quarkonia show a non-linear behavior with
the nuclear thickness function, uncharacteristic of the Glauber
distribution as well.  

Present data from d+Au collisions at forward rapidity already shows a
suppression of correlated pions\cite{Adare:2011sc} in a manner
consistent with the CGC. Further theoretical analysis will be
necessary to differentiate this interpretation from other nuclear
effects. The analysis could also be complicated by the presence of
hadron pairs arising from multiparton interactions
(MPI)\cite{Strikman:2010bg} in which case the pairs 
made by this mechanism would not be probing the gluons at low-x.
In addition, PHENIX data
on the J/$\psi$ already indicate that cold nuclear matter effects are
non-linear. Such effects may be due to final state effects (absorption
and energy loss), or initial state effects (e.g. the gluon
PDF)\cite{Adare:2010fn}.

\begin{figure}[hbt]
  \begin{center}
    \hspace*{-0.12in}
    \includegraphics[width=0.6\linewidth]{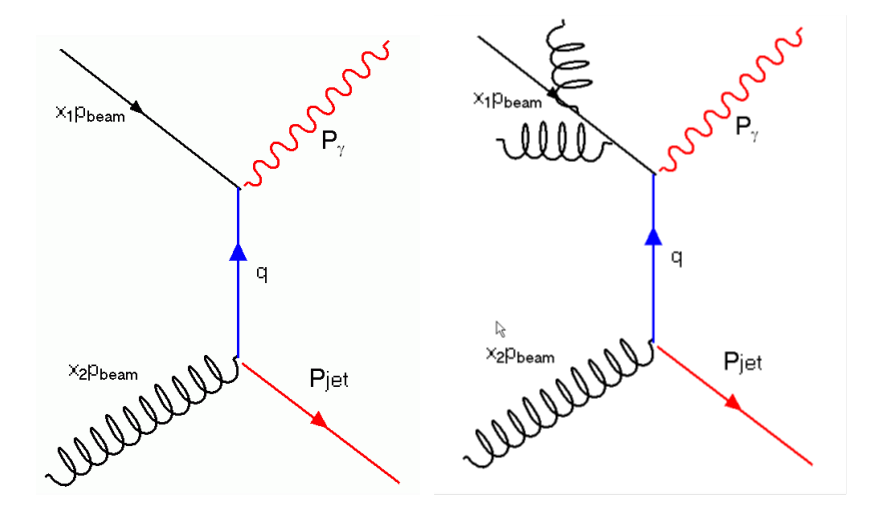}
  \end{center}\vspace*{-0.12in}
  \caption{\label{fig:mpcexcnm3} Diagrams for the production of direct
    photons in hadron-hadron collisions. To the left is a first order
    diagram, the right shows an example of a higher order diagram.  }
\end{figure}

Measurements of hadrons have an ambiguity since they involve a
fragmentation function.  Direct photons originating from the primary
vertex should clarify the situation.  Figure~\ref{fig:mpcexcnm3}, left
shows the basic first-order production diagram for direct photons at
forward rapidities.  The primary interaction is between a quark in the
deuteron and the gluon of interest in the gold nucleus, producing an
outgoing photon and jet.

\begin{figure}[hbt]
  \begin{center}
    \hspace*{-0.12in}
    \includegraphics[width=0.6\linewidth]{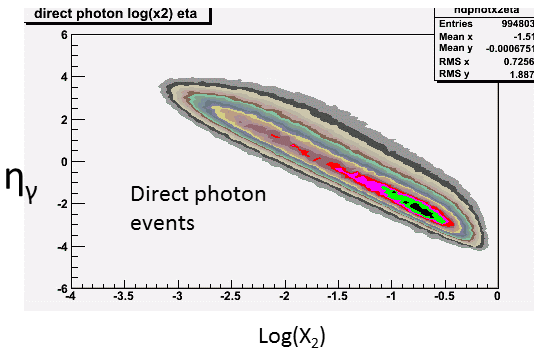}
  \end{center}\vspace*{-0.12in}
  \caption{\label{fig:dphot_eta_x2.png} Direct photon events:
    $\eta_\gamma$ vs log(x$_2$), showing the correlation between the
    pseudorapidity of the photon and x$_2$. In this figure no
    correlated hadron is required.  }
\end{figure}

\begin{figure}[hbt]
  \begin{center}
    \hspace*{-0.12in}
    \includegraphics[width=0.6\linewidth]{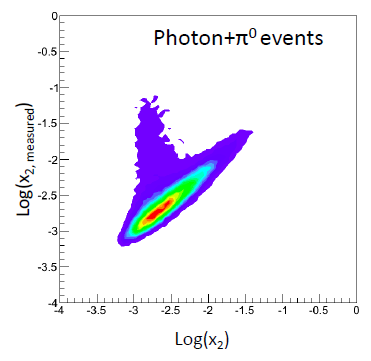}
  \end{center}\vspace*{-0.12in}  \caption{\label{fig:mpcexcnm5} 
x$_{measured}$ vs x$_2$ as described in the text. The non-diagonal
portion of the plot corresponds to cases in which the detected $\pi^0$
was did not give a good estimate of the direction of the outgoing jet,
as in the case in which the detected pion was not the leading
particle. Such events form about 10\% of the events which had a
detected photon and $\pi^0$ in the MPC-EX.}
\end{figure}

Figure~\ref{fig:dphot_eta_x2.png} shows that the rapidity of the direct
photon is directly related to the x$_2$ of the gluon.  Once the direct
photon is observed the x$_2$ can be more accurately determined by
including a correlation with a $\pi^0$ originating from the opposite
side jet.  If one assumes that the pseudorapidity of the pion is the
same as the pseudorapidity of the jet, one can deduce the x$_2$ of the
gluon to leading order through the relationship
$$x_2=p_{T\gamma}(e^{-\eta_{\gamma}}+e^{-\eta_{\pi}})/\sqrt{s}$$ where
$p_{T\gamma}$ and $\eta_{\gamma}$ refer to the direct photon,
$\eta_{\pi}$ is the pseudorapidity of the $\pi^0$ and $\sqrt{s}$ is
the nucleon-nucleon center of mass energy. We are currently exploring
our capability to measure the complete jet to improve the resolution
on x$_{gluon}$. Figure~\ref{fig:mpcexcnm5} shows that the measured value
of x$_2$ is nicely correlated with the true x$_2$ assuming that the
first order scattering diagram dominates.

While not simulated for this proposal, correlations of photons and either hadrons of jets 
will then allow us to vary the x$_2$ of the gluon in the following
manner.  We first require that the direct photon be in the positive
rapidity MPC-EX.  To reach the lowest values of x, we require the
correlated pion to be in the same MPC-EX (and be opposite in azimuth). To
reach moderate values of x, we will require a hadron to be stopped in
the positive rapidity muon arm (note that we only need the rapidity of
the pion, and not its momentum). To reach yet higher values of x, we
will require that the pion be in the VTX or central arms. We also plan
also to measure the jet angle, using the MPC-EX on both sides, and the
new silicon detectors - the VTX at mid-rapidity (installed in 2010)
and and the FVTX at forward rapidity (installed in late 2011) to
cover essentially the full range in x.


\subsection{Transverse Momentum Dependent Gluon distributions}
An exciting new development \cite{Dominguez:2011wm} has been made in
understanding the transverse momentum dependent (TMD) gluon
distributions at low-x.  Measuring direct photon-jet process in d+Au
collisions at low-x, i.e. in the forward direction will give the
MPC-EX the opportunity to measure these distributions. These TMD
distributions have been shown to be equivalent to the distributions
obtained in the CGC frame work. The relationship between these TMD
distributions and the spin dependent TMD distributions described in
section~\ref{sec:spin}, is analogous to the relationship between the ordinary
partons distribution functions, e.g. xG(x) and the spin dependent
$g^p_1$. Hence a unified picture is emerging. The MPC-EX is can access
both the spin dependent TMD PDFs and the spin independent TMD
PDFs. The spin independent TMD PDFs at low-x can be identified as those
obtained in the framework of the CGC - i.e. there is a ``duality'' between
the two methods of calculation.  This is briefly described
in what follows.

Recently work has been done in trying to understand the gluon
distributions in cold nuclear matter taking into account the k$_T$
dependence \cite{Dominguez:2011wm}. Models such as EPS09, which we are
using to benchmark the measurement, assume ``collinear
factorization'', i.e. that the physical description of the processes
depend only on x$_2$, the fraction of the proton momentum carried by a
parton. This assumes that physical processes do not depend on k$_T$,
the transverse momentum of the partons with respect to the nucleon.
The hope was that a similar procedure could be applied to physical
processes in which the k$_T$ was an important factor, e.g. in
exclusive channels, such as di-jet production, and that cross sections
could be factorized into two pieces. The first piece is the hard
parton scattering cross section which can be calculated using
pQCD. The second piece is the non-pertubative part - the
``unintegrated'' parton distributions dependent on both x$_2$ and k$_T$.
These would come from measurements. One of the important assumptions
is that the unintegrated PDFs are universal i.e., that the PDFs are
the same for all process in question. This ``TMD factorization'' is
analogous to the collinear factorization assumed in the standard k$_T$
independent analysis. Recently it has been shown that TMD
factorization is violated in a variety of process (e.g. di-jet
production)\cite{Vogelsang:2007jk}.

In the past decade, these so called unintegrated gluon distributions have been
studied in several contexts\cite{Gelis:2010nm}.  The CGC model assumes
that for small-x gluons, a regime is reached characterized by the
saturation scale Q$_{sat}$, below which the process could be calculated
semi-classically. The scale Q$_{sat}$ is the typical transverse
momentum of the small-x gluons and is related to the transverse
color-charge density - thereby leading to a ``condensate'' extending
over a large transverse portion of nuclear target. Since thick
targets, e.g. Au, would lead to a larger transverse charge density,
the transition happens at higher-x or lower energy in proton-heavy
nucleus collisions than in p+p collisions.

\begin{table}
\centering
\label{Tab:G1G2}
    \begin{tabular}{|l|c|c|c|c|c|c|}
        \hline
        ~         & DIS and DY & SIDIS    & hadron in pA & $\gamma$-Jet in pA & Dijet in DIS & Dijet in pA \\ \hline
        G$^{(1)}$ & $\times$   & $\times$ & $\times$     & $\times$           & $\surd$      & $\surd$     \\ \hline
        G$^{(2)}$ & $\surd$    & $\surd$  & $\surd$      & $\surd$            & $\times$     & $\surd$     \\
        \hline
    \end{tabular}
\caption{Processes which are sensitive to G$^{(1)}$ and G$^{(2)}$. Direct photon - jet events in pA collisions are sensitive
to G$^{(2)}$ and dijet events in pA collisions are sensitive to G$^{(1)}$. Taken from \cite{Dominguez:2011wm}. Check marks indicate the gluon distributions relevant to the given process.}
\end{table}

Recent progress\cite{Dominguez:2011wm} indicates that TMD
factorization can be recovered in the low-x limit if one considers two
different unintegrated gluon distributions, G$^{(1)}$ and
G$^{(2)}$. G$^{(1)}$ can be interpreted as the gluon
density. G$^{(2)}$ is the dipole gluon distribution and does not have
an easily understood physical interpretation. This gluon distribution
can be related to the color-dipole cross section evaluated from a
dipole of size r$_\perp$ scattering on the nuclear target.  It is
G$^{(2)}$ that enters into most of the processes of interest - for
instance the total cross section (or the structure functions) in DIS,
single inclusive hadron production in DIS and pA collisions and
Drell-Yan lepton pair production in pA collisions.  G$^{(1)}$ can be
measured in dijet final states of proton-nucleus collisions, while
G$^{(2)}$ can be measured in photon-jet final states, thus it is
crucial to measure G$^{(2)}$, which can be done by the MPC-EX.
Table~\ref{Tab:G1G2} shows a variety of processes and the relevant
gluon distributions.  The MPC-EX will be able measure both of these
distributions since the di-jet final state is also within its
capabilities.

\subsection{Measurements Simulated in this Proposal}
\begin{figure}[hbt]
  \begin{center}
    \hspace*{-0.12in}
    \includegraphics[width=0.8\linewidth]{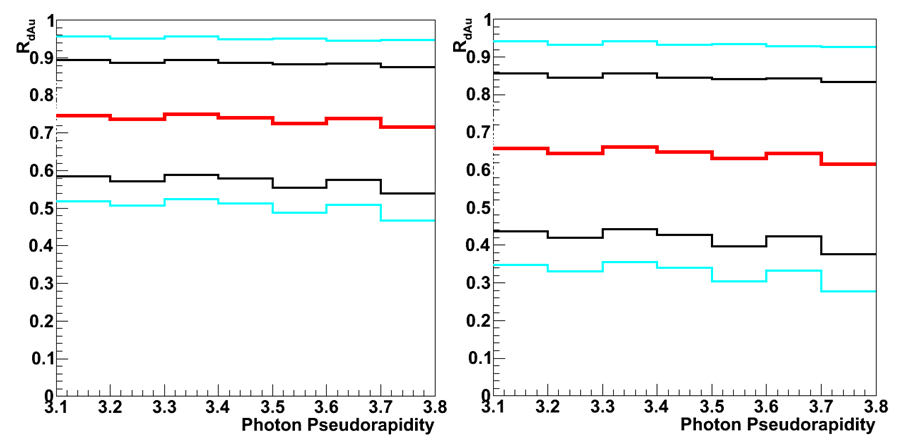}
  \end{center}\vspace*{-0.12in}
  \caption{\label{fig:rdaufragplot} Left: R$_{dAu}$ as simulated in
    minimum bias events in the MPC-EX vs $\eta$ of the photon, where
    no attempt is made to suppress fragmentation photons. Red: central
    value of EPS09. Black: R$_{dAu}$ obtained when using the least and
    most suppressed values of the nuclear gluon PDF.  Light blue: the
    envelope of all parametrization to give a one sigma range. Right:
    same for 0-20\% central events.  }
\end{figure}

\begin{figure}[hbt]
  \begin{center}
    \hspace*{-0.12in}
    \includegraphics[width=0.8\linewidth]{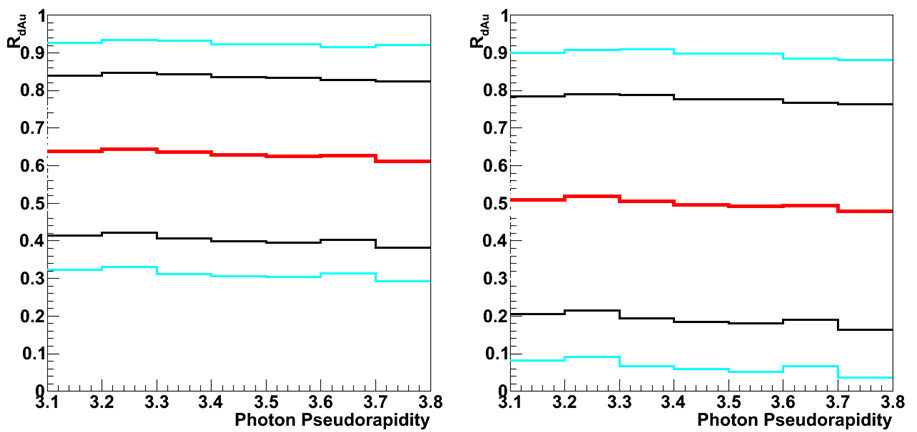}
  \end{center}\vspace*{-0.12in}
  \caption{\label{fig:rdauplot} Left: R$_{dAu}$ as simulated in minimum
    bias events in the MPC-EX vs $\eta$ of the photon assuming no
    contamination from fragmentation photons. Red: central value of
    EPS09. Black: R$_{dAu}$ obtained when using the least and most
    suppressed values of the nuclear gluon PDF.  Light blue: the
    envelope of all parametrization to give a one sigma range. Right:
    same for 0-20\% Central events.  }
\end{figure}

We have simulated the performance for the basic observable for this physics,
namely the single direct photon in forward d+Au events. 
First
we assume that the gluon distributions in Au (Figure~\ref{fig:mpcexcnm1}),
with the addition of a Glauber model will give us the centrality
dependence.  Figure~\ref{fig:dphot_eta_x2.png} shows that we will be
dominated by events where x$_2\sim10^{-3}$. We simulate the measurement
of R$_{dAu}$. In a realistic measurement, one has a contamination of,
among other things, fragmentation photons - i.e. photons which
fragment off of the outgoing quark legs of the initial hard
interaction. These of course, can be reduced using appropriate cuts,
however, for completeness we show distributions both with and without
these additional sources of photons. A detailed description will be
given in Section~\ref{sim:dphot}.  Figure~\ref{fig:rdaufragplot} shows R$_{dAu}$ for
minimum bias events (left) and central events (right), where we have
assumed that there is no attempt to suppress fragmentation
photons. The red line shows results where we have assumed
the central value of EPS09, the black line
shows the results where we have used the parametrization from EPS09 giving
the lowest and highest values or R$_{dAu}$. Recalling that all the
possible pasteurizations given by the EPS09 fits are equally good, we
take the envelope of all parametrization to give a one sigma range,
shown in light blue.  Figure~\ref{fig:rdauplot} shows the same
plot, where we have assumed that all fragmentation photons could be
eliminated. The final result will lie somewhere between
Figure~\ref{fig:rdaufragplot} and Figure~\ref{fig:rdauplot}.

It must be emphasized, as we conclude this section, that the interpretation
of our results will need to be done in close coordination with theorists
as in any measurement of a PDF, since, in reality the diagram shown in
Figure~\ref{fig:mpcexcnm3}, left, is only a first order diagram, and
higher orders (e.g Figure~\ref{fig:mpcexcnm3},right) will
contribute. What these measurements will give, however, are data to to
clarify out understanding of cold nuclear matter and to constrain the
initial condition leading to the formation of the sQGP.

\subsection{Other Experiments}
\subsubsection{STAR}
STAR will not be able to extract direct photon measurements from the
Run-8 d+Au data they have already taken with the Forward Pion
Detector (FPD), mainly due to the fact that the tower-to-tower gain
variations were too large to allow effective triggering. The STAR FPD
covers a similar kinematic region as the MPC-EX upgrade and can
distinguish $\pi^0$ from photon showers up to 50~GeV based on the size
of the crystals in the FPD and location from the interaction point. The
MPC-EX uses a finely segmented Si-W preshower detector to enable the
direct reconstruction of $\pi^0$s up to energies $>$80GeV. In many ways
the STAR FPD and PHENIX MPC-EX are complimentary and will make
complimentary and competitive measurements using different approaches
in future d+Au running.

It should be noted that the MPC-EX adds the ability to detect charged
particles as well, making possible improved isolation cuts and the
correlation of $\pi^0$s with respect to a charged cluster axis that is
sensitive to the Collins effect in spin-polarized pp collisions. In
this way the PHENIX MPC-EX adds significant new capabilities beyond
the existing STAR detector. 

\subsubsection{LHC Experiments}

\begin{figure}[hbt]
  \begin{center}
    \hspace*{-0.12in}
    \includegraphics[width=0.8\linewidth]{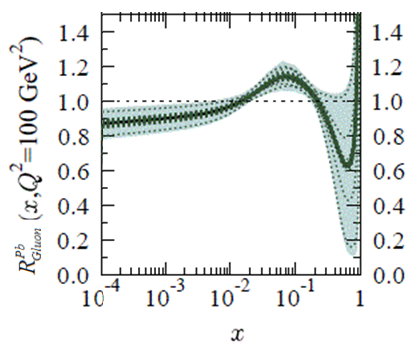}
  \end{center}\vspace*{-0.12in}
  \caption{\label{fig:Gx_lhc} EPS09 fits to R$^{Pb}_{Gluon}$ at a 
   Q$^2$ of 100 GeV$^2$. The suppression seen at lower Q$^2$ is no longer present. }
\end{figure}

Recently the LHC took a short run with p+Pb collisions. 
Both ATLAS and CMS
have electromagnetic calorimeters in the forward region. However a
crucial element of the d+Au measurement is the ability of the MPC-EX to
measure relatively low p$_T$ direct photons to measure R$_{G}$ at low
Q$^2$. Figure~\ref{fig:Gx_lhc} shows that for a Q$^2$ of 100 GeV$^2$
(p$_T$ ~ 10 GeV/c) the suppression of the gluon structure function in
nuclei prominent at low Q$^2$ is absent. The $\gamma/\pi^0$ ratio
at LHC energies even at a p$_T$ of 10 GeV/c is greater than 100,
making it essentially impossible to measure the direct photon signal.

ALICE does not have electromagnetic calorimeters in the relevant region. 
Upgrade plans call for the construction of a forward Calorimeter (FOCAL) 
which may be able to make measurements at low Q$^2$. The timescale of 
the ALICE FOCAL is after the MPC-EX is scheduled to take physics data. 

\clearpage
  \label{spin}

\section[Nucleon Spin Structure]{Nucleon Spin Structure}
\label{sec:spin}

\subsection{Nucleon Structure: Transverse Spin Physics}

\begin{figure}[htbp]
\centerline{\includegraphics[width=0.6\linewidth]{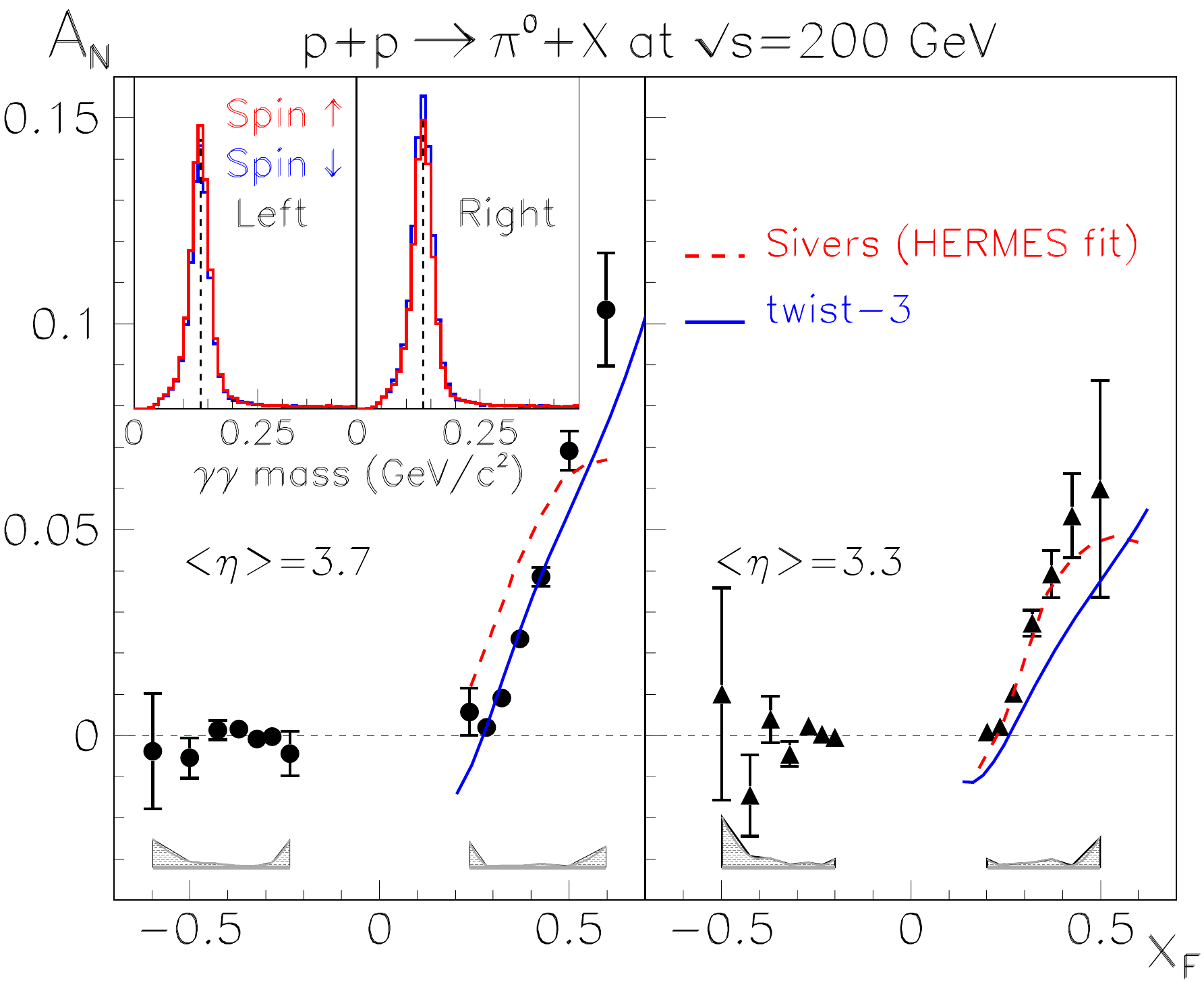}}
\caption{Single spin asymmetry $A_N$ from $\pi^0$ mesons at two different forward
rapidity bins ($\langle\eta\rangle = 3.3, 3.7$) as a function of Feynman
$x_F$, measured at the STAR experiment from transversely polarized
$p+p$ collisions at $\sqrt{s} = 200$~GeV~\protect\cite{spin:STAR2008}.
 The calculations are: i) a fit~\protect\cite{spin:UMBERTO2004} of quark Sivers function from HERMES proton Sivers results, ii) a twist-3 calculation~\protect\cite{spin:TWIST32006}
 as described later in the
text. The inset shows examples of the spin-sorted invariant mass
distributions. The vertical lines mark the $\pi^0$ mass.}
\label{fig:NCC10}
\end{figure}

\begin{figure}[htbp]
\centerline{\includegraphics[width=0.6\linewidth]{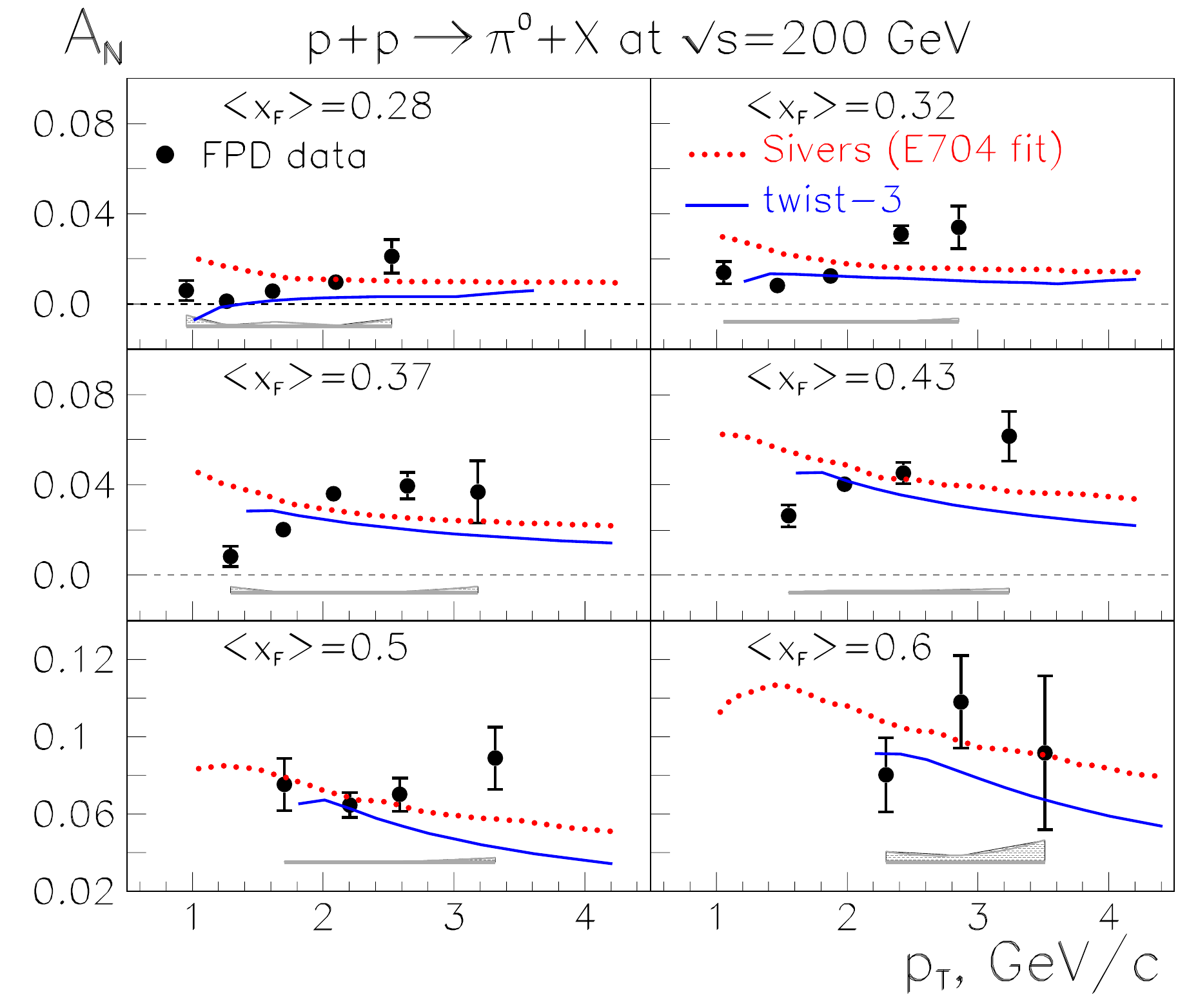}}
\caption{Data from STAR: Transverse
momentum ($p_T$) dependence of Single spin asymmetry $A_N$  in fixed $x_F$ bins of $\pi^0$ mesons production 
in $p+p$ collisions at $\sqrt{s} = 200$~GeV. 
~\protect\cite{spin:STAR2008}.
 The calculations are: i) a fit~\protect\cite{spin:UMBERTO2004} of quark Sivers function from HERMES proton Sivers results, ii) a twist-3 calculation~\protect\cite{spin:TWIST32006}
 as described later in the
text.}
\label{fig:NCC10_2}
\end{figure}

\begin{figure}[htbp]
\centerline{\includegraphics[width=0.8\linewidth]{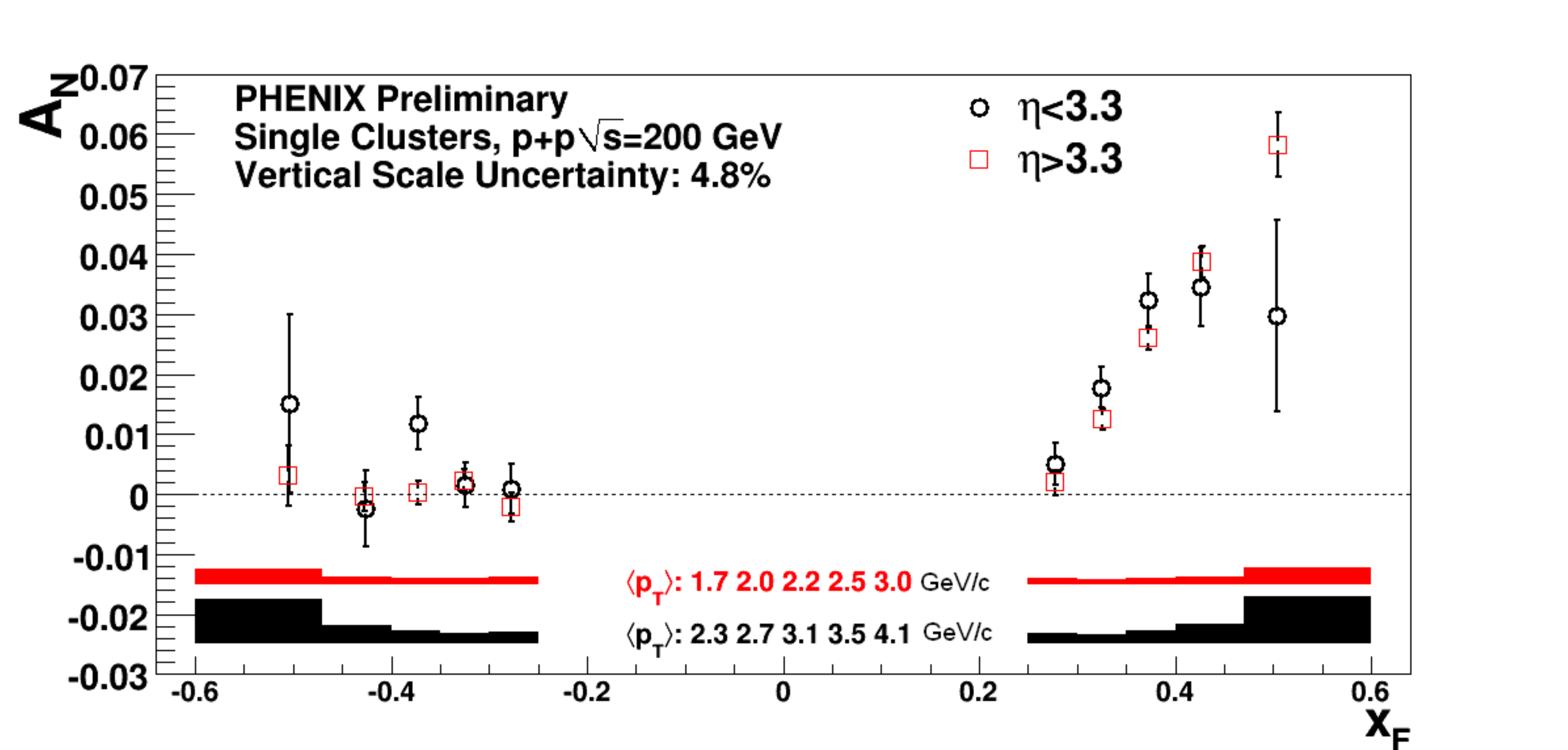}}
\caption{PHENIX preliminary results of single spin asymmetry $A_N$ vs $x_F$ of MPC single-cluster hits (presumably $\pi^0$s)
in $p+p$ collisions at $\sqrt{s} = 200$~GeV. 
}
\label{fig:NCC10_3}
\end{figure}

\begin{figure}[htbp]
\centerline{\includegraphics[width=1.0\linewidth]{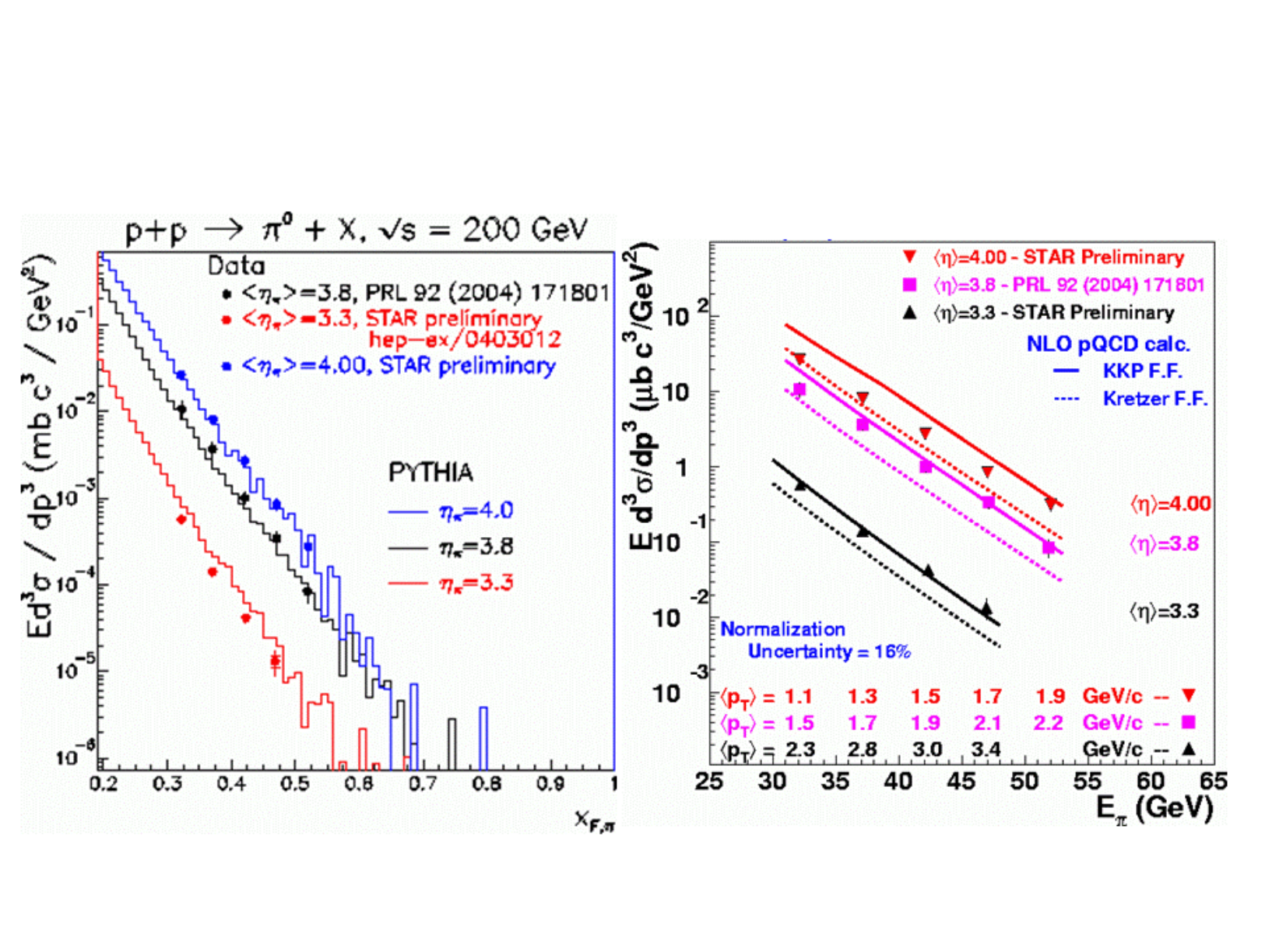}}
\caption{Forward inclusive $\pi^0$ cross sections measured at the STAR 
experiment from transversely polarized
$p+p$ collisions at $\sqrt{s} = 200$~GeV~\protect\cite{spin:STAR2008}; 
the average pseudorapidity
is $\langle\eta\rangle = 3.8$.
In the left panel, these results are compared to
predictions using PYTHIA~\protect\cite{Sjostrand:2000wi} 
as a function of Feynman x; in the right panel 
they are compared to NLO pQCD 
calculations as a function of the pion
energy.}
\label{fig:NCC11}
\end{figure}

Since the observation of surprisingly large single transverse spin 
asymmetries (SSAs) in $p^\uparrow + p\to \pi + X$ at Fermilab during
1980s and 1990s~\cite{spin:E704}, the exploration of the physics behind the
observed SSAs has become a very active research branch in hadron
physics, and has played an important role in our efforts to
understand QCD and nucleon structure. The field of  transverse spin physics has now become one of the hot spots in high energy nuclear physics,
 generating tremendous excitement on both theoretical and experimental fronts.
Fermilab E704's observation
of large SSA~\cite{spin:E704} initially presented a challenge for QCD theorists and contradicted the general expectation from pQCD of vanishingly
small SSA assuming it is originated from a helicity flip of a collinear parton.  
It was even more startling that the SSA discovered by E704
at $\sqrt{s}$ = 20 GeV did not vanish at all, as expected from pQCD, at the much higher 
$\sqrt{s}$ of 62.4 GeV and 200 GeV from the BRAHMS~\cite{spin:BRAHMS2008} and the STAR~\cite{spin:STAR2008} experiments. 
The surprisingly large SSA of $\pi^0$ mesons observed at STAR,
as a function of Feynman $x$, is shown in Figure~\ref{fig:NCC10}. Although theory calculations based 
on a fit~\cite{spin:UMBERTO2004} of Sivers 
Transverse Momentum Dependent parton distributions (TMD) and a twist-3 calculation~\cite{spin:TWIST32006} roughly 
described the $x_F$ dependencies of SSAs, they failed to describe the trend of transverse momentum ($p_T$) 
dependencies of SSA, as shown in Figure~\ref{fig:NCC10_2}.  PHENIX preliminary results of forward  ``single-cluster'' MPC hits (presumably  $\pi^0$s) SSA
$A_N$, as in Figure~\ref{fig:NCC10_3}, also showed similar large size asymmetries. 
 One might
question whether the forward reactions are hard enough to apply perturbative
QCD, but as shown in Figure~\ref{fig:NCC11} the cross sections of $p + p\to \pi^0 + X$ are reasonably 
described by NLO pQCD~\cite{spin:STARNLO} as well as 
by PYTHIA simulations~\cite{Sjostrand:2000wi}.
The existence of large single spin asymmetries at very forward rapidities at RHIC, along with the
good theoretical understanding 
of the unpolarized cross-sections gives hope that
transverse spin phenomena in polarized $pp$ collisions at RHIC 
can be used as a tool to probe the correlation between parton's transverse motion and the nucleon's spin 
in order to provide a 3-dimensional dynamical image of the nucleon.

In order to explain these large single-spin asymmetry phenomena associated with transversely polarized $p+p$ collisions, 
three basic mechanisms have been introduced (although they can not be clearly separated from each other 
in inclusive hadron SSA measurements): 

\begin{enumerate}

\item The ``Collins Effect'': a quark's transverse spin~\cite{spin:transversity} (transversity) generates 
a left-right bias during the (spin-dependent) quark fragmentation process~\cite{spin:Collins}. 

\item The ``Sivers Effect'': a parton's transverse motion generates a left-right bias~\cite{spin:Sivers}. \\ 
The existence of the parton's Sivers distribution functions ($f^\perp_{1T}$),
 one of the eight leading order Transverse Momentum Dependent parton distributions (TMDs),
 which is naive T-odd and describes the correlation between parton's transverse momentum and the nucleon's transverse spin, allows a left-right bias 
to appear in the final hadron's azimuthal distribution.  This ``TMD factorization approach'' is valid in 
the low $p_T$ region ($ p_T \sim \Lambda_{QCD} \ll Q$).

\item The so-called ``twist-3 colinear factorization approach'', valid in high $p_T$ region 
($p_T \gg \Lambda_{QCD}$): a higher twist (twist-3) mechanism in the initial and/or final 
state~\cite{spin:KANG2011_1} that describes SSA in terms of twist-3 transverse-spin-dependent correlations between quarks and gluons. 
It was shown theoretically that in the intermediate $p_T$ region ($\Lambda_{QCD} \ll p_T \ll Q$) that overlap between 
the TMD factorization approach and the twist-3 approach, as in the case of SSAs measured at RHIC $p+p$ collisions, 
both methods describe the same physics such that
a link between the moments of twist-3 three-parton correlation function $T_{q,F}(x,x)$, and the quark Sivers distribution $f_{1T}^{\perp q}(x)$
can be established~\cite{spin:KANG2011_1}.
\end{enumerate}

The Collins and the Sivers effects, although not possible to be separated in inclusive hadron SSA in $p+p$ collisions, 
can be clearly separated through azimuthal angle dependence of SSA measured in semi-inclusive deep-inelastic 
scattering (SIDIS) reactions. It has been a world-wide effort over 
the last several years to measure SSA in SIDIS reactions.
The HERMES experiment at DESY carried out the first SSA measurement in SIDIS reaction on a transversely polarized proton target~\cite{spin:HERMES2005,spin:HERMES2010}.
The COMAPSS experiment at CERN carried out similar SSA measurements on transversely polarized deuteron and proton targets~\cite{spin:COMPASS2009,spin:COMPASS2010}.
Most recently, Jefferson Lab Hall A published results of SSA measurements on a transversely polarized neutron ($^3$He) target~\cite{spin:HALLA2011}.   

\begin{figure}[htbp]
\centerline{\includegraphics[width=0.8\linewidth]{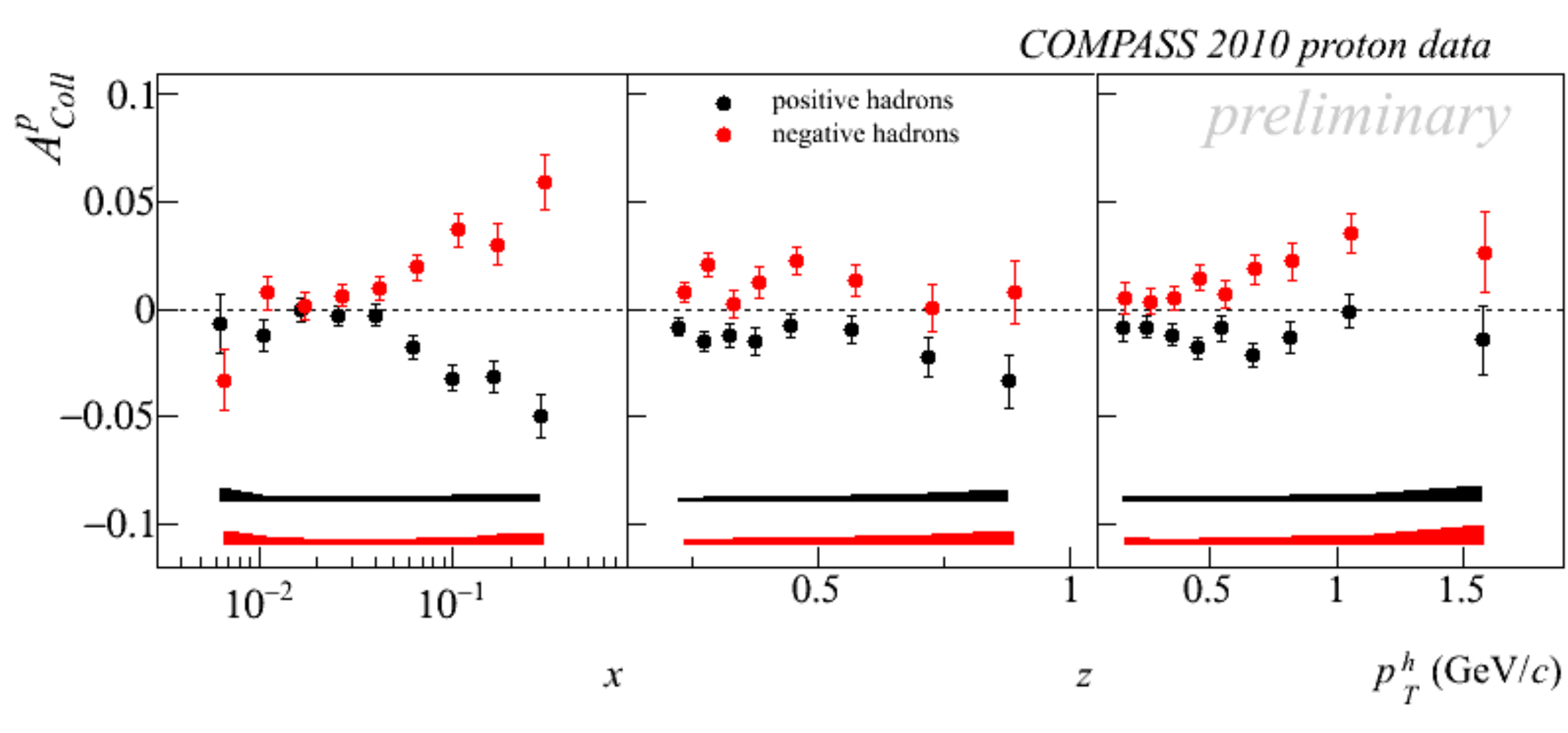}}
\caption{The COMPASS Collaboration's preliminary Collins single spin asymmetry results in semi-inclusive deep-inelastic scattering on a transversely polarized proton target
~\protect\cite{spin:COMPASS2011}.
}
\label{fig:COMPASS2011_Collins}
\end{figure}

In the recent Transversity-2011 Workshop,  the COMPASS Collaboration presented their new preliminary data of high statistic SSA results of 2010-run on a 
transversely polarized proton target~\cite{spin:COMPASS2011}, as shown in Figure~\ref{fig:COMPASS2011_Collins}. 
The Collins SSA of proton for COMPASS and HERMES agree reasonably well in the overlapping kinematic region, 
and show clear non-zero SSA for both positively and negatively charged hadrons with opposite signs of asymmetries.

The observed non-zero Collins asymmetry in SIDIS, which is related to the convolution products of the chiral-odd quark transversity distribution~\cite{spin:transversity} with another chiral-odd object the ``Collins Fragmentation Function'', strongly indicated that both the quark transversity as well 
as the quark to hadron Collins fragmentation functions are non-vanishing. 
The similar amplitudes and the opposite signs of positive-hadron SSA relative to that 
of the negative hadron indicated that the the up-quark transversity is opposite 
to that of down-quark, but similar in amplitudes, and the ``unfavored'' Collins fragmentation function is opposite in sign to that of the ``favored'' one, 
perhaps with an even larger amplitude.  
Independently, effects of non-zero Collins fragmentation function have been observed by the BELLE Collaboration
~\cite{spin:BELLE} in $e^+e^-$ annihilation and the quark to hadron Collins fragmentation function have been first extracted from these data~\cite{spin:Anselmino2009}. 

%
%

\begin{figure}[htbp]
\centerline{
\includegraphics[width=0.5\linewidth]{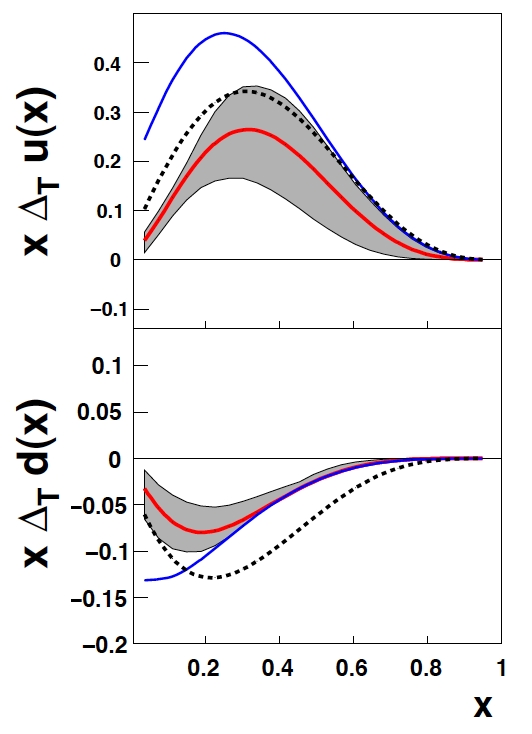}
\includegraphics[width=0.48\linewidth]{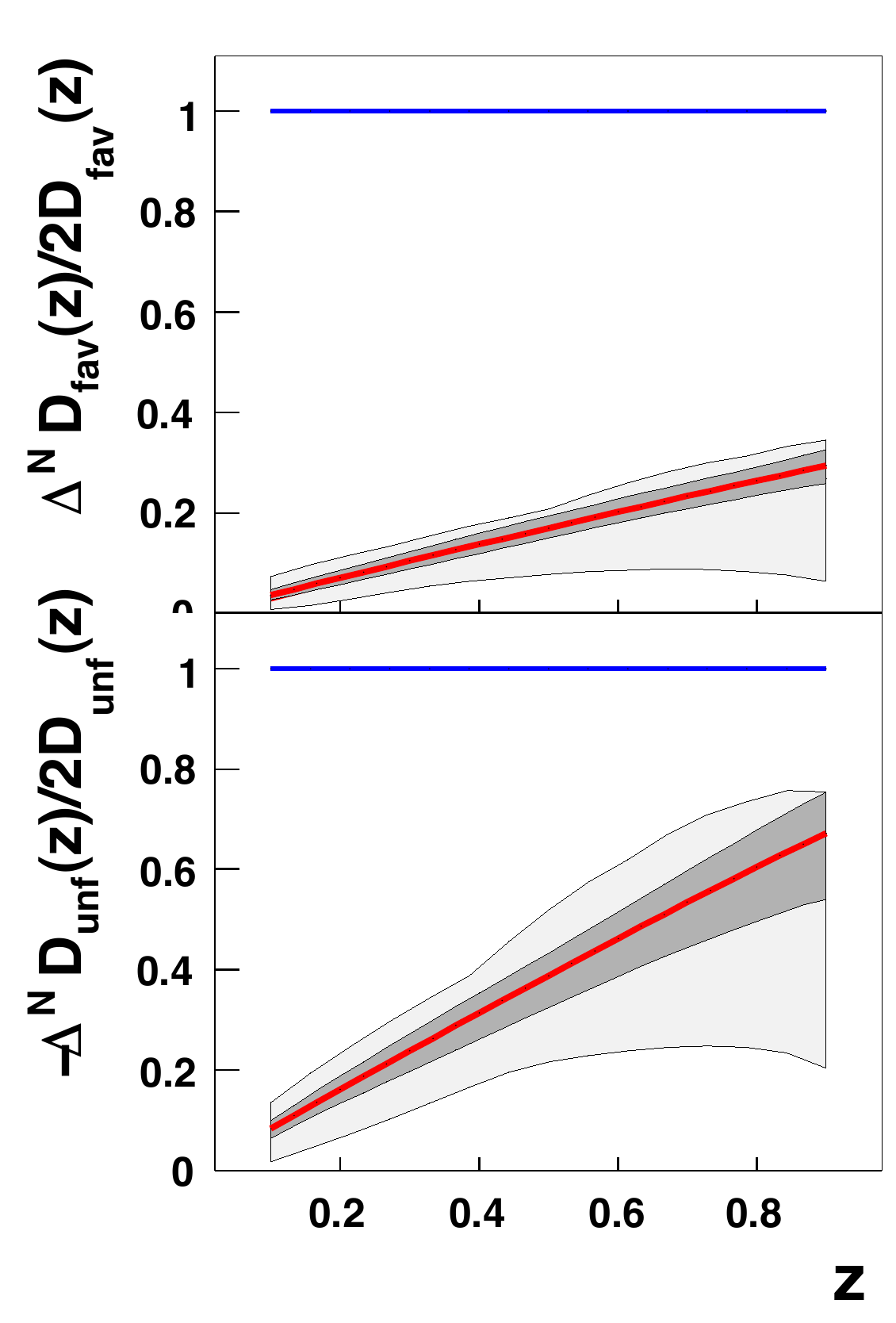}
}
\caption{The quark transversity (left) distributions, and the Collins fragmentation functions (right) as extracted from SIDIS and $e^+e^-$ data. 
In both cases the solid red curve indicates the distributions as determined by the global best fit to the data. The gray bands
are an indication of the uncertainty in the extraction. In the left panel, the extracted
transversity (solid line) is compared with the helicity distribution (dashed
line) at $Q^2=2.4$ GeV$^2$ and the Soffer positivity bound (blue solid line). In the right panel, the favored and the unfavored Collins
fragmentation functions, at $Q^2$ =
2.4 GeV$^2$; are compared with the positivity bound and the (wider)
uncertainty bands obtained in an earlier fit. 
}
\label{fig:transversity}
\end{figure}

The existence of non-zero Collins fragmentation function
allows the extraction of the quark transversity distributions inside the nucleon.
Transversity or $\delta q_f(x)$, is one of the three leading order quark distributions which survive the integration of 
quark transverse momentum. They are: quark momentum distribution $f_q(x)$, helicity distribution $\Delta f_q(x)$ 
and transversity distribution $\delta q_f(x)$.   
Quark transversity is a measure of the quark's spin-alignment
along the nucleon's transverse spin direction, and it is different from that of helicity distribution
since operations of rotations and boosts do not commute.
The $0^{th}$-moment of transversity, $\sum_f \int_0^1 \delta q_f(x) dx$, yields nucleon's tensor-charge as one of the 
fundamental properties of the nucleon just like its charge and magnetic moment. 
Transversity requires a helicity change of 1-unit between the initial and the final state of the parton such that 
gluons, which have spin-1, are not allowed to have transversity.
Therefore, quark transversity distribution is sensitive only to the
valence quark spin structure, and its $Q^2$ evolution follows that of non-singlet densities which do not 
couple with any gluon related quantities, a completely different behavior compared to that of 
the longitudinal spin structure. 
These attributes provide an important test of our understanding of the
anti-quark and gluon longitudinal spin structure functions, especially
with regard to relativistic effects.
Quark transversity distributions and quark spin-dependent Collins fragmentation functions have been extracted from a QCD global fit~\cite{spin:Anselmino2009}
of published HERMES proton and COMPASS deuteron SIDIS Collins asymmetries in 
conjunction with the BELLE $e^+e^-$ data. The results are shown in Figure~\ref{fig:transversity}.

\begin{figure}[htbp]
\centerline{\includegraphics[width=0.8\linewidth]{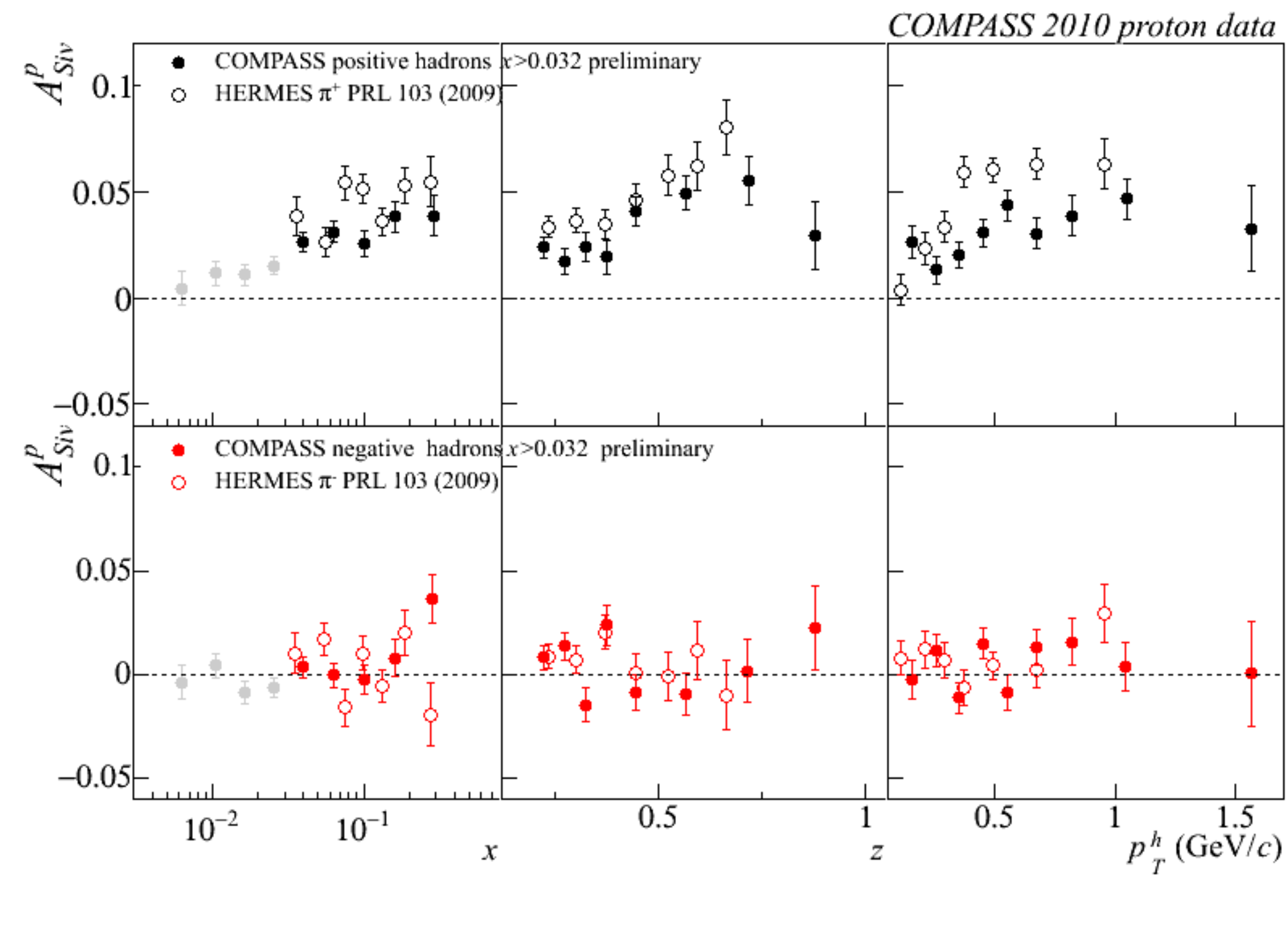}}
\caption{The COMPASS Collaboration's preliminary Sivers single spin asymmetry results in semi-inclusive deep-inelastic scattering on a transversely polarized proton target
~\protect\cite{spin:COMPASS2011} compared with that of published HERMES data~\protect\cite{spin:HERMES2005}.
}
\label{fig:COMPASS2011_Sivers}
\end{figure}

The ``Sivers effect'', and the quark Sivers distributions as a completely different mechanism, 
was thought to be forbidden since early 1990s due to its odd nature under the ``naive'' time-reversal operation. 
It was only in 2002 when Brodsky {\it et al.}~\cite{spin:BRODSKY2002}
demonstrated that 
when quark's transverse motion is considered a left-right biased quark Sivers distribution is not only allowed, it could also be large enough 
to account for the large observed inclusive hadron 
SSAs in $p+p$ collisions.  
Subsequent SIDIS measurements have shown the existence of such non-zero Sivers SSAs, as summarized in Figure\ref{fig:COMPASS2011_Sivers} for a comparison of 
proton Sivers SSA of preliminary COMPASS run-2010 data and the published HERMES data.
 Clear non-zero Sivers SSA are observed in the positive hadron ($\pi^+$ in HERMES) production, 
while the negative hadron ($\pi^-$ in HERMES) SSA are consistent with zero,  along with the
 COMPASS deuteron~\cite{spin:COMPASS2010} $\pi^+$ and $\pi^-$ Sivers SSA, 
indicating that up-quark and down-quark Sivers distributions are opposite in sign. 
Such pronounced flavor dependence of the quark Sivers functions were also indicated
by a phenomenological fit~\cite{Anselmino:2005ea} of the published proton and deuteron Sivers SSA data.

\begin{figure}[htbp]
\centerline{
\includegraphics[width=0.45\linewidth]{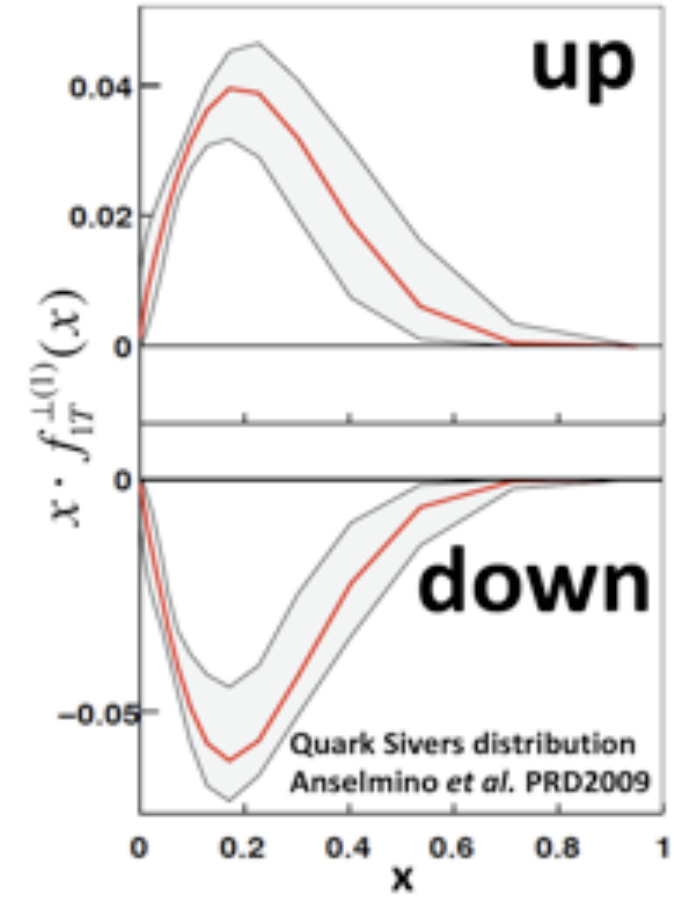}
\includegraphics[width=0.34\linewidth]{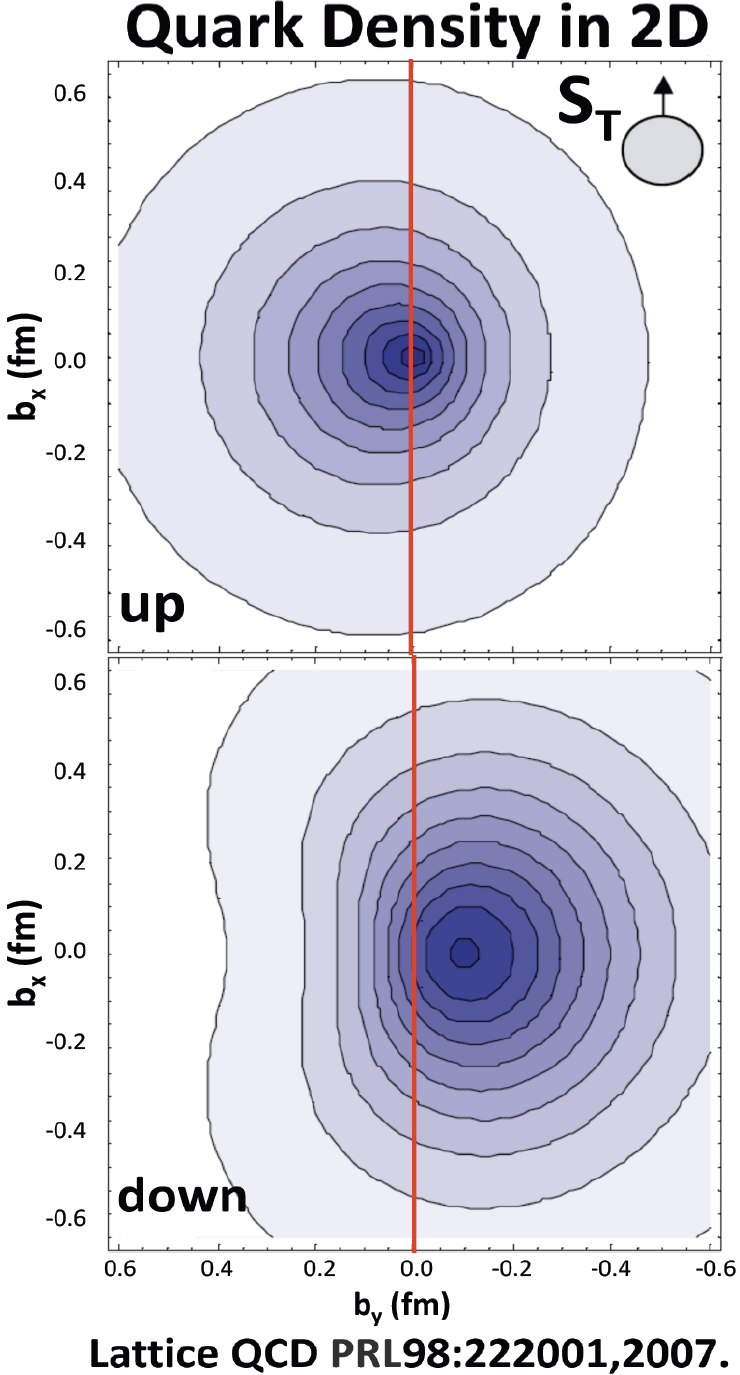}
}
\caption{The quark Sivers distributions (left plot), as extracted from published proton and deuteron target SIDIS data, for up-quark (top) and down-quark (bottom).  The gray bands
are an indication of the uncertainty in the extraction. A Lattice-QCD calculation of quark 2-dimensional 
density distribution in the impact parameter space ($b_x$ vs $b_y$) for up-quark and down-quark is shown (right plot) 
with the nucleon polarized in the transverse direction.} 
\label{fig:Anselmino}
\end{figure}

Since the Sivers SSA is related to the convolution products of  the 
quark Sivers distributions $f^{\perp}_{1T}$ and the ``regular-type'' spin-independent quark to hadron fragmentation function, 
which are reasonably well-known through 
$e^+e^-$ annihilation and SIDIS hadron production data, quark Sivers distributions have been
extracted through global QCD fits~\cite{Anselmino:2005ea} of existing proton and deuteron targets SIDIS data, as shown in Figure~\ref{fig:Anselmino}. An illustration of quark 2D density distribution from a Lattice-QCD calculation is also shown, indicating a left-right imbalance of quark density in a transversely polarized nucleon.
Sivers function $f^{\perp}_{1T}$ represents a correlation between
the nucleon spin and the quark transverse momentum, and it corresponds
to the imaginary part of the interference between light-cone wave function 
components differing by one unit of orbital angular momentum~\cite{spin:BRODSKY2002}.
A nonzero $f_{1T}^{\perp}$ arises due to initial (ISI) and/or final-state interactions (FSI) 
between the struck parton and the remnant of the polarized nucleon~\cite{spin:BRODSKY2002}. It was further demonstrated through 
gauge invariance that the same Sivers function, originates from 
a gauge link, would lead to SSAs in SIDIS from FSI and in Drell-Yan from ISI
but with an opposite sign~\cite{Collins_sivers,brodsky_sivers}. 
This ``modified universality'' of quark Sivers distribution is an important test of the QCD gauge-link 
formalism, and the underline assumption of QCD factorization used to calculate these initial/final state colored 
interactions.
A direct test of such a fundamental QCD prediction of Sivers function sign change between SIDIS and Drell-Yan
has become a major challenge to spin physics, and it has been designated an 
DOE/NSAC milestone. 
Polarized Drell-Yan experiments are currently under preparation 
at COMPASS and at RHIC IP2, and in the planning stage for both STAR and PHENIX upgrades at RHIC and possibly for 
a fixed target Drell-Yan experiment at Fermilab.  
The existence of non-zero quark Sivers distributions is now generally accepted and well defined. 
Quark Sivers distribution provides an interesting
window into the transverse structure of the nucleon, and provides constraints to quark's 
orbital angular momentum, although currently only in a model-dependent fashion.
Recently, using a lattice-QCD ``inspired'' assumption that links quark Sivers distribution with quark Generalized Parton Distributions $E$,
 quark total angular momentum ($J^q$) has been quantified~\cite{spin:BACCHETTA2011} for the first time as: $J^u = 0.266 \pm 0.002^{+0.009}_{-0.014}$ and $J^d = -0.012 \pm 0.003^{+0.024}_{-0.006}$.
 
\begin{figure}[htbp]
\centerline{
\includegraphics[width=0.45\linewidth]{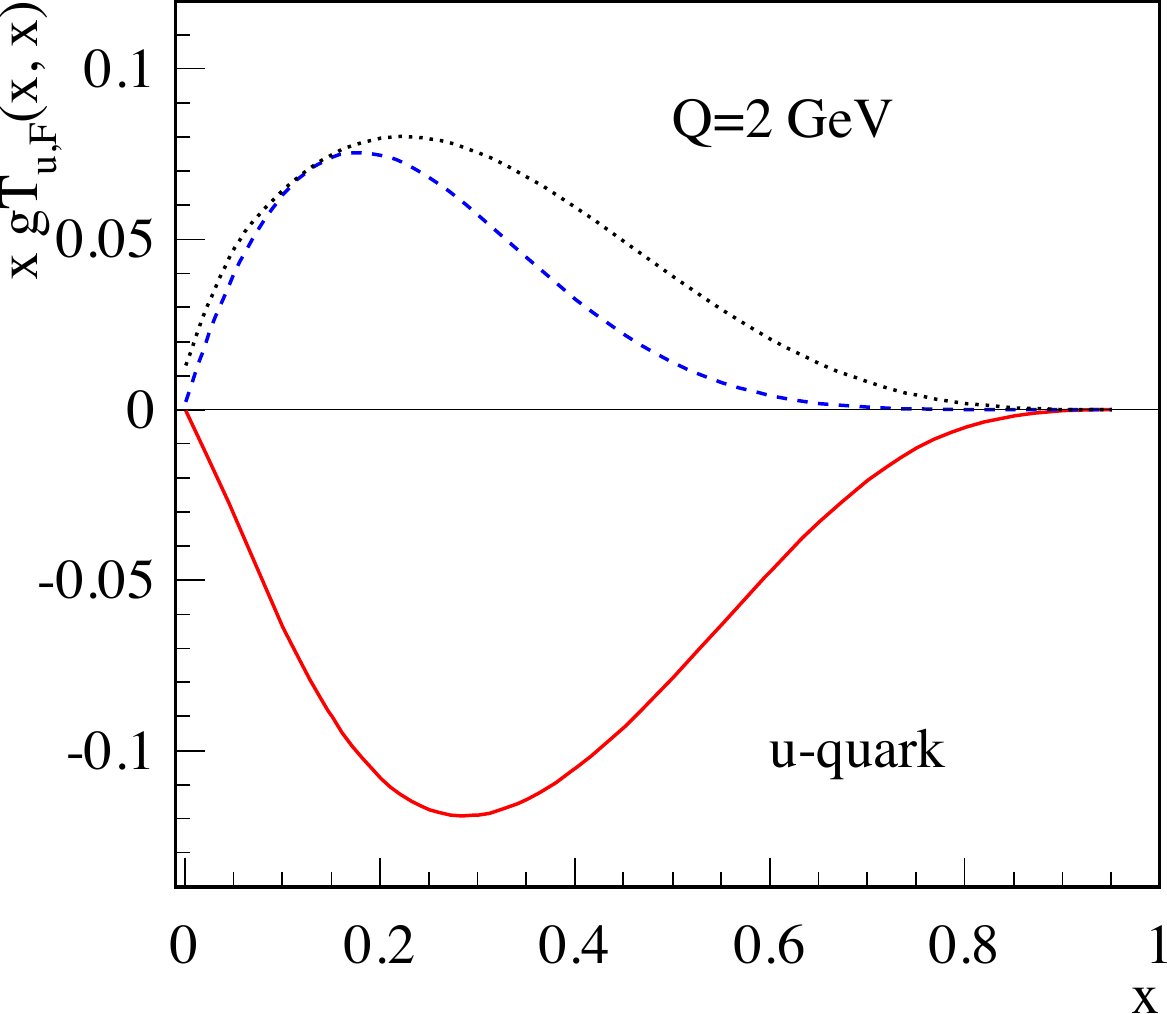}
\includegraphics[width=0.45\linewidth]{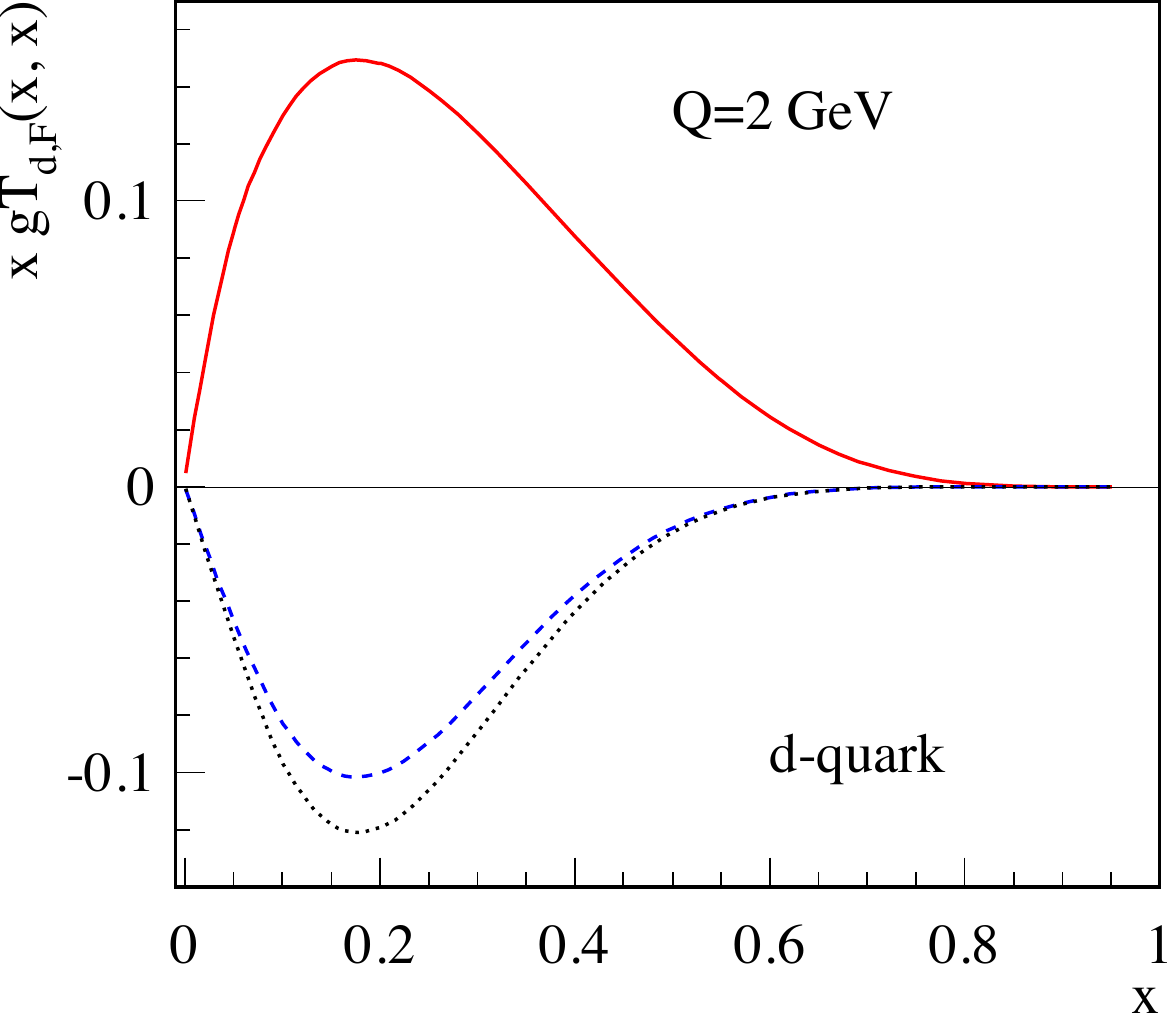}
}
\caption{The
quark-gluon correlation function $gT_{q,F}(x, x)$ as a function of
momentum fraction $x$ for $u$-quarks (left) and $d$-quarks (right).
The solid lines represent ``direct extraction'' from $p+p$ inclusive SSA data in the twist-3 approach, 
while the dashed and dotted lines represent Sivers functions extracted from SIDIS data assuming two different functional forms.
}
\label{fig:Kang2011_twist3}
\end{figure}

Linking the Sivers effect with the twist-3 colinear factorization approach, 
the twist-3 transverse-spin-dependent quark-gluon correlation function $T_{q,F}(x,x)$ 
extracted from $p+p$ inclusive SSA data was shown to be directly 
related to the moments of Sivers functions, 
thus provide an independent check of our understanding of SSA phenomena in SIDIS and in $p+p$. 
However, very recent studies by Kang et al. showed that {\bf the quark Sivers function moments extracted by these two methods are similar in size, 
but  opposite in sign}~\cite{spin:KANG2011_1}, as shown in Figure~\ref{fig:Kang2011_twist3} for the up-quark (left) and the down-quark (right). 
The solid lines represent twist-3 approach 
``direct extraction'' from $p+p$ inclusive SSA data, 
while the dashed and dotted lines represent Sivers functions extracted from published 
SIDIS data assuming two different functional forms.  
This controversy of Sivers function sign ``mismatch'' indicates either a serious flaw in our understanding of transverse spin phenomena, or alternatively
drastic behaviors~\cite{spin:BOER2011} of 
quark Sivers function in high momentum fraction ($x$) or in high transverse momentum ($k_t$).
Given the facts that the existing SIDIS measurements are limited to $x \le 0.35$, 
high precision $p+p$ SSA measurements at very forward rapidity 
are urgently needed to provide constraints in the high-$x$ region. 
 
\begin{figure}[htbp]
\centerline{
\includegraphics[width=0.6\linewidth]{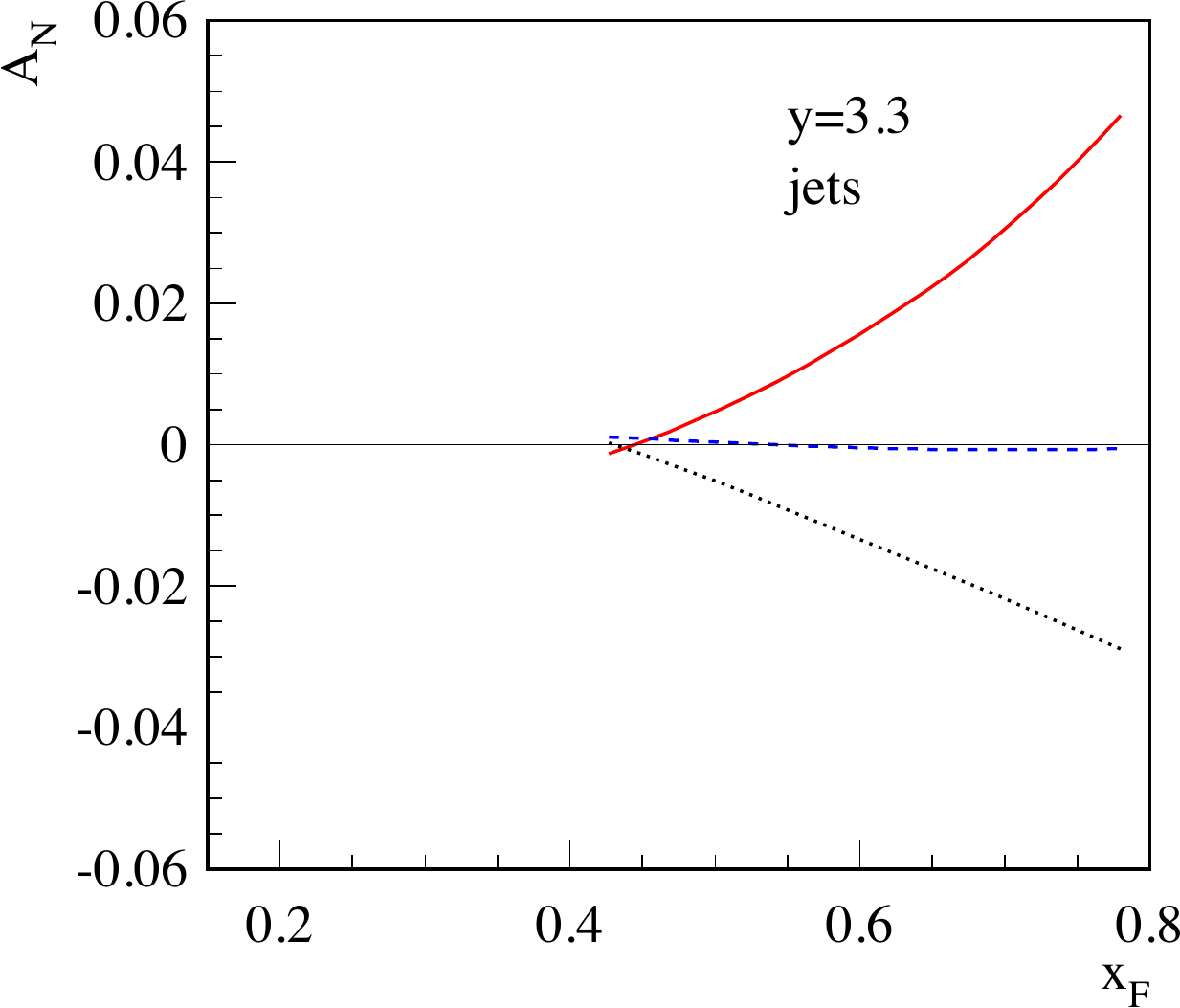}
}
\caption{The SSAs for inclusive jet production $A_N^{jet}$  in
$p^\uparrow p$ collisions~\protect\cite{spin:KANG2011_1} at $\sqrt{S}=200$ GeV, as functions of 
$x_F$ for rapidity $y=3.3$. 
The solid lines represent ``direct extraction'' from $p+p$ inclusive SSA data in the twist-3 approach, 
while the dashed and dotted lines represent Sivers functions extracted from SIDIS data assuming two different functional forms.
}
\label{fig:jetAn}
\end{figure}

\begin{figure}[htbp]
\centerline{
\includegraphics[width=0.6\linewidth]{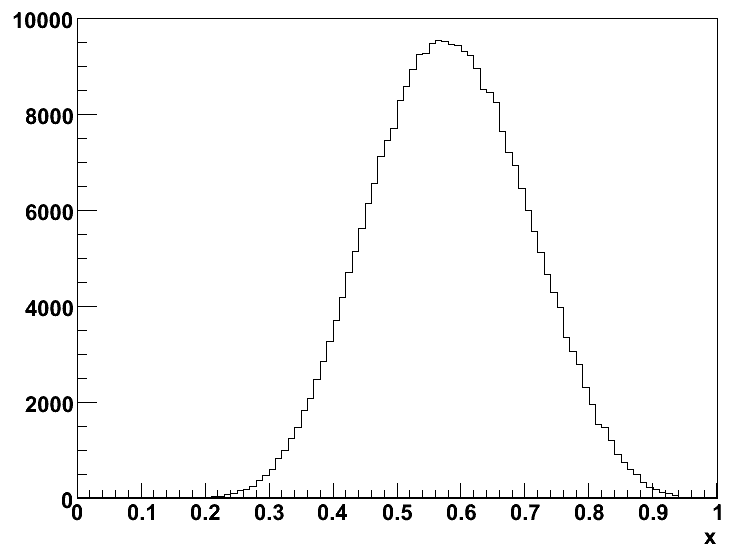}}
\caption{Bjorken-x distribution in polarized proton for PYTHIA events with a hadron scattered into $3.1<\eta<3.8$ for $x_{F}>0$. A substantial fraction of the data is at $x_{B}>0.35$, where the DIS data ends.}
\label{fig:x_fract}
\end{figure}

Unlike polarized SIDIS reactions, SSA effects in forward hadron production 
in transversely polarized $p+p$ collisions are somewhat
more complicated to interpret since both the Final State Interactions and the Initial State Interactions exist. 
From past observations, the single-spin effects in $p+p$ are typically larger than those of SIDIS, 
thus are much easier for experiments to measure. 
The main goal of these types of $p+p$ measurements must be to clearly 
isolate individual effects in SSAs in order to gain a 
deeper understanding of the fundamental physics.
The MPC-EX, along with the Muon Piston Calorimeter (MPC)
and the standard PHENIX central and muon-arm detectors,
will allow a series of transverse spin 
measurements to be carried out
at PHENIX. Especially, with the capability to reconstruct ``jet-like'' structures at forward rapidity, 
two kinds of SSA observables are of particular interest:

\begin{enumerate}

\item {\bf Hadron azimuthal distribution asymmetry  inside a jet ($A_N^{h~in-jet}$) arises purely from the Collins effect.} \\ 
The quark's transverse spin (transversity) can generate a left-right bias inside a jet. 
A measurement of $A_N^{h~in-jet}$ will provide constraints on the product of 
quark transversity distributions and the Collins fragmentation function 
Specifically for MPC-EX, the left-right asymmetry of $\pi^0$ inside a jet ($A_N^{\pi^0~in-jet}$) is a pure Collins effect.
The experimental observable in MPC-EX would be the azimuthal distribution of $\pi^0$ yields around the jet axis reconstructed with the MPC-EX, 
and the azimuthal angle $\phi_S$ is between the proton spin direction $\vec{S}_p$ and the transverse momentum $\vec{k}_T$ of the pion 
with respect to the jet axis, $\vec{p}_{jet}$
One advantage that such a measurement would have over existing SIDIS measurements would be that the $x$ range measured for the transversity distribution
would be substantially higher than that reached in SIDIS, see Figure~\ref{fig:x_fract}.  
While the next generation SIDIS experiments at JLab-12GeV will extend to high-$x$ region starting in FY-2015, 
the current SIDIS data do not exceed beyond $x_{Bj}=0.35$,

\item {\bf The azimuthal asymmetry of inclusive jet ($A_N^{jet}$) arises purely from the Sivers effect.} \\
The Collins effect does not contribute to $A_N^{jet}$ as it averages out 
in the integration over the azimuthal angle of hadrons inside the jet. 
A measurement of $A_N^{jet}$ will provide information on the product of quark Sivers distributions and 
the well-known spin-independent fragmentation functions. 
Predictions of $A_N^{jet}$ in the MPC-EX acceptance are at a few $\%$ level with a large range of variations reflecting our 
lack of knowledge on quark Sivers functions at high-$x$, as shown in Figure~\ref{fig:jetAn}
The measurement of $A_N^{jet}$ can be carried out with the 
MPC-EX by recording the jet yields for the different transverse proton spin
orientations and constructing the relative luminosity corrected asymmetries
between the yields for the up versus down proton spin orientations.

\end{enumerate}

The most critical experimental performance parameters for these type of MPC-EX
 measurements would include the angular resolution for the direction of the jet axis and the resolution in the hadron momentum fraction $z$. 
Uncertainties in knowing the jet axis will dilute the amplitude of the azimuthal Collins asymmetry and uncertainties in measuring hadron's energy fraction 
($z=E_h/E_{jet}$) will 
smear the spin analyzing power of the Collins fragmentation function in the stage of data interpretation.
The latter of these two is very important, given that the Collins fragmentation function has a strong $z$-dependence, see Figure~\ref{fig:transversity}.




\subsection{Other possible SSA measurements with MPC-EX}

\begin{figure}[htbp]
\centerline{
\includegraphics[width=0.8\linewidth]{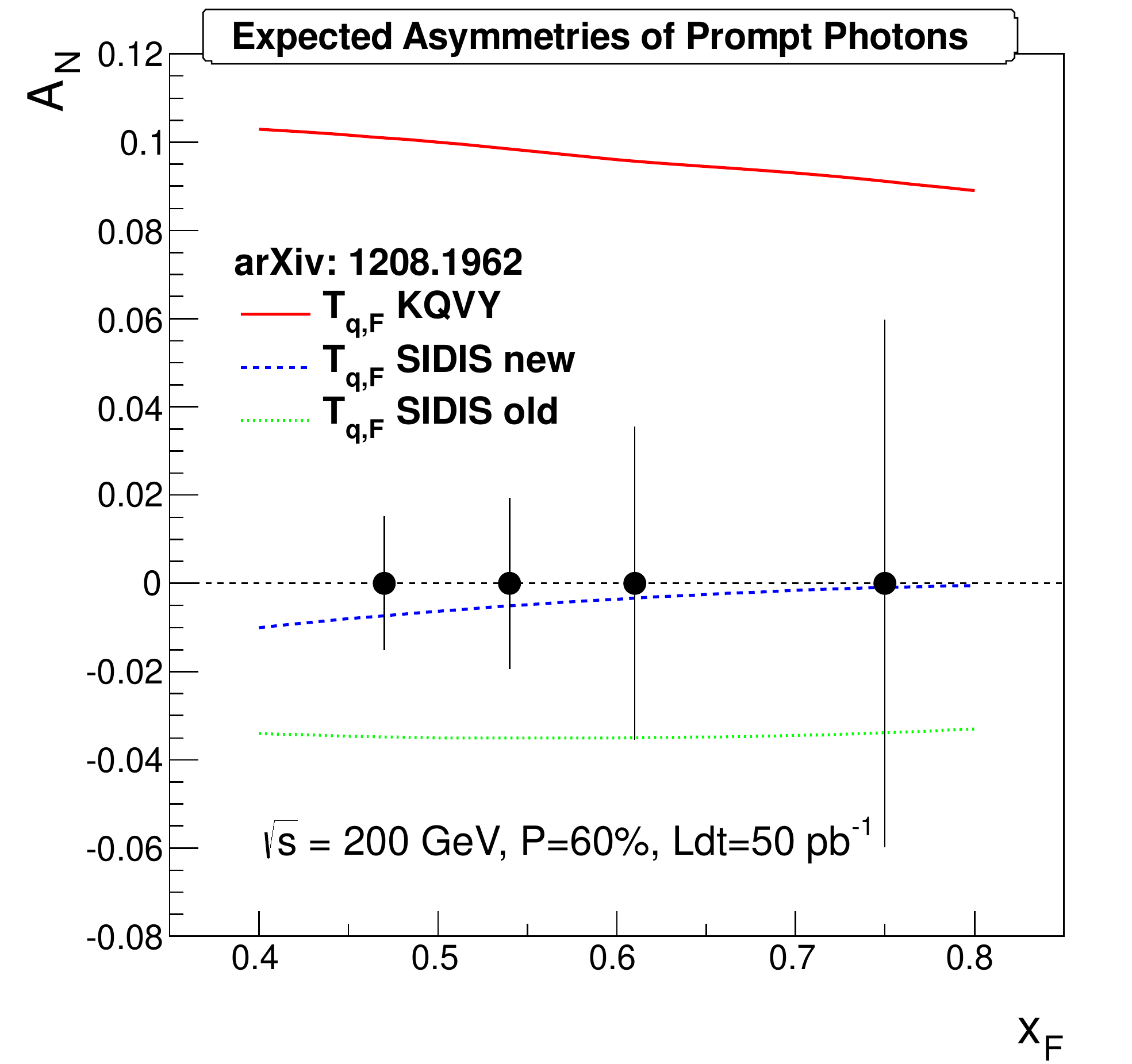}
}
\caption{The SSAs for prompt photon production $A_N^{\gamma}$ in
$p^\uparrow p$ collisions~\protect\cite{spin:KANG2011_1} at $\sqrt{S}=200$ GeV, as functions of 
$x_F$ for rapidity $y=3.5$. The solid lines represent ``direct extraction'' from $p+p$ inclusive SSA data in the twist-3 approach, 
while the dashed and dotted lines represent Sivers functions extracted from SIDIS data assuming two different functional forms. The statistical 
error bars include statistical errors as well as uncertainties introduced by subtraction of background photon SSA originating from meson decays. 
See Section~\ref{sim:photon_AN} for details.}
\label{fig:photonAn}
\end{figure}

In addition, not elaborating on the details, we list here other possible SSA measurements with MPC-EX:

\begin{enumerate}

  \item Prompt photon SSA ($A_N^\gamma$), which purely arises from the Sivers effect. The expected measurement statistical 
precision of an MPC-EX measurement 
(data points from Monte Carlo simulations, see Section~\ref{sim:photon_AN}) are shown in Figure \ref{fig:photonAn}, with theory predictions of prompt photon $A_N^{\gamma}$ of Kang {\it et al.}\cite{spin:KANG2011_1}, which includes contributions from direct and fragmentation photons. Different assumptions for the quark Sivers functions lead to predictions of opposite signs for $A_N^{\gamma}$. 

  \item SSA of back-to-back di-hadrons and back-to-back di-jets.

  \item SSA of back-to-back $\gamma$-jet~\cite{spin:Bacchetta} and back-to-back photon-pairs.

\end{enumerate}

\subsection{Measurements Simulated in this Proposal}

In order to demonstrate the capabilities of the MPC-EX, we have chosen to simulate a particular observable 
in detail, namely the correlation of $\pi^{0}$ mesons with the axis of a jet. Such a correlation would 
yield information about the Collins fragmentation function and transversity within the nucleon. This observable exercises 
two main features of the MPC-EX: the ability to identify charged tracks, and the ability to reconstruct
$\pi^{0}$ mesons at very large momentum. These simulations are described in Section~\ref{sim:collins}. 

Of course, without full jet reconstruction the MPC-EX cannot measure the full $z$ dependence of the Collins fragmentation
function (it can, however, yield measurements in ``low-$z$'' and ``high-$z$'' samples by selecting $\pi^{0}$ momentum regions). 
The main goal of this measurement will be to quantify what fraction of the inclusive $\pi^{0}$ $A_N$ results from the 
Collins fragmentation function and transversity, and by inference, what is the role played by the Sivers effect. In this
sense this measurement with the MPC-EX can be considered a ``pathfinder'' measurement that will point the way to future
experiments at RHIC with complete forward spectrometers. 

\newpage

\subsection{Summary: MPC-EX and the Study of Nucleon's Transverse Spin Structure}

The goal of nucleon spin structure studies is to understand how the nucleon spin 
is composed of the spin and orbital angular momenta of the quarks and gluons inside the nucleon. 
With the MPC-EX we will address the following fundamental questions regarding
the nucleon's intrinsic spin structure and the color-interactions that hold together the nucleon's building blocks: 

\begin{enumerate}
  \item Is a quark's spin aligned with nucleon spin in the transverse direction ?   
  \item What is the role of quark's transverse spin (transversity) during fragmentation  ?
  \item What is the role of parton's transverse motion and its correlation with nucleon spin ? 
  \item What is the role of the color-interactions between a hard-scattering parton and the remnant of the nucleon ?
\end{enumerate}

Specifically, with the new experimental capabilities provided by the MPC-EX, we will make precision measurements  
that provide clear answers to the following questions:

{\bf 
When a transversely polarized proton produces a very forward jet in a
high energy $p+p$ collision, relative to the direction of proton's spin,

\begin{itemize}

  \item Would a $\pi^0$ particle favor the left side or the right side within the jet (Collins + Transversity)?

  \item Would the jet itself favor the left side or the right side of the collision (Sivers)?

\end{itemize}

}

\cleardoublepage

\resetlinenumber

\cleardoublepage

\resetlinenumber

  \chapter{The MPC-EX Preshower Detector}
  \label{design}

\section[The MPC-EX]{The MPC-EX Detector}
\label{dsn:design}

The MPC-EX detector system includes both the existing Muon Piston Calorimeters (MPCs) and the proposed extensions which are two, nearly identical, W-Si preshower segments located upstream of the north and south MPCs respectively. This pairing will share the available space inside the PHENIX muon magnet piston pit.  Their functionality is largely complementary. The new preshower will
\begin{itemize}
\item {Improve the quality of measurements of electromagnetic showers in the MPC aperture by reducing the longitudinal leakage of energy,}
 
\item {Improve the discrimination between electromagnetic and hadronic 
showers,}

\item {Enable the reconstruction of $\pi^0$'s via an effective mass measurement and shower shape analysis to the $p_T$ extent allowed by the calorimeter 
acceptance and RHIC luminosity,}

\item {Measure jet 3-vectors with a precision sufficient to allow a correlation with $pi^{0}$ meson to measure the Collins asymmetry in polarized proton-proton collisions,}

\item {Assist in measuring energies inside jet cone around high-$p_T$ lepton candidates for isolation testing.}
\end{itemize}

\begin{figure}[htbp]
\centerline{\includegraphics[width=0.8\linewidth]{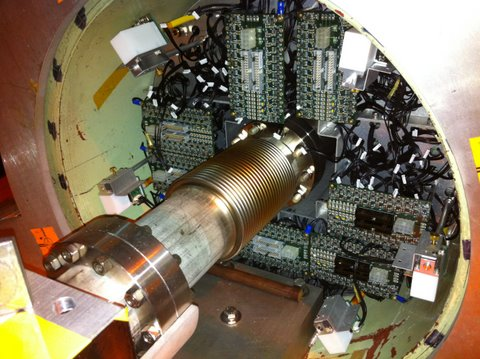}}
\caption{\label{fig:MPCInThePit} 
A beam view of the North Muon Piston with MPC installed. Signal cables removed.}
\end{figure}

The current MPC's\cite{chiu:2007MPC} (see Fig.~\ref{fig:MPCInThePit}) were installed in 2006 and have already produced a wealth of physics results. With the aim to further extend the physics reach of the existing PHENIX forward spectrometers we have designed extensions (a preshower) to complement the existing MPC's.  By themselves, the MPCs are highly segmented total absorption detectors with a depth of $\sim$18$X_{0}$. The preshower converts photons and will track and measure the energy deposited in the active Si layers by photons and by charged particles.  Additionally, the preshower will count and classify hits (as originating from electromagnetic or hadronic showers), measure hit-to-hit separations, and reconstruct effective masses from hit pairs, which can be further used to extract $\pi^0$ yields. By measuring the $\pi^0$ yields in the same detector, a direct photon extraction can be performed in a self-consistent way, without using extrapolated data with often unknown systematics for background subtraction.

The MPC-EX's are located $\sim$210\,cm from the nominal collision point north and south of the PHENIX central magnet. The MPC alone is capable of resolving close hits with similar energies down to a separation of the order of 3\,cm; this effectively limits the $\pi^0$ reconstruction range to momenta below $\sim$15\,GeV/$c$. To extend that range towards the $\pi^0$ luminosity limit in the forward direction ($\sim$100\,GeV) the preshower is designed as a sampling structure of tungsten and active pixelated silicon layers with readout integrated with silicon in the form of micromodules.  Silicon provides for versatility of segmentation, while tungsten has a small Moli\`{e}re radius (9.3\,mm) so the showers in the preshower are very compact. Tungsten also has an excellent ratio of radiation and absorption lengths, well matching that of PbWO4 (MPC crystals) which is important for electromagnetic energy measurements in the presence of a large hadronic background.  The preshower is comprised of eight sampling layers each consisting of 2\,mm thick W plate and 3\,mm deep readout. The total depth of the preshower ($\sim$$4X_0$) is chosen to allow both photons from a $\pi^0\rightarrow\gamma\gamma$ decay to convert and be reliably measured in at least two X and two Y sampling layers.

The granularity of the preshower is chosen to match the expected two photon separation in $\pi^0$  decays. A $p$\,=\,100\,GeV/$c$ $\pi^0$ produced at the nominal collision point will generate two hits in the preshower separated by $\sim$1\,cm (compare this to the Moli\`{e}re radius of the detector $\sim$2\,cm).  To match both the shape of the MPC towers and the minimal two photon separation requirement, the silicon pixels are rectangular in shape and have a transverse size of $\sim$1.8$\times$15\,mm$^2$.  The signal from each pixel is split with a ratio of 1:30 with individual copies sent to two independent SVX4 chips. 

The ideal location for this preshower would be flush with the front face of crystals in MPC to minimize large-angle spray fluctuations at the boundary. Unfortunately, this is precluded by the earlier decision to locate the MPC readout (APD's and signal drivers) upstream of crystals. The actual preshower location on the beam line is also constrained by concerns about additional background to muon tracker station 1 from inside of the Muon Piston pit.  This concern will be ultimately decided upon upon completion of integration study of utilities and cable routing which is currently being pursued for the MPC-EX upgrade.

\begin{figure}[htbp]
\centerline{\includegraphics[width=0.8\linewidth]{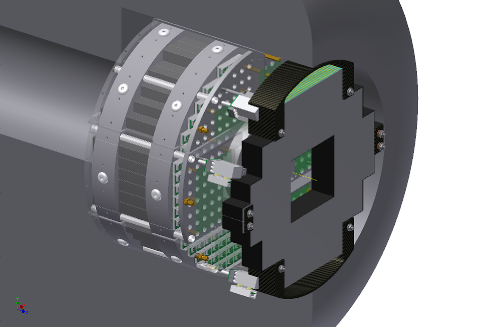}}
\caption{\label{fig:MPC-EX-3D} 
3D rendering of the Muon Piston Pit with fully installed MPC-EX detector components.}
\end{figure}

Figure~\ref{fig:MPC-EX-3D} shows a three dimensional model of the MPC-EX system  installed into the pit of the muon piston.  Both components of the system perform calorimetry-style measurements of the energy deposited by charged and neutral particles inside its active volume (crystals in case of MPC and Si in case of preshower).  The total sampling depth of the combined detectors (4 $X_0$ in the preshower and 18 $X_0$ in the MPC) will contribute to the energy measurement.

The pit has a diameter of $450$\,mm and a depth of $\sim$43\,cm.  Its opening in front of the MPC is occupied by the sparsely installed MPC signal and power cables  (see Fig.~\ref{fig:MPCInThePit}), cooling lines for the MPC, and fixtures supporting beam pipe.  A conflict arises between the preshower and MPC monitoring system (distribution boxes), which will be resolved by redesigning MPC cable routing and MPC LED light distribution boxes to illuminate fibers with back-scattered light.

Details of the MPC design can be found in \cite{chiu:2007MPC}. The mechanical design of the preshower, and its electronics chain and readout, are described in the following sections.

\clearpage
  \label{dsn:det_design}

\section[Detector Design]{Detector Design}
\label{dsn:det_design}

The physics program described in the first chapter of this proposal requires excellent calorimetry being available very close to beam pipe I both forward directions in PHENIX. The calorimeters must provide good photon energy resolution, reliable hit counting under conditions of extreme occupancy, and two shower resolving power never before implemented in the electromagnetic calorimetry. 
The Muon Piston Calorimeter (MPC) which covers rapidity range 3.1$<\eta<$4.2 solves this problem only partially. Its transverse momentum range is limited by granularity to $\sim2.5 GeV/c$, it has no resolving power between single photons and  $\pi^0$ at momenta above $\sim15GeV/c$ and its resolution is seriously degraded by the presence of sparsly distributed material in front of the detector (beam pipe,  BBC, cabling and readout electronics) and radiation damage to crystals. In PHENIX the particles emitted in the very forward direction travel mostly along the direction of magnetic field lines.  There are no tracking detectors in the acceptance of the muon piston.  High particle multiplicities (especially in the jet events) futher limit the ability of the MPC to address the physics of forward produced direct photons and  $\pi^{0}$'s. A meaningful upgrade to the very forward calorimetry  required to bring forward jet and direct photon physics within the reach of PHENIX is impossible without major improvement in  shower resolving power (a precondition for extaction of the direct photon signal) and single particle tracking in calorimeter (a precondition for jet extraction).
 
Given the space constraints of the muon piston bore, space is an issue for any new detector component. The space available for the preshower detector is no exception. There is only few cm depth between the tip of muon piston and area already occupied by readout cables and buffer amplifiers of MPC. With this limitation and extreme granularity requirements for detector which must resolve electromagnetic showers as close as 5 mm, a Si based detector becomes the only practical choice for preshower detector. 
In the past few years our efforts have been primarily directed towards the simulation and R\&D of the preshower detector.  We opted for a W/Si ionization device so the effect of varying environmental conditions (temperature, humidity)  on signal proportionality to energy deposited in readout layers can be either neglected or is easy to monitor with charge injection.   In this section we discuss issues such as basic detector geometry and calibration and monitoring schemes. 

\begin{figure}[htbp]
\centerline{\includegraphics[width=0.8\linewidth]{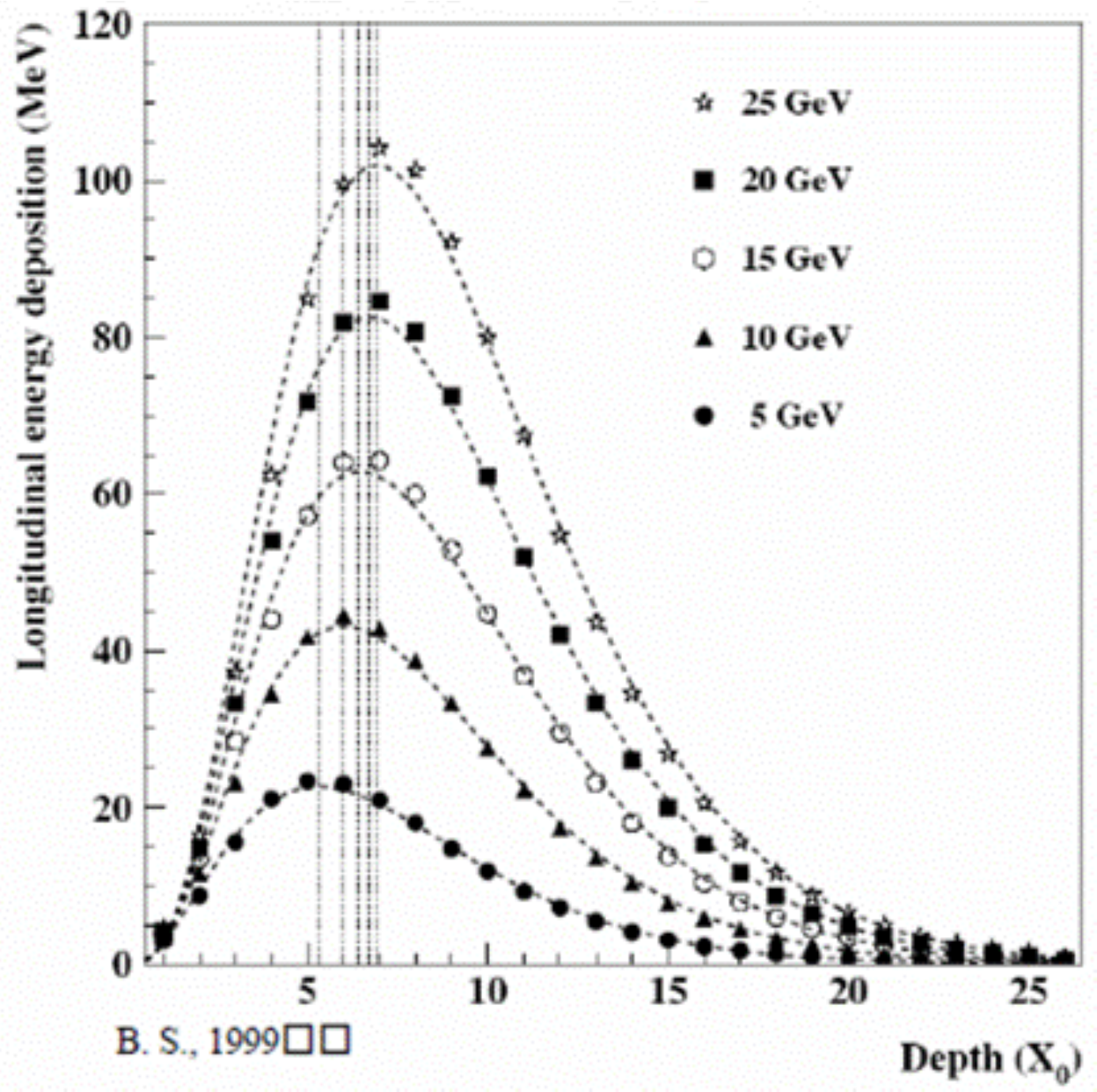}}
\caption{\label{fig:EMLongShProfile}
Longitudinal energy deposition (dE/dX) for electrons in the energy range 5 to 25 GeV as function of the depth in calorimeter measured in units of $X_0$. Lines indicate shower maximum positions at different energies.}
\end{figure}

The depth of the Preshower detector is chosen equal to 4.6$X_0$ based upon the following considerations.  In the geometry of MPC-EX most  electrons (photons) will beging showering in the first one (two)  radiation lengths in the preshower . It takes one more radiation length in depth to insure reasonable probability for both photons from high energy $\pi^{0}$ decay to convert and become separately  measurable entities in preshower. We add one more $X_0$ to the total preshower depth  to make sure that both electrons and photons deposit substantial part of their energies in the preshower detector (see Fig.~\ref{fig:EMLongShProfile} which shows longitudinal shower profile \cite{BerndSurrow} for electrons of different energies).

\begin{figure}[htbp]
\centerline{\includegraphics[width=0.8\linewidth]{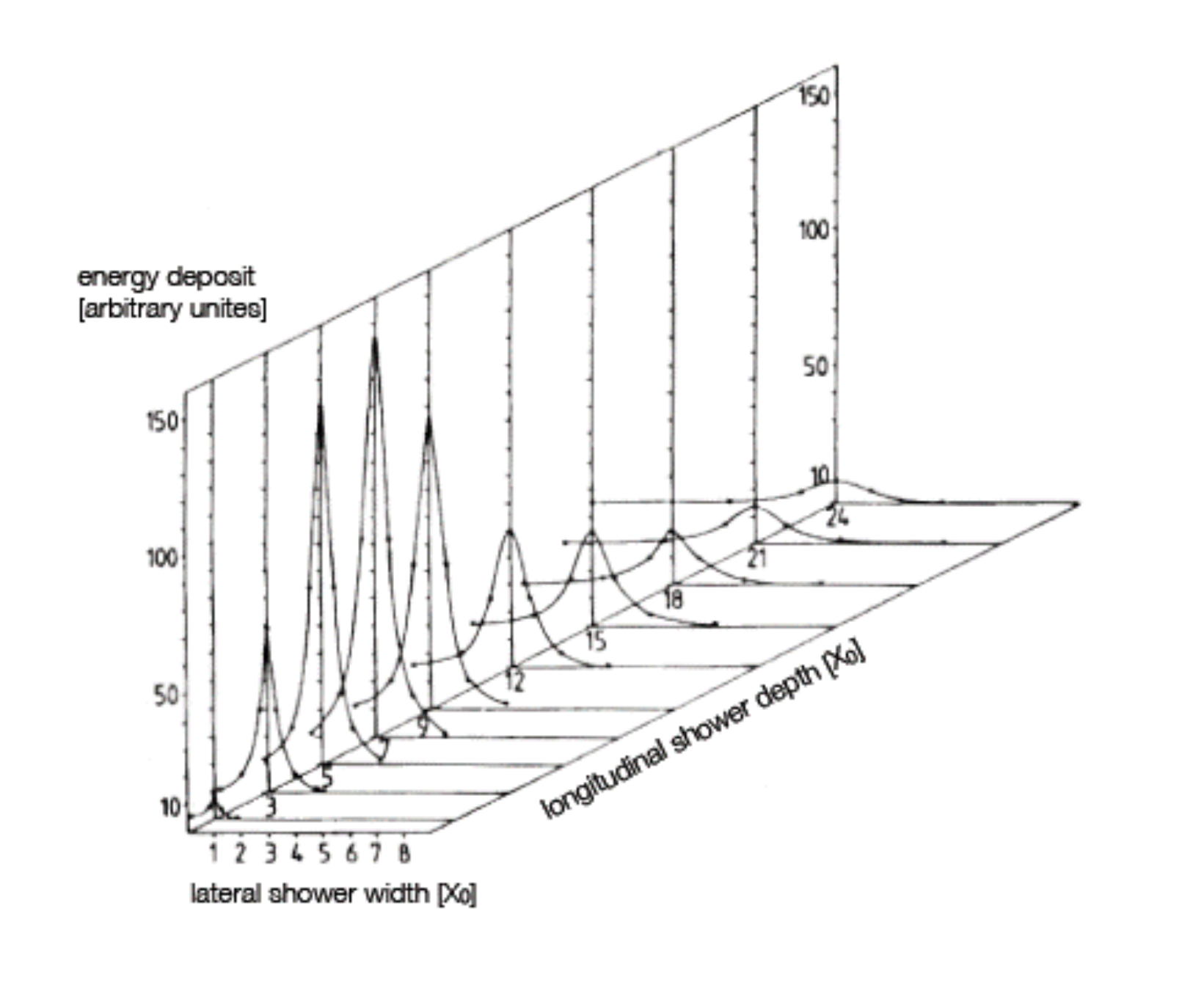}}
\caption{\label{fig:EMLatShProfile}
Lateral energy deposition  for electrons as function of the distance from shower axis  in calorimeter measured in units of Molier radius.}
\end{figure}

Within the first three (four) radiation length of material electrons (photons)  deposit on average about 30\% of their energy. 
The choice of depth is nearly optimal in terms of  detector sensitivity to electromagnetic  vs. hadron  variations in the longitudinal shower profile (critical mainly for hadron rejection) and for its resolving power which is based upon its ability to locate individual maxima in the lateral profiles (see Figure~\ref{fig:EMLatShProfile}\cite{HCSC})  of electromagnetic showers and to measure shower to shower separation.

The preshower consists of 8 sampling cells each built of ~2mm thick W plate and fine position resolution Si readout layer. The Si dtectors are structured  into  $1.8\times15mm^2$ minipads. The minipad orientation in sequential sampling cells alternates between X and Y to avoid cluster shadowing  and allow for separation measurements (see next chapter for details).  The over-all granularity of the two-layer XY pair is thus about $2\times2 mm^2$.
The minipad shaped diodes are implemented on $62\times62 mm^2$ silicon wafers $\sim 500\mu$m thick each. Each sensor is laminated with a sensor readout control board (micromodule) which carries 2 readout chips (SVX4) together with a number of passive components and two precision positioned low height microcontact connectors used to connect the  micromodule to readout bus on a sensor carrier board.  Experience with PC board manufacturing houses shows that given due diligence the positioning of connectors both on a carrier board and SRC can be made to better then $50\mu$m precision resulting in the contribution to uncertainty in hit position of the order of $75\mu$m (compared to $\sim0.6mm$ intrinsic position resolution of the minipad measurements). The carrier boards are glued to $2mm$ W plates held together by presision bolts penetrating whole depth of the preshower in the areas free of silicon. Alignment between the MPC-EX preshower and MPC will rely on MIP hits in both detectors (measuring edge positions of the shadow of MPC towers in preshower plane). 

The preshower is  a sampling calorimeter and essentially counts the number of charged particles passing through the silicon. The particles used to calibrate MPC today are mostly pions with momenta of a few GeV. This means they are nearly minimum ionizing, for simplicity we will refer to them as MIPs. We will use the same particles selected in the MPC to reach design goal accuracy of the minipad-to-minipad intercalibration of 5\%. After in-situ calibration and measurement of the noise in every minipad , an estimation of the signal to noise ratio for MIP will be made for every individual minipad. The design specification (confirmed in the CERN beam test) for this value is $\sim 10$. We expect on average 32 minipads contributing to shower energy measurements in preshower resulting in the total noise value (pedestal width) of the order of 200 MeV (compare to $\sim 6GeV$ of energy deposited in preshower by a 20 GeV photon).  

\subsection[Sensor Radiation Dose]{Sensor Radiation Dose}

The MPC-EX preshower will certainly be exposed to high radiation doses. Albedo from the MPC
and surrounding PHENIX components, electromagnetic showers, and charged and neutral hadrons from primary collisions 
are the sources of the radiation in the MPC-EX.

During PHENIX running the muon piston bore is filled with a fairly homogeneous distribution of albedo neutrons with a 
logarithmic energy spectrum which peaks roughly around 1 MeV (see Section~\ref{dsn:impact}). 
We will use D0 estimates for the neutron flux density of $\phi_{n}\ =\ 1.2\times 10^4\ cm^2 s^{-1}$ at a
luminosity of $\mathcal{L}\ =\ 10^{32} cm^{-2} s^{-1}$ \cite{D0-preshower}. Using a conversion factor of
$1\ neutron/cm^2\ =\ 1.8\times10^{-9} rad$, the dose from neutrons is given by $\frac{dD_n}{dt}\ =\ 2.2\times 10^{-5}rad/s$.

The charged particle flux at rapidity of 4 (corresponding to inner radius of MPC-EX,  which is equal to a radius of ~7 cm) 
computed assuming an inelastic cros-section of 50mb and an average multiplicity of 4 per unit of rapidity 
is $\phi_{\pm}\ \sim 0.5\times 10^5\ cm^{-2} s^{-1}$.

Assuming a factor $1/2$ difference between radiation damage due to neutrons and charged particles, taking into account a factor of two for low
$p_T$ looping tracks, a factor of 10 for particle showering in the calorimeter and an extra factor of 5 for $\pi^0$'s, an 
upper limit for the dose rate related to collision produced particles is given 
by $\frac{dD_{coll}}{dt} = c_{E}\times \phi_{tot} \times \frac{1}{\rho} \frac{dE}{dx}\ = \sim 7.6\times 10^{-4}$. 
Adding the rates from neutrons and an upper estimate from collision related particles the total dose rate at $\mathcal{L}\ =\ 10^{32} cm^{-2} s^{-1}$ 
equals to $\sim 7.8\times 10^{-4}$ rad/s. The total dose rate accumulated at the highest pseudorapidity edge of MPC-EX preshower detector 
in one year ($10^7\ s$) is not expected to exceed 10~krad, or 100~krad for a 10 year running period.  This dose rate will 
result in only minimal radiation damage to silicon sensors.  
Any related increase in a leakage current is taken care of by decoupling the sensor from the readout chip (see Section~\ref{dsn:electronics}) .

\clearpage
  \label{dsn:mech}
\section[Mechanical Design]{Mechanical Design}
\label{dsn:mech}

The MPC-EX uses the digital sum of pixel energies measured in a region of interest around a vector pointing from the collision point to a shower found and measured in the MPC.  The energy from successive $1.8\times15$\,mm$^2$ pixels in the preshower are added to form $\sim$$15\times 15$\,mm$^2$ towers, with both X and Y pixels allowed to be combined into correlated (partially overlapping)
sets of towers.  Consequently, defined towers are shower-position dependent and thus could be distinct for different showers, even those which are closely spaced.  Their size can be varied depending on the shower width, greatly improving the quality of energy sharing between individual objects. Configured towers are pointing and have energies, positions, hit counts, and object width measured in every sampling layer so both particle identification and particle tracking are simplified and improved. The short (15\,mm) length of the pixel makes its energy measurement robust against the adverse effects of occupancy (each layer has $\sim$2500 pixels compared to $\sim$200 crystals in MPC).
The advantages of this ``configure on the go'' approach will be especially important for forward jet measurements which in the case of the MPC-EX system could use both jet definitions based on hit counting in the preshower and the total electromagnetic energy measurements associated with hits in a hybrid preshower/MPC calorimeter.

The radial dimensions and geometry of the preshower were chosen to fit within the envelope defined by the muon piston front face (see Fig.~\ref{fig:MPCInThePit}) coupled the the reorganized MPC signal cables -- the last foot of cable length is unjacketed, and the cables will be restrained on the pit wall close to the diver boards.  This provides the best match between the preshower and the existing MPC acceptance, resulting in an approximately annular configuration with a central opening of 124$\times$150\,mm$^2$ to accommodate the beam pipe flanges and support. Note that the actual shape of W absorbers is defined by a 62$\times$62\,mm$^2$ transverse footprint of the individual Si micromodules.

 The preshower is constructed as 2\,mm W plates interleaved with readout layers -- to allow for micromodule installation the readout layer depth is set to 3.0\,mm. G10 carrier boards (0.5\,mm thick) are glued to the W plates by conductive tape creating a nearly-perfect Faraday cage for the silicon sensors which are embedded into micromodules pluggable into carrier boards. In designing the micromodules, we decided on a very unconventional design. The sensors are laminated between a 0.4\,mm ceramic tile and a 0.4\,mm thick sensor readout card (SRC) carrying dual RC network which is used to split the signals and AC decouple silicon diodes from SVX4 input circuitry. The SRC carries two SVX4 chips which combine both analog amplifiers and storage and digitizers and carry two separate grounds (analog and digital). The unconventional part of this design is a presence of digital signals on the traces immediately above the silicon sensors so we went to the extreme to minimize the pickup of digital activity signals on Si. Fortunately calorimetry is forgiving of the additional material in readout layers and a good ground layer between sensor and first layer with traces was sufficient to keep noise level related to digital activity on the board well within SVX4 pedestal width.

We have chosen to use the FNAL-developed SVX4 128 channels pipelined chips as a base for our readout system. 

 A number of ongoing R{\&}D projects aimed at building similar calorimeters for experiments at a future electron-positron linear collider are considering the option to digitize signals from every pixel in all sampling layers.  The  proposed solutions are all in their preliminary stages, have a number of constraints (range, power etc), and are expensive. We believe that we have found a unique if not perfect solution to this problem based upon inexpensive and commercially available components which is equally applicable to calorimetry in all kinds of collider experiments. The MPC-EX preshower is the first ever built calorimetry detector with pluggable silicon micromodules and on-detector digital conversion of the analog signals generated by particles passing layers of silicon detectors. 

The main design parameters of the MPC-EX preshower can be found in Table~\ref{tbl:PSParameters}. Details of the readout electronics can be found in Section~\ref{dsn:electronics}.

\begin{table}[htbp]
\begin{center}
\caption{\label{tbl:PSParameters}
MPC-EX Preshower design features. All counts are for a single unit.}
\begin{tabular}{|p{87pt}|p{70pt}|p{100pt}|p{142pt}|}
\hline
\multicolumn{2}{|p{158pt}|}{Parameter} &  Value &  Comment \\
\hline
\multicolumn{2}{|p{158pt}|}{Distance from collision vertex} & 220\,cm &   \\
\hline
\multicolumn{2}{|p{158pt}|}{Radial coverage} &  $\sim 18$\,cm &    \\
\hline
\multicolumn{2}{|p{158pt}|}{Geometrical depth} &  $\sim 5$\,cm &   \\
\hline
\multicolumn{2}{|p{158pt}|}{Absorber} & W (2mm plates)  & $\sim$$0.5\ X_0$ or $\sim$$2\%\ L_{abs}$  \\
\hline
\multicolumn{2}{|p{158pt}|}{Readout} & {Si pixels (1.8$\times$15\,mm$^2$)} & \\
\hline
\multicolumn{2}{|p{158pt}|}{Sensors } & $62\times62$\,mm$^2$ & 192 ($1.8\times15$\,mm$^2$ minipixels)  \\
\hline
\multicolumn{2}{|p{158pt}|}{Pixel count } & 24576 & \\
\hline
\multicolumn{2}{|p{158pt}|}{SVX4's} & 384 & \\
\hline
\end{tabular}
\end{center}
\end{table}

\clearpage
  \label{dsn:electronics}

\section[Electronics and Readout]{Electronics and Readout}
\label{dsn:electronics}

The MPC-EX detector system is composed of eight identical readout layers arranged around the beam pipe in front of the MPC detector. The enclosure diameter is 44 cm. Each layer consists of two identical carrier boards, attached to the tungsten absorber plates. Each carrier board contains 12 plug-in modules with silicon sensors and readout ASICs. The technology for the sensors will be p-on-n detectors with narrow mini-pads 15.0$\times$1.8\,mm. The sensors will be orthogonally oriented in alternate layers. To provide a high dynamic range, the signal from each mini-pad is split with ratio 30:1 using a capacitive divider and it is sent to different ASICs. 

The electronics unit counts for the MPC-EX, per arm, are:

\begin{tabular} { l c r }
number of readout planes: & 8 \\
number of minipad modules: & 192 \\
number of minipads: & 24576 \\
number of readout chips: & 384 \\
number of carrier boards: & 16 \\ 
number of FEMs: & 8 \\
\end{tabular}

\begin{figure}[!ht]
  \begin{center}
    \hspace*{-0.12in}
    \includegraphics[width=0.6\linewidth]{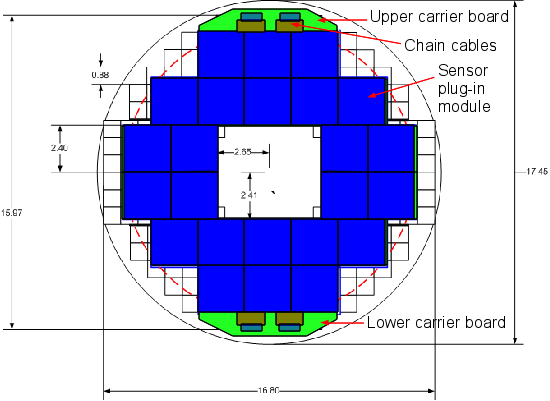}
  \end{center}\vspace*{-0.12in}
  \caption{\label{fig:fee_location.png} Location of the MPC-EX readout electronics in front of the MPC (dimensions are in inches).
    }
\end{figure}

\begin{figure}[!ht]
  \begin{center}
    \hspace*{-0.12in}
    \includegraphics[width=0.6\linewidth]{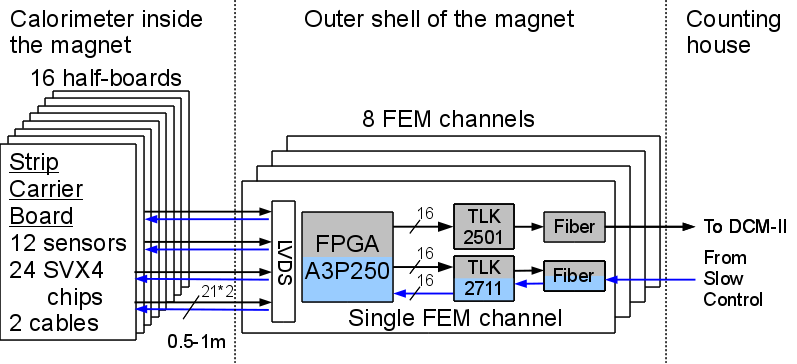}
  \end{center}\vspace*{-0.12in}
  \caption{\label{fig:fee_readout_diagram.png} Block diagram of the MPC-EX readout electronics components. The blue area - front-end clock domain, the grey area - back-end clock domain.
    }
\end{figure}

The data from the readout ASICs will go to PHENIX DCMs through FEM boards as indicated on Figure~\ref{fig:fee_readout_diagram.png}. The FEM will reside on the outer shells of the muon piston magnet and will perform the functions of converting the continuous stream of commands from the control optical fiber into the SVX4 control signals, collecting the data of several SVX4 chains, serializing it and sending it out on data optical fiber to the PHENIX DCMs. 

\subsection[Strip Readout Module]{Strip Readout Module}

The design goal of the readout plane is to keep it is as thin as possible to minimize the transversal expansion of the particle shower in the absorber-free areas. The sensor plane consists of two carrier boards (upper and lower) which are conductively attached to the tungsten absorber plates. The carrier board is thin PCB, which has low-profile (0.9\,mm thick) connectors where the minipad modules will be plugged in.

The readout card is mounted on top (p+ side) of the sensor, it is wire bonded to the sensor pads at the edge of the sensor using 25$\mu$ Al wires. The positive bias voltage is applied to the backside (n- side) of the sensor using flexible leaf of gold-plated fabric. A thin (0.4\,mm) ceramic cover is attached to the backside of the sensor, which provides mechanical rigidity to the assembly. 

\begin{figure}[!ht]
  \begin{center}
    \hspace*{-0.12in}
    \includegraphics[width=0.6\linewidth]{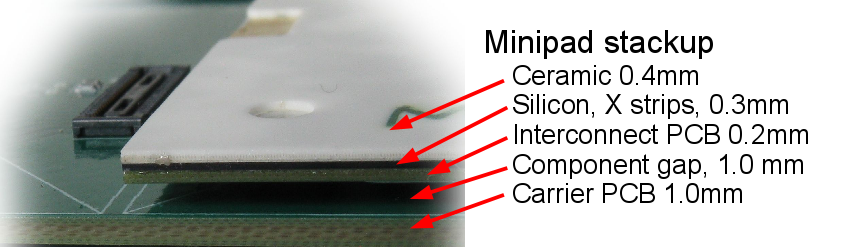}
  \end{center}\vspace*{-0.12in}
  \caption{\label{fig:minipad_stackup.png} Stack-up of the minipad module.
    }
\end{figure}

The signals from each of the minipads are routed to two SVX4 ASICs through different decoupling capacitors. The high-gain leg SVX4 is optimized for measuring MIP signals, the low-gain leg SVX4 - for measuring large energy deposition at the center of the shower. The expected energy deposition of the MIP particle in one minipad is 80 KeV, the energy deposition in the central minipad from the 50 GeV electromac shower is expected to be 40 MeV. The ratio between the two legs is 30:1, and is chosen to ensure that the maximal signal in the high-gain leg will, at the same time, be detectable in the low-gain leg. 
\begin{figure}[H]
  \begin{center}
    \hspace*{-0.12in}
    \includegraphics[width=0.6\linewidth]{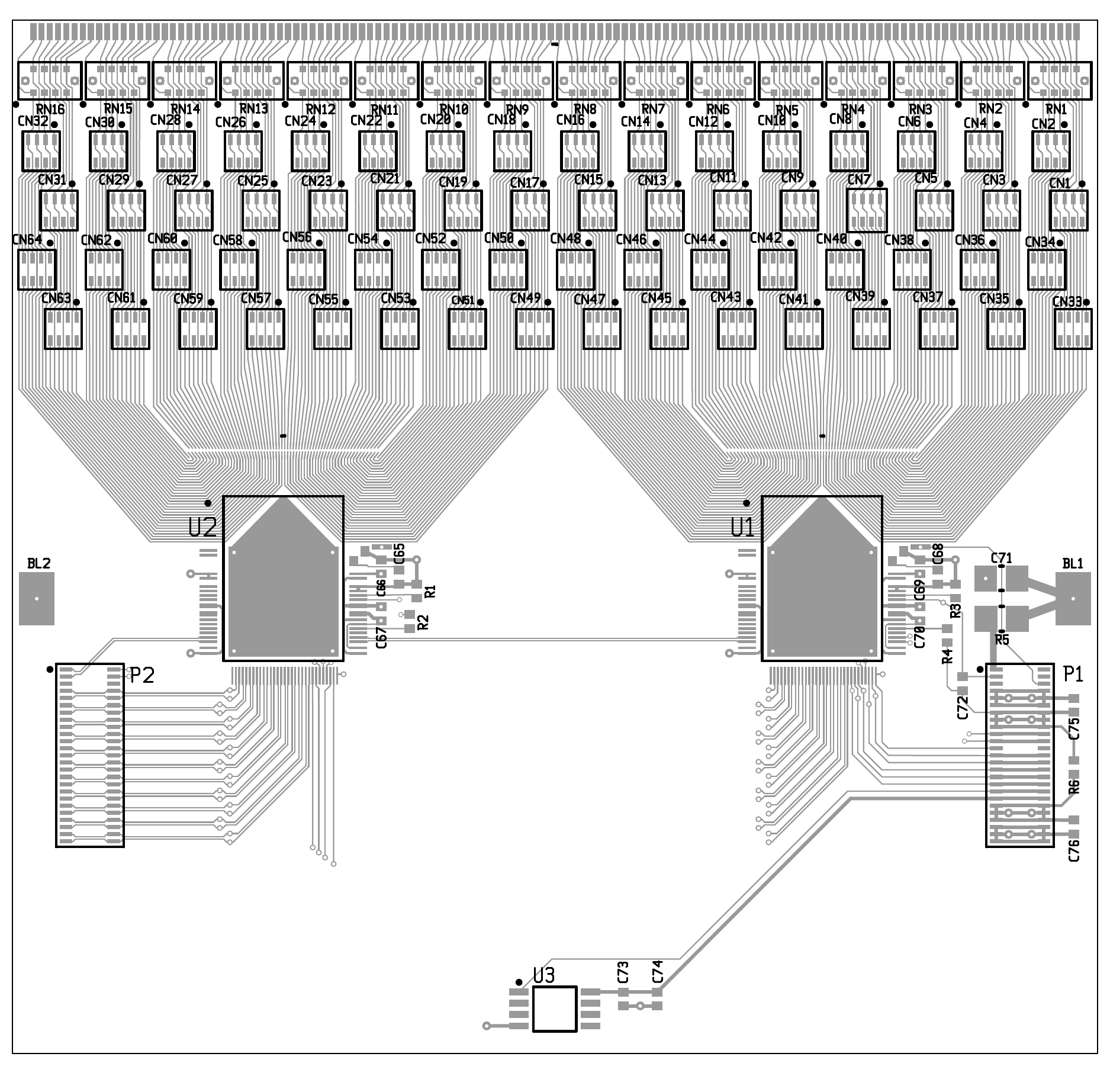}
  \end{center}\vspace*{-0.12in}
  \caption{\label{fig:src_arrangement.png} Layout and design of the dual-gain SRC. The minipad sensor connections are made at the top of the board, through a set of decoupling capacitor arrays, and then into the SVX4 ASICs. }
\end{figure}

\begin{figure}[H]
  \begin{center}
    \hspace*{-0.12in}
    \includegraphics[width=0.6\linewidth]{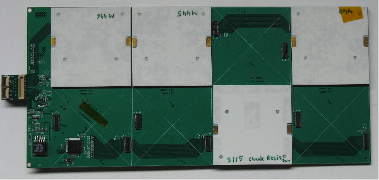}
  \end{center}\vspace*{-0.12in}
  \caption{\label{fig:carrier_board.png} Prototype of the carrier board (green) with four installed minipad modules (white).
    }
\end{figure}

Each carrier board provides two readout chains with 6 modules per chain. Each chain is connected to an FEM using off-the-shelf low profile flex cable assemblies (JF04 from JAE). All signals in the cable are LVDS, the carrier boards have receivers to convert SVX4 control signals from LVDS to LVCMOS levels. The total thickness of the readout layer is 3.0\,mm. The space between adjacent sensors is 0.5\,mm.
The prototype of the carrier board (see Fig.~\ref{fig:carrier_board.png}) has been designed and tested successfully.

\subsection[Dual SVX4 Readout]{Dual SVX4 Readout}

\begin{figure}[!ht]
  \begin{center}
    \hspace*{-0.12in}
    \includegraphics[width=0.6\linewidth]{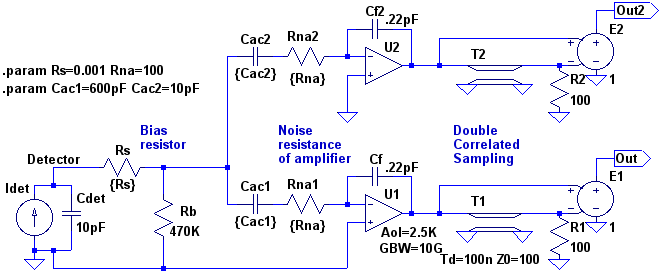}
  \end{center}\vspace*{-0.12in}
  \caption{\label{fig:dual_svx4_sch.png} SPICE model of the strip readout channel.
    }
\end{figure}

The dual SVX4 readout has been simulated using LTSpice, the schematics of which is shown in Fig.~\ref{fig:dual_svx4_sch.png}. 

The sensor strip is presented as a current source with realistic strip capacitance of 10\,pF, the bias resistance is the highest available in a small package. The open gain loop (Aol) of the amplifier is from the specs of the SVX4. The unity gain bandwidth (GBW) was selected to match the published rise time of the SVX4 with the fastest setting. The effective series resistance (Rna) was estimated by matching its contribution to the published ENC versus Cdet dependence. The shaping in the SVX4 is done using a double correlating sampling technique, simulated using an ideal transmission line and a subtractor.

If we assume the infinite open loop gain (Aol) of the operational amplifiers, then  the gain of legs Out and Out2 are 

\textit{G1 = 1/Cf * Cac1/(Cdet+Cac1+Cac2), G2 = 1/Cf * Cac2/(Cdet+Cac1+Cac2)}.

It can be shown that the S/N at Out is proportional to 1/Cdet and it does not depend on its decoupling capacitor Cac1. 

\textit{SN1 $\sim$ 1/(Cdet+Cac2), similarly, SN2 $\sim$ 1/(Cdet+Cac1)}.

\begin{figure}[!ht]
  \begin{center}
    \hspace*{-0.12in}
    \includegraphics[width=0.6\linewidth]{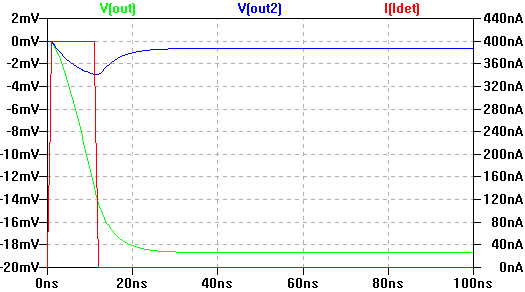}
  \end{center}\vspace*{-0.12in}
  \caption{\label{fig:dual_svx4_transient.png} Response to one MIP (4.4 fC) charge injection. I(Idet) is the current pulse generated on the detector, V(out) and V(Out2) are the voltage output on the high-gain channel and the low gain channel respectively.
    }
\end{figure}

\begin{figure}[H]
  \begin{center}
    \hspace*{-0.12in}
    \includegraphics[width=0.6\linewidth]{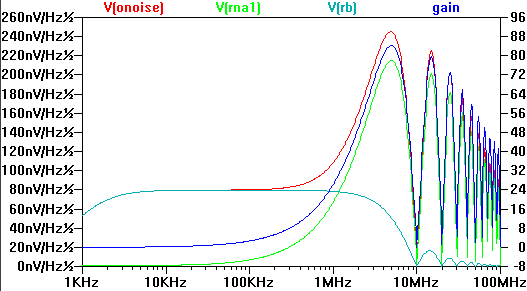}
  \end{center}\vspace*{-0.12in}
  \caption{\label{fig:dual_svx4_noise.png} Noise analysis. Shown are the gain and spectral densities of the system (V(onoise)) and contributions to it from the preamplifier (V(rna1)) and from the bias resistor (V(rb)). 
    }
\end{figure}

To have the SN1 small, we need to choose Cac2 to be as small as possible, but controllable.  The reasonable choice is 10\,pF. If we select the gain of the low leg, G2 = 1/30 of G1 then the Cac1 should be 300\,pF.
The simulation, which includes the finite Aol and GBW shows that the G1/G2 = 30 is achieved when Cac2 = 10\,pF and Cac1 = 600\,pF. The results of the simulation are shown in Figs.~\ref{fig:dual_svx4_transient.png}~and~\ref{fig:dual_svx4_noise.png}.

  The signal amplitude of the high-gain leg is 18.68\,mV, of the low-gain leg it is 0.64\,mV.  The main noise contribution above 1\,MHz comes from the preamplifier, below 1\,MHz - from the bias resistor. 

  \textbf{For the high-gain leg, the total RMS noise at Out is 1.17\,mV, this corresponds to ENC of 0.28\,fC or 1730 electrons.} The contribution from Rna1 is 1.04\,mV, from Rb is 0.17\,mV. If serial resistance of input traces (Rs) is 40\,$\Omega$, then the total RMS noise is 1.21\,mV. 
  \textbf{We can conclude that the noise contributions from the bias resistor and from the input traces are not significant}.

  \textbf{For the low-gain leg, the total RMS noise at Out2 is 0.76\,mV, ENC\,=\,5.2\,fC or 32600 electrons}, this is slightly larger than 1~MIP but still less than one ADC count.

The saturation level of the pipeline cell of the SVX4 is $\sim$ 100\,fC, the saturation level of its preamp is at $\sim$200\,fC. 

  \textbf{With charge division of 1/30 between two legs we can achieve the following S/N. 
In the high-gain leg, S/N\,=\,16 for 1~MIP and saturation occurs at 22~MIP or 1.8\,MeV deposited energy.
In the low-gain leg, saturation occurs at 660~MIP or 32\,MeV of deposited energy}.

One important feature of this design is that the gain of both legs depends very weakly on the varying detector capacitance.

\subsection[FEM]{Front End Module (FEM)}

The FEM services up to eight SVX4 chains and serializes them through one fiber link to the PHENIX DCM. The zero suppression of data on SVX4 will be turned off. SVX4 has a unique feature: robust suppression of the common mode noise in real time (RTPS), this will be used to reduce the low frequency noise originating from power supplies and electromagnetic interference.  For each trigger every SVX4 generates 129 of 2-byte words. The FPGA in the FEM strips off the channel number byte, selects either the low-gain or high-gain value for output from the two SVX4 and streams the result to the serializer. The input stream of 8 of 16-bit data words @40\,MHz is reduced by a factor of four and the resulting stream is serialized with nominal DCM data rate of 1600 Mbps. The first factor of two of reduction is due to the removal of channel bits from the data word, the second factor of two comes from reading out only one of two legs. The leg bits, representing which of the legs was selected for output, are embedded into the output streams (2 bytes of leg bits after 16 ADC bytes). 

\begin{figure}[!ht]
  \begin{center}
    \hspace*{-0.12in}
    \includegraphics[width=0.6\linewidth]{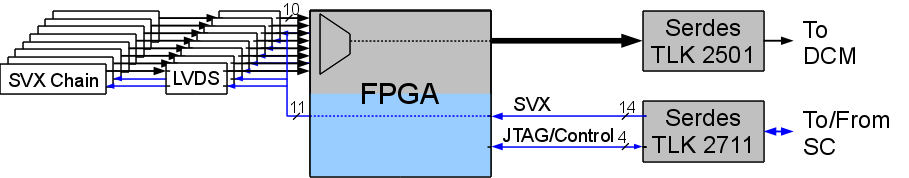}
  \end{center}\vspace*{-0.12in}
  \caption{\label{fig:FEM.png} Block diagram of FEM board.
    }
\end{figure}

There are two clock domains in the system as shown on Fig.~\ref{fig:FEM.png}: the front-end clock and the back-end clock. The front-end clock, synchronous with the beam crossing, is provided by the PHENIX GTM and it is trasferred to the FEM through the optical link from the Serial Control module. The back-end clock is local to the FEM it synchronizes the data transfer to DCM. 

The readout is dead-time free and fully pipelined, the SVX4 can store up to four samples in its input FIFO. Digitization of all channels with 40\,MHz front-end clock takes 4.0\,$\mu$s. The readout time of one SVX4 at 40\,MHz front-end clock is approximately 3.4\,$\mu$s. All 8 chains with 12 SVX4s in each can be received into the FEMs FIFO in 45\,$\mu$s, the transfer to the DCM can start immediately after the digitization and it will take the same 45\,$\mu$s to transfer output data to DCM.

The FEM has a very transparent architecture, divided into two, practicaly independent partitions, corresponding to the clock domains - front-end (shown as blue in Fig.~\ref{fig:FEM.png}) and back-end (grey). The back-end partition streams out to the data fiber link whatever it receives from the SVX4 chains. The front-end partition simply transfers the SVX4 signals from the  Serial Control fiber link to SVX4 chains. 

The FEM de-serializes the 16-bit commands coming with the rate of 80\,MHz from that link, synchronously with the beam clock. The allocation of parellel bits is shown in Table~\ref{dsn:electronics:scbits}. 

Four bits of the command word (ADDR*) are used to address the FEMs. Three bits (CTRL*) are reserved for FPGA control: initialization and reset of the beam clock counters are encoded here. Six bits of the command word (SVX*) are translated directly into the signals on SVX chain according to Tables~\ref{dsn:electronics:triggerbits}~and~\ref{dsn:electronics:modebits}.

Three bits of the command word and one bit from the SerDes receiver (JTAG*) constitute the JTAG interface. The main purpose of the JTAG interface is the programmatic control of the FPGA in real time, this is implemented using UJTAG macro in the FPGA. The JTAG is also used to re-configure the FPGA firmware. The SerDes for  Serial Control connection is the small-footprint TLK2711 working at 1.6\,Gbps, the SerDes for DCM connection is TLK2501.

\begin{table}
\centering
	\caption{\label{dsn:electronics:scbits} Serial Control bit assignment.}
	\begin{tabular} {| l | l | l |}
\hline
Bit & In & Out \\
\hline
1	&In[0]	&ADDR\_CS[0]\\
2	&In[1]	&ADDR\_CS[1]\\
3	&In[2]	&ADDR\_CS[2]\\
4	&In[3]	&ADDR\_CS[3]\\
5	&In[4]	&CTRL\_Cmd[0]\\
6	&In[5]	&CTRL\_Cmd[1]\\
7	&In[6]	&CTRL\_Cmd[2]\\ \cline{3-3}
8	&In[7]	&SVX\_FEClk	\\
9	&In[8]	&SVX\_Trig[0]\\
10	&In[9]	&SVX\_Trig[1]\\
11	&In[10]	&SVX\_Mode[0]\\
12	&In[11]	&SVX\_Mode[1]\\
13	&In[12]	&SVX\_Readout\\ \cline{3-3}
14	&In[13]	&JTAG\_TMS\\
15	&In[14]	&JTAG\_TCK\\ \cline{2-2}
16	&JTAG\_TDO	&JTAG\_TDI\\
\hline
	\end{tabular}
\end{table}

\begin{table}
\centering
	\caption{\label{dsn:electronics:triggerbits} Trigger[1:0] encoding}
	\begin{tabular}{l l l}
	\hline
	Code&Action	&SVX signals\\
	\hline
	0	&no action	&\\
	1	&Trigger	&L1A\\
	2	&Abort gap	&PARst,PRD2\\
	3	&Calibration	&CalSR\\
	\hline
	\end{tabular}
\end{table}

\begin{table}
\centering
	\caption{\label{dsn:electronics:modebits} Mode[1:0] encoding}
	\begin{tabular}{l l l}
	\hline
	Code&Action	&SVX signals\\
	\hline
	0	&Configuration	&FEMode=0\\
	1	&Reserved	&\\
	2	&Acquire	&FEMode=1, BEMode=0\\
	3	&Acquire\&Digitization 	&FEMode=1, BEMode=1\\
	\hline
	\end{tabular}
\end{table}

The power consumption required for one arm is approximately 110\,W for all 16 carrier boards and 20\,W for the 4 FEMs. The details are shown 
in Table~\ref{dsn:electronics:power}. 

\begin{table}
\centering
\caption{\label{dsn:electronics:power} Power budget for the FEM and carrier boards.}
\begin{tabular} { l l c c c }
\hline
Board	&Line	&Voltage	&Current	&Wattage\\
\hline
Carrier Board & AVDD SVX4 & 2.5V & 2.0A & 5W\\
	& DVDD SVX4 & 2.5V & 0.5A & 1.3W\\
	& DVDD LVDS & 2.5V & 0.2A & 0.5W\\
Total & & & & 6.75W\\
\hline
FEM & DVDD LVDS & 2.5V & 1.0A & 3.8W\\
	& FPGA Core & 1.5V & 0.6A & 0.9W\\
	& FPGA IO & 2.5V & 0.2A & 0.4W \\
Total & & & & 5.1W\\
\hline
\end{tabular}
\end{table}

The JF04 cable assembly between the FEM and the carrier board carries 21 LVDS pairs and also a ground plane and 9 extra lines -- which can be used to provide power to the carrier board. The powering of the carrier boards from the FEMs through the signal cable simplifies the cable routing in the tight area of the muon piston magnet but it may have an impact on the noise figure of the system and should be tested before the final installation in PHENIX. 

The current FEM channel design, serving 4 of the SVX4 chains has been successfully implemented on a Virtex-II XILINX FPGA. The full design for 8 chains will be implemented using more radiation hard A3P1000 ACTEL FPGA. 

\subsection[Serial Control]{Serial Control}

The  Serial Control module is responsible for the following:
 
\begin{tabular} { l }
	distributes the front-end clock from the PHENIX GTM to the FEMs, \\
	generates trigger and SVX4 control signals from mode bits of the PHENIX GTM, \\
	provides run control of the FEMs, \\
	provides configuration of the SVX4 chains, \\
	provides configuration for FPGA in FEMs, \\
	monitors the status of the FEMs \\
\end{tabular}

All this information is sent to and from FEMs through the optical fibers. The  Serial Control FPGA contains several serial transceivers, one transceiver is used to emulate the fixed-latency GLINK protocol of the GTM, the rest are used to connect to FEMs. Communication with the external world over ethernet is provided by a micro-processor unit Digi ConnectMe 9210 from Digi International, which is embedded into the modular ethernet jack.

\begin{figure}[H]
  \begin{center}
    \hspace*{-0.12in}
    \includegraphics[width=0.6\linewidth]{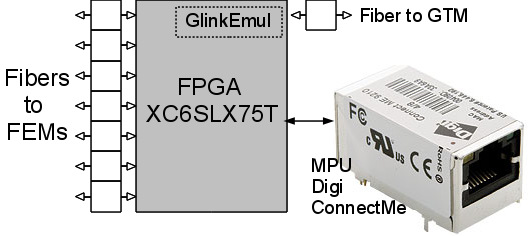}
  \end{center}\vspace*{-0.12in}
  \caption{\label{fig:serial_control.png} Block diagram of the  Serial Control module.
    }
\end{figure}

The communication protocol between MPU and FPGA as well as graphical user interface to the Serial Control have been developed and tested on the FEM prototype.

\subsection[Radiation Environment and Component Selection]{Radiation Environment and Component Selection}

The FPGA used in FEM is FLASH-based, the same FPGA family as used in PHENIX SVTX and FVTX subsystems. It is immune to configuration loss due to neutron irradiation (firm errors). 

FLASH memories exhibit dissipation of the charge on the floating gate after 20kRad of
integrated dose. The dissipation is not permanent damage and is remediated by reprogramming
the device. Flash memories also displayed SEE problems when programmed during radiation
exposure that included gate punch-through, a destructive effect. These types of SEEs are
avoided by not programming the FLASH under radiation exposure conditions, namely during
machine operation. The Single Event Upsets (SEU) will be mitigated using Triple Modular Redunduncy (TMR) technique.

\clearpage
  \label{dsn:impact}
\section[Impact of the MPC-EX on PHENIX]{Impact of the MPC-EX on Existing PHENIX Detector Systems}
\label{dsn:impact}

\subsection{Neutrons}

The addition of dense material in the muon piston hole can potentially have severe 
effects on other detector subsystems.
The hole or cut-out in the muon piston was originally motivated by neutron studies 
in order to move the origin of spallation products far back into the iron yoke.
This way, the yoke itself serves as an effective shielding of the active detector 
components of the muon tracker stations gainst secondary particles from the 
spallation process.
Furthermore, the MPC is now located inside the piston hole.
Several radiation legths of material in front of the detector can cause increased 
radiation damage in the PbWO$_{4}$ crystals and large fake signals in the read-out
electronics.

We study the effect of additional tungsten layers in the muon piston hole with
a GEANT4 based Monte-Carlo simulation.
This simulation was developed especially for the investigation of thermal neutrons
from spallation processes in the context of a new steel absorber in the PHENIX muon 
arms.
The geometry includes a full representation of the south arm tracker stations with 
a slighly reduced acceptance compared to the north hemisphere 
($12^{\deg}<\theta<30^{\deg}$).
The simulation uses primary particles from PYTHIA generated events in a forward
direction of $\theta<45^{\deg}$.
Consequently, the central magnet iron yoke with both copper coils is included in
the setup together with the copper nose cone and the copper flower pot with lead
end-caps.
Also, more importantly for the current studies, both the BBC quartz and the MPC 
crystals are represented by a cylindrical mock-up geometry.
All materials are constructed with the complete isotope composition on top of the 
chemical structure.
The correct isotope mix is important for the thermalization process of neutrons 
when they scatter elastically from nuclei.
It also can change the neutron absorption cross section significantly over a wide 
range of energies.
We use the QGSP\_BERT\_HP package in GEANT4 for interaction and processes with its
default settings instead of tuning all particle and material cut-offs manually.
This package has been tuned for the thermalization of neutrons and includes 
electromagnetic and hadronic interactions down to a few eV with neutron capture
modelled from world data.
While the total flux of particles from spallation may be off by a factor of two 
or so, we currently only use the simulation to compare changes in the setup.
The absolute normalization can be infered from measured data in the years 2010 
and 2011.

\begin{figure}[!ht]
  \begin{center}
    \includegraphics[width=1.0\linewidth]{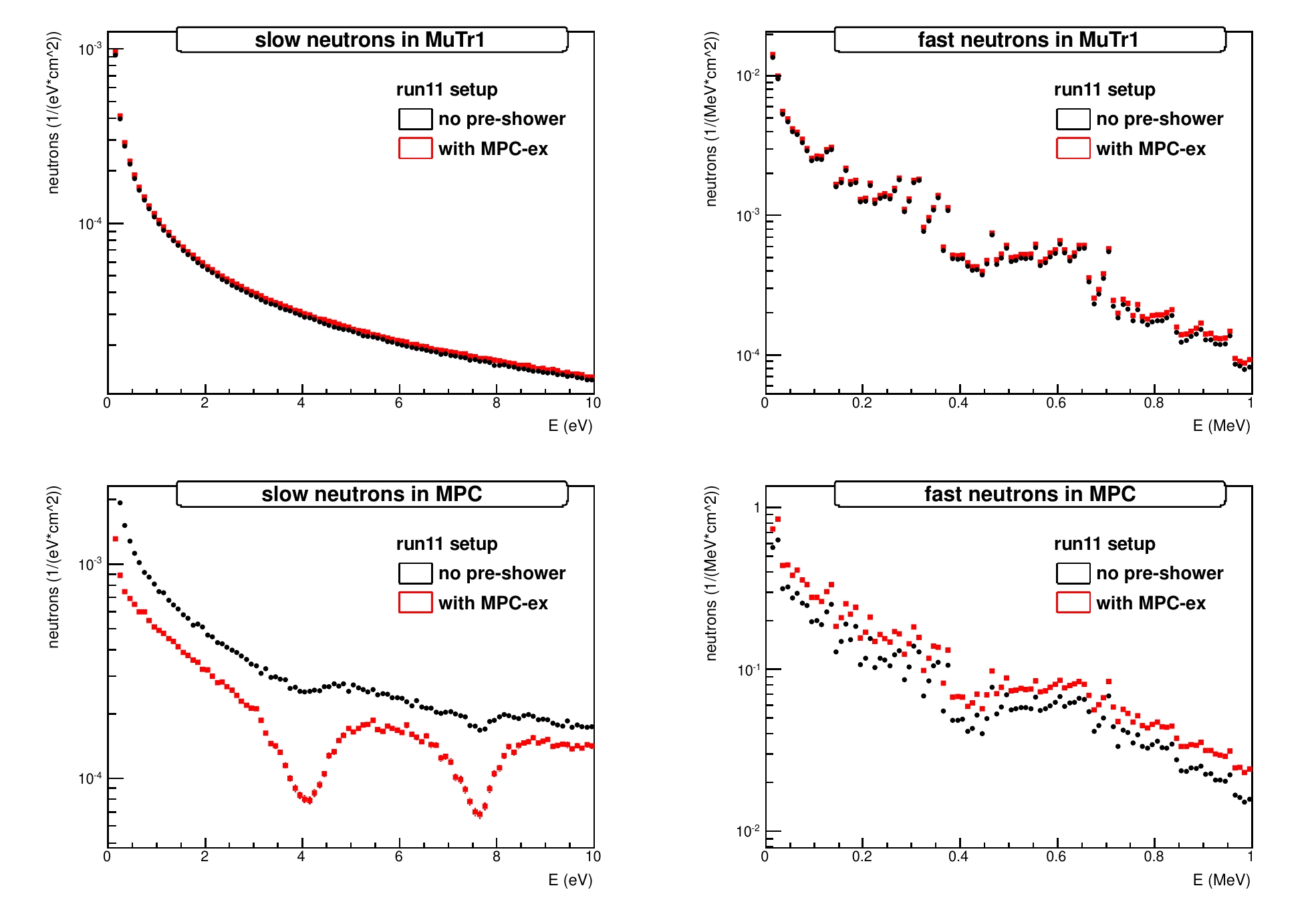}
  \end{center}
  \caption{\label{fig:neutron_flux}
    Slow (left) and fast (right) neutron fluxes in the location of muon tracker 
    station 1 (top) and the read-out of the existing MPC (bottom).
    }
\end{figure}

\begin{figure}[!ht]
  \begin{center}
    \includegraphics[width=1.0\linewidth]{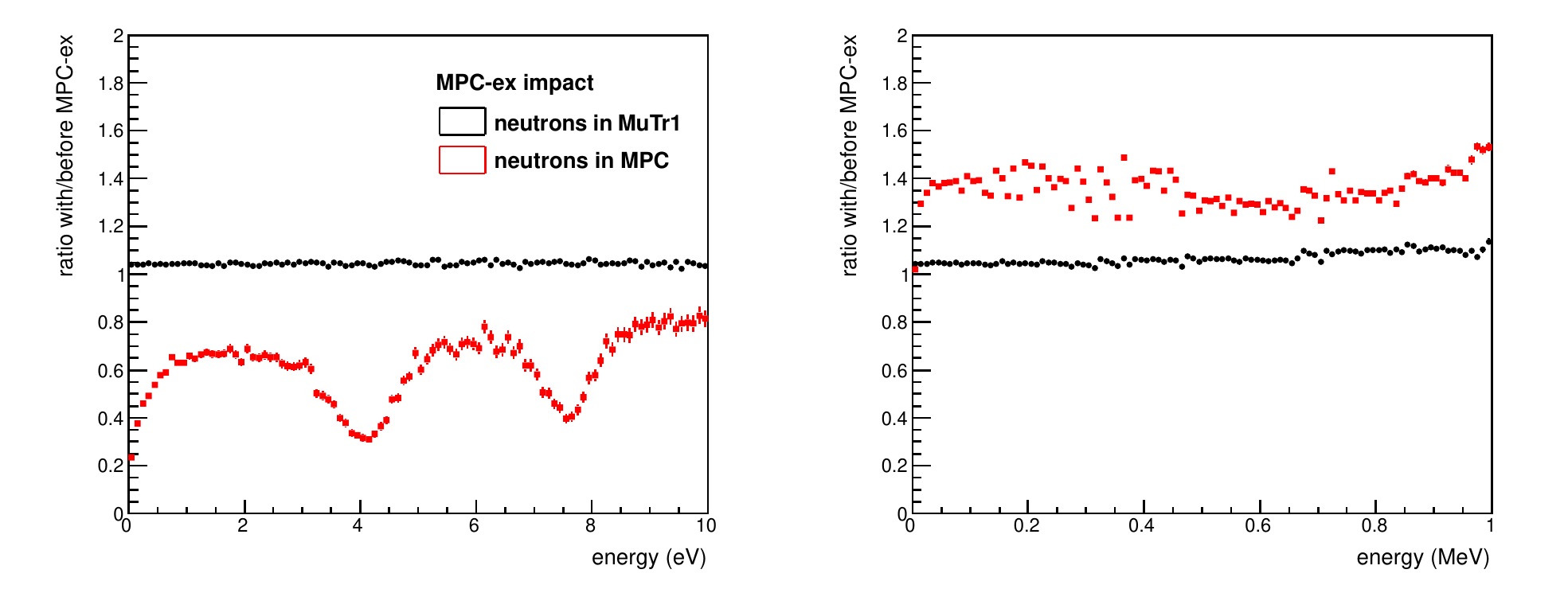}
  \end{center}
  \caption{\label{fig:impact_mpcex}
    Impact of the tungsten layers of the pre-shower on the slow and fast neutron flux
    on the muon tracker station 1 and the MPC read-out (APDs and pre-amps).
    }
\end{figure}

For previous studies, we focused on the three muon tracker stations where a big
pulse background has been observed with a strong timely correlation to proton
proton collisions (heavy ion collisions, respectively).
This background was assumed to be from thermal neutron capture and subsequent
gamma emmission in material close to the active gas volume of the muon tracker
stations.
Although the quantitative time structure of the thermal neutrons could not be 
reproduced in the simulation in detail, it was shown that the gamma emmission 
was indeed consistent with the measurement.
Medium and low energy neutrons are typically behaving very much like a gas in 
the muon tracker volume.
Neutrons from the main spallation source in the back part of the muon piston
hole can easily reach into the muon trackers, either by diffusion through the 
iron piston or by expanding and moving around the hole opening.
Thermalization typically happens after the neutrons have already left the
constrained volume of the piston hole.
Those neutrons can pass through the active detector parts many times before
being captured.
In summary, differences between the neutron exposure of the three muon tracker
layers are mainly scaling with the general geometry, i.e. distance and size.
In the following we use the first station (MuTr1 in Figures~\ref{fig:neutron_flux} 
and \ref{fig:impact_mpcex}) as a (worst case) proxy for the whole muon tracker 
volume.

For the purposes of thermal neutrons in the muon trackers, the MPC has 
previously been a minor obstruction in the muon piston hole that shifts the
main origin of spallation a little towards the opening (upstream).
In order to also estimate the impact of the new pre-shower on the MPC, we 
include an additional active detector in place of the APDs and the 
pre-amplifiers in front of the PbWO$_{4}$ crystals.

The neutron flux results of the simulation are presented in Figure~\ref{fig:neutron_flux}.
We choose two energy ranges to illustrate the behavior of the neutron
spectra depending on the location of the detector.
The slow neutrons behave very gas-like, the steep incline towards the 
lower end of the spectrum here ($E_{neutron}<0.4$ eV) are thermal 
neutrons.
The fast neutrons are not very different from high energetic neutrons in
that their tracks are rather straight as compared to the gas-like thermal
neutrons.
The shape of the spectrum looks very ragged, but a general fall-off with 
energy can be seen that is almost exponential all the way up to a few GeV.
Therefore, the fast neutrons can be also be taken as a good proxy of 
variations in the total neutron flux where most of the energy is contained 
which is the important parameter for the radiation damage in the MPC.

As expected, the slow neutrons in the muon tracker stations (top left 
in Figure~\ref{fig:neutron_flux} are very little affected by the inclusion 
of additional tungsten layers in the muon piston hole.
Although the main spallation happens in the tungsten now, most of those
high energetic neutrons leave the piston hole before they thermalize.
Thermalization times are on the order of 30 $\mu$s and it happens either 
the muon piston steel or (more likely) in hydrogen rich materials in the
muon tracker volume (plastic in the electronics read-out).
We like to point out that the piston hole still serves its purpose in that 
there is ample free space both in front and behind the pre-shower tungsten
layers.
If the whole volume was filled with heavy material, the neutron flux in 
the muon tracker stations would likely be significantly increased from 
neutrons emerging from the front of that material.
As a side note, the slow neutrons in the MPC are actually reduced by the
tungsten in front of it, see bottom left of Figure~\ref{fig:neutron_flux}.
We see two very broad absorption resonances which have been confirmed from
other data, too (and a hint of them is already observed from the tunsten
content in the crystals of the MPC themselves).
Again, this is evidence that the thermalization takes place outside of the
piston hole; the neutrons are then captured in the tungsten before they 
can reach the MPC.

The spectra do include statistical errors, so the variations we see in 
the fast neutron spectra are real, right side of Figure  
~\ref{fig:neutron_flux}.
They do correspond to different absorption resonances in the material
mix in the whole detector setup.
Generally, the spectra in the muon tracker and the MPC show the same
distinct features, e.g. the two rising spikes at $E_{neutron}\approx 0.3$ MeV 
and the flat area around 0.5 MeV.
Minor variations are expected in the proximity of different elements.
In summary, the inclusion of additional tungsten in the piston hole 
leads to a slight increase in the fast neutron flux in the muon tracker
volume.
This probably is due to secondary spallation from an upstream shifted 
primary spallation.
Not surprisingly, the fast neutron flux in the MPC is increased 
significantly.

Figure \ref{fig:impact_mpcex} is a quantitative summary of the 
effects of additional tungsten layers as part of a pre-shower
in front of the MPC.
For the moun tracker, the thermal neutrons are the main concern
because they likely lead to the big pulse background that has 
been observed in real data.
From the simluations we conclude, that we do not an increase
in this background by more than 5\%.
Both the fast neutrons and the total neutron flux can lead to a 
significant performance reduction of the MPC.
We estimate this effect to be on the order of 30-40\%.
This increase is not negligible, but it is on the order of what
we expect from a high luminosity and high energy environment, 
$\sqrt{s}=500$ GeV instead of 200 GeV.

\subsection{Effect of the MPC-EX on MPC APDs}

The MPC is read out with avalanche photodiodes (Hamamatsu model S8664-55) mounted on the 
front face of the MPC crystals. The original motivation for this location for the APD's (as opposed to the 
rear of the crystals) was in part 
to avoid the effect of leakage of charged particles from the back of the MPC crystals into the APD's. 
Charged particles depositing ionization in the silicon of the APD prior to the 
avalanche region will create a signal that will be amplified by the same gain as the light from 
the crystals. When the MPC-EX is installed in front of the MPC the electromagnetic showers will
be partially developed at the location of the MPC APD's and they will see a significant charged
particle flux. This could potentially affect the energy calibration and resolution of the MPC. 

In order to measure this effect we first estimate the number of charged particles at the location of the APD's
for a sample of photon showers in the PISA simulation with energies between 5-100GeV. The number of charged particles
at the location of the APD's is estimated by summing the ionization energy in the last layer of the 
the MPC-EX silicon and dividing by the energy deposited by a minimum ionizing particle (MIP) in a 300$\mu$m 
thickness of Si ($\sim$ 116 keV). This is almost certainly an over-estimate, as it does not account for the 
fact that some shower particles will be very low energy and will stop in the silicon. 
For low-energy electromagnetic showers (5-10 GeV) the average number
of MIPS at the longitudinal location of the APD's is 6 MIPs/GeV of incident energy. At higher energy this 
average drops a bit, down to 5.4 MIPs/GeV at shower energies of 35\,GeV.

The MPC APD's cover a smaller area than the MPC crystals, 25$mm^2$ compared to 400$mm^2$. Typically the 
central tower in the MPC contains $\sim$40\% of the shower energy. The CMS collaboration has studied the 
effect of minimum ionizing particles on similar Hamamatsu APD's in a test beam by exposing a test setup 
to high momentum muons~\cite{cms_apd}. As expected, the effect depends on the size of the depleted region
in front of the avalanche region. For an APD gain approximately equal to 50 the equivalent response to a 
MIP in the APD's was between 145 and 520MeV, depending on the terminal capacitance of the Hamamatsu
APD. The MPC APD's are operated at a gain of 25, which means that a single MIP in the 
MPC APD's should contribute an equivalent signal between 73 and 260MeV.

Based on the above we can estimate the additional energy due to the flux of charged particles through
the MPC APD's as:

\begin{eqnarray*}
E_{charged} &  = & 6 MIPs/GeV \times 0.7 \times (25mm^2/400mm^2) \times (145 - 520 MeV/MIP) \\
            &  = & 38-136 MeV/GeV     
\end{eqnarray*}

which is equivalent to between 4-14\% additional energy measured by the APD's. 
This can be easily accounted 
for in the calibration of the energy response of the combined MPC-EX and MPC detectors and should
not significantly influence the energy resolution. Fluctuations
due to the shower development in the MPC-EX be accounted for by using the MPC-EX track location relative to the 
APD's and the depth of the starting point of the shower, as well as an estimate of the number of MIPs exiting the MPC-EX
shower from the energy deposition in the last layer. Therefore, we conclude that it while the effect of the MPC-EX
will need to be included in calibrating the MPC, it will not be necessary to redesign the MPC detector to move the
APDs.

The MPC-EX collaboration is investigating the possibility of testing the MPC APD response in a  
test beam at FNAL in late 2012 in order to verify the CMS measurements.

\cleardoublepage

\resetlinenumber

\cleardoublepage

\resetlinenumber

  \chapter{Simulations and Physics Observables}
  \label{sim}

To demonstrate the physics capabilities of the MPC-EX we have performed extensive simulations of two key physics observables.
The first, the measurement of direct photons in d+Au collisons, is extremely challenging experimentally but offers a 
sensitive measurement of the suppression of gluons in nuclei at small-x.  The second, a measurement of the azimuthal 
asymmetry of hadrons within jets in polarized proton-proton collisions, can elucidate the origin of the single-spin 
asymmetries observed at large $x_F$ in hadron collisions. 

We begin by first describing the reconstruction method and performance in the MPC-EX in identifying $\pi^{0}$ mesons, 
photons and charged hadrons.  This is followed by specific descriptions of the measurement for direct photons and 
$\pi^{0}$ correlations in jets.  Our strategy is to use a full simulation of the PHENIX detector including the MPC-EX 
to simulate the observables as completely as possible. From these simulations we extract the expected experimental 
sensitivity we for given sample of integrated luminosity.

\clearpage
  \label{sim:reco}

\label{sim:reco}

\section[Electromagnetic Shower Reconstruction]{Reconstruction of electromagnetic showers.}
\label{sim:photonrecomethod}

\subsection{Overview of Section}

The method used for reconstructing electromagnetic showers in the MPC-EX
is detailed in this section.  Specifically, a discussion
of the reconstruction in general terms is given, with little consideration 
of the source of the e-m shower (i.e. $\gamma$ or $\pi^{0}$).

The detector, as described previously, uses
eight layers of silicon strips to determine a track vector and its energy
from localized clusters of hits.  The overall aim of this
section is to describe how the clusters are formed in the MPC-EX, how the energy 
of the preshower is determined, and finally how the preshower is
connected to the existing MPC cluster reconstruction.
This section does not
discuss the cluster-finding algorithms employed in the MPC reconstruction, but focuses
solely on the preshower.

\subsection{Preshower Cluster Reconstruction}

The methodology used for the cluster-finding in the preshower is quite
simple. Tracks in the MPC-EX preshower are determined in ``Hough'' space for 
ease of matching.  In Hough
tracking, the coordinates of the track are converted into ``Hough''
parameters.  For the case of neutral particles, or those which do not
deviate from a straight line trajectory, the Hough parameters are the
slope ($x$/$z$ or $y$/$z$) and intercept.  This is far more convenient
as these parameters are the same for each layer of silicon in
the preshower {\em and} the same for the MPC itself.  By contrast,
tracking or track matching with Cartesian coordinates would be more
difficult as the $x$ and $y$ positions change with the $z$ position of
the layer.  When many points of reference are available, localized
slopes and intercepts between each pair can be created allowing both
the slope and intercept to be used for matching.  In our case, the
number of layers is small, so the Hough parameters are formed using
the intercept (vertex position) and each point.  Thus, only the slope
parameter is used in the track finding in the MPC-EX and later track matching
with the MPC.

\begin{figure}[hbt]
  \begin{center}
    \hspace*{-0.12in}
    \includegraphics[trim=5cm 2cm 1cm 1cm,angle=90, width=0.24\linewidth]{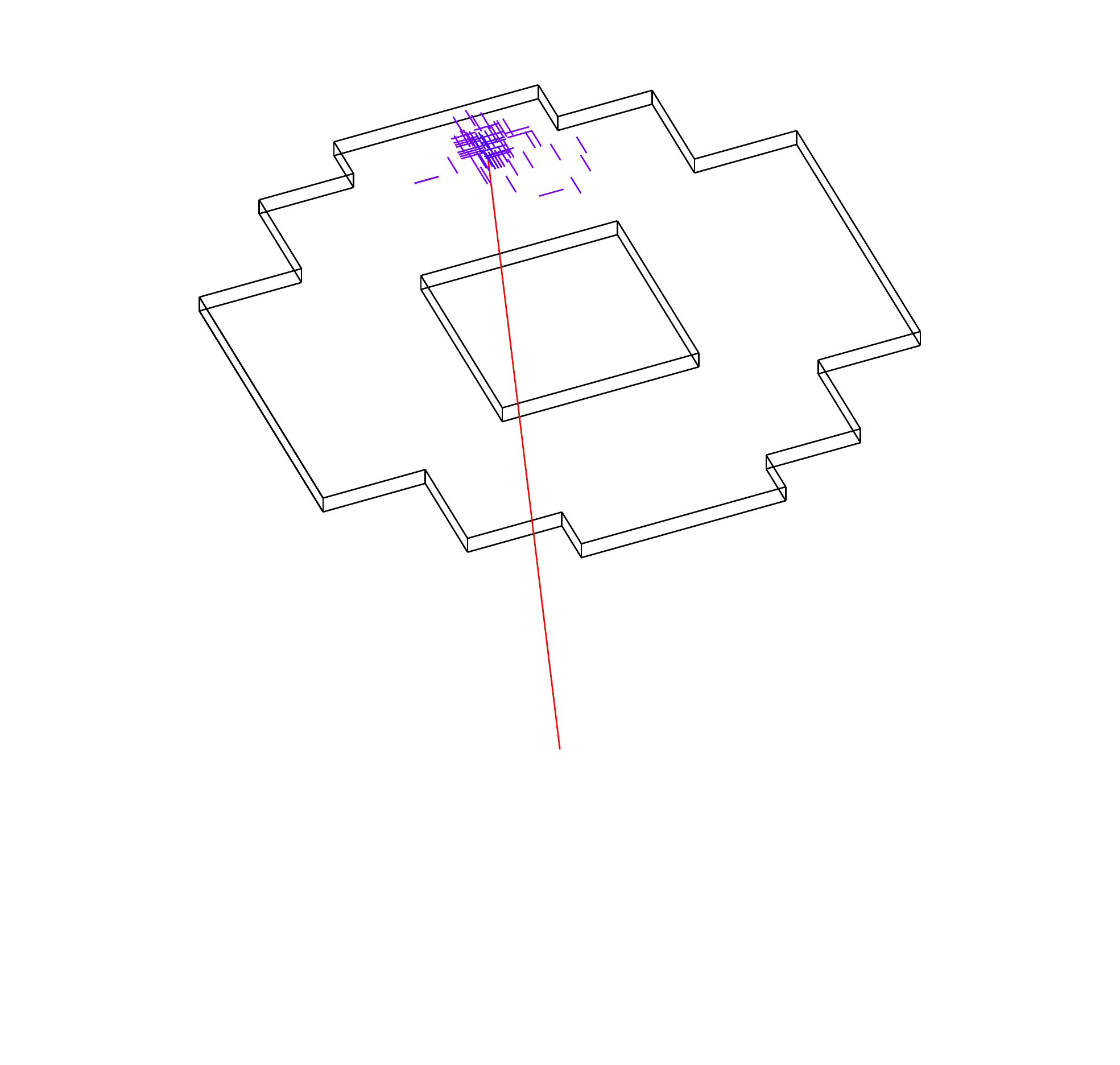}
    \includegraphics[trim=5cm 2cm 1cm 1cm,angle=90, width=0.24\linewidth]{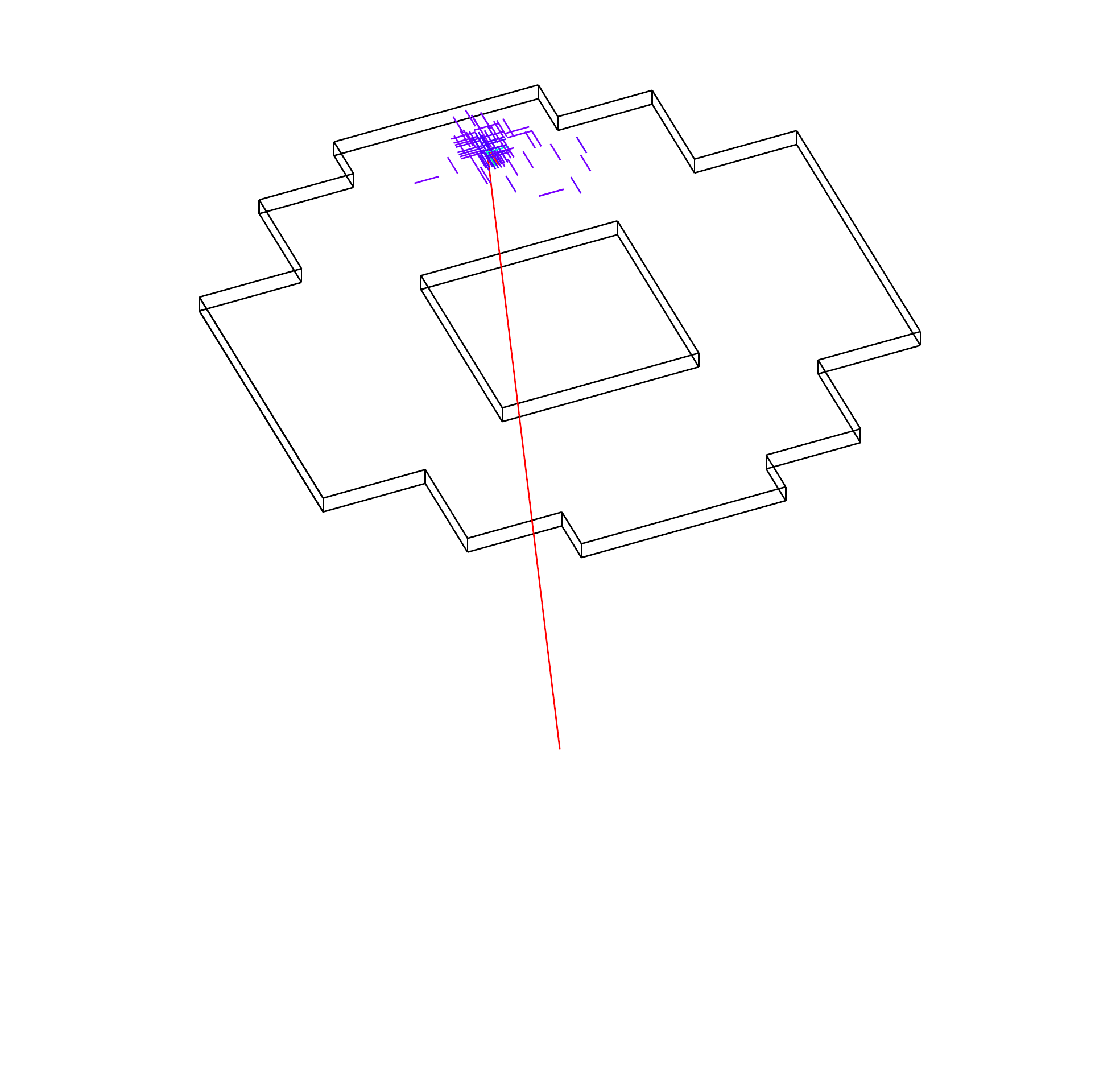}
    \includegraphics[trim=5cm 2cm 1cm 1cm,angle=90, width=0.24\linewidth]{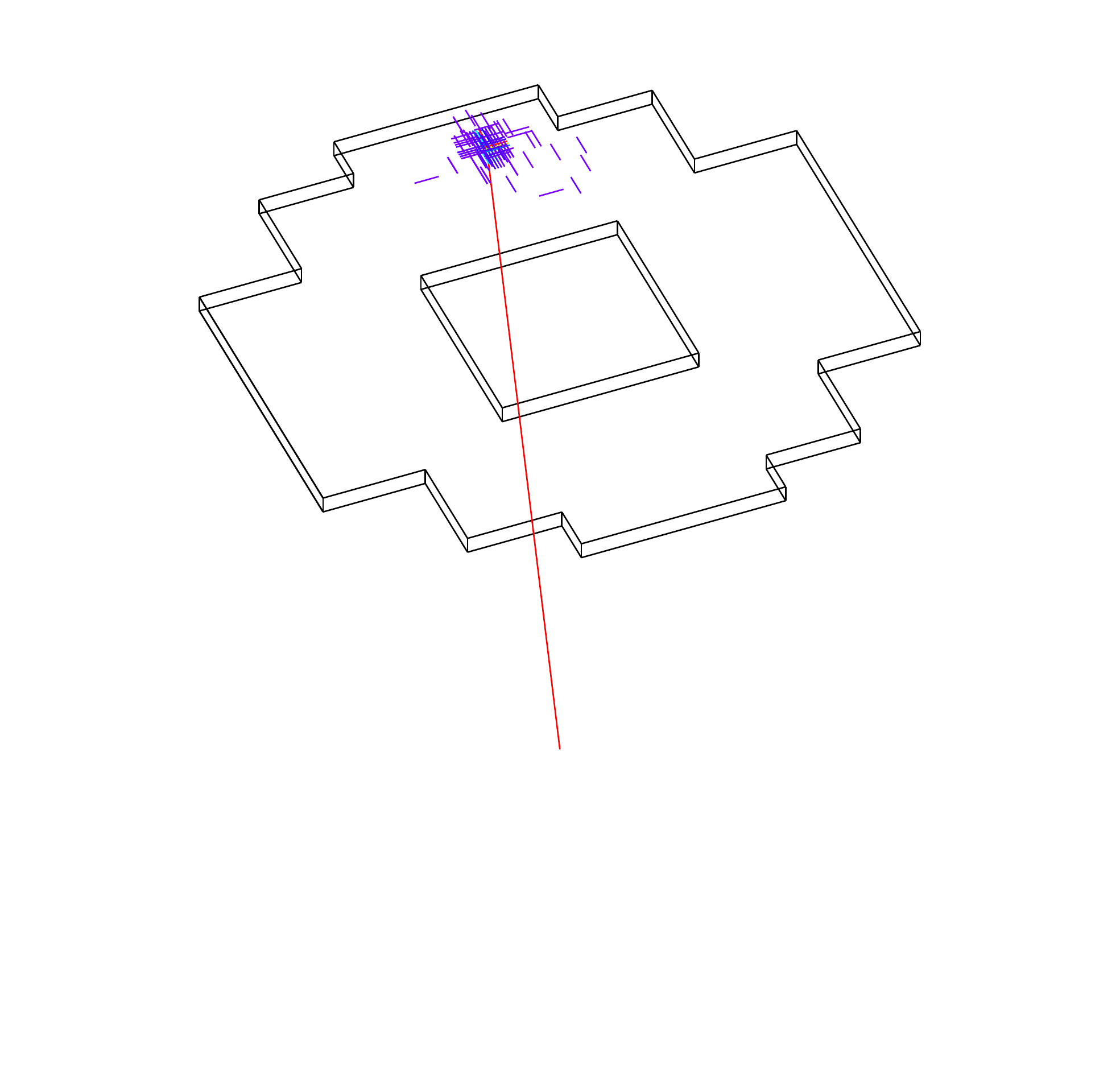}
    \includegraphics[trim=5cm 2cm 1cm 1cm,angle=90, width=0.24\linewidth]{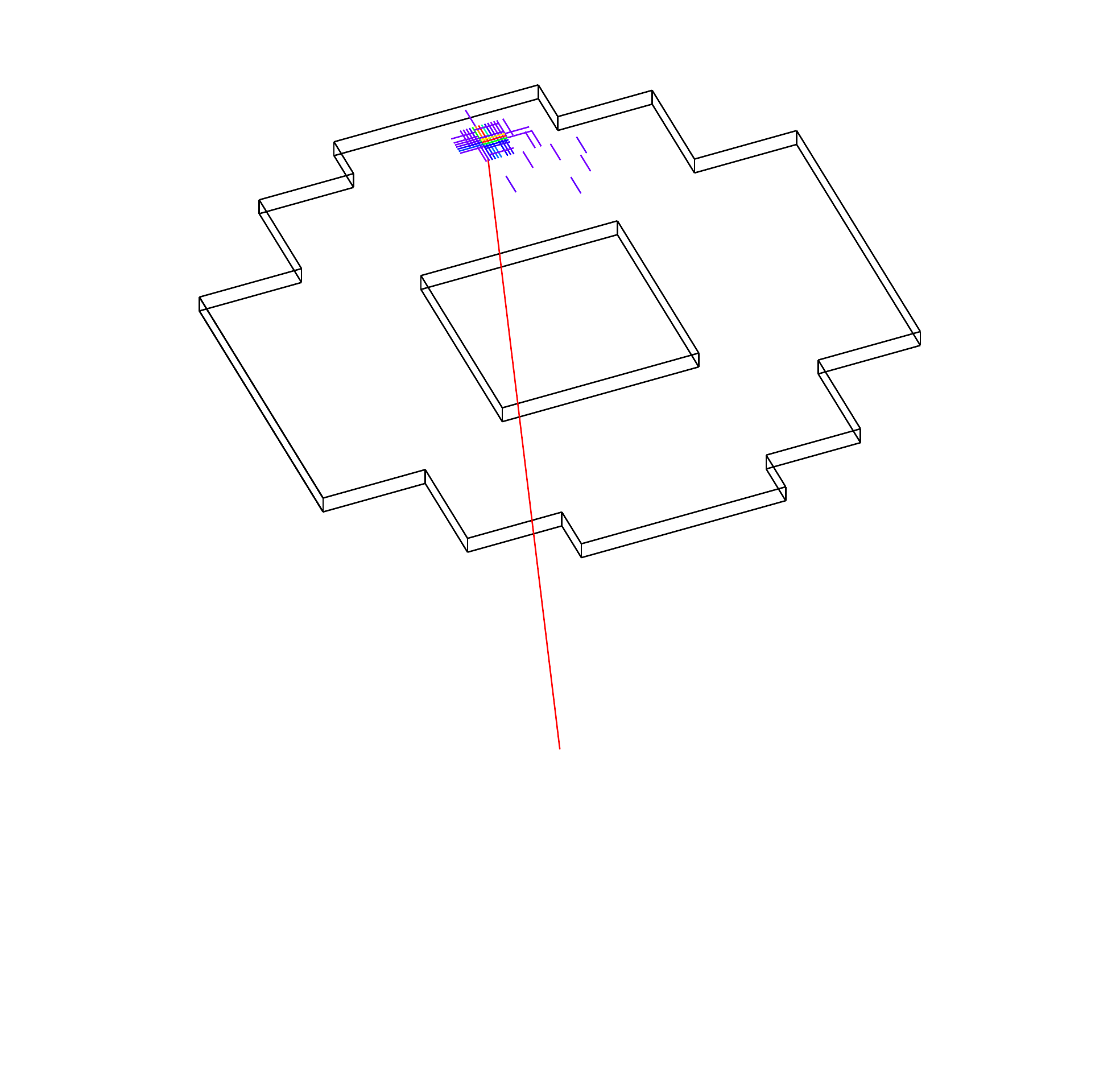}
  \end{center}\vspace*{-0.12in}
  \caption{\label{fig:Hits} Hits in successive layers 0\&1, 2\&3, 4\&5, and 6\&7 ((left to right).  Colors represent the energy deposited in a particular minipad -- purple/blue are low energy, yellow/red are higher energies.}
\end{figure}

\begin{figure}[hbt]
  \begin{center}
    \hspace*{-0.12in}
    \includegraphics[trim=5cm 2cm 1cm 1cm,angle=90, width=0.24\linewidth]{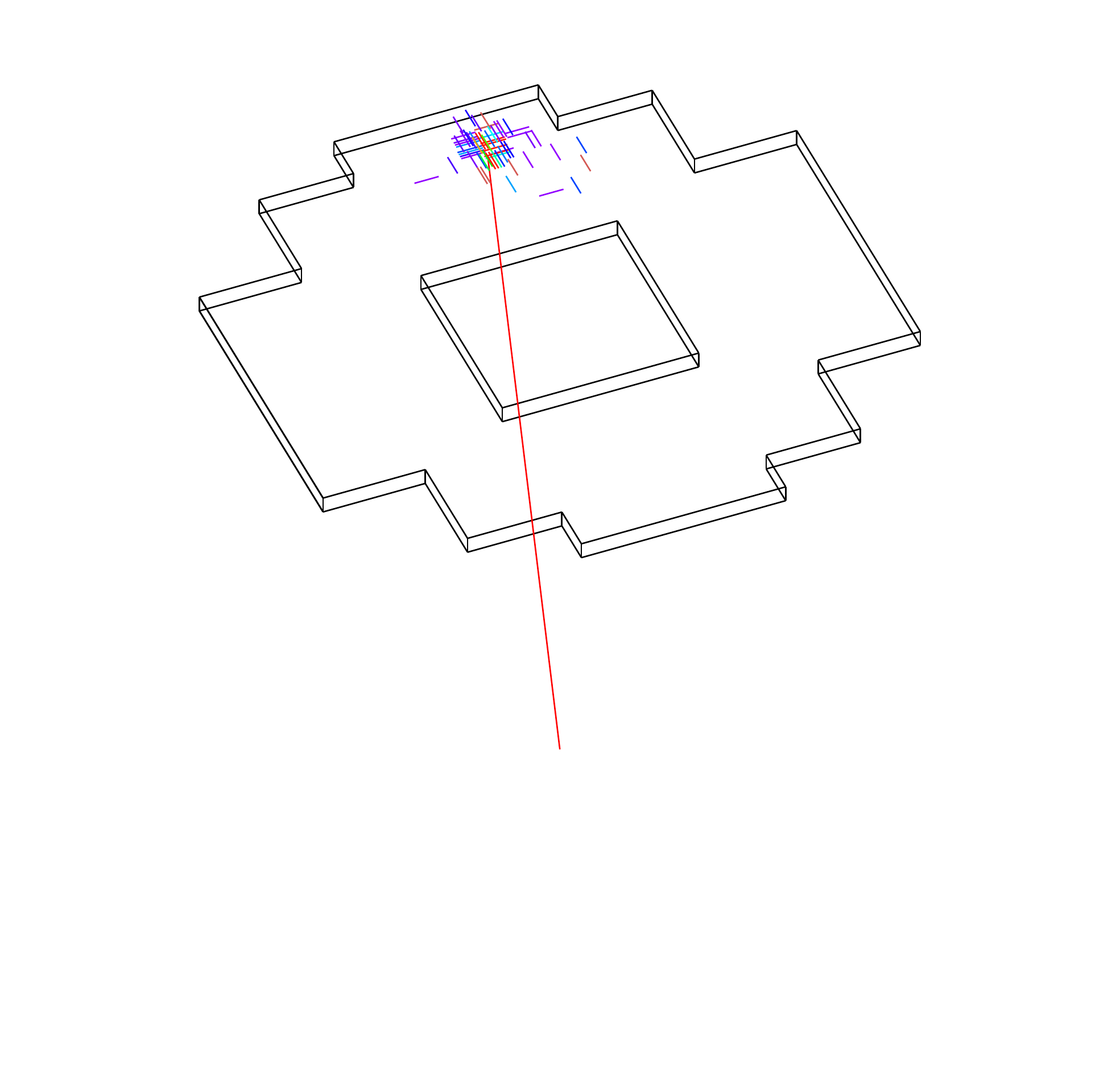}
     \includegraphics[angle=0, width=0.24\linewidth]{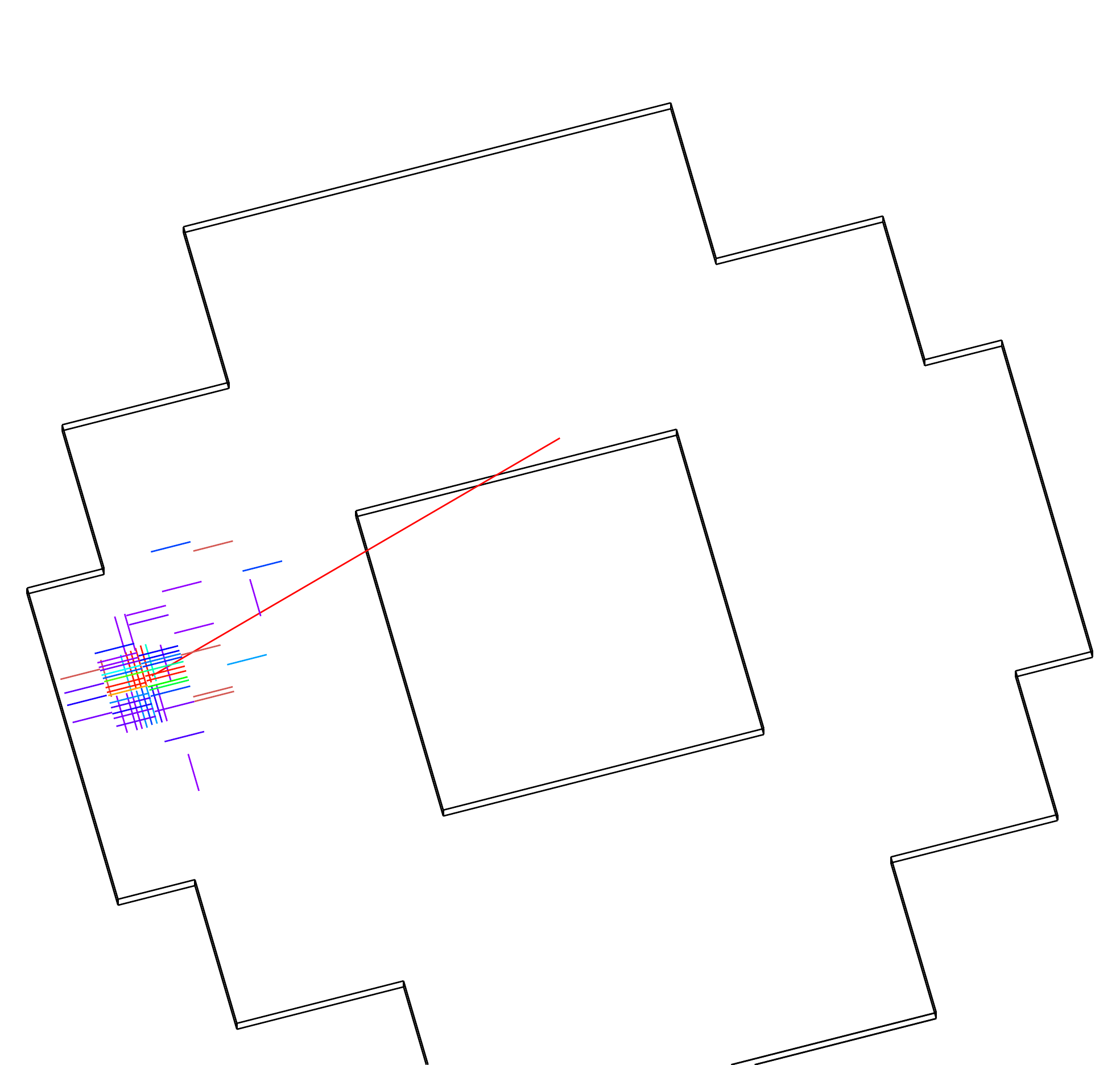}
 \end{center}\vspace*{-0.12in}
  \caption{\label{fig:HitTowers} Two views of the energy deposited in ``towers'' from the same event in Figure~\ref{fig:Hits}.  Colors represent the energy deposited in a particular minipad -- purple/blue are low energy, yellow/red are higher energies. }
\end{figure}

Towers of minipads are formed in Hough space from the sum of minipad
energies in successive layers.  The $x$ and $y$ oriented
minipads are independently summed.  Each tower consists of four minipads (one per layer) which
have the same $x$ and $y$ Hough parameters, but different $z$.
Figure~\ref{fig:Hits} shows hits in layers 0\&1 (for $x$ and $y$ oriented
layers respectively), 2\&3, 4\&5, and 6\&7 (left to right).  The colors
represent the amount of energy deposited in the minipads: purple/blue
are low energy hits, yellow/red are high energy hits.  The summed energies
are shown in Figure~\ref{fig:HitTowers} as towers.  The two views show
different perspectives of the same event.

\begin{figure}[hbt]
  \begin{center}
    \hspace*{-0.12in}
    \includegraphics[angle=90, width=0.99\linewidth]{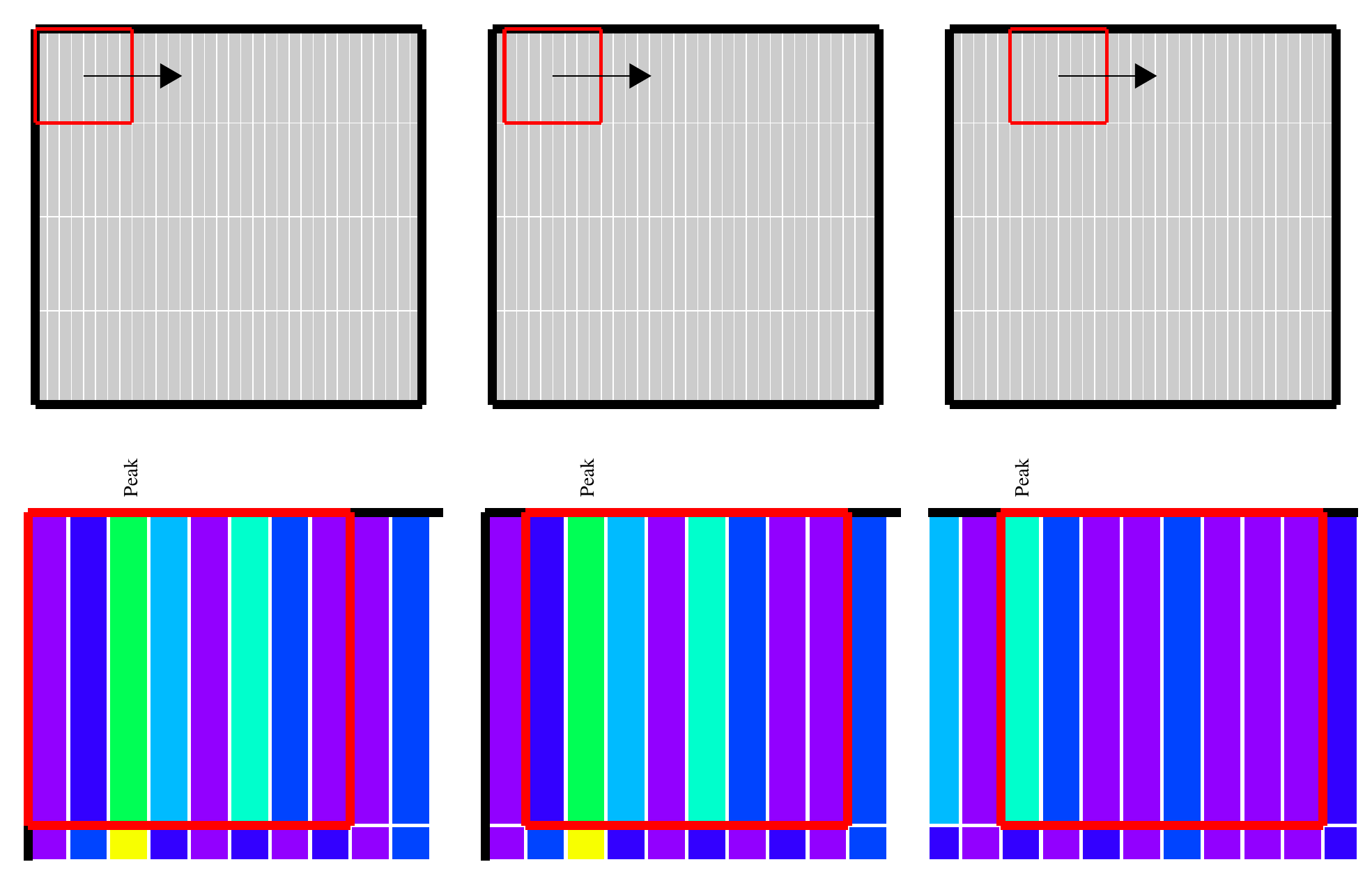}
  \end{center}\vspace*{-0.12in}
  \caption{\label{fig:RecoCartoon} Cartoon showing the reconstruction
procedure.  The upper panels depict the full (32$\times$4) sensor, the lower are zoomed
in on the region in question for that particular step.  The colors in the
lower panel show the energy recorded in a particular minipad (purple/blue
are low energies, yellow/red are high energy).  The leftmost
column shows the starting point of the reconstruction, the center shows
the next step, i.e. moved over one minipad.  The rightmost column shows a
few steps later.  The red outline shows the current search window.  A single
peak is found in each window corresponding to the highest energy deposited
in a single minipad.}
\end{figure}

Next, a search for all possible peaks of energy in the minipad towers
performed using the tower energies (see Figure~\ref{fig:RecoCartoon}
for a cartoon of this).
A search window of eight towers is defined.  Within the search window, 
the tower with the highest energy is considered to be a candidate
for forming preshower track; only one peak per window is allowed.
In the next step, the search window is shifted by one tower, i.e. seven original
towers, plus one new one.  A new peak is sought.  If the peak is the
same as a previous peak, it is not added to the list of candidates.
The whole detector is scanned to find all possible peaks.
The top row of Figure~\ref{fig:RecoCartoon} shows one sensor (32$\times$4
minipads) being searched for peaks.  The bottom row shows more details
of the corresponding search windows, with energy deposits illustrating the found
peaks.  Purple/blue colors represent small energy deposits in the tower,
whilst yellow/red colors represent large energy deposits in that region.  In the first
column, the peak is found in the third tower.  In the second search
window, the same peak is found.  In this example, a second ``peak'' 
is not added to the candidate list until the third search window is reached. (That peak would have
been defined in the fourth search window trial, but this is not shown
in the cartoon.)

\begin{figure}[hbt]
  \begin{center}
    \hspace*{-0.12in}
    \includegraphics[angle=90, width=0.7\linewidth]{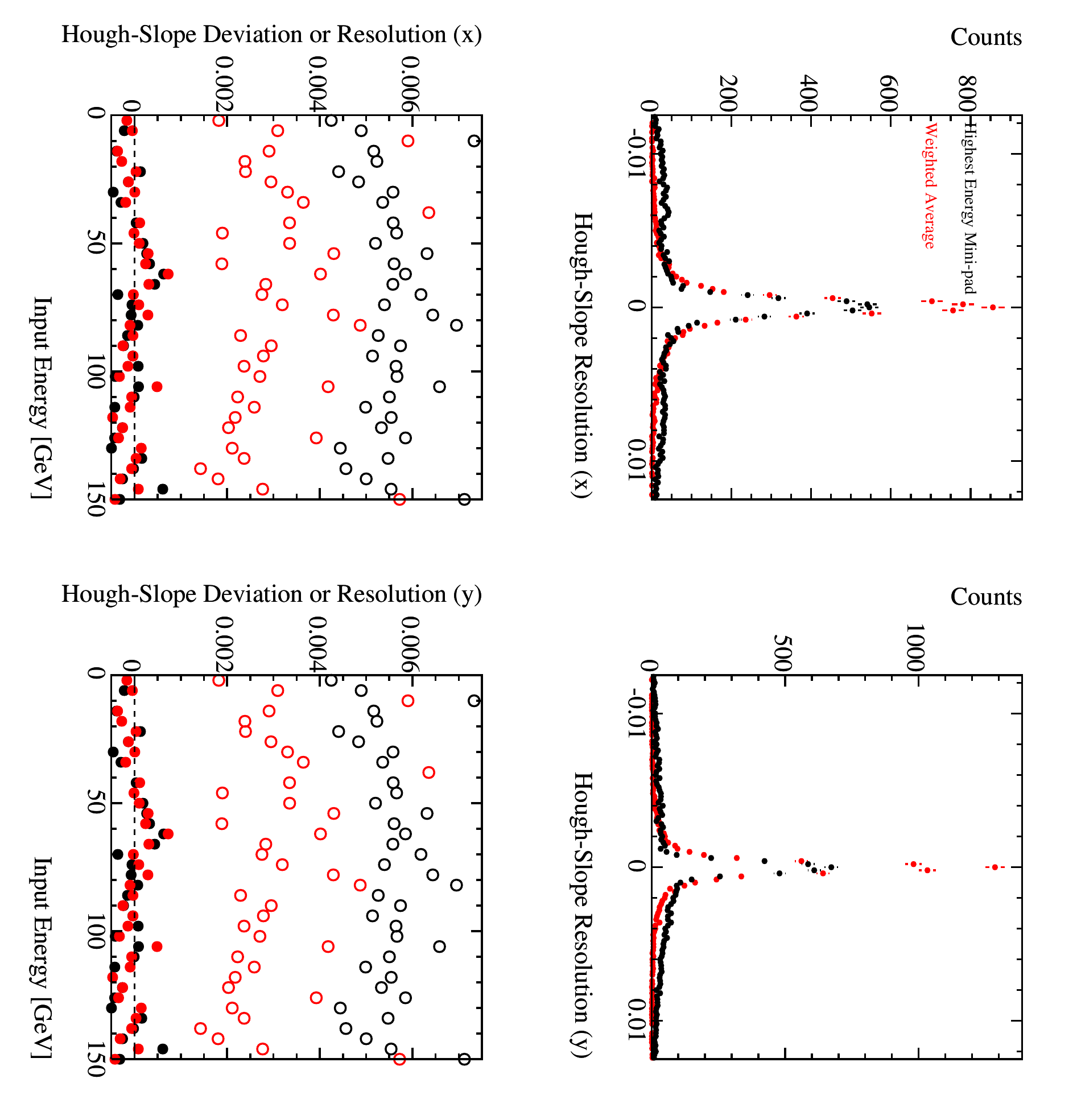}
  \end{center}\vspace*{-0.12in}
  \caption{\label{fig:PosResolutionComp} Comparison between the resolution
determined from the highest energy minipad (trial peak position - black
symbols) and that from the energy-weighted average (red) for single input $\gamma$s. The top panels
show the the difference between the reconstructed and true Hough slope
for the $x$- and $y$-directions, left and right figures respectively.  The
lower panels show the mean deviation (solid symbols) and resolution (open)
as a function of true energy.}
\end{figure}

For each preshower track candidate, the track energy and energy-weighted
track-vector is calculated as the sum of energy deposited in a 
region that is 48 strips wide in the ``short'' dimensions of the strips, and 
three strips wide in the long dimension.  This yields an average position over many minipads, rather
than the position of the highest-energy hit.
Figure~\ref{fig:PosResolutionComp} shows a comparison of the Hough space resolution
between the highest energy minipad and the energy-weighted
average.  The latter has a superior resolution.
A new energy (summed over the central and surrounding towers) and an improved energy-weighted track-vector
is recalculated for added precision.

\subsection{MPC Cluster Pointing Resolution}

\begin{figure}[hbt]
\hspace*{-0.12in}
\begin{minipage}[b]{0.5\linewidth}
\centering
\includegraphics[angle=90, width=0.95\linewidth]{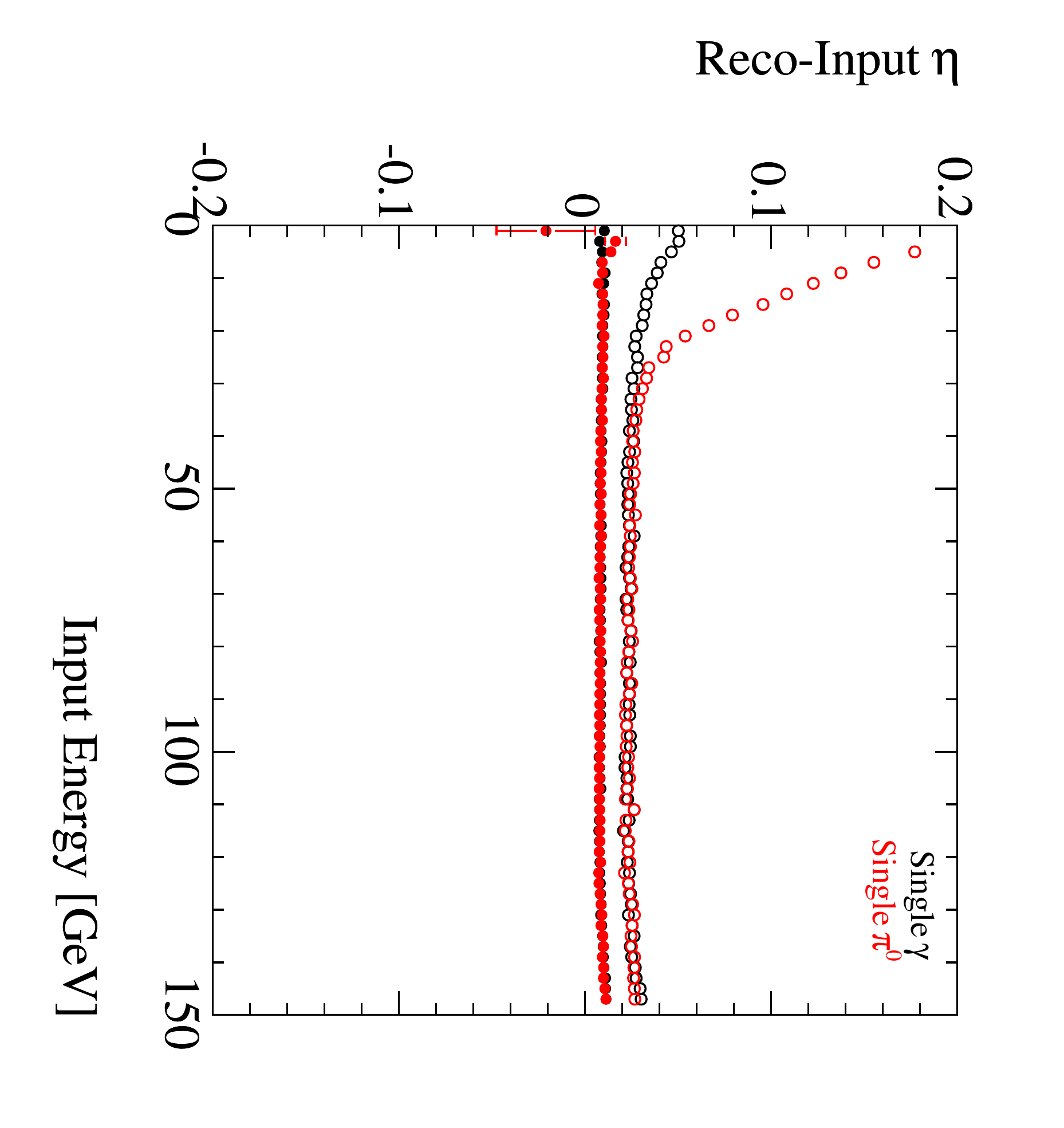}
\end{minipage}
\hspace{0.5cm}
\begin{minipage}[b]{0.5\linewidth}
\centering
\includegraphics[angle=90, width=0.95\linewidth]{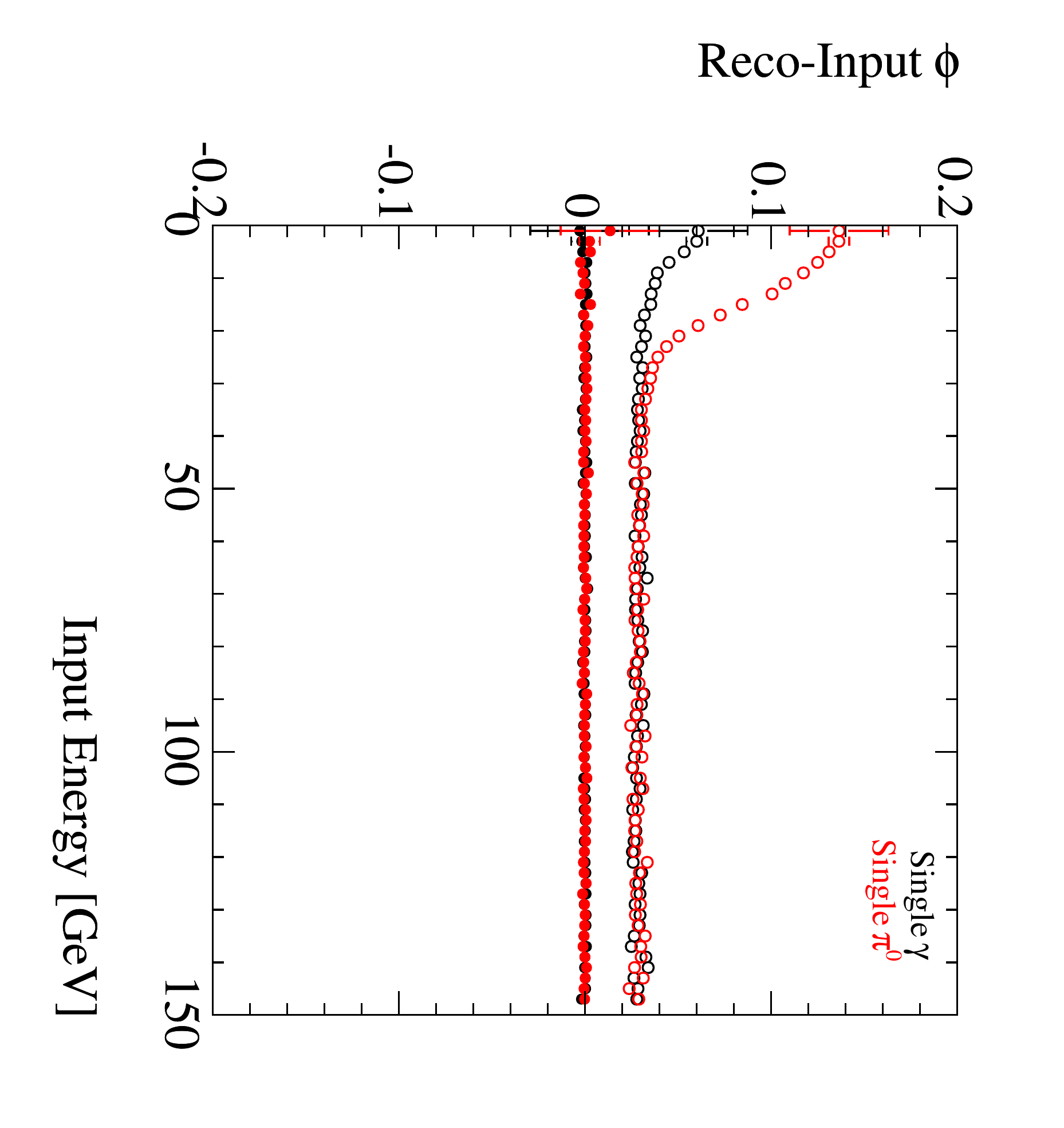}
\end{minipage}
\vspace*{-0.12in}
\caption{\label{fig:MpcPointingResolution} Pointing resolution of the 
MPC for single-$\gamma$ (black) and single-$\pi^{0}$s (red) versus the true energy
of the particle.  The left (right)
panel shows the $\eta$ ($\phi$) resolution (open symbols) and offset
(closed).  The deviation between $\pi^{0}$s and $\gamma$s observed at
low energies is due to two distinct MPC clusters being found. }
\end{figure}

The MPC has an intrinsic limitation in its pointing resolution due its tower size.
Figure~\ref{fig:MpcPointingResolution} shows the $\eta$ (left panel) 
and $\phi$ (right) resolution for MPC Clusters, versus the true
energy of the particle.  A resolution of $\Delta\phi(\eta)$$<$0.04
is observed for high momentum track (i.e. $E_{Input}$$>$30\,GeV).
Tracks at lower momenta have a worse resolution owing to the diminished
energy available for showering and (for $\pi^{0}$s) the deflection of
decay $\gamma$s from the original direction.

\begin{figure}[hbt]
\hspace*{-0.12in}
\begin{minipage}[b]{0.5\linewidth}
\centering
\includegraphics[angle=0, width=0.95\linewidth]{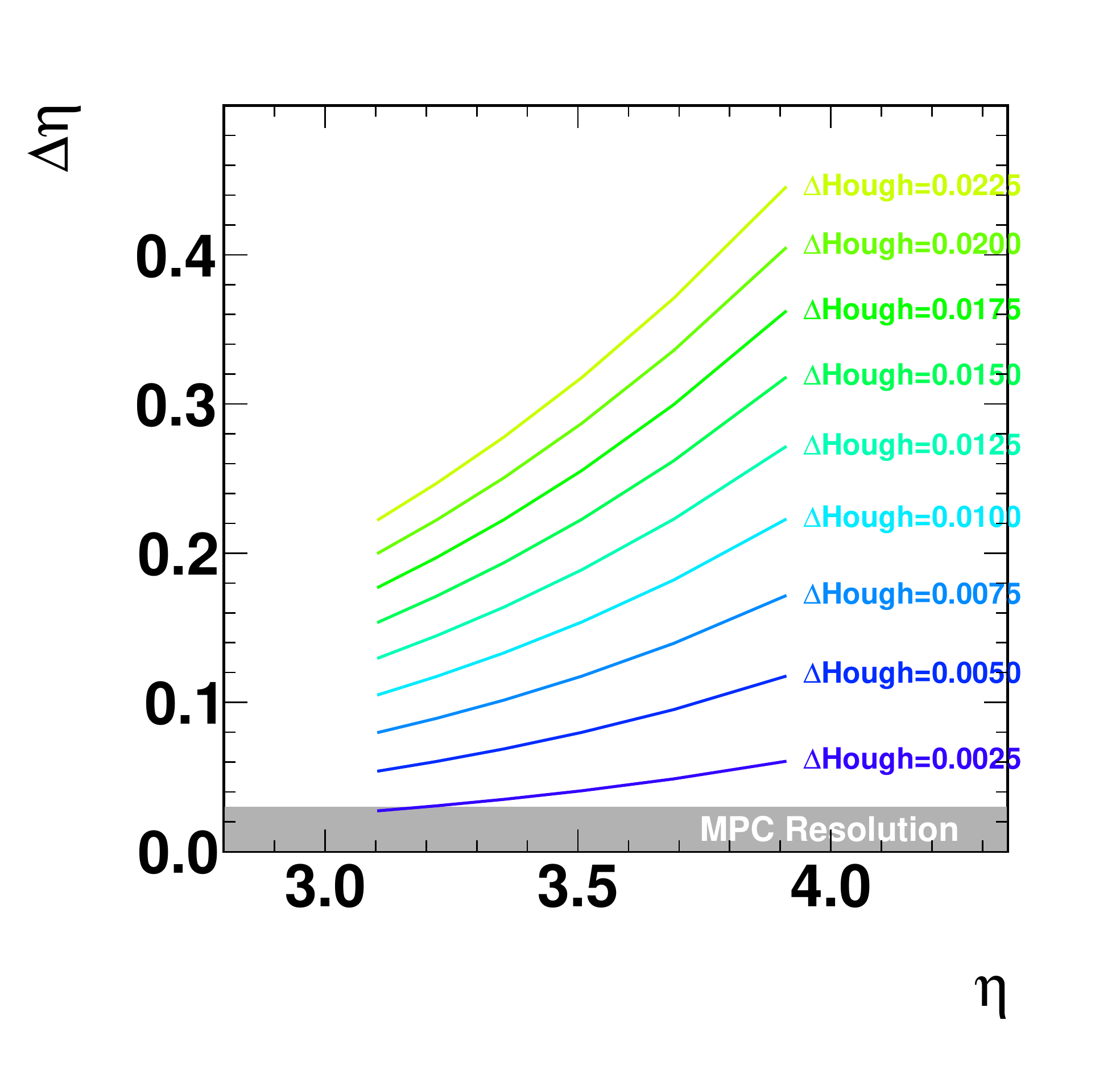}
\end{minipage}
\hspace{0.5cm}
\begin{minipage}[b]{0.5\linewidth}
\centering
\includegraphics[angle=0, width=0.95\linewidth]{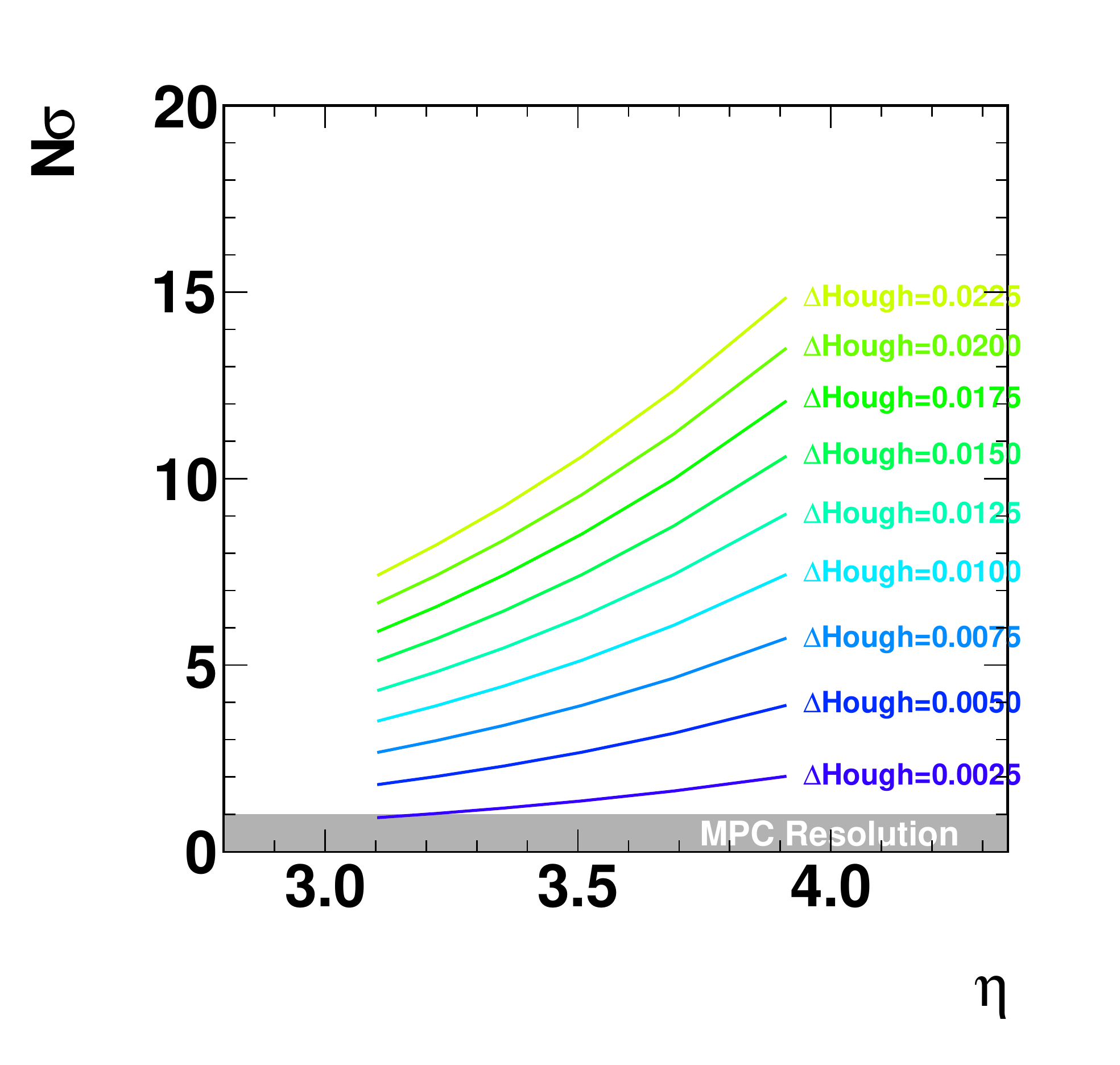}
\end{minipage}
\vspace*{-0.12in}
\caption{\label{fig:MpcPointingResolutionHough} Hough slope difference cut
compared to the pointing resolution of the MPC.  The left panel shows the
absolute $\Delta\eta$ for a prospective difference in Hough slope between
the MPC and the pre-shower.  The right panel shows the same data, but
divided by the approximate resolution of the MPC (0.03).
Colored lines represent the limit of difference between the MPC and 
the pre-shower.  The grey box shows the approximate MPC resolution.}
\end{figure}

In terms of the resolution in Hough space,
Figure~\ref{fig:MpcPointingResolutionHough} shows various prospective 
Hough slope differences between the MPC and the pre-shower, representing
a possible quality control cut (see below).  The resolution of the MPC is
good enough such that a very tight cut on this parameter can be used,
reducing contamination from additional particles.

\subsection{Track Matching}

\begin{figure}[hbt]
  \begin{center}
    \hspace*{-0.12in}
    \includegraphics[angle=0, width=0.8\linewidth]{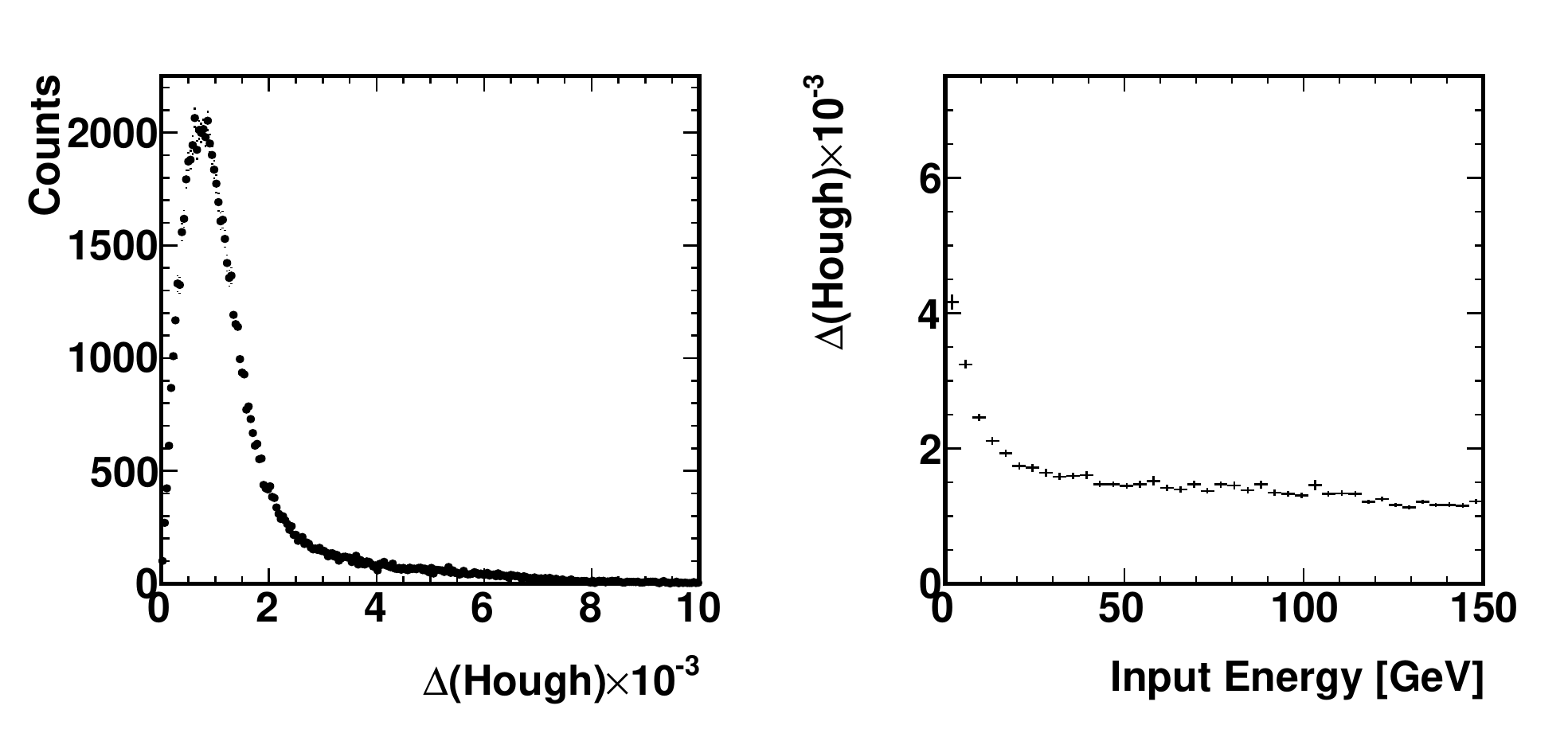}
  \end{center}\vspace*{-0.12in}
  \caption{\label{fig:TrackMatchResolution} The left panel shows the
$\Delta(Hough)$ distribution in a single-$\gamma$ simulation; the right
panel shows the dependence on the true energy.}
\end{figure}

The cluster-finding procedure of the MPC is completely independent of the one used in the 
preshower.  To join the two systems, the track-vector found for each preshower track candidate
is compared to each cluster found in the MPC.  In Hough space, the closest
matched MPC cluster is assigned to a preshower candidate.
As this allows 
multiple preshower candidates to be associated to a single MPC cluster.  
A scan through all candidates with the same MPC cluster is made to determine
which preshower candidate is the closest.
Figure~\ref{fig:TrackMatchResolution} shows the difference in Hough slope
found between the closest MPC cluster to a given MPC-EX track candidate.
The right panel shows the energy dependence of the mean of $\Delta(Hough)$
distribution (the distance in the 2-dimensional Hough space).
One reason for multiple tracks is fluctuations in
the showering process which can form a spur of hits in multiple layers
that happen to line-up in Hough-space.  Multiple tracks can also be formed
from the decay of particles, for example $\pi^{0}\rightarrow\gamma\gamma$.
These are treated in a second peak-finding algorithm.  

\begin{figure}[hbt]
  \begin{center}
    \hspace*{-0.12in}
    \includegraphics[angle=0, width=0.8\linewidth]{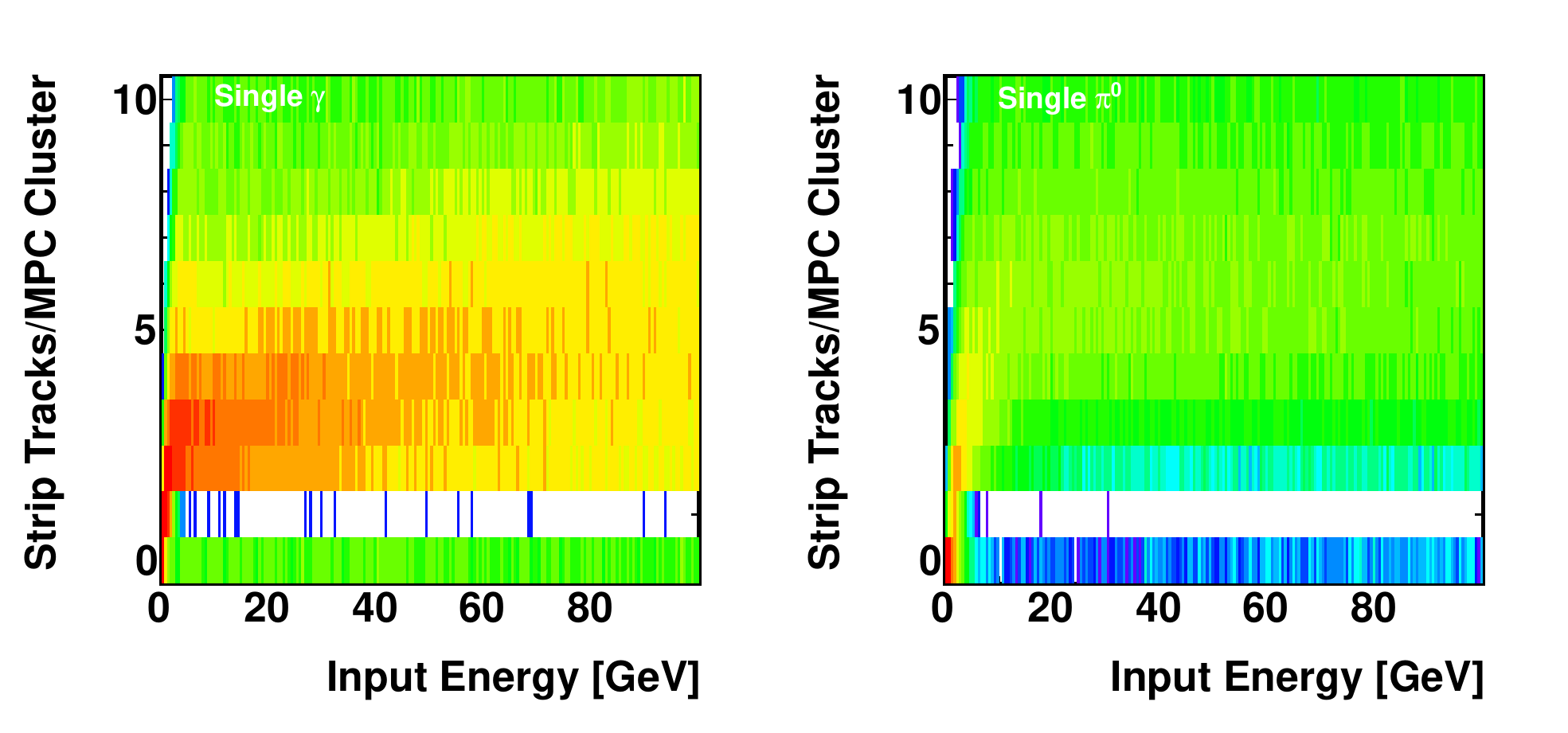}
  \end{center}\vspace*{-0.12in}
  \caption{\label{fig:NumTrackPerMPCCluster} Number of minipad-tracks per
MPC cluster.  The left panel shows a simulation of single-$\gamma$s, the
right shows single-$\pi^{0}$s.  The color scale is logarithmic.}
\end{figure}

For illustration,
Figure~\ref{fig:NumTrackPerMPCCluster} shows the number of MPC-EX track
candidates found per MPC cluster found for single-$\gamma$s (left) and
single-$\pi^{0}$s (right). Once the closest MPC-EX track to an MPC cluster is determined, 
a clustering 
algorithm is used to recombine all tracks within a radius of 0.0175 in Hough 
space.  This track is then associated with the MPC cluster. 
The constituent tracks of the cluster are eliminated from the track list.  

\subsection{Energy Recalibration}

When working with real data, the whole MPC-EX detector will be calibrated
as a single unit.  This can be performed in the low-energy region
iteratively using pairs of clusters to form an invariant mass. A
comparison between the known mass of the $\pi^{0}$ and the $\eta$-meson
will facilitate the calibration.  For high-energy clusters, the
single-track approach (discussed in Section~\ref{sim:photonreco}) can
be used to verify (or fine-tune) the calibration.  For the purposes of this document, a full calibration procedure
has not been developed.  Rather, a re-calibration of clusters produced
using the existing infrastructure is performed.  This procedure is
necessary as the current calibration assumes there is no impediment to 
photons prior to the MPC, whereas the MPC-EX represents approximately
four electromagnetic radiation lengths.

\begin{figure}[hbt]
\hspace*{-0.12in}
\begin{minipage}[b]{0.5\linewidth}
\centering
\includegraphics[angle=90, width=0.95\linewidth]{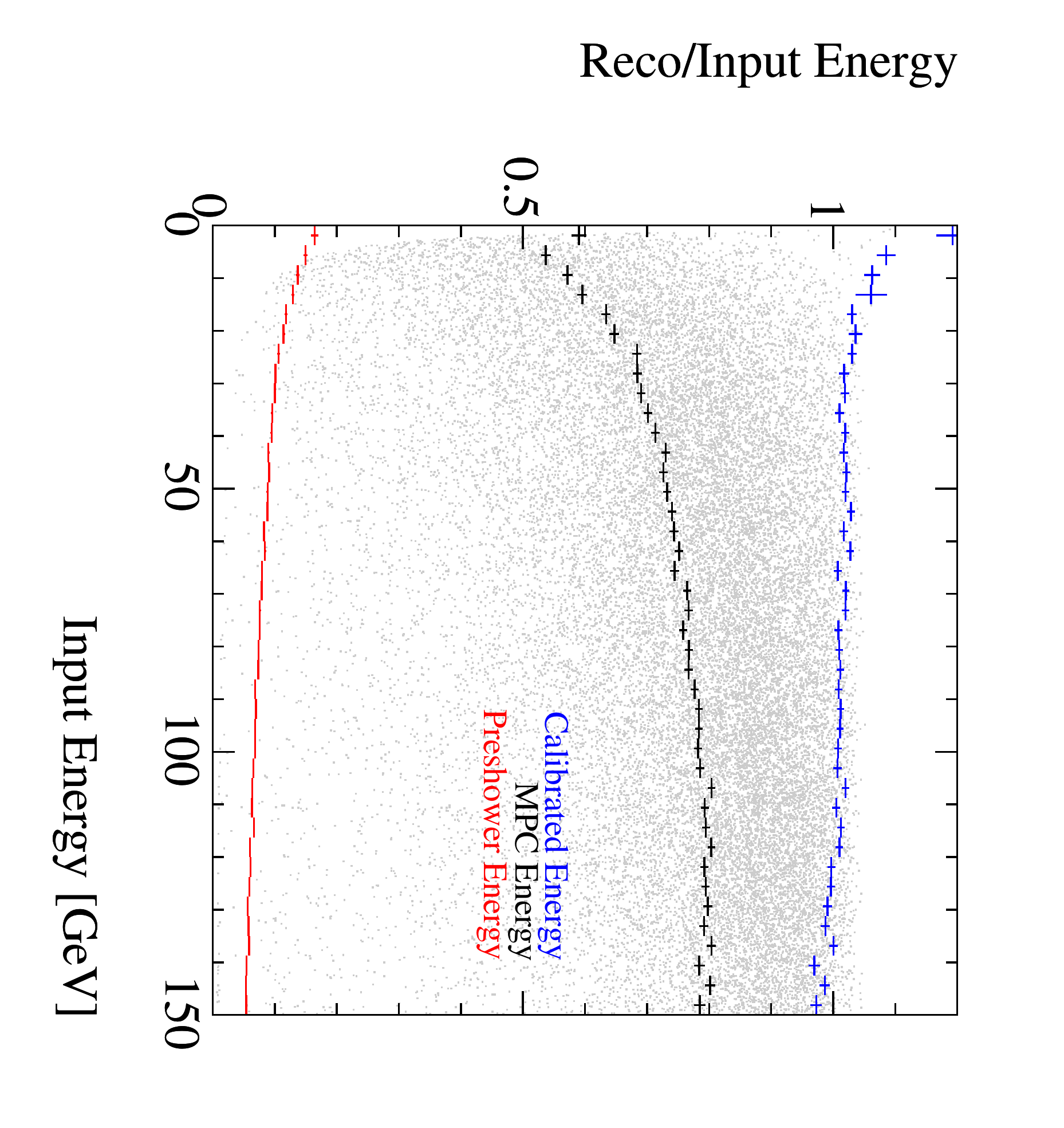}
\end{minipage}
\hspace{0.5cm}
\begin{minipage}[b]{0.5\linewidth}
\centering
\includegraphics[angle=90, width=0.95\linewidth]{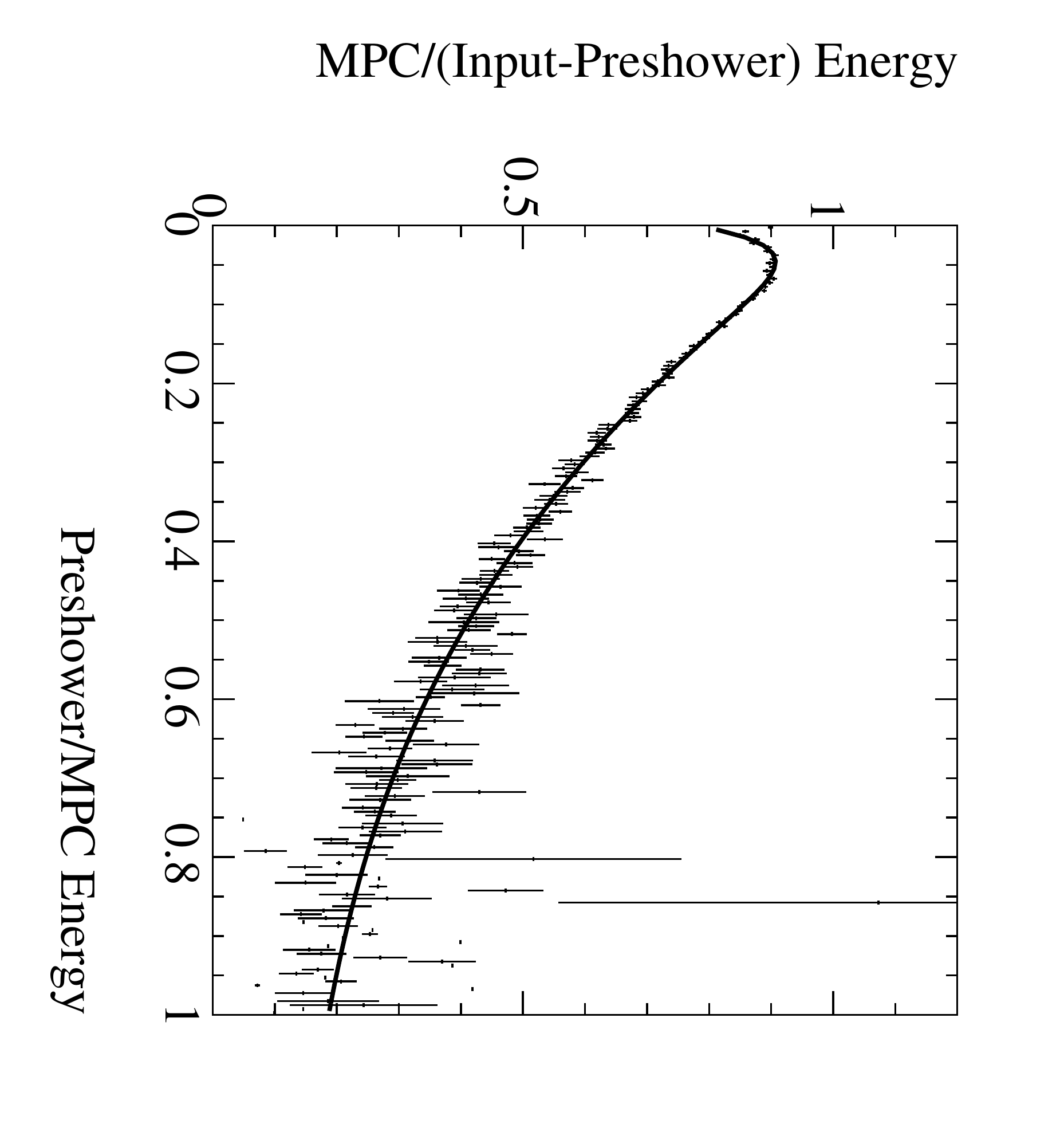}
\end{minipage}
\vspace*{-0.12in}
\caption{\label{fig:TrackRecalibration} The left panel shows the energy
deposition in the preshower (red) and MPC (black profile, grey symbols).
The blue histogram represents the calibrated data.  Most energy is
deposited in the MPC, with a diminishing amount in the preshower for
larger true energies.  The right panel shows the calibration method used,
see text for details.}
\end{figure}

The reconstructed energy is formed independently from the silicon
preshower and the MPC crystals.  Figure~\ref{fig:TrackRecalibration}
(left panel) shows the amount of energy typically deposited in each
section as a function of the true energy.  Less than 20\% of the
total true energy is reconstructed in the preshower.  To estimate
the total energy, one needs to recalibrate the energy reconstructed
in the MPC.  
Here, we simply
apply an additional calibration to the reconstructed energy in the MPC
from the existing reconstruction algorithms.

The calibration itself is a two-pass procedure. The relevant scaling variable
for the first pass is the ratio of the preshower to the MPC energy,
$fRat$.  The calibration is made from a sample of single $\gamma$s at
various energies.  The measured MPC energy is divided by the known
true energy of the $\gamma$ minus the measured energy from the
preshower.  This difference then represents the amount of energy which
needs to be reconstructed in the MPC.  Figure~\ref{fig:TrackRecalibration}
(right panel) shows this ratio as a function of $fRat$.
The distribution is fit with a ninth order polynomial 
in the region 0.1$<fRat<$1. The total energy $E_{TOT}$ is calculated by summing the 
MPC-EX energy and the corrected MPC energy. The total energy is then subject to a 
second-order scaling correction based on the reconstructed total energy.  
The parameters used in the first and second-pass calibrations are in
Table~\ref{tbl:EnRecalibration}.  For cases when the preshower energy
exceeds that of the MPC, the energy calibration is fixed to a constant
value (corresponding to $fRat=1$), due to statistical limitations in the
calibration of the simulation data.  The final calibration can be seen in the left panel
of Figure~\ref{fig:TrackRecalibration} (blue symbols).  The reconstructed energy in the 
region of interest ($E>$20\,GeV) is in good agreement with the
true energy.

\begin{table}
\centering
\caption{Parameters for first- and second-pass energy calibration.}
\label{tbl:EnRecalibration}
\begin{tabular}{|l|c|c|}
\hline
\multirow{2}{*}{Function} & Parameter & \multirow{2}{*}{Value} \\
 & Number &  \\
\hline
\multirow{9}{*}{First Pass (MPC Energy), Polynomial ($fRat$)} & 0 & 1.00492 \\
& 1 & 0.571689 \\
& 2 & -14.6354 \\
& 3 & 371.281 \\
& 4 & -2608.27 \\
& 5 & 8990.66 \\
& 6 & -1.67062 $\times 10^{4}$ \\
& 7 & 1.6094 $\times 10^{4}$ \\
& 8 & -6327.06 \\
\hline
\multirow{7}{*}{Second Pass (Total Energy), Polynomial ($E_{TOT}$)} & 0 & 1.04137 \\
& 1 & -0.009866 \\
& 2 &  0.000506 \\
& 3 & -1.2361 $\times 10^{-5}$ \\
& 4 & 1.595451 $\times 10^{-7}$ \\
& 5 & 1.02758 $\times 10^{-8}$ \\
& 6 & 2.5889 $\times 10^{-12}$ \\
\hline
\end{tabular}
\end{table}

\begin{figure}[hbt]
  \begin{center}
    \includegraphics[width=0.8\linewidth]{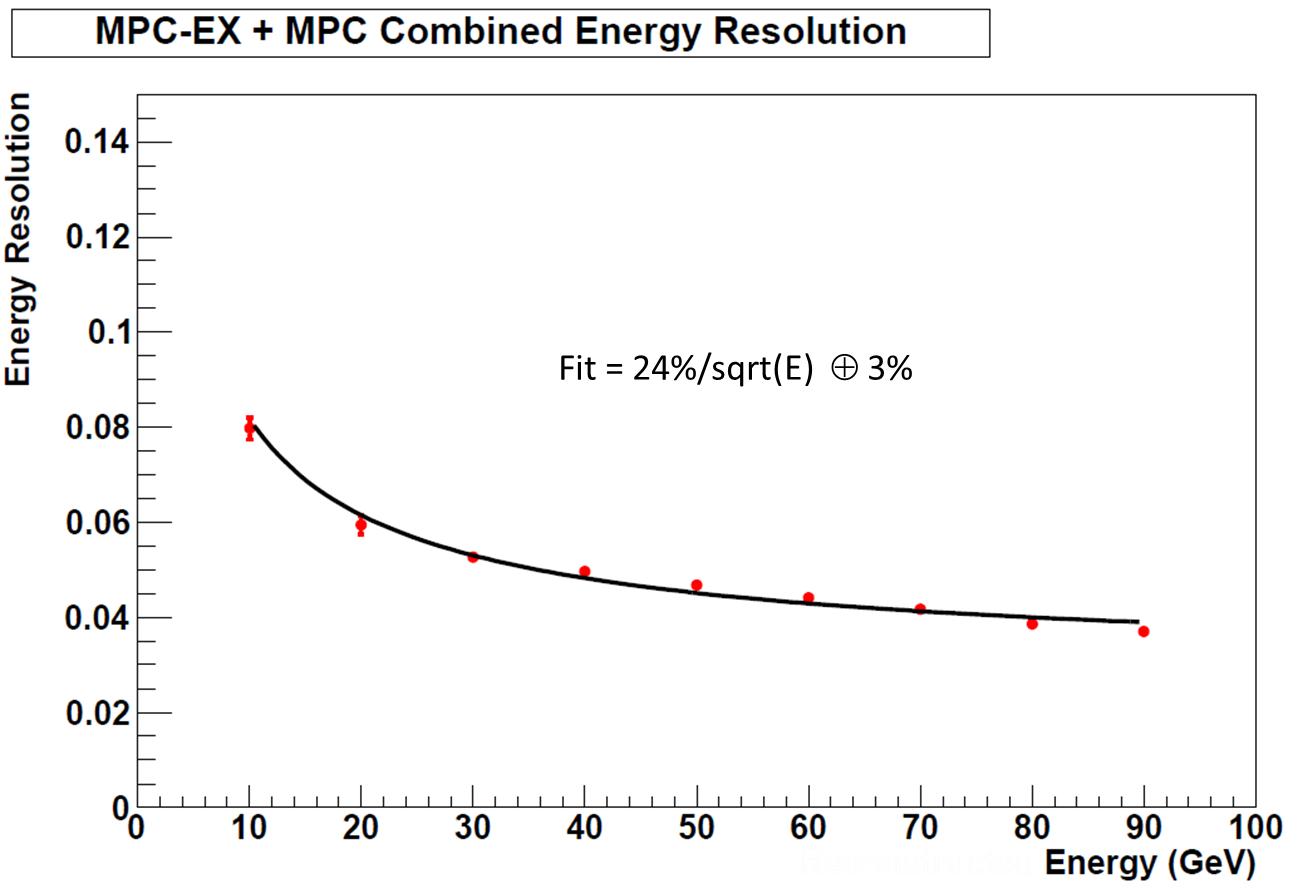}
  \end{center}\vspace*{-0.12in}
  \caption{\label{fig:EnergyRes} Energy resolution of the MPC-EX + MPC resonstruction
in simulated events, using the calibration procedure outlined in the text. The large constant
term in the resolution arises from combining MPC clusters with MPC-EX tracks. In the analysis of real
data, a combined reconstruction will be used and it is anticipated that the energy resolution will improve. }
\end{figure}

The energy resolution of the combined MPC-EX + MPC after calibration of the simulations is shown in 
Figure~\ref{fig:EnergyRes}. In the energy range of interest for the direct photon ($E>$20\,GeV) the
energy resolution is below 6\%. The large constant term in the fit to the resolution 
is likely a result of the imperfect method used to combine the MPC-EX and MPC detectors, 
and will certainly be improved in real data when a combined reconstruction procedure is used.
As a comparison, the MPC alone achieved and energy resolution of roughly $14\%/\sqrt{E}$ in 
a test beam.  

\clearpage
\section[EM Shower Reconstruction Performance]{EM Shower Reconstruction Performance: $\gamma$ and $\pi^{0}$.}
\label{sim:photonreco}

\subsection{Overview of Section}

In this section, the reconstruction performance is evaluated and discussed
in the context of single $\gamma$ and $\pi^{0}$ simulations.  This is followed
by a brief discussion of the performance in a simulation of the full $p+p$
cross-section using {\sc Pythia}.

\subsection{Basic cuts used in the reconstruction}

In this analysis, a series of ``basic cuts'' are applied to all data to
clean-up the event sample prior to more detailed analysis.  These cuts
are aimed at reducing effects due to poorly reconstructed electromagnetic
showers, due to errors in the reconstruction or acceptance effects.
The final product should be a sample of events which include
all single-$\gamma$s and clusters of merged showers from $\pi^{0}$
decays.

The first cut
separates tracks with a reconstructed energy of at least 20\,GeV.
Below this limit the number of $\pi^{0}$s which form two distinct MPC clusters
increases, thus most tracks below this energy are single-$\gamma$'s and not overlapping
$\pi^{0}$s.
Such tracks can be used to reconstruct $\pi^{0}$s using a ``two-track'' method
similar to that used in the current MPC-only analysis. Tracks with reconstructed energies
greater than 20\,GeV are subject to an additional analysis in a region of interest around
the track to determine if the track is consistent with a single shower (a photon) or 
an overlap between photon showers (a $\pi^{0}$). This procedure is described in 
Section ~\ref{sec:SingleTrackInvMass}.

Only a preshower track that is tagged as ``closest'' to an MPC cluster
is considered in the direct photon analysis.  It is also required that the
preshower track vector and the direction vector formed from the MPC
cluster agree to within $\Delta({\rm Hough})$$<$0.0025 (distance in Hough space), see
Figure~\ref{fig:TrackMatchResolution}.

\begin{figure}[hbt]
\hspace*{-0.12in}
\centering
\includegraphics[angle=90, width=0.65\linewidth]{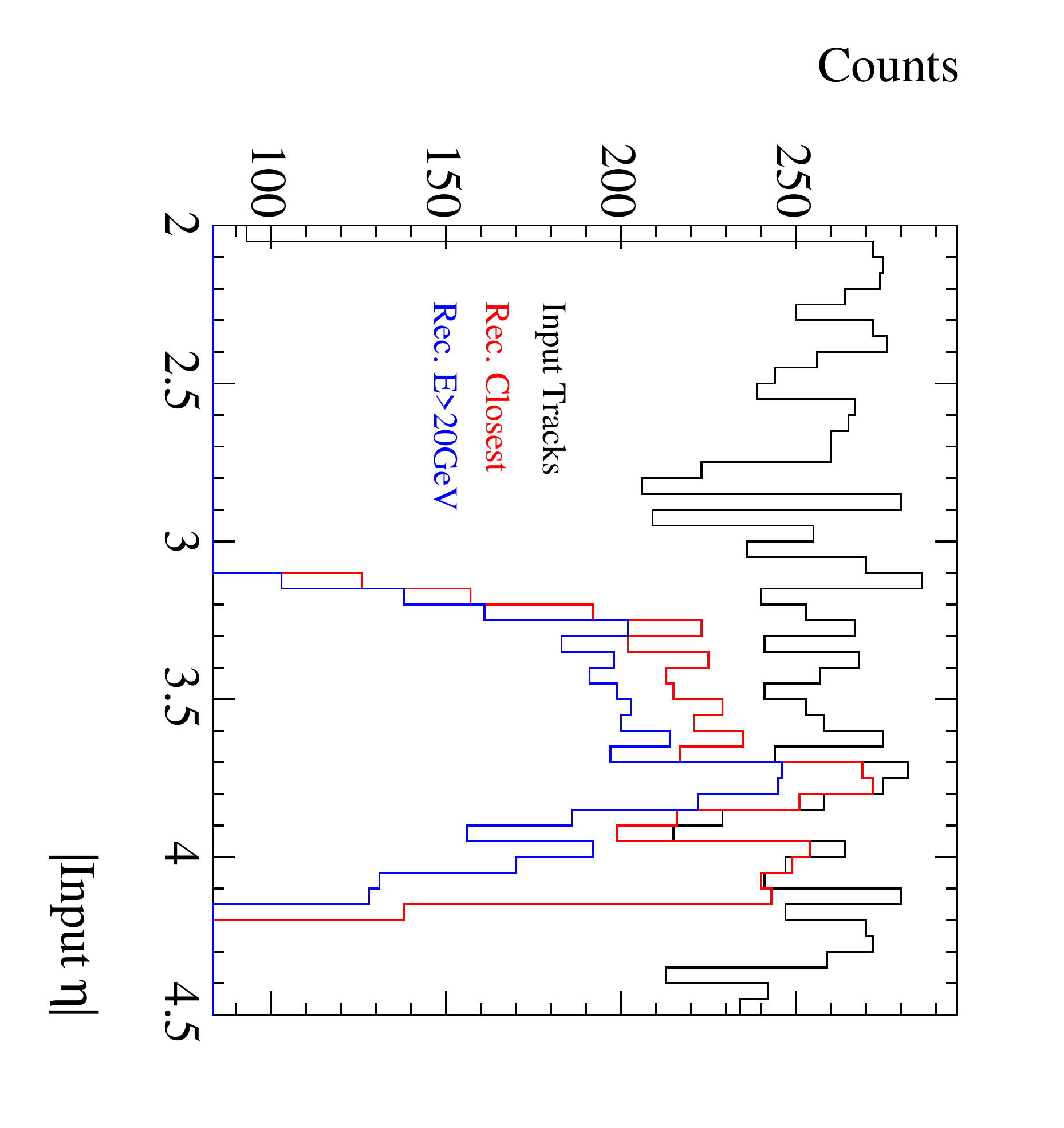}
\vspace*{-0.12in}
\caption{\label{fig:RecoEtaFullInputRange} $\eta$ distribution of
single-$\gamma$s from a simulation where the input distibution is wider than
the acceptance of the detector (black).  The reconstructed closest tracks (red)
and those with a reconstructed energy greater than 20\,GeV (blue) are shown.}
\end{figure}

Finally, a fiducial cut in $\eta$-space rejects tracks reconstructed
close to the edge of the preshower.  These are typically malformed
and may, for example, have missing or distorted energy signals due to shower leakage.
Figure~\ref{fig:RecoEtaFullInputRange} shows the pseudorapidity dependence
of reconstructed tracks in a single-$\gamma$ simulation.  The
input $\eta$ was thrown over 2$<$$|\eta|$$<$4.5 (black histogram), and the full reconstruction
was run.  By selecting tracks with both an MPC and preshower component,
with the latter being ``closest'', the pseudorapidity coverage was found to
be $\sim$3.1$<$$\eta$$<$$\sim$4.2 (red).  The requirement of a minimum energy
($E_{\rm Rec}$$>$20\,GeV) serves only to reduce statistics, as expected (blue).

\begin{figure}[hbt]
\hspace*{-0.12in}
\begin{minipage}[b]{0.5\linewidth}
\centering
\includegraphics[angle=0, width=0.9\linewidth]{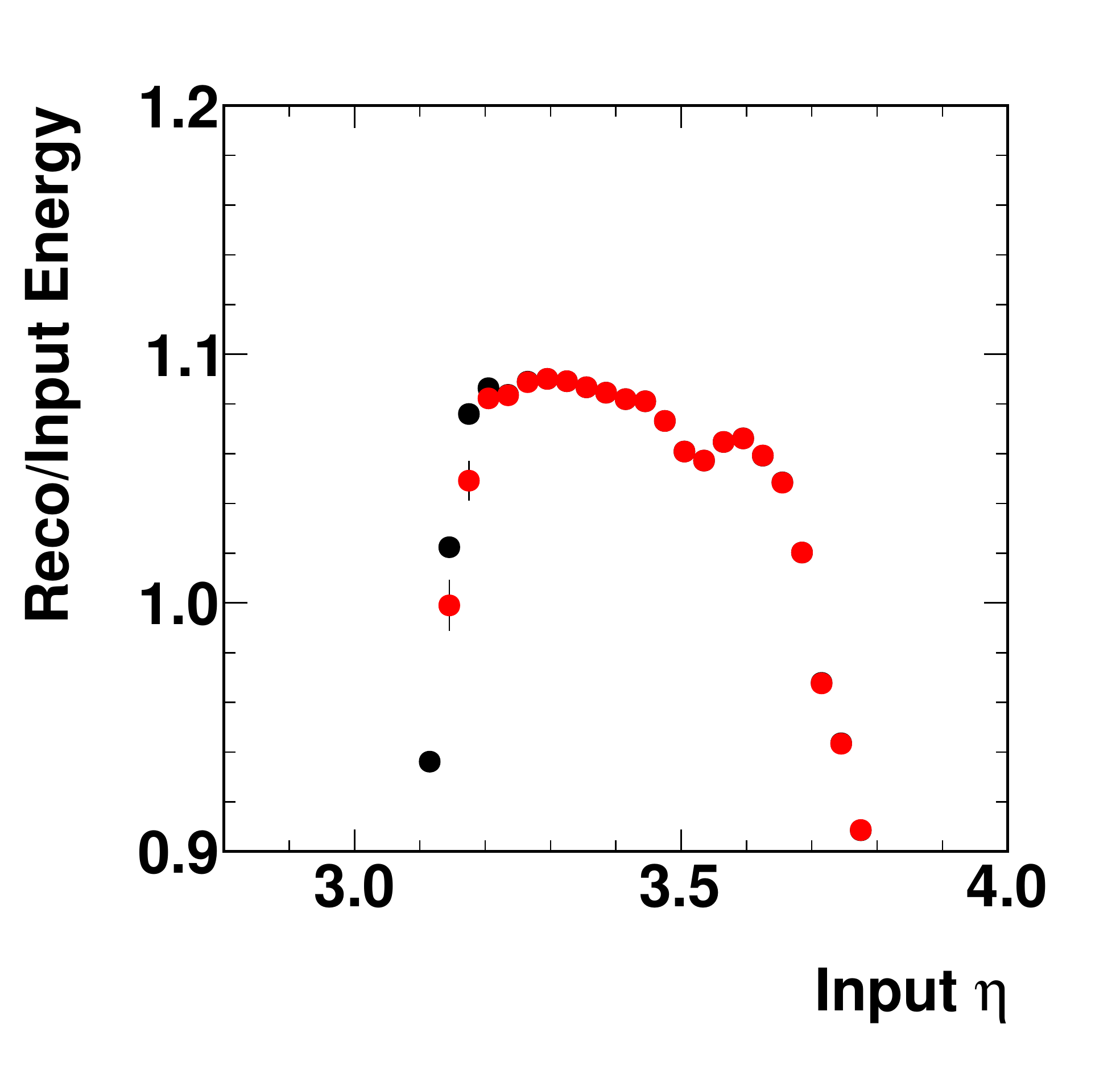}
\end{minipage}
\hspace{0.5cm}
\begin{minipage}[b]{0.5\linewidth}
\centering
\includegraphics[angle=0, width=0.9\linewidth]{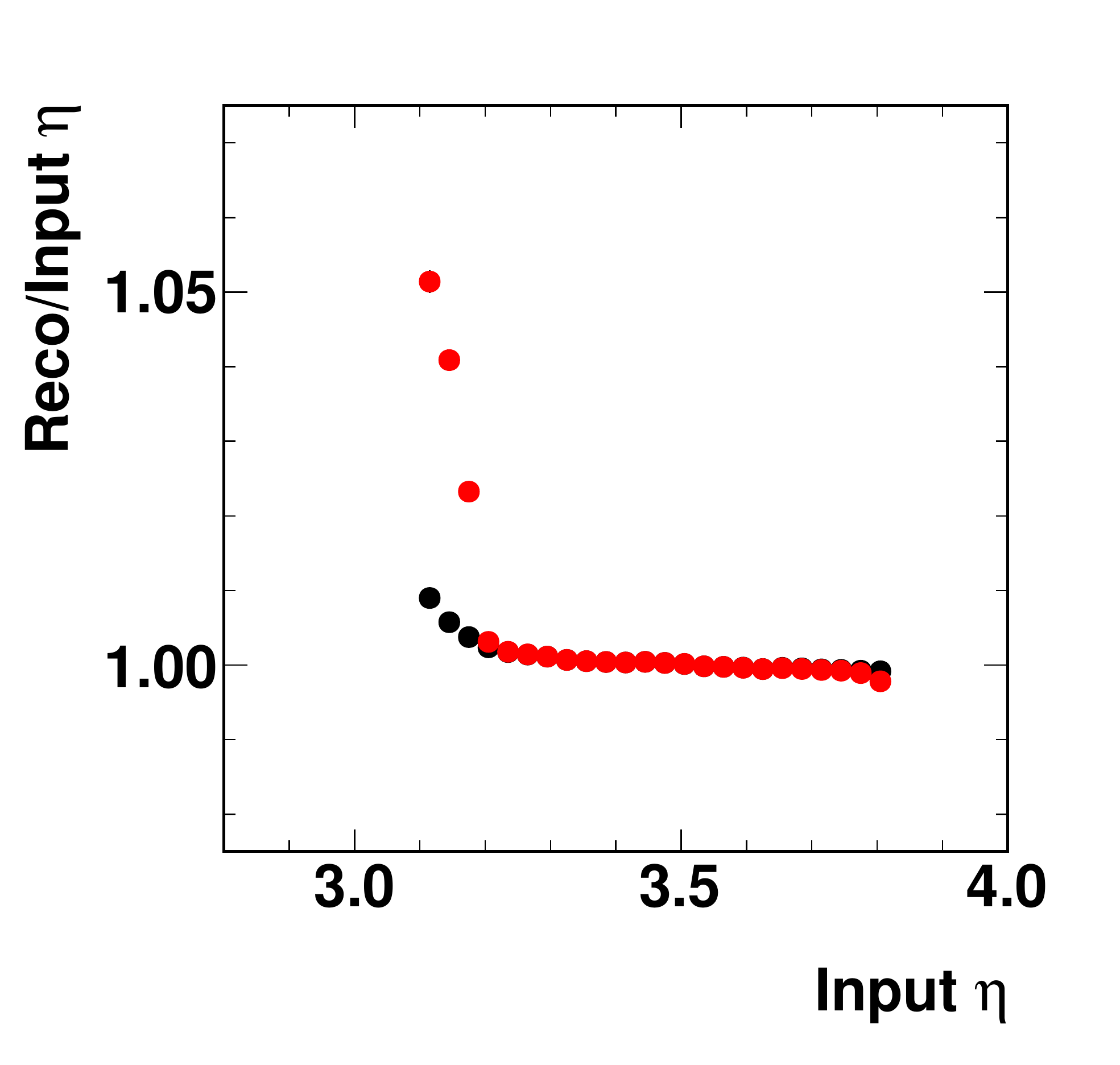}
\end{minipage}
\begin{minipage}[b]{0.5\linewidth}
\centering
\includegraphics[angle=0, width=0.9\linewidth]{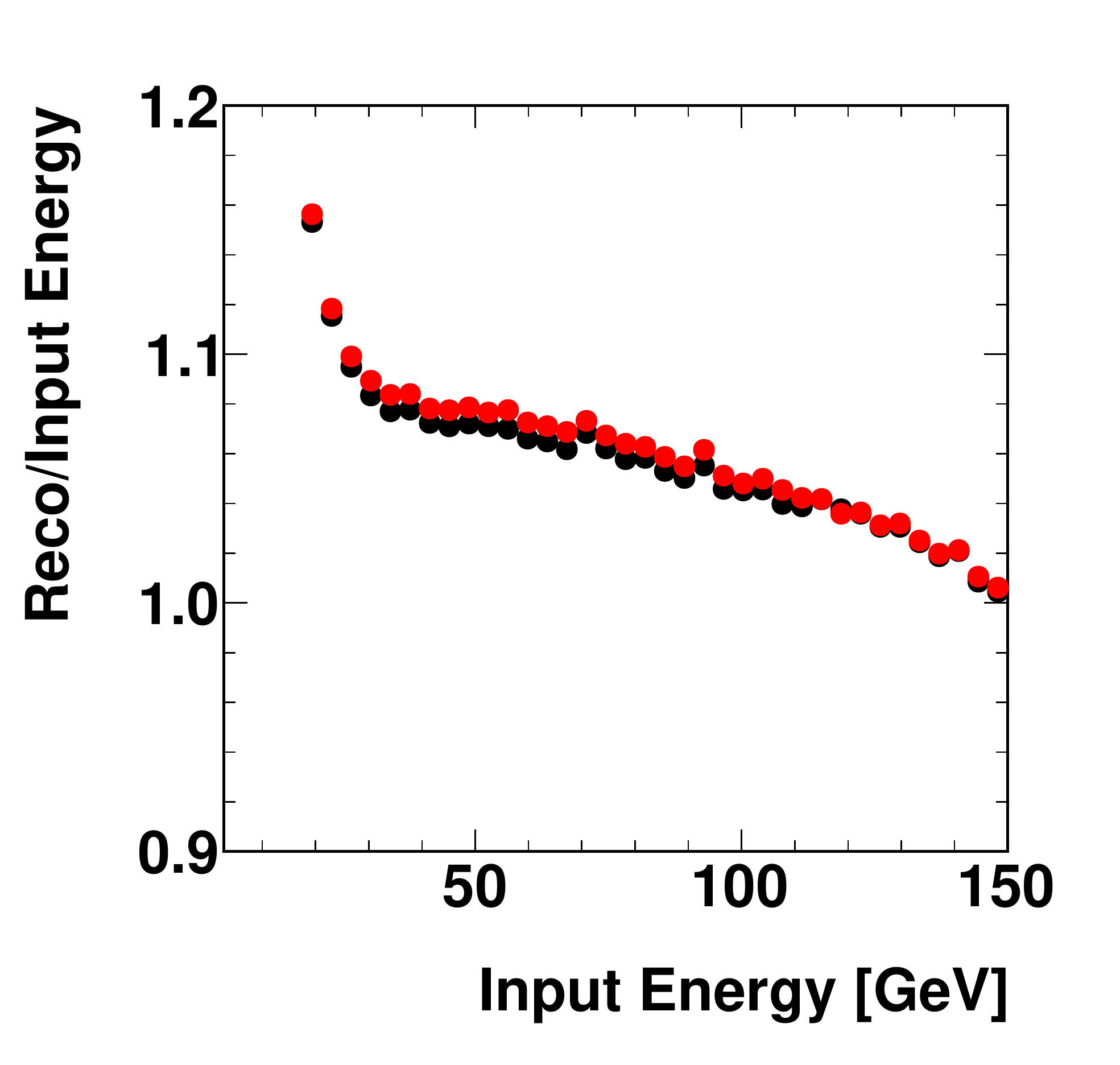}
\end{minipage}
\hspace{0.5cm}
\begin{minipage}[b]{0.5\linewidth}
\centering
\includegraphics[angle=0, width=0.9\linewidth]{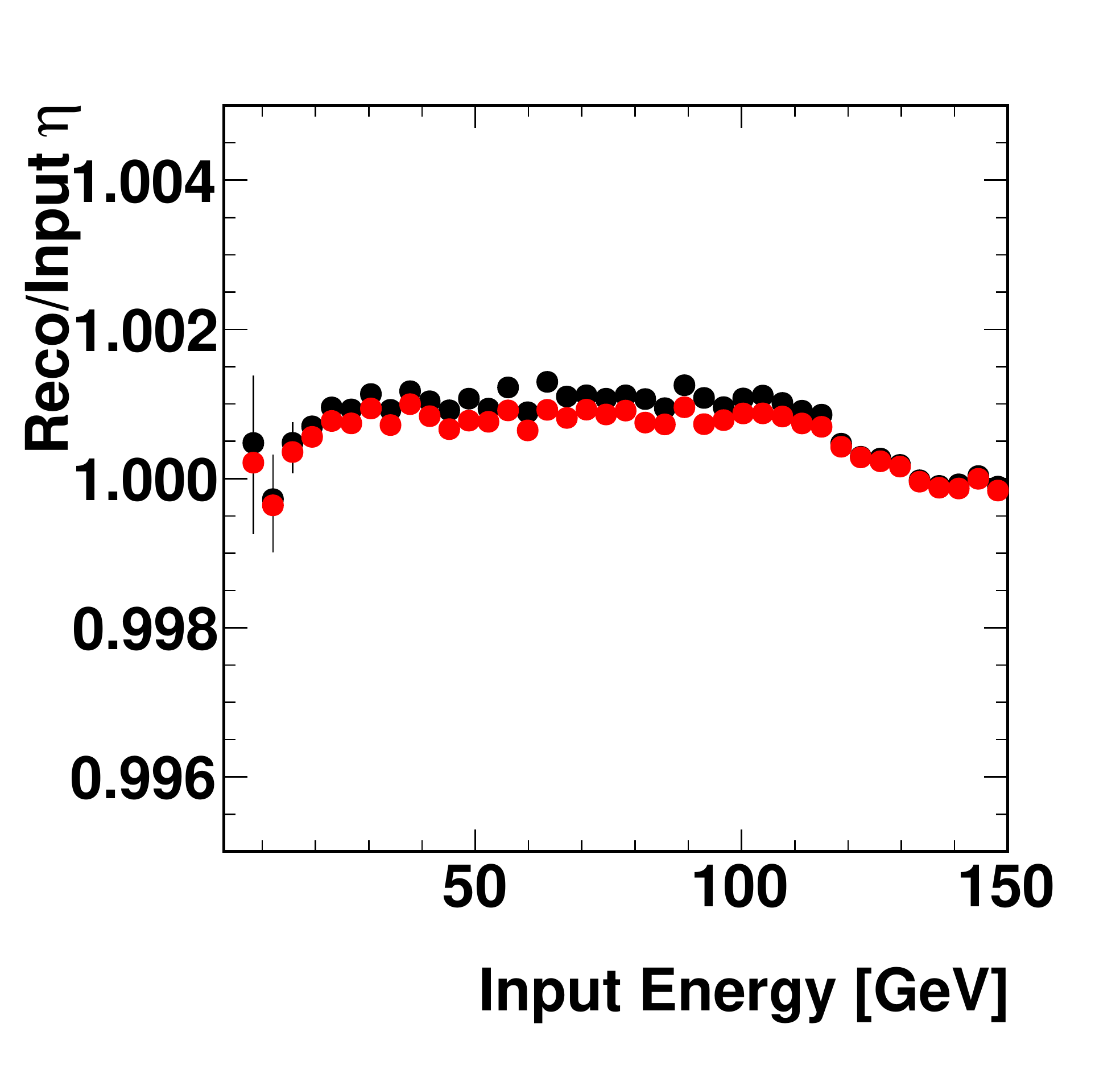}
\end{minipage}
\vspace*{-0.12in}
\caption{\label{fig:TrackCuts} Reconstructed energy (left figures) and
reconstructed $\eta$ (right) as a function of $\eta$ (top) and energy
(bottom). }
\end{figure}

Figure~\ref{fig:TrackCuts} shows the reconstructed energy (left panels)
and $\eta$ (right panels) relative to the input value.  The upper panels
depict this versus the input $\eta$ for each event, and the lower panels show
this versus the input energy.  The black (red) histogram shows the
distribution for $\gamma$s ($\pi^{0}$s).  The accuracy of the\
reconstructed $\eta$ is found to degrade for low $\eta$ (worse for
$\pi^{0}$s), whilst the energy reconstruction degrades only as a function
of energy.

\subsection{Single-track invariant mass calculation}\label{sec:SingleTrackInvMass}

Once an electromagnetic track candidate is found, it is tested to determine whether it
is consistent with a single-electromagnetic shower or two close
showers, similar to that expected from $\pi^{0}\rightarrow\gamma\gamma$
decays.  Only high energy tracks (E$>$20\,GeV) and the closest track to a
single MPC cluster are considered.  The process begins with a detailed examination
of the minipad hits in a region of interest around the MPC-EX track. This region 
of interest is fixed at 0.0175 in Hough space in both the $x$ and $y$ directions, or
a width of $\sim$41 minipads. A histogram of the minipad energies is then divided into 
two halves (performed independently
for $x$- and $y$-minipads).  In a first pass, the dividing line is 
the center of gravity of the energy distribution in each direction.  
This is subsequently changed in an iterative fashion
until a small change in the split-point results in no change in the energy assigned to 
each track, i.e. a stable point is found.  Typically, the stable point is
found at the first or second iteration.

Once a stable point is found, the energy in the MPC-EX for each sub-track is determined, and 
the total energy of the track is shared between the subtracks based in these energy fractions. 
The subtracks can then be combined to form an invariant mass, which is associated with the 
MPC-EX track object. If all goes well, a single $\pi^{0}$ will be reconstructed into 
two subtracks, and those subtracks will reconstruct to the $\pi^{0}$ invariant mass. 



It is important to note that the calculation of the invariant mass serves
as both a method to identify and measure $\pi^{0}$s  as well as a method 
to exclude $\pi^{0}$s from the direct photon analysis.  If no invariant mass
is found, then this is more likely to be a $\gamma$ candidate, rather than
a $\pi^{0}$.  However, a two-track decay does not necessarily produce
two discernable tracks.  This will be discussed further in the next section.
The reconstruction of the invariant mass is quite aggressive at rejecting
all candidates which do not appear as $\pi^{0}$s.  This helps in
reconstructing $\pi^{0}$s, and forces all ``failed'' invariant mass
reconstruction into one of two categories.

The first category deals with candidates which have too few minipad 
hits associated with the track.  Many tracks in this category fail due to a failure 
of the track matching between the MPC
and MPC-EX, rather than due to too few hits in the pre-shower detector.
Because of the resolution of the
MPC it is possible to select a track which is a fluctuation
from the main shower, but this happens to be the closest track in Hough
space.  In this scenario, there are too few hits in the MPC-EX to
continue reconstruction simply from matching to the wrong track. This 
mode is greatly reduced by clustering tracks associated with a given MPC shower 
as described previously.

The second point of failure occurs when the algorithm cannot divide the
energy up enough to separate two distinct peaks in the $x$- {\b or}
$y$-directions, i.e. no stable dividing point is found.  The reason for this could be a real single-$\gamma$
(for example a signal photon), an asymmetrically decayed $\pi^{0}$, a
track-matching failure whereby only one track (from a $\pi^{0}$ decay) falls in the acceptance.


As will be discussed in later sections, the performance of this algorithm on a given 
$\pi^{0}$ track depends greatly on the asymmetry of the event. For large asymmetries, the lower-energy photon 
can be buried under the shower from the higher energy photon. This results in an essentially symmetric shower, 
and the algorithm will either fail to converge on a dividing point in the shower, or a dividing point will be selected 
with a very small opening angle between the subtracks, resulting in a very low invariant mass. 
A shower due to a single photon will also be symmetric, and will reconstruct in a similar fashion, with 
the subtracks being determined by fluctuations in the photon shower. For this reason, we can define photon 
candidates as tracks that either fail the invariant mass reconstruction, or reconstruct to a very low mass. 
This is not a ``true'' invariant mass of the photon, but an artifact of the reconstruction. 
A photon sample defined in this way will be subject to contamination from $\pi^{0}$s that will have to 
be removed with additional cuts.

\subsection{Reconstruction Performance}

\subsubsection{Position Resolution}
\begin{figure}[hbt]
\hspace*{-0.12in}
\begin{minipage}[b]{0.5\linewidth}
\centering
\includegraphics[angle=0, width=0.95\linewidth]{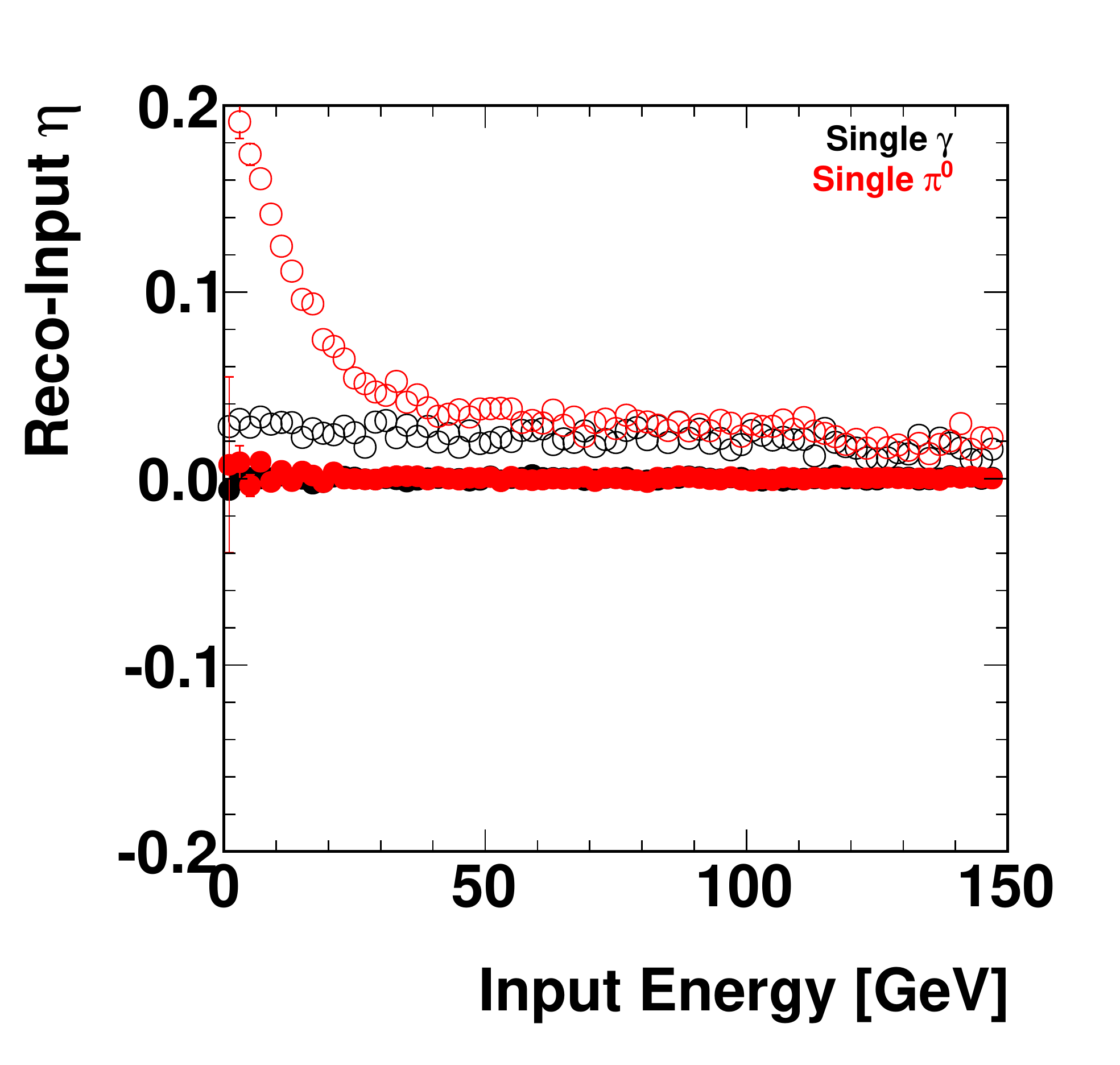}
\end{minipage}
\hspace{0.5cm}
\begin{minipage}[b]{0.5\linewidth}
\centering
\includegraphics[angle=0, width=0.95\linewidth]{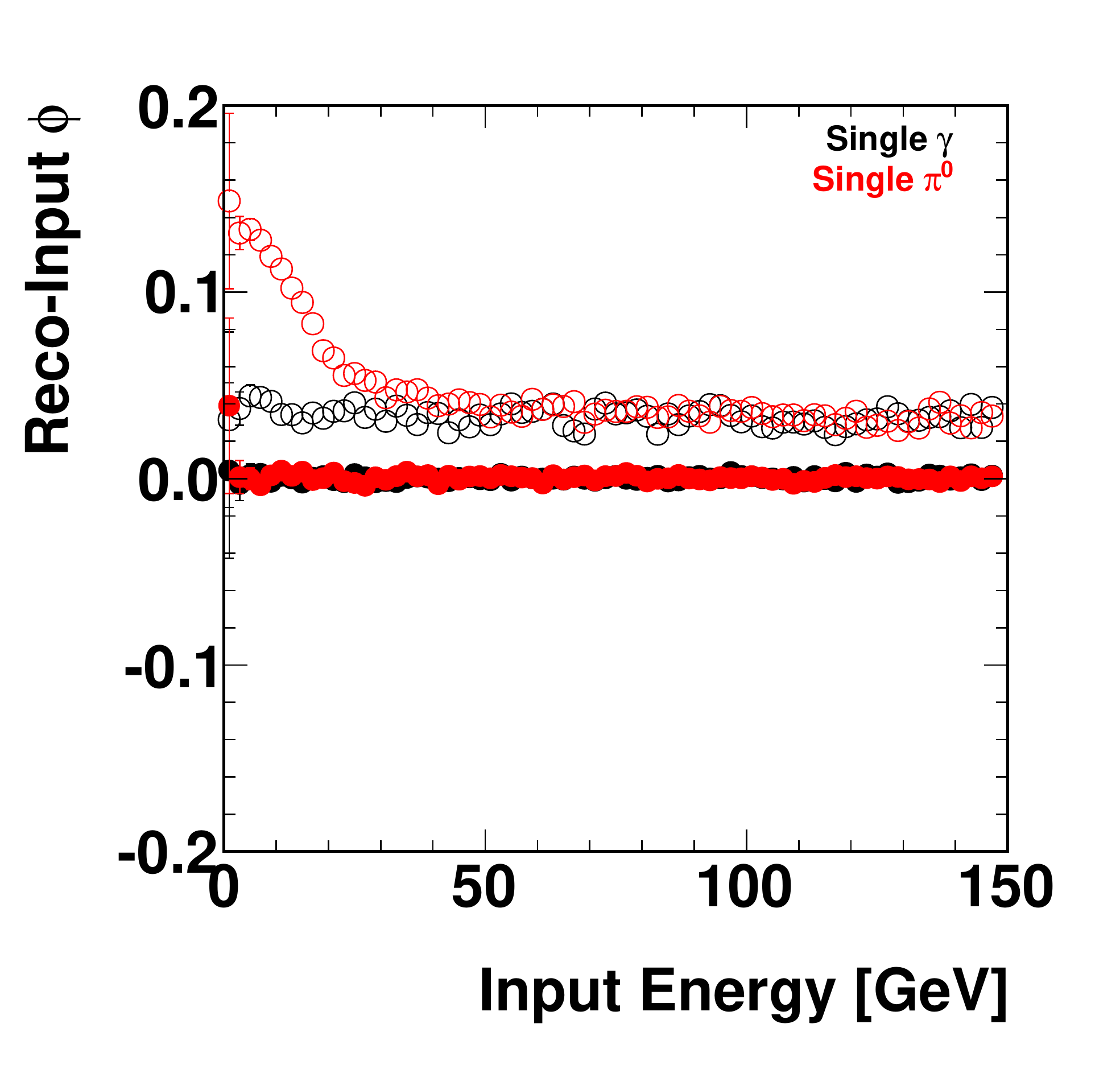}
\end{minipage}
\vspace*{-0.12in}
\caption{\label{fig:PositionResolution} $\eta$ (left panel) and $\phi$ (right)
resolution (open symbols) and offset (closed) for single-$\gamma$ (black)
and single-$\pi^{0}$s.  The deviation between $\pi^{0}$s and $\gamma$s
observed at low energies is due to the reference $\eta$ (or $\phi$)
following the original $\pi^{0}$ -- not the decay $\gamma$ which has a
different vector. }
\end{figure}

The resolution and offset to the reconstructed track vector, relative to the input vector of the particle, for a
single-$\gamma$ and single-$\pi^{0}$ simulations is shown in
Figure~\ref{fig:PositionResolution}.  The left panel shows the $\eta$
resolution (open symbols) and offset (closed) for single-$\gamma$ and
single-$\pi^{0}$ simulations.

%

\subsubsection{Invariant Mass Reconstruction}

Candidate tracks in full events originate from a variety of sources.  The dominant source
is $\pi^{0}$s which can appear as two independent tracks in the detector
(two-track $\pi^{0}$s) or merged as a single cluster (as seen in the MPC).
The aim of the analysis in this document is to measure the production
of direct-$\gamma$s, whose dominant background source is the $\pi^{0}$.
Thus, $\pi^{0}$s must be measured to enable the extraction of
direct-$\gamma$s.  The measurement of the $\pi^{0}$ cross-section is
not less important, but is not the focus here.  Other sources, for example
charged hadrons, $\eta$ mesons, and other decay photon sources are
discussed in more detail in Section~\ref{sim:dphotcuts}.

\begin{figure}[hbt]
\hspace*{-0.12in}
\centering
\includegraphics[angle=0, width=0.95\linewidth]{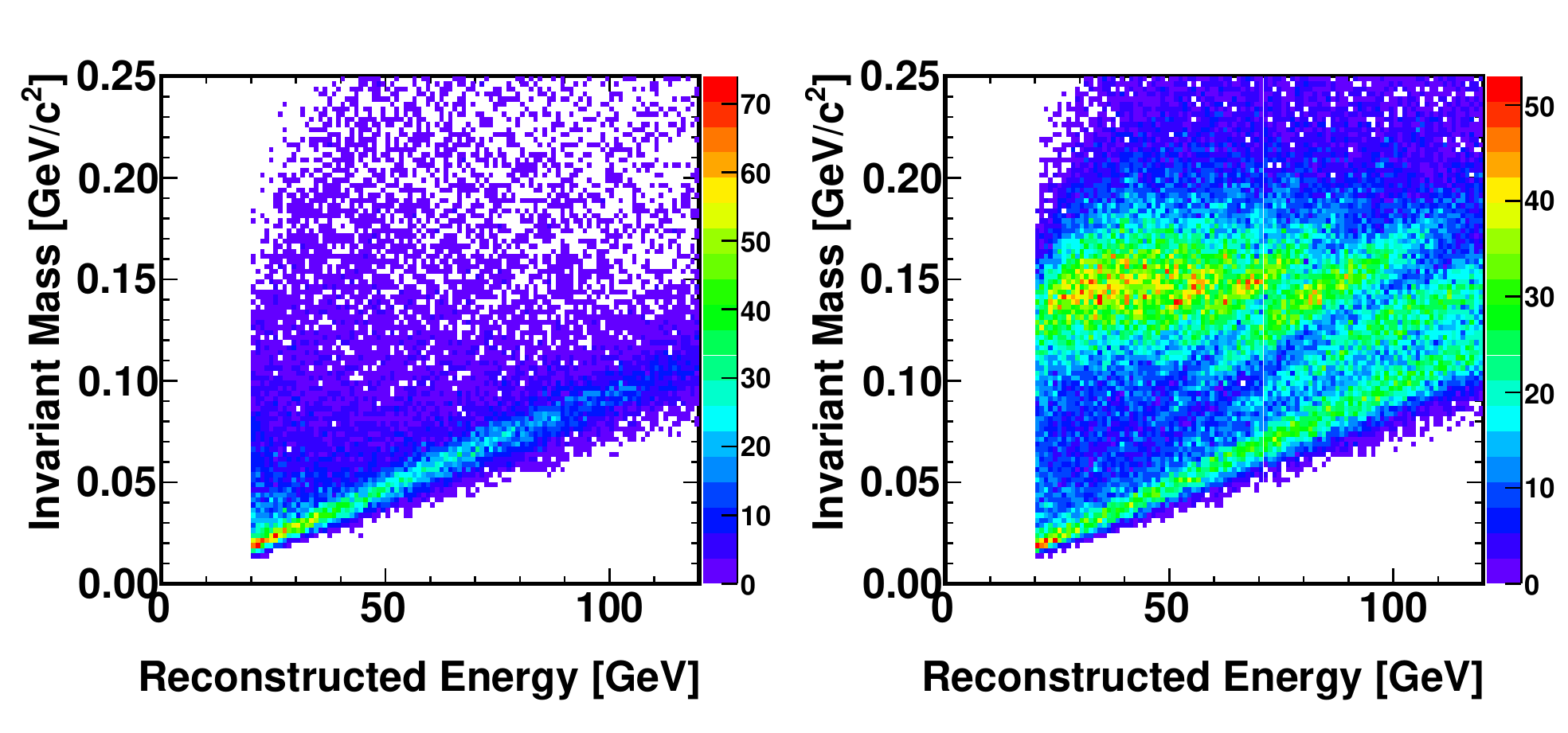}
\vspace*{-0.12in}
\caption{\label{fig:RecoAllEnergies} The left (right) panel shows the
reconstructed invariant mass versus reconstructed energy for $\gamma$s
($\pi^{0}$s).}
\end{figure}
\begin{figure}[hbt]
\hspace*{-0.12in}
\begin{minipage}[b]{0.5\linewidth}
\centering
\includegraphics[angle=0, width=0.95\linewidth]{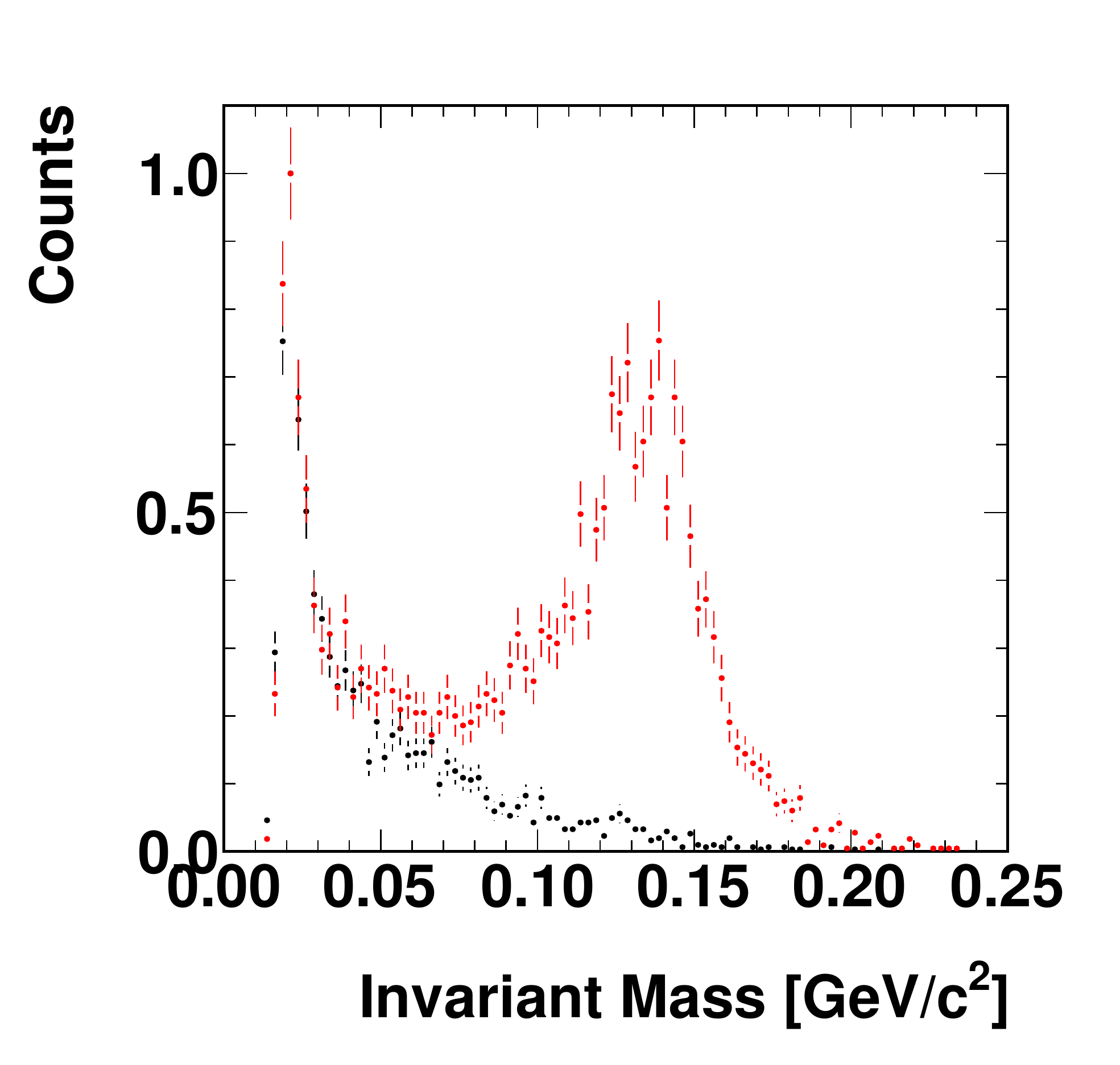}
\end{minipage}
\hspace{0.5cm}
\begin{minipage}[b]{0.5\linewidth}
\centering
\includegraphics[angle=0, width=0.95\linewidth]{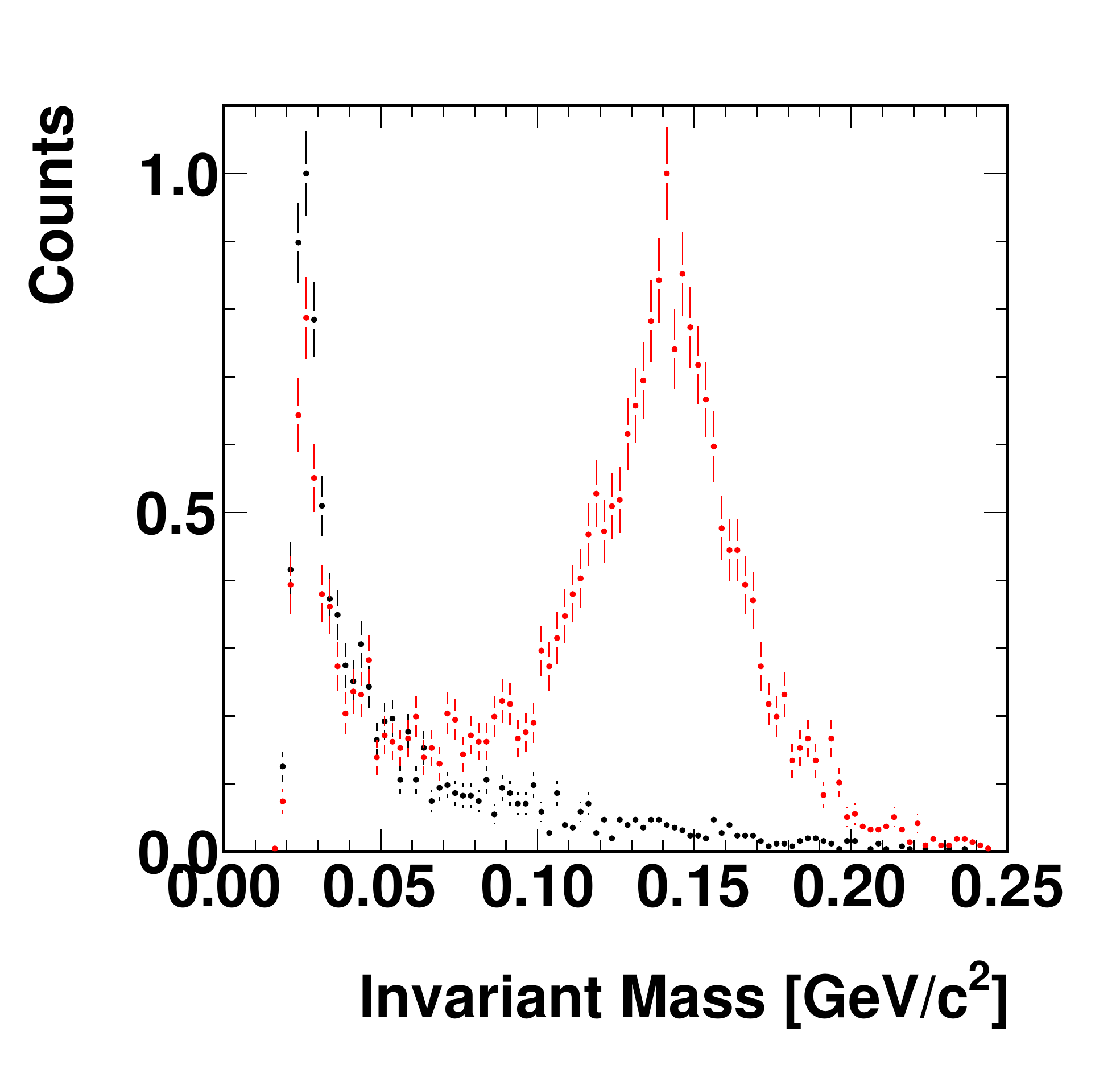}
\end{minipage}
\vspace*{-0.12in}
\caption{\label{fig:Reco20_30} Comparison of the invariant mass distribution
for $\gamma$s (black) and $\pi^{0}$s (red) for the 20$<$E$<$25\,GeV
(25$<$E$<$30\,GeV) energy range. The normalization of the two histograms
was chosen to fix the maximum height to be the same.}
\end{figure}
\begin{figure}[hbt]
\hspace*{-0.12in}
\begin{minipage}[b]{0.5\linewidth}
\centering
\includegraphics[angle=0, width=0.95\linewidth]{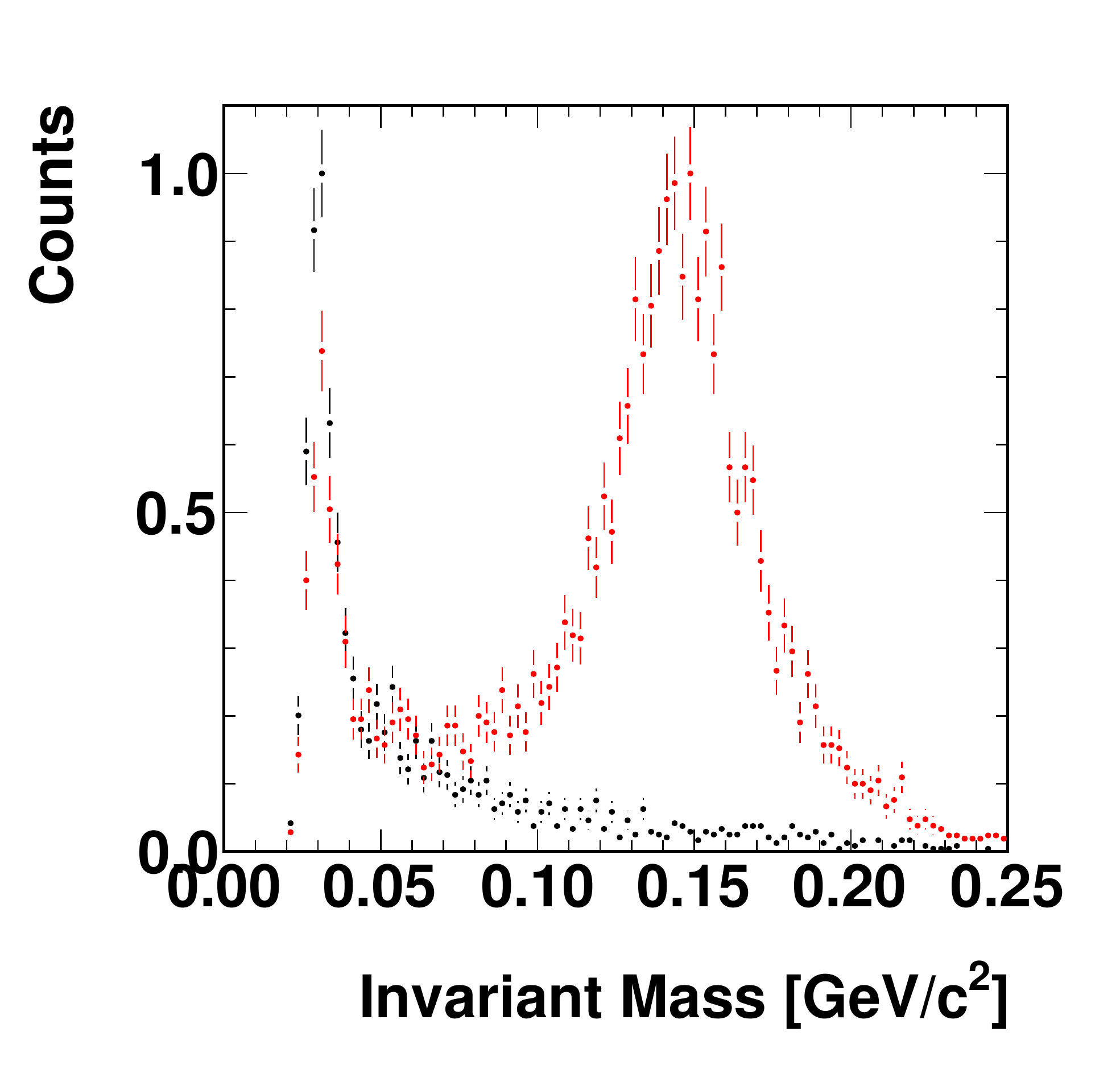}
\end{minipage}
\hspace{0.5cm}
\begin{minipage}[b]{0.5\linewidth}
\centering
\includegraphics[angle=0, width=0.95\linewidth]{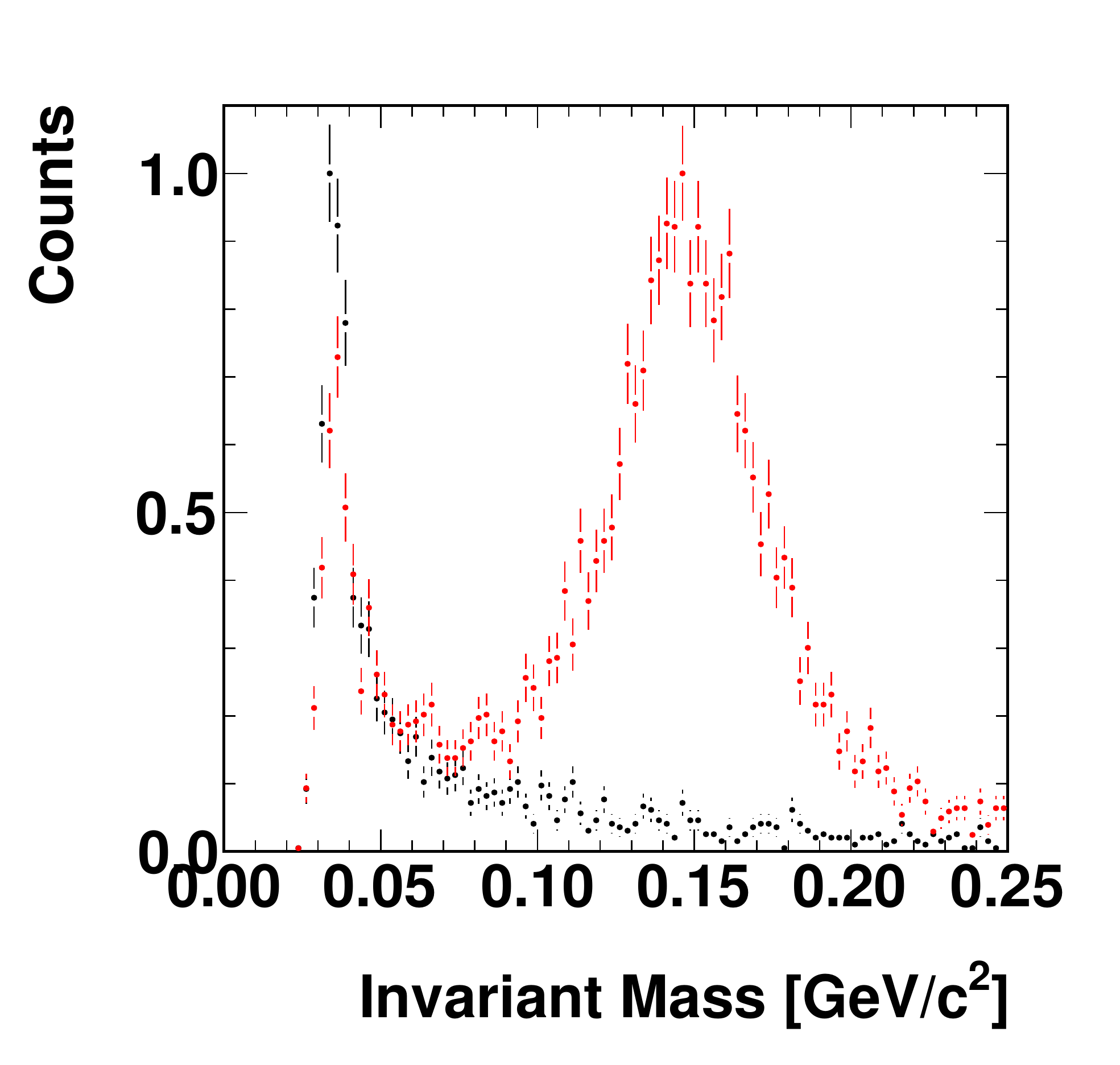}
\end{minipage}
\vspace*{-0.12in}
\caption{\label{fig:Reco30_40} Comparison of the invariant mass distribution
for $\gamma$s (black) and $\pi^{0}$s (red) for the 30$<$E$<$35\,GeV
(35$<$E$<$40\,GeV) energy range.}
\end{figure}
\begin{figure}[hbt]
\hspace*{-0.12in}
\begin{minipage}[b]{0.5\linewidth}
\centering
\includegraphics[angle=0, width=0.95\linewidth]{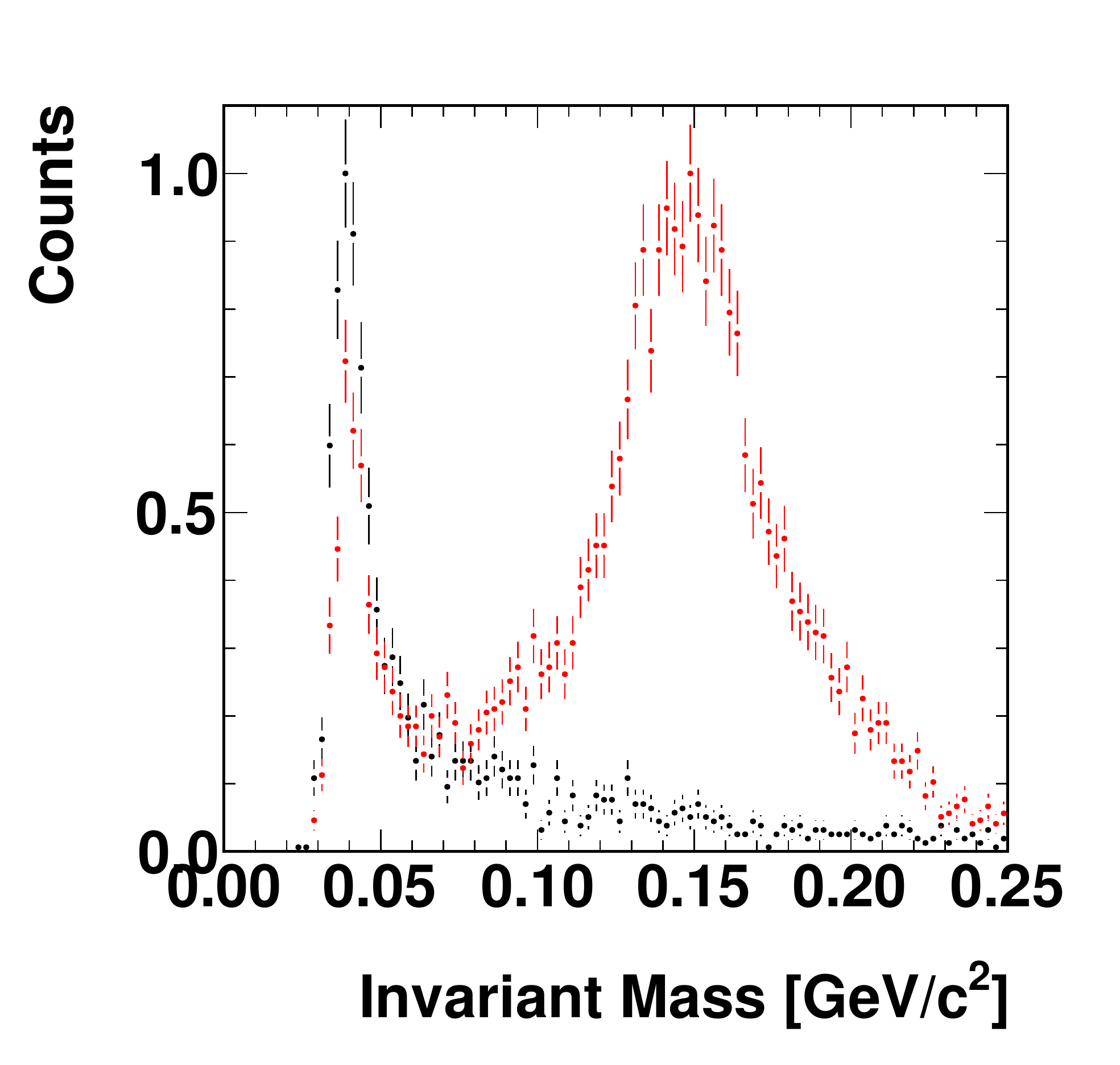}
\end{minipage}
\hspace{0.5cm}
\begin{minipage}[b]{0.5\linewidth}
\centering
\includegraphics[angle=0, width=0.95\linewidth]{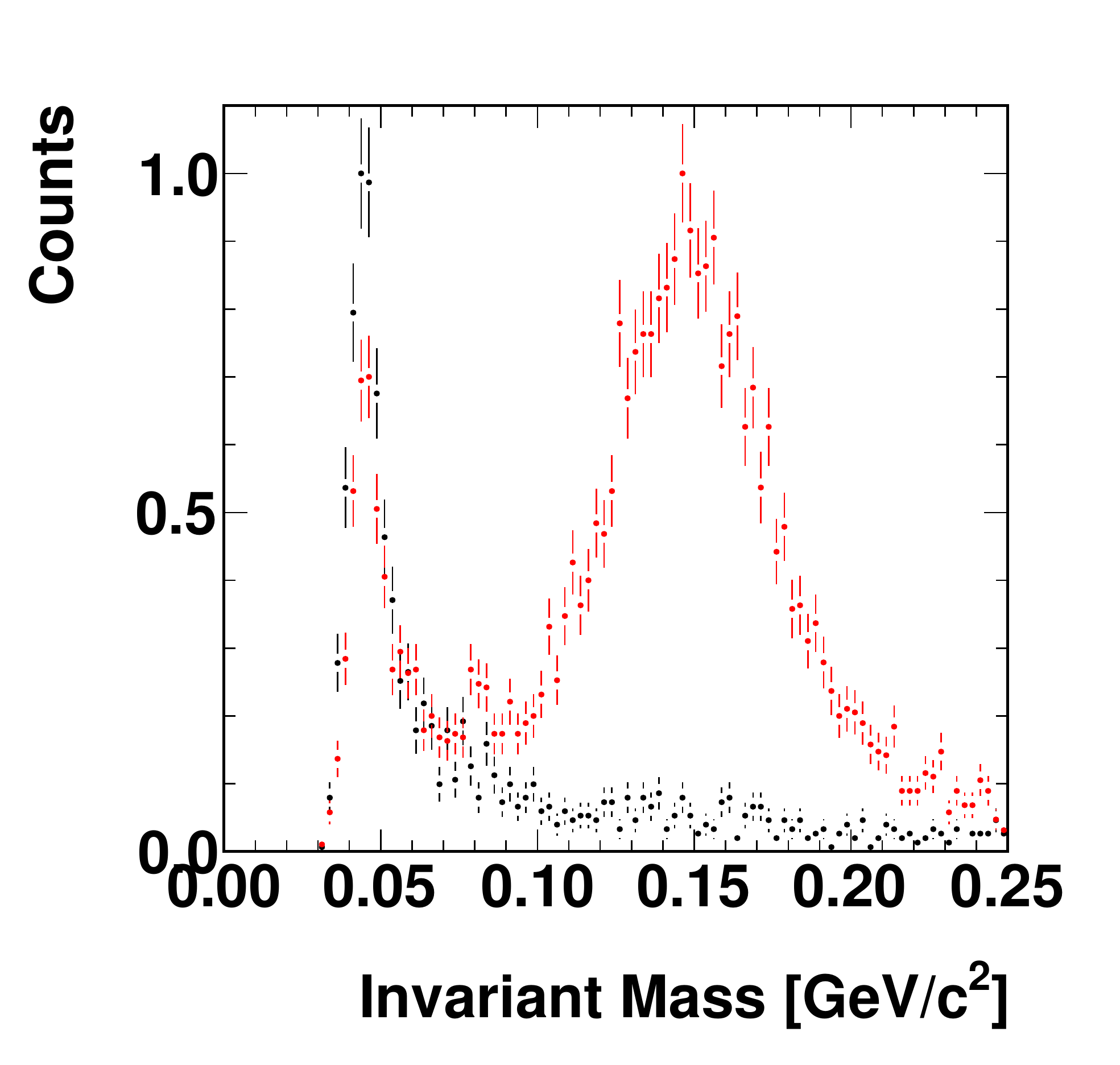}
\end{minipage}
\vspace*{-0.12in}
\caption{\label{fig:Reco40_50}  Comparison of the invariant mass distribution
for $\gamma$s (black) and $\pi^{0}$s (red) for the 40$<$E$<$45\,GeV
(45$<$E$<$50\,GeV) energy range.}
\end{figure}
\begin{figure}[hbt]
\hspace*{-0.12in}
\begin{minipage}[b]{0.5\linewidth}
\centering
\includegraphics[angle=0, width=0.95\linewidth]{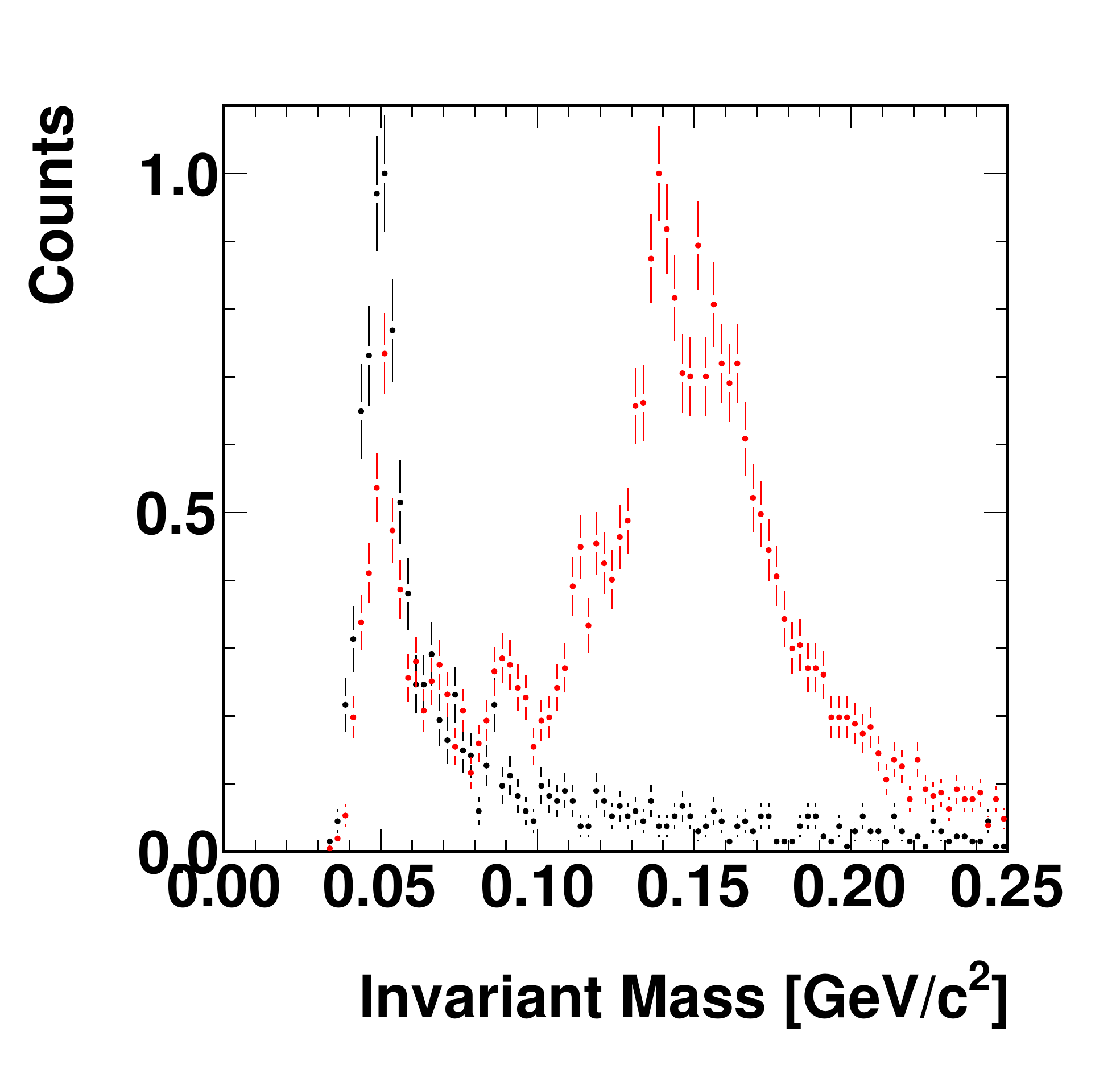}
\end{minipage}
\hspace{0.5cm}
\begin{minipage}[b]{0.5\linewidth}
\centering
\includegraphics[angle=0, width=0.95\linewidth]{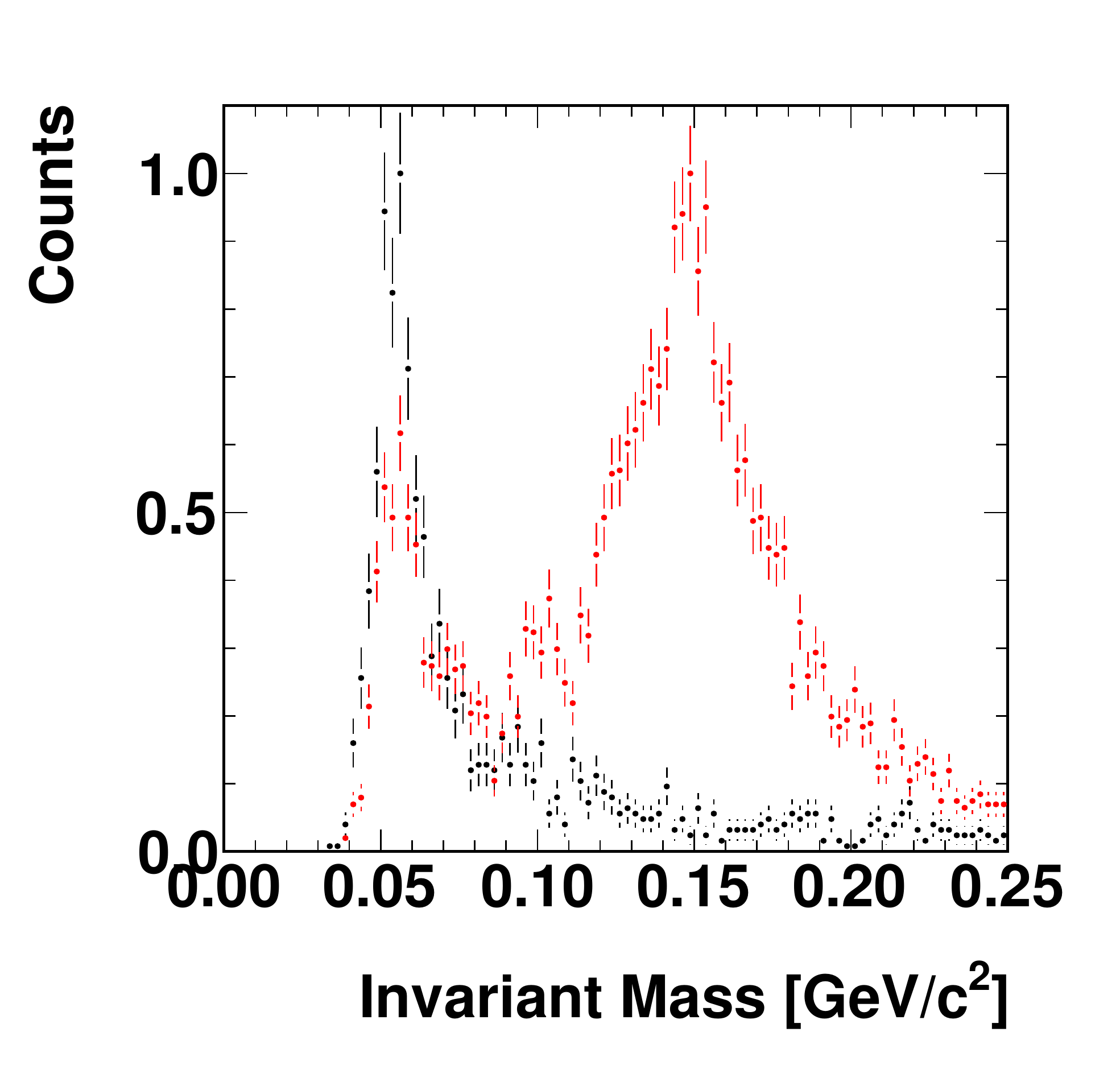}
\end{minipage}
\vspace*{-0.12in}
\caption{\label{fig:Reco50_60}  Comparison of the invariant mass distribution
for $\gamma$s (black) and $\pi^{0}$s (red) for the 50$<$E$<$55\,GeV
(55$<$E$<$60\,GeV) energy range.}
\end{figure}
\begin{figure}[hbt]
\hspace*{-0.12in}
\begin{minipage}[b]{0.5\linewidth}
\centering
\includegraphics[angle=0, width=0.95\linewidth]{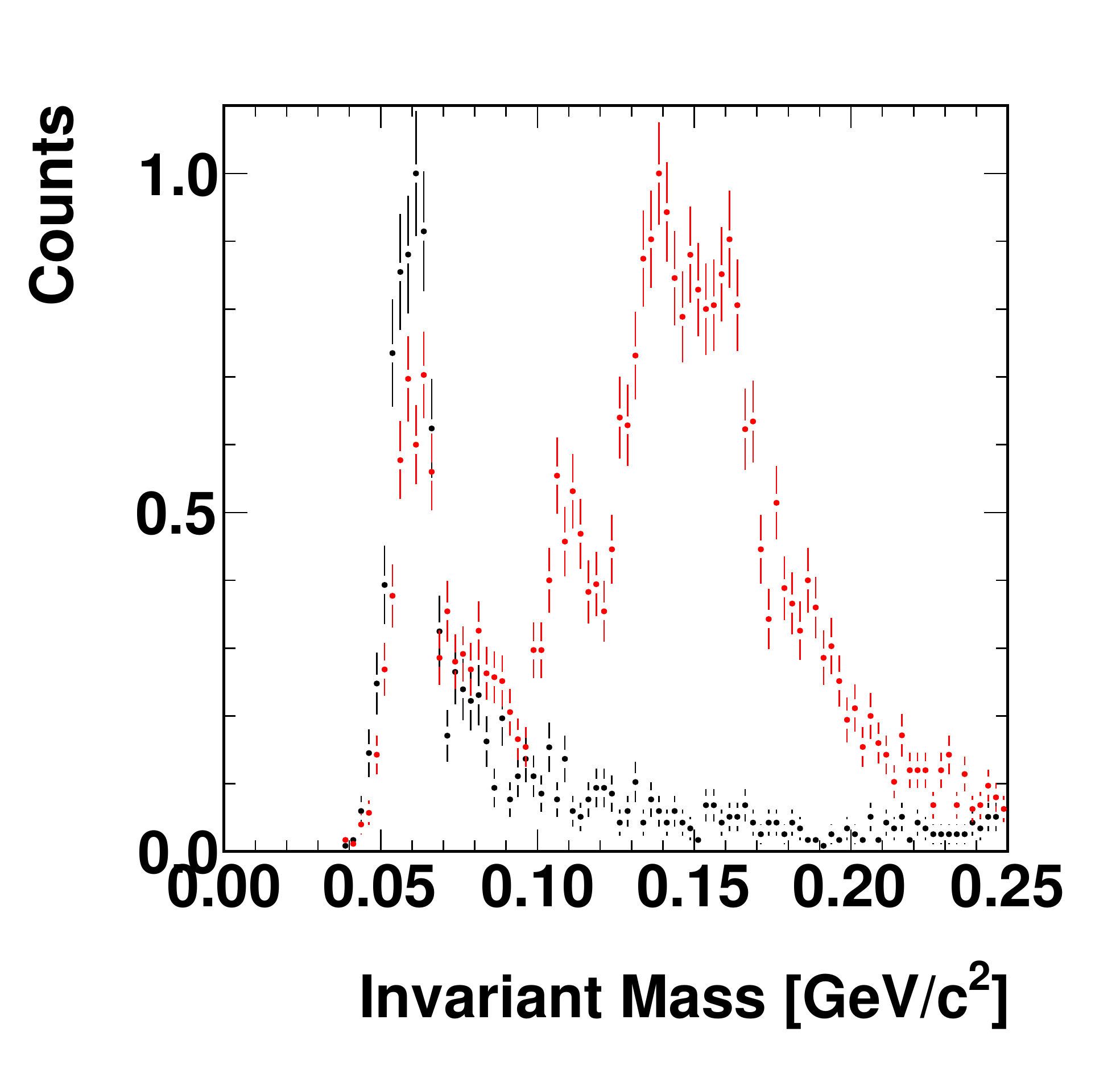}
\end{minipage}
\hspace{0.5cm}
\begin{minipage}[b]{0.5\linewidth}
\centering
\includegraphics[angle=0, width=0.95\linewidth]{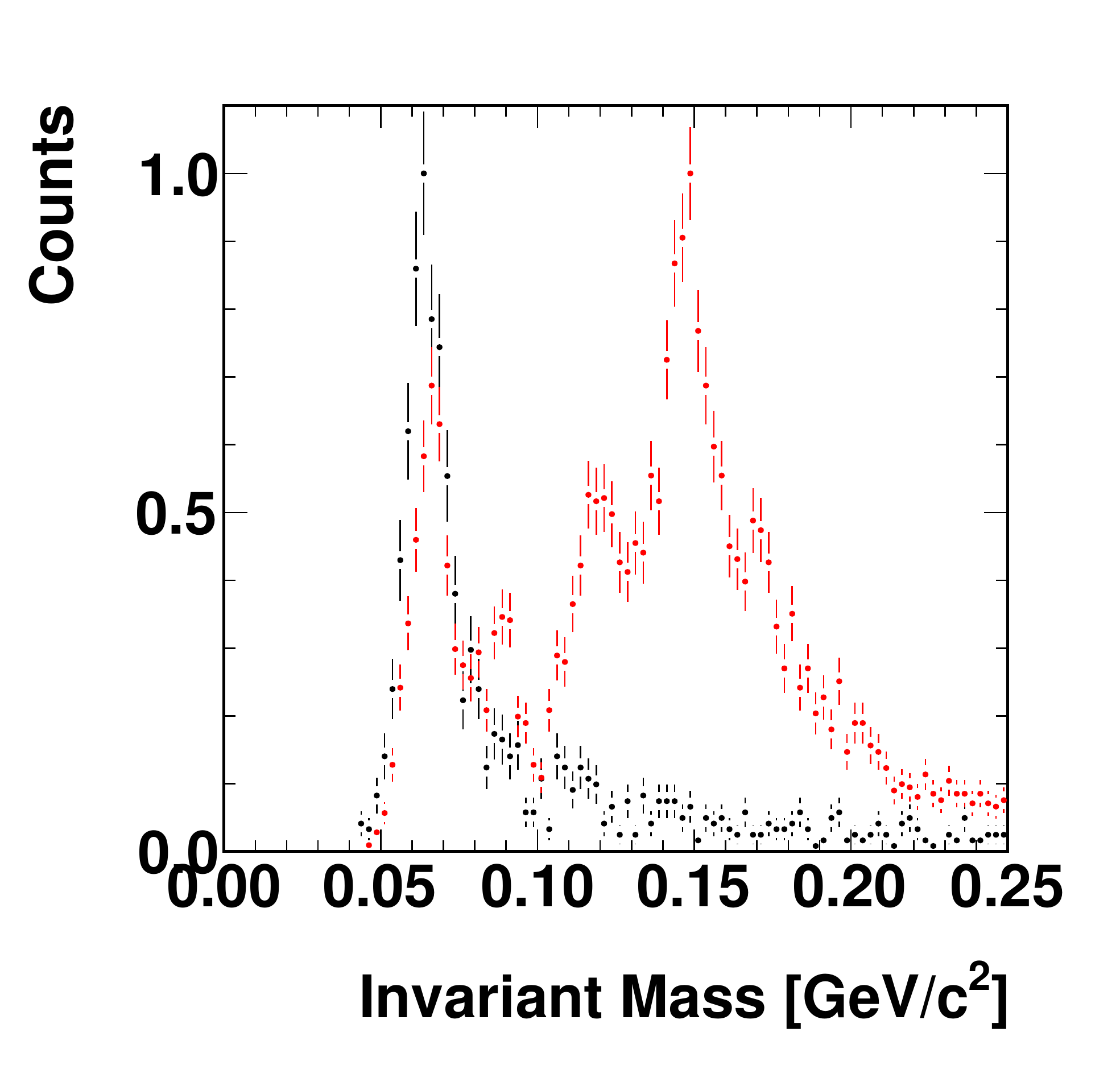}
\end{minipage}
\vspace*{-0.12in}
\caption{\label{fig:Reco60_70}  Comparison of the invariant mass distribution
for $\gamma$s (black) and $\pi^{0}$s (red) for the 60$<$E$<$65\,GeV
(65$<$E$<$70\,GeV) energy range.}
\end{figure}
\begin{figure}[hbt]
\hspace*{-0.12in}
\begin{minipage}[b]{0.5\linewidth}
\centering
\includegraphics[angle=0, width=0.95\linewidth]{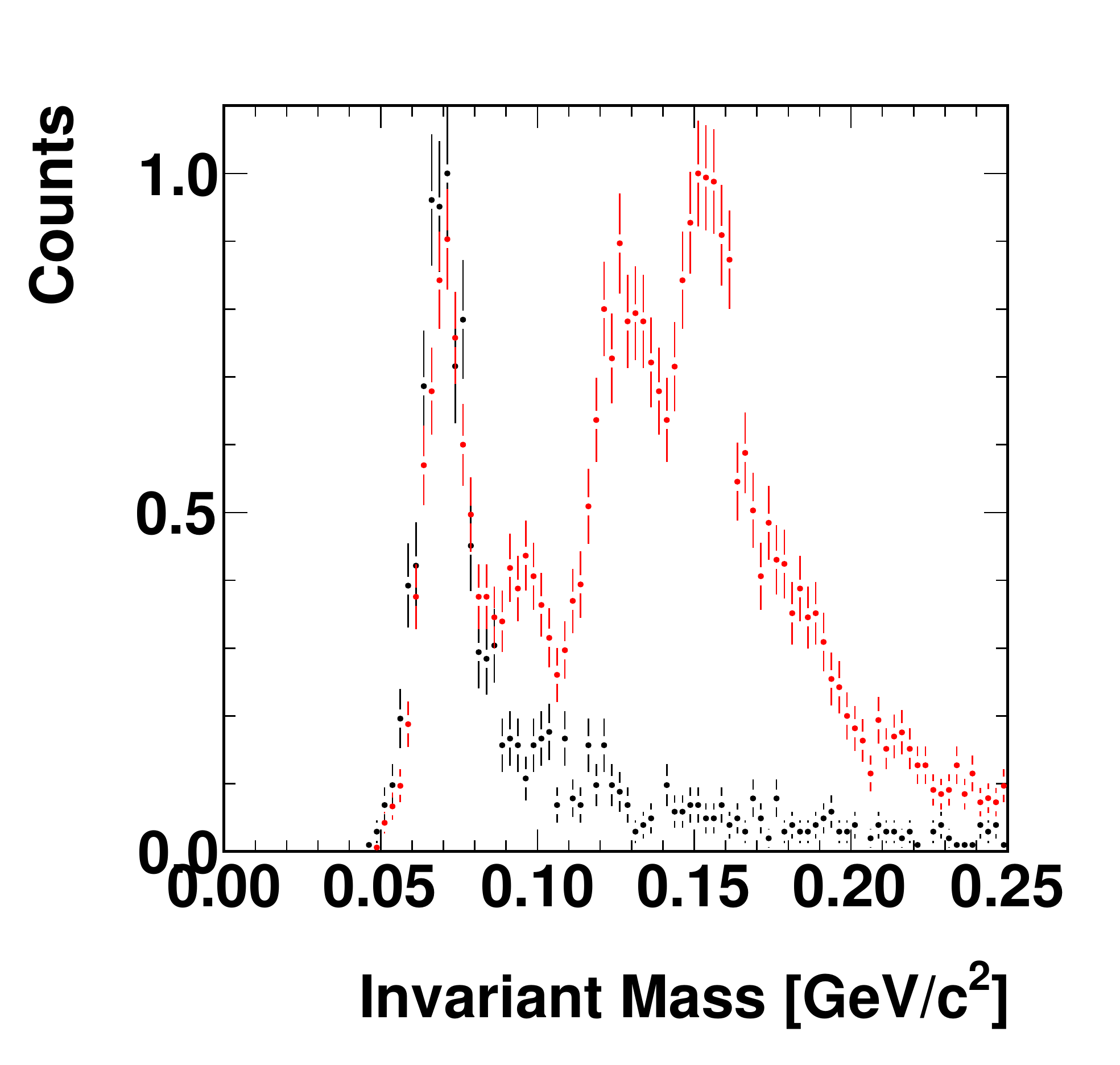}
\end{minipage}
\hspace{0.5cm}
\begin{minipage}[b]{0.5\linewidth}
\centering
\includegraphics[angle=0, width=0.95\linewidth]{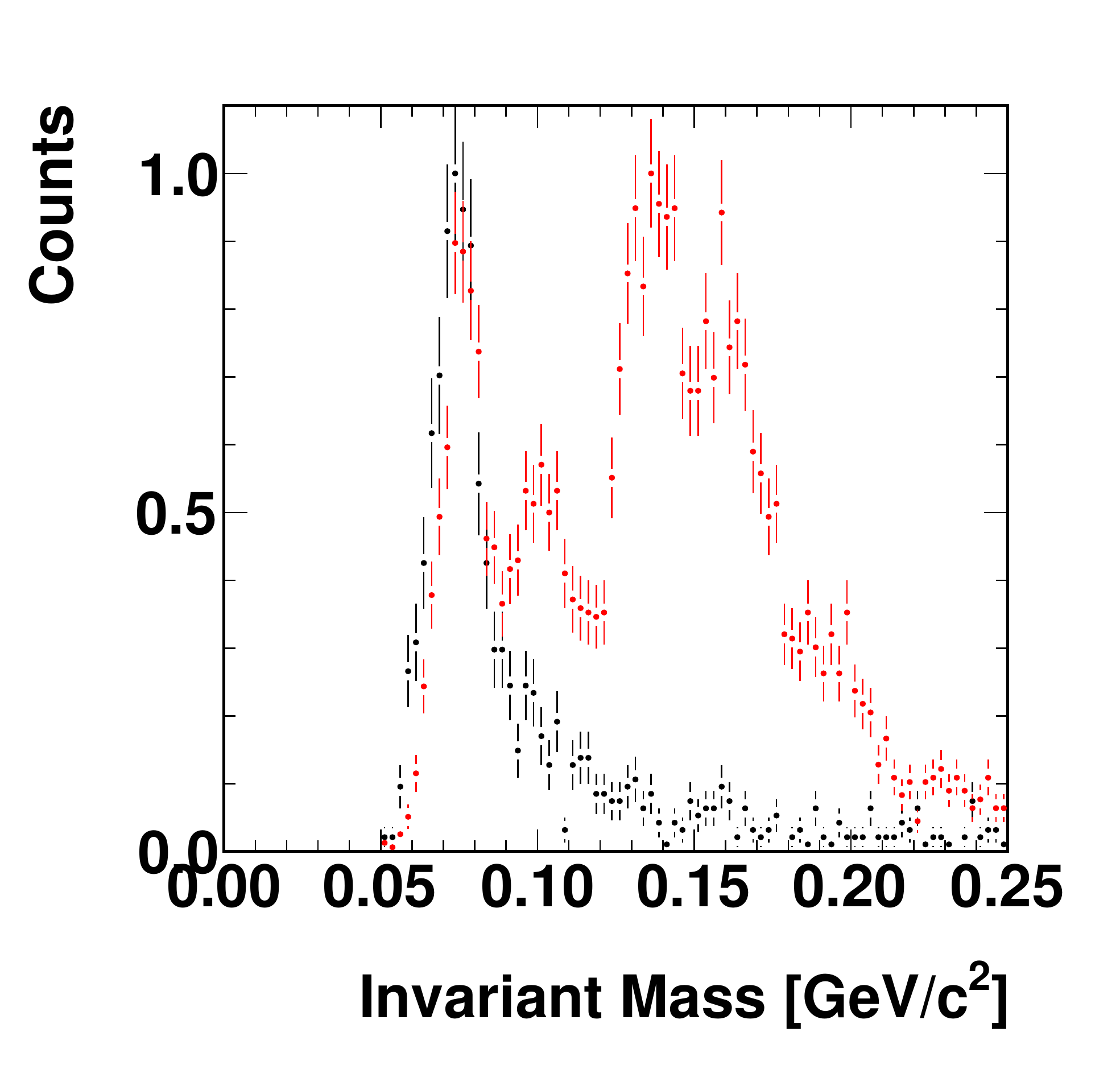}
\end{minipage}
\vspace*{-0.12in}
\caption{\label{fig:Reco70_80}  Comparison of the invariant mass distribution
for $\gamma$s (black) and $\pi^{0}$s (red) for the 70$<$E$<$75\,GeV
(75$<$E$<$80\,GeV) energy range.}
\end{figure}
\begin{figure}[hbt]
\hspace*{-0.12in}
\begin{minipage}[b]{0.5\linewidth}
\centering
\includegraphics[angle=0, width=0.95\linewidth]{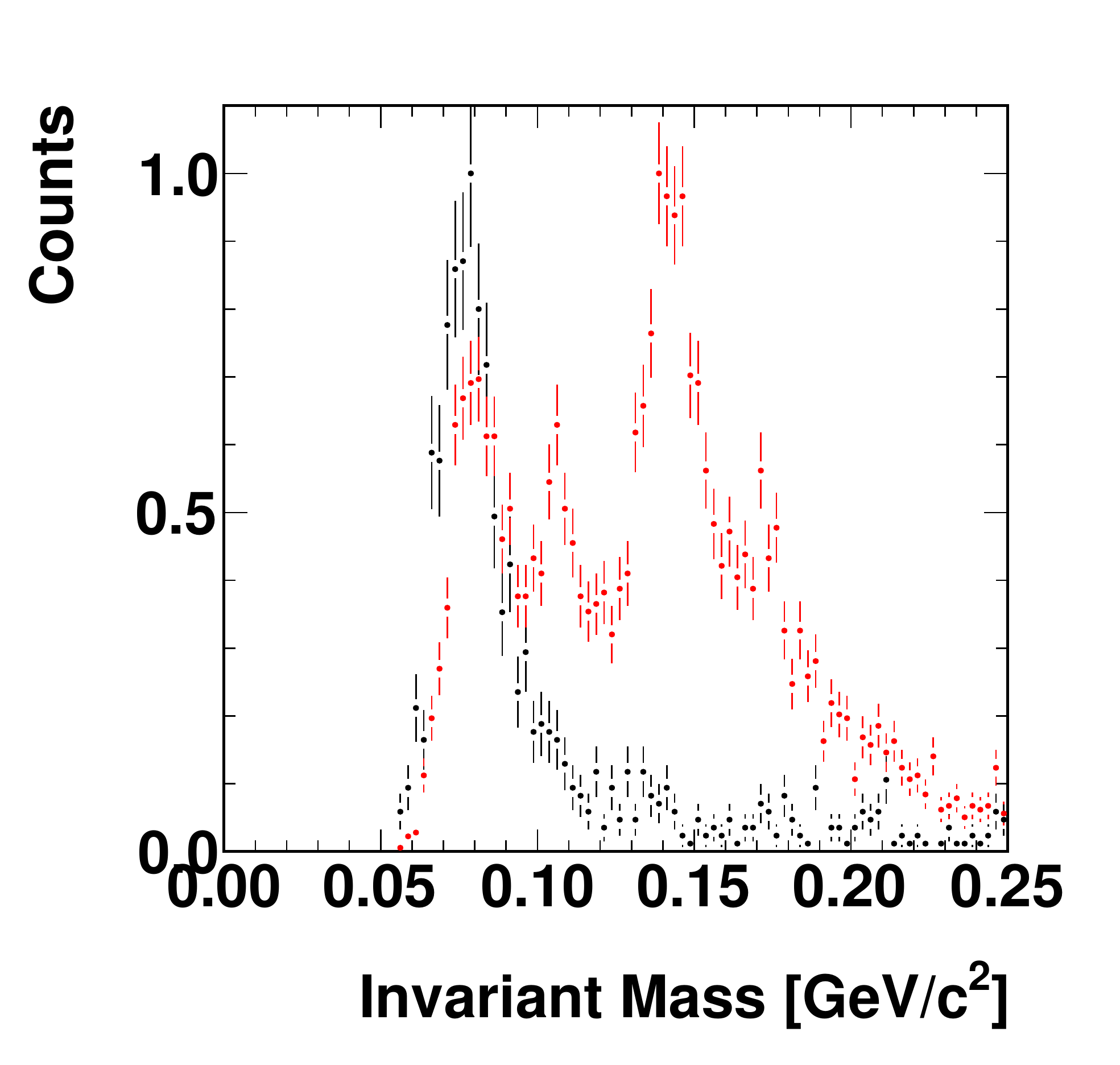}
\end{minipage}
\hspace{0.5cm}
\begin{minipage}[b]{0.5\linewidth}
\centering
\includegraphics[angle=0, width=0.95\linewidth]{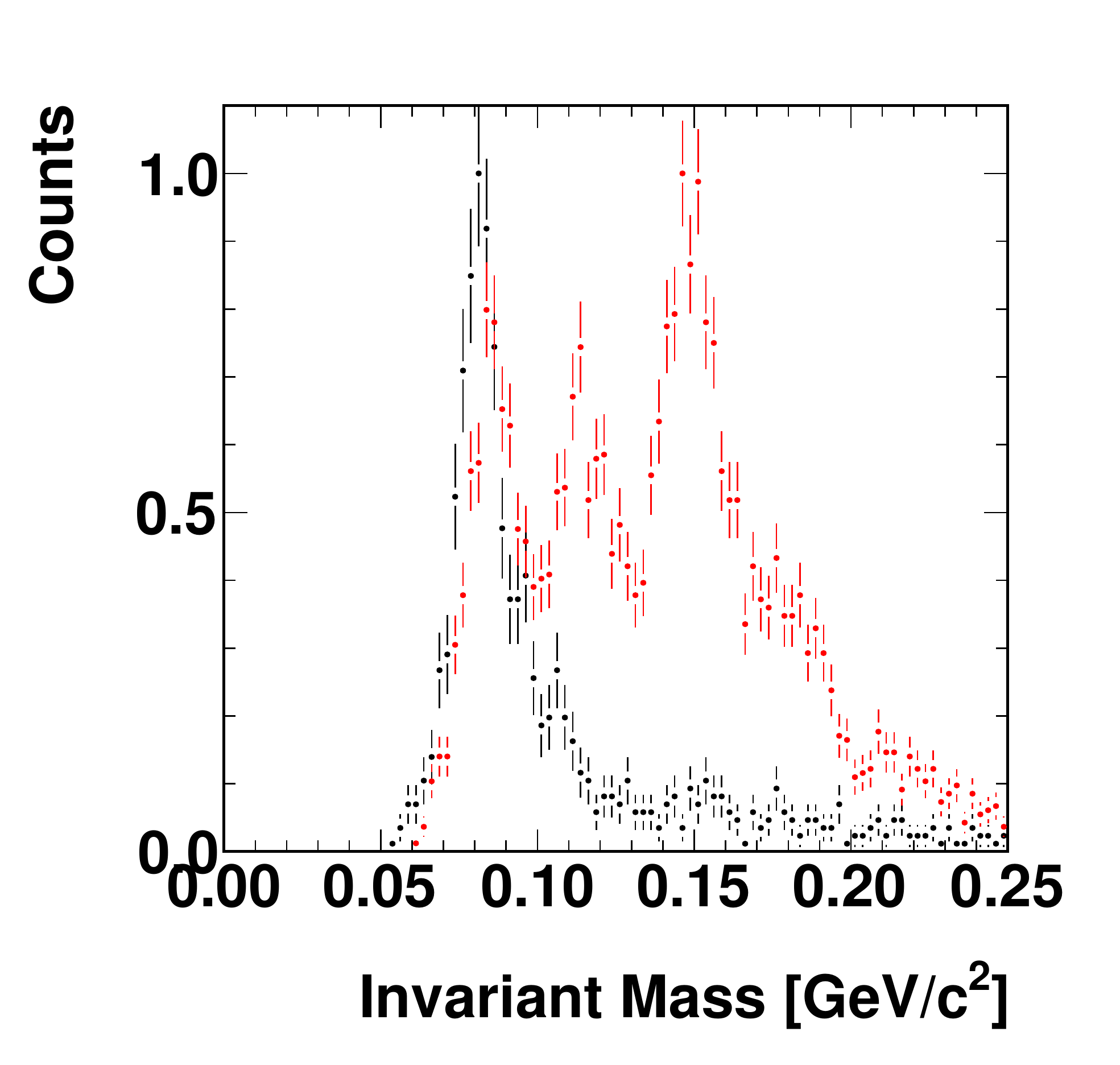}
\end{minipage}
\vspace*{-0.12in}
\caption{\label{fig:Reco80_90}  Comparison of the invariant mass distribution
for $\gamma$s (black) and $\pi^{0}$s (red) for the 80$<$E$<$85\,GeV
(85$<$E$<$90\,GeV) energy range.}
\end{figure}

Figure~\ref{fig:RecoAllEnergies} shows the reconstructed invariant mass
versus the reconstructed energy for single input $\gamma$s (left) and
$\pi^{0}$s (right).  A cut in the reconstruction at $E$\,=\,20\,GeV is
applied, as below this energy two distinct tracks are typically found in the
detector.  A more detailed view of this
can be found in Figs.~\ref{fig:Reco20_30}~to~\ref{fig:Reco80_90}, where
slices in reconstructed energy are made in $\Delta E$\,=\,5\,GeV bins.
Figure~\ref{fig:Reco20_30} (left) shows the 20$<$$E_{\rm Rec}$$<$25\,GeV bin.
The red histogram shows single-$\pi^{0}$s, where a small correct-mass
peak is observed, black shows single-$\gamma$s reconstructed at the 
same energy.  A low-mass tail is observed for
single-$\pi^{0}$s which is due to the reconstruction algorithm picking
up a fluctuation in energy from a single shower and thus reconstructing
a mass based from one $\gamma$.  The normalization of the two histograms
was chosen to fix the maximum height to be the same. 
Figure~\ref{fig:Reco20_30} (right) shows the 25$<$$E_{\rm Rec}$$<$30\,GeV
bin.  A more prominant correct-mass peak is observed.  At higher
reconstructed energies, the correct-mass peak becomes dominant and the
low-mass peak shrinks (relatively).  The overall low mass peak distribution,
however, still retains the same shape as that from single-$\gamma$s.

\clearpage

\subsubsection{Single-Particle Reconstruction Efficiencies}

\begin{table}
\centering
\caption{Single-Particle Reconstruction Efficiency.  The first column are
the number of MC particles thrown into MPC-EX with the stated energy.
The second column are the number in the reconstruction which passed all
the cuts and have a reconstructed mass, note that there is no restriction
on the incoming particle energy applied, so in- and out-flow are possible.
The percentages in this column are relative to the input tracks.  The next
columns are the number of tracks at low/high reconstructed mass
the percentages are ``per reconstructed track'', i.e. ``passed''.}
\label{tbl:SingleParticleRecoI006}
\begin{tabular}{|l|c|c|c||c|c|}
\hline
Energy Range & & Thrown & Passed & 0$<$ Inv. Mass $<$0.06 & Inv. Mass $>$0.06 \\
\hline
\multirow{4}{*}{$>$20\,GeV}  & \multirow{2}{*}{$\gamma$} & 147527 & 25071 & 9902 & 15169 \\
 & & 100\% & 16.9\% & 39.4\% & 60.5\% \\\cline{2-6}
 & \multirow{2}{*}{$\pi^{0}$} & 148927 & 92298 & 8080 & 84218 \\
 & & 100\% & 61.9\% & 8.7\% & 91.2\% \\\cline{2-6}
\hline
\multirow{4}{*}{20-25\,GeV}  & \multirow{2}{*}{$\gamma$} & 9329 & 2611 & 1877 & 734 \\
 & & 100\% & 27.9\% & 71.8\% & 28.1\% \\\cline{2-6}
 & \multirow{2}{*}{$\pi^{0}$} & 9247 & 5037 & 1456 & 3581 \\
 & & 100\% & 54.4\% & 28.9\% & 71\% \\\cline{2-6}
\hline
\multirow{4}{*}{25-30\,GeV}  & \multirow{2}{*}{$\gamma$} & 9268 & 2322 & 1623 & 699 \\
 & & 100\% & 25\% & 69.8\% & 30.1\% \\\cline{2-6}
 & \multirow{2}{*}{$\pi^{0}$} & 9123 & 5615 & 1126 & 4489 \\
 & & 100\% & 61.5\% & 20\% & 79.9\% \\\cline{2-6}
\hline
\multirow{4}{*}{30-35\,GeV}  & \multirow{2}{*}{$\gamma$} & 9147 & 2074 & 1351 & 723 \\
 & & 100\% & 22.6\% & 65.1\% & 34.8\% \\\cline{2-6}
 & \multirow{2}{*}{$\pi^{0}$} & 9234 & 6084 & 976 & 5108 \\
 & & 100\% & 65.8\% & 16\% & 83.9\% \\\cline{2-6}
\hline
\multirow{4}{*}{35-40\,GeV}  & \multirow{2}{*}{$\gamma$} & 9200 & 1847 & 1126 & 721 \\
 & & 100\% & 20\% & 60.9\% & 39\% \\\cline{2-6}
 & \multirow{2}{*}{$\pi^{0}$} & 9338 & 6332 & 941 & 5391 \\
 & & 100\% & 67.8\% & 14.8\% & 85.1\% \\\cline{2-6}
\hline
\multirow{4}{*}{40-45\,GeV}  & \multirow{2}{*}{$\gamma$} & 9204 & 1727 & 975 & 752 \\
 & & 100\% & 18.7\% & 56.4\% & 43.5\% \\\cline{2-6}
 & \multirow{2}{*}{$\pi^{0}$} & 9318 & 6476 & 858 & 5618 \\
 & & 100\% & 69.4\% & 13.2\% & 86.7\% \\\cline{2-6}
\hline
\multirow{4}{*}{45-50\,GeV}  & \multirow{2}{*}{$\gamma$} & 9210 & 1618 & 862 & 756 \\
 & & 100\% & 17.5\% & 53.2\% & 46.7\% \\\cline{2-6}
 & \multirow{2}{*}{$\pi^{0}$} & 9385 & 6331 & 789 & 5542 \\
 & & 100\% & 67.4\% & 12.4\% & 87.5\% \\\cline{2-6}
\hline
\end{tabular}
\end{table}

\begin{table}
\centering
\caption{Continuation of Tbl.~\ref{tbl:SingleParticleRecoI006}.}
\label{tbl:SingleParticleRecoII006}
\begin{tabular}{|l|c|c|c||c|c|}
\hline
Energy Range & & Thrown & Passed & 0$<$ Inv. Mass $<$0.06 & Inv. Mass $>$0.06 \\
\hline
\multirow{4}{*}{50-55\,GeV}  & \multirow{2}{*}{$\gamma$} & 9148 & 1495 & 751 & 744 \\
 & & 100\% & 16.3\% & 50.2\% & 49.7\% \\\cline{2-6}
 & \multirow{2}{*}{$\pi^{0}$} & 9348 & 6292 & 706 & 5586 \\
 & & 100\% & 67.3\% & 11.2\% & 88.7\% \\\cline{2-6}
\hline
\multirow{4}{*}{55-60\,GeV}  & \multirow{2}{*}{$\gamma$} & 9164 & 1535 & 612 & 923 \\
 & & 100\% & 16.7\% & 39.8\% & 60.1\% \\\cline{2-6}
 & \multirow{2}{*}{$\pi^{0}$} & 9238 & 6258 & 590 & 5668 \\
 & & 100\% & 67.7\% & 9.4\% & 90.5\% \\\cline{2-6}
\hline
\multirow{4}{*}{60-65\,GeV}  & \multirow{2}{*}{$\gamma$} & 9441 & 1453 & 391 & 1062 \\
 & & 100\% & 15.3\% & 26.9\% & 73\% \\\cline{2-6}
 & \multirow{2}{*}{$\pi^{0}$} & 9197 & 6068 & 383 & 5685 \\
 & & 100\% & 65.9\% & 6.3\% & 93.6\% \\\cline{2-6}
\hline
\multirow{4}{*}{65-70\,GeV}  & \multirow{2}{*}{$\gamma$} & 9103 & 1399 & 192 & 1207 \\
 & & 100\% & 15.3\% & 13.7\% & 86.2\% \\\cline{2-6}
 & \multirow{2}{*}{$\pi^{0}$} & 9312 & 5923 & 169 & 5754 \\
 & & 100\% & 63.6\% & 2.8\% & 97.1\% \\\cline{2-6}
\hline
\multirow{4}{*}{70-75\,GeV}  & \multirow{2}{*}{$\gamma$} & 9166 & 1355 & 81 & 1274 \\
 & & 100\% & 14.7\% & 5.9\% & 94\% \\\cline{2-6}
 & \multirow{2}{*}{$\pi^{0}$} & 9375 & 6007 & 66 & 5941 \\
 & & 100\% & 64\% & 1\% & 98.9\% \\\cline{2-6}
\hline
\multirow{4}{*}{75-80\,GeV}  & \multirow{2}{*}{$\gamma$} & 9142 & 1290 & 38 & 1252 \\
 & & 100\% & 14.1\% & 2.9\% & 97\% \\\cline{2-6}
 & \multirow{2}{*}{$\pi^{0}$} & 9423 & 5677 & 15 & 5662 \\
 & & 100\% & 60.2\% & 0.2\% & 99.7\% \\\cline{2-6}
\hline
\multirow{4}{*}{80-85\,GeV}  & \multirow{2}{*}{$\gamma$} & 9270 & 1222 & 13 & 1209 \\
 & & 100\% & 13.1\% & 1\% & 98.9\% \\\cline{2-6}
 & \multirow{2}{*}{$\pi^{0}$} & 9415 & 5470 & 5 & 5465 \\
 & & 100\% & 58\% & 0\% & 99.9\% \\\cline{2-6}
\hline
\multirow{4}{*}{85-90\,GeV}  & \multirow{2}{*}{$\gamma$} & 9320 & 1144 & 10 & 1134 \\
 & & 100\% & 12.2\% & 0.8\% & 99.1\% \\\cline{2-6}
 & \multirow{2}{*}{$\pi^{0}$} & 9305 & 5141 & 0 & 5141 \\
 & & 100\% & 55.2\% & 0\% & 100\% \\\cline{2-6}
\hline
\end{tabular}
\end{table}

\clearpage

\subsection{Shower Shape}

Candidate tracks in the MPC-EX are compared to a shower library in order to provide an additional 
discrimination between true photon showers, and sources of background (such as high
energy $\pi^{0}$s). A shower library was developed from a sample of single-particle photon showers
in the MPC-EX, and indexed by energy and by the layer of first interaction in the MPC-EX. This depth information
is important because the MPC-EX is seeing the electromagntic shower in its infancy, and a shower that
developed late in the MPC-EX will have a very different profile than a shower that starts in the 
first layers of the MPC-EX, and it is in this way that depth information from the MPC-EX is
incorporated into the reconstruction. The shower profiles are kept separately for the $x$ and $y$ oriented minipads.

MPC-EX tracks are compared to the shower profiles using the Kolmogorov test \cite{ktest}, which is a 
statistical method for determining if two distribution are consistent with being drawn from the same 
underlying probability distribution. The Kolmogorov test returns this information as a ``distance'' value that 
characterizes the comparison - a large distance indicates a smaller probability that two two distributions are compatible. 

\begin{figure}
\hspace*{-0.12in}
\begin{minipage}[b]{0.5\linewidth}
\centering
\includegraphics[angle=0, width=0.95\linewidth]{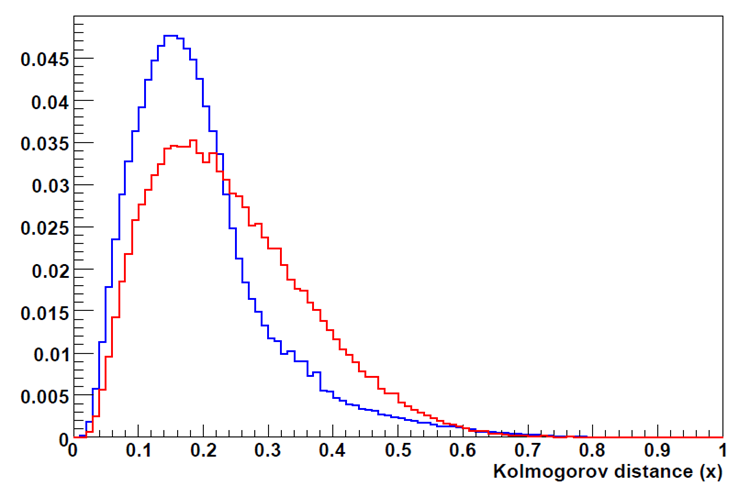}
\end{minipage}
\hspace{0.5cm}
\begin{minipage}[b]{0.5\linewidth}
\centering
\includegraphics[angle=0, width=0.95\linewidth]{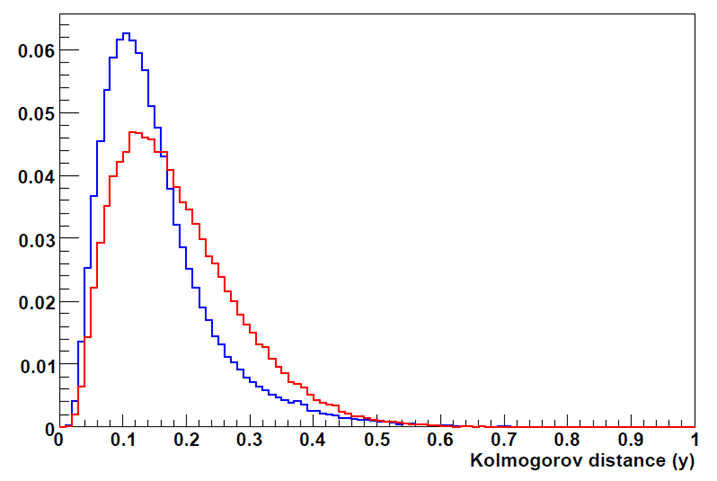}
\end{minipage}
\vspace*{-0.12in}
\caption{\label{fig:KTest} The Kolmogorov distance distributions for $\pi^{0}$s (red) and photons (in blue). Both distributions
are normalized to the same area (number of events). The difference in two projections is driven by the fact that the $y$ projection
minipads are always after the $x$ projection minipads.}
\end{figure}

Every MPC-EX track with $E>$20\,GeV is compared to the reference shower library for the initial shower depth and 
reconstructed energy, and the corresponding Kolmogorov distances are recorded with the track object. Figure~\ref{fig:KTest} 
shows the Kolmogorov distances for the $x$ and $y$ minipads for a sample of single $\pi^{0}$ and photons
showers in the MPC-EX. It can be seen that the shower shape comparison can provide some discrimination between 
photon and high energy $\pi^{0}$ showers, especially when the distance in both projections are combined (see the 
direct photon analysis, Section~\ref{sec:pi0_back}). Of course, the power of a shape cut is limited in the MPC-EX 
because it only sees a fraction of the electromagnetic shower development.  Additional cuts on the MPC shower shape
are also used in the direct photon analysis (see Section~\ref{sec:pi0_back}). 

\subsection{$\chi^{2}$ Calibration}

The $\chi^{2}$ calibration is an important piece of the electromagnetic
shower reconstruction.  The purpose is to remove contamination from
charged hadrons, which may masquerade as electromagnetic showers.  Electromagnetic showers are
distinct as, for a given energy, the shower profile is somewhat
predictable.  For charged hadrons, the shower profile is different,
sometimes a shower develops, sometimes only MIPs are seen in the silicon. Here we 
are primarily concerned with removing charged hadrons that begin showering 
in the MPC-EX. 

\begin{figure}[hbt]
\hspace*{-0.12in}
\centering
\includegraphics[angle=0, width=0.99\linewidth]{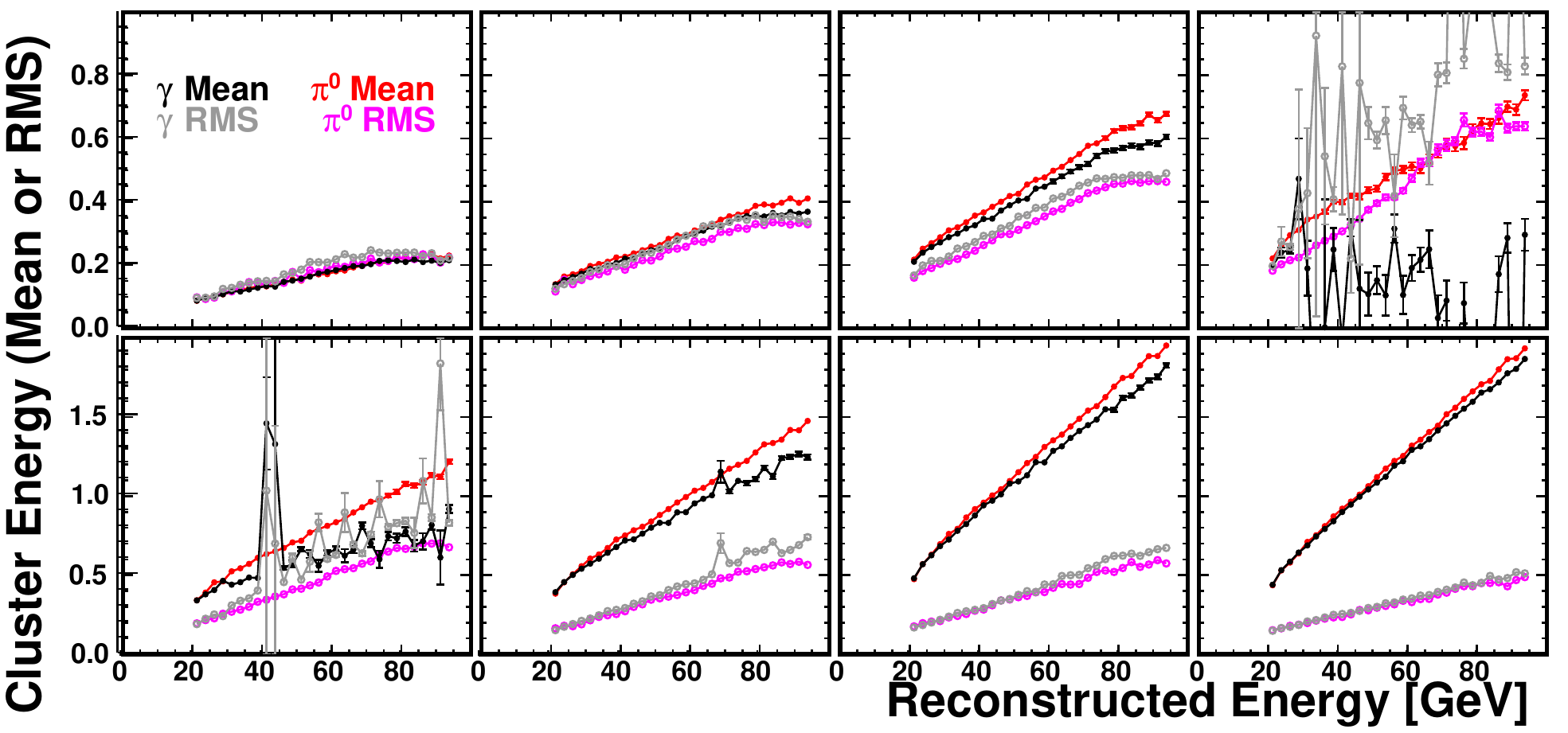}
\vspace*{-0.12in}
\caption{\label{fig:EnChi2} Calibration of the energy deposition profile
for a reconstructed track.  The calibration is performed using
single-$\gamma$s and single-$\pi^{0}$s.  The top row (left to right) show
the layers 0 through 3, the bottom row 4 through 7.  The black/grey symbols
show the mean/RMS of the single-$\gamma$ distribution.  The red/purple symbols
show the mean/RMS of the single-$\pi^{0}$ distribution.}
\end{figure}

\begin{figure}[hbt]
\hspace*{-0.12in}
\includegraphics[angle=0, width=0.99\linewidth]{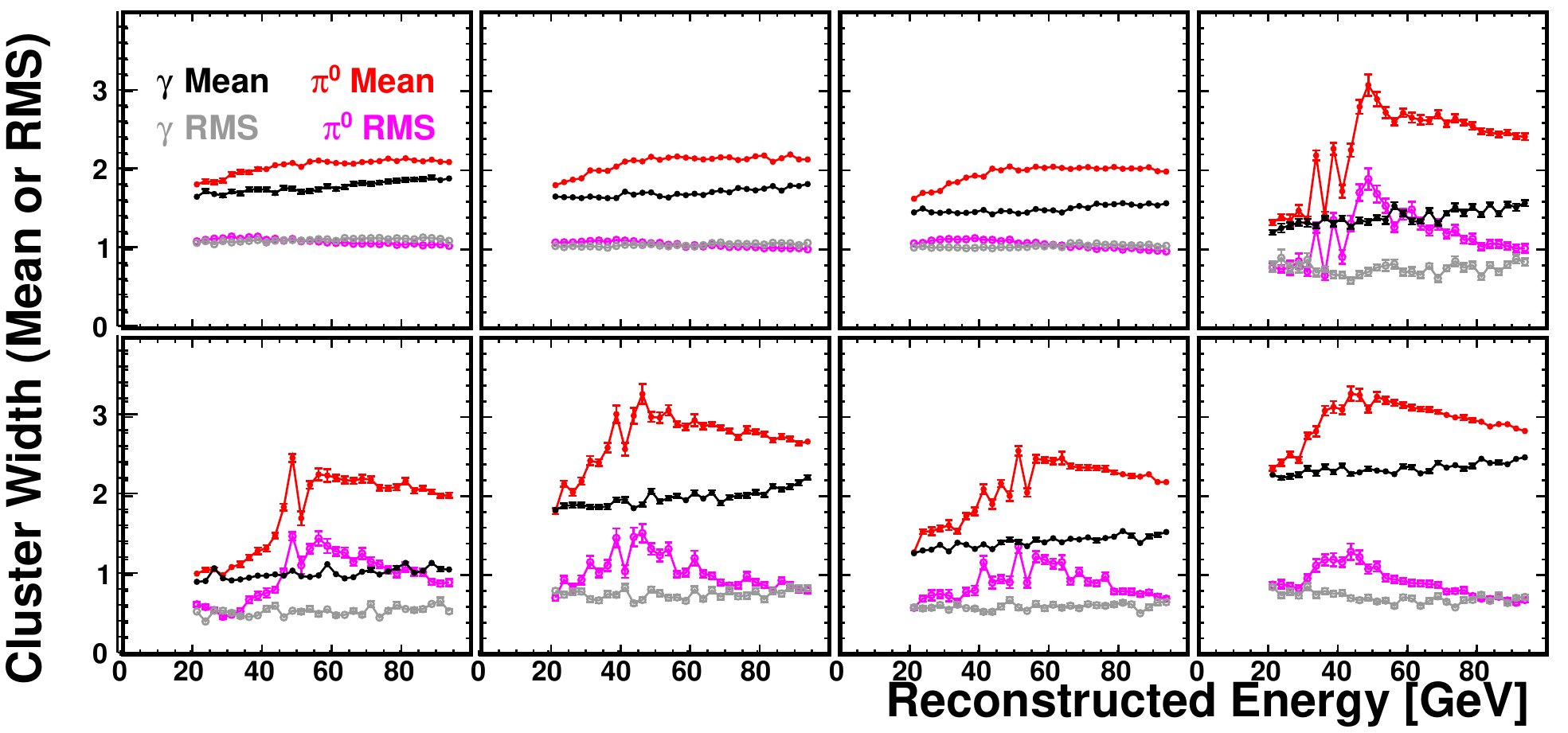}
\vspace*{-0.12in}
\caption{\label{fig:WidthChi2} Calibration of the transverse profile
at each layer for a reconstructed track.  The calibration is performed using
single-$\gamma$s and single-$\pi^{0}$s.  The top row (left to right) show
the layers 0 through 3, the bottom row 4 through 7.  The black/grey symbols
show the mean/RMS of the single-$\gamma$ distribution.  The red/purple symbols
show the mean/RMS of the single-$\pi^{0}$ distribution.}
\end{figure}

A $\chi^{2}$ is formed to characterize the shower profile at each layer 
in the silicon with the aim to reduce the charged hadron contamination.
Both the energy deposited and the transverse profile of the shower in
each layer are measured for many events, then the mean and RMS of those
distributions are found as references.  
Figures~\ref{fig:EnChi2}~and~\ref{fig:WidthChi2} show the mean and RMS of the 
calibration, versus collision energy, for the energy deposition profile
and transverse profile respectively.  The calibration is performed twice,
once for single-$\gamma$s, and again as a cross-check with
single-$\pi^{0}$s.  The energy deposition, as expected, is very similar for
the two cases, as both are electromagnetic showers.  However, the width is distinct,
as the $\pi^{0}$ showers are from two separated $\gamma$s.  For higher
reconstructed energies, those distributions start to approach each other,
as the angular separation diminishes.

\begin{figure}[hbt]
\hspace*{-0.12in}
\includegraphics[angle=0, width=0.24\linewidth]{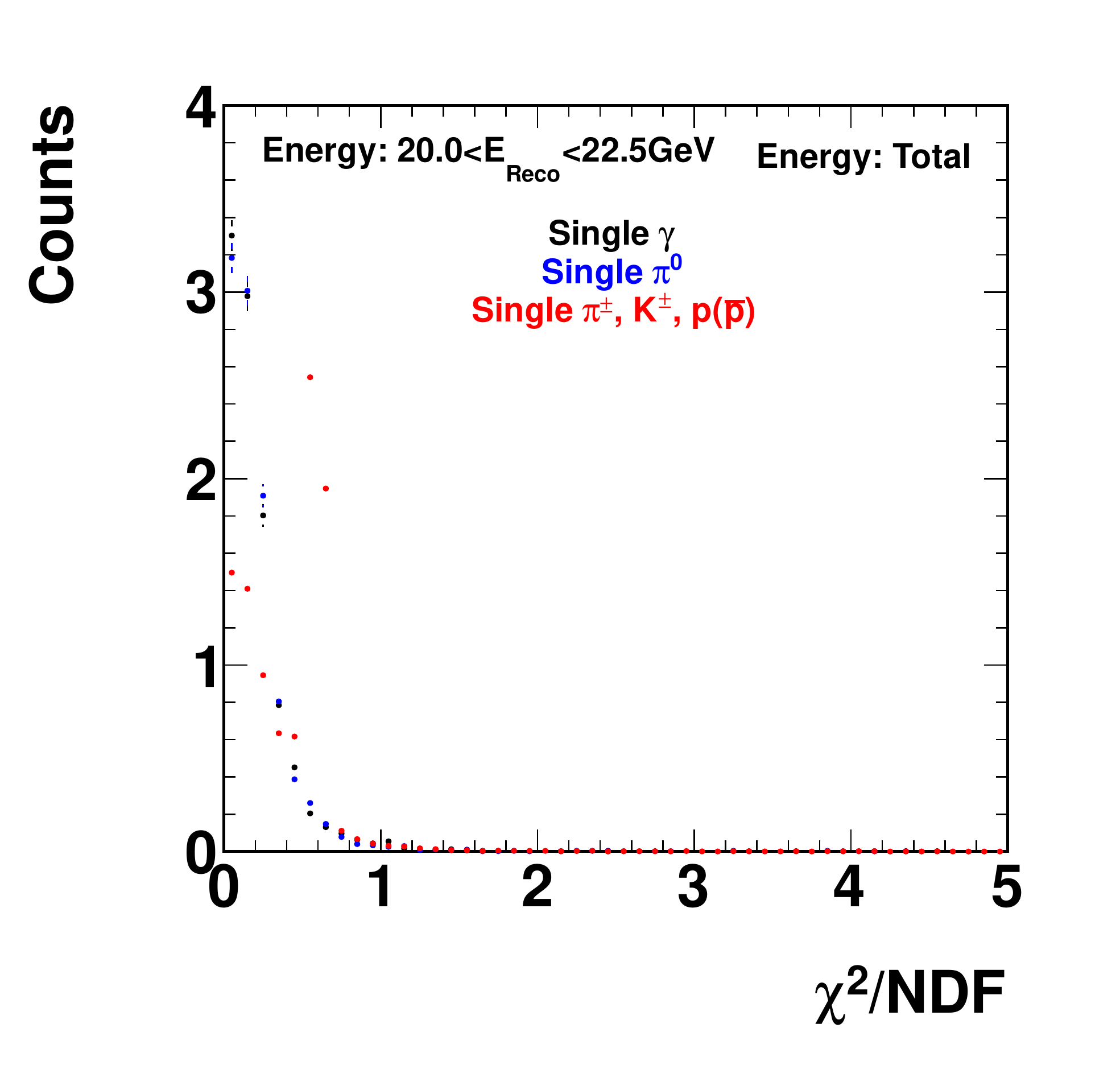}
\includegraphics[angle=0, width=0.24\linewidth]{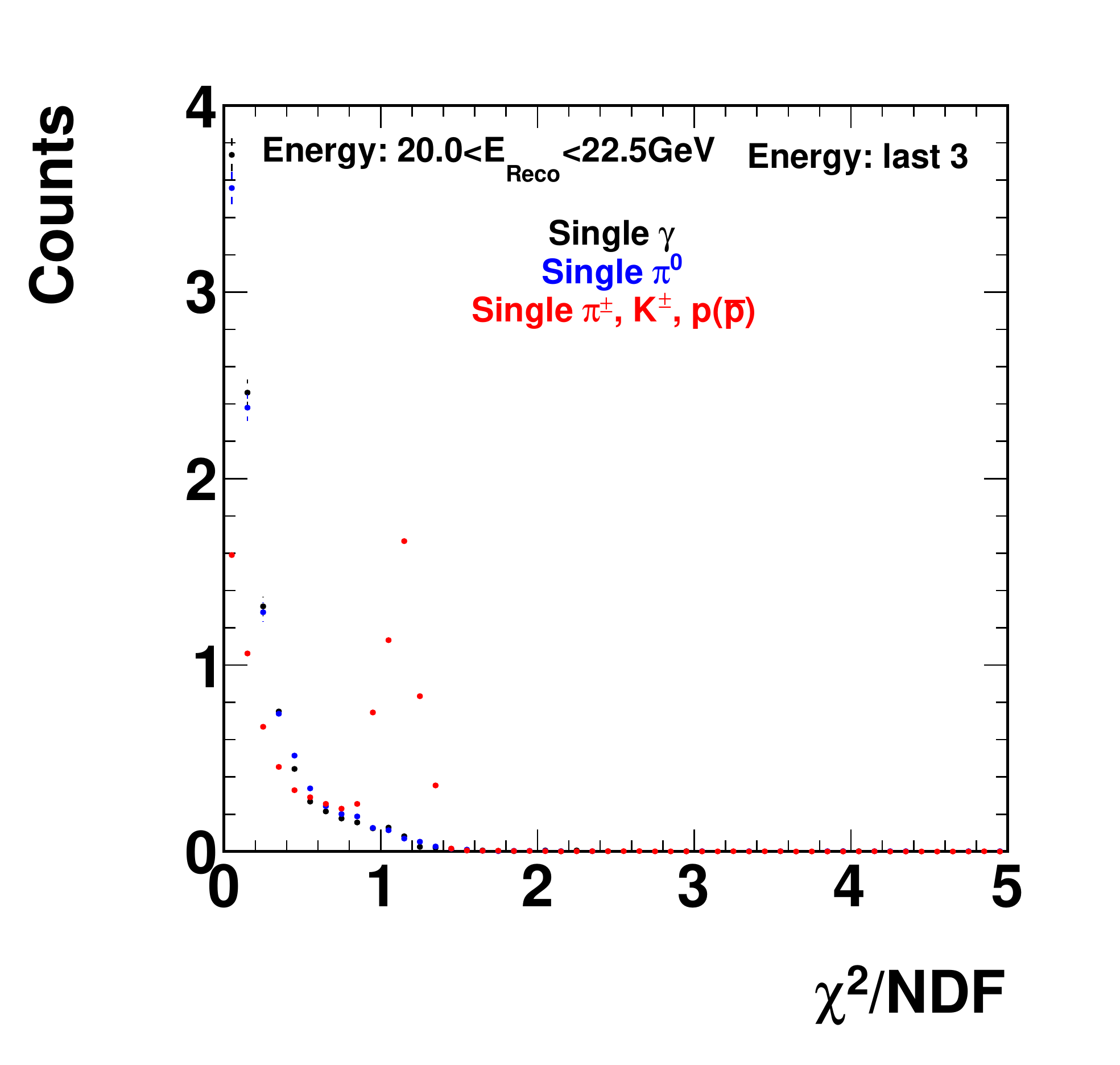}
\includegraphics[angle=0, width=0.24\linewidth]{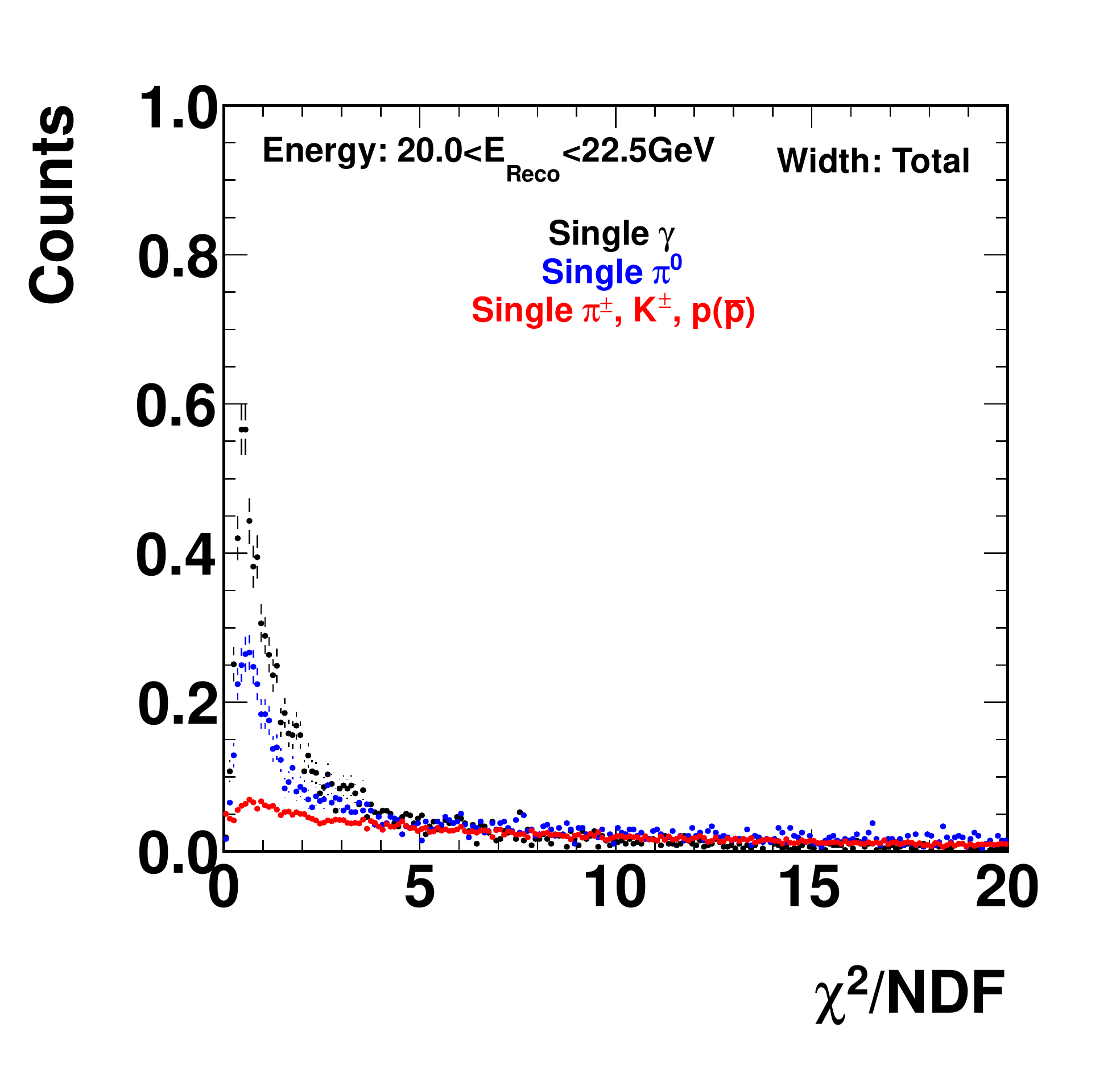}
\includegraphics[angle=0, width=0.24\linewidth]{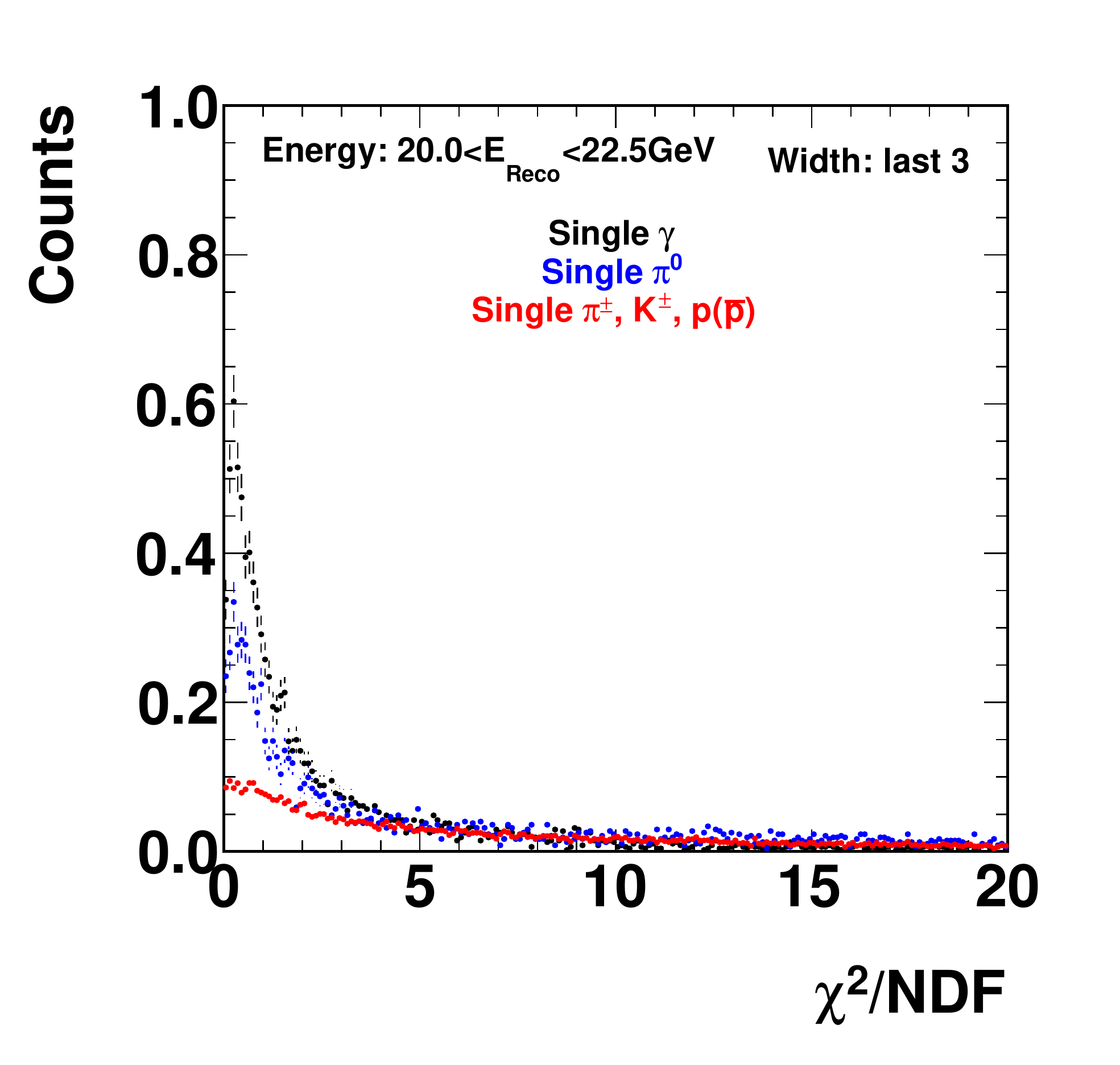}
\vspace*{-0.12in}
\caption{\label{fig:Chi2LayerCalib_20GeV} $\chi^{2}$ calibration applied
to single-$\gamma$s (black), single-$\pi^{0}$s (red), and single-charged
hadrons (mixed $\pi^{\pm}$, $K^{\pm}$, and $p(\overline{p})$), with 
reconstructed energy of 20$<$$E_{\rm Rec}$$<$22.5\,GeV.  Left to
right, the panels show the calibration using energy (all layers), energy
(last 3 only), width (all layers), and width (last 3 only). }
\end{figure}

\begin{figure}[hbt]
\hspace*{-0.12in}
\includegraphics[angle=0, width=0.24\linewidth]{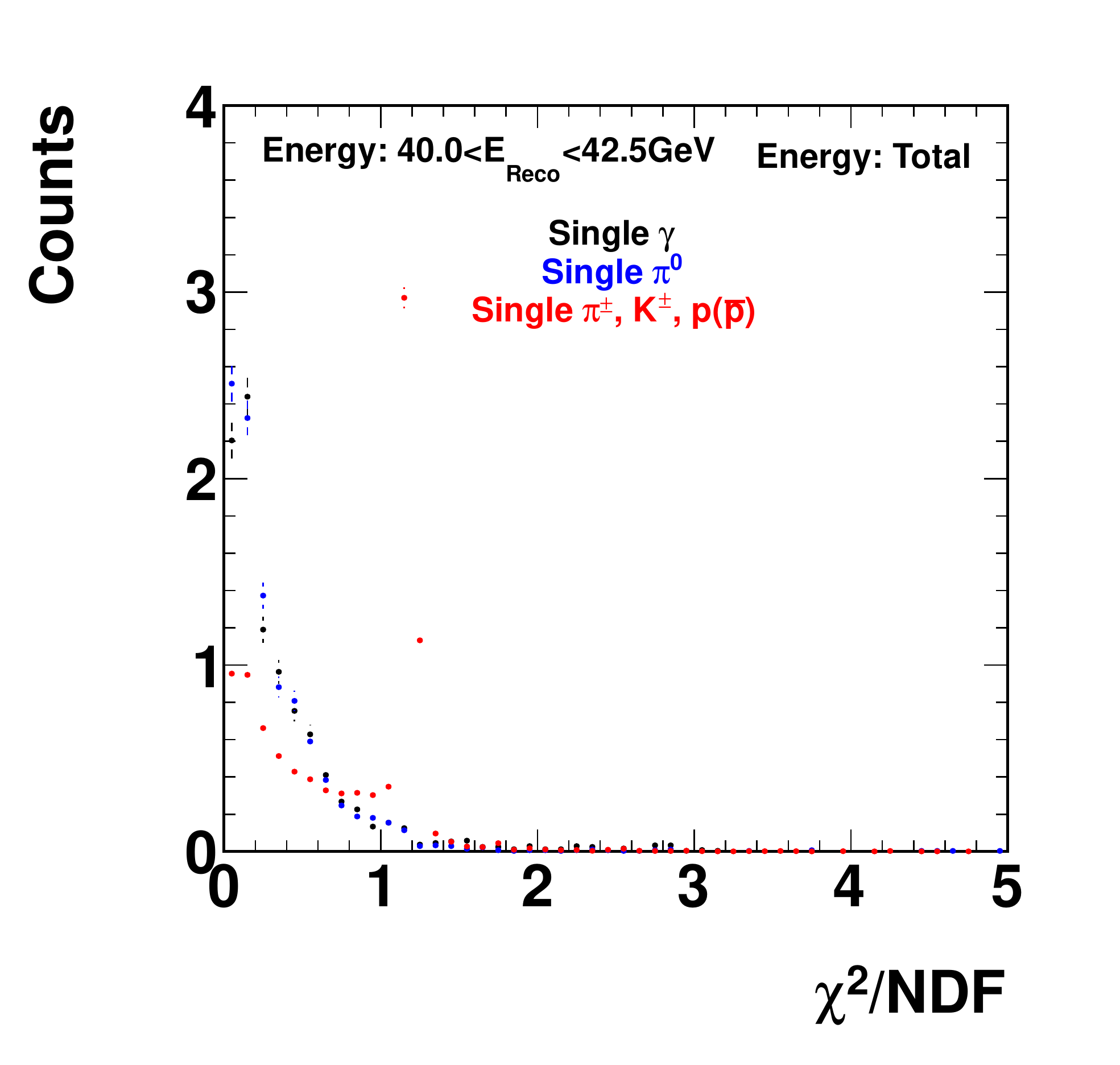}
\includegraphics[angle=0, width=0.24\linewidth]{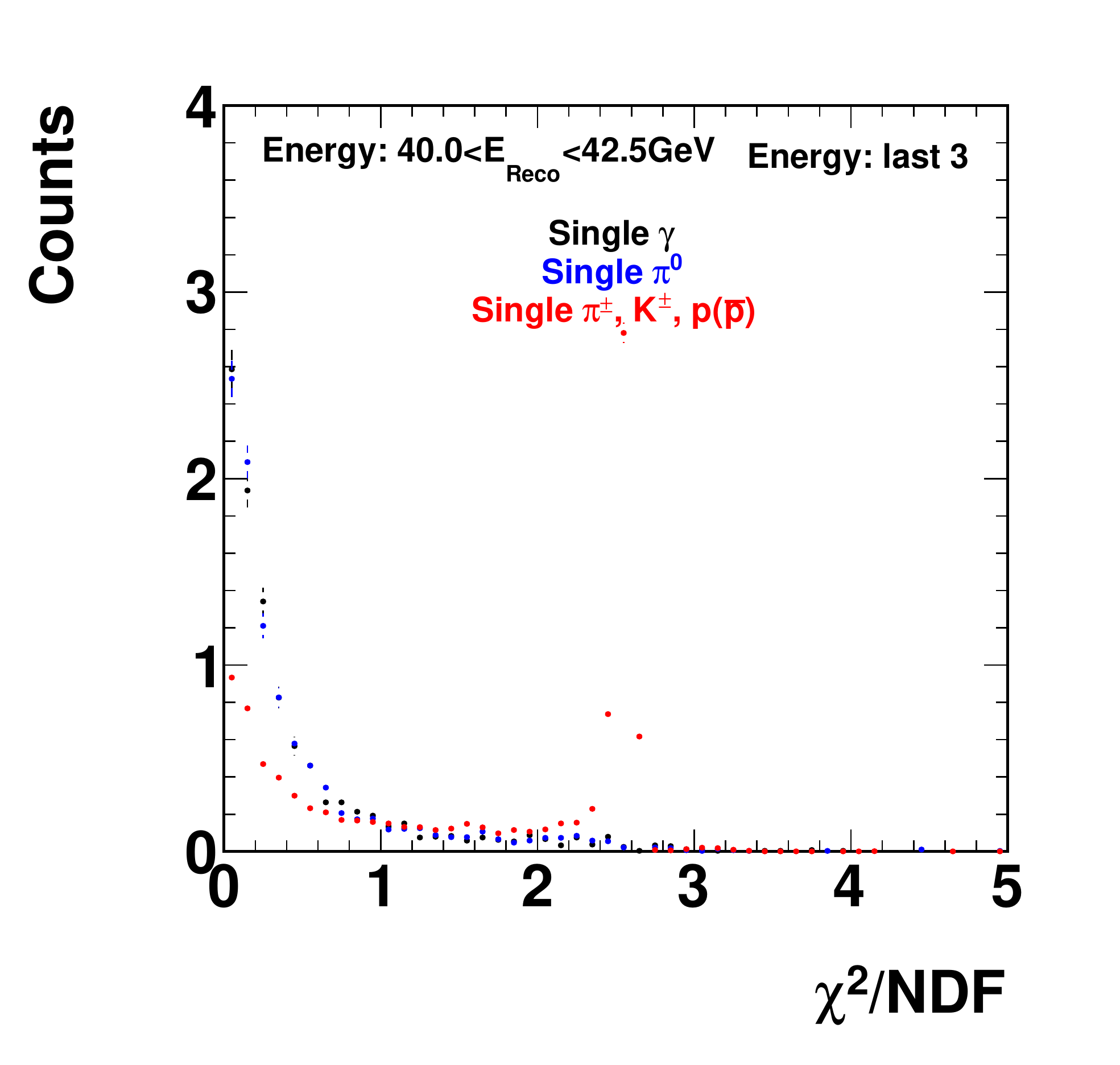}
\includegraphics[angle=0, width=0.24\linewidth]{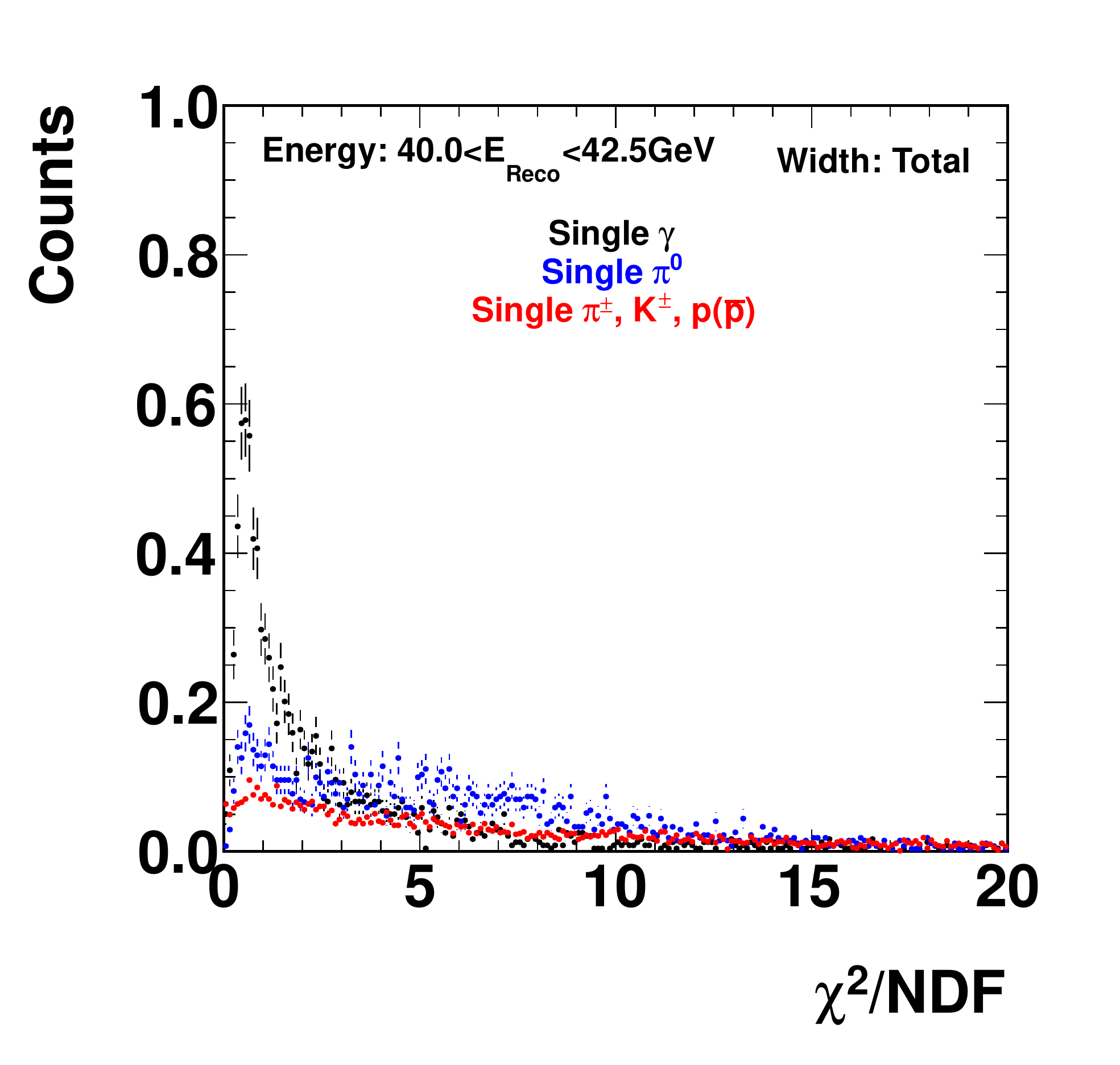}
\includegraphics[angle=0, width=0.24\linewidth]{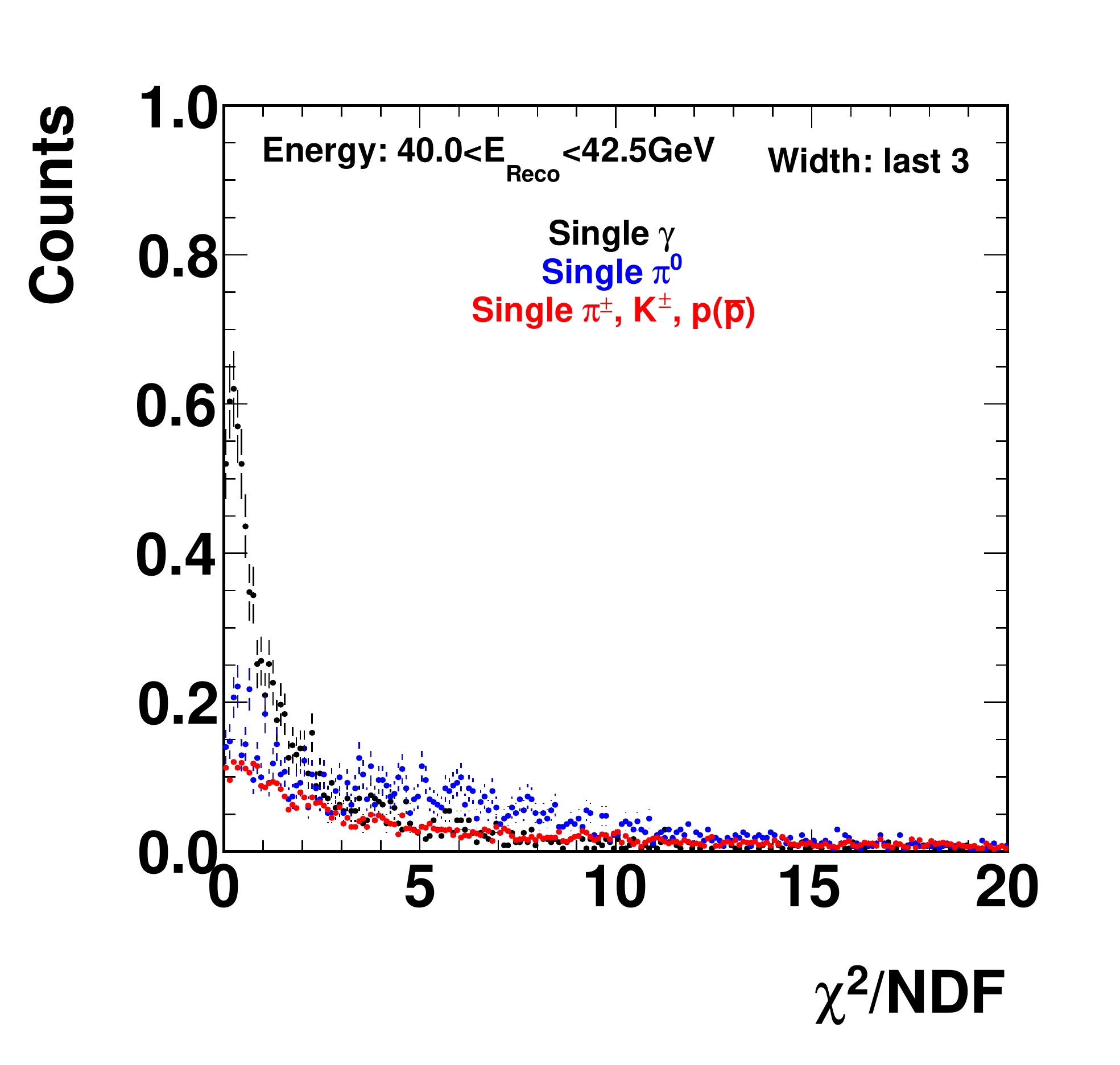}
\vspace*{-0.12in}
\caption{\label{fig:Chi2LayerCalib_40GeV} $\chi^{2}$ calibration applied
to single-$\gamma$s (black), single-$\pi^{0}$s (red), and single-charged
hadrons (mixed $\pi^{\pm}$, $K^{\pm}$, and $p(\overline{p})$), with 
reconstructed energy of 40$<$$E_{\rm Rec}$$<$42.5\,GeV.  Left to
right, the panels show the calibration using energy (all layers), energy
(last 3 only), width (all layers), and width (last 3 only). }
\end{figure}

\begin{figure}[hbt]
\hspace*{-0.12in}
\includegraphics[angle=0, width=0.24\linewidth]{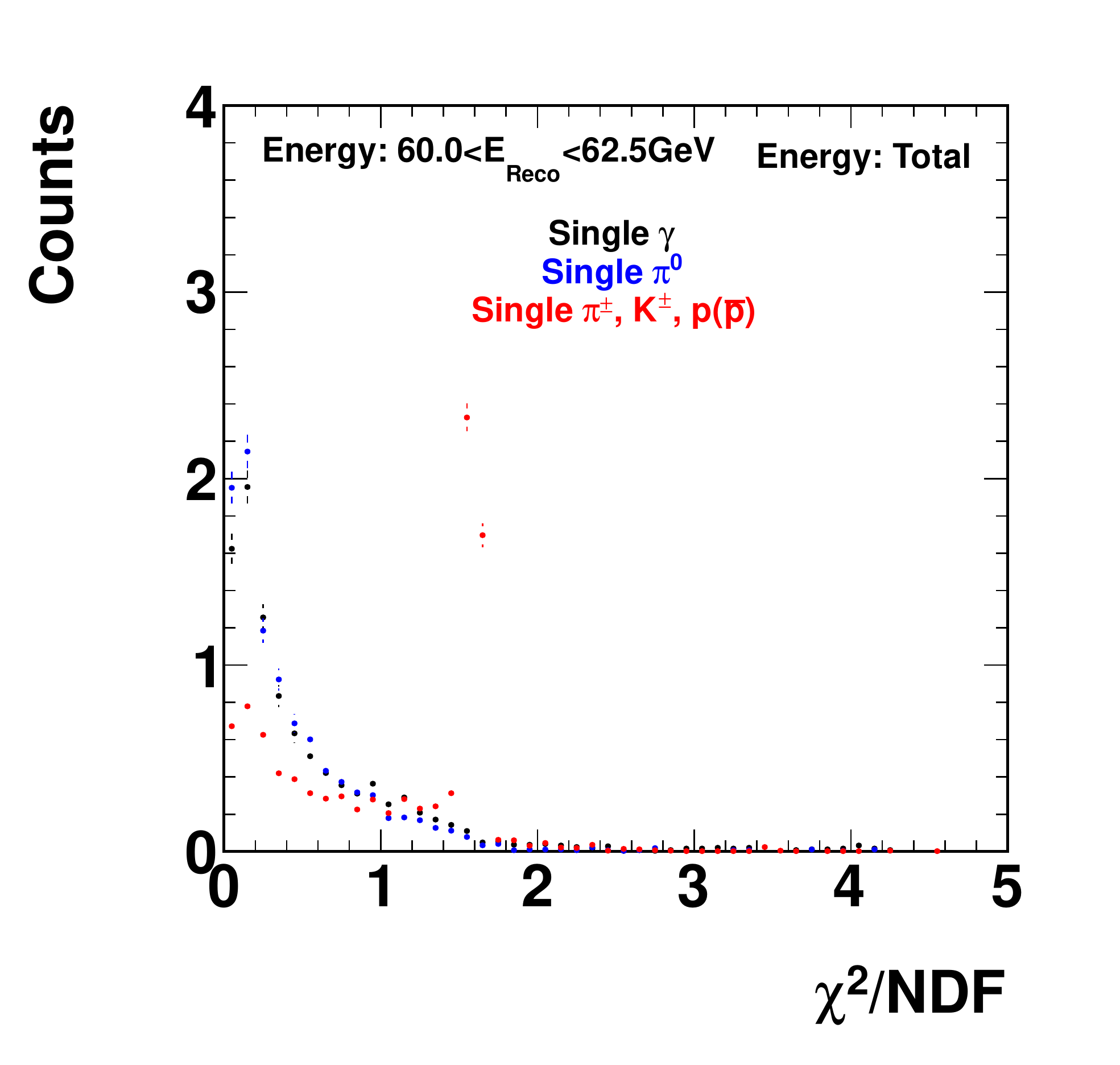}
\includegraphics[angle=0, width=0.24\linewidth]{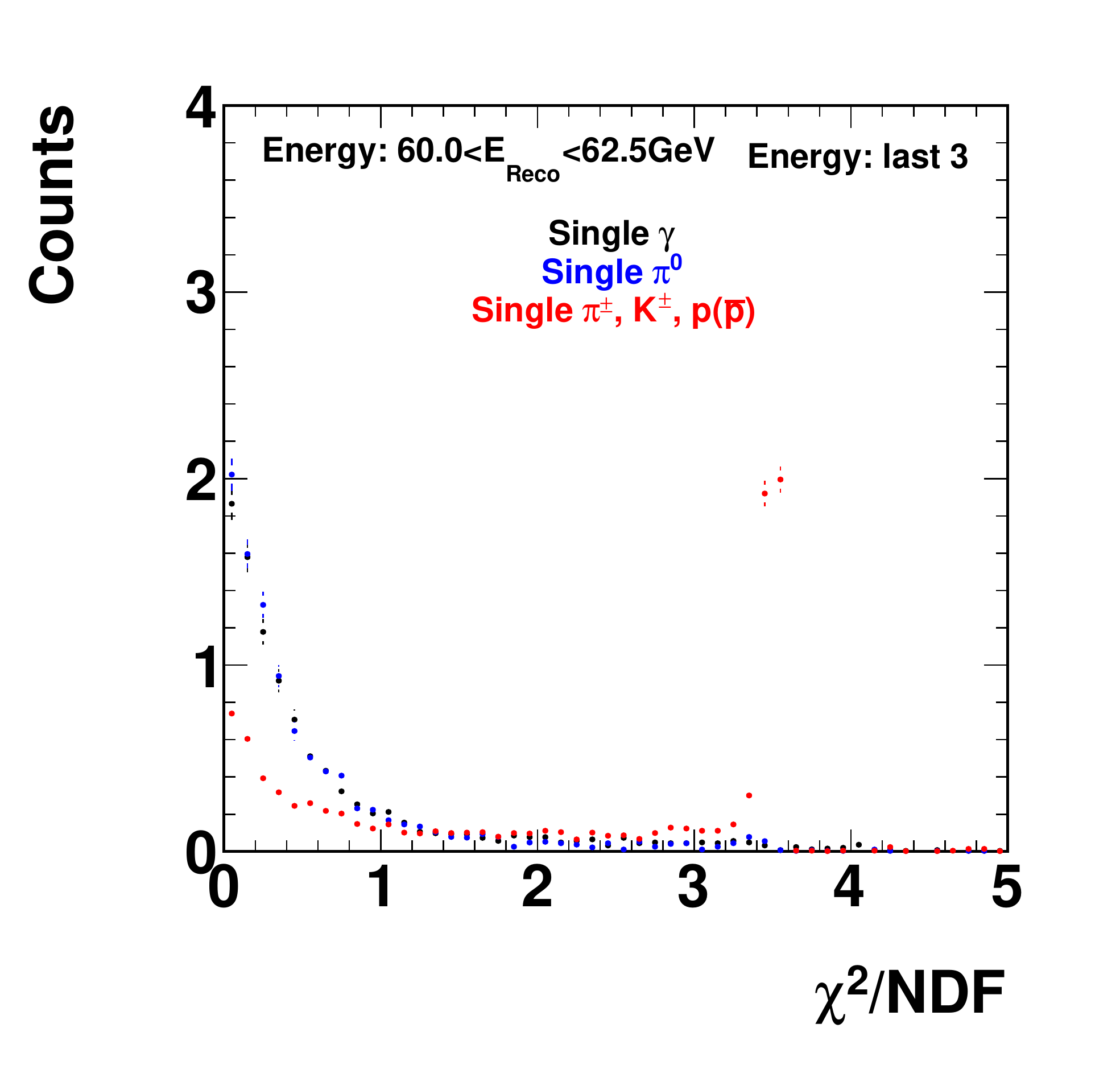}
\includegraphics[angle=0, width=0.24\linewidth]{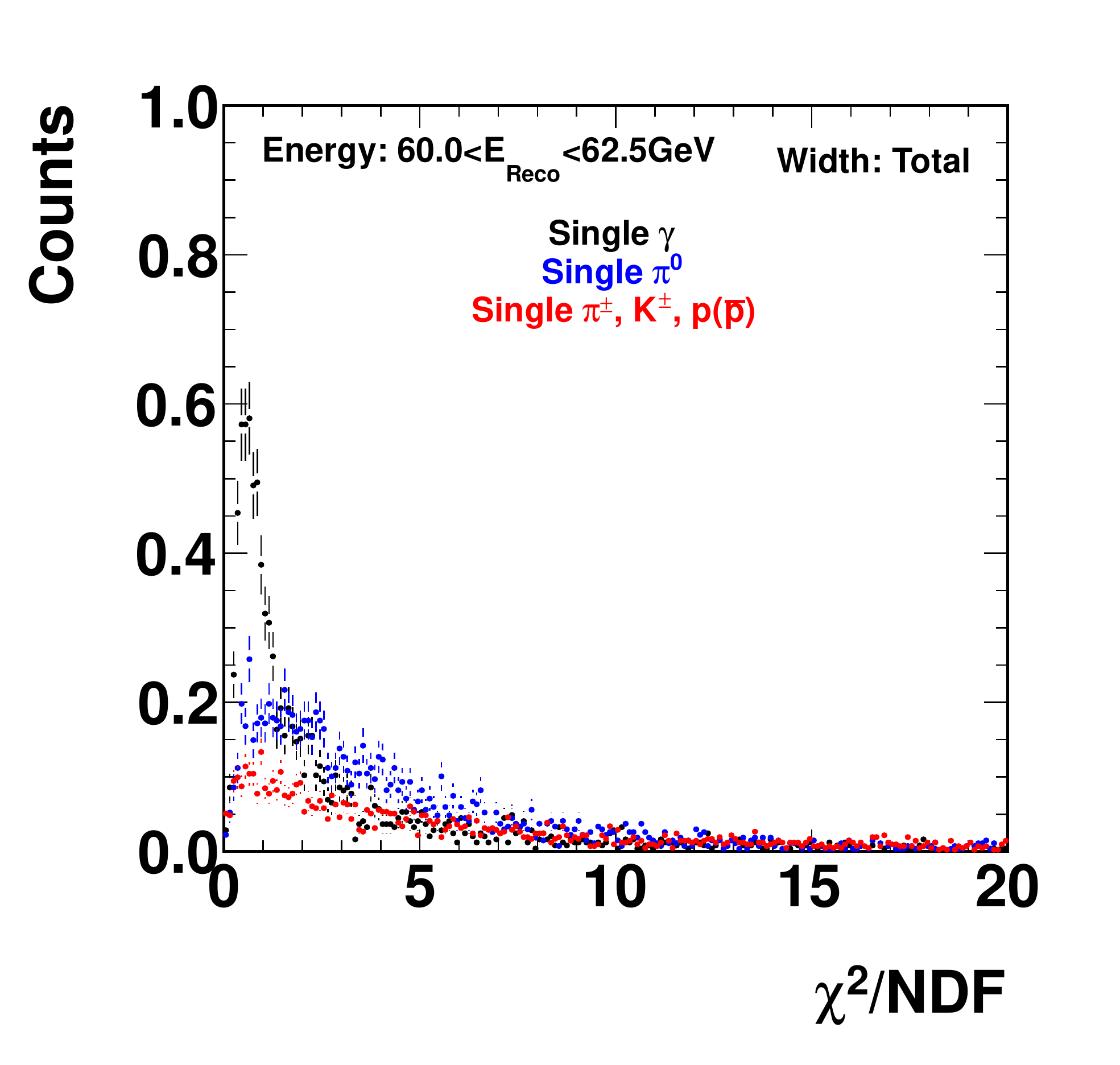}
\includegraphics[angle=0, width=0.24\linewidth]{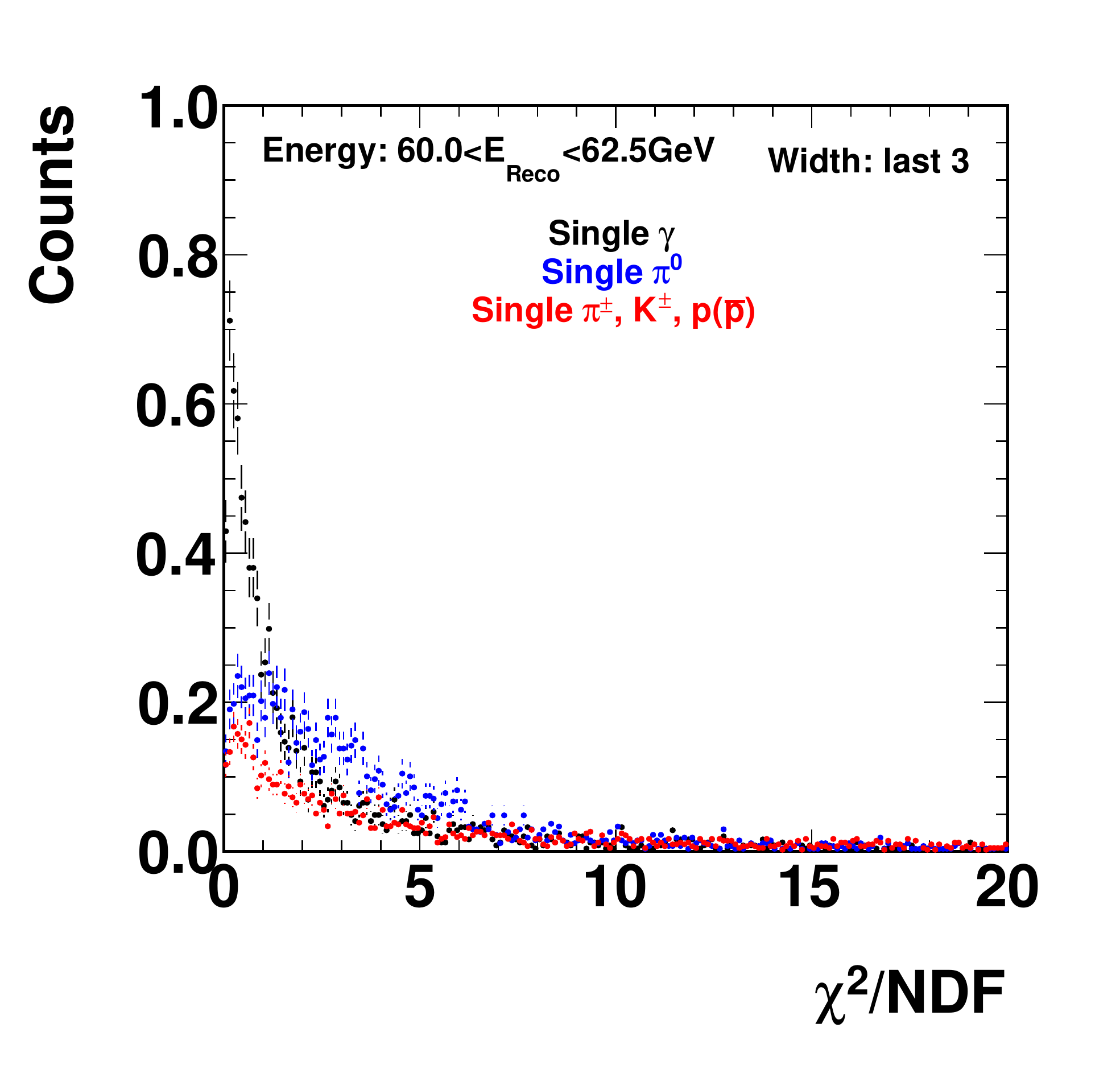}
\vspace*{-0.12in}
\caption{\label{fig:Chi2LayerCalib_60GeV} $\chi^{2}$ calibration applied
to single-$\gamma$s (black), single-$\pi^{0}$s (red), and single-charged
hadrons (mixed $\pi^{\pm}$, $K^{\pm}$, and $p(\overline{p})$), with 
reconstructed energy of 60$<$$E_{\rm Rec}$$<$62.5\,GeV.  Left to
right, the panels show the calibration using energy (all layers), energy
(last 3 only), width (all layers), and width (last 3 only). }
\end{figure}

\begin{figure}[hbt]
\hspace*{-0.12in}
\includegraphics[angle=0, width=0.24\linewidth]{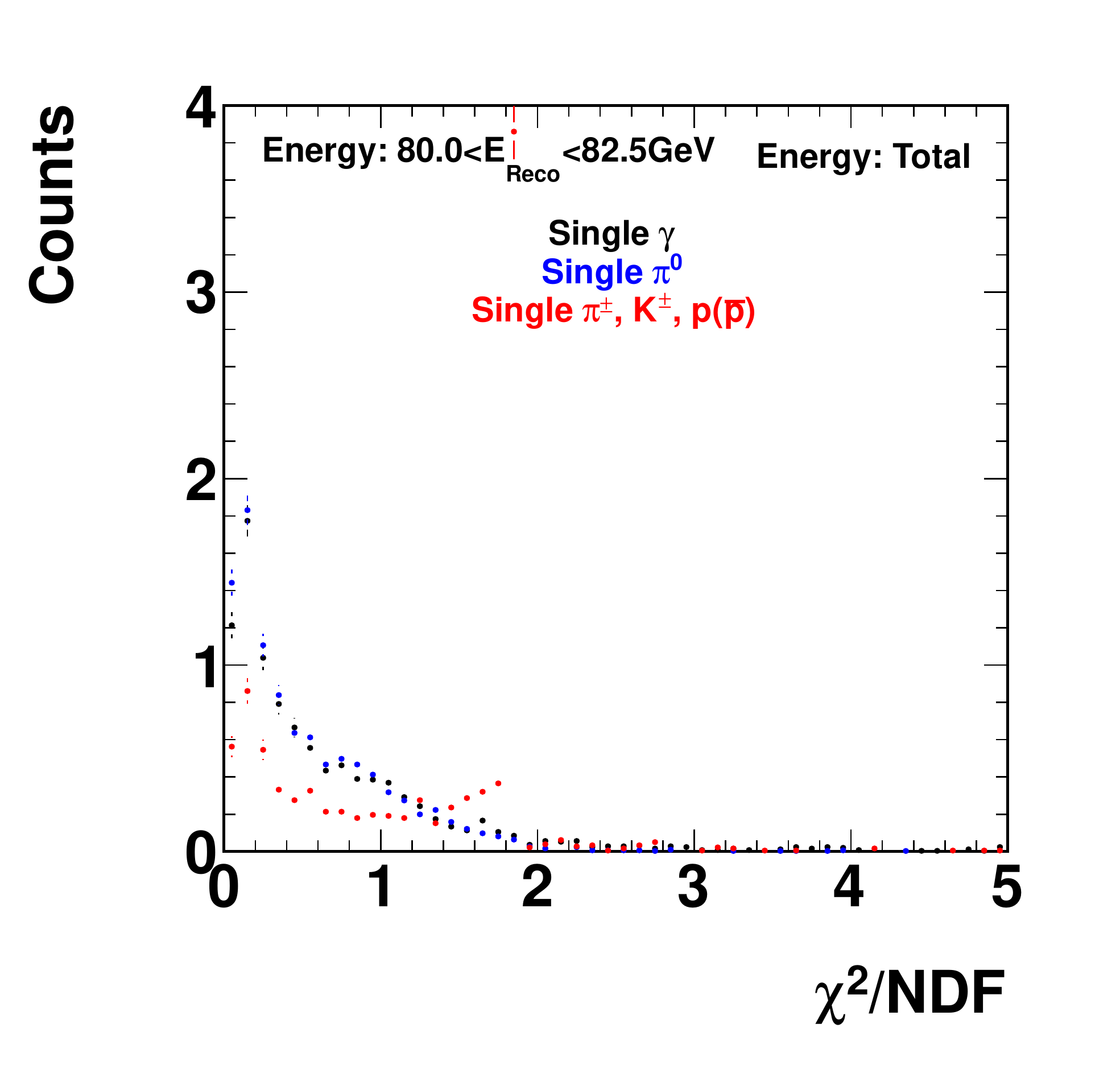}
\includegraphics[angle=0, width=0.24\linewidth]{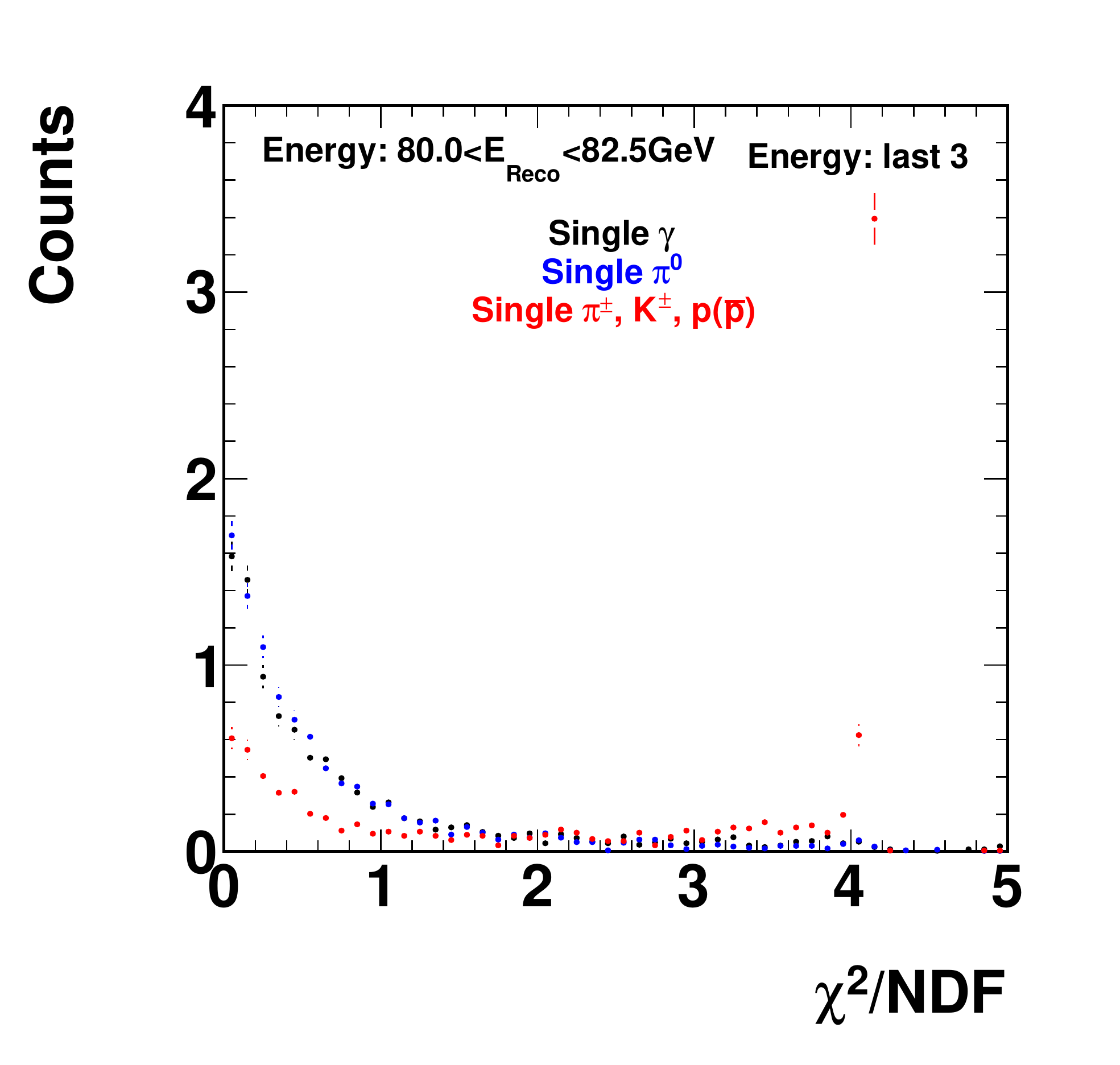}
\includegraphics[angle=0, width=0.24\linewidth]{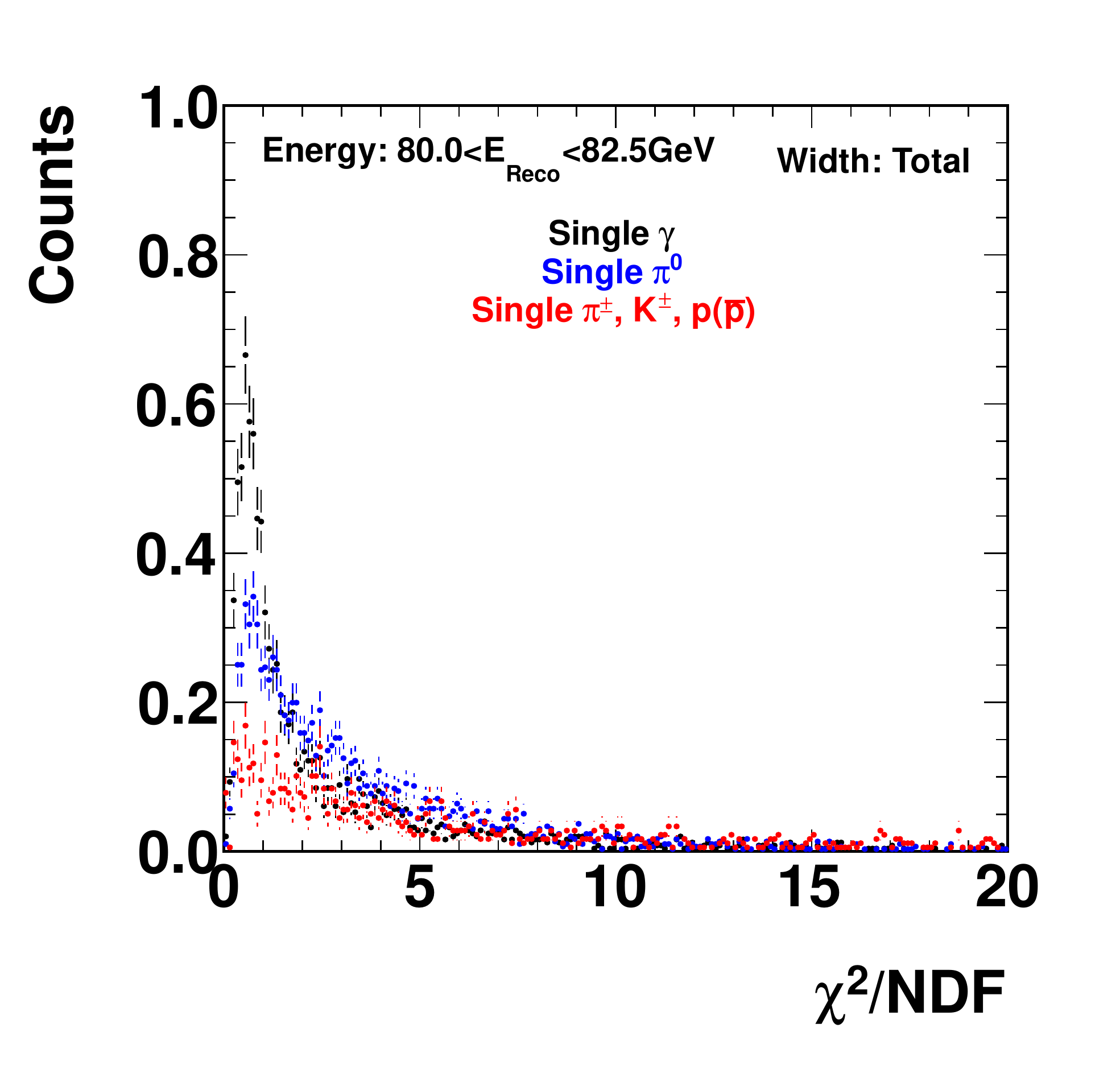}
\includegraphics[angle=0, width=0.24\linewidth]{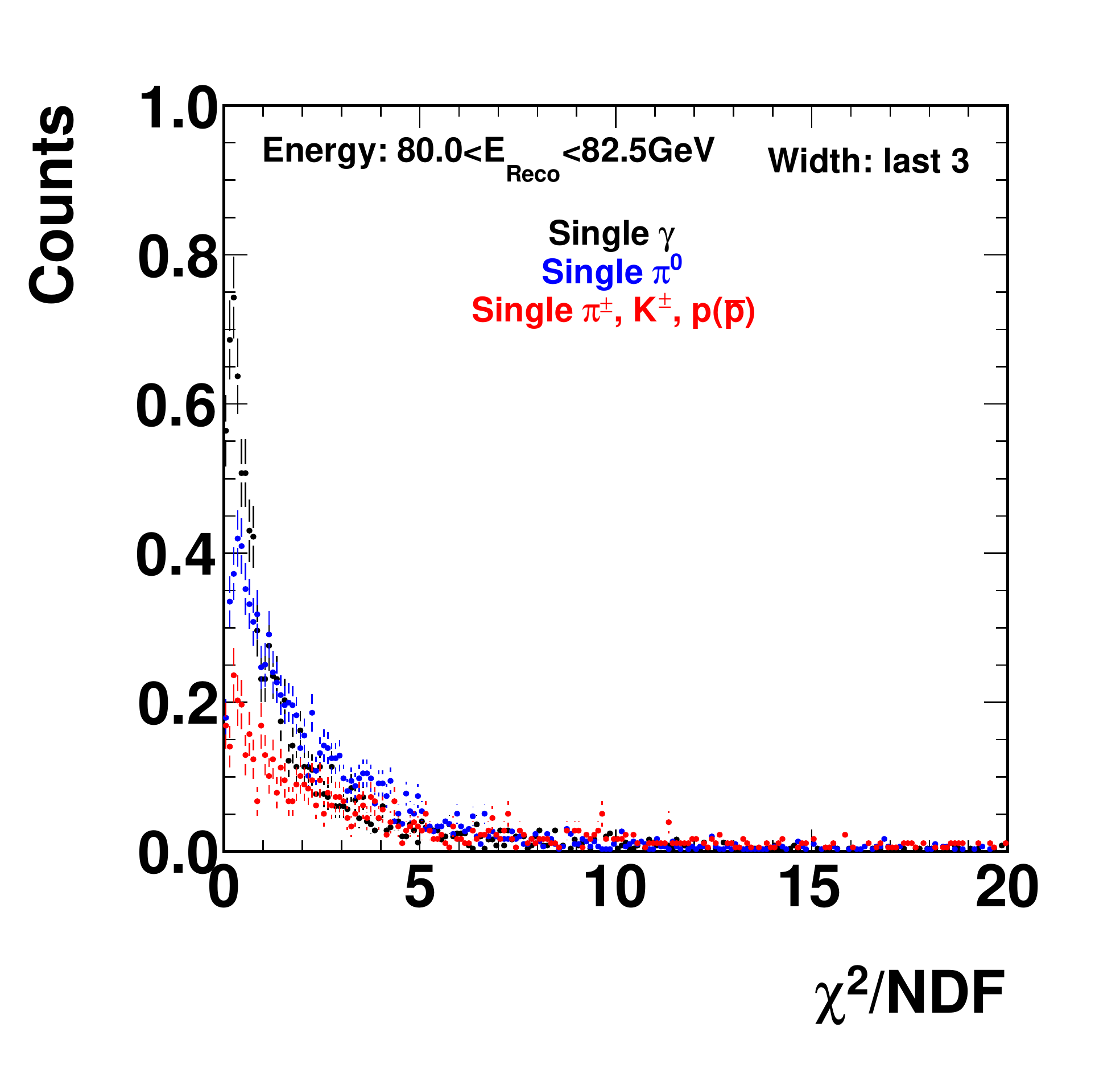}
\vspace*{-0.12in}
\caption{\label{fig:Chi2LayerCalib_80GeV} $\chi^{2}$ calibration applied
to single-$\gamma$s (black), single-$\pi^{0}$s (red), and single-charged
hadrons (mixed $\pi^{\pm}$, $K^{\pm}$, and $p(\overline{p})$), with 
reconstructed energy of 80$<$$E_{\rm Rec}$$<$82.5\,GeV.  Left to
right, the panels show the calibration using energy (all layers), energy
(last 3 only), width (all layers), and width (last 3 only). }
\end{figure}

Four final $\chi^{2}$ methods have been developed for this analysis, with varying
degrees of success.  Two (one in energy, the other in width) are based
on the $\chi^{2}$ formed from all layers, and a second pair is derived
from only the last 3 layers, where the showers are more definitely defined.
As an example, Fig~\ref{fig:Chi2LayerCalib_20GeV} shows the four cases
for the lowest energy considered (20.0$<$$E_{\rm Rec}$$<$22.5\,GeV).
Energy bins 40.0$<$$E_{\rm Rec}$$<$42.5\,GeV,
60.0$<$$E_{\rm Rec}$$<$62.5\,GeV, and
80.0$<$$E_{\rm Rec}$$<$82.5\,GeV are shown in
Figs.~\ref{fig:Chi2LayerCalib_40GeV},~\ref{fig:Chi2LayerCalib_60GeV},~and~\ref{fig:Chi2LayerCalib_80GeV} respectively.

\begin{figure}[hbt]
\centering
\hspace*{-0.12in}
\includegraphics[angle=0, width=0.55\linewidth]{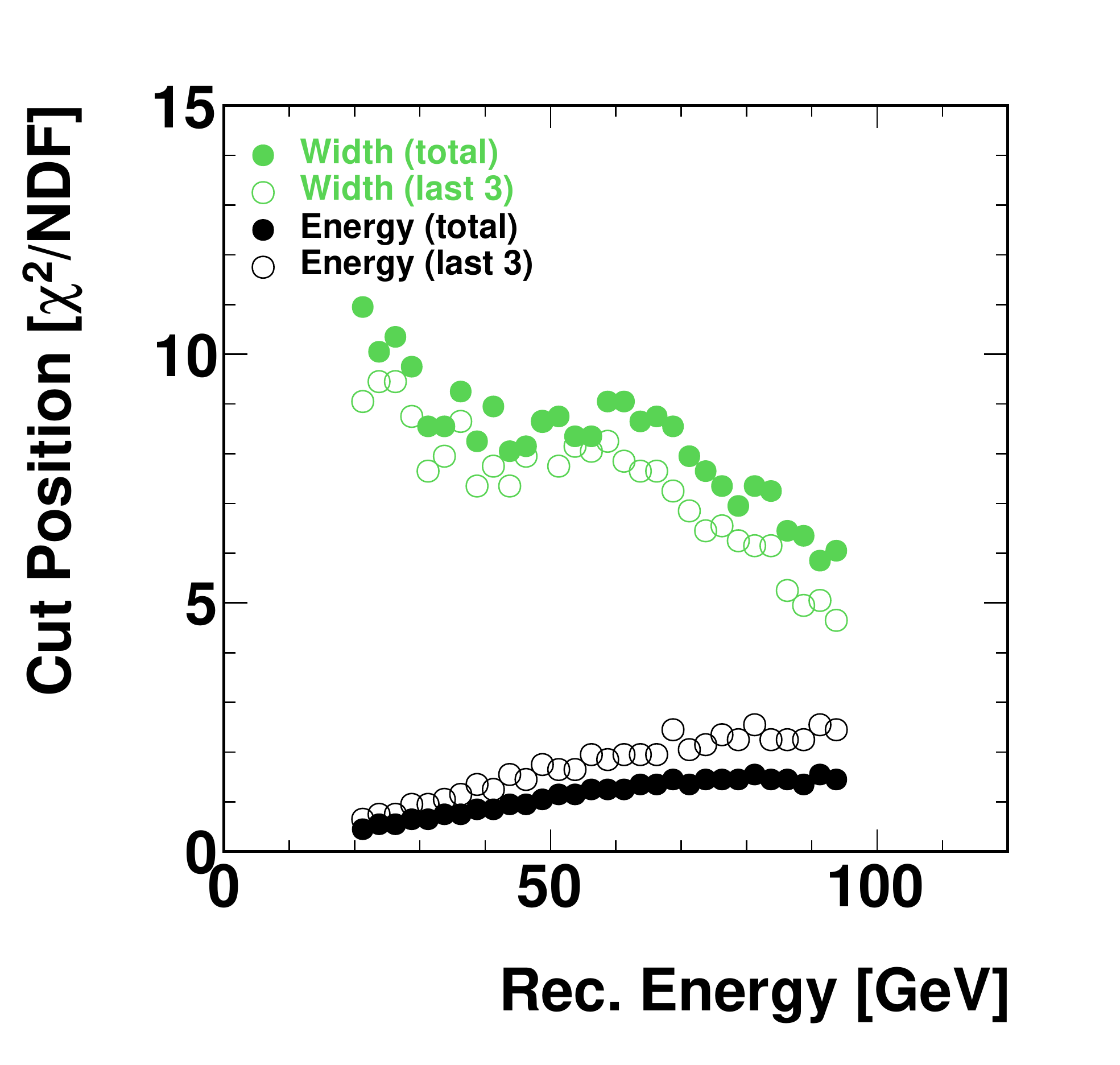}
\vspace*{-0.12in}
\caption{\label{fig:Chi2CutPosition_90pc} $\chi^{2}$ cut positions, determined
to retain 90\% of single-$\gamma$s for the energy (black) and width (green)
methods.  ``Total'' and ``last 3'' are depicted as closed and open symbols
respectively.}
\end{figure}

\begin{figure}[hbt]
\centering
\hspace*{-0.12in}
\includegraphics[angle=0, width=0.49\linewidth]{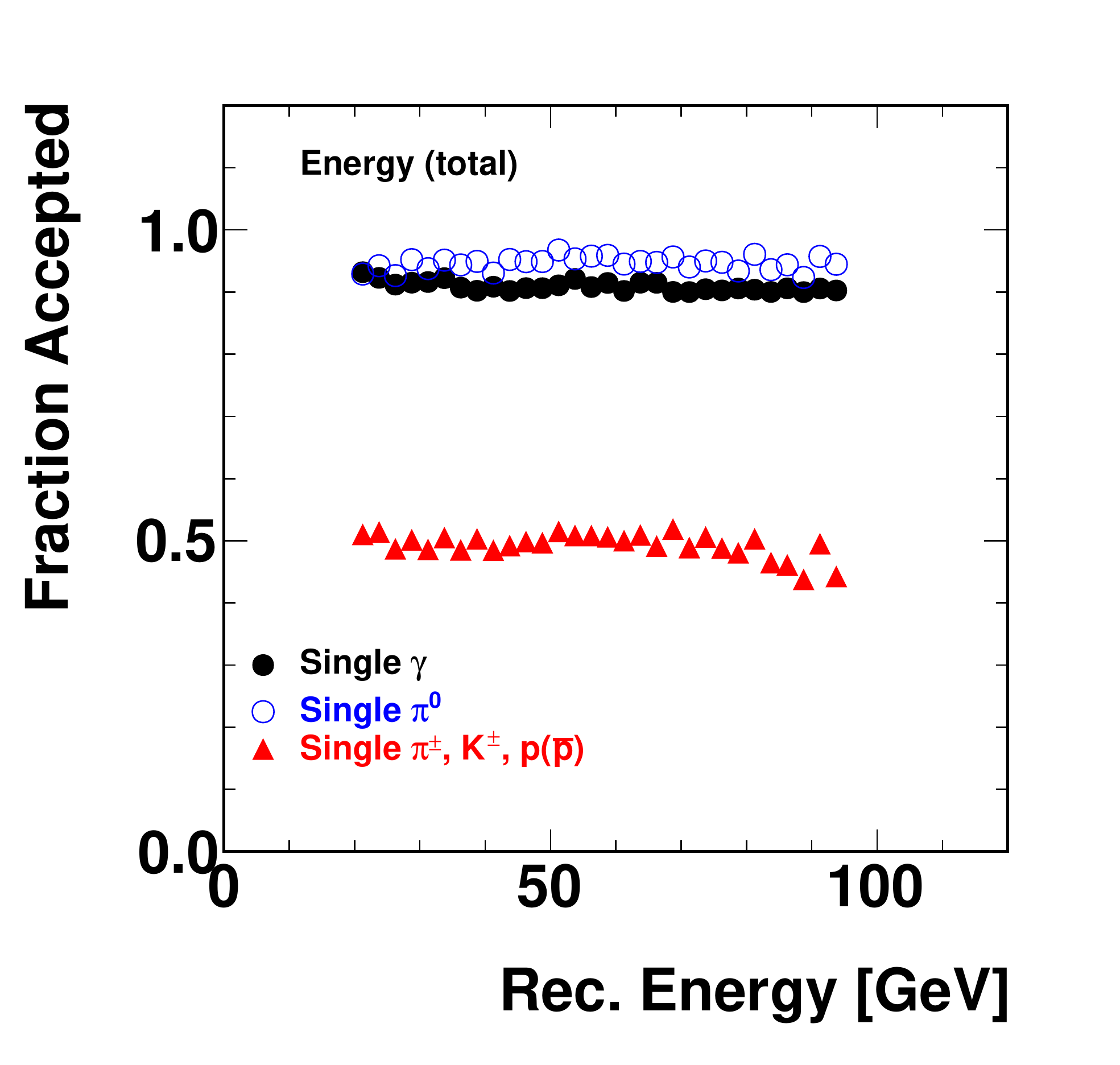}
\includegraphics[angle=0, width=0.49\linewidth]{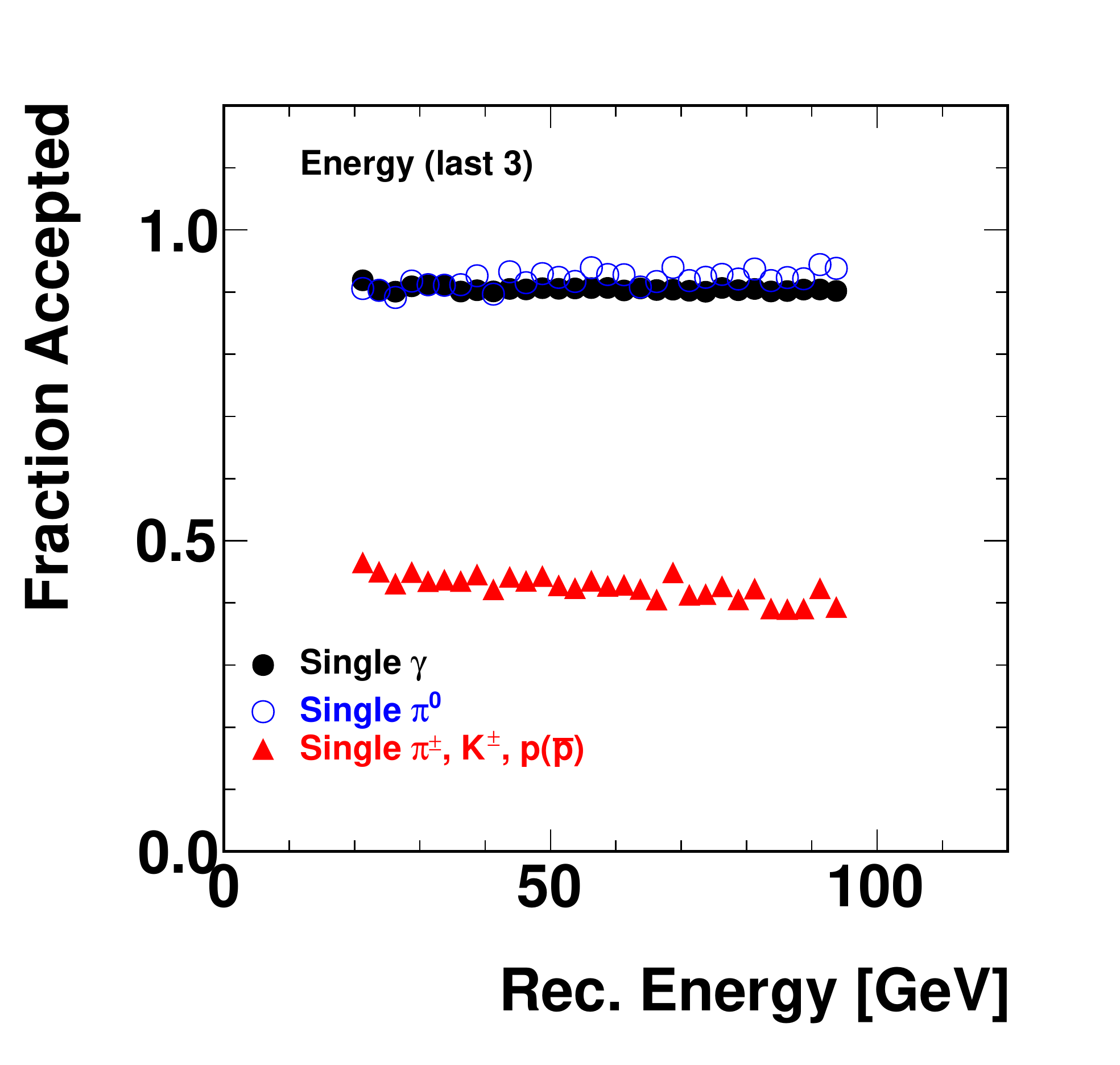}
\vspace*{-0.12in}
\caption{\label{fig:Chi2FracAcc_E_90pc} Fraction of single-$\gamma$s (black),
single-$\pi^{0}$ (blue), and charged hadrons (red) retained by the energy
total (right panel) and last 3 method (left).  The cut position was set
such that 90\% of single-$\gamma$ passed the cuts.}
\end{figure}

\begin{figure}[hbt]
\centering
\hspace*{-0.12in}
\includegraphics[angle=0, width=0.49\linewidth]{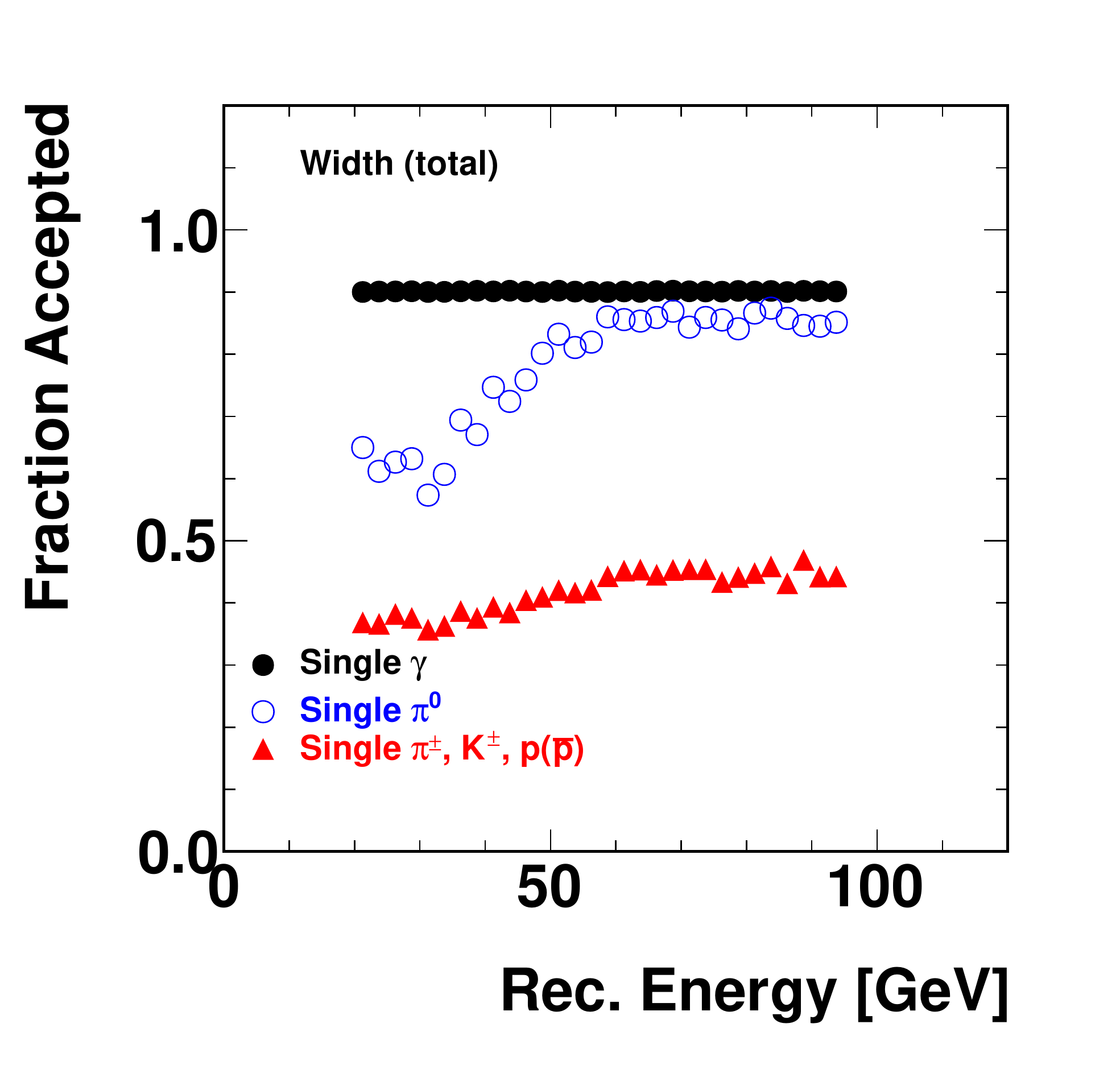}
\includegraphics[angle=0, width=0.49\linewidth]{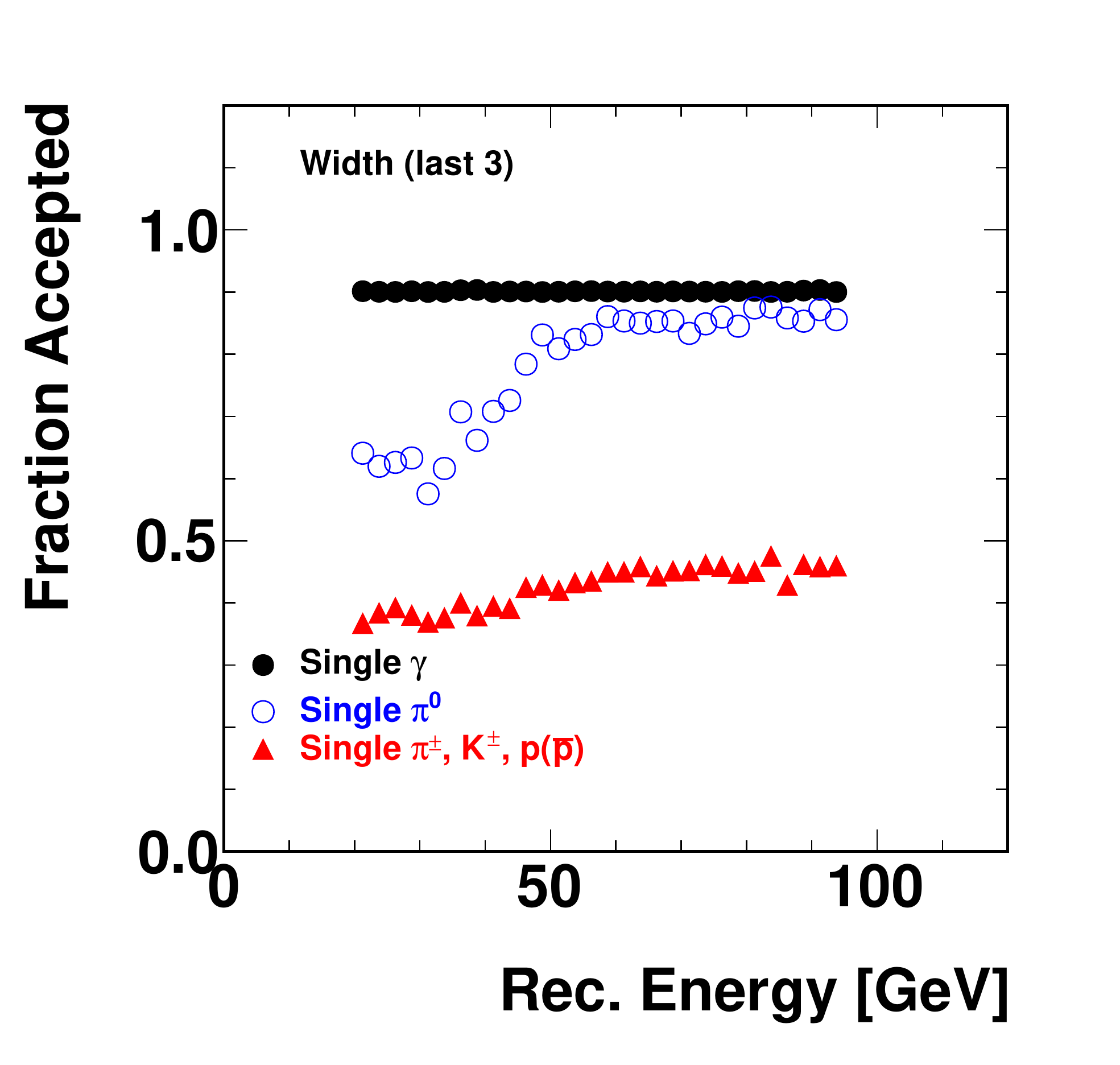}
\vspace*{-0.12in}
\caption{\label{fig:Chi2FracAcc_W_90pc} Fraction of single-$\gamma$s (black),
single-$\pi^{0}$ (blue), and charged hadrons (red) retained by the width
total (right panel) and last 3 method (left).  The cut position was set
such that 90\% of single-$\gamma$ passed the cuts.}
\end{figure}

Using these calibrations, one can define a $\chi^{2}$ cut for each of the
methods.  Figure~\ref{fig:Chi2CutPosition_90pc} shows the cut positions,
as a function of reconstructed energy, which keeps about 90\%
of the single~$\gamma$s.  The effect applying these cut to the
single-$\pi^{0}$s, and single-charged hadrons is shown in
Figure~\ref{fig:Chi2FracAcc_E_90pc} (energy $\chi^{2}$) and
Figire~\ref{fig:Chi2FracAcc_W_90pc} (width $\chi^{2}$) as a fraction of the total which were
reconstructed.

\begin{figure}[hbt]
\centering
\hspace*{-0.12in}
\includegraphics[angle=0, width=0.55\linewidth]{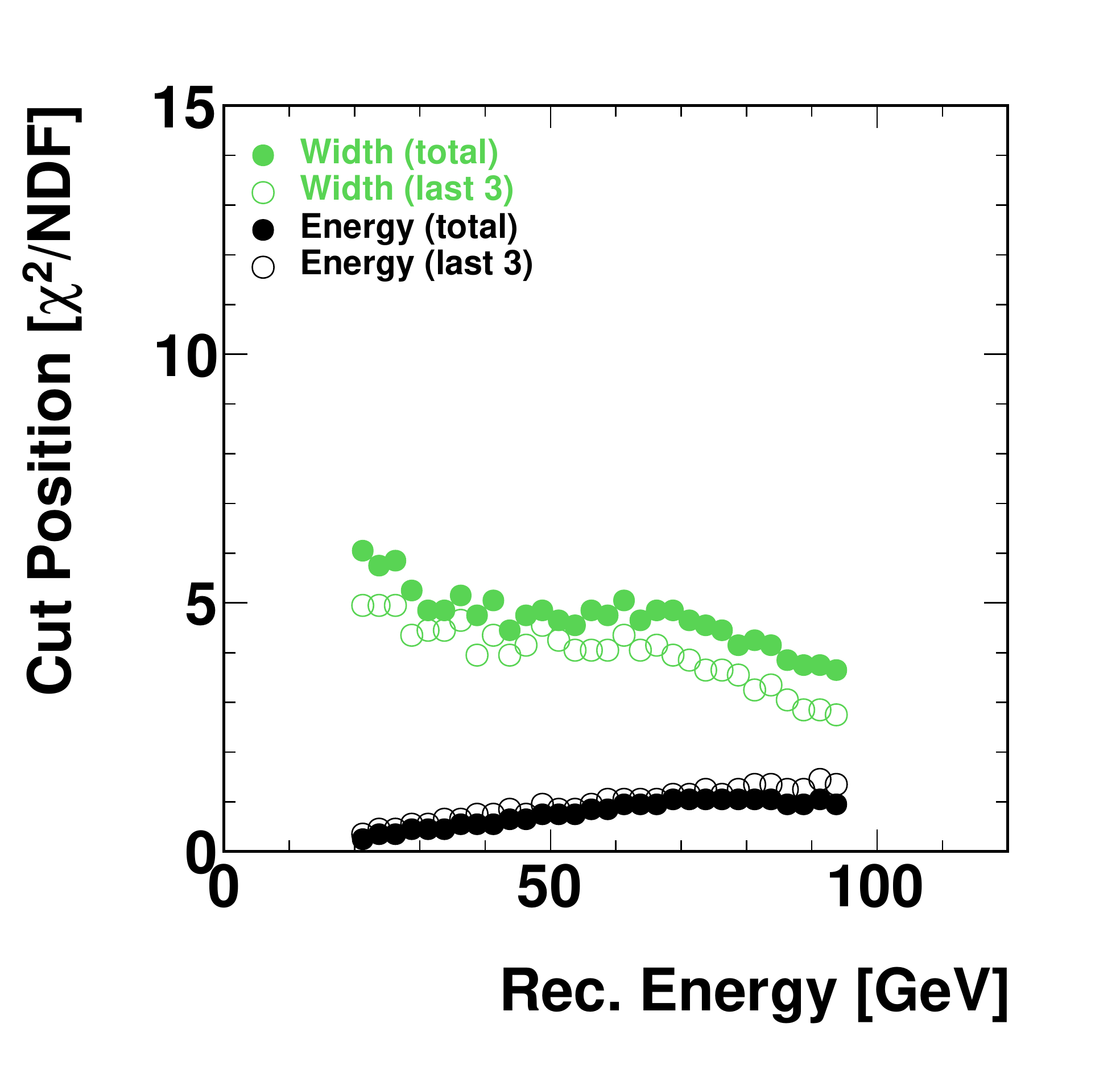}
\vspace*{-0.12in}
\caption{\label{fig:Chi2CutPosition_80pc} $\chi^{2}$ cut positions, determined
to retain 80\% of single-$\gamma$s for the energy (black) and width (green)
methods.  ``Total'' and ``last 3'' are depicted as closed and open symbols
respectively.}
\end{figure}

\begin{figure}[hbt]
\centering
\hspace*{-0.12in}
\includegraphics[angle=0, width=0.49\linewidth]{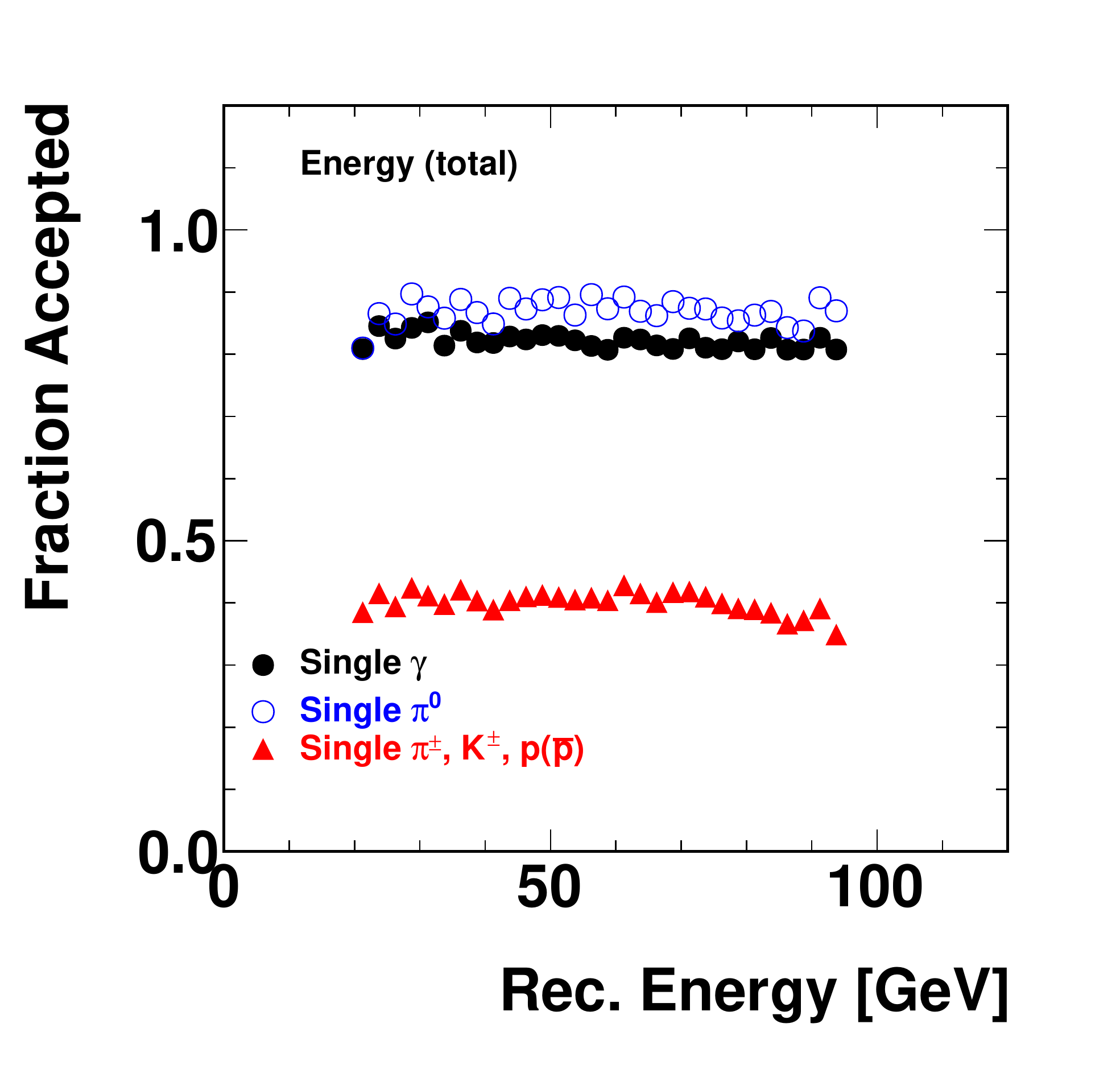}
\includegraphics[angle=0, width=0.49\linewidth]{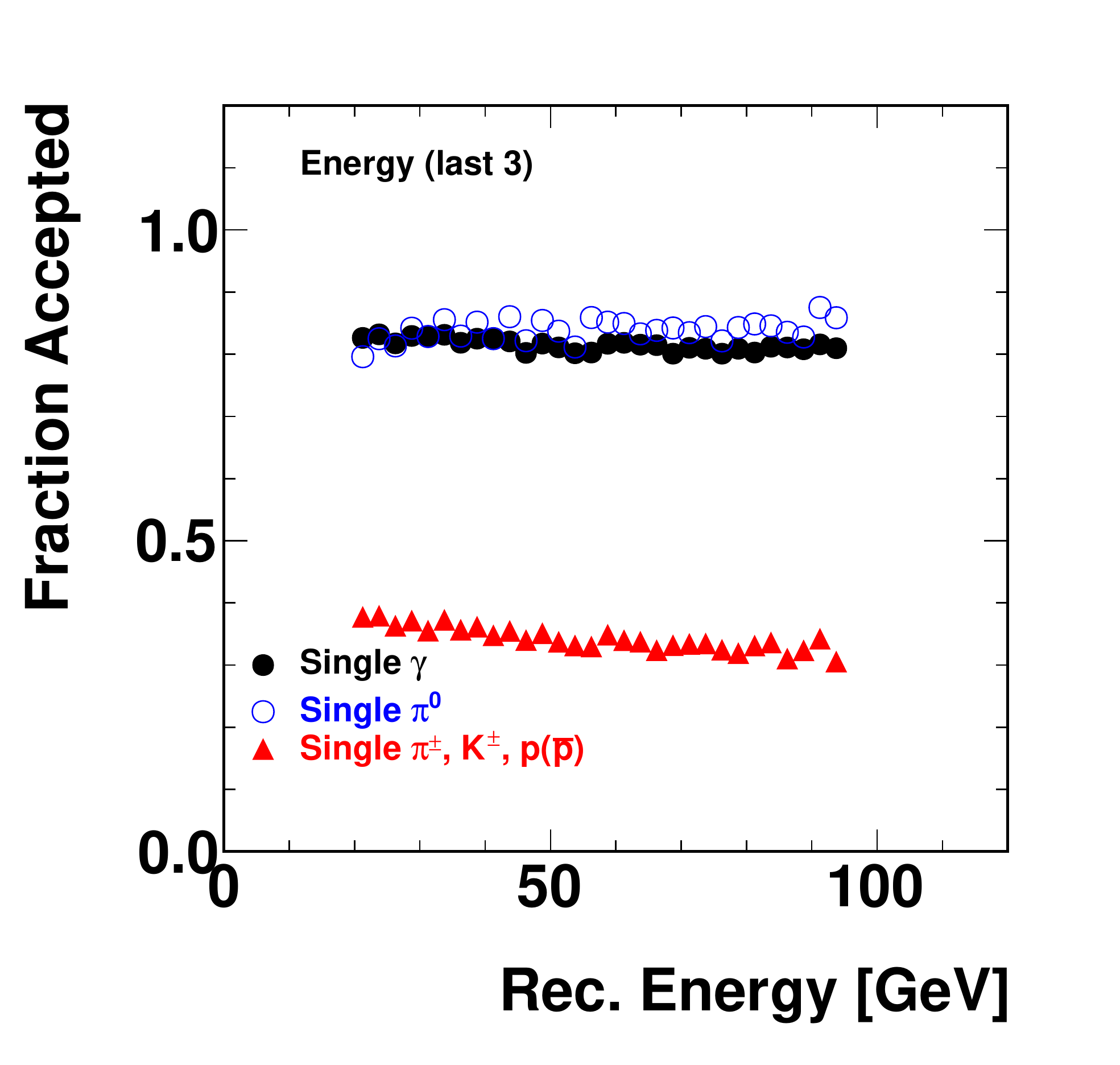}
\vspace*{-0.12in}
\caption{\label{fig:Chi2FracAcc_E_80pc} Fraction of single-$\gamma$s (black),
single-$\pi^{0}$ (blue), and charged hadrons (red) retained by the energy
total (right panel) and last 3 method (left).  The cut position was set
such that 80\% of single-$\gamma$ passed the cuts.}
\end{figure}

\begin{figure}[hbt]
\centering
\hspace*{-0.12in}
\includegraphics[angle=0, width=0.49\linewidth]{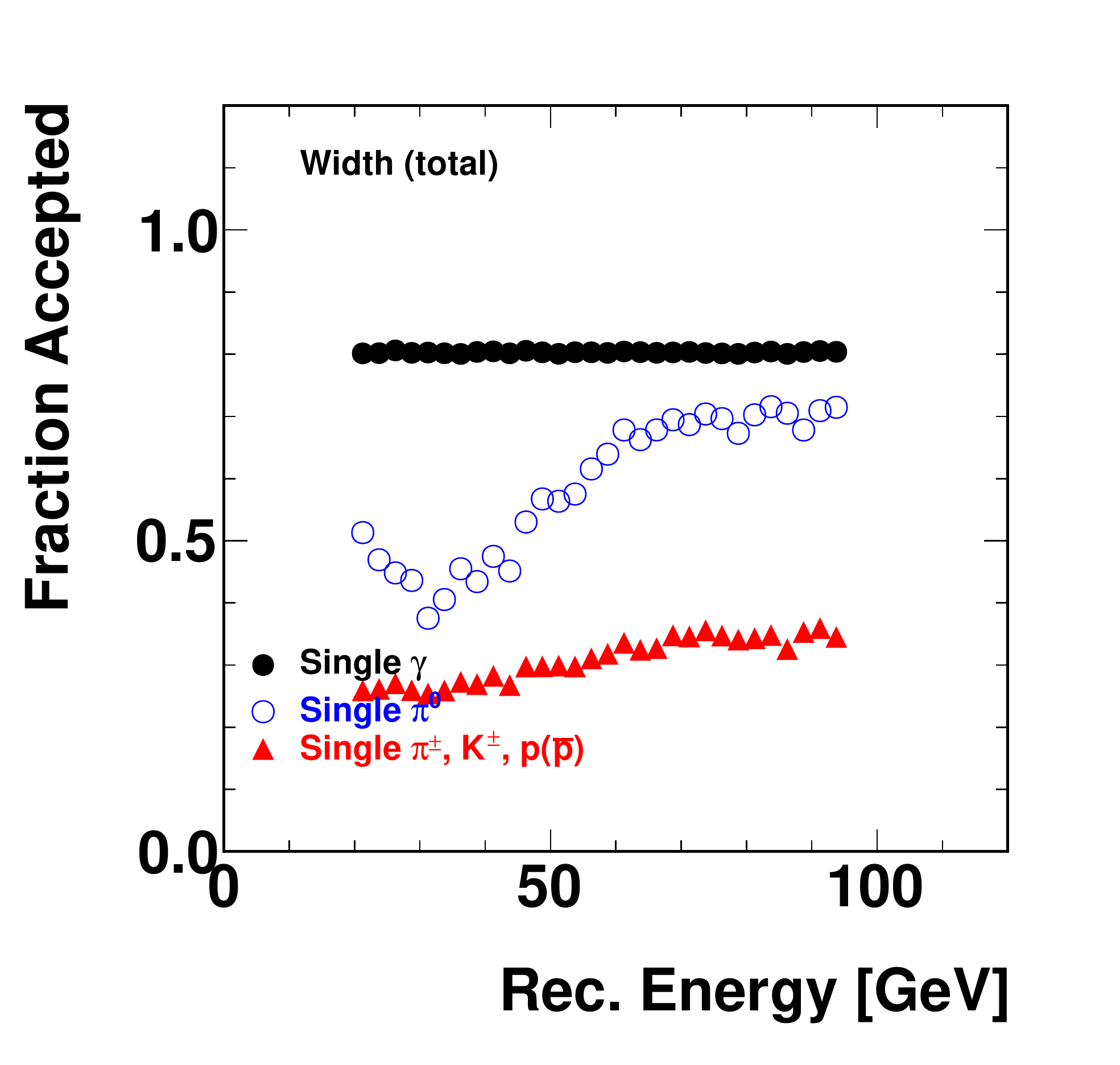}
\includegraphics[angle=0, width=0.49\linewidth]{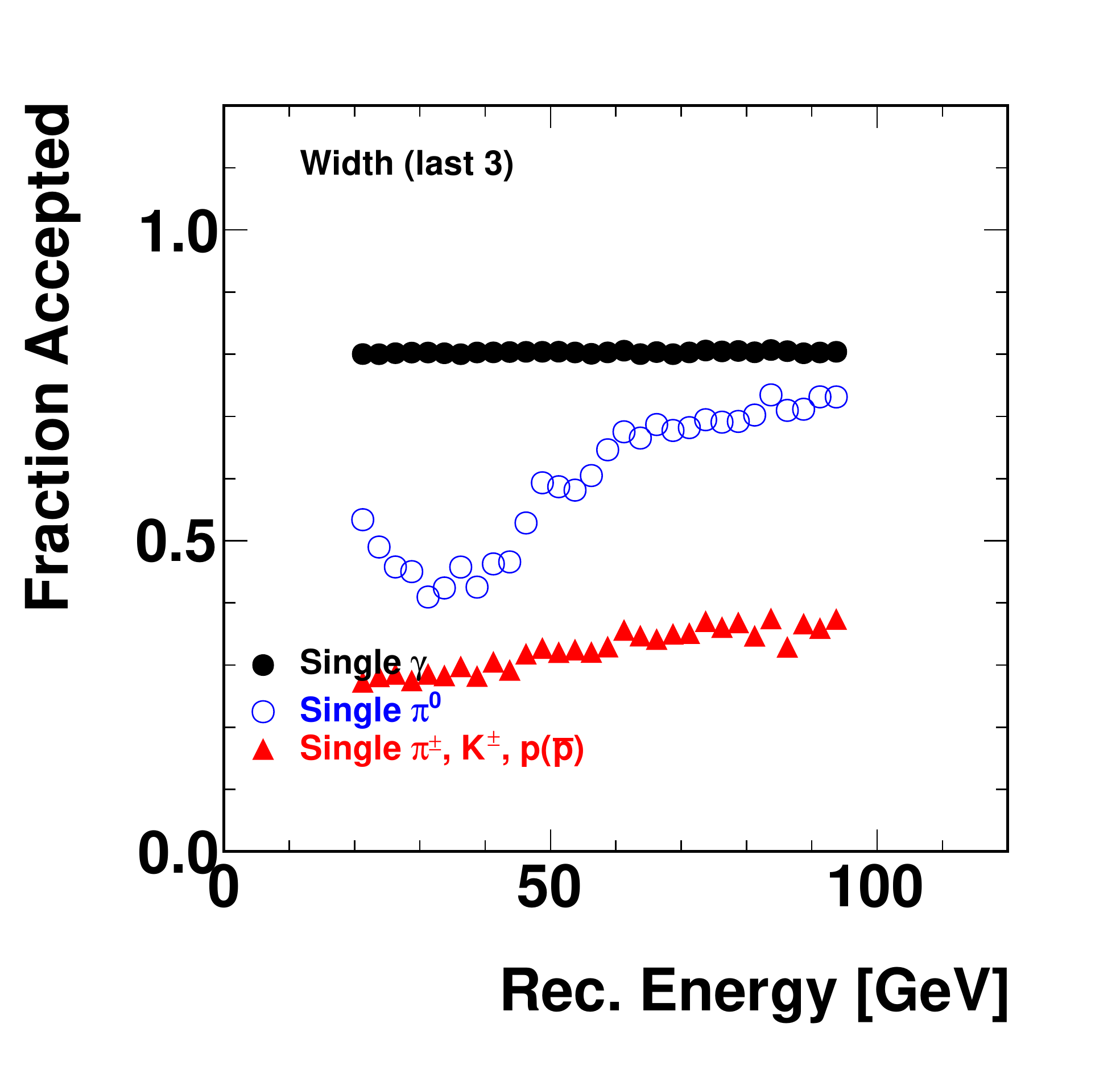}
\vspace*{-0.12in}
\caption{\label{fig:Chi2FracAcc_w_80pc} Fraction of single-$\gamma$s (black),
single-$\pi^{0}$ (blue), and charged hadrons (red) retained by the width
total (right panel) and last 3 method (left).  The cut position was set
such that 80\% of single-$\gamma$ passed the cuts.}
\end{figure}

The $\chi^{2}$ for any event sample is then calculated
relative to those references, and summed over each layer.

\clearpage
\subsection{Reconstruction in {\sc Pythia} events}

\begin{figure}[hbt]
\hspace*{-0.12in}
\centering
\includegraphics[angle=0, width=0.75\linewidth]{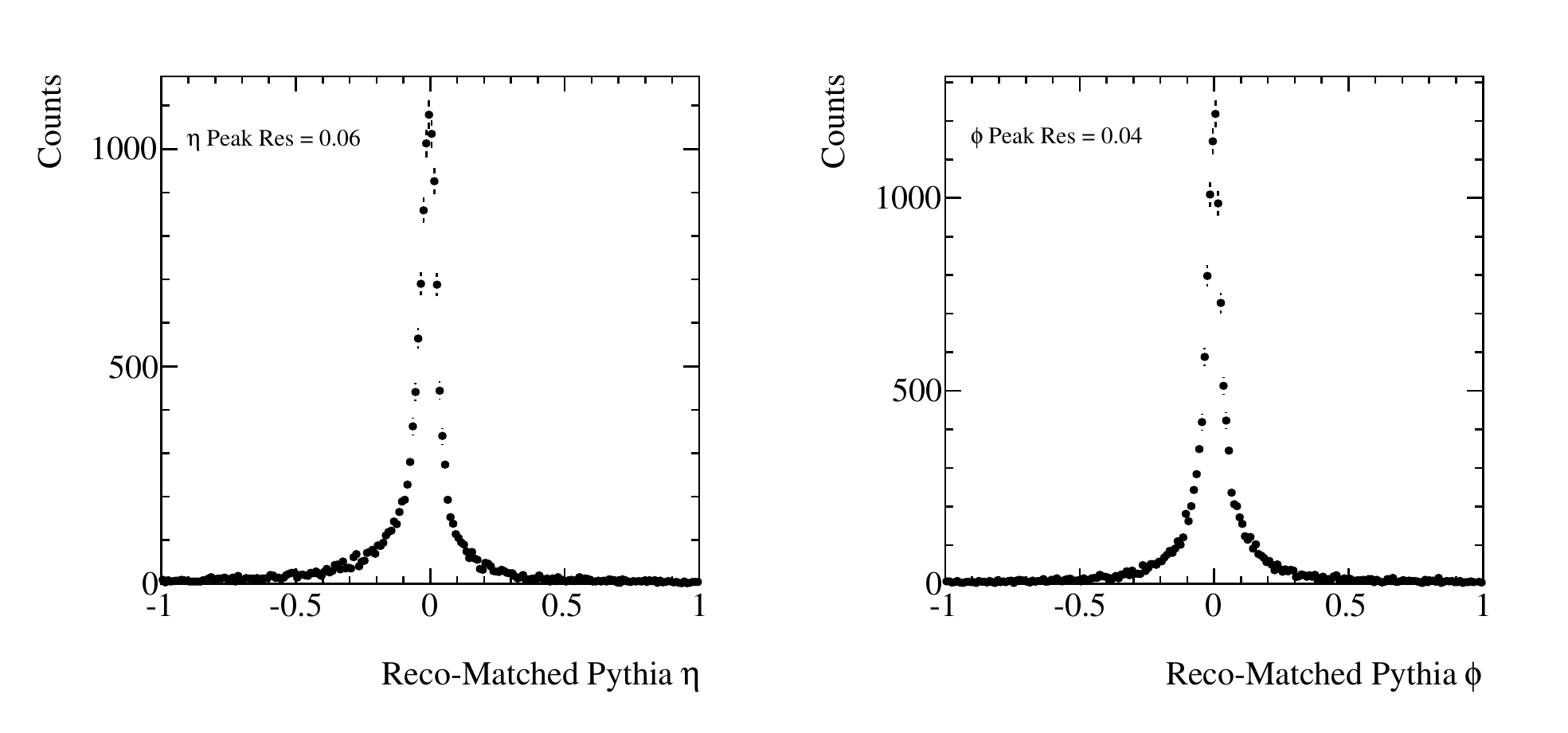}
\vspace*{-0.12in}
\caption{\label{fig:PyMatchReso} Resolution of matching between reconstructed
and {\sc Pythia} primary particles.  The left panel shows the $\delta\eta$
resolution and the right shows the $\delta\phi$ resolution.  The text states
the Gaussian sigma of a fit to a region close to the peak, to exclude the 
outliers.}
\end{figure}

Moving on from single-particle simulations, a full {\sc Pythia} simulation
was made to check for reconstruction features in a $p+p$ scenario.
To gain a full understanding of the reconstruction, we match the closest
match particle produced from {\sc Pythia} and also the next-closest (to
study effects due to track merging).
The matching resolution between the reconstruction and
{\sc Pythia} is 0.13 in $\delta\eta$ and 0.05 in $\delta\phi$, see
Figure~\ref{fig:PyMatchReso}.

\subsubsection{$\pi^{0}$ Reconstruction}

The reconstruction of $\pi^{0}$s is critical in determining the
direct-$\gamma$ yields, as this is the largest contribution to the 
background.  The reconstruction algorithms are well understood in 
single-particle events and are observed to find high-$p_T$ $\pi^{0}$s.
In {\sc Pythia}, the addition of more particles, thus more energy in 
the MPC and preshower, could cause the algorithm to fail.  This section
discusses this behavior of the reconstruction method in the environment 
of a full {\sc Pythia} event. 

\begin{figure}[hbt]
\hspace*{-0.12in}
\centering
\includegraphics[angle=0, width=0.99\linewidth]{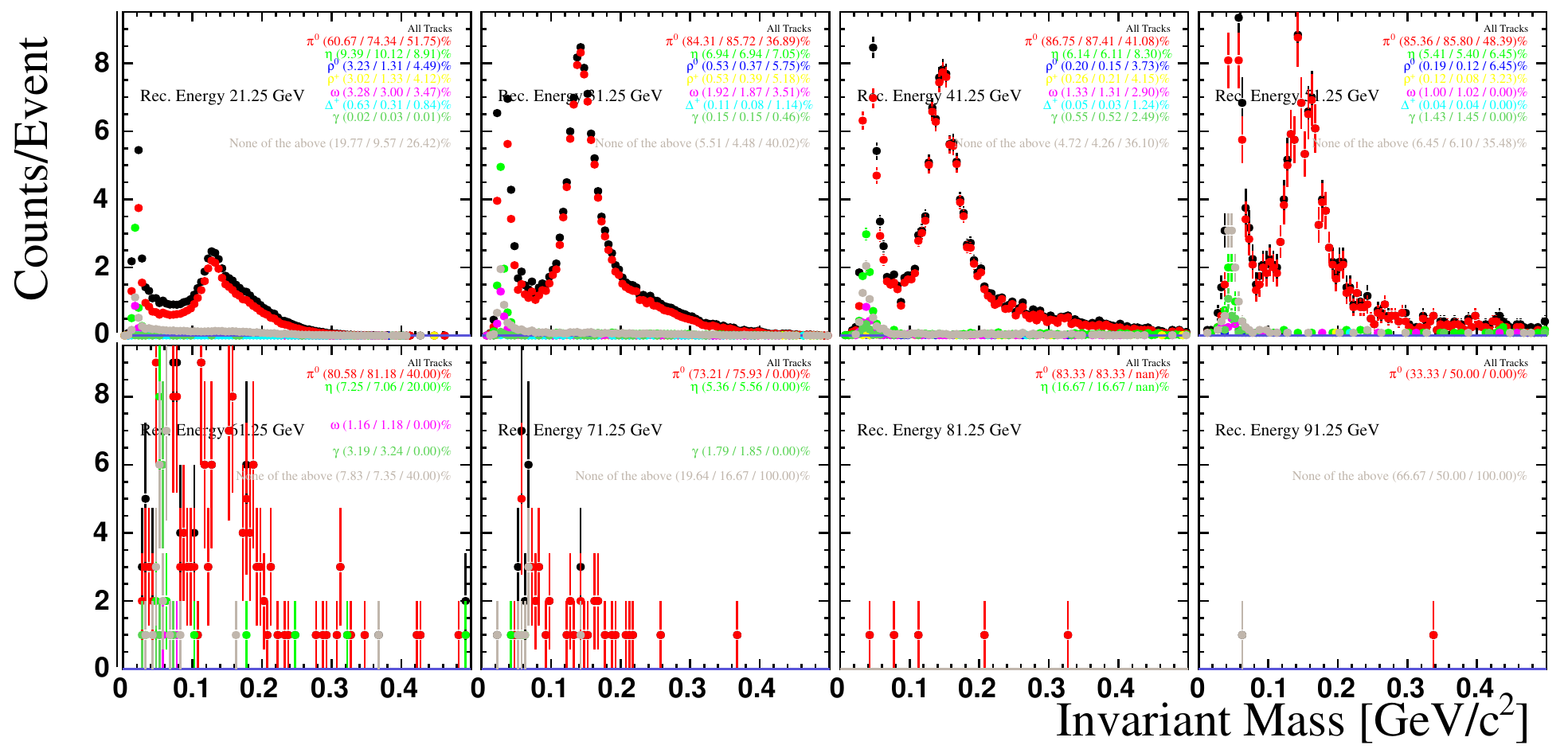}
\vspace*{-0.12in}
\caption{\label{fig:PyReco_breakdown} Reconstruction of all particles, showing the
well reconstructed mass range.  Numbers in the caption are: fraction of total
/ InvMass$>$0 (single track reconstruction converged) / InvMass$<$0 (single tracks
reconstruction did not converge). The different panels show the energy dependence.}
\end{figure}

Figure~\ref{fig:PyReco_breakdown} shows the reconstructed invariant
mass in several bins of $p_T$; no quality-control cuts have been
applied to the data -- this is discussed later.  In each panel, the
black histogram shows the reconstructed data without knowing the
origin of the particle which makes that track.  Tracks which were
created from $\pi^{0}$s are shown in red and, as expected, dominate
the data.  The legend in the figure denotes the fraction of all
events in the given $p_T$ bin.  The first number is the fraction
which passed all reconstruction cuts, but ignores the value determined
for the invariant mass.  The second number shows the fraction of that
particular particle with a well-reconstructed invariant mass tracks
(i.e. $Inv. Mass$$>$0, indicating that the reconstruction did not succeed).  
The final number shows the fraction of that
particular particle which did not produce a well-reconstructed
invariant mass.  The numbers are summarized in
Table~\ref{tbl:PyReco_breakdown}.  Further in the Figure, more minor 
contributions to the total histogram are shown, specifically, the $\eta$
and low-mass vector mesons contribute a significant fraction and are
a further source of background to the direct-$\gamma$s.

\begin{table}
\centering
\caption{Breakdown by particle type of the contribution to the invariant mass spectrum.  No
cut refers to the all data (irrespective of whether the invariant mass was reconstructed or not.
($<$0) $\ge$0 are tracks with a (not) well reconstructed invariant mass.  ``--'' denotes too 
low statistics to be reliable. No quality-control cuts are applied to this data.}
\label{tbl:PyReco_breakdown}
\begin{tabular}{|l|c|c|c|c|c|c|c|c|c|}
\hline
Energy & Invariant & $\pi^{0}$ & $\eta$ & $\rho^{0}$ & $\rho^{+}$ & $\omega$ & $\Delta^{+}$ & $\gamma$ & Something \\
GeV & Mass & \% & \% & \% & \% & \% & \% & \% &  Else (\%) \\
\hline
\multirow{3}{*}{20-22.5} & No Cut    & 60.7 & 9.4 & 3.2  & 3.0  & 3.3 & 0.63 & 0.02 & 19.8 \\
                            & $\ge$0 & 74.3 & 10.1& 1.3  & 1.3  & 3.0 & 0.31 & 0.03 & 9.6  \\
                            & $<$0   & 51.8 & 8.9 & 4.5  & 4.1  & 3.5 & 0.84 & 0.01 & 26.4 \\
\hline
\multirow{3}{*}{30-32.5} & No Cut    & 84.3 & 6.9 & 0.53 & 0.53 & 1.9 & 0.11 & 0.15 & 5.5  \\
                            & $\ge$0 & 85.7 & 6.9 & 0.37 & 0.39 & 1.9 & 0.08 & 0.15 & 4.5  \\
                            & $<$0   & 39.9 & 7.1 & 5.8  & 5.2  & 3.5 & 1.4  & 0.46 & 40.0 \\
\hline
\multirow{3}{*}{40-42.5} & No Cut    & 86.8 & 6.1 & 0.20 & 0.26 & 1.3 & 0.05 & 0.55 & 4.7  \\
                            & $\ge$0 & 87.4 & 6.1 & 0.15 & 0.21 & 1.3 & 0.03 & 0.52 & 4.3  \\
                            & $<$0   & 41.1 & 8.3 & 3.7  & 4.2  & 2.9 & 1.2  & 2.5  & 36.1 \\
\hline
\multirow{3}{*}{50-52.5} & No Cut    & 85.4 & 5.4 & 0.19 & 0.12 & 1.0 & 0.04 & 1.43 & 6.5  \\
                            & $\ge$0 & 85.8 & 5.4 & 0.12 & 0.08 & 1.0 & 0.04 & 1.45 & 6.1  \\
                            & $<$0   & 48.4 & 6.5 & 6.5  & 3.2  & --  & --   & --   & 35.5 \\
\hline
\multirow{3}{*}{60-62.5} & No Cut    & 80.6 & 7.3 & --   & --   & 1.2 & --   & 3.19 & 7.8  \\
                            & $\ge$0 & 81.2 & 7.1 & --   & --   & 1.2 & --   & 3.24 & 7.4  \\
                            & $<$0   & --   & --  & --   & --   & --- & --   & --   & --   \\
\hline
\multirow{3}{*}{70-72.5} & No Cut    & 73.2 & --  & --   & --   & --  & --   & 1.8  & 19.6 \\
                            & $\ge$0 & 75.9 & --  & --   & --   & --  & --   & 1.9  & 16.7 \\
                            & $<$0   & --   & --  & --   & --   & --  & --   & --   & --   \\
\hline
\multirow{3}{*}{80-82.5} & No Cut    & 83.3 & --  & --   & --   & --  & --   & --   & --   \\
                            & $\ge$0 & 83.3 & --  & --   & --   & --  & --   & --   & --   \\
                            & $<$0   & --   & --  & --   & --   & --  & --   & --   & --   \\
\hline
\multirow{3}{*}{90-92.5} & No Cut    & 33.3 & --  & --   & --   & --  & --   & --   & --   \\
                            & $\ge$0 & 50.0 & --  & --   & --   & --  & --   & --   & --   \\
                            & $<$0   & --   & --  & --   & --   & --  & --   & --   & --   \\
\hline
\end{tabular}
\end{table}

\begin{figure}[hbt]
\hspace*{-0.12in}
\centering
\includegraphics[angle=0, width=0.75\linewidth]{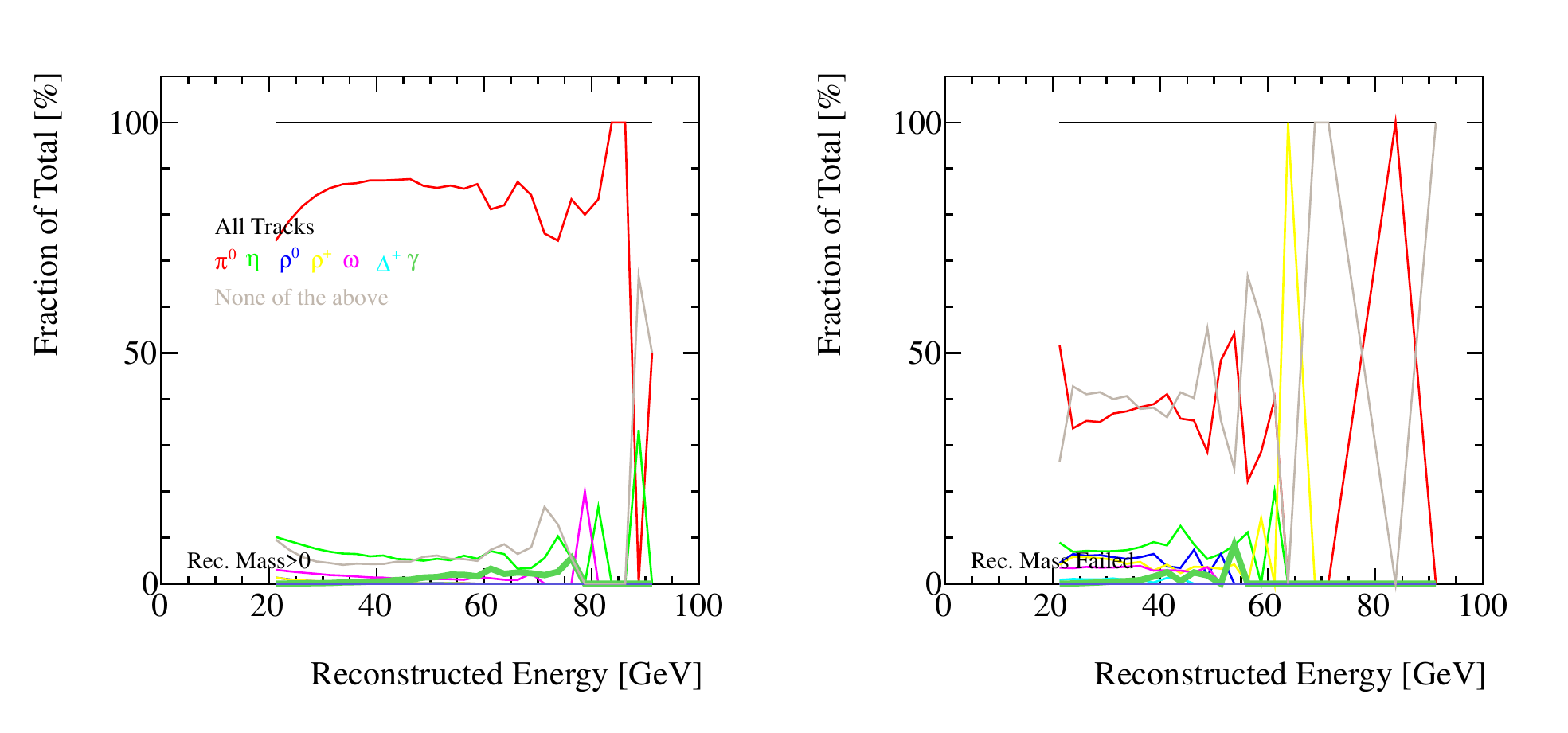}
\vspace*{-0.12in}
\caption{\label{fig:pyReco_breakdownVsEn} Energy dependence of fraction of
each particle type, no cuts except fiducial cut.  }
\end{figure}

\begin{figure}[hbt]
\hspace*{-0.12in}
\centering
\includegraphics[angle=0, width=0.75\linewidth]{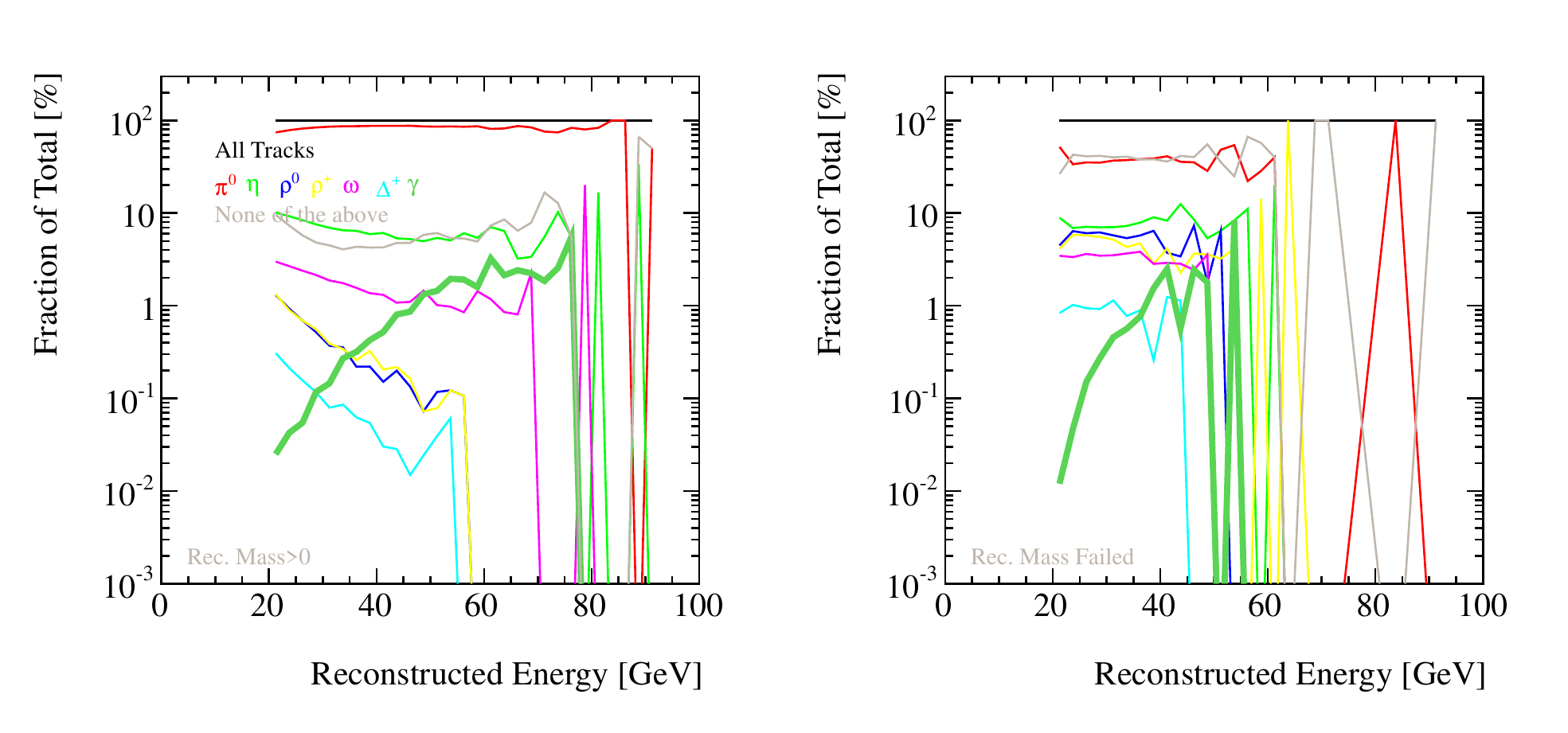}
\vspace*{-0.12in}
\caption{\label{fig:pyReco_breakdownVsEnLog} Same as Figure~\ref{fig:pyReco_breakdownVsEn}, but logarithmic scale on the y-axis for detail. }
\end{figure}

Figures~\ref{fig:pyReco_breakdownVsEn}
(linear scale)~and~\ref{fig:pyReco_breakdownVsEnLog} (logarithmic scale)
show a summary of the simulated data from
Figure~\ref{fig:PyReco_breakdown} / Table~\ref{tbl:PyReco_breakdown}.
The left panels show the fraction of well reconstructed invariant mass
data relative to the total number of tracks reconstructed.
The right show the fraction that do not reconstruct to an invariant mass.  The data at high energy are
statistics starved; additionally, there are no quality-control cuts
applied.  The fraction of $\gamma$ events rises steadiily from low-to-high
energy as the cross-section of the hadronic sources drops.

\subsubsection{$\pi^{0}$ Reconstruction In Detail}

\begin{figure}[hbt]
\hspace*{-0.12in}
\centering
\includegraphics[angle=0, width=0.75\linewidth]{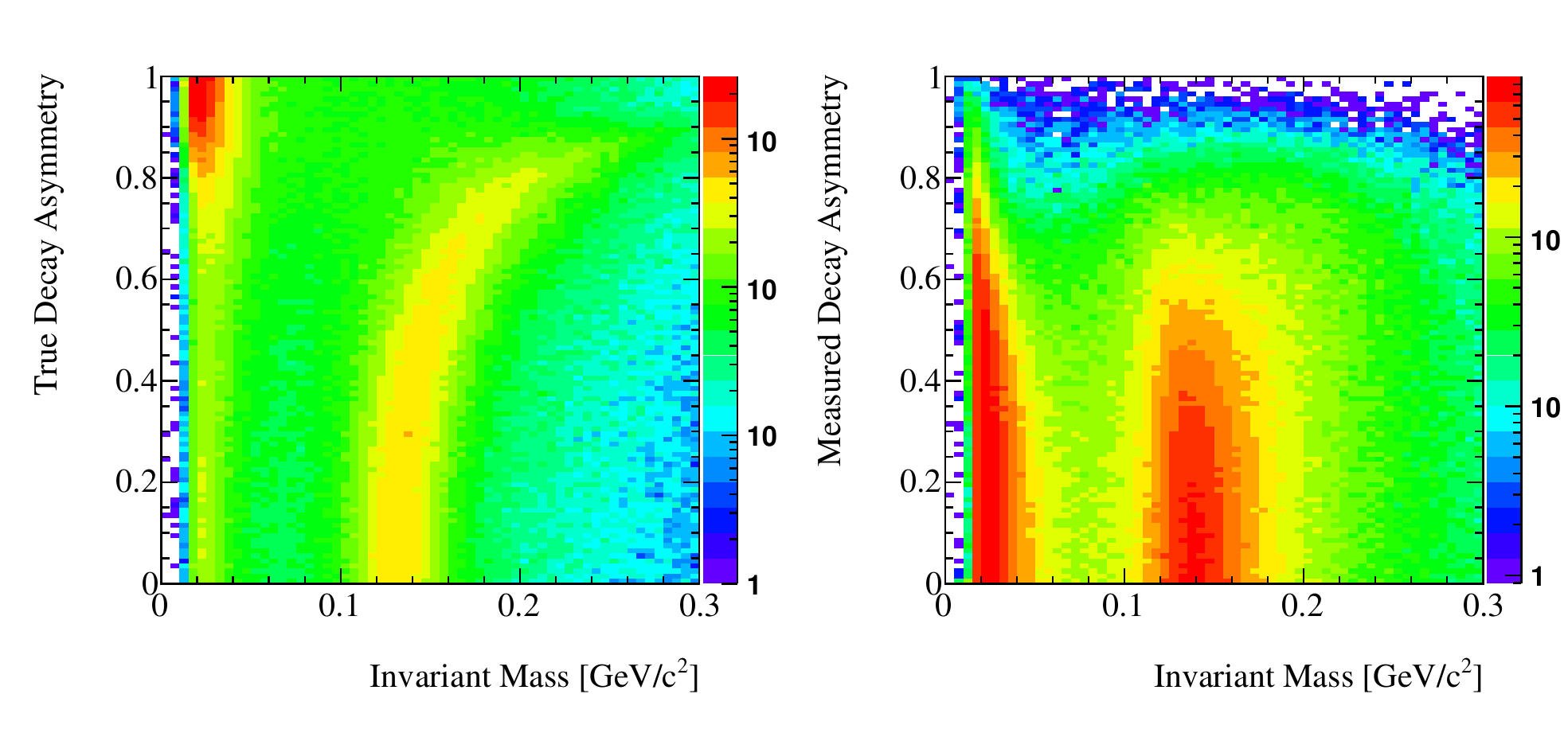}
\vspace*{-0.12in}
\caption{\label{fig:pyReco_asymm} True (left) and measured (right)
asymmetry versus the reconstructed invariant mass.}
\end{figure}

The main problem in the analysis of the $\pi^{0}$s is a kinematic property of
the decay of the $\pi^{0}$.  In the rest frame of the $\pi^{0}$, the decay
angle is evenly distributed.  As such, the energy of each daughter $\gamma$,
when boosted into the collision rest frame, is not equally.  In fact, it is
unusual for an equal energy for each $\gamma$.  This is not a particular
problem for the two-track analyses, but for single tracks, where the distance
between tracks is small, the lesser-energy track can be overwhelmed by 
its counterpart to the degree that the signal is lost in the
tails of the larger-energy $\gamma$.  This can be illustrated by looking
at the asymmetry of the decay versus the reconstructed invariant mass.
Figure~\ref{fig:pyReco_asymm} shows this for the true decay asymmetry (left
panel) and the estimated decay asymmetry based on the energy in the minipads
associated to each track.  For the true case, very asymmetric tracks mostly
reconstruct as a 'single $\gamma$' at low invariant mass.  For tracks with
an asymmetry of 50\% or less (zero is equally-shared energy), the mass is
amost always well reconstructed.  The estimated asymmetry from the
reconstructed mini-pad energy associated to each $\gamma$ does not reflect
this.  This is due to the small sampling fraction coupled with the
optimization within the algorithm to separate the daughter $\gamma$s.

\begin{figure}[hbt]
\hspace*{-0.12in}
\centering
\includegraphics[angle=0, width=0.9\linewidth]{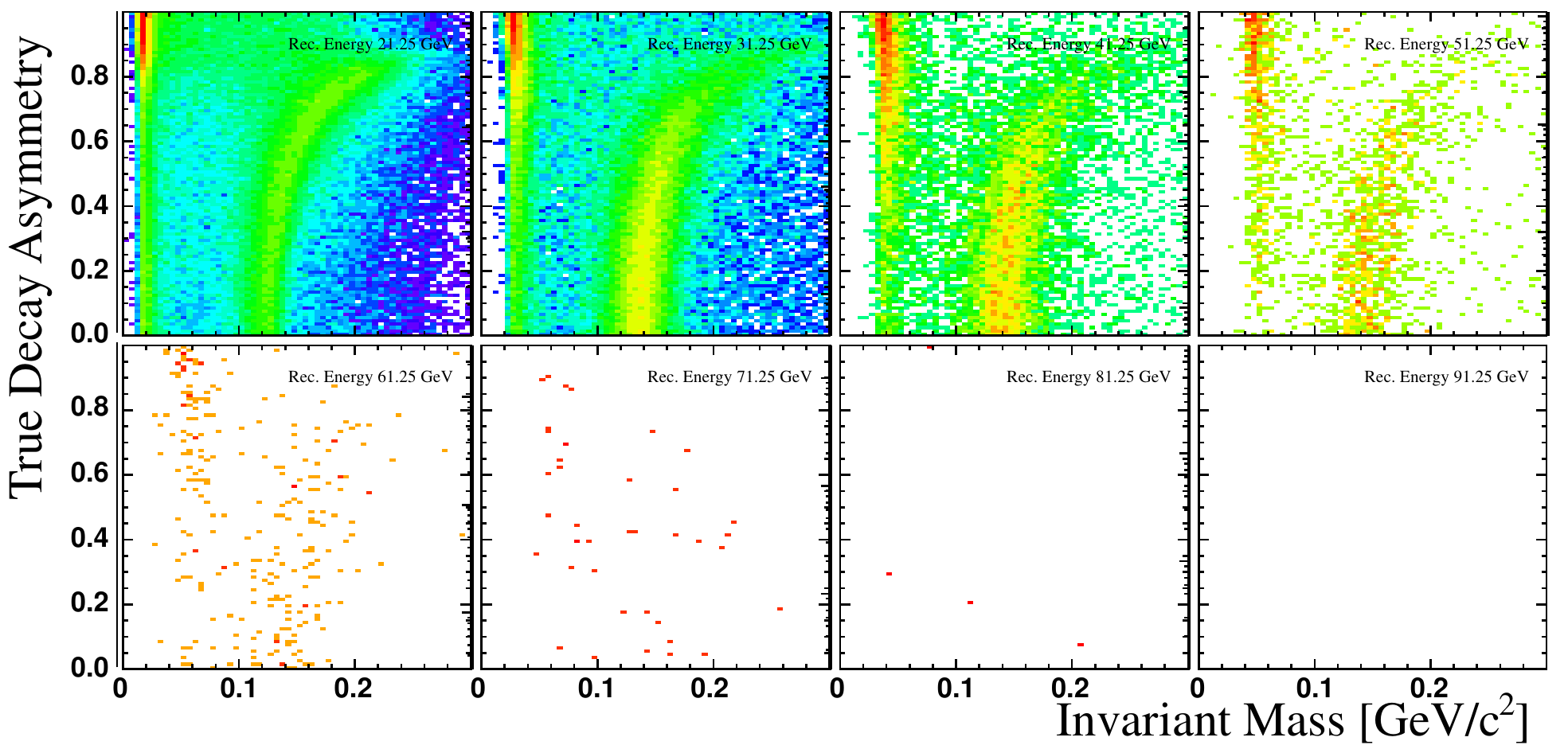}
\vspace*{-0.12in}
\caption{\label{fig:pyReco_asymmVsEn} Energy dependence of true asymmetry
versus invariant mass. }
\end{figure}

One can look at this in more detail with an energy dependence of the 
asymmetry, see Figure~\ref{fig:pyReco_asymmVsEn}.  One finds that the
fraction of mis-reconstructed mass is highest in the low-energy bins,
whilst at higher energy, the mass becomes better defined, even for
very asymmetry decays.

\clearpage
\subsection{Track Selection Cuts in {\sc Pythia} events}
\label{sim:dphotcuts}

The {\sc Pythia} simulation is analyzed to identify direct photon candidates.  
The event vertex must be within $\pm50$\,cm.  Tracks must fall within the MPC-EX 
$\eta$ acceptance (3.1$<$$|\eta|$$<$3.8), match to a unique MPC cluster, and have an 
MPC-to-MPC-EX hough separation below 0.005.  These requirements ensure that all 
surviving tracks have a similar acceptance, are properly reconstructed in the MPC-EX 
and are well matched to the MPC.  The studies presented in this section apply these 
cuts to ensure that the sample contains to only well reconstructed tracks. 

\begin{figure}[hbt]
  \hspace*{-0.12in} 
  \begin{minipage}[b]{0.5\linewidth}
    \centering
    \includegraphics[width=0.95\linewidth]{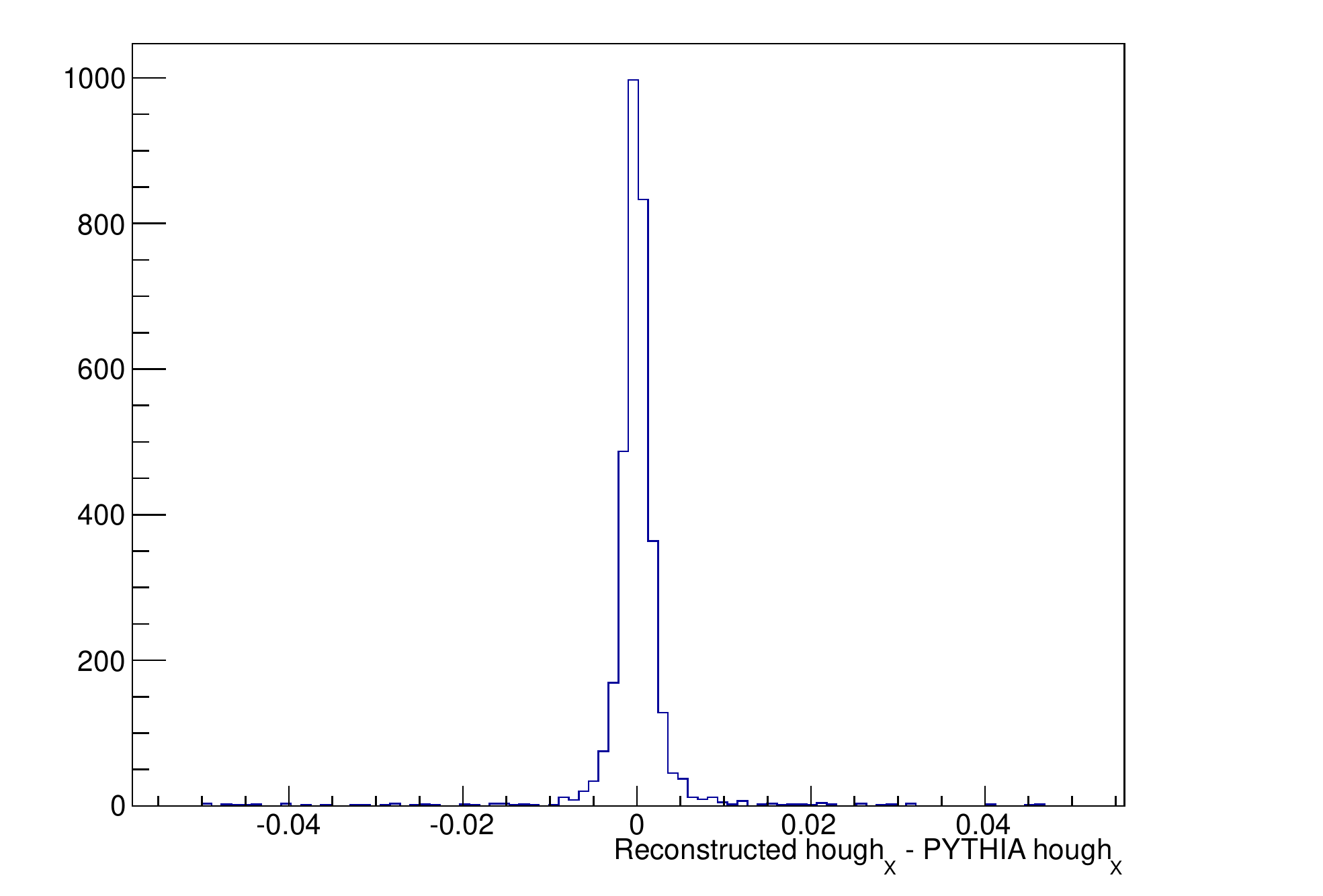}
  \end{minipage}
  \hspace{0.5cm}
  \begin{minipage}[b]{0.5\linewidth}
    \centering
    \includegraphics[width=0.95\linewidth]{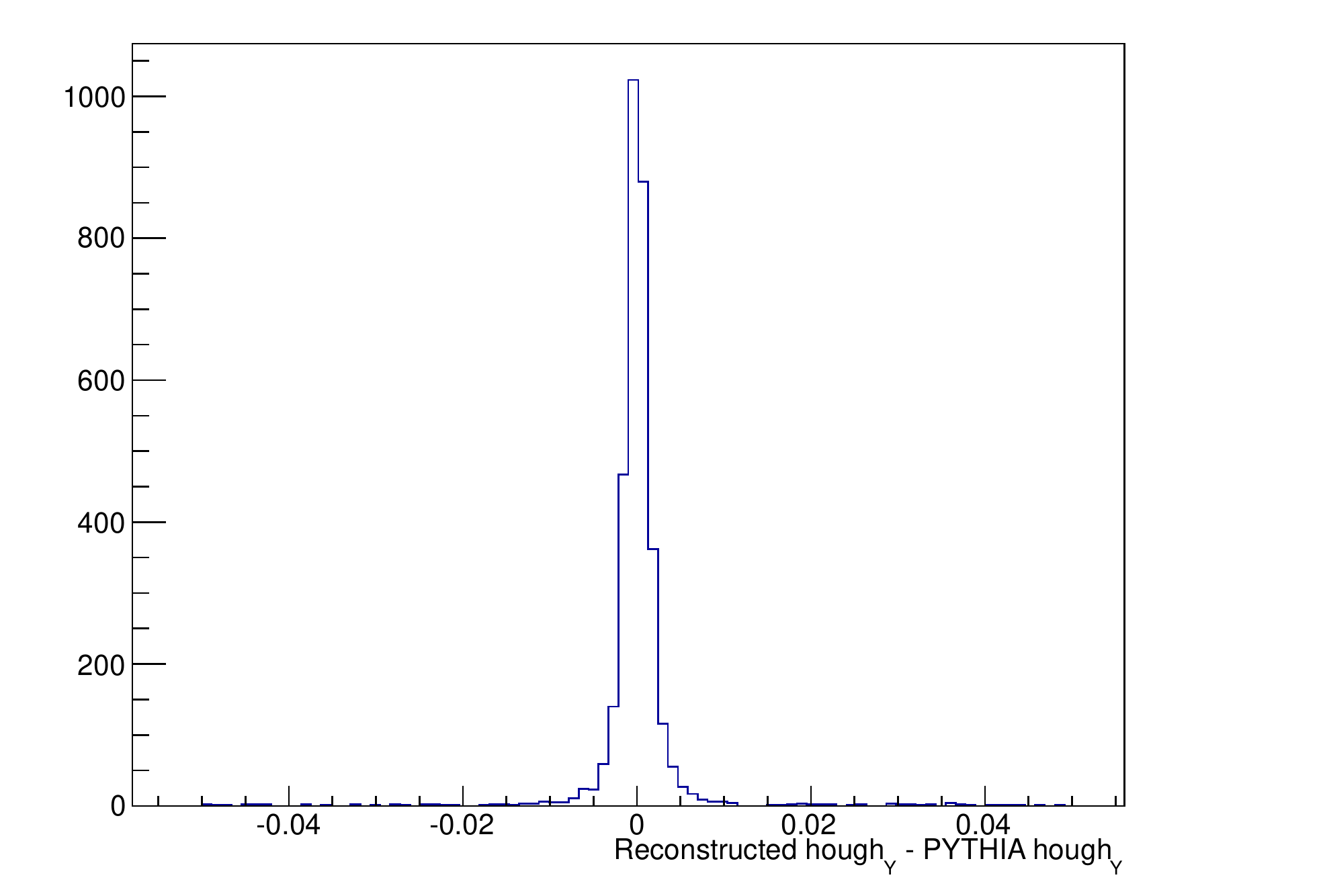}
  \end{minipage}
  \vspace*{-0.12in}
  \caption{\label{Fig:dhough} The separation between reconstructed tracks and 
    the {\sc Pythia} associated tracks in Hough space for the $x$- (left) 
    and $y$-coordinate (right) in a subset of the simulated data sample.}
\end{figure}

{\sc Pythia} primary information is used to identify the source of a photon 
candidate tracks.  Reconstructed tracks are associated to hadrons, $\pi^{0}$'s, 
decay photons, and direct photons.  The {\sc Pythia} association requires 
that the {\sc Pythia} primary and reconstructed track are in the same arm and 
have a Hough space separation of less than 0.02.  The Hough space separation 
between reconstructed tracks and the {\sc Pythia} primaries in the $x$ and $y$ 
coordinates are shown in Figure~\ref{Fig:dhough}.  Tracks 
that fail to match to a {\sc Pythia} primary are retained and classified 
as unassociated tracks.  After hadron removal cuts are applied unassociated tracks 
are less than one percent of the remaining yield.

Using the {\sc Pythia} association, photons can be separated 
into one of two categories: direct photons from the initial hard 
scattering and fragmentation photons from outgoing quark fragmentation.  
For the remainder of this section, the term ``signal photons'' refers to the 
sum of these sources.  There is also a third source of photons from QED 
radiation off of the incoming quark lines.  These are produced in the 
{\sc Pythia} simulation.  However, the amount of QED radiation is much larger 
than what is expected from NLO rate calculations. 
As a result we have excluded this contribution from our analysis.  We are 
working with theorists to understand the proper reweighting of the {\sc Pythia} 
components to ensure agreement with NLO calculations.  This is discussed
in more detail in Section~\ref{sim:dphot_pythia}.

The direct photon measurement must contend with large backgrounds from hadrons,
$\pi^{0}$'s and decay photons.  Decay photons primarily consist of $\eta$ decays, 
with a small contribution from $\omega$ and $\eta$' decays.  The first step in 
this analysis is to remove as much of the hadron, $\pi^{0}$ and decay photon 
backgrounds as possible.  The remaining backgrounds are removed through a double 
ratio calculation.  The double ratio method is discussed in detail in Section~\ref{sim:dphot}.

The goal of the direct photon measurement is to determine the gluon contribution 
to direct photons using the measured $R_{dA}$ suppression.  Direct photons 
from the initial hard interaction are more sensitive to suppression in the 
quark-gluon production mechanism than fragmentation photons radiated later in 
the collision.  Isolations cuts and tighter cuts on the MPC and 
MPC-EX variables reduce the fragmentation photon contribution to the signal 
photon measurement.

In this section, we present the cuts used to identify direct photon candidates.  
These cuts remove the hadronic tracks, reduce the $\pi^{0}$ and decay photon 
backgrounds, and lessen the fragmentation photon contribution.  The relevant 
variables are discussed and their distributions and cut efficiencies are shown.  
The efficiencies of the photon candidate tracks with all of the direct photon 
identification cuts applied are given at the end of this section.

\subsubsection{Hadronic background}
\begin{figure}[hbt]
  \hspace*{-0.12in}
  \centering
  \includegraphics[width=0.5\linewidth]{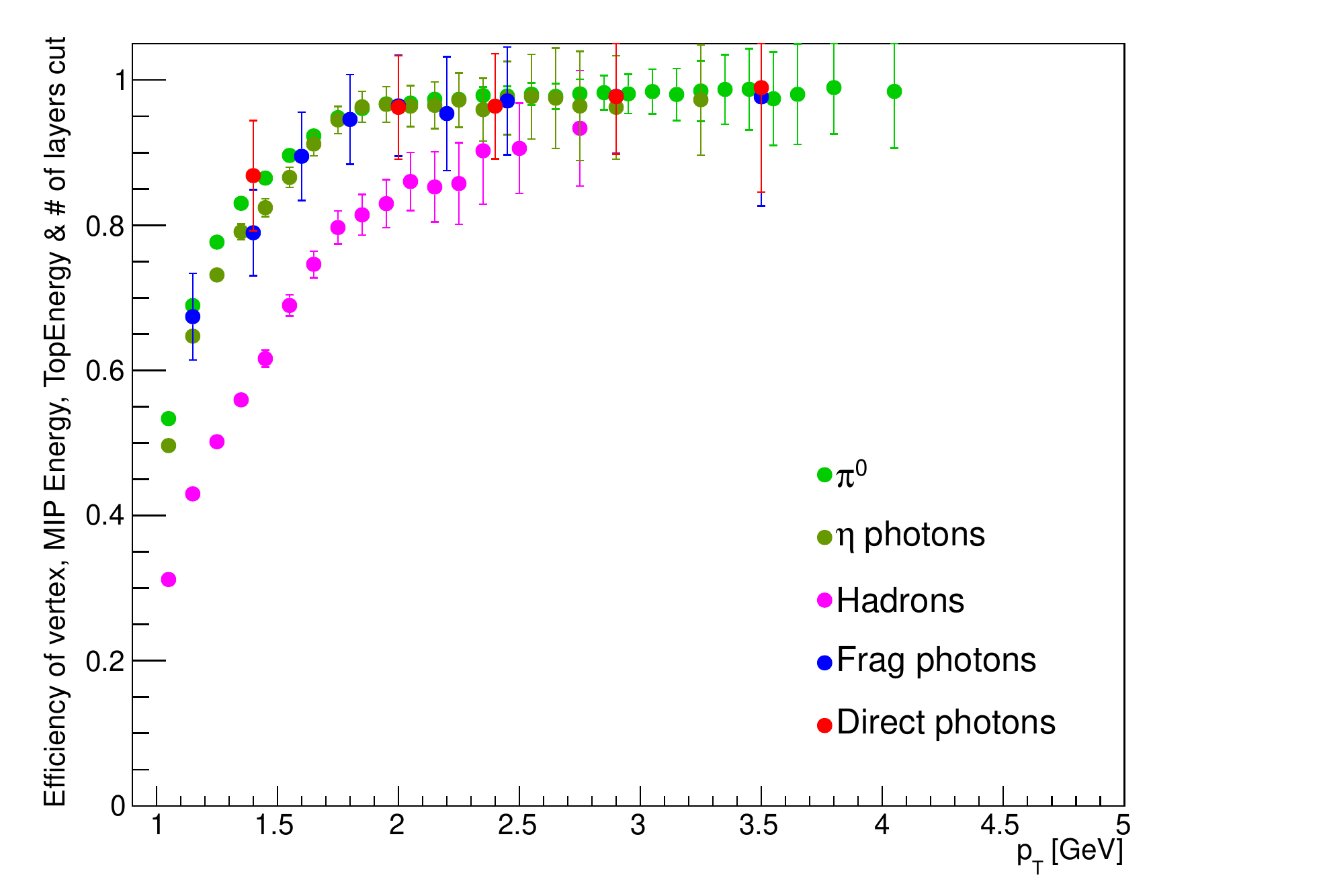}
  \vspace*{-0.12in}
  \caption{\label{Fig:Effi_good2}The efficiency distribution of the higest energy, 
    MIP energy, vertex and number of MPC-EX layers cuts as a funtion of $p_{T}$. 
    $\pi^{0}$ and $\eta$ are shown in bright and olive green respectively.  
    Hadrons are shown in pink.  Direct photons from the initial hard interaction 
    are in red and fragmentation photons are in dark blue. }
\end{figure}

The hadronic background is reduced using event characteristics and cuts that 
distinguish between hadronic tracks and electromagnetic showers in the MPC-EX.
In a direct photon event, the direct photons are the highest energy particles 
in the event.  By considering only the highest energy track in each event, 
81.6\% of the hadron tracks are removed with only a 3\% reduction in the signal 
photon yields.  Since the backgrounds are largest at low $p_{T}$, we require the 
$p_{T}$ to be greater than 3\,GeV and reduce the background contribution in the 
sample significantly.  This is particularly effective for the hadron and $\pi^{0}$ 
backgrounds both of which drop by over 99.5\%.  As charged hadrons pass through 
the MPC-EX, they deposit a small amount of their energy in the layers of the 
MPC-EX.  To remove these tracks we reject all candidates that deposit less than 
the minimum ionizing energy of 0.07\,GeV in a narrow region of interest and tracks 
with fewer than two layers of the MPC-EX hit.  The MIP requirement removes 62.9\%
of hadrons and 2.7\% of signal photons.  Figure \ref{Fig:Effi_good2} shows the 
efficiency of all of the hadron rejection cuts as a function of $p_{T}$.  Over 
the entire $p_{T}$ range hadrons and unassociated tracks are reduced by 92.7\% 
and 85.9\%. $\pi^{0}$, $\eta$ and other decays drop by 65.2\%, 66.3\% and 
64.5\% respectively.  Only 5.1\% of direct photons are affected.

The remainder of this section focuses on removing $\pi^{0}$ and fragmentation 
photons.  The hadron rejection cuts are applied throughout.  Cuts designed to 
remove $\pi^{0}$ and fragmentation photons further reduce the small 
backgrounds from hadronic tracks.

\subsubsection{$\pi^{0}$ background} \label{sec:pi0_back}

The $\pi^{0}$ background is removed with a variety of cuts on the energy 
characteristics and shower widths in the MPC-EX and MPC.  The track's shower 
width is considered and a shower shape comparison to a single particle 
distribution is made using a Kolmogrov test~\cite{ktest}.  The ratio of 
the track's energy to the amount of energy deposited in a cone surrounding 
the track is used to separate isolated tracks from hadrons and $\pi^{0}$ 
in a jet.  These variables and their distributions are presented below.  
The specific variable cut ranges are determined by using a multivariate 
analysis presented later in this section.

\begin{figure}[hbt]
  \hspace*{-0.12in}
  \begin{minipage}[b]{0.5\linewidth}
    \centering
    \includegraphics[width=0.95\linewidth]{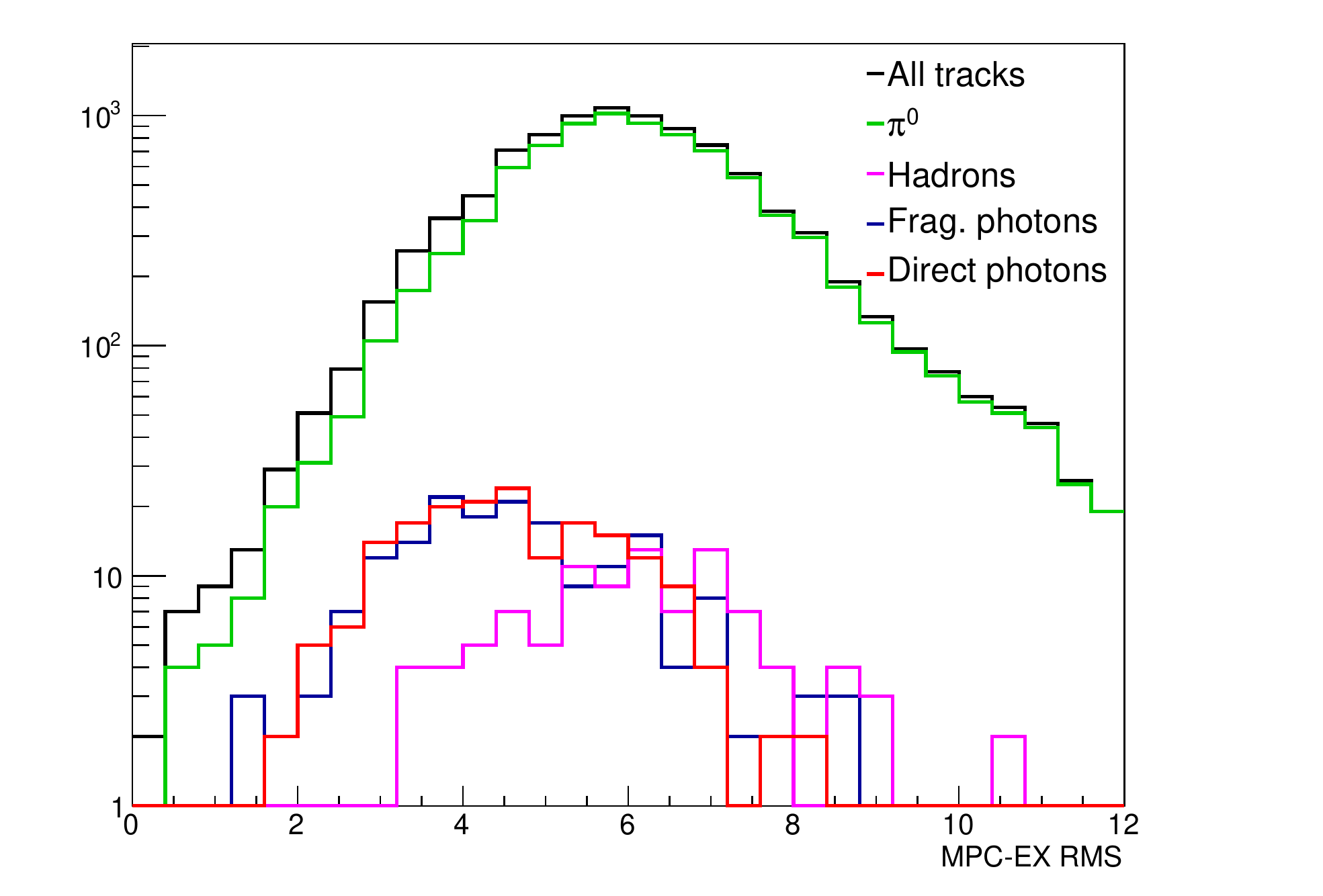}
  \end{minipage}
  \hspace{0.5cm}
  \begin{minipage}[b]{0.5\linewidth}
    \centering
    \includegraphics[width=0.95\linewidth]{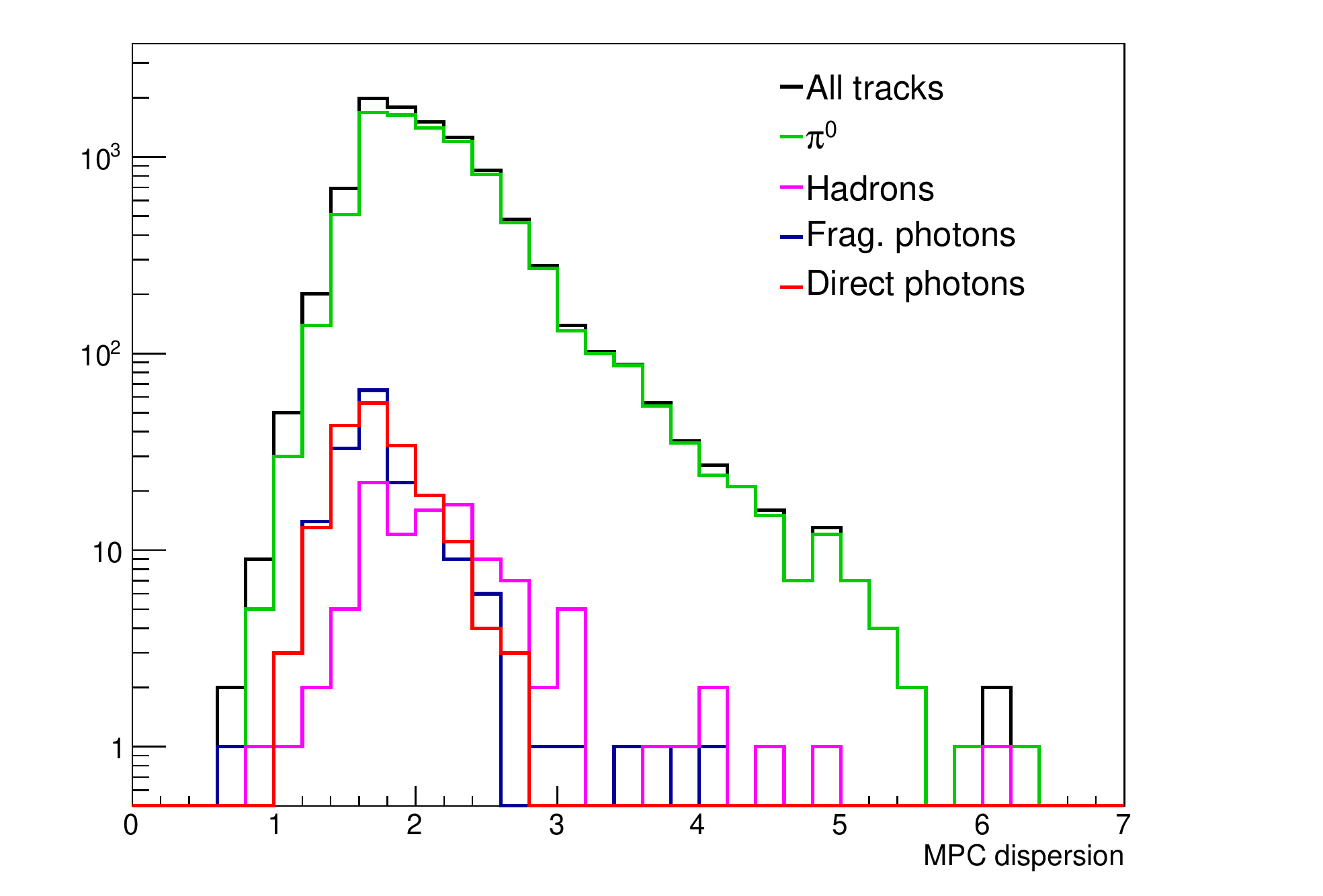}
  \end{minipage}
  \vspace*{-0.12in}
  \caption{\label{Fig:Width} The the MPC-EX RMS and MPC dispersion 
    distributions.  The right panel shows the RMS and the left 
    panel shows the dispersion.  Direct photon are shown in red and 
    fragmentation photons are in blue.  Hadrons and $\pi^{0}$ 
    tracks are in pink and green respectively.  The sum of all 
    tracks is shown in black.}
\end{figure}

The shower width is characterized in the MPC-EX using the root mean square 
(RMS) of the energy distribution.  The energy distributions are considered 
separately for the combined x and y layers and are summed in quadrature, 
$RMS = \sqrt{RMS_{X}^{2} + RMS_{Y}^{2}}$, where $RMS_{X}$ and $RMS_{Y}$ are 
the RMS of the showers in the x and y layers.  In the MPC, shower widths are 
determined using the dispersion of the shower shape in the MPC clusters.  
The MPC dispersion in the $x$ and $y$ directions are combined, 
$disp = \sqrt{(\log(dispx))^{2} + (\log(dispy))^{2}}$.  Figure \ref{Fig:Width} presents 
the MPC-EX RMS and MPC dispersion distributions for charged hadrons, $\pi^{0}$ 
and direct and fragmentation photons.  Hadrons and $\pi^{0}$'s peak at and 
extend out to higher values than the direct photons in both variables.  
This shift allows us to separate the $\pi^{0}$ and remaining hadrons from 
the direct photons.  A small shift is seen in the narrow peak of the $\pi^{0}$ 
dispersion distribution with a shoulder at $p_{T}$ greater than 2\,GeV.  
The separation between the direct photons and $\pi^{0}$'s  is more pronounced 
in the RMS distribution however the distribution is also a wider.

\begin{figure}[hbt]
  \hspace*{-0.12in}
  \centering
  \includegraphics[width=0.5\linewidth]{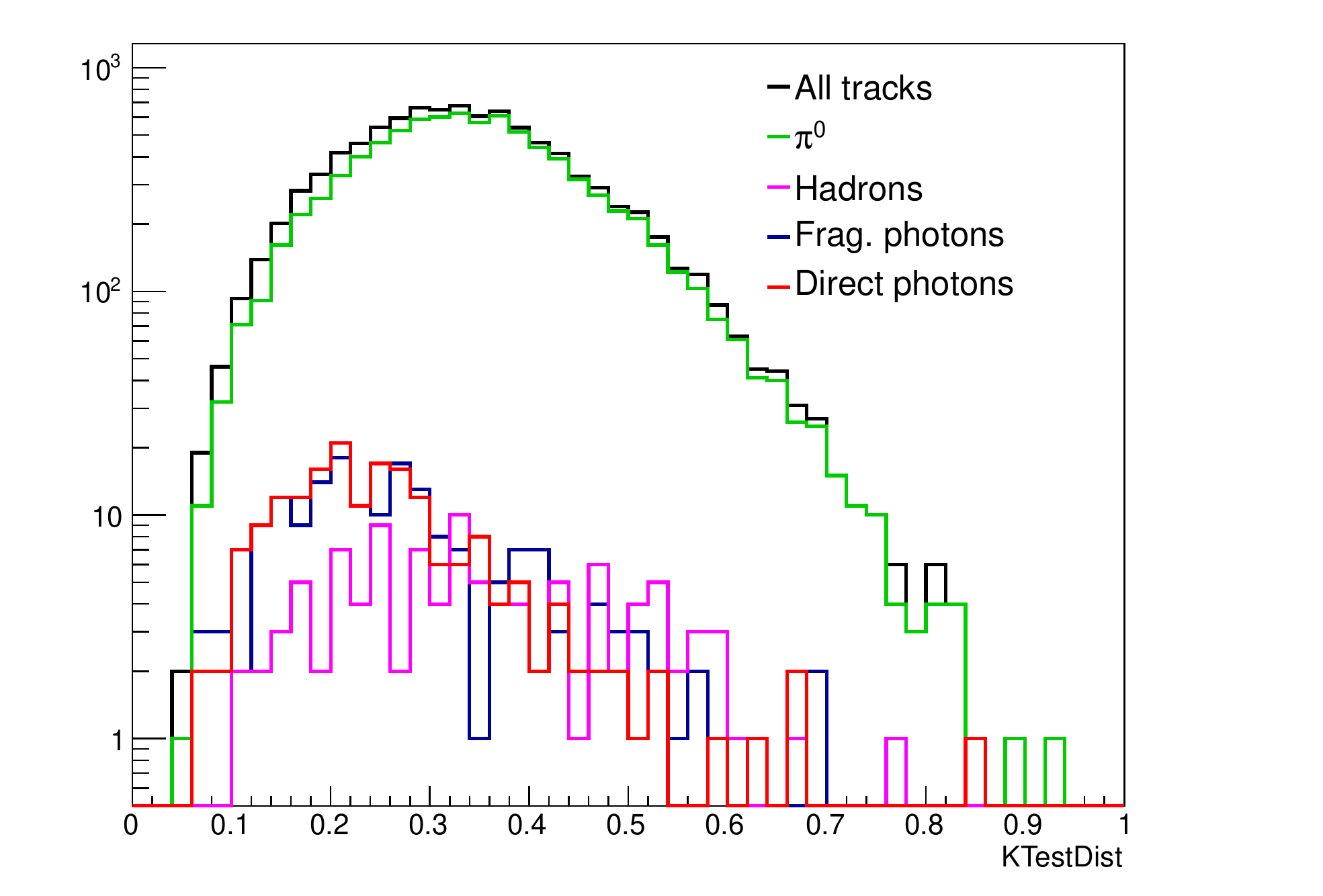}
  \vspace*{-0.12in}
  \caption{\label{Fig:KTest} The distribution of the Kolmogorov test
    variable (KTestDist) as a function of $p_{T}$.  
    The right panel shows the efficiency for with MPC energies 
    and the left panel MPC-EX energies.  Direct photon are shown 
    in red and fragmentation photons are in blue.  Hadron and $\pi^{0}$
    tracks are in pink and green respectively.  The sum of 
    all tracks is shown in black.}
\end{figure}

The Kolmogorov test compares the expected shower shape of the candidate track 
with those from single particle $\pi^{0}$ simulations by calculating the 
distance between the two energy profiles.  The Kolmogorov test is performed 
separately on the combined x layers and y layers.  These values are restricted 
between zero and one and are summed in quadrature creating the KTestDist 
variable.  Figure~\ref{Fig:KTest} displays the KTestDist distribution for all 
tracks, $\pi^{0}$, hadrons and fragmentation and direct photons.  The direct 
and fragmentation photon distributions are peaked near a value of 0.2 and 
the $\pi^{0}$ distribution peaks at a value above 0.3. 

\begin{figure}[hbt]
  \hspace*{-0.12in}
  \begin{minipage}[b]{0.5\linewidth}
    \centering
    \includegraphics[width=0.95\linewidth]{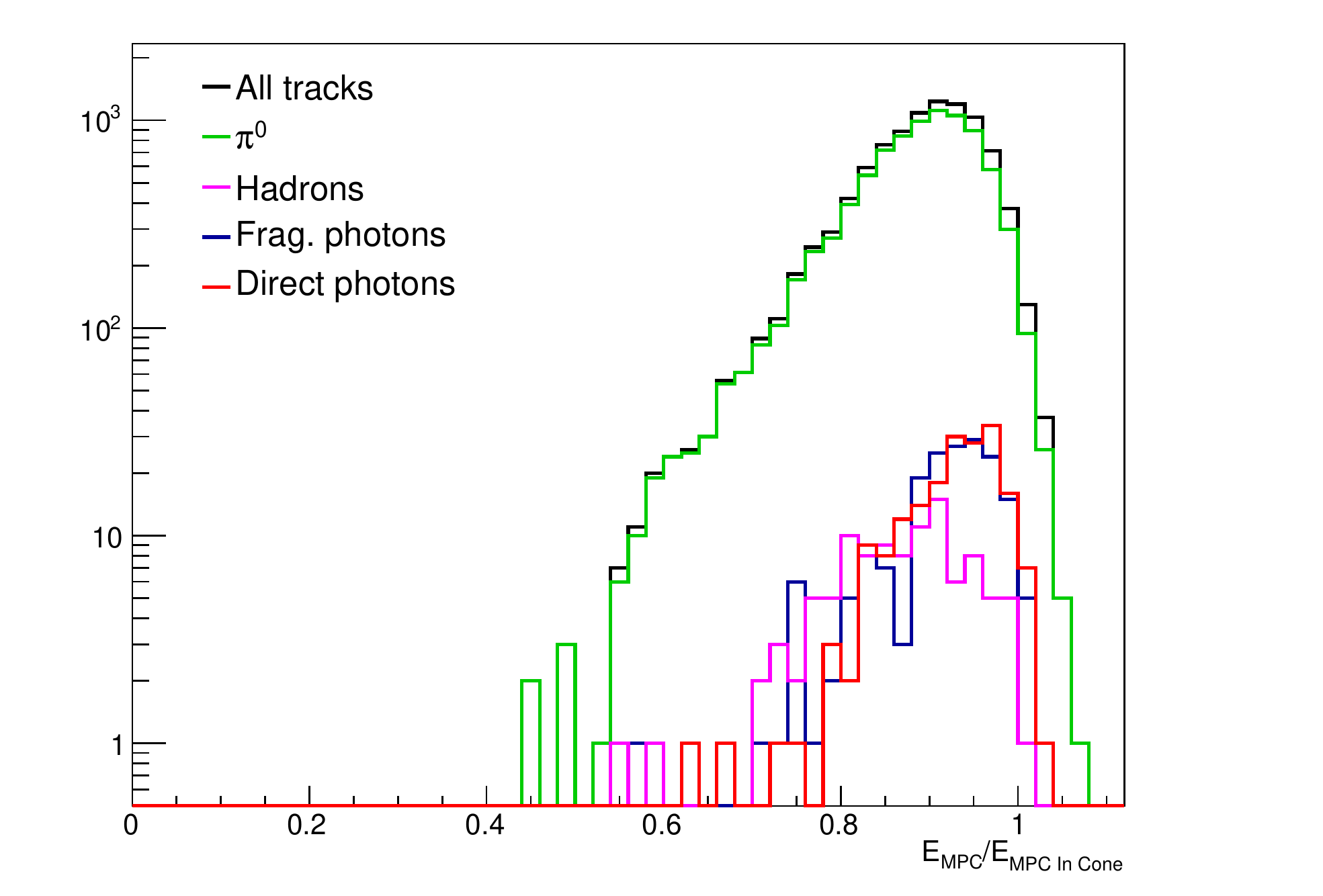}
  \end{minipage}
  \hspace{0.5cm}
  \begin{minipage}[b]{0.5\linewidth}
    \centering
    \includegraphics[width=0.95\linewidth]{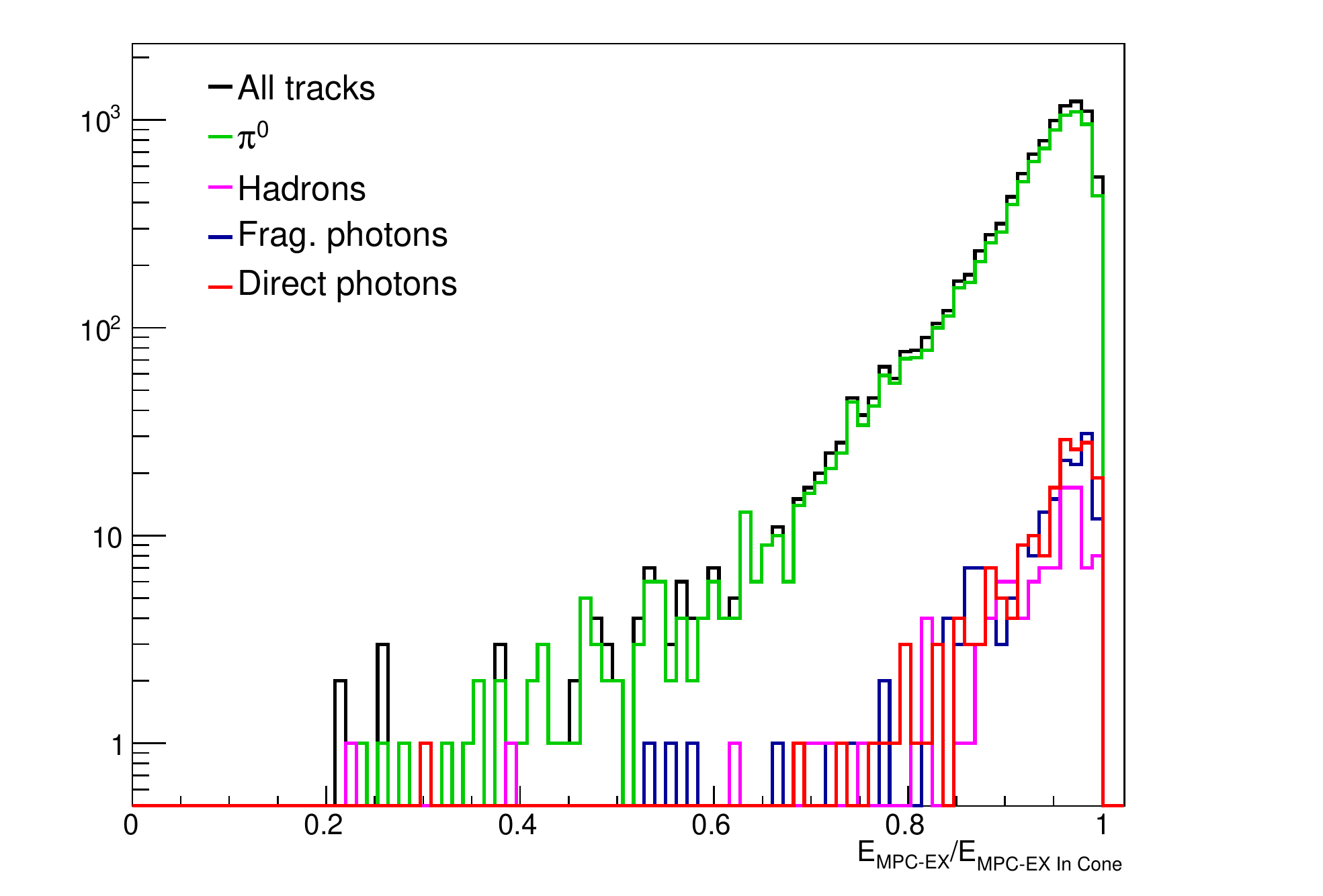}
  \end{minipage}
  \vspace*{-0.12in}
  \caption{\label{Fig:EIC} The distribution of the energy ratio in cone 
    cuts with the energy in the MPC and MPC-EX as a function of $p_{T}$.  
    The right panel shows the efficiency for with MPC energies and the 
    left panel MPC-EX energies.  Direct photon are shown in red and 
    fragmentation photons are in blue.  Hadrons and $\pi^{0}$ are shown 
    in pink and green respectively.  The sum of all tracks is 
    shown in black.}
\end{figure}

Ratios of the track's energy to the amount of energy deposited 
in a one radius cone in $\eta$-$\phi$ space surrounding the track
are calculated separately for the MPC and MPC-EX energies.  
Figure \ref{Fig:EIC} shows both of these distributions for all 
tracks, $\pi^{0}$, hadrons and fragmentation and direct photons.
These ratios are peaked near one for all particles.  Values of one 
occur when the photon candidate track is isolated and is the only 
source of energy in the cone.  The distribution of the MPC ratio 
peaks slightly below one and extends to values above one.  This is
because the MPC energy of the track is the MPC cluster energy, 
while the MPC energy in the cone is uncalibrated and unclustered.

\begin{figure}[hbt]
  \hspace*{-0.12in}
  \begin{minipage}[b]{0.5\linewidth}
    \centering
    \includegraphics[width=0.95\linewidth]{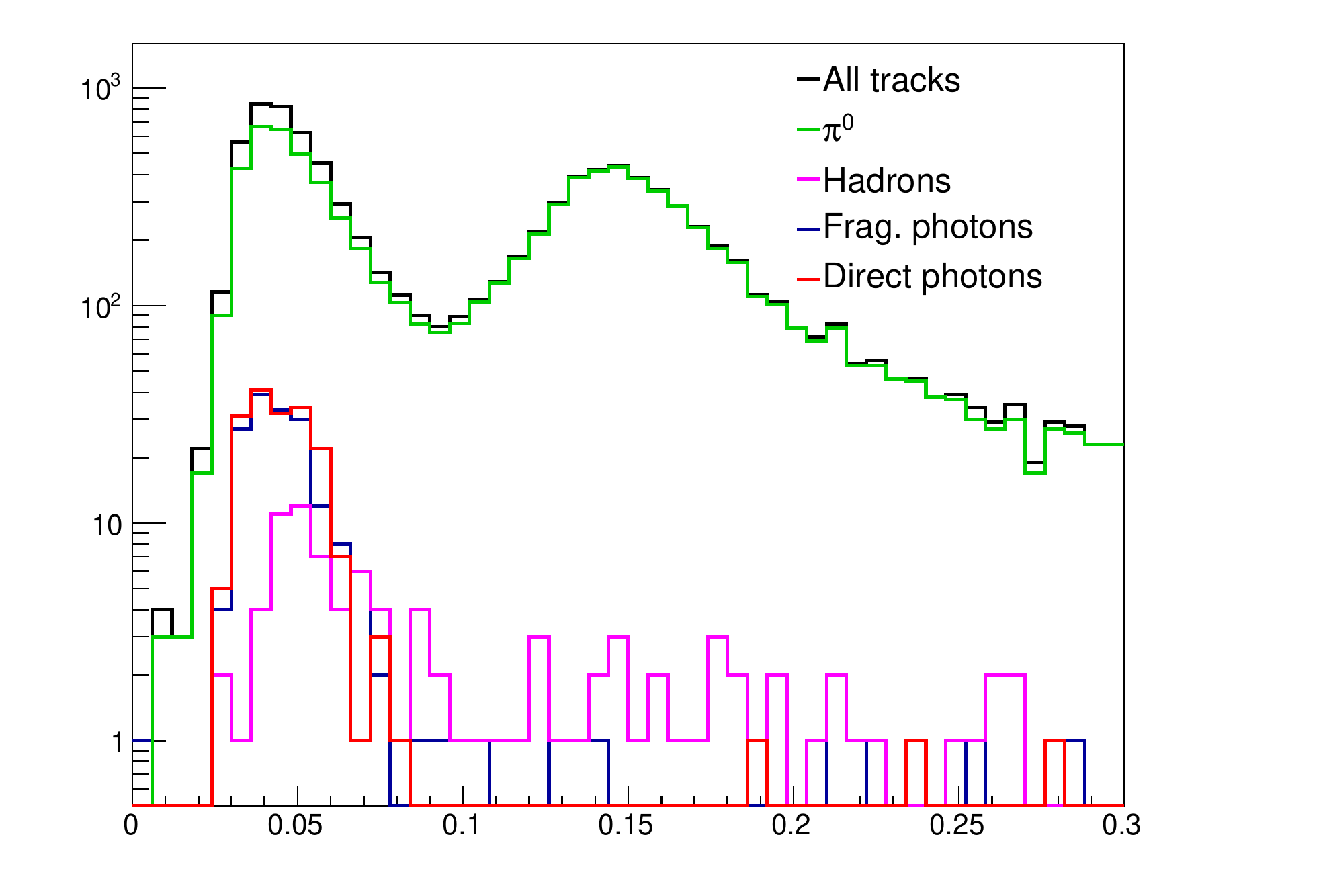}
  \end{minipage}
  \hspace{0.2cm}
  \begin{minipage}[b]{0.5\linewidth}
    \centering
    \includegraphics[width=0.95\linewidth]{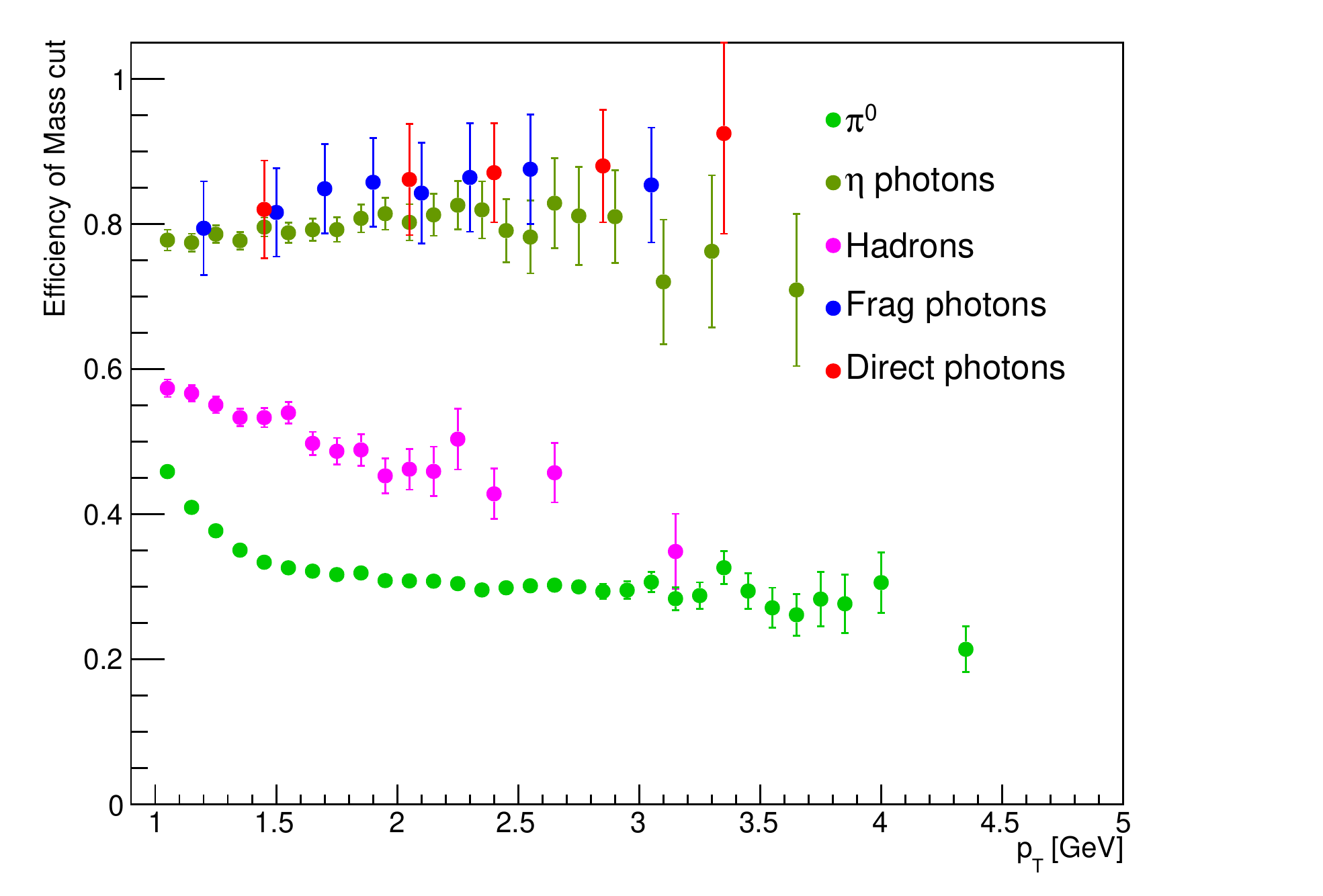}
  \end{minipage}
  \vspace*{-0.12in}
  \caption{\label{Fig:Mass}The single track invariant mass distributions 
    of all high $p_{T}$ good tracks on the left and the efficiency of 
    a $m < 0.06$\,GeV cut on the right.  $\pi^{0}$ and $\eta$ are in light
    and olive green.  Hadrons are in pink.  Direct photons are in red and 
    fragmentation photons are in blue. }
\end{figure}

Before cutting on these variables, we first remove known
$\pi^{0}$'s from our sample.  The single track invariant
mass variable provides this rejection.  The calculation 
of the single track invariant mass is discussed in detail 
in Section~\ref{sec:SingleTrackInvMass}.  The mass 
distributions for $\pi^{0}$'s, hadrons and fragmentation 
and direct photons are shown in Figure~\ref{Fig:Mass}.  
The $\pi^{0}$'s have a clear peak at the pion mass and a 
low mass peak at 0.04\,GeV.  This low mass peak occurs 
when the mass reconstruction algorithm identifies an energy 
fluctuation as the basis for the second photon in the 
$\pi^{0}$ decay.  The low mass peak is seen in all particle 
types.  A cut requiring a single track mass below 0.06\,GeV
removes the properly reconstructed $\pi^{0}$ while retaining
the direct photons in the low mass peak.  The efficiency of 
this cut is also shown in Figure~\ref{Fig:Mass} and listed 
in Table~\ref{Tab:CutsEffi}.  Over 70.9\% of high $p_{T}$ 
$\pi^{0}$'s are removed and high $p_{T}$ direct photons 
survive with 90.3\% efficiency.

\begin{figure}[hbt]
  \hspace*{-0.12in}
  \centering
  \includegraphics[width=0.5\linewidth]{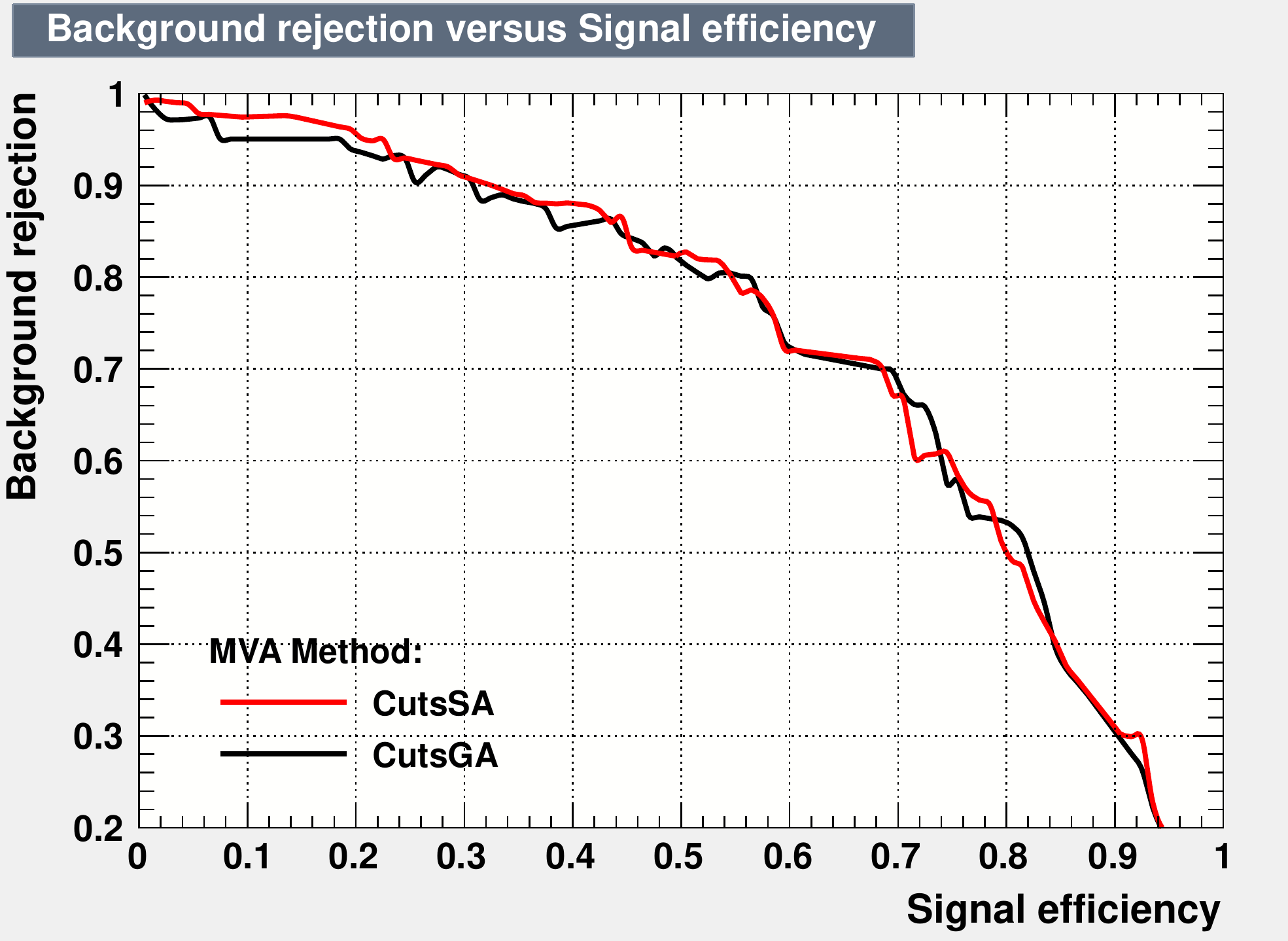}\\
  \caption{\label{Fig:TMVA}The signal efficiency versus background rejection
    curves (ROC curves) for the TMVA analysis of the direct and fragmentation 
    photon signal and the $\pi^{0}$ background.  The projected signal efficiencies 
    and background rejections for the CutsSA and CutsGA curves are drawn in red and 
    black respectively.}
\end{figure}

A multi-variate analysis tool, TMVA, is used to determine the optimal cut 
locations of the MPC and MPC-EX variables to remove the $\pi^{0}$ background 
and maintain the direct photon signal \cite{TMVA}.  This is performed on the 
$p_{T} > 3$\,GeV {\sc Pythia} sample after the invariant mass and hadron cuts 
are applied.  TMVA characterizes each of the variable distributions and 
dependencies for both the signal and background tracks.  It determines optimal 
cuts for signal efficiencies between zero and one hundred percent with a one 
percent step size.  Two rectangular cut algorithms are attempted in this study, 
the simulated annealing algorithm (CutsSA) and the genetic algorithm (CutsGA).   
Early studies also considered a Monte Carlo generation algorithm but this 
algorithm produced lower background rejections.  The $\pi^{0}$ background is 
the only background considered in this TMVA analysis; the signal consists of 
both direct and fragmentation photons.  Figure~\ref{Fig:TMVA} presents the 
projected direct and fragmentation photon efficiencies versus $\pi^{0}$ 
rejection (ROC curve) for both algorithms.  The CutsSA algorithm is used in 
this analysis.  It consistently produces cuts within reasonable ranges and a 
more reliable and higher predictions of the signal efficiencies.  The cuts 
determined by the CutsSA algorithm with a projected direct and fragmentation 
photon efficiency of $28\%$ and a projected $85\%$ rejection of $\pi^{0}$ are 
used and the corresponding acceptable variable ranges are listed in 
Table~\ref{Tab:pi0Cuts}.  

\begin{table}
\centering
\caption{Acceptable ranges for variables as defined by TMVA to remove the $\pi^{0}$ background.}
\label{Tab:pi0Cuts}
\begin{tabular}{|l||c|} \hline 
Variable                        &      Range       \\ \hline
dispersion                      &   0.812 - 1.631  \\
RMS                             &   0.663 - 6.118  \\
KTestDistX                      &   0.105 - 0.556  \\
$E_{MPC-EX}/E_{MPC-EX In Cone}$ &   $>$ 0.767      \\
$E_{MPC}/E_{MPC In Cone}$       &   0.727 - 1.013  \\  \hline
\end{tabular}
\end{table}

\begin{table}
\centering 
\caption{Efficiencies of the cuts in Table \ref{Tab:pi0Cuts} for $\pi^{0}$, hadrons, direct and fragmentation photons for yields at $p_{T} > 3$\,GeV.}
\label{Tab:CutsEffi}
\begin{tabular}{|l||c|c|c|c|} \hline 
Variable & $\epsilon_{\pi^{0}}$ & $\epsilon_{hadrons}$ & $\epsilon_{\gamma frag}$ & $\epsilon_{\gamma direct}$ \\ \hline
Mass                            &  $29.1\%$  &  $37.5\%$  &  $84.2\%$  &   $90.3\%$  \\  
dispersion                      &   $8.6\%$  &   $9.6\%$  &  $35.0\%$  &   $35.5\%$  \\  
RMS                             &  $51.9\%$  &  $48.1\%$  &  $81.4\%$  &   $84.4\%$  \\  
KTestDist                       &  $48.3\%$  &  $58.4\%$  &  $61.0\%$  &   $71.5\%$  \\  
$E_{MPC-EX}/E_{MPC-EX In Cone}$ &  $48.9\%$  &  $46.2\%$  &  $64.4\%$  &   $74.2\%$  \\ 
$E_{MPC}/E_{MPC In Cone}$       &  $49.2\%$  &  $47.1\%$  &  $65.5\%$  &   $74.2\%$  \\ \hline 
All $\pi^{0}$ cuts applied      &   $2.9\%$  &   $6.1\%$  &  $24.3\%$  &   $31.2\%$  \\ \hline
\end{tabular}
\end{table}

\begin{figure}[hbt]
  \hspace*{-0.12in}
  \begin{minipage}[b]{0.5\linewidth}
    \centering
    \includegraphics[width=0.95\linewidth]{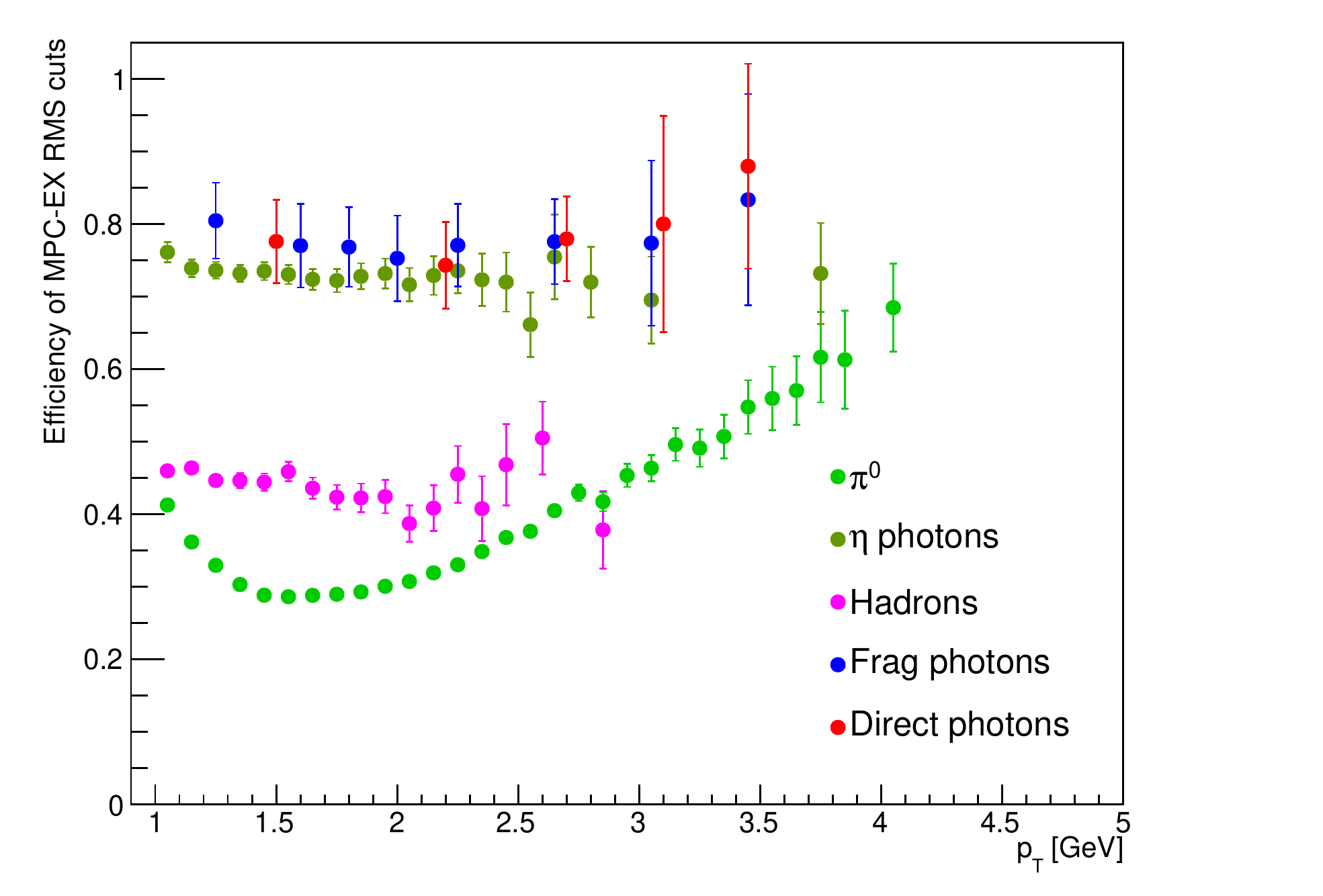}
  \end{minipage}
  \hspace{0.5cm}
  \begin{minipage}[b]{0.5\linewidth}
    \centering
    \includegraphics[width=0.95\linewidth]{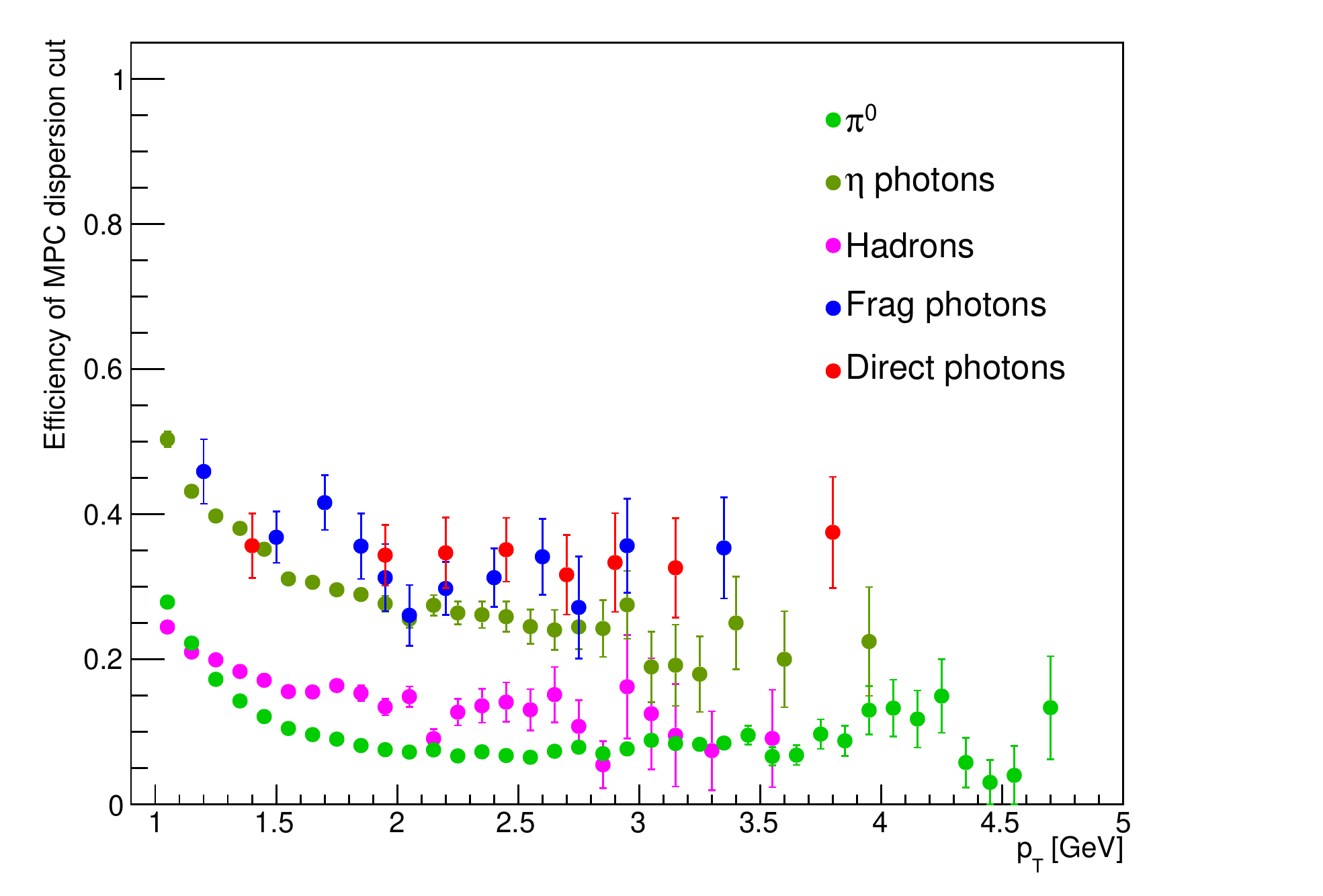}
  \end{minipage}
  \vspace*{-0.12in}
  \caption{\label{Fig:Effi_width}The efficiency of the MPC-EX RMS and MPC 
    dispersion cuts as a function of $p_{T}$.  The right panel shows the 
    efficiency for the MPC-EX RMS cut and the left panel the MPC dispersion.  
    $\pi^{0}$ and $\eta$ decays are shown in bright and olive green respectively.  
    Hadrons are shown in pink. Direct photons are shown
    in red and fragmentation photons in dark blue. }
\end{figure}

\begin{figure}[hbt]
  \hspace*{-0.12in}
  \centering
  \includegraphics[width=0.5\linewidth]{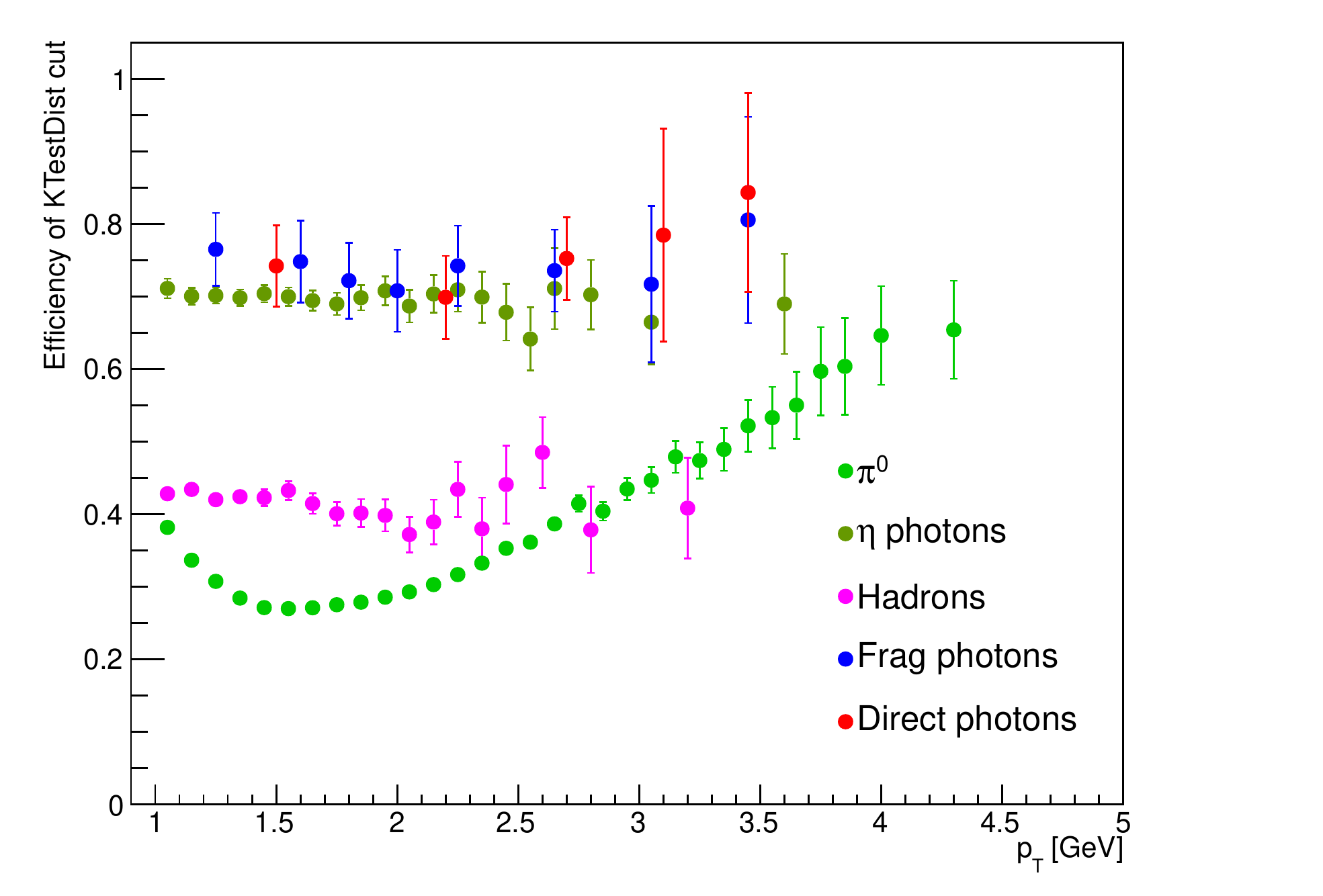}
  \vspace*{-0.12in}
  \caption{\label{Fig:Effi_KT}The efficiency of the KTestDist cut as a function of 
    $p_{T}$.  $\pi^{0}$ and $\eta$ decays are shown in bright and olive 
    green respectively.  Hadrons are shown in pink.  Direct photons are shown
    in red and fragmentation photons in dark blue. }
\end{figure}

\begin{figure}[hbt]
  \hspace*{-0.12in}
  \begin{minipage}[b]{0.5\linewidth}
    \centering
    \includegraphics[width=0.95\linewidth]{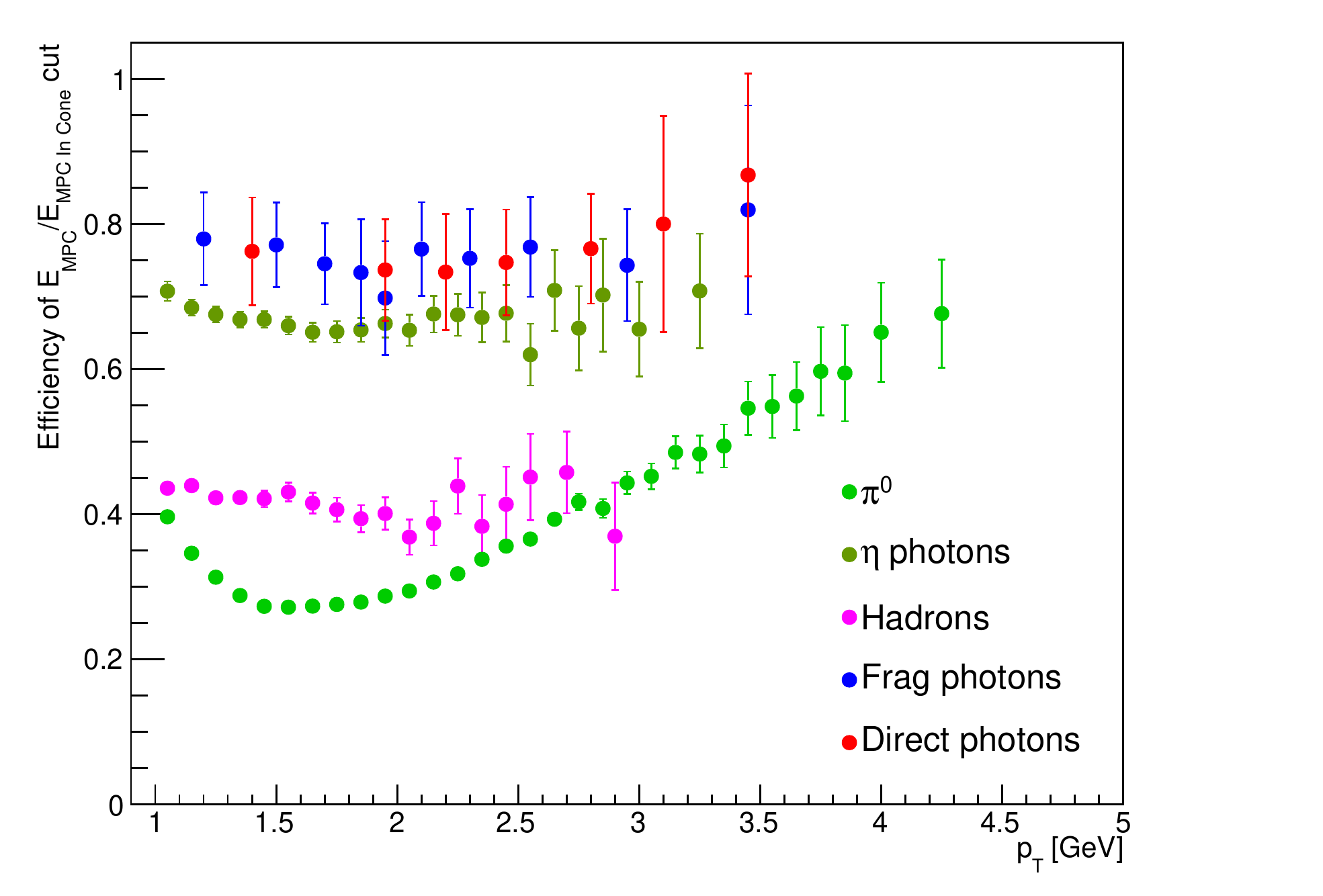}
  \end{minipage}
  \hspace{0.2cm}
  \begin{minipage}[b]{0.5\linewidth}
    \centering
    \includegraphics[width=0.95\linewidth]{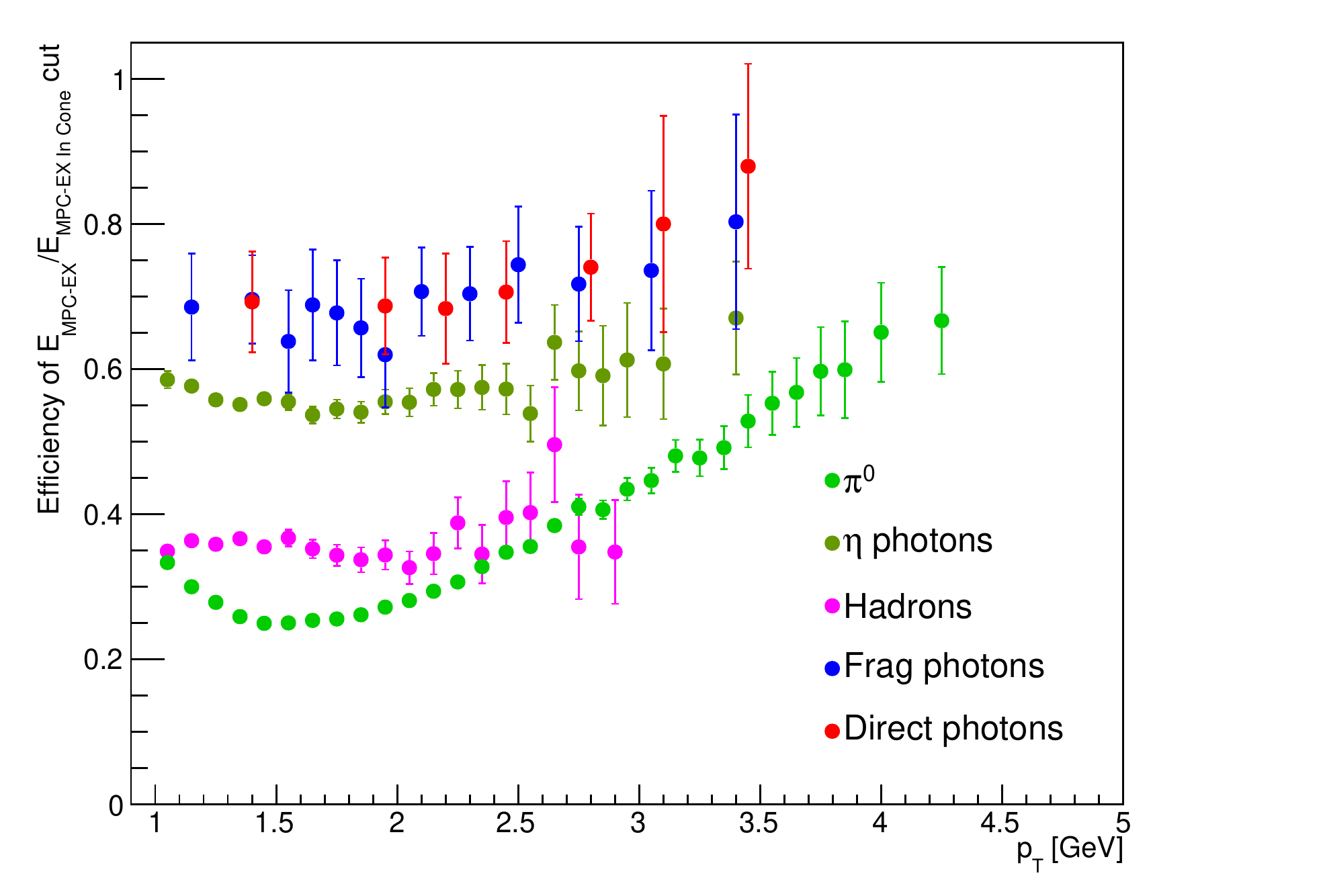}
  \end{minipage}
  \vspace*{-0.12in}
  \caption{\label{Fig:Effi_EIC}The efficiency of the energy ratio in cone cuts 
    considering the energies in the MPC and MPC-EX as a function of $p_{T}$.  
    The right panel shows the efficiency using the MPC energies and the left 
    panel shows the ratio with MPC-EX energies.  $\pi^{0}$ and $\eta$ decays 
    are shown in bright and olive green respectively.  Hadrons are shown in 
    pink.  Direct photons are in red and fragmentation photons in dark blue. }
\end{figure}

Table \ref{Tab:CutsEffi} presents the efficiencies of each cut for $\pi^{0}$, 
hadrons, and direct and fragmentation photons above 3\,GeV in $p_{T}$.
Figures~\ref{Fig:Effi_width},~\ref{Fig:Effi_KT} and~\ref{Fig:Effi_EIC}, show 
the efficiencies for each of these cuts as a function of $p_{T}$ for $\pi^{0}$,
hadrons and direct and fragmentation photons.  All five variables are able to 
separate $\pi^{0}$ from direct and fragmentation photons.  The dispersion cut
has the lowest efficiency for both the signal photons and $\pi^{0}$.  The 
mass cut provides a large separation between the $\pi^{0}$ background 
and signal photons while maintaining a high direct photon efficiency.

\begin{figure}[hbt]
  \hspace*{-0.12in}
  \centering
  \includegraphics[width=0.5\linewidth]{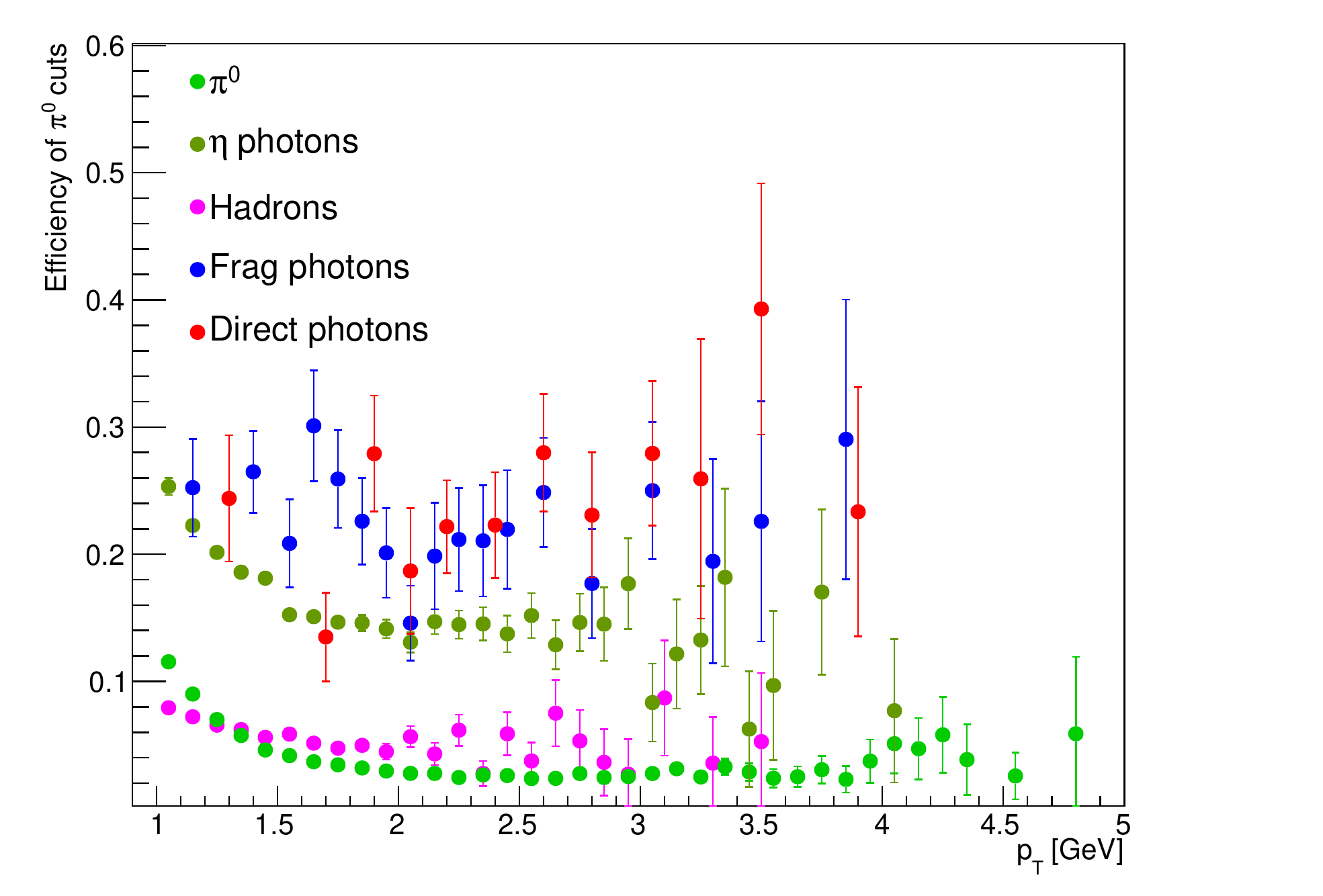}
  \vspace*{-0.12in}
  \caption{\label{Fig:Effi_pi0}The efficiency of the photon candidate cuts 
    as a function of $p_{T}$.  $\pi^{0}$ and $\eta$ are shown in bright and 
    olive green.  Hadrons are shown in pink.  Direct photons are in red and 
    fragmentation photons in dark blue.}
\end{figure}

Table~\ref{Tab:pi0CutsYieldEffi} presents the yields and efficiencies 
for all particle types in this analysis.  Figure~\ref{Fig:Effi_pi0} 
shows the efficiency distribution of the photon candidates as a function 
of $p_{T}$.  Direct photons are the most efficient particle type, with an 
efficiency of $31.2\%$ at a $p_{T}$ above 3~GeV.  Fragmentation photons 
have a slightly lower efficiency of $23.2\%$ at high $p_{T}$.  Hadrons and 
$\pi^{0}$ background efficiencies are $5.1\%$ and $2.8\%$ respectively.  
These cuts separate hadrons and $\pi^{0}$ backgrounds from the direct photon 
signal.  

\begin{table}
\centering 
\caption{Particle yields and efficiencies with all of the cuts in Table \ref{Tab:pi0Cuts} applied.}
\label{Tab:pi0CutsYieldEffi}
\begin{tabular}{|l||c|c|} \hline 
Particle Type                &   Yield  &  Efficiency  \\ \hline
Charged hadrons              &       6  &    6.1$\%$   \\
\hspace{5mm} $\pi$           &       5  &   11.4$\%$   \\
\hspace{5mm} K               &       0  &    0.0$\%$    \\ 
\hspace{5mm} p               &       0  &    0.0$\%$    \\ 
\hspace{5mm} neutrons        &       0  &    0.0$\%$    \\
\hspace{5mm} other           &       1  &    1.9$\%$    \\
$\pi^{0}$                    &     234  &    2.9$\%$    \\
$\eta$ decay                 &      51  &   11.5$\%$    \\
Other decays                 &      11  &    9.6$\%$    \\
\hspace{5mm} $\omega$        &       9  &    9.6$\%$    \\
\hspace{5mm} $\eta$'         &       1  &    5.3$\%$    \\
\hspace{5mm} other           &       1  &  100.0$\%$    \\
Signal photons               &     101  &   27.8$\%$    \\ 
\hspace{5mm} frag photons    &      43  &   24.3$\%$    \\ 
\hspace{5mm} direct photons  &      58  &   31.2$\%$    \\ \hline 
\end{tabular}
\end{table}

\subsubsection{Fragmentation photons} \label{sec:frag_back}

While fragmentation photons are included in NLO calculations 
within the MPC-EX acceptance, PHENIX measurements of fragmentation 
photons~\cite{ppg095} indicate that fragmentation photons are not a 
significant fraction of the signal photon yield at midrapidity.  
Despite the uncertainty in the expected levels of fragmentation 
photon production, we design cuts to increase the relative 
contribution of direct photons to the signal and reduce the 
fragmentation photons.  This allows the direct photon measurement 
to more easily access the gluon contribution to the measured photon 
signal since fragmentation photons lower the sensitivity of the signal 
photon $R_{dA}$, as seen in Section~\ref{sec:cnm}.  In this section, 
we consider our ability to separate fragmentation and direct photons 
with the MPC-EX detector.

For the direct photon measurement, both the fragmentation and direct 
photons are measured as signal.  The analysis in Section~\ref{sec:pi0_back} 
focuses primarily on removing the $\pi^{0}$ background and results in 
101 signal photons, 58 of which are direct photons.  A direct photon 
concentration of 57.4\%.  This is an increase from the roughly equal 
contribution of fragmentation and direct photons before the $\pi^{0}$ 
rejection cuts are applied.  The contribution from fragmentation photons 
is reduced even further by tightening the MPC and MPC-EX variables 
introduced in Section~\ref{sec:pi0_back} and using new variables to 
differentiate between fragmentation and direct photons.  Here we present 
two additional analyses to highlight our ability to remove fragmentation photons.

Direct photons are produced with no neighboring particles while fragmentation 
photons are within a jet of correlated particles.  Isolation variables, such as 
the number of tracks within a cone around the candidate track and the ratio of the
track's energy to the amount of energy deposited in a cone around the track,
are able remove fragmentation photons.  These isolation variables are designed 
to reject fragmentation photons, but they can also reduce $\pi^{0}$ and hadrons 
from jets. The energy ratio in cone variables, $E_{MPC-EX}/E_{MPC-EX In Cone}$ 
and $E_{MPC}/E_{MPC In Cone}$, shown in Figure \ref{Fig:EIC}, successfully 
reduce $\pi^{0}$ and hadrons with the cuts presented in Section~\ref{sec:pi0_back}.
By tightening these energy ratios, fragmentation photons can also be removed from 
the sample.

\begin{figure}[hbt]
  \hspace*{-0.12in}
  \begin{minipage}[b]{0.5\linewidth}
    \centering
    \includegraphics[width=0.95\linewidth]{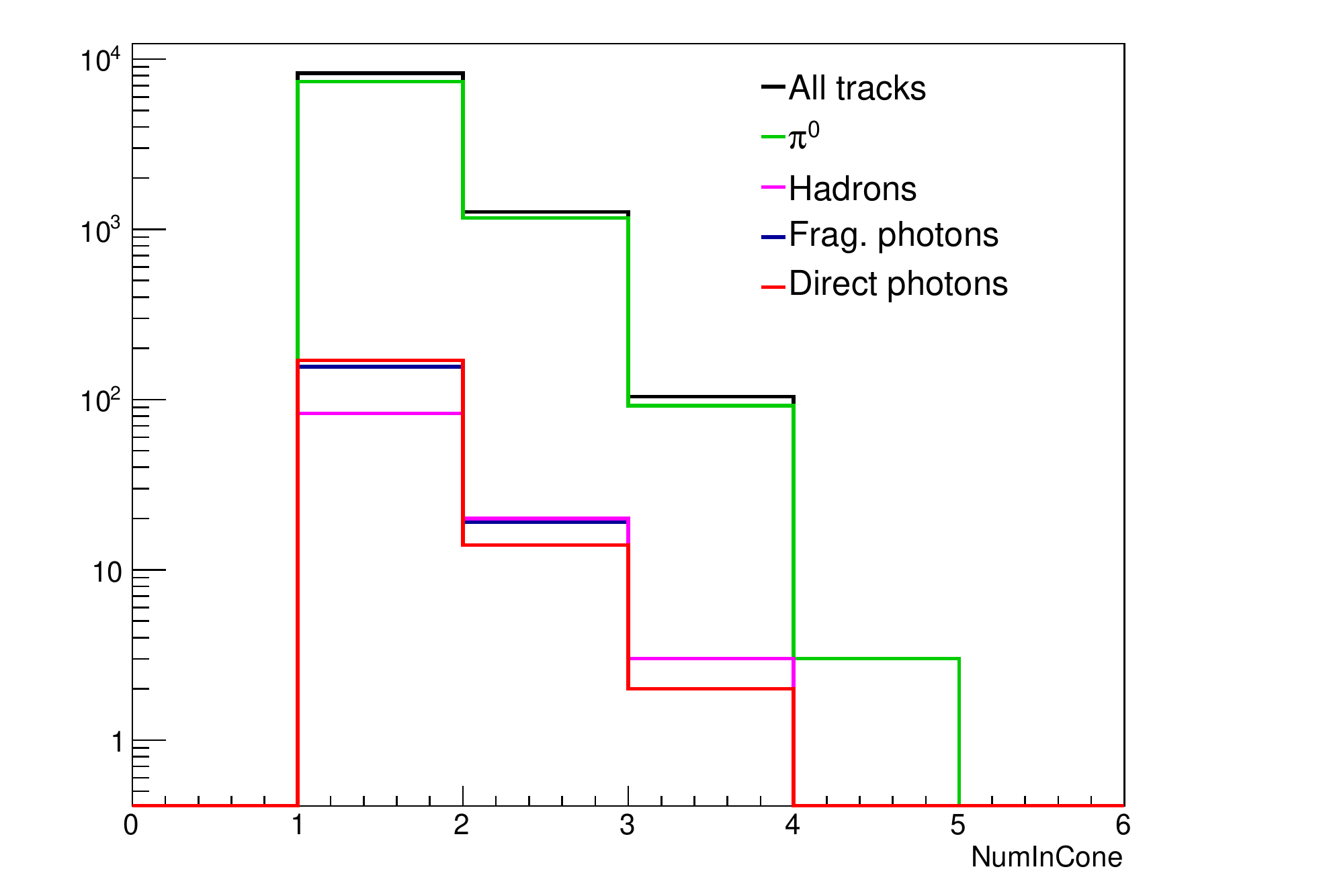}
  \end{minipage}
  \hspace{0.5cm}
  \begin{minipage}[b]{0.5\linewidth}
    \centering
    \includegraphics[width=0.95\linewidth]{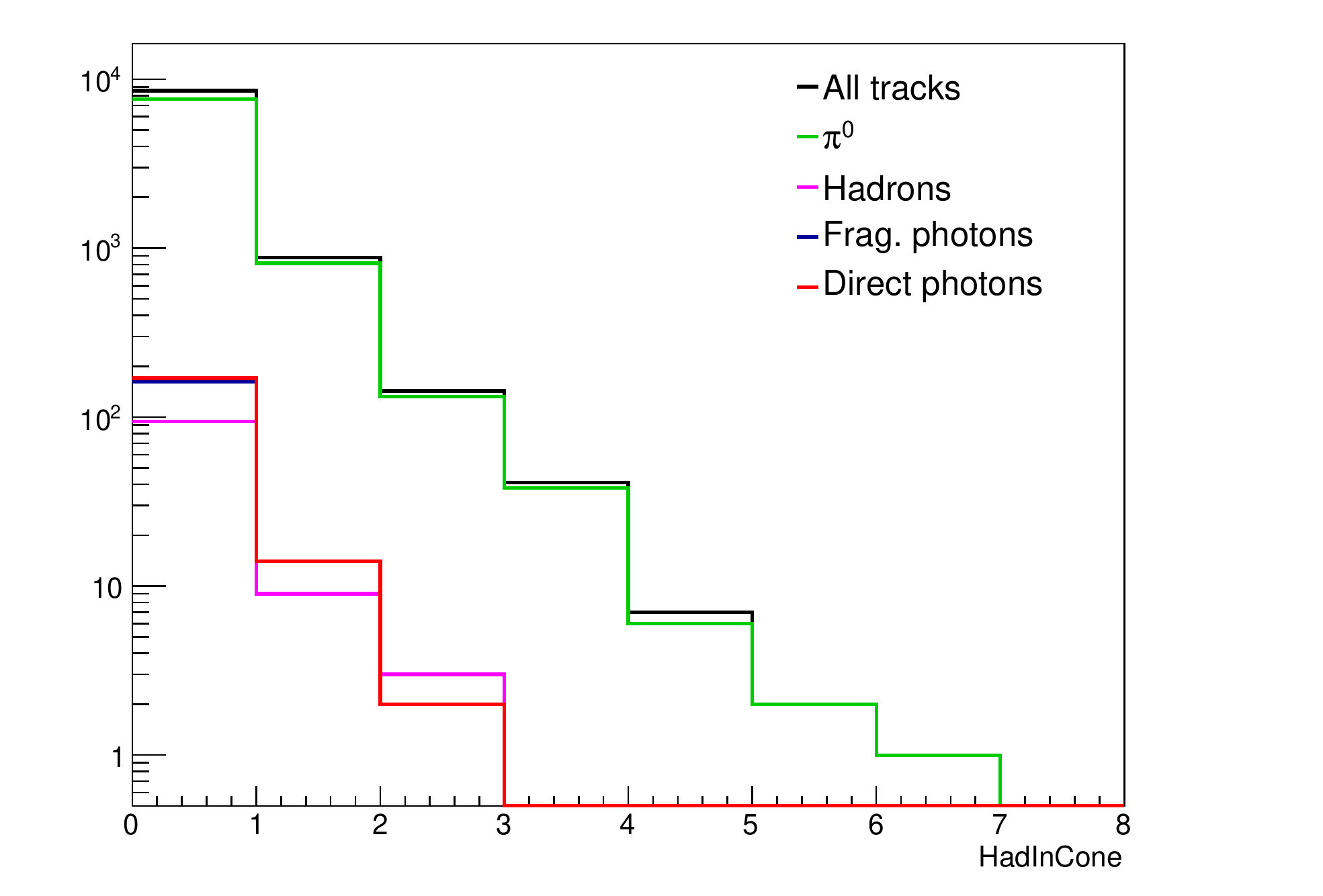}
  \end{minipage}
  \vspace*{-0.12in}
  \caption{\label{Fig:NumHadIC} The NumInCone and HadInCone distributions are 
    shown on the left and right respectively.  Pions are shown in green and hadrons
    are shown in pink.  The direct photon are shown in red and fragmentation 
    photons are in blue.  The sum of all tracks is shown in black.}
\end{figure}

The number of electromagnetic tracks (NumInCone) and hadron-like tracks (HadInCone) 
in a one radius cone in $\eta$-$\phi$ space around the photon candidate track are 
additional measures of the candidate track's isolation.  The hadron-like tracks 
used in HadInCone are defined by requiring narrow MIP deposits in all layers of 
the MPC-EX with poor non-unique matching to the MPC.  The electromagnetic tracks 
are only required to have a unique match to an MPC cluster.  The NumInCone variable 
has a minimum value of one since the cone contains the photon candidate track.  
Figure \ref{Fig:NumHadIC} shows the NumInCone and HadInCone distributions for 
$\pi^{0}$, hadrons and direct and fragmentation photons.  Both the NumInCone and 
HadInCone distributions are largest at their minimum values with the yields falling 
steeply as the value increases.  At the lowest values for each variable, the direct 
photon yields to the fragmentation photon contribution.  At higher values of 
NumInCone a small separation can be seen with the direct photon yield lower than 
the fragmentation photon yield.  

\begin{figure}[hbt]
  \hspace*{-0.12in}
  \begin{minipage}[b]{0.5\linewidth}
    \centering
    \includegraphics[width=0.95\linewidth]{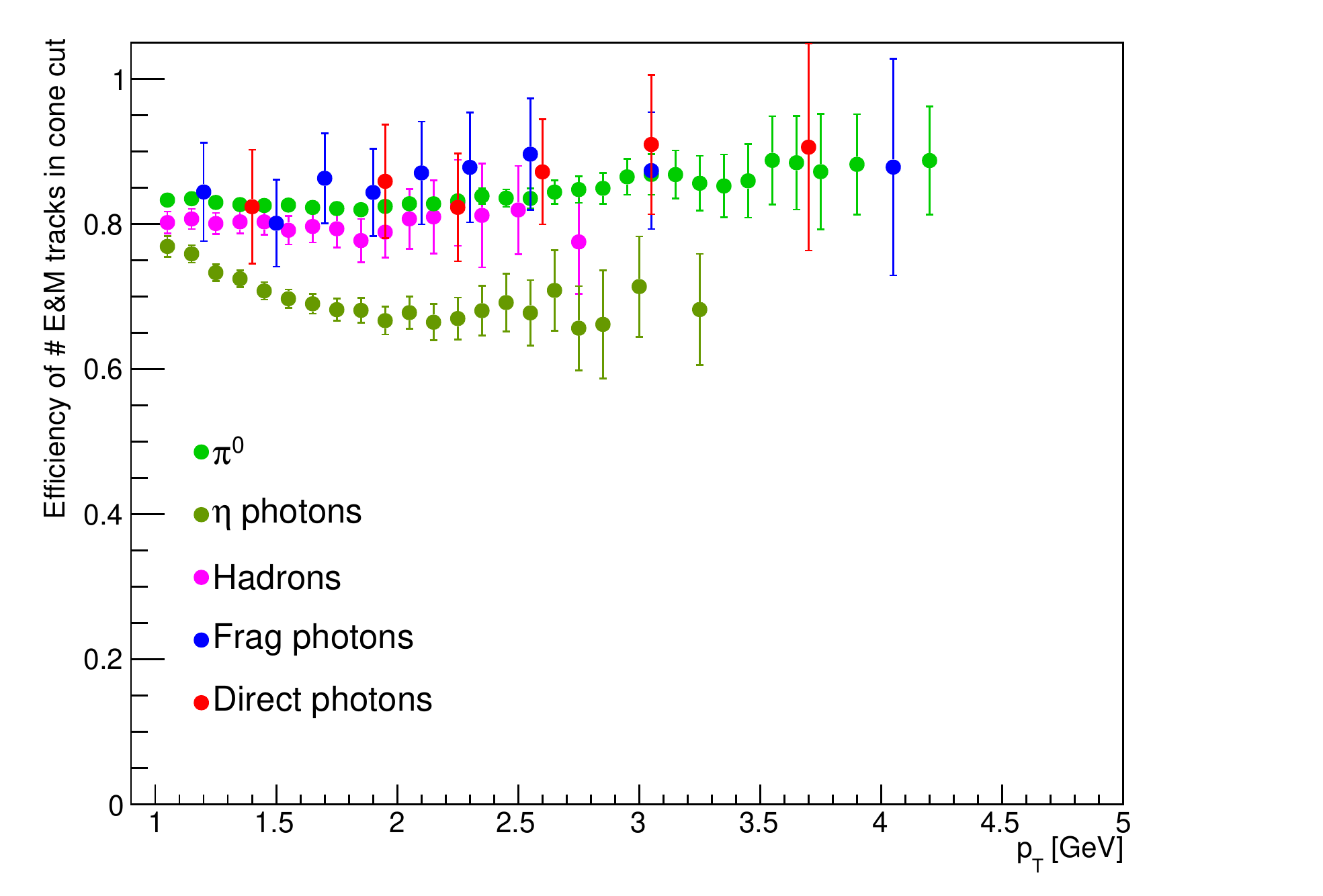}
  \end{minipage}
  \hspace{0.5cm}
  \begin{minipage}[b]{0.5\linewidth}
    \centering
    \includegraphics[width=0.95\linewidth]{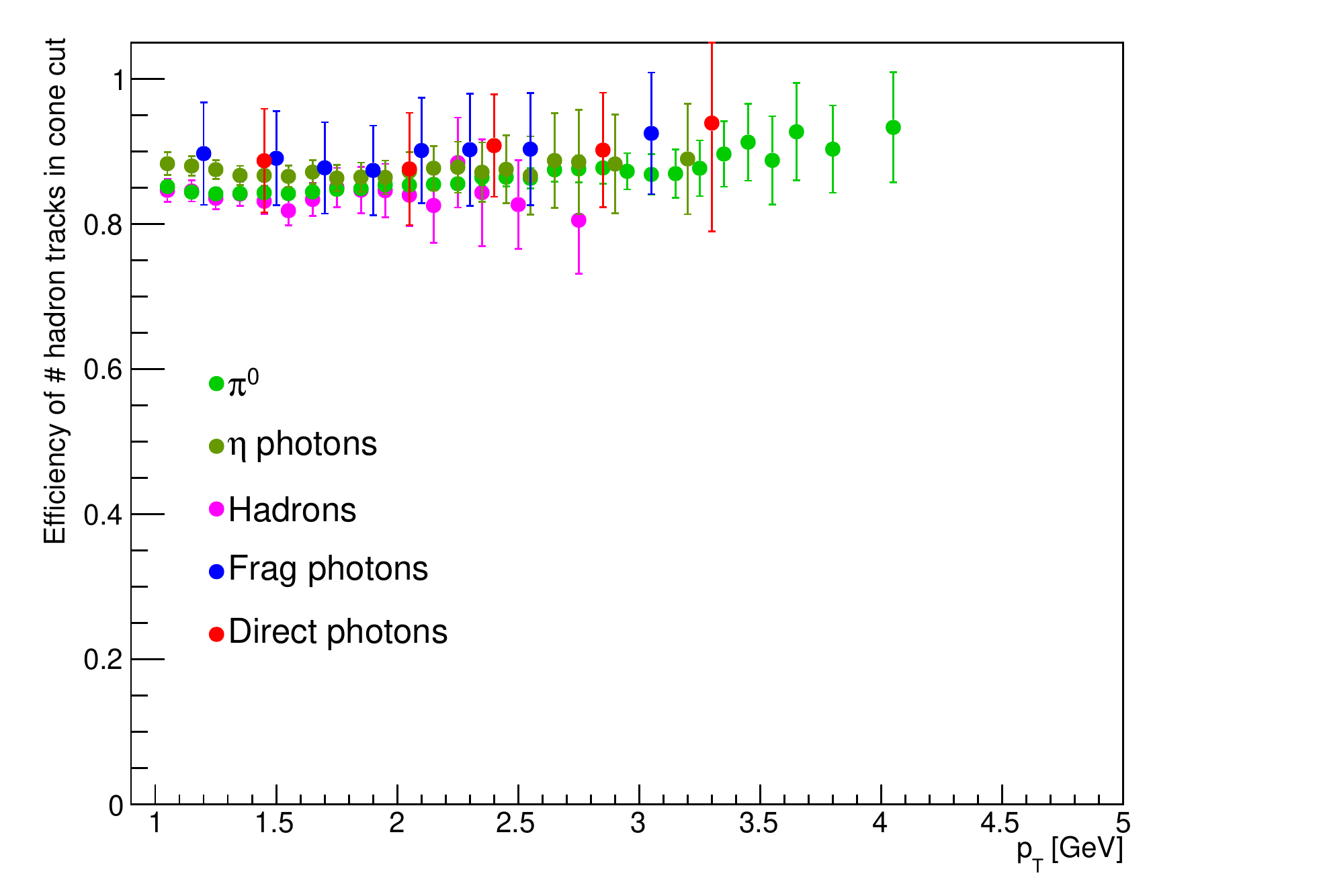}
  \end{minipage}
  \vspace*{-0.12in}
  \caption{\label{Fig:Effi_NumHadIC} The efficiency of the track isolation cuts 
    as a function of $p_{T}$.  On the left, the NumInCone of one cut efficiency 
    is shown and on the right the HadInCone of zero cut efficiency distributions 
    is shown.  $\pi^{0}$ and $\eta$ are shown in light and olive green.  Hadrons 
    are shown in pink.  Direct photons are shown in red and fragmentation photons 
    are in blue.}
\end{figure}

Isolated tracks are defined as tracks where NumInCone is one and HadInCone 
is zero.  The efficiencies of these requirements is plotted in 
Figure~\ref{Fig:Effi_NumHadIC}.  Unfortunately the HadInCone and NumInCone 
isolation cuts do not yet provide a substantial separation between direct
and fragmentation photons on their own.  Photons from $\eta$ decay are 
reduced by the NumInCone equal to one requirement, but the desired reduction 
of fragmentation photons is limited.  At $p_{T}$ greater than 3\,GeV, direct 
photons have an efficiency of 91.4\% and fragmentation photons have an 
efficiency of 87.6\%.  Similar efficiencies for the HadInCone cuts are 
91.4\% and 91.0\% for direct and fragmentation photons respectively.
This limits our ability to remove the fragmentation photon contribution with 
these cuts alone.  However, when these cut are applied in conjunction with 
other tightened cuts fragmentation photons are removed from the sample.

\begin{figure}[hbt]
  \hspace*{-0.12in}
  \centering
  \includegraphics[width=0.5\linewidth]{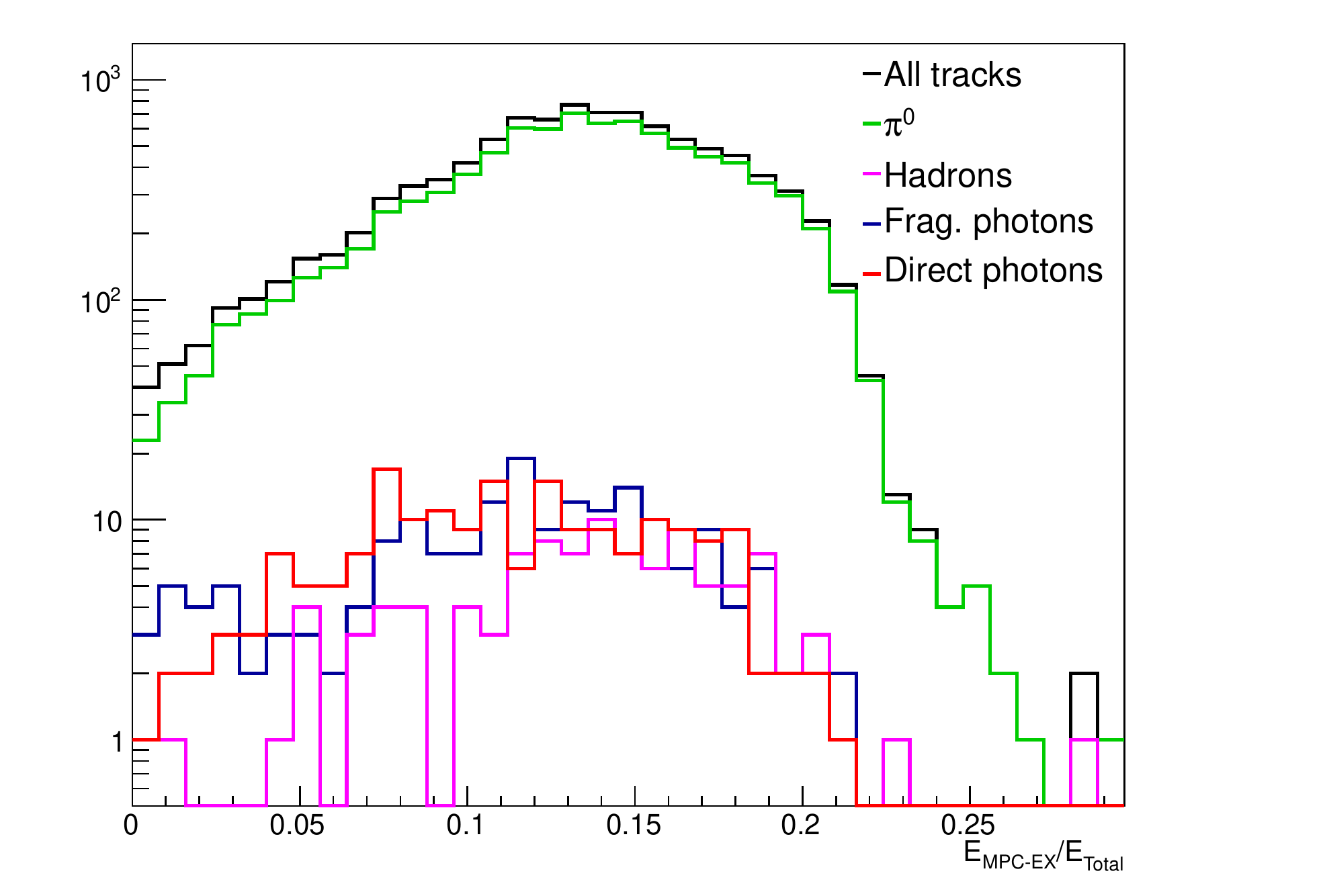}
  \vspace*{-0.12in}
  \caption{\label{Fig:Eratio} The ratio of the MPC-EX energy over the
    total energy in the MPC and MPC-EX as a function of $p_{T}$.  
    Direct photon are shown in red and fragmentation photons are in 
    blue.  Hadron and $\pi^{0}$ tracks are in pink and green 
    respectively.  The sum of all tracks is shown in black.}
\end{figure}

We also consider an additional variable, the ratio of the energy in the MPC-EX to 
the total energy from both the MPC and MPC-EX; this distribution is shown in 
Figure~\ref{Fig:Eratio}.  A shape difference in the direct photon distribution 
compared to the $\pi^{0}$ is seen.  Like many of the distributions shown 
previously, it is difficult to identify a difference in the direct and fragmentation 
photon distributions with the low statistics available in the $p_{T} > 3$ region.

To separate the direct photons from fragmentation photons a second 
TMVA analyses is completed.  Direct photons are identified as the signal and 
fragmentation photons are the background.  The TMVA sample is restricted to 
isolated tracks with $p_{T}$ above 3\,GeV that pass the mass cut, the 
hadron cuts and the KTestDist cut from the Table \ref{Tab:pi0Cuts}.  
These cuts are applied to restrict the TMVA phase space to those tracks 
surviving the hadron and limited $\pi^{0}$ cuts detailed previously.
The results from the CutsSA algorithm at two points on the ROC curve are considered.  
The first is located at a projected direct photon efficiency of 57\% and a 
projected fragmentation photon rejection of 60.2\%; this is refered to Frag Cuts 1.  
The second is located at a projected direct photon efficiency of 47\% and a 
projected fragmentation photon rejection of 78.9\%; this is referred to Frag Cuts 2.
Both of these cuts are listed in Table \ref{Tab:fragCuts}.  

\begin{table}
\centering
\caption{Acceptable ranges for variables as defined by TMVA to remove fragmentation photons.}
\label{Tab:fragCuts}
\begin{tabular}{|l||c|c|} \hline 
Variable                  &    Frag Cuts 1  &  Frag Cuts 2   \\ \hline
$E_{MPC-EX}/E_{Total}$    &   0.034 - 0.214 & 0.017 - 0.172  \\ 
dispersion                &   1.154 - 2.645 & 1.530 - 2.544  \\
RMS                       &   3.075 - 6.847 & 2.479 - 8.242  \\
KTestDist                 &   0.153 - 0.378 & 0.115 - 0.385  \\
$E_{MPC-EX}/E_{MPC-EX In Cone}$ & $>$ 0.868 &   $>$ 0.869    \\
$E_{MPC}/E_{MPC In Cone}$ &   0.832 - 1.021 & 0.793 - 0.960  \\  \hline
\end{tabular}
\end{table}

A comparsion of the acceptable variable ranges between Frag Cuts 1 and Frag Cuts 2 
shows that the minimum on the dispersion and the maximum on the RMS are increased 
as the fragmentation photon rejection rises.  The $E_{MPC-EX}/E_{Total}$ 
and KTestDist variables are lowered and the maximum on the $E_{MPC}/E_{MPC In Cone}$ 
is reduced.  The $E_{MPC-EX}/E_{MPC-EX In Cone}$ range is unaffected.  
Comparison of the cuts in Table~\ref{Tab:fragCuts} and those in Table~\ref{Tab:pi0Cuts} 
shows that the RMS and dispersion variables are similarly constrained but 
with an lower ranges than seen in the $\pi^{0}$ cuts.  The energy in cone ratios 
and KTestDist are more tightly restricted when removing fragmentation photons.  
The $\pi^{0}$ and fragmentation photon rejection cut sets are combined so that 
the intersection of the cut regions are applied.  By retaining only the overlap 
of the allowable variable ranges background removal is maximized.  The downside 
to this is that the efficiency of the direct photons are reduced.  The combined 
cuts are listed in Table~\ref{Tab:pi0fragCuts} and Cuts 1 efficiencies are listed in 
Table~\ref{Tab:CutsEffi_pi0frag}.  
 
\begin{table}
\centering
\caption{Acceptable ranges for variables as defined by TMVA to remove both $\pi^{0}$ and fragmentation photons.}
\label{Tab:pi0fragCuts}
\begin{tabular}{|l||c|c|} \hline 
Variable                  &       Cuts 1    &      Cuts 2      \\ \hline
$E_{MPC-EX}/E_{Total}$    &   0.034 - 0.214 &   0.017 - 0.172  \\ 
dispersion                &   1.154 - 1.631 &   1.530 - 1.631  \\
RMS                       &   3.075 - 6.118 &   2.479 - 6.118  \\
KTestDist                 &   0.153 - 0.378 &   0.115 - 0.385  \\
$E_{MPC-EX}/E_{MPC-EX In Cone}$ & $>$ 0.868 &   $>$ 0.869      \\
$E_{MPC}/E_{MPC In Cone}$ &   0.832 - 1.013 &   0.793 - 0.960  \\ \hline
\end{tabular}
\end{table}

\begin{table}
\centering 
\caption{Efficiencies using the Cuts 1 ranges in Table \ref{Tab:pi0fragCuts} for $\pi^{0}$, hadrons, direct and fragmentation photons for yields at $p_{T}>3$\,GeV.}
\label{Tab:CutsEffi_pi0frag}
\begin{tabular}{|l||c|c|c|c|} \hline 
Variable & $\epsilon_{\pi^{0}}$ & $\epsilon_{hadrons}$ & $\epsilon_{\gamma frag}$ & $\epsilon_{\gamma direct}$ \\ \hline
Mass                            &  $29.1\%$  &  $37.5\%$  &  $84.2\%$  &   $90.3\%$  \\  
$E_{MPC-EX}/E_{Total}$          &  $48.6\%$  &  $45.2\%$  &  $62.2\%$  &   $73.1\%$  \\ 
dispersion                      &   $8.4\%$  &   $7.7\%$  &  $34.5\%$  &   $34.4\%$  \\  
RMS                             &  $50.2\%$  &  $47.1\%$  &  $66.1\%$  &   $74.7\%$  \\  
KTestDist                       &  $33.6\%$  &  $35.6\%$  &  $39.5\%$  &   $54.8\%$  \\  
$E_{MPC-EX}/E_{MPC-EX In Cone}$ &  $45.9\%$  &  $44.2\%$  &  $58.2\%$  &   $68.3\%$  \\ 
$E_{MPC}/E_{MPC In Cone}$       &  $44.7\%$  &  $37.5\%$  &  $59.3\%$  &   $71.0\%$  \\ 
NumInCone                       &  $86.7\%$  &  $78.8\%$  &  $87.6\%$  &   $91.4\%$  \\  
HadInCone                       &  $88.6\%$  &  $89.4\%$  &  $91.0\%$  &   $91.4\%$  \\  \hline 
All Cuts 1 applied              &   $1.4\%$  &   $1.0\%$  &   $7.3\%$  &   $15.1\%$  \\ \hline
\end{tabular}
\end{table}

\begin{figure}[hbt]
  \hspace*{-0.12in}
  \begin{minipage}[b]{0.5\linewidth}
    \centering
    \includegraphics[width=0.95\linewidth]{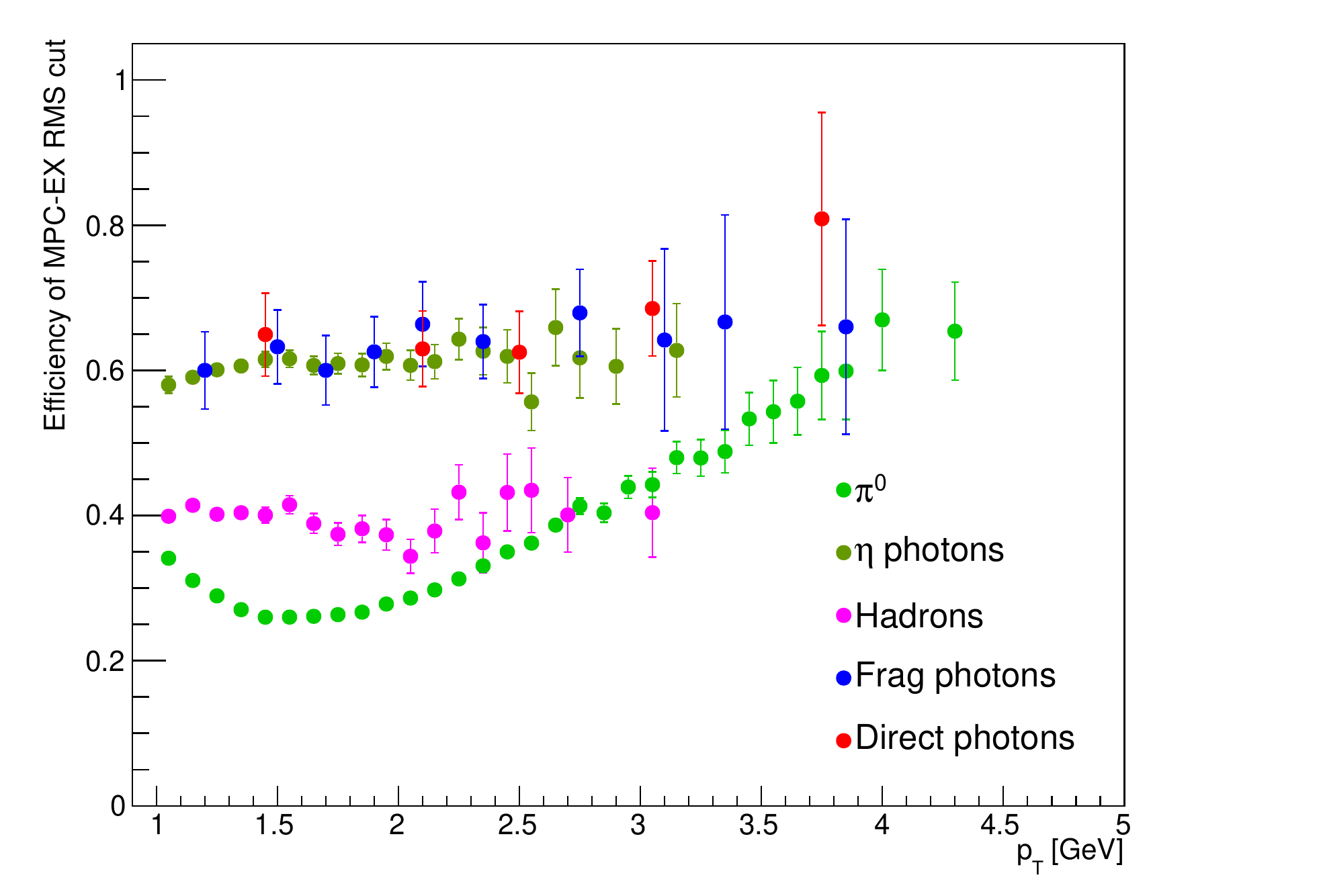}
  \end{minipage}
  \hspace{0.2cm}
  \begin{minipage}[b]{0.5\linewidth}
    \centering
    \includegraphics[width=0.95\linewidth]{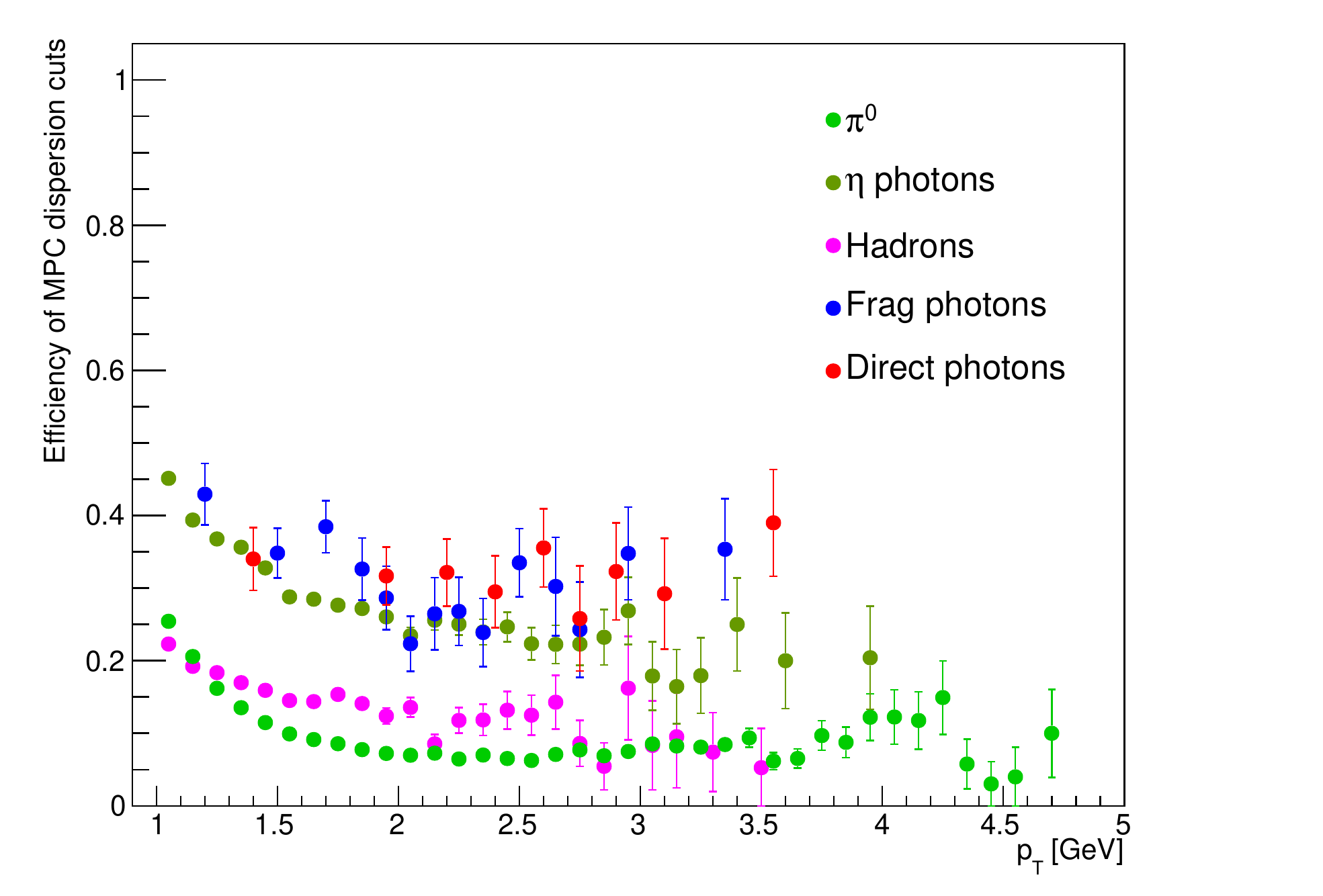}
  \end{minipage}
  \vspace*{-0.12in}
  \caption{\label{Fig:Effi_Width_frag}The efficiency of the MPC-EX RMS 
    and the MPC dispersion cuts in the Cuts 1 range as a function of $p_{T}$.  
    The right panel shows the MPC-EX RMS cut efficiency and the
    left panel shows the efficiency of the MPC dispersion cut.  $\pi^{0}$ and 
    $\eta$ are shown in bright and olive green.  Hadrons are shown in pink.  
    Direct photons are shown in red and fragmentation photons in blue. }
\end{figure}

\begin{figure}[hbt]
  \hspace*{-0.12in}
  \begin{minipage}[b]{0.5\linewidth}
    \centering
    \includegraphics[width=0.95\linewidth]{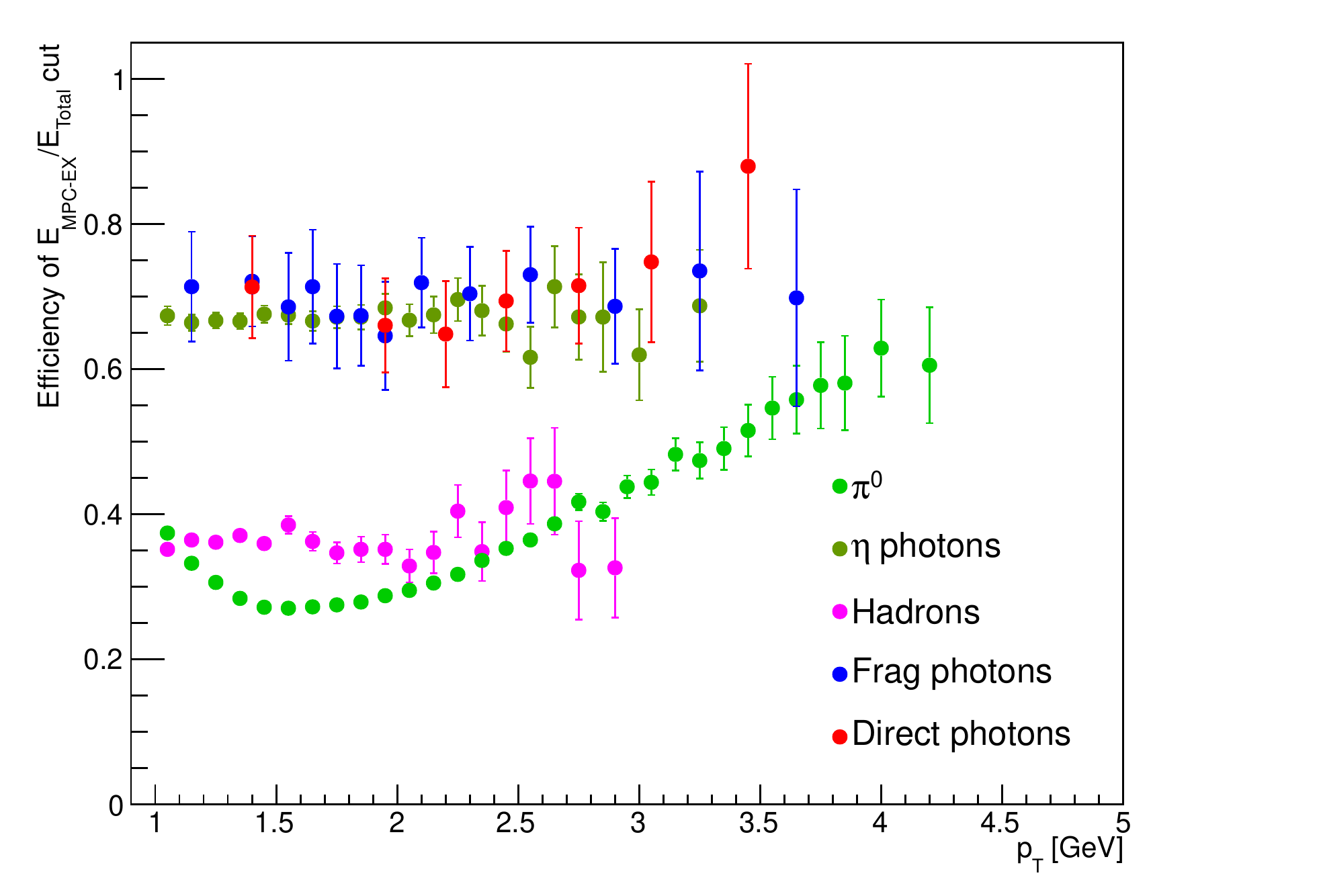}
  \end{minipage}
  \hspace{0.2cm}
  \begin{minipage}[b]{0.5\linewidth}
    \centering
    \includegraphics[width=0.95\linewidth]{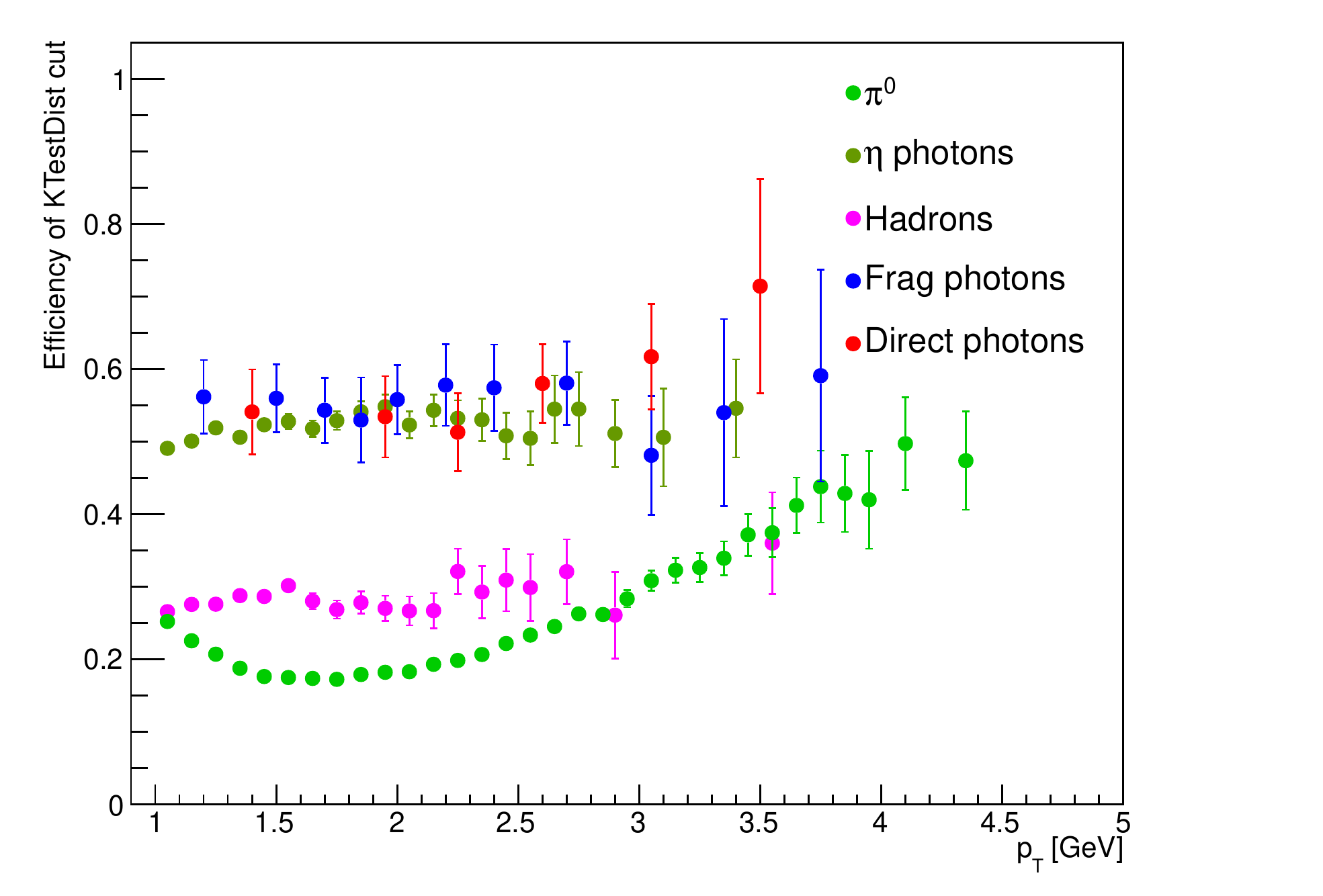}
  \end{minipage}
  \vspace*{-0.12in}
  \caption{\label{Fig:Effi_Eratio}The efficiency of the $E_{MPC-EX}/E_{Total}$ 
    and KTestDist cuts in the Cuts 1 range as a function of $p_{T}$.  
    The right panel shows the $E_{MPC-EX}/E_{Total}$ cut efficiency and the
    left panel shows the efficiency of the KTestDist cut.  $\pi^{0}$ and 
    $\eta$ are shown in bright and olive green.  Hadrons are shown in pink.  
    Direct photons are shown in red and fragmentation photons in blue. }
\end{figure}

\begin{figure}[hbt]
  \hspace*{-0.12in}
  \begin{minipage}[b]{0.5\linewidth}
    \centering
    \includegraphics[width=0.95\linewidth]{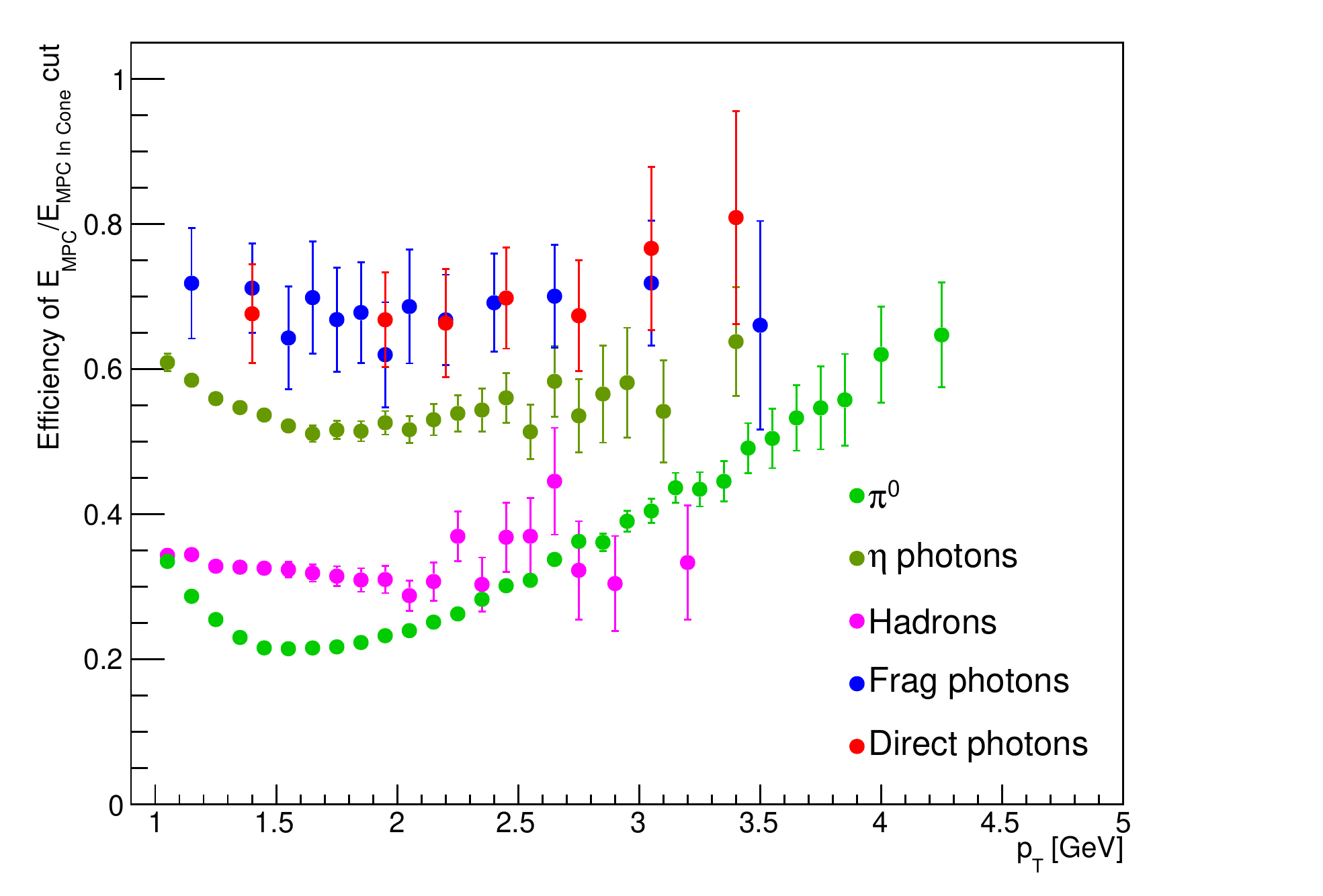}
  \end{minipage}
  \hspace{0.2cm}
  \begin{minipage}[b]{0.5\linewidth}
    \centering
    \includegraphics[width=0.95\linewidth]{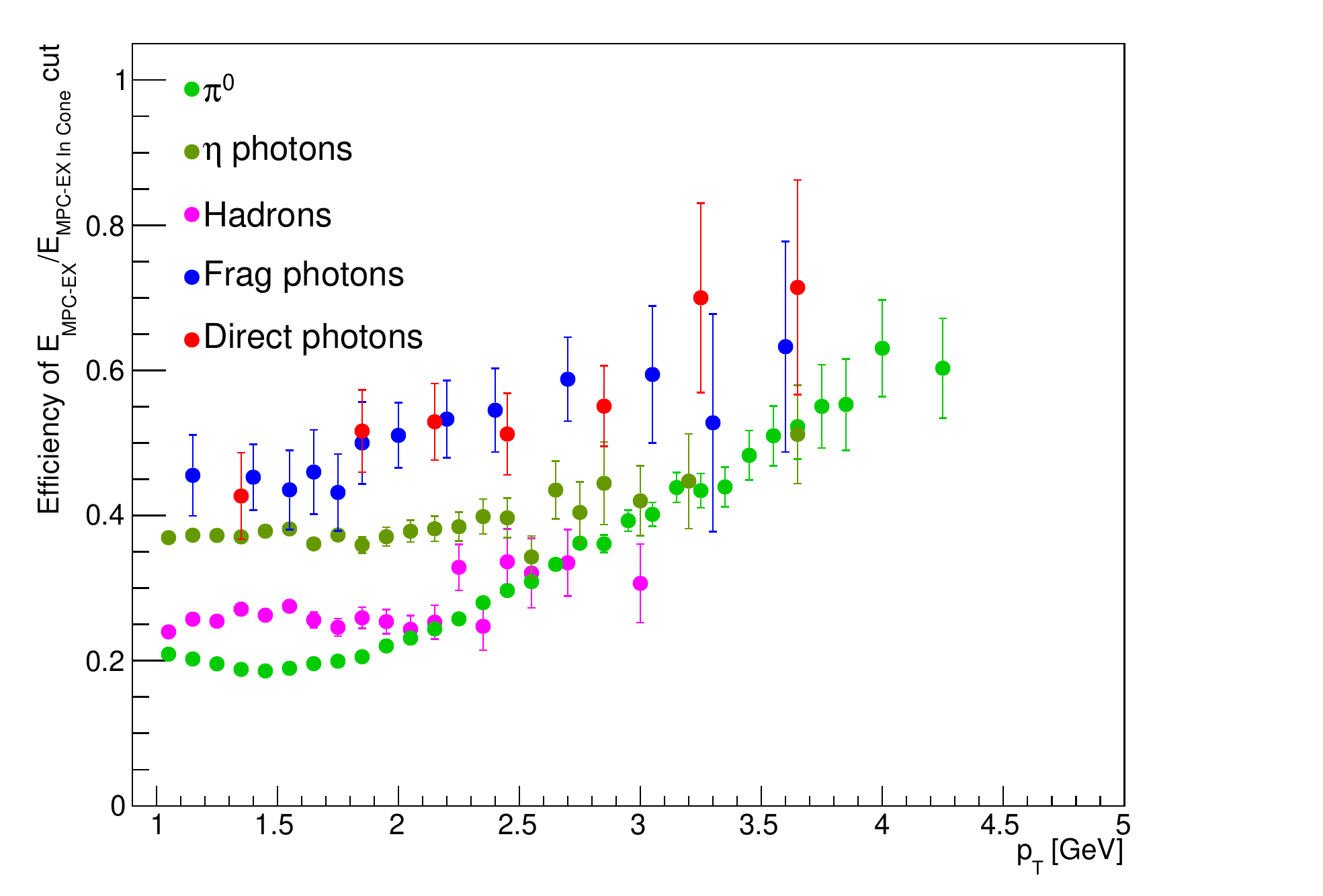}
  \end{minipage}
  \vspace*{-0.12in}
  \caption{\label{Fig:Effi_EICfrag}The efficiency of the energy ratio in cone cuts 
    considering the energies in the MPC and MPC-EX as a function of $p_{T}$.  
    The right panel shows the efficiency using the MPC energies and the left 
    panel shows the ratio with MPC-EX energies.  $\pi^{0}$ and $\eta$ decays 
    are shown in bright and olive green.  Hadrons are shown in pink.  
    Direct photons are shown in red and fragmentation photons in blue. }
\end{figure}

The hadrons and $\pi^{0}$ backgrounds are rejected by the mass, 
$E_{MPC-EX}/E_{Total}$, dispersion, KTestDist, and energy ratio in 
cone variables with $\pi^{0}$ efficiencies of less than $50\%$.
The $E_{MPC-EX}/E_{Total}$, KTestDist and energy ratio in cone variables 
also distinguish between fragmentation and direct photons with a separation 
in direct and fragmentation photons efficiencies of more than 10\%.
Figures~\ref{Fig:Effi_Width_frag}, ~\ref{Fig:Effi_Eratio} and~\ref{Fig:Effi_EICfrag} 
show the efficiencies as a function of $p_{T}$ for the variables: 
dispersion, RMS, $E_{MPC-EX}/E_{Total}$, KTestDist, and the energy in cone 
ratios with the Cuts 1 ranges.  
Comparing the Cut 1 efficiencies with the RMS, dispersion, KTestDist and 
energy ratio in cone variables with the $\pi^{0}$ cut efficiencies in 
Figures~\ref{Fig:Effi_width},~\ref{Fig:Effi_KT} and~\ref{Fig:Effi_EIC} 
shows an overall decrease in the cut efficiencies except for the dispersion cut
and a minimal affect seen in KTestDist.  

\begin{figure}[hbt]
  \hspace*{-0.12in}
  \centering
  \includegraphics[width=0.5\linewidth]{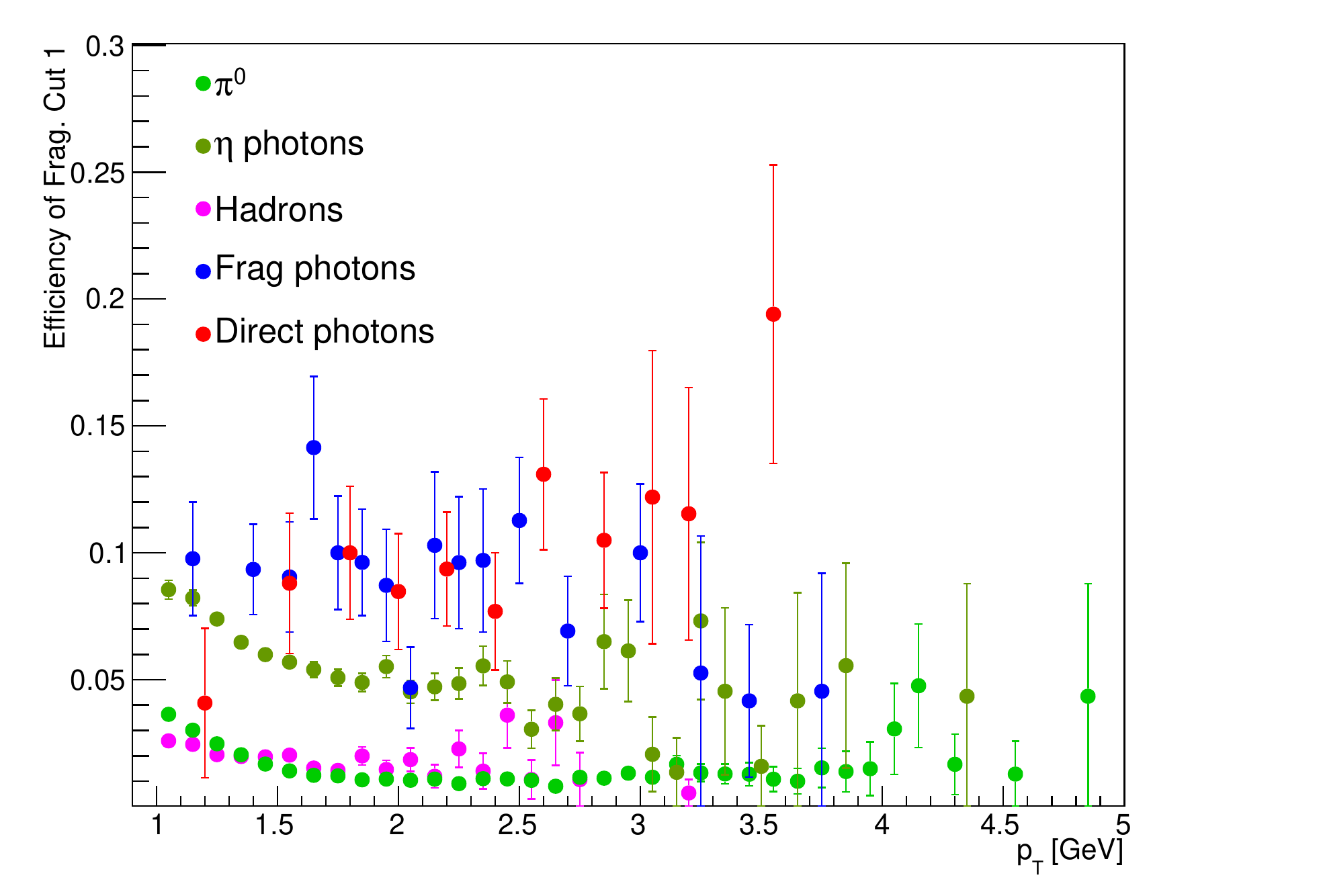}
  \vspace*{-0.12in}
  \caption{\label{Fig:Effi_fragcuts} The total efficiency of the Cuts 1 ranges as 
    a function of $p_{T}$.  Direct photon are shown in red and fragmentation 
    photons are in blue.  Hadrons, $\pi^{0}$ and $\eta$ are in pink, light 
    and olive green respectively.}
\end{figure}   

Comparing the efficiencies in Tables~\ref{Tab:CutsEffi} 
and~\ref{Tab:CutsEffi_pi0frag} confirms that the RMS, KTestDist and energy 
in cone ratio cuts show a lower direct photon efficiency and an increased 
separation between the direct and fragmentation photon efficiencies; the 
$\pi^{0}$ efficiencies remain roughly the same.  The mass cut and resulting 
efficiencies are unchanged in both cases.  The dispersion efficiencies are 
only minimally effected despite the reduced range.  This is because the 
dispersion distribution is sparingly populated below 1.154.  

As expected, the efficiencies in the Cuts 1 analysis are reduced compared 
to the $\pi^{0}$ cuts analysis.  Figure~\ref{Fig:Effi_fragcuts} shows the 
efficiency in the Cuts 1 analysis as a function of $p_{T}$.  A comparison to
Figure~\ref{Fig:Effi_pi0} shows an increased separation between the direct
and fragmentation photons at high $p_{T}$.  Table~\ref{Tab:AllCutsYield} 
shows the yields for the $\pi^{0}$, Cuts 1 and Cuts 2 analyses and when only 
the hadron cuts are applied.  The resulting $\pi^{0}$ and direct photons 
efficiencies and yields in the Cuts 1 case are approximately half, 48\%, of the 
$\pi^{0}$ cuts case.  Fragmentation photons are down by roughly a third, 30\%.  
The Cuts 2 analysis reduces the direct photons and $\pi^{0}$'s by 19\% and 
15\% compared to the $\pi^{0}$ rejection analysis; fragmentation photons are 
reduced by 7\%.  By tightening cuts and ensuring tracks are isolated we can 
increase the direct photon concentration while maintaining or increasing 
the direct photon-to-$\pi^{0}$ ratio.

\begin{table}
\centering
\caption{Particle yields with and with out various photon cuts.}
\label{Tab:AllCutsYield}
\begin{tabular}{|l||c|c|c|c|} \hline 
Particle Type &   Yield with &  Yield with    &  Yield with & Yield with \\ 
              & hadron cuts  & $\pi^{0}$ cuts &     Cuts 1  &  Cuts 2   \\ \hline
Charged hadrons              &      99  &   6 &       1     &   0   \\ 
\hspace{5mm} $\pi$           &      44  &   5 &       1     &   0   \\ 
\hspace{5mm} K               &       0  &   0 &       0     &   0   \\ 
\hspace{5mm} p               &       0  &   0 &       0     &   0   \\ 
\hspace{5mm} neutrons        &       2  &   0 &       0     &   0   \\
\hspace{5mm} other           &      53  &   1 &       0     &   0   \\
$\pi^{0}$                    &    8013  & 234 &     112     &  35   \\
$\eta$ decay                 &     445  &  51 &      16     &   8   \\
Other decays                 &     114  &  11 &       4     &   2   \\
\hspace{5mm} $\omega$        &      94  &   9 &       4     &   2   \\
\hspace{5mm} $\eta$'         &      19  &   1 &       0     &   0   \\
\hspace{5mm} other           &       1  &   1 &       0     &   0   \\
Signal photons               &     363  & 101 &      41     &  14   \\ 
\hspace{5mm} frag photons    &     177  &  43 &      13     &   3   \\ 
\hspace{5mm} direct photons  &     186  &  58 &      28     &  11   \\ \hline 
\end{tabular}
\end{table}

Before the $\pi^{0}$ cuts are applied, direct and fragmentation 
photons have a roughly equal contribution to the signal photons 
in the {\sc Pythia} sample.  After applying the $\pi^{0}$ cuts, 
direct photons are 57.4\% of the signal photons.  With the 
Cuts 1 analysis, the relative contribution of direct photons to 
the signal photon measurement increases to 68.3\%.  By 
applying the tighter Cuts 2 cuts, the relative contribution of 
the direct photons to 78.6\% of the signal photons.  These 
results show that with a MPC-EX detector we are able to adjust 
the concentration of direct photons in the measured signal.

In Section~\ref{sim:dphot} the direct photon measurement using 
the double ratio method is discussed using the results from the 
$\pi^{0}$ cuts analysis.  The measurements and the corresponding 
systematics for the Cuts 1 and Cuts 2 analyses with the 
increased relative direct photon contribution are also discussed 
for comparison.


\clearpage
  \label{sim:dphot_pythia}
\section[Direct Photons in the {\sc Pythia} Monte Carlo]{Direct Photons in the {\sc Pythia} Monte Carlo}
\label{sim:dphot_pythia}

A {\sc Pythia} Monte Carlo is used to benchmark the MPC-EX's ability to measure direct 
photons\cite{Sjostrand:2006za}.  While {\sc Pythia} is a simulation of p+p events, it serves 
as a good proxy for d+Au collisions in the deuteron-going direction.  {\sc Pythia} is well 
tuned to measured cross sections at the Tevatron and fixed-target energies. Studies with the 
event generator HIJING indicate that the additional multiplicity in a d+Au collision is 
localized at low energies, below the high energy direct photons emitted in the forward direction.

\begin{figure} 
  \begin{center}
  \includegraphics[width=0.8\linewidth]{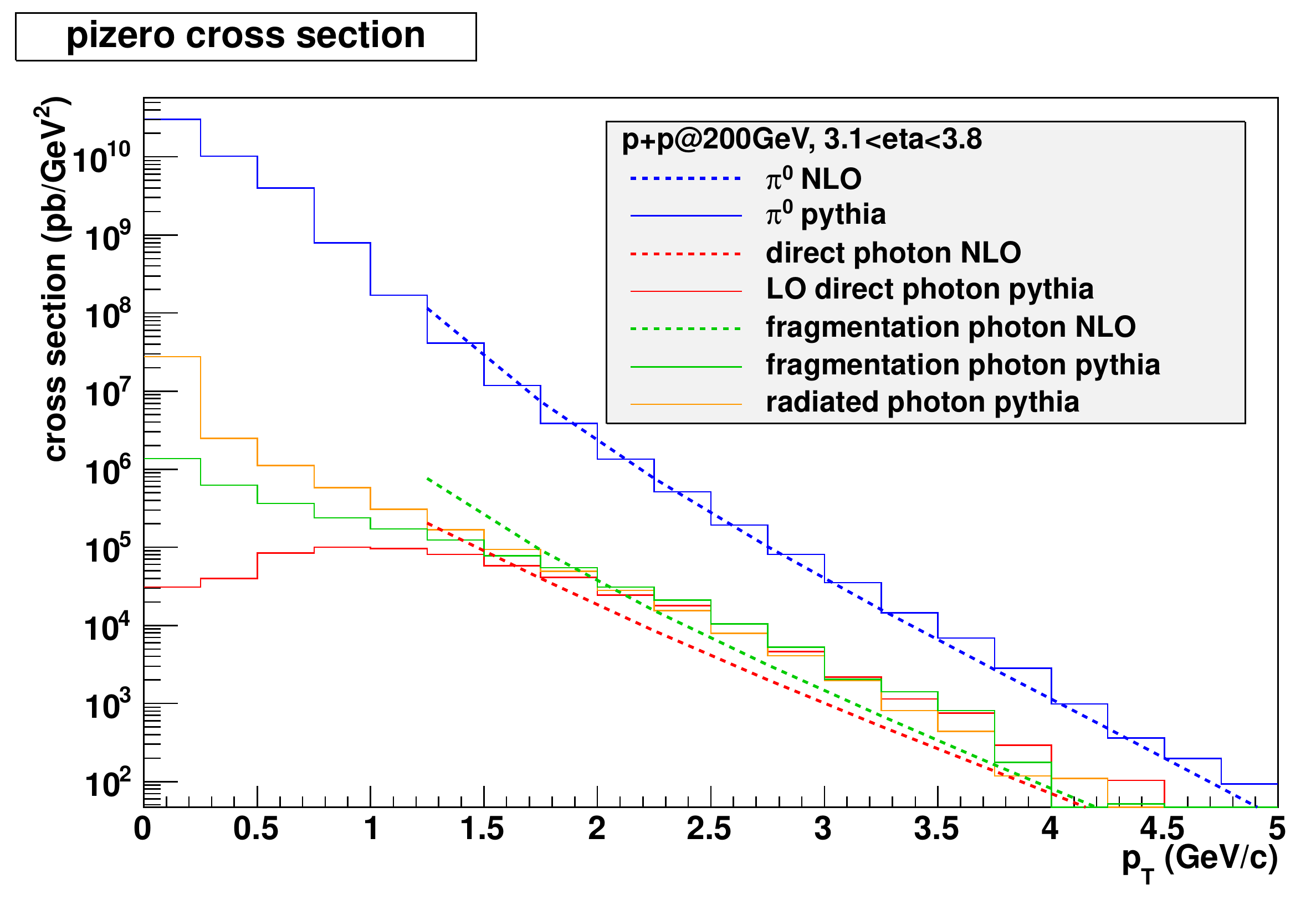}
  \end{center}
  \caption{\label{fig:Werner_comp} NLO cross sections fom Werner Vogelsang compared the cross sections extracted from {\sc Pythia} for $\pi^{0}$'s and direct
photons.  Direct photons from {\sc Pythia} are selected only from processes that produce a photon at the hard scattering vertex, fragmentation photons and 
QED radiation from incoming quarks are shown as separate entries.}
\end{figure}

In addition to the Monte Carlo generator {\sc Pythia}, the MPC-EX collaboration also solicited 
NLO calculations of the direct photon cross section from Werner Vogelsang for p+p collisions 
at 200~GeV~\cite{Werner_NLO}.  A comparison between the NLO calculations and {\sc Pythia} is 
shown in Figure~\ref{fig:Werner_comp}.  The NLO cross sections identify two photon sources: 
photons from partonic processes and QED radiation from incoming partons, which we refer to as 
``direct'' photons, and photons from parton fragmentation, which we refer to as ``fragmentation'' 
photons.  In the NLO calculation it is not possible to cleanly separate direct photons into 
their partonic and QED radiated components because the cross section calculation involves 
interference between different amplitudes.  The {\sc Pythia} Monte Carlo calculates partonic 
processes at leading order (LO) and QED radiated photons from incoming quarks are implemented 
in a separate process.  In {\sc Pythia} these two different sources of photons can 
be distinguished.  

Before proceeding to use {\sc Pythia} to evaluate the performance of the MPC-EX, it is important to understand if {\sc Pythia} 
properly reproduces the expected cross sections for direct and fragmentation photons, and $\pi^0$ mesons (the dominant 
background).  In Figure~\ref{fig:Werner_comp}, we compare the NLO cross sections for direct and fragmentation photons in the 
MPC-EX acceptance with the cross sections from the {\sc Pythia} Monte Carlo model.  A comparison of the $\pi^0$ cross section 
is also included.  The $\pi^0$ cross sections are a very good match between {\sc Pythia} and the NLO calculations.  The 
fragmentation photon cross section is similar between the {\sc Pythia} and NLO calculations, agreeing to within a factor of 
approximately two.  The LO photons in {\sc Pythia} are a good comparion in overall magnitude to the NLO direct photon cross 
sections (again, within roughly a factor of two).  However, this is not an apples-to-apples comparison.  The {\sc Pythia} 
calculation is LO and does not include QED radiation from quarks incoming to the hard scattering vertex, while the NLO 
calculation includes this radiation and additional amplitudes.  The cross section in {\sc Pythia} for photons from QED radiation 
is comparable to the {\sc Pythia} LO cross section.  If the photons from incoming quark radiation are included in the MPC-EX analysis, 
the effective cross section will be much larger than the NLO cross section, potentially overestimating the sensitivity of the 
measurement. 

\begin{figure}
  \begin{center}
    \centering
    \subfloat[]{
      \label{fig:Werner_dphot_PDF_A}
      \includegraphics[width=0.4\linewidth]{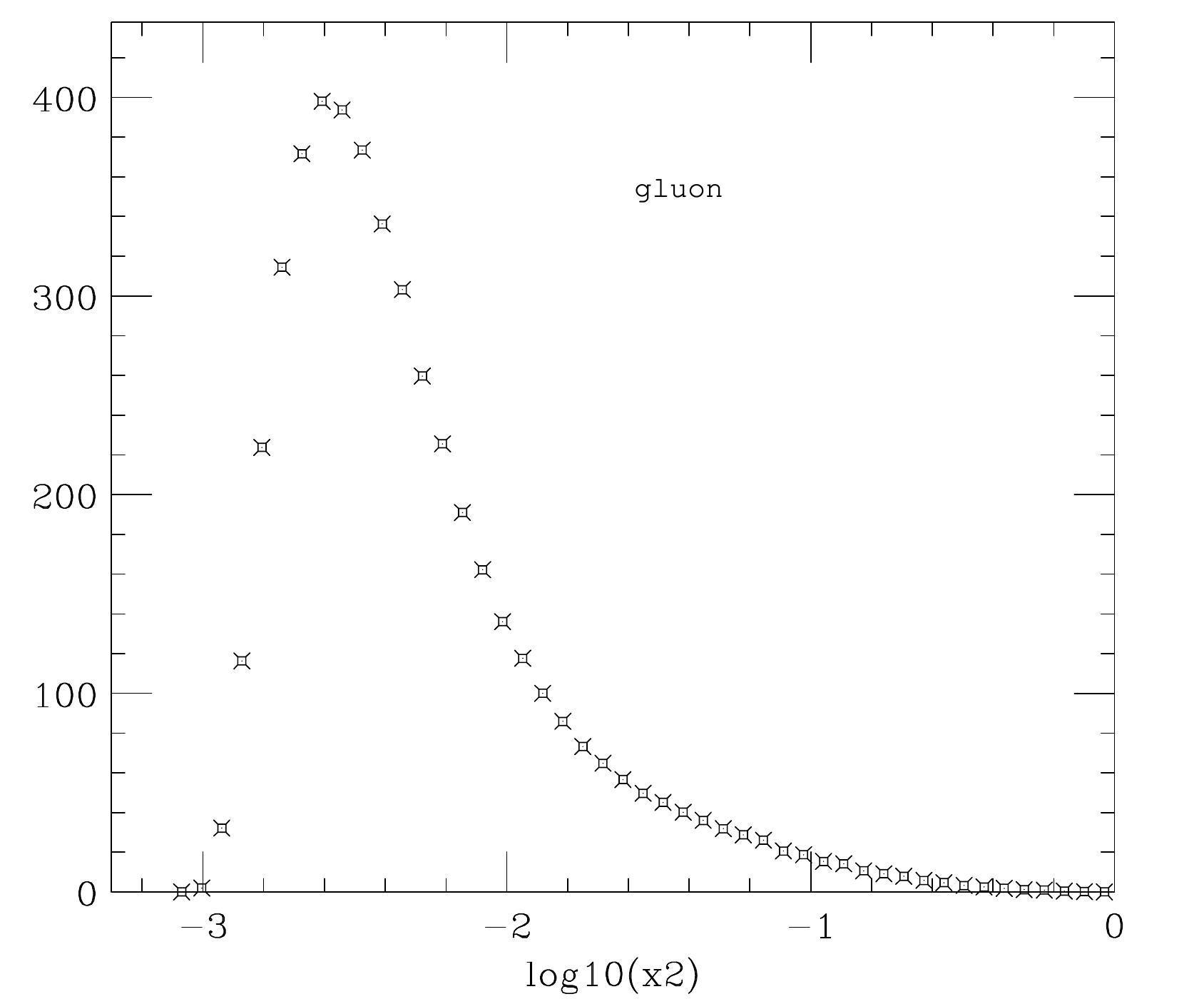}
    }
    \subfloat[]{
      \label{fig:Werner_dphot_PDF_B}
      \includegraphics[width=0.4\linewidth]{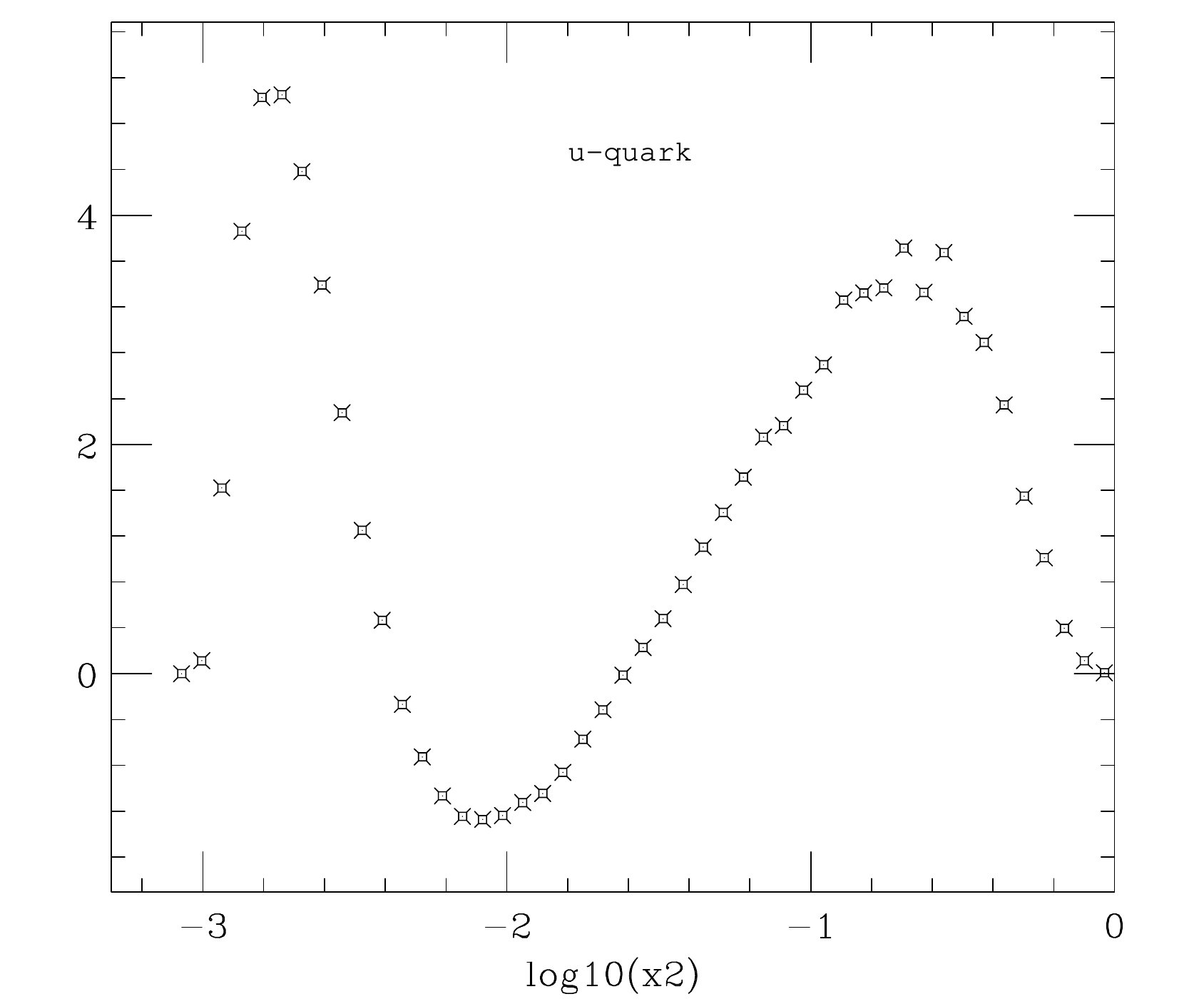}
    }
    \subfloat[]{
      \label{fig:Werner_dphot_PDF_C}
      \includegraphics[width=0.4\linewidth]{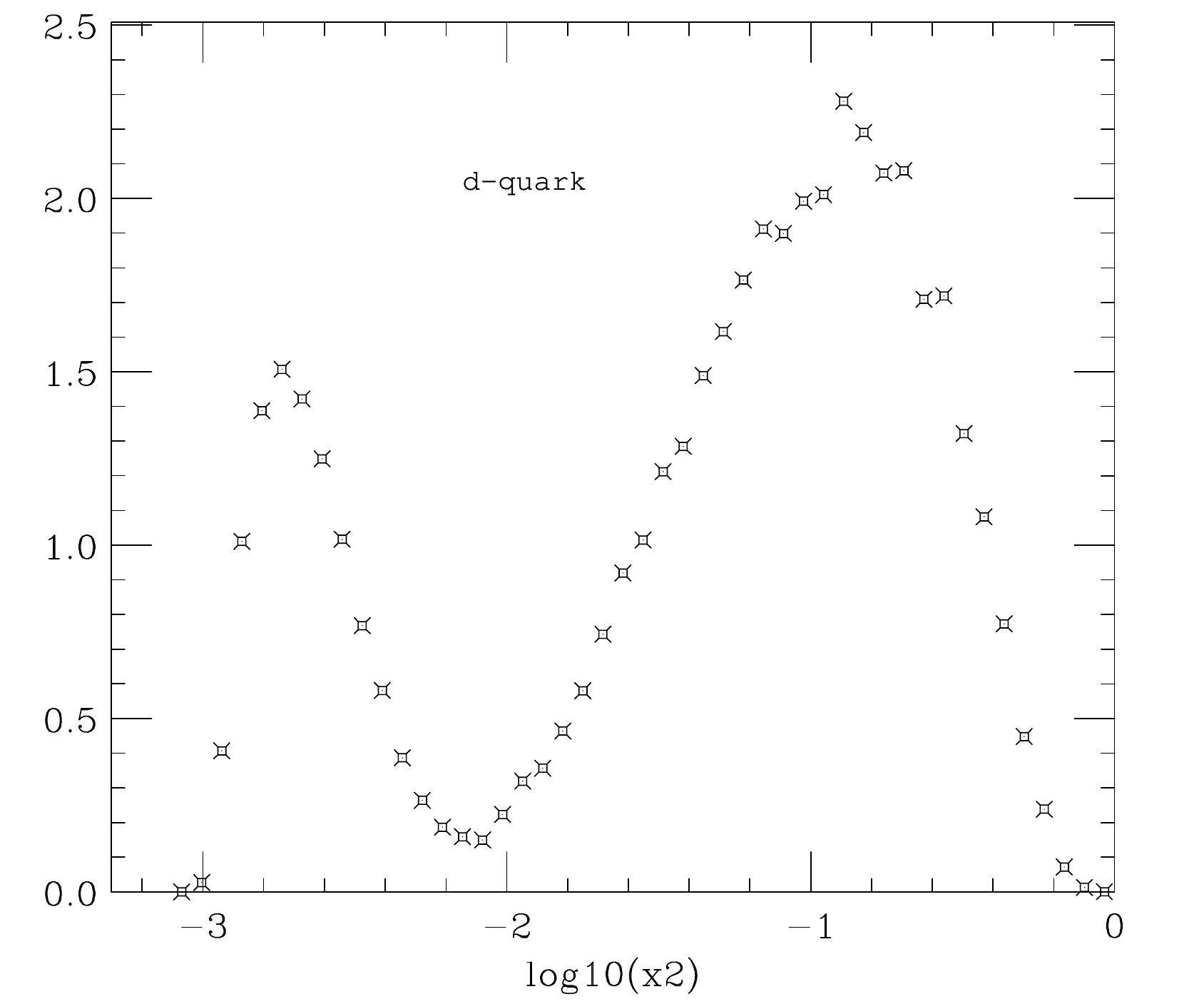}
    }
    \subfloat[]{
      \label{fig:Werner_dphot_PDF_D}
      \includegraphics[width=0.4\linewidth]{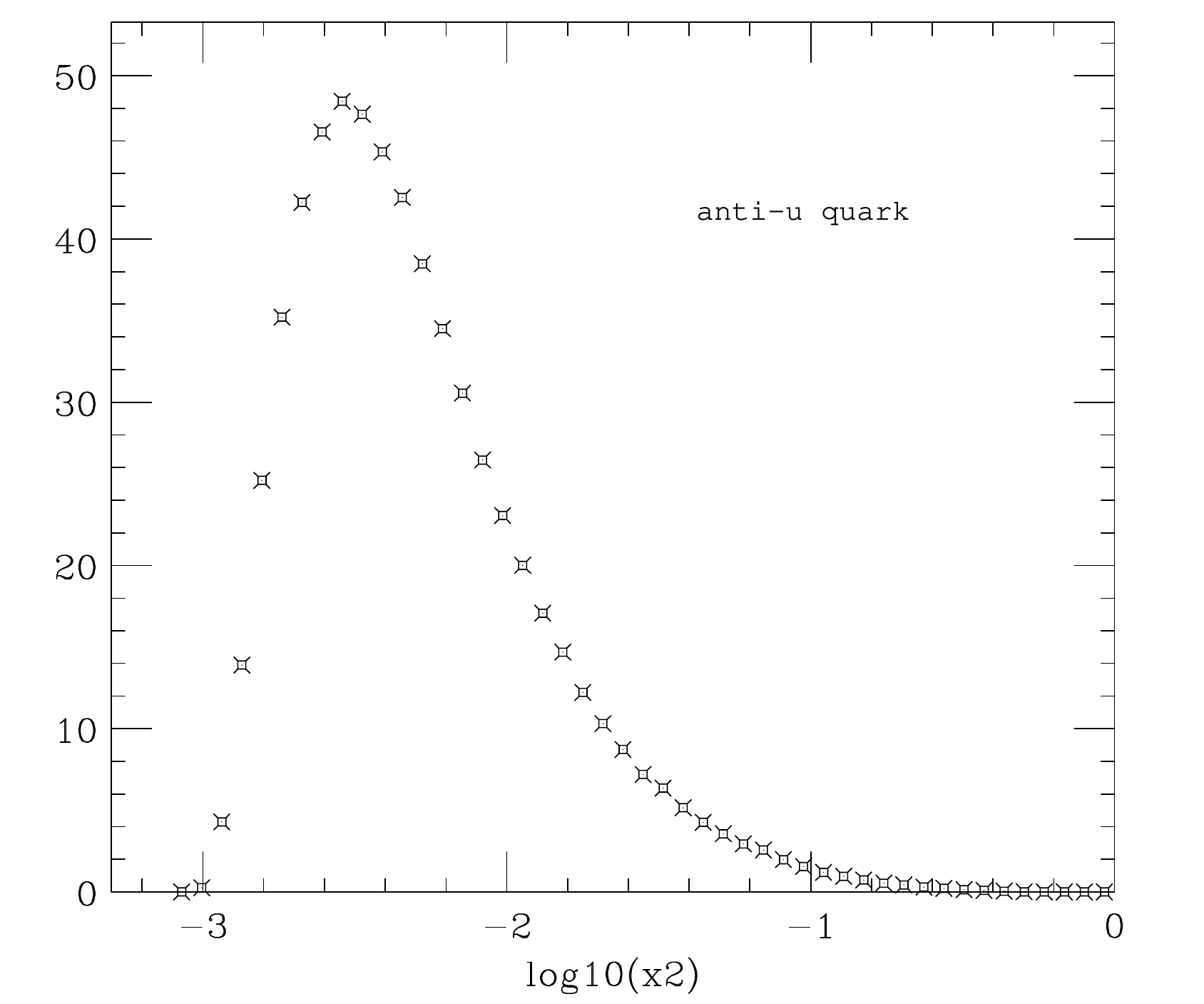}
    }
    \subfloat[]{
      \label{fig:Werner_dphot_PDF_E}
      \includegraphics[width=0.4\linewidth]{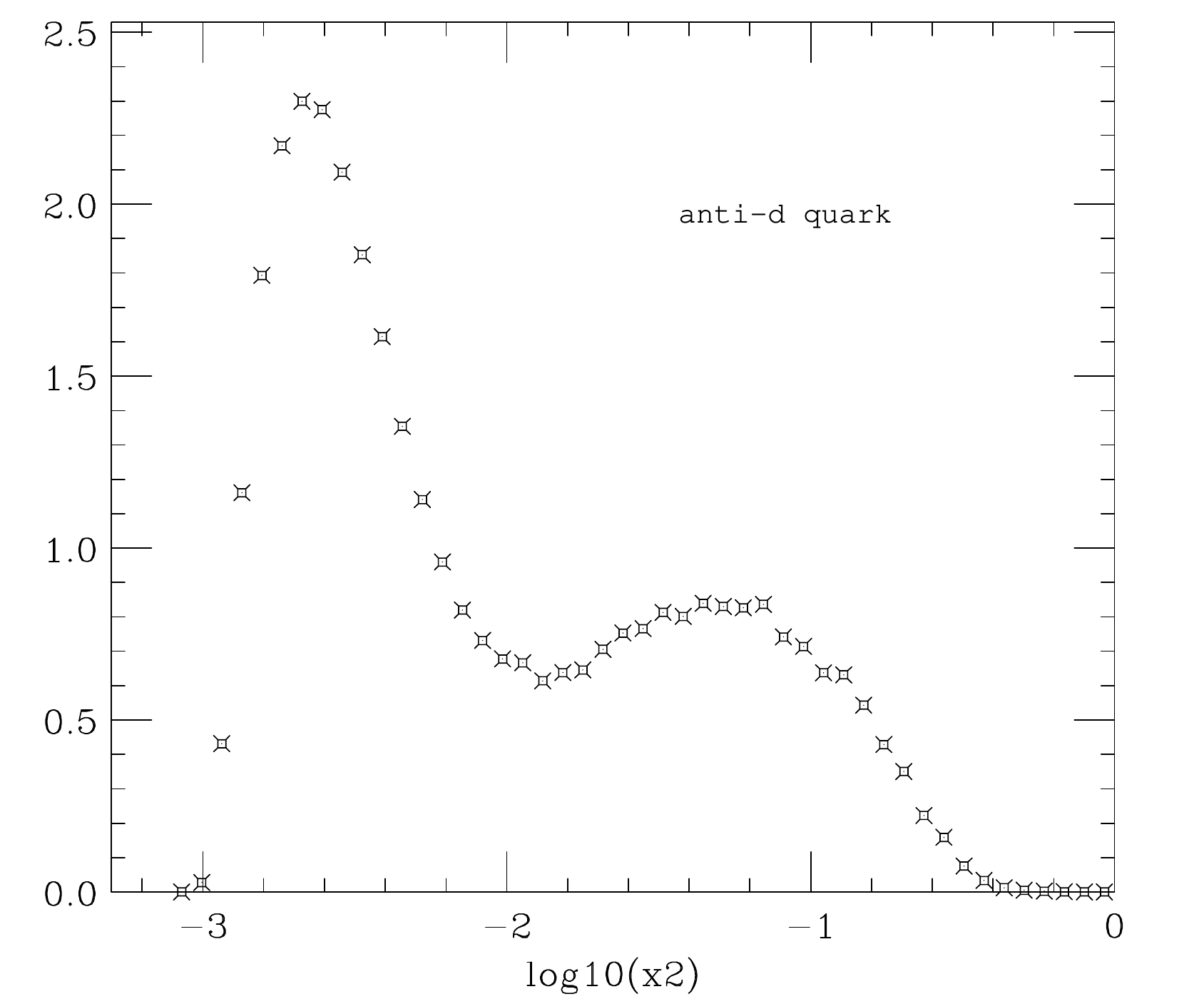}
    }
    \subfloat[]{
      \label{fig:Werner_dphot_PDF_F}
      \includegraphics[width=0.4\linewidth]{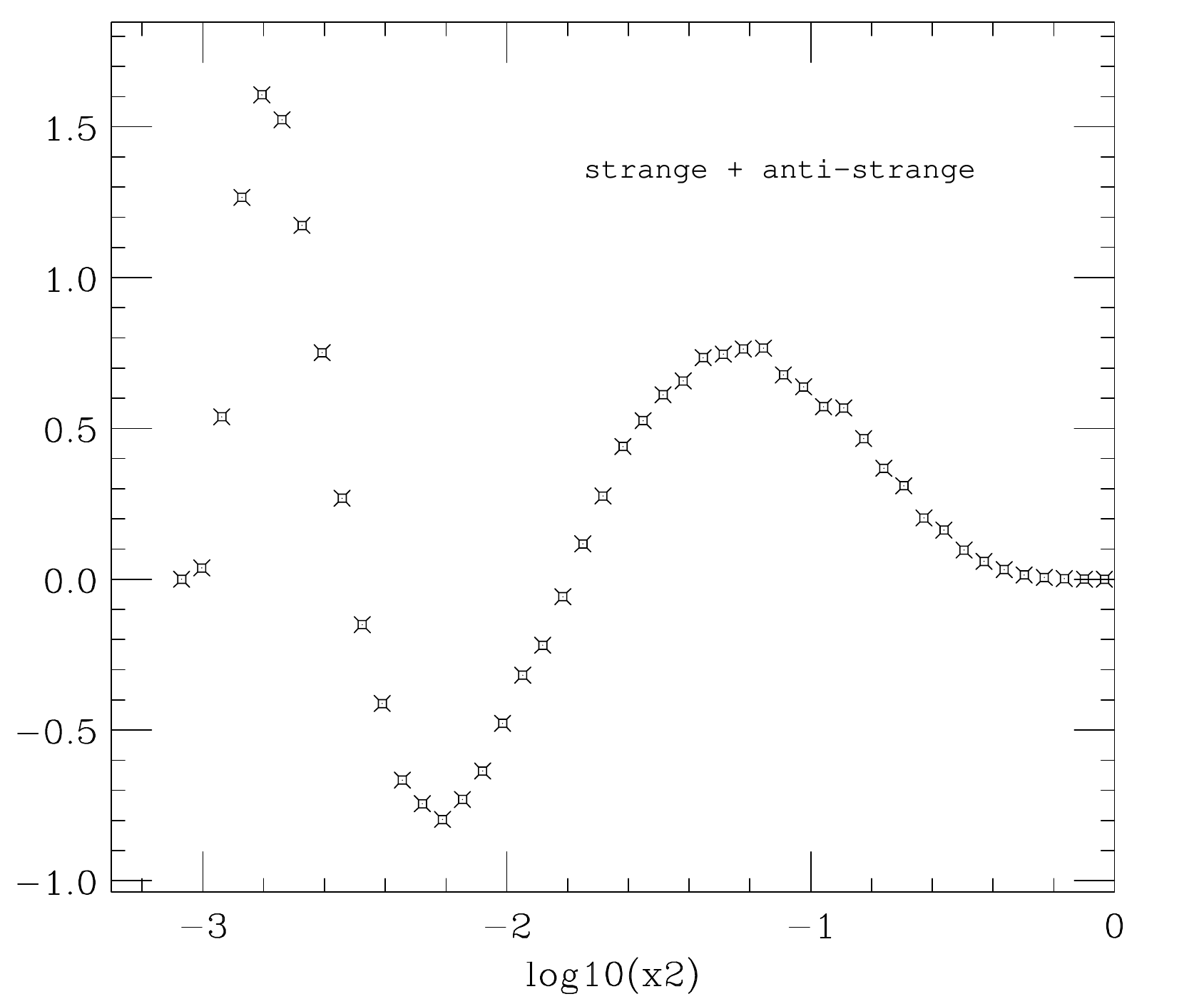}
    }
  \end{center}
  \caption{\label{fig:Werner_dphot_PDF} Differential NLO cross sections fom Werner Vogelsang as a function of $x_2$ for gluons (a),
u-quarks (b), d-quarks (c), anti-up quarks (d), anti-down quarks (e) and the combination of strange and anti-strange 
quarks (f). The vertical axis is $d\sigma/d\log(x_2)$ in the pseudorapidity range $3.1<\eta<3.8$.}

\end{figure}

Due to the nature of the NLO calculation, we cannot directly compare the QED radiated components between the NLO 
and {\sc Pythia} calculations.  However, in the MPC-EX acceptance the hard-scattering contribution to the direct 
photon cross section is dominated by gluon Compton scattering, which samples the gluon distribution in the Au 
nucleus.  The ability to access the Au nuclei's gluon distribution is the focus of the direct photon measurement.  
The portion of the cross section resulting from QED radiation samples the quark and antiquark parton distribution 
functions.  These PDFs access a different region in Bjorken x.  The NLO cross sections can be calculated as 
differential quantities in the parton $x_2$ by parton flavor to 
estimate the relative effect of direct photons from the hard scattering vertex (dominated by gluons) or from 
QED radiation from incoming quarks (dominated by the quark PDF's).  These differential cross sections are plotted 
in Figure~\ref{fig:Werner_dphot_PDF}, which shows that interactions involving the gluon in the target nucleon 
dominate the direct photon cross section by a large factor.  Interactions involving quarks and antiquarks are 
substantially smaller, and result from interactions that produce direct photons (via processes other than gluon 
Compton) and proceses that produce direct photons by QED radiation from the incoming quarks.

From the NLO differential cross sections in Figure~\ref{fig:Werner_dphot_PDF}, we conclude that the contribution of direct photons from 
QED radiation is small in the NLO calculation, and the {\sc Pythia} Monte Carlo dramatically overestimates the contribution of these 
photons.  Based on this, we exclude {\sc Pythia} events from our analysis when direct photons are generated by this process.  The 
remaining {\sc Pythia} yield of LO direct and fragmentation photons, as well as the relative yield of these photons to $\pi^0$ mesons, 
more closely approximates the NLO calculations.

\clearpage
  \label{sim:dphot}
\section[Direct Photons]{Direct Photons in d+Au Collisions}
\label{sim:dphot}


The direct photon can be measured with the MPC-EX detector at $p_{T} > 3$\,GeV.  When 
the simulation is scaled up to the expected event rates, the $R_{dAu}$ measurement 
for signal photons at $p_{T}>3$~GeV is statistically precise.  The measured $R_{dA}$'s 
ability to constrain existing models of the gluon nuclear parton distribution function 
depends on the systematic errors on the $R_{dAu}$ ratio.

At $p_{T} > 3$\,GeV, $\pi^{0}$ and $\eta$ decays are the primary backgrounds after 
applying the photon identification cuts.  Signal photons are extracted from the 
$\pi^{0}$ and $\eta$ backgrounds using the double ratio method.  First the simulation 
and the resulting signal photon and background yields are presented.  Then the double
ratio method and the corresponding systematic errors are discussed.  Finally these
values are used to demonstrate the sensitivity available with the MPC-EX detector to
restrict the viable region of the EPS09 gluon modification distribution\cite{Eskola:2009uj}.

\subsection{Simulation yields}
The MPC-EX detector's ability to measure direct photons in d+Au collisions 
is determined using a minimum bias 200 GeV p+p {\sc Pythia} simulation with a realistic 
vertex distribution.  At high transverse momenta, where the direct photon measurement is 
performed, binary-scaled p+p events provide a good approximation of d+Au events.  
Of the approximately 868 million simulated events, 270 million 
events satisfy the trigger requirement of a track with a MPC-EX energy above 16.5 GeV.  

\begin{table}[hbtp!]
  \begin{center} 
    \caption{Projected direct photon candidate yields in 49 $pb^{-1}$ p+p (12 weeks) and 0.35 $pb^{-1}$ d+Au (12 weeks) collisions using the Table~\ref{Tab:pi0Cuts} photon identification cuts.}
    \begin{tabular}{|l||c|c|} \hline
      Candidate Sources           & p+p       &  d+Au \\ \hline
      Hadrons                     &  14293.6  &    847.4    \\ 
      \hspace{5mm} $\pi$          &  11911.3  &    706.2    \\
      \hspace{5mm} K              &      0.0  &      0.0    \\
      \hspace{5mm} p              &      0.0  &      0.0    \\ 
      \hspace{5mm} neutrons       &      0.0  &      0.0    \\ 
      \hspace{5mm} other          &   2382.3  &    141.2    \\
      $\pi^{0}$                   & 557448.4  &  33048.7    \\
      $\eta$ decay                & 121495.2  &   7202.9    \\
      Other hadron decays         &  26204.8  &   1553.6    \\
      \hspace{5mm} $\omega$       &  21440.3  &   1271.1    \\
      \hspace{5mm} $\eta$'        &   2382.3  &    141.2    \\
      \hspace{5mm} other          &   2382.3  &    141.2    \\
      Signal photons              & 240608.1  &  14264.6    \\ 
      \hspace{5mm} frag. photons  & 102437.1  &   6073.1   \\ 
      \hspace{5mm} direct photons & 138171.0  &   8191.6   \\ \hline
    \end{tabular}
    \label{Tab:real_yields}
  \end{center}
\end{table}

The direct photon analysis yields using the {\sc Pythia} simulation is adjusted to 
give the total expected yields in p+p and d+Au collisions.  Table \ref{Tab:real_yields} 
presents the projected yields assuming a total integrated luminosity of 49 $pb^{-1}$ 
in 12 weeks of $\sqrt{200}$\,GeV p+p collisions and 0.35 $pb^{-1}$ in 12 weeks of 
$\sqrt{200}$\,GeV d+Au collisions.  These luminosities are calculated in Appendix~\ref{sec:A}.  
The simulated results using the $\pi^{0}$ rejection cuts from Table~\ref{Tab:pi0Cuts} 
in Section~\ref{sec:pi0_back} are used in this projection.  With these projected yields 
the $R_{dAu}$ measurement is statisically precise with a relative statistical error 
of approximatly 1.16\%.  The significance of the signal photon $R_{dA}$ measurement 
hinges on the systematic errors discussed later in this section.

\begin{figure}[hbt]
  \hspace*{-0.12in}
  \centering
  \includegraphics[width=0.7\linewidth]{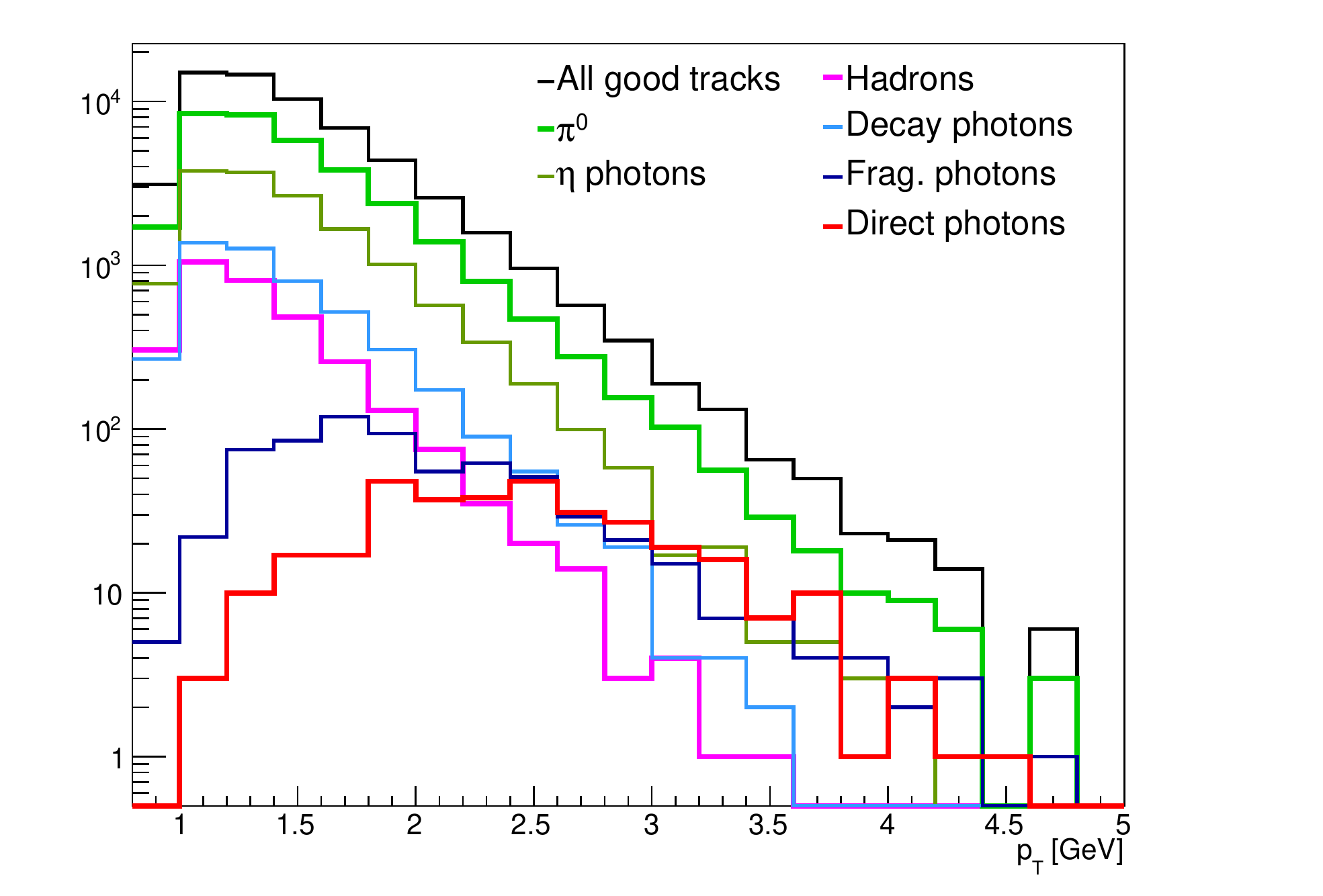}
  \vspace*{-0.12in}
  \caption{\label{Fig:PtDistrib}The $p_{T}$ distribution of photon candidate tracks.  
    All of the photon candidate tracks are shown in black. The spectrum of photon
    candidates from $\pi^{0}$ (bright green) and $\eta$ (olive green) decays are 
    shown separately.  Remaining decays, primarily from $\omega$ and $\eta$', are 
    in light blue.  Signal photons including those produced by fragmentation and 
    the initial hard interactions are in blue and red respectively.  Hadrons are in pink.} 
\end{figure}

\begin{figure}[hbt]
  \hspace*{-0.12in} 
  \begin{minipage}[b]{0.5\linewidth}
    \centering
    \includegraphics[width=0.95\linewidth]{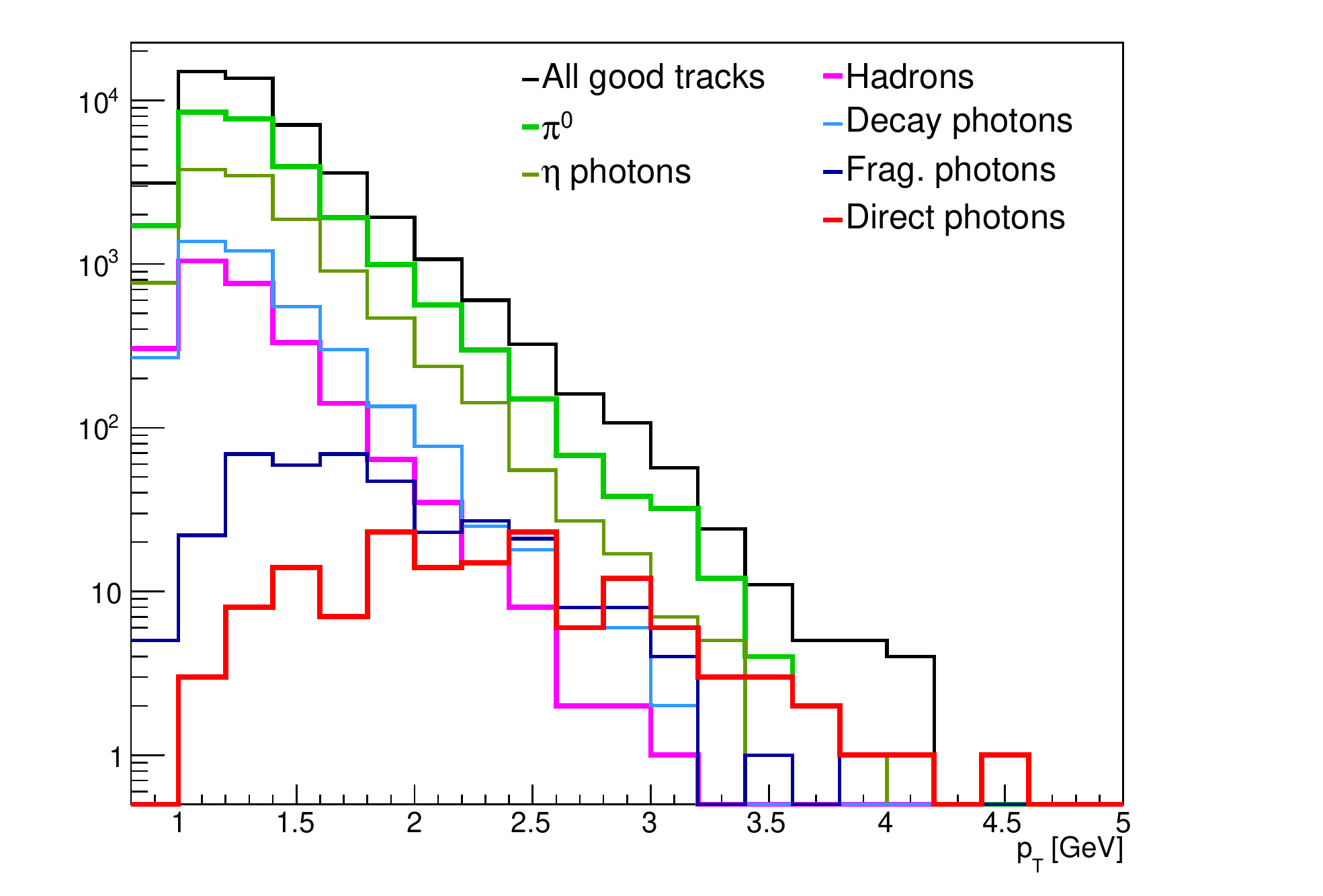}
  \end{minipage}
  \hspace{0.5cm}
  \begin{minipage}[b]{0.5\linewidth}
    \centering
    \includegraphics[width=0.95\linewidth]{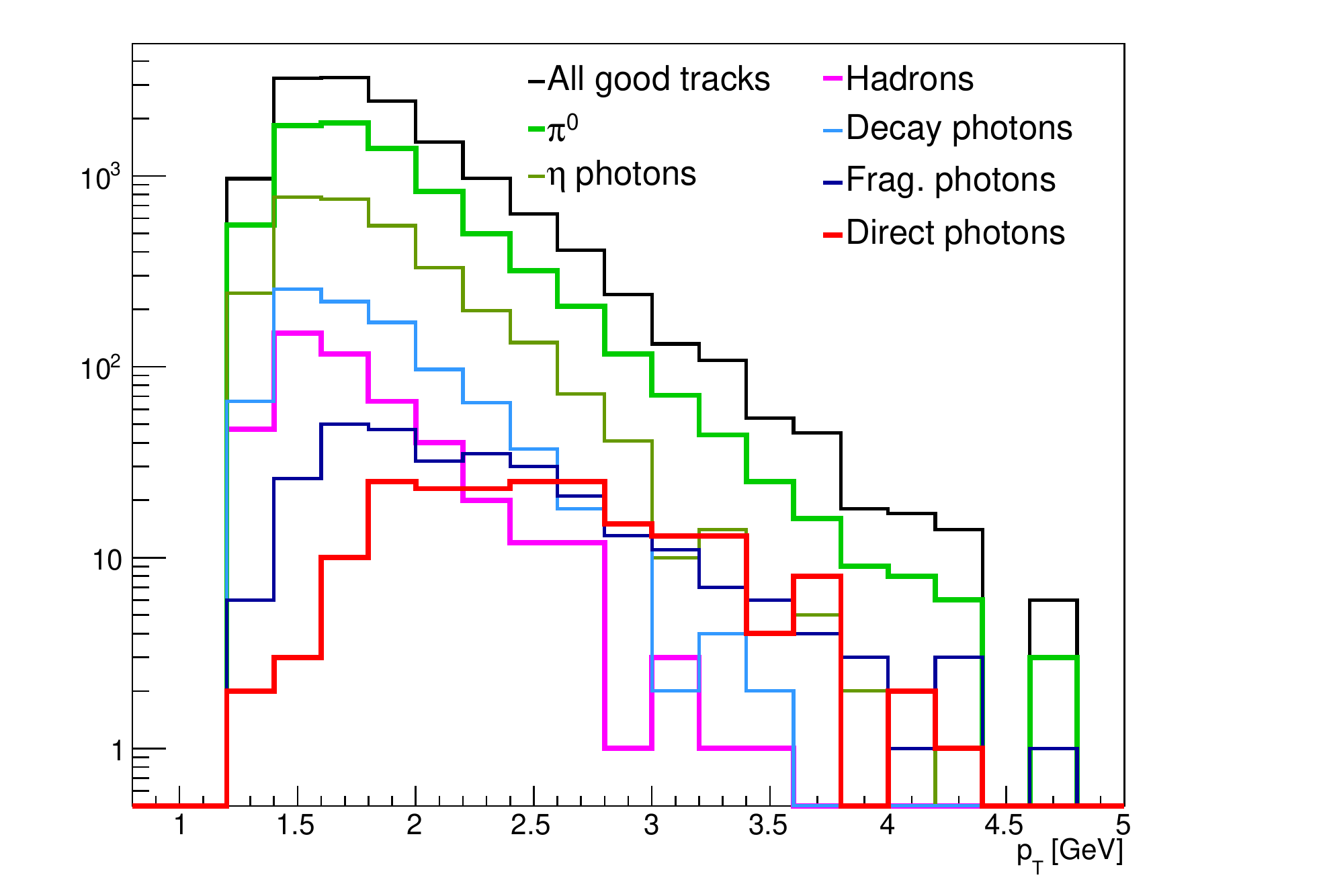}
  \end{minipage}
  \vspace*{-0.12in}
  \caption{\label{Fig:PtDistrib_eta}The $p_{T}$ distribution of photon candidate tracks in 
    inner and outer $\eta$ ranges.  The right panel shows the $p_{T}$ distribution with 
    $\eta$ between 3.1 and 3.45.  The left panel shows the $\eta$ distribution with $\eta$ 
    between 3.45 and 3.8.  All of the photon candidate tracks are shown in black, the 
    spectrum of photon candidates from $\pi^{0}$ and $\eta$ decay are shown in bright 
    and olive green.  Direct photons from the initial hard interaction are shown in red 
    and fragmentation photons are in dark blue. The direct and fragmentation photons are 
    further broken down according to their interaction type, $q+q$, $q+\bar{q}$, $q+g$ 
    and $g+g$ with various linestyles.}
\end{figure}

Signal photon candidates are selected with the $\pi^{0}$ rejection 
cuts described in Table~\ref{Tab:pi0Cuts} in Section~\ref{sim:dphotcuts}.  
Figure \ref{Fig:PtDistrib} presents the $p_{T}$ distributions for the 
direct photons and backgrounds surviving these cuts.  $\pi^{0}$ and 
photons from $\eta$ decays are the largest remaining components.  The 
direct photon signal remains below the $\pi^{0}$ contribution at all 
$p_{T}$ values.  At $p_{T}$ greater than 3\,GeV, the direct photons 
yield is larger than the background photons from $\eta$ decays.  Hadrons 
and other decay photons are relevant at lower $p_{T}$'s but are negligible 
compared to the direct photon contributions in the high $p_{T}$ range.  
Table \ref{Tab:breakdownPt} lists the particle contributions in four 
$p_{T}$ ranges with varied lower limits between 2.5 and 4\,GeV.  The 
direct photon candidates are separated into two $\eta$ ranges, an inner 
range of $3.1 < \eta < 3.45$ and an outer range of $3.45 < \eta < 3.8$, 
in Figure \ref{Fig:PtDistrib_eta} and Table \ref{Tab:breakdownEta}.

\begin{figure}[hbt]
  \hspace*{-0.12in}
  \centering
  \includegraphics[width=0.5\linewidth]{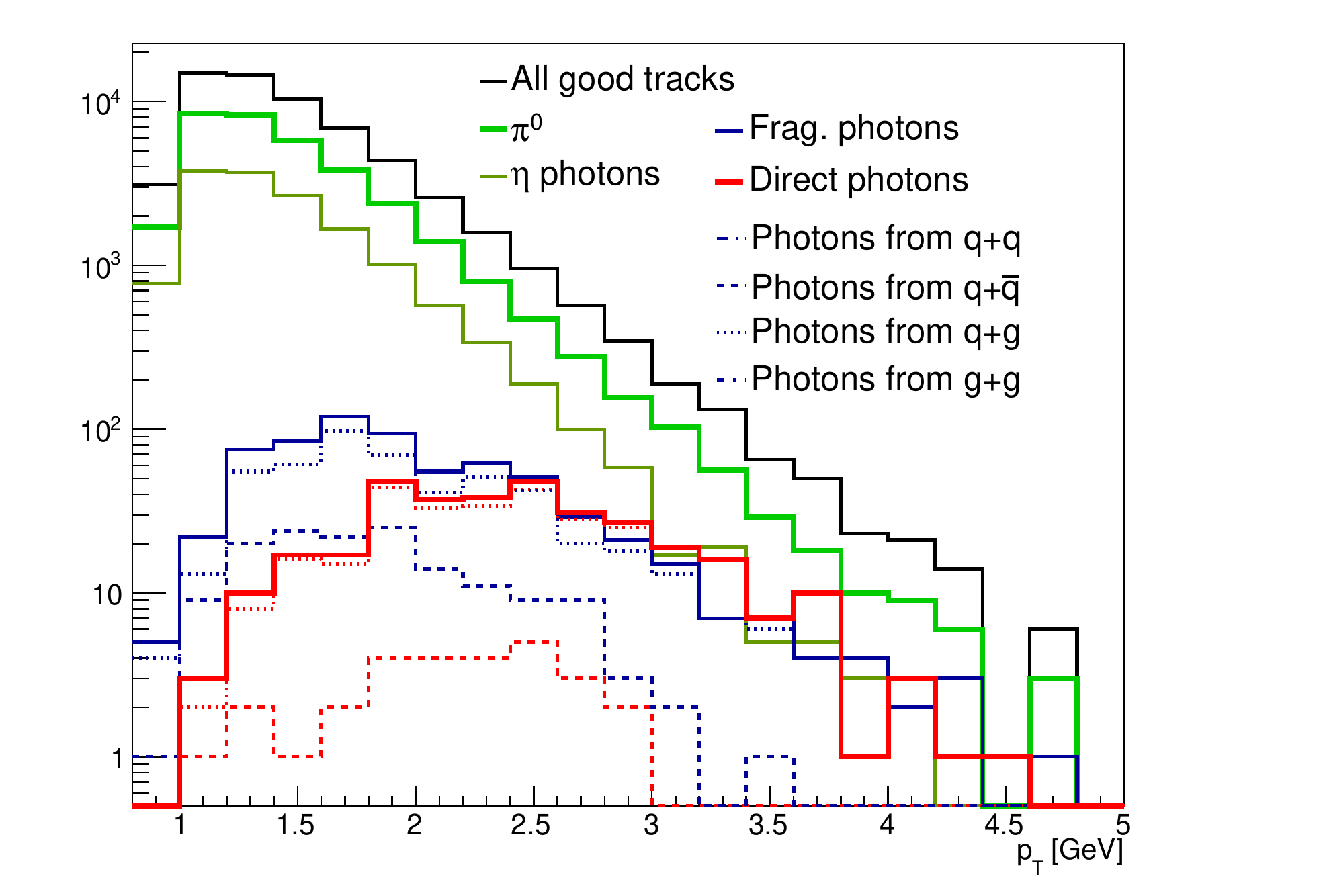}
  \vspace*{-0.12in}
  \caption{\label{Fig:DistribsQG}The $p_{T}$ distribution of the surviving signal photons
    separated by interaction type.  All of the photon candidate tracks are shown in black.
    Direct photons from the initial hard interaction are shown in red and fragmentation 
    photons are in dark blue. The signal photons are separated according to their 
    interaction type, $q+q$, $q+\bar{q}$, $q+g$ and $g+g$ with the various linestyles, 
    dashed-dot, dashed, dotted, and dash-dot-dot.}
\end{figure}

Signal photons are separated into direct photons from the initial hard scattering 
and photons from outgoing quark fragmentation.  They are further divided by their 
initial hard production mechanisms.  Direct photons from the initial hard scattering 
are produced by quark-gluon Compton scattering, quark-antiquark annihilation and 
gluon fusion. 
Fragmentation photons are produced in the following initial hard processes:
\begin{itemize} 
\item $q + q \rightarrow q + q $
\item $q + \bar{q} \rightarrow q + \bar{q}$ and $q + \bar{q} \rightarrow g + g$
\item $q + g \rightarrow q + g$
\item $g + g \rightarrow q + \bar{q}$ and $g + g \rightarrow g + g$
\end{itemize}
Direct photons from the initial hard scattering are divided into $q+\bar{q}$, 
$q+g$ and $g+g$ interaction types.  Fragmentation photons are separated into 
$q+q$, $q+\bar{q}$, $q+g$ and $g+g$ interaction types.  Figure \ref{Fig:DistribsQG} 
presents the $p_{T}$ distributions for the signal photon contributions, direct 
photons (red) and fragmentation photons (dark blue).
The signal photon contributions are further separated by their hard interaction 
type, $q+q$ (dash-dot-dotted), $q+\bar{q}$ (dashed), $q+g$ (dotted) and $g+g$ (dash-dotted).
The signal photon components separated by their hard interaction types are also 
detailed in Tables \ref{Tab:breakdownPt} and \ref{Tab:breakdownEta}.
Both the fragmentation photons and direct photons are dominated 
by the $q+g$ interaction.  $q+\bar{q}$ interactions provide small contributions 
to fragmentation photons.  There are no remaining $q+q$ or $g+g$ interactions. 

\begin{table}[hbtp!]
  \begin{center} 
    \caption{Breakdown of direct photon candidates with various minimum $p_{T}$'s }
    \begin{tabular}{|l||c|c|c|c|} \hline
      Candidate Sources   & $p_{T } > 2.5$ & $p_{T} > 3.0$ & $p_{T} > 3.5$ & $p_{T} > 4.0$ \\ \hline
      Charged hadrons             &   29      &    6       &      1        &        0      \\ 
      \hspace{5mm} $\pi$          &   25      &    5       &      1        &        0      \\
      \hspace{5mm} K              &    0      &    0       &      0        &        0      \\
      \hspace{5mm} p              &    0      &    0       &      0        &        0      \\ 
      \hspace{5mm} neutrons       &    1      &    0       &      0        &        0      \\ 
      \hspace{5mm} other          &    3      &    1       &      0        &        0      \\
      $\pi^{0}$                   &  857      &  234       &     57        &       18      \\
      $\eta$ decay                &  293      &   51       &     13        &        2      \\
      Other hadron decays         &   75      &   11       &      2        &        1      \\
      \hspace{5mm} $\omega$       &   59      &    9       &      2        &        1      \\
      \hspace{5mm} $\eta$'        &   15      &    1       &      0        &        0      \\
      \hspace{5mm} other          &    1      &    1       &      0        &        0      \\
      Signal photons              &  261      &  101       &     36        &       11      \\ 
      \hspace{5mm} frag. photons  &  117      &   43       &     17        &        6      \\ 
      \hspace{10mm} $q+q$         &    0      &    0       &      0        &        0      \\
      \hspace{10mm} $q+\bar{q}$   &   19      &    3       &      0        &        0      \\
      \hspace{10mm} $q+g$         &   98      &   40       &     17        &        6      \\
      \hspace{10mm} $g+g$         &    0      &    0       &      0        &        0      \\ 
      \hspace{5mm} direct photons &  144      &   58       &     19        &        5      \\ 
      \hspace{10mm} $q+\bar{q}$   &    4      &    0       &      0        &        0      \\
      \hspace{10mm} $q+g$         &  137      &   58       &     19        &        5      \\
      \hspace{10mm} $g+g$         &    0      &    0       &      0        &        0      \\ \hline
    \end{tabular}
    \label{Tab:breakdownPt}
  \end{center}
\end{table}

\begin{table}[hbtp!]
  \begin{center} 
    \caption{Breakdown of direct photon candidates at $p_{T} > 3$\,GeV with various $\eta$ ranges }
    \begin{tabular}{|l||c|c|c|c|} \hline
      Candidate Sources      & $3.1 < |\eta| < 3.8$ & $3.1 < |\eta| < 3.45$ & $3.45 < |\eta| < 3.8$ \\ \hline
      Hadrons                     &    6       &      4        &        2      \\ 
      \hspace{5mm} $\pi$          &    5       &      4        &        1      \\
      \hspace{5mm} K              &    0       &      0        &        0      \\
      \hspace{5mm} p              &    0       &      0        &        0      \\ 
      \hspace{5mm} neutrons       &    0       &      0        &        0      \\ 
      \hspace{5mm} other          &    1       &      1        &        0      \\
      $\pi^{0}$                   &  234       &    182        &       52      \\
      $\eta$ decay                &   51       &     37        &       14      \\
      Other hadron decays         &   11       &      8        &        2      \\
      \hspace{5mm} $\omega$       &    9       &      1        &        1     \\
      \hspace{5mm} $\eta$'        &    1       &      1        &        1      \\
      \hspace{5mm} other          &    1       &      0        &        0      \\
      Signal photons              &  101       &     77        &       24      \\ 
      \hspace{5mm} frag. photons  &   43       &     36        &        7      \\ 
      \hspace{10mm} $q+q$         &    0       &      0        &        0      \\
      \hspace{10mm} $q+\bar{q}$   &    3       &      3        &        0      \\
      \hspace{10mm} $q+g$         &   40       &     33        &        7      \\
      \hspace{10mm} $g+g$         &    0       &      0        &        0      \\ 
      \hspace{5mm} direct photons &   58       &     41        &       17       \\ 
      \hspace{10mm} $q+\bar{q}$   &    0       &      0        &        0      \\
      \hspace{10mm} $q+g$         &   58       &     41        &       17      \\
      \hspace{10mm} $g+g$         &    0       &      0        &        0      \\ \hline
    \end{tabular}
    \label{Tab:breakdownEta}
  \end{center}
\end{table}


%
%

\begin{figure}[hbt]
  \hspace*{-0.12in}
  \centering
  \includegraphics[width=0.5\linewidth]{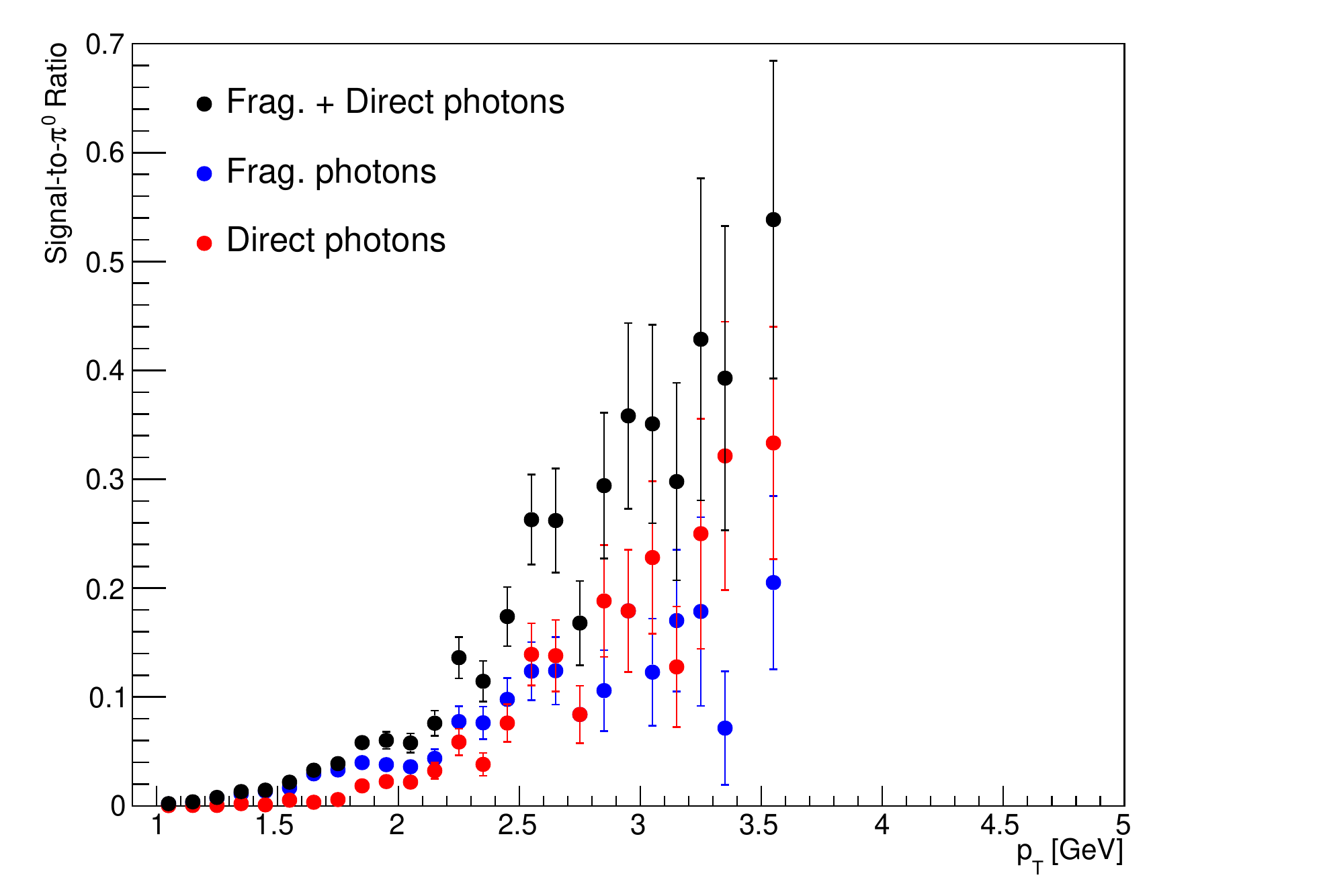}
  \vspace*{-0.12in}
  \caption{\label{Fig:DistribSigPi0}The signal-to-$\pi^{0}$ ratio versus $p_{T}$ 
    for photon candidate tracks. The combined direct and fragmentation photon signal
    is shown in black.  Fragmentation photons are in blue and direct photons are in red.}
\end{figure}

Figure \ref{Fig:DistribSigPi0} shows the signal-to-$\pi^{0}$ ratio as a function of $p_{T}$
for the photon candidates from fragmentation and direct photons using the $\pi^{0}$
rejection cuts.  The signal-to-$\pi^{0}$ for the combined direct and fragmentation photon 
signal is shown and the ratios for the direct (red) and fragmentation (blue) photons are
shown independently.  As the $p_{T}$ increases, the signal-to-$\pi^{0}$ ratio also rises
leading to a ratio of 0.43 $\pm$ 0.05 (stat) at $p_{T}$ greater than 3\,GeV.  
Table~\ref{Tab:sigtopi0} presents the signal-to-$\pi^{0}$ ratio for high $p_{T}$ signal 
photons with various photon identification cuts applied.  The combined fragmentation and direct
photon signal-to-$\pi^{0}$ ratio is unaffected by changing the $\eta$ range. 
The signal-to-$\pi^{0}$ ratios when
the different cuts are applied ($\pi^{0}$ rejection vs. Cuts 1 vs. Cuts 2) are consistent with
no effect on the ratio.  This comparison is limited in
by the large statistical errors in the simulation particularly for the Cuts 2 results.
In Figure~\ref{Fig:DistribSigPi0}, 
the increase in the signal to background at high $p_{T}$ is in part a result of the decrease in the $\pi^{0}$
yields at high $p_{T}$ but it also reflects the fact that our $\pi^{0}$ rejection cuts are 
tailored to enhance the signal-to-$\pi^{0}$ ratio at high $p_{T}$.
The direct photon signal-to-$\pi^{0}$ ratio increases at a steeper rate than then the 
fragmentation photon ratio.  This shows the potential to increase the direct photon 
concentration in the measured signal at high $p_{T}$.  The importance of the direct
photon concentration and the ability to increase this quantity in the MPC-EX is presented 
in Sections~\ref{sim:frag_phot} and~\ref{sim:R}. 

\begin{table}[hbtp!]
  \begin{center} 
    \caption{Signal-to-$\pi^{0}$ ratio of direct photon candidates with $p_{T} > 3$\,GeV and various cuts applied.  
      The simulation's statistical errors in these values are also presented.}
    \begin{tabular}{|l||c|c|c|} \hline 
      Cuts used     &  Frag + direct & Frag. & Direct \\ \hline
      $\pi^{0}$ rejection (Table \ref{Tab:pi0Cuts}),  &    0.43 $\pm$ 0.05   &   0.18 $\pm$ 0.03  &   0.25 $\pm$ 0.04  \\
      $3.1 < |\eta| < 3.8$     &    &   &  \\
      $\pi^{0}$ rejection (Table~\ref{Tab:pi0Cuts}),    &    0.42 $\pm$ 0.06   &   0.20 $\pm$ 0.04  &   0.22 $\pm$ 0.04  \\
      $3.1 < |\eta| < 3.45$     &    &   &  \\
      $\pi^{0}$ rejection (Table~\ref{Tab:pi0Cuts}),   &    0.46 $\pm$ 0.11   &   0.13 $\pm$ 0.05  &   0.33 $\pm$ 0.06  \\
      $3.45 < |\eta| < 3.8$     &    &   &  \\
      Cuts 1 frag. rejection (Table~\ref{Tab:pi0fragCuts}), &    0.37 $\pm$ 0.07   &   0.12 $\pm$ 0.03  &   0.25 $\pm$ 0.05  \\
      $3.1 < |\eta| < 3.8$     &    &   &  \\
      Cuts 2 frag. rejection (Table~\ref{Tab:pi0fragCuts}), &    0.40 $\pm$ 0.13   &   0.09 $\pm$ 0.05  &   0.31 $\pm$ 0.11  \\ 
      $3.1 < |\eta| < 3.8$     &    &   &  \\ \hline
    \end{tabular}
    \label{Tab:sigtopi0}
  \end{center}
\end{table}


\subsubsection{Remaining fragmentation photons} 
\label{sim:frag_phot}

Direct photons are more strongly suppressed 
by the nuclei's gluon distribution in a d+Au collision.  Reducing the
number of measured fragmentation photons and understanding their
contribution to the signal photon measurement may increase the significance
of the $R_{dA}$ measurement and thereby increase the EPS09 exclusion regions.  
The effect of some fragmentation photon contamination in the signal photon 
$R_{dA}$ is simulated and shown in Section \ref{sec:cnm}.

\begin{figure}[hbt]
  \hspace*{-0.12in}
  \begin{minipage}[b]{0.5\linewidth}
    \centering
    \includegraphics[width=0.95\linewidth]{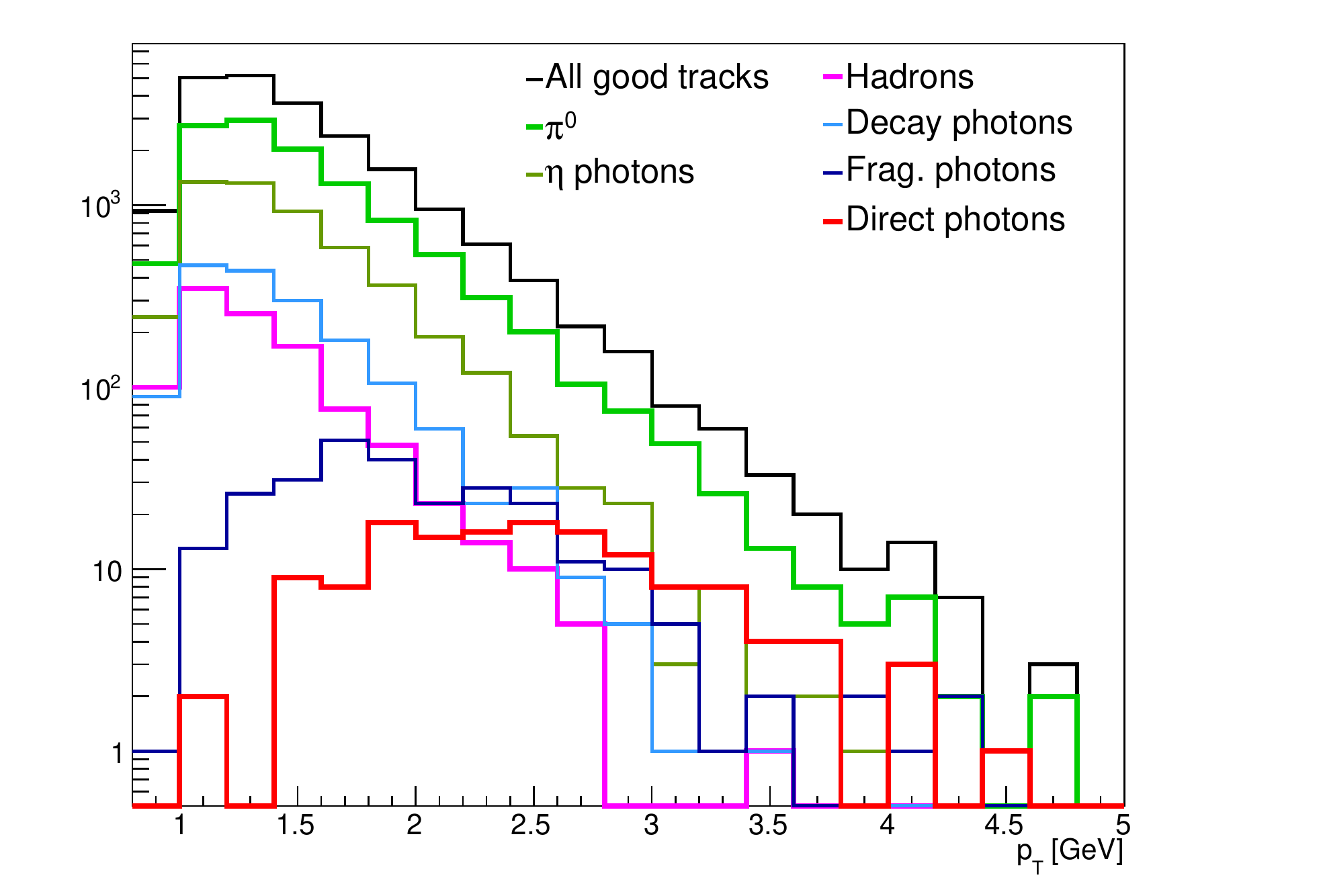}
  \end{minipage}
  \hspace{0.5cm}
  \begin{minipage}[b]{0.5\linewidth}
    \centering
    \includegraphics[width=0.95\linewidth]{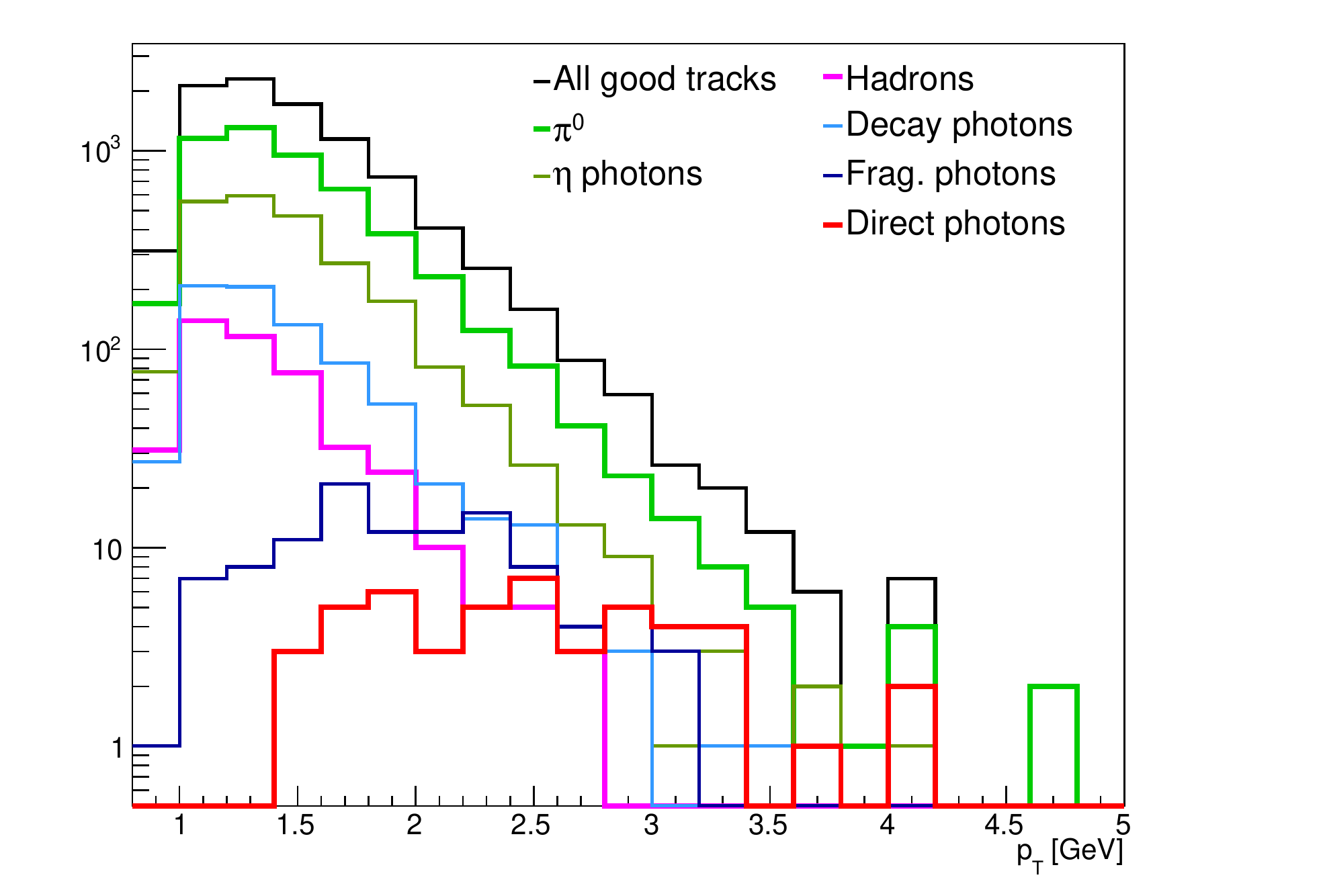}
  \end{minipage}
  \vspace*{-0.12in}
  \caption{\label{Fig:PtDistrib_frag}The $p_{T}$ distribution of photon candidate tracks 
    using the Cuts 1 and Cuts 2 analyses presented in Section~\ref{sim:dphotcuts}.  
    The right panel shows the $p_{T}$ distribution with the Cuts 1 ranges applied.
    The left panel shows the $p_{T}$ distribution with the Cuts 2 ranges applied.
    All of the photon candidate tracks are shown in black, the spectrum of photon 
    candidates from $\pi^{0}$ and $\eta$ decay are shown in bright and olive green.  
    Other decays and hadrons are in light blue and pink respectively.
    Direct photons from the initial hard interaction are shown in red 
    and fragmentation photons are in dark blue.}
\end{figure}

Before photon identification cuts are applied, direct and 
fragmentation photons have roughly equal contributions
to the signal photons in the {\sc Pythia} sample.  As mentioned 
in Section \ref{sim:dphotcuts}, the $\pi^{0}$ cuts increase the 
direct photon contribution to 57.4\% of the signal photon measurement.
With the less efficient Cuts 1 and Cuts 2 analyses, which are designed 
to reduce the fragmentation photon contribution, direct photons are 68.3 
and 78.6\% of the signal measurement respectively.  These cuts are 
presented in Table~\ref{Tab:pi0fragCuts} in Section~\ref{sim:dphotcuts}.  
As a comparison two additional analyses are completed with these tighter 
fragmentation cuts, Cuts 1 and Cuts 2, applied.  
Figure~\ref{Fig:PtDistrib_frag} shows the $p_{T}$ distributions of the 
surviving photon candidate tracks from the Cuts 1 and Cuts 2 analyses.  
The photon candidate yields decrease considerably.  The simulation yields 
with the Cuts 1 and Cuts 2 ranges applied are listed in 
Table~\ref{Tab:AllCutsYield} in the Section~\ref{sim:dphotcuts}.  

\begin{table}[hbtp!]
  \begin{center} 
    \caption{Systematic errors of direct photon candidates in various $\eta$ ranges. For the final exclusion plots the full $\eta$ range is used.}
    \begin{tabular}{|l||c|c|c|} \hline 
                          & $3.1 < |\eta| < 3.8$ & $3.1 < |\eta| < 3.45$ & $3.45 < |\eta| < 3.8$ \\ \hline
      direct-to-signal ratio                     &    57.4\%   &   58.3\%  &   56.6\%  \\
      $\Delta \gamma_{Incl}/ \gamma_{Incl}$      &    0.65\%   &   0.85\%  &    0.21\% \\
      R$_\gamma$                                 &    1.34     &   1.34    &    1.35   \\ 
      $\Delta$ R$_\gamma$/ R$_\gamma$            &    7.22\%   &   7.24\%  &    7.24\% \\ \hline
    \end{tabular}
    \label{Tab:sysErrR}
  \end{center}
\end{table}

\begin{table}[hbtp!]
  \begin{center} 
    \caption{Systematic errors of direct photon candidates in various $p_{T}$ ranges. For the final exclusion plots the full $p_{T} > 3.0$\,GeV range is used.}
    \begin{tabular}{|l||c|c|c|c|} \hline 
                                         & $p_{T} > 2.5$ & $p_{T} > 3.0$ & $p_{T} > 3.5$ & $p_{T} > 4/0$ \\ \hline
      direct-to-signal ratio                  & 43.7\%   &   57.4\%    &  52.8\%  &  45.5\%   \\
      $\Delta \gamma_{Incl}/ \gamma_{Incl}$   &  0.85\%  &    0.65\%   &   0.93\% &   0.00\%  \\
      R$_\gamma$                              &  1.21    &    1.34     &   1.50   &   1.52    \\ 
      $\Delta$ R$_\gamma$/ R$_\gamma$         &  7.25\%  &    7.22\%   &   7.21\% &   7.21\%  \\ \hline
    \end{tabular}
    \label{Tab:sysErrR_pt}
  \end{center}
\end{table}

\begin{table}[hbtp!]
  \begin{center} 
    \caption{Systematic errors of direct photon candidates with varied fragmentation photon contributions. For the final exclusion plots the $\pi^0$ cuts are used.}
    \begin{tabular}{|l||c|c|c|} \hline
                                                 & $\pi^{0}$ Cuts &  Cuts 1 &  Cuts 2  \\ \hline
      direct-to-signal ratio                     &    57.4\%      &  68.3\% &  78.6\%  \\
      $\Delta \gamma_{Incl}/ \gamma_{Incl}$      &    0.65\%      &  0.46\% &   0.00\% \\
      R$_\gamma$                                 &    1.34        &  1.31   &   1.31   \\ 
      $\Delta$ R$_\gamma$/ R$_\gamma$            &    7.22\%      &  7.22\% &   7.21\% \\ \hline
    \end{tabular}
    \label{Tab:sysErrR_frag}
  \end{center}
\end{table}

\subsection{Calculation of R$_{dAu}$ and  Systematic errors}
\label{sim:R}

The direct photon yield is calculated using a double ratio method
that allows for the cancellation of many of the systematic
errors.  This method is used in many other analyses published by the
PHENIX collaboration \cite{Adler:2005ig}.  The ratio, R$_\gamma$, 
is calculated using the measured inclusive photon and $\pi^{0}$
spectra compared to the known contributions from hadronic decays
relative to the $\pi^{0}$.  Equation \ref{Eq:R} presents the 
formula for R$_\gamma$,
\begin{equation}
R_\gamma = \frac{ \left( \frac{\gamma_{Incl}}{\pi^{0}} \right)_{Meas} }{ \left( \frac{\gamma_{Incl}}{\pi^{0}} \right)_{Sim}}
\label{Eq:R}
\end{equation}
where $\gamma_{Incl}$ and $\pi^{0}$ are the inclusive photon and
$\pi^{0}$ spectra respectively.  The numerator is the measured
inclusive photon-$\pi^{0}$ ratio and the denominator is a simulated
ratio based on the known decay particle yields.  Direct photons in the
measured inclusive photon spectrum result in R$_\gamma$ values greater
than one.  In turn the R$_\gamma$ value is used to determine the
direct photon contribution from the inclusive photon spectrum
according to Equation \ref{Eq:DoubleRDirect}.
\begin{equation}
\gamma_{Direct} = \gamma_{Incl} * \left(1 - 1/R_\gamma\right)
\label{Eq:DoubleRDirect}
\end{equation}
We estimate R$_\gamma$ using our simulated analysis.  First the 
hadron contributions are subtracted to determine the measured
$\gamma_{Incl}$ spectrum.  The simulated $\gamma_{Incl}^{sim}$ 
spectrum is approximated by summing the various photon decay 
contributions, $\gamma_{Incl}^{sim} = \pi^{0} + \eta + \eta' + \omega +$other
hadronic decays.  The R$_\gamma$ value is calculated from the 
yields in Table \ref{Tab:breakdownEta} producing an average value 
of 1.34 over the MPC-EX's $\eta$ acceptance.

Based on the MPC-EX's capabilities and the experience with the
MPC\cite{Meredith:2011a}, the $\pi^{0}$ cross-section at energies
above 20~GeV can be measured to within an estimated systematic error
of about 6\%, including the effects of an uncertainty in the absolute
energy scale of 1 to 2\%.  The error in $\gamma_{incl}$ is also
dominated by the energy scale. The effects of these energy scale
uncertainties in the final results are quantified later in this
section. In the MPC analysis at low $p_{T}$ the main systematic
uncertainty arose from the combinatorial background subtraction.  
This is not an issue for single-track analysis in the MPC-EX. 
The MPC-EX will be unable to provide detailed information on false
photon candidates created by charged hadrons interacting in the 
detector.  We expect that this background will be model dependent,
and assign as 20\% systematic error to this component.  Finally there
is an additional systematic error of 3\% on $R_{dAu}$ from the
determination of $\langle N_{coll}\rangle$ in the Glauber model (under
the assumption that the systematic error in the p+p cross section
cancels in the $R_{dAu}$ ratio)\cite{PHENIX_GLAUBER}.

The systematic error in R$_\gamma$ is calculated according to Equation \ref{Eq:deltaR}
\begin{equation}
\Delta R_\gamma = \frac{\delta R_\gamma}{\delta \gamma_{Incl}}\Delta \gamma_{Incl} \oplus \frac{\delta R_\gamma}{\delta \pi^{0}} \Delta \pi^{0} \oplus \frac{\delta R_\gamma}{\delta MC_{sim}} \Delta MC_{sim} 
\label{Eq:deltaR}
\end{equation}
where $MC_{sim}$ is the simulated inclusive photon-to-$\pi^{0}$ ratio.
  Equation \ref{Eq:deltaRR} gives the relative systematic error of R$_\gamma$.
\begin{equation}
\frac{\Delta R_\gamma}{R_\gamma} = \frac{\Delta \gamma_{Incl}}{\gamma_{Incl}} \oplus \frac{\Delta \pi^{0}}{\pi^{0}} \oplus \frac{\Delta MC_{sim}}{ MC_{sim}}
\label{Eq:deltaRR}
\end{equation}
Tables \ref{Tab:sysErrR}, \ref{Tab:sysErrR_pt} and \ref{Tab:sysErrR_frag} 
present the systematic errors calculated for the R$_\gamma$ assuming 6\% 
relative error in the $\pi^{0}$ as mentioned previously, 4\% relative error 
in the simulated $\gamma_{Inc}$-to-$\pi^{0}$ ratio, and 20\% relative error 
in the hadron subtraction from the inclusive photons.  Table~\ref{Tab:sysErrR}
displays the R$_\gamma$ and its systematic error for the $\pi^{0}$ rejection 
analysis over the entire MPC-EX $\eta$ acceptance and separated into inner 
and outer $\eta$ ranges.  Table~\ref{Tab:sysErrR_pt} presents the same information
for various $p_{T}$ lower limits ranging from 2.5 to 4\,GeV.  In these tables, 
the R$_{\gamma}$ appears to increase at higher $p_{T}$ and is unaffected by the separate $\eta$ ranges.  
The relative systematic error in R$_\gamma$ remains stable at values around 7.22\%.
Table~\ref{Tab:sysErrR_frag} is particularly valuable since it allows us to 
compare the R$_\gamma$ and the corresponding systematic errors for various 
direct photon concentrations.  The R$_\gamma$ values are flat within the 
systematic errors and there is little to no improvedment in the R relative 
systematic error.  However, the statistical errors in the Cuts 1 and Cuts2 
analyses are larger as a result of the reduced yields and low signal efficiencies 
of these cuts as seen in Section~\ref{sec:frag_back}.  We continue
with the preferred $\pi^{0}$ cuts analysis because of the larger signal 
efficiency and the lack of a substantial effect on R$_\gamma$ as a result 
of the fragmentation photon contribution.

We now have R$_\gamma$, $\gamma_{incl}$ and their associated systematic errors
for both p+p and d+Au. The statistical errors on the yields as input to the final 
result are calculated using Appendix~\ref{sec:rates} and the efficiencies given 
in Table~\ref{Tab:pi0CutsYieldEffi}.  R$_{dAu}$ is then calculated in each bin
of $\eta$ as 
\begin{equation} R_{dAu} =\frac{1}{\langle
    N_{coll}\rangle}\frac{ \gamma_{Incl}^{dAu} * \left(1 -
    1/R_{\gamma}^{dAu}\right)}{\gamma_{Incl}^{pp} * \left(1 -
    1/R_{\gamma}^{pp}\right)}
\label{Eq:RdAu}
\end{equation}.

\subsection{EPS09 Exclusion Plot}

\begin{figure}[hbt]
  \hspace*{-0.12in}
  \centering
  \includegraphics[width=0.5\linewidth]{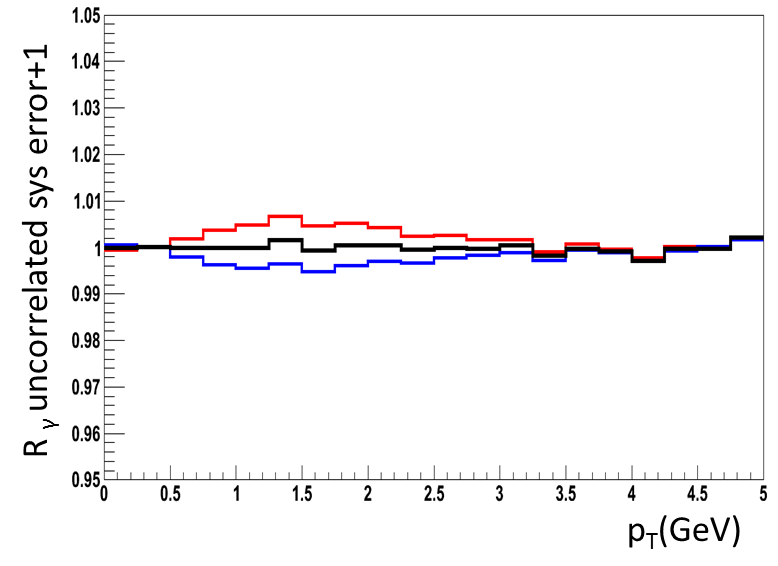}
  \vspace*{-0.12in}
  \caption{\label{Fig:pi0syserr} The ratio of R$_{\gamma}$ in blue
    (red) where we have allowed the energy scale to be increased
    (decreased) by 2\% over the nominal value. The black line is when
    no shift is applied and indicates the statistics of the simulation.
    The systematic error in $R_{\gamma}$ due to the energy scale less
    than 1\%.}
\end{figure}

The MPC-EX's ability to constrain the viable region of the EPS09 gluon 
modification, R$_{G}$, is presented as an exclusion plot.  Here we detail
how the exclusion plot is calculated and how the 1$\sigma$ and 90\% 
confidence level bands are determined.  The gluon suppression factor from 
EPS09 is assumed.  Our simulated photon events are weighted 
according to their x$_2$ and Q$^2$ values and their sources, either direct photons, 
fragmentation photons, hadron decays or $\pi^0$ backgrounds. We then
account for both statistical and systematic errors, and make an exclusion plot for the
various R$_G$ distributions given by EPS09\cite{Eskola:2009uj}.  We assume that the
systematic error in R$_{\gamma}^{pp}$ and R$_{\gamma}^{dAu}$, which are
dominated by the energy scale, are largely correlated.  If the spectra p+p
and d+Au spectra were identical, they would be completely correlated.
To estimate the error from any difference between the p+p and d+Au spectra, 
a toy model of the raw spectra in p+p and d+Au is made.  This gives the 
correct value of R$_{dAu}$ for $\pi^0$s as measured in \cite{Meredith:2011a,Adare:2011sc}.  
The 2\% energy scale error is propagated to R$_{\gamma}$. Figure
\ref{Fig:pi0syserr} shows the ratio between the value of R$_{\gamma}$
with a shifted energy scale to the R$_{\gamma}$ without the energy scale
shift.  The error in R$_{\gamma}$ is less than 1\%.  A similar
procedure is followed to find the error in the ratio
$\gamma_{inc}^{dAu}/\gamma_{inc}^{pp}$ resulting in an error value of 2\%.

\begin{figure}[hbt]
  \hspace*{-0.12in}
  \centering
  \includegraphics[width=0.9\linewidth]{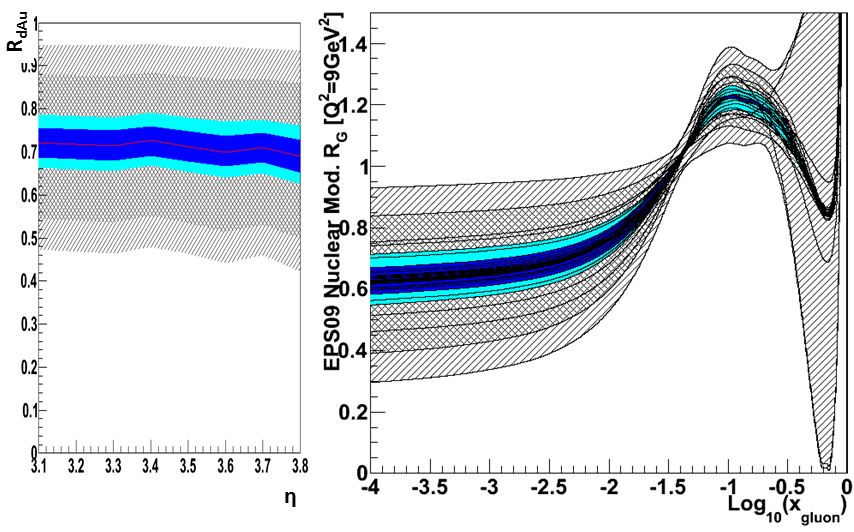}
  \vspace*{-0.12in}
  \caption{\label{Fig:rf_limits} EPS09 exclusion plots in R$_{dAu}$
    (left) and R$_G$ (right). The outer hatched lines are the 90\%
    confidence level envelope of all the EPS09 curves. The light blue
    areas represent the 90\% confidence level limits of the simulated
    measurement, while the dark blue represent the 1$\sigma$
    limits.  The nominal value is taken as the central EPS09 curve.}
\end{figure}

We vary all of the systematic errors over three standard deviations in 
each direction, a standard PHENIX procedure \cite{Adare:2008cg}.  
The systematic errors input into this calculation are summarized 
in Table \ref{Tab:sysErrExclusion}.  For each of the EPS09 R$_G$ 
suppression functions we find the set of R$_\gamma^{dAu}$, R$_\gamma^{pp}$, 
$\gamma_{incl}^{dAu}$ and $\gamma_{incl}^{pp}$ values within 
three standard deviation of the nominal value located at the $\chi^2$ minimum.  
For each R$_G$ curve, a value of the $\chi^2$ consistent with the simulated
data is calculated.  The associated $\chi^{2}$ values are used to identify the 
R$_G$ curves that are within 1$\sigma$ of the simulated data.  The R$_G$ curves 
consistent at the 90\% confidence level are also identified.  The exclusion plots 
of R$_{dAu}$ and the corresponding EPS09 suppression factors are shown in Figure
\ref{Fig:rf_limits}.  The hatched and light blue regions show the 
exclusion at the 90\% confidence level for all of the EPS09 curves and 
using the simulated MPC-EX measurement respectively.  The dark blue region 
represents the MPC-EX exclusion at the 1$\sigma$ limit.  The black lines 
within the dark blue band are the EPS09 R$_{G}$ curves that are consistent 
with the nominal values of the simulated data.  The exclusion plots 
corresponding to Cuts 1 and 2 results are also calculated and are 
consistent with these results.  

\begin{table}[hbtp!]
  \begin{center}
    \caption{Quantities used for the calculation of the exclusion plots.}
    \begin{tabular}{|p{9cm}||p{1cm}|p{3.5cm}|} \hline
      Quantity                                     & value      & reference \\ \hline\hline
      R$_{\gamma}$                                    &    1.34    &   Table \ref{Tab:sysErrR_frag}        \\ \hline 
      systematic error $\Delta$ R$_\gamma$/ R$_\gamma$ correlated between dAu and pp   &    7.2\%     &   Table \ref{Tab:sysErrR_frag}      \\ \hline 
      systematic error $\Delta$ R$_\gamma$/ R$_\gamma$ uncorrelated between dAu and pp &    1\%     & Fig. \ref{Fig:pi0syserr}   \\ \hline
      systematic error $\Delta \gamma_{Incl}/\gamma_{Incl} $ uncorrelated between dAu and pp &    2\%     &  from calculation similar to Fig. \ref{Fig:pi0syserr}   \\ \hline
      relative systematic error on $\langle N_{coll}\rangle$      &     3\%     & \cite{PHENIX_GLAUBER}  \\ \hline
    \end{tabular}
    \label{Tab:sysErrExclusion}
  \end{center}
\end{table}

\begin{figure}[hbt]
  \hspace*{-0.12in}
  \centering
  \includegraphics[width=0.9\linewidth]{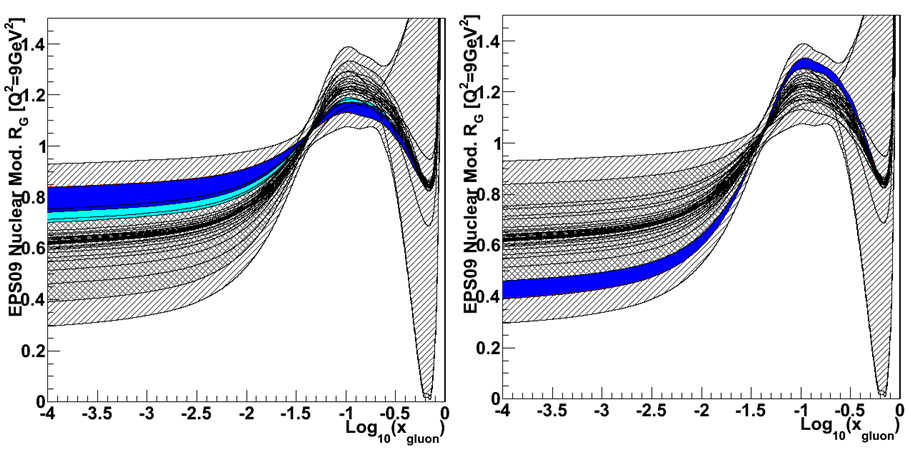}
  \vspace*{-0.12in}
  \caption{\label{Fig:limits_lower} EPS09 exclusion plots in R$_G$
    where the nominal value is taken as the EPS09 curve corresponding to the
    least (left) or greatest (right) amount of suppression.  The outer
    hatched lines are the 90\% confidence level envelope of all the
    EPS09 curves. The light blue line represents the 90\% confidence
    level limits of the simulated measurement, while the dark blue
    represent the 1$\sigma$ limits. No light blue region is visible if the
    90\% CL and the 1$\sigma$ level coincide to within the resolution given
    by the EPS09 theoretical curves available.}
\end{figure}

The central value of EPS09 is taken as the nominal value in Figure
\ref{Fig:rf_limits}.  This is arbitrary.  All values in the EPS09
range are equally probable and consistent with the world's data.
Figure \ref{Fig:limits_lower} shows the excluded regions in two other
cases where the upper and lower values are taken as the gluon 
suppression factor.  In both of these instances, the viable region is 
reduced compared to the nominal EPS09 case.  The exclusion region is 
particularly large when the lower value is used.  Present PHENIX 
$\pi^0$ data shows a suppression at low-x indicating that lower values 
of EPS09 may be favored.  However, this suppression could also be 
explained by the low $p_{T}$ of the measured $\pi^0$s.  
The MPC-EX direct photon measurement will clarify the effects of gluon 
suppression at low-x.  It is important to note that EPS09 serves as the 
basis of comparison for this proposal.  These exclusion plots illustrate 
the sensitivity of the MPC-EX direct photon measurement in a the context 
of a variety of theoretical pictures.

\clearpage
  \label{sim:photon_AN}
\section[Direct Photon $A_{N}$]{Direct Photon $A_{N}$}
\label{sim:photon_AN}

In a prompt (direct + fragmentation) photon measurement with the MPC-EX detector, the signal of prompt photons ($S$) and 
the background of hadron-decay photons ($B$) are mixed, with a signal to background ratio of $r=S/B$.  
For the MPC-EX,  according to Monte Carlo simulations described in preceeding sections, with a cut of $p_T^\gamma > 3$ GeV/c
the value of $r$ typically is $\approx 0.4-0.5 $.  
In measurements of a prompt photon SSA in 200~GeV spin polarized p+p collisions, the background photon events might carry a non-zero SSA, such as from $\pi^0$ 
or $\eta$ decay photons. The signal asymmetry ($A_S$) can be extracted from the measured asymmetry ($A_{meas}$ ) 
and the independently measured background asymmetry ($A_B$) according to:

\begin{equation*}
A_S = (1+{\frac{1}{r}}) A_{meas} - \frac{1}{r} A_B.
\end{equation*}

\begin{equation*}
(\delta A_S)^2 = (1+{\frac{1}{r}})^2 (\delta A_{meas})^2  +  (\frac{1}{r})^2 (\delta A_B)^2.
\end{equation*}

Corresponding to a total luminosity of 49 pb$^{-1}$  and a cut of $p_T^\gamma > 3$ GeV/c,  from Monte Carlo simulations, 
0.75 million photon events will be observed. The Monte Carlo events are split into 4 $p_T$-bins  that  corresponding to 
central values of  ($p_T$,  $x_F$, number of events) as:  (3.2,  0.47,  400k),  (3.6, 0.54, 250k), (4.0,  0.61, 75k) and 
(4.4, 0.75, 25k). 

\begin{figure}[htbp]
\centerline{
\includegraphics[width=0.8\linewidth]{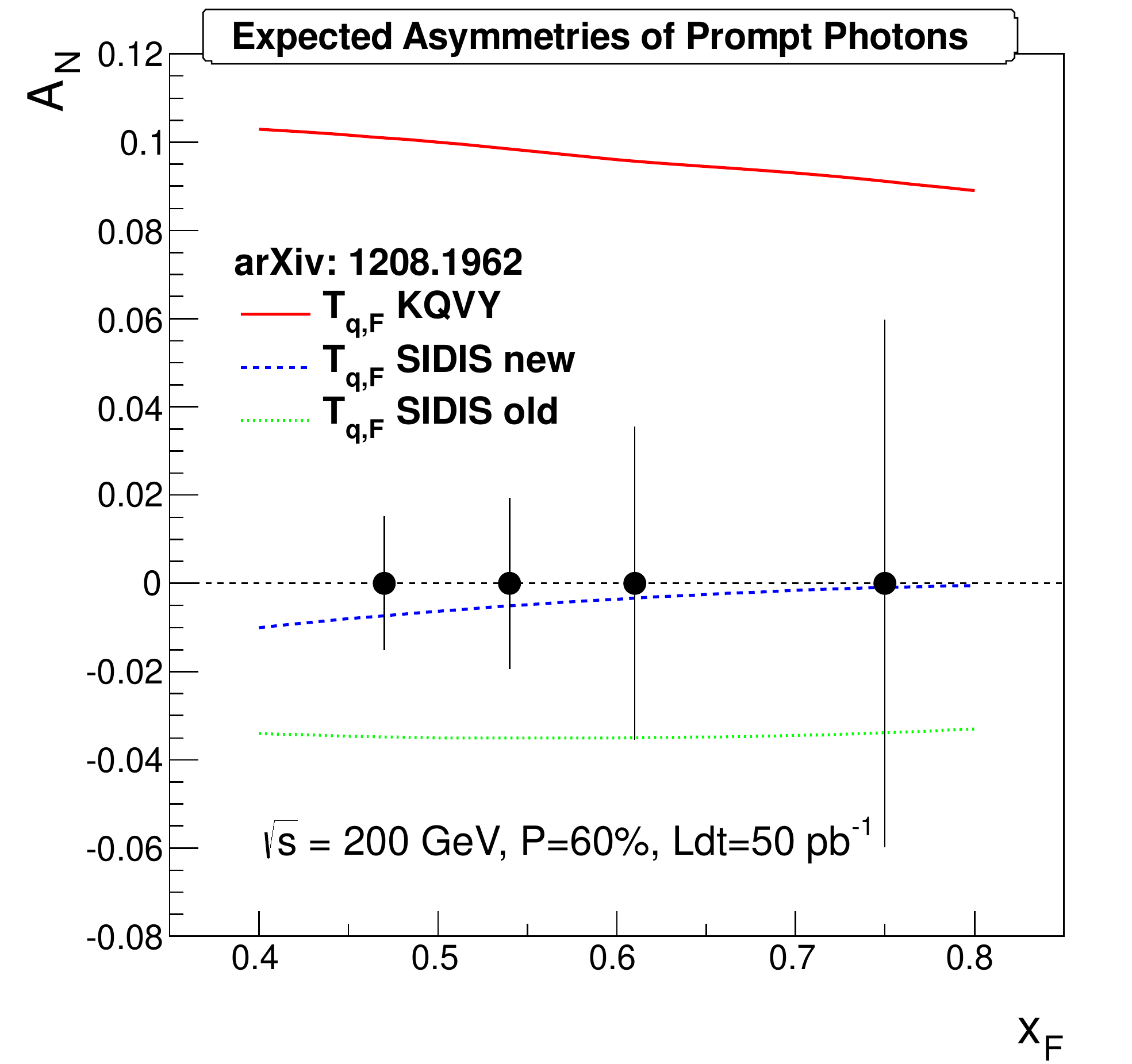}
}
\caption{Projected sensitivity for the prompt photon single spin asymmetry with the MPC-EX assuming an integrated luminosity of 
50 pb$^{-1}$ and 60\% beam polarization at $\sqrt{s}=200$ GeV. The sensitivities are shown compared to calculations in the 
collinear factorized approach ~\protect\cite{spin:KANG2011_1,Gamberg:2012iq} using a direct extraction of the 
quark-gluon correlation function from polarized $p+p$ data (upper solid curve), compared to the correlation
function derived from SIDIS extractions (lower dotted and dashed curves).
} 
\label{fig:photonAn2}
\end{figure}

We assume that MPC-EX can independently measure SSA of mesons, such as $\pi^0$ and $\eta$  to a precision at least ``twice 
as precise'' as  the photon SSA in each bin.  We also assume a proton beam polarization of $65\%$.  The estimated precision 
of  $A_N^\gamma$ is shown in Figure \ref{fig:photonAn2}, with theory predictions of prompt photon $A_N^{\gamma}$ of 
Kang {\it et al.}\cite{spin:KANG2011_1,Gamberg:2012iq}.

\clearpage
  \label{sim:collins}

\section[Determination of Jet Axis Proxy]{Determination of a Jet Axis Proxy via Charged Particles}
\label{sim:jet_axis}

\subsection{A Simple Toy Model}
\label{sim:jaxis_toy}

Because of the dual-gain readout capabilities of the MPC-EX the detector can be sensitive to both full-energy 
electromagnetic showers, as well as the passage of single minimum-ionizing particles. We propose to use this 
latter capability to reconstruct the axis of a jet in the MPC-EX acceptance. Such a measurement poses 
several challenges. 

First of all, there is very little magnetic bend between the event vertex and the MPC-EX, so 
we cannot determine the momentum of charged particles. In generating a proxy for the jet axis from a selection of charged tracks the
only possibility available is to equally-weight the charged tracks to reconstruct the jet axis, but we will 
have no information on the jet energy. Information on the jet energy could be obtained by combining 
a charged particle reconstriction with an energy measurement from the MPC-EX and MPC (or by just using the 
energy measurement alone), but this would 
bias the axis used to determine the asymmetry of $\pi^{0}$'s (see Section \ref{sim:collins}). This approach will 
limit the resolution of the jet axis direction that can be achieved. 

Second, the presence of an electromagnetic shower creates a ``dead zone'' in the MPC-EX where we will not be able to 
find isolated charged tracks as the energy from the shower will dominate the minipad response in this region. This
can potentially bias the determination of the jet axis proxy. In this section we demonstrate the method with a simple 
Monte Carlo model of a jet. In this model a ``jet'' consists of a high energy $\pi^{0}$, which defines the jet direction, and 
a set of four positive and negative muons that are distributed around the $\pi^{0}$ direction according to a Gaussian with 
$\sigma=0.4$ in pseudorapidity $\eta$ and azimuthal angle $\phi$. The same set of events is generated in two ways. In the first 
set the $\pi^{0}$ is ignored in the simulation. This first sample allows us to examine the ``jet'' axis reconstruction without 
the destructive effects of an electromagnetic shower. In the second sample, the $\pi^{0}$ is included in the event, which 
allows an examination of the effect of the $\pi^{0}$ shower on the reconstructed ``jet'' direction. Both samples of events 
are put through the full MPC-EX simulation and reconstruction. 

\begin{figure}
\centering
\includegraphics[angle=90, width=0.85\linewidth]{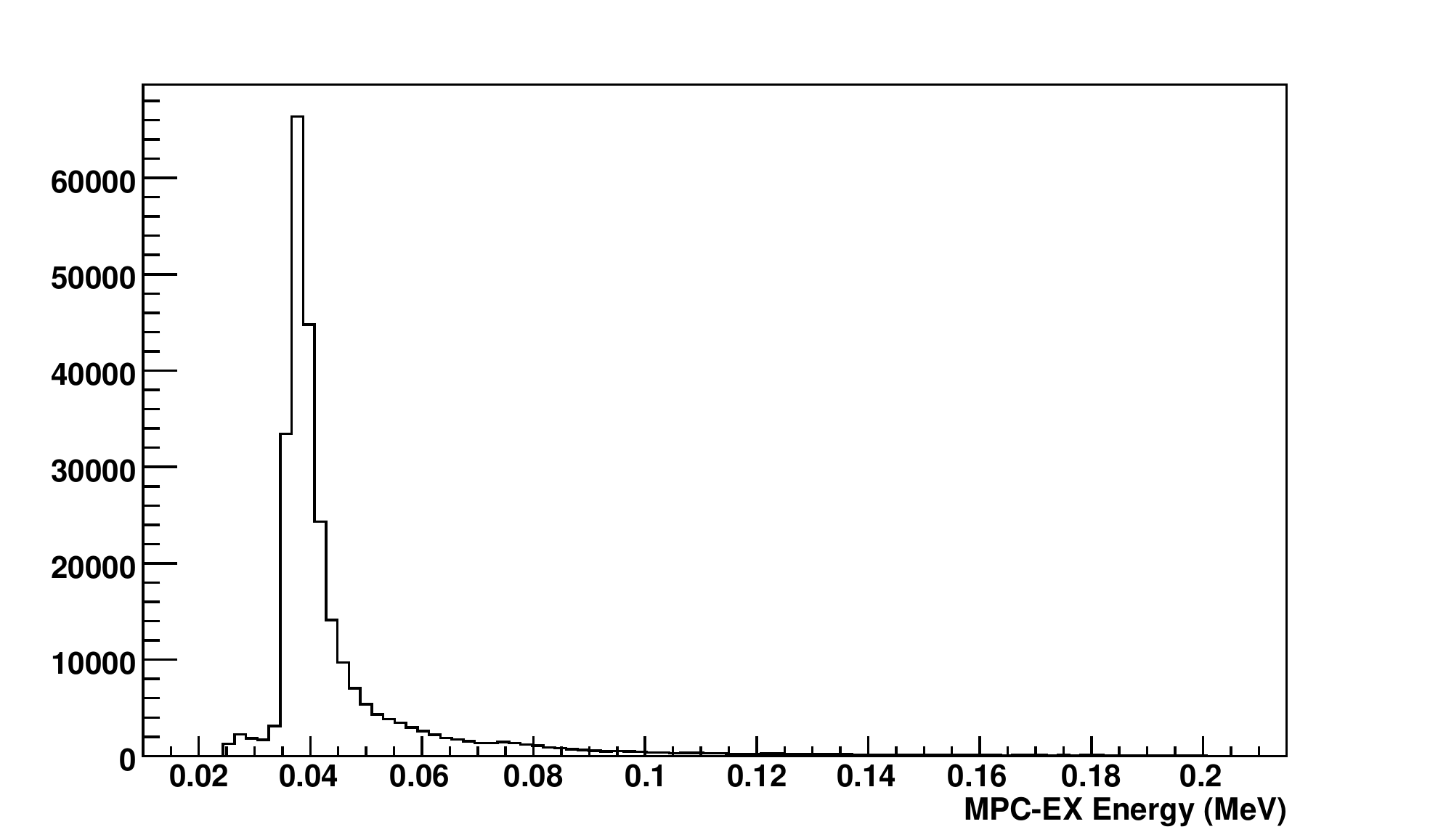}
\caption{\label{fig:MIPS_jets} Energy distribution of MPC-EX tracks from single charged particles.  
The MIP peak at 36~MeV is clearly visible. In the analysis a cut is placed at $<$70~MeV to select charged particle tracks.}
\end{figure}

Charged particles in the MPC-EX are reconstructed via a loose set of cuts. A charged particle track is identified  
by demanding that the MPC-EX track have a hit in all eight layers, and that the energy within a narrow window around the 
tracks (+/-10 strips centered on the track) is less than 70~MeV (see Figure ~\ref{fig:MIPS_jets}). 
No check is made on the association of the MPC-EX track to a cluster in the 
MPC. A more restrictive cut has been explored that required that the RMS of the distribution of hits in each later be less than one 
strip, but the addition of this cut reduces the track finding efficiency and does not significantly reduce backgrounds. 

As we have no momentum measurement for charged tracks, all charged tracks are arbitrarily assigned an equal momentum of 1~GeV. 
The jet cluster algorithm is a seeded cone algorithm that uses every particle in the track list as seed for a cluster cone.
The cluster cone is taken as a fixed radius in $\eta$ and $\phi$ space of 0.7 units. 
For each selection of a seed track, the cone algorithm and cluster axis are iterated until further iteration produces 
no change in the cluster axis. This cluster is recorded and the next seed is analyzed. Finally, from the list of all found clusters 
the cluster that contains the largest number of tracks is returned as a jet proxy for each MPC-EX arm.

\begin{figure}[hbt]
\hspace*{-0.12in}
\begin{minipage}[b]{0.5\linewidth}
\centering
\includegraphics[angle=90, width=0.95\linewidth]{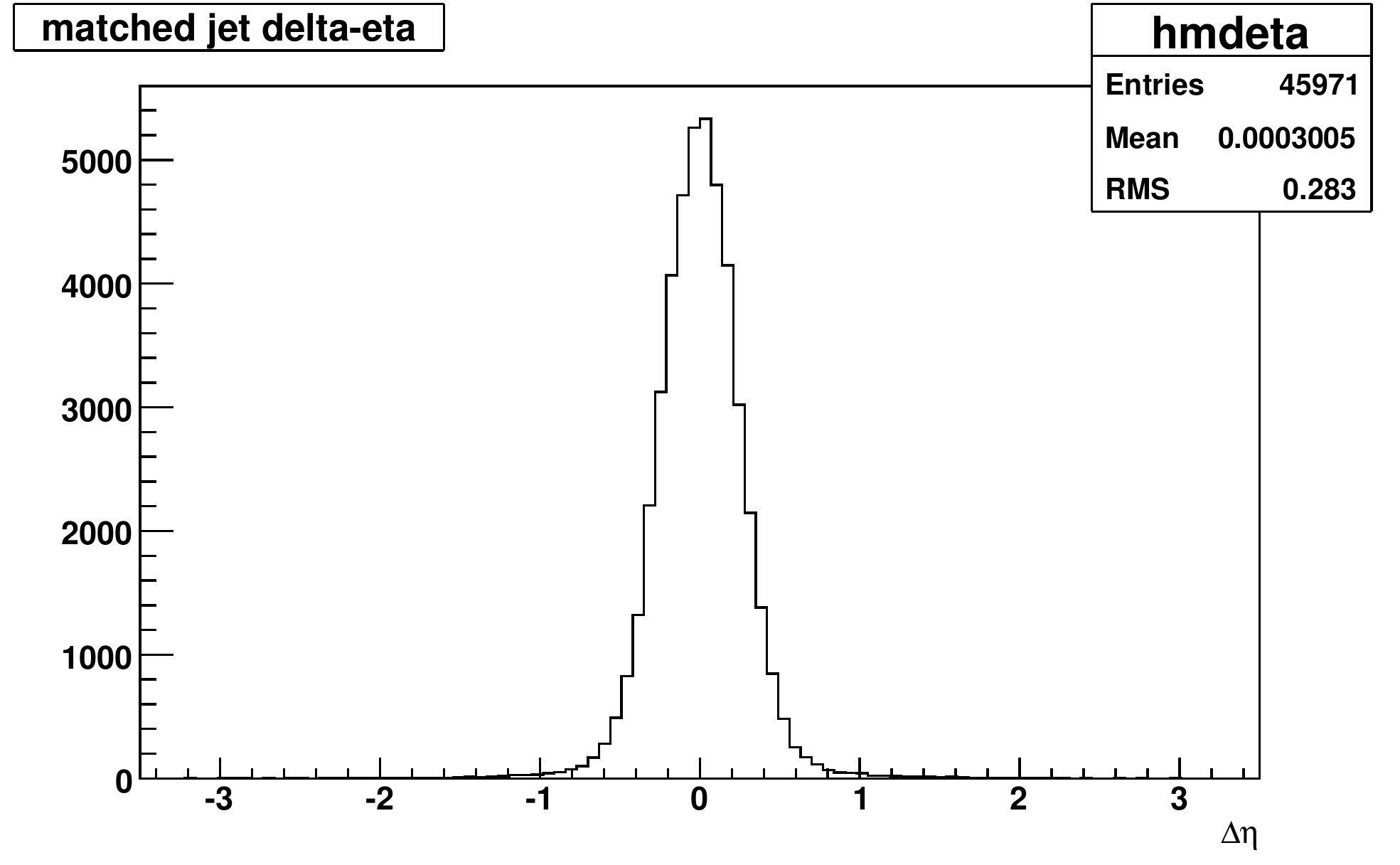}
\end{minipage}
\hspace{0.5cm}
\begin{minipage}[b]{0.5\linewidth}
\centering
\includegraphics[angle=90, width=0.95\linewidth]{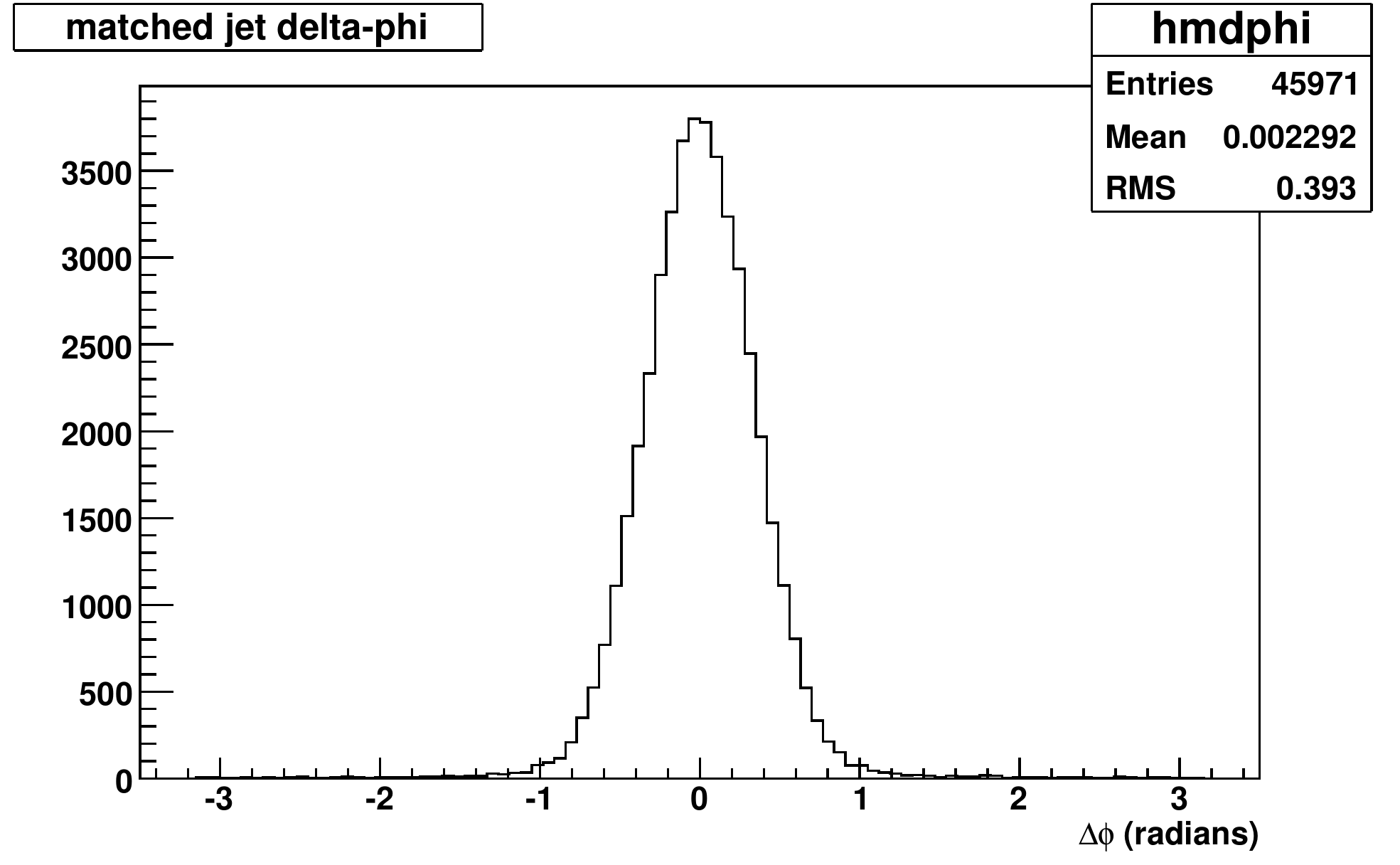}
\end{minipage}
\begin{minipage}[b]{0.5\linewidth}
\centering
\includegraphics[angle=90, width=0.95\linewidth]{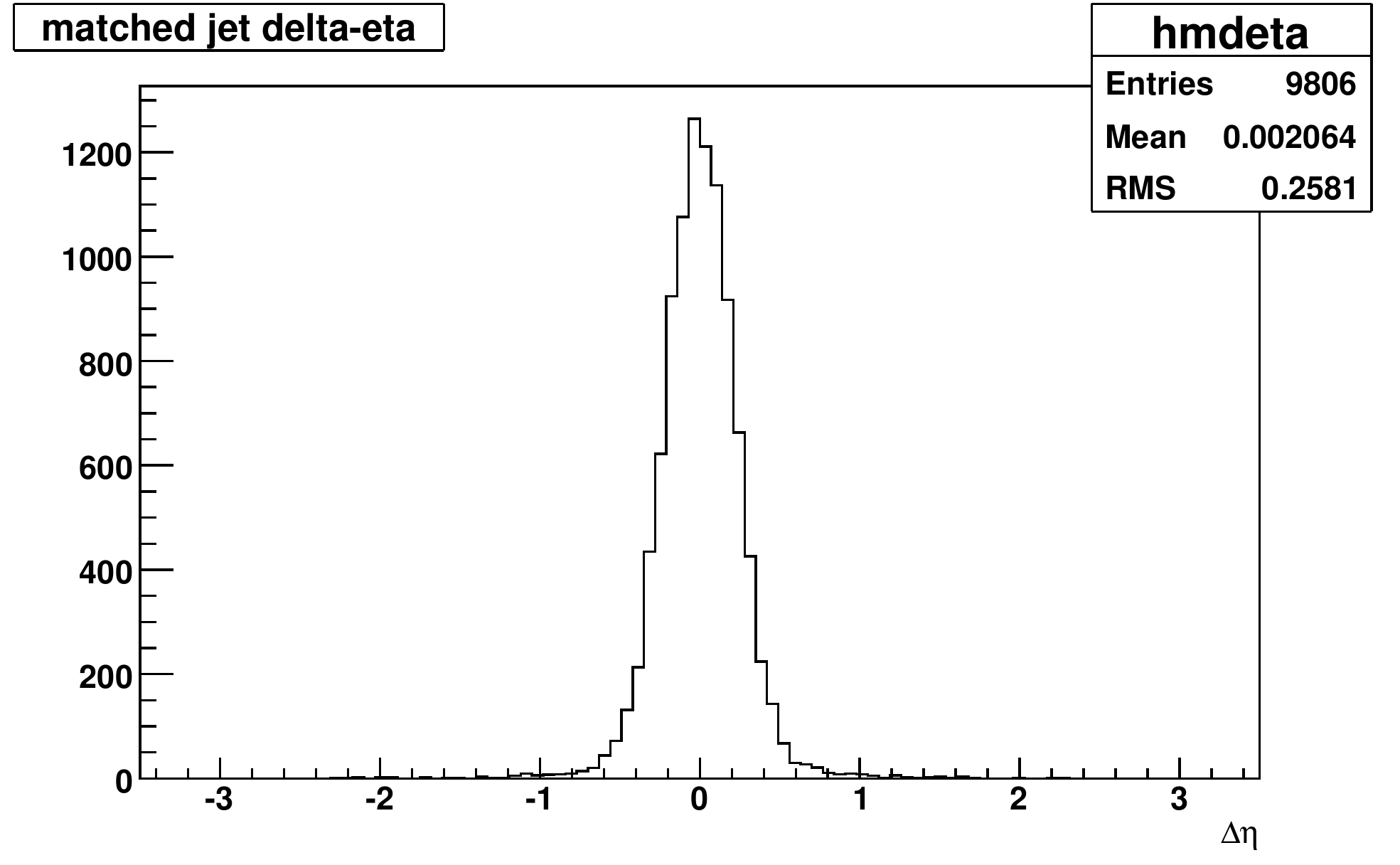}
\end{minipage}
\hspace{0.5cm}
\begin{minipage}[b]{0.5\linewidth}
\centering
\includegraphics[angle=90, width=0.95\linewidth]{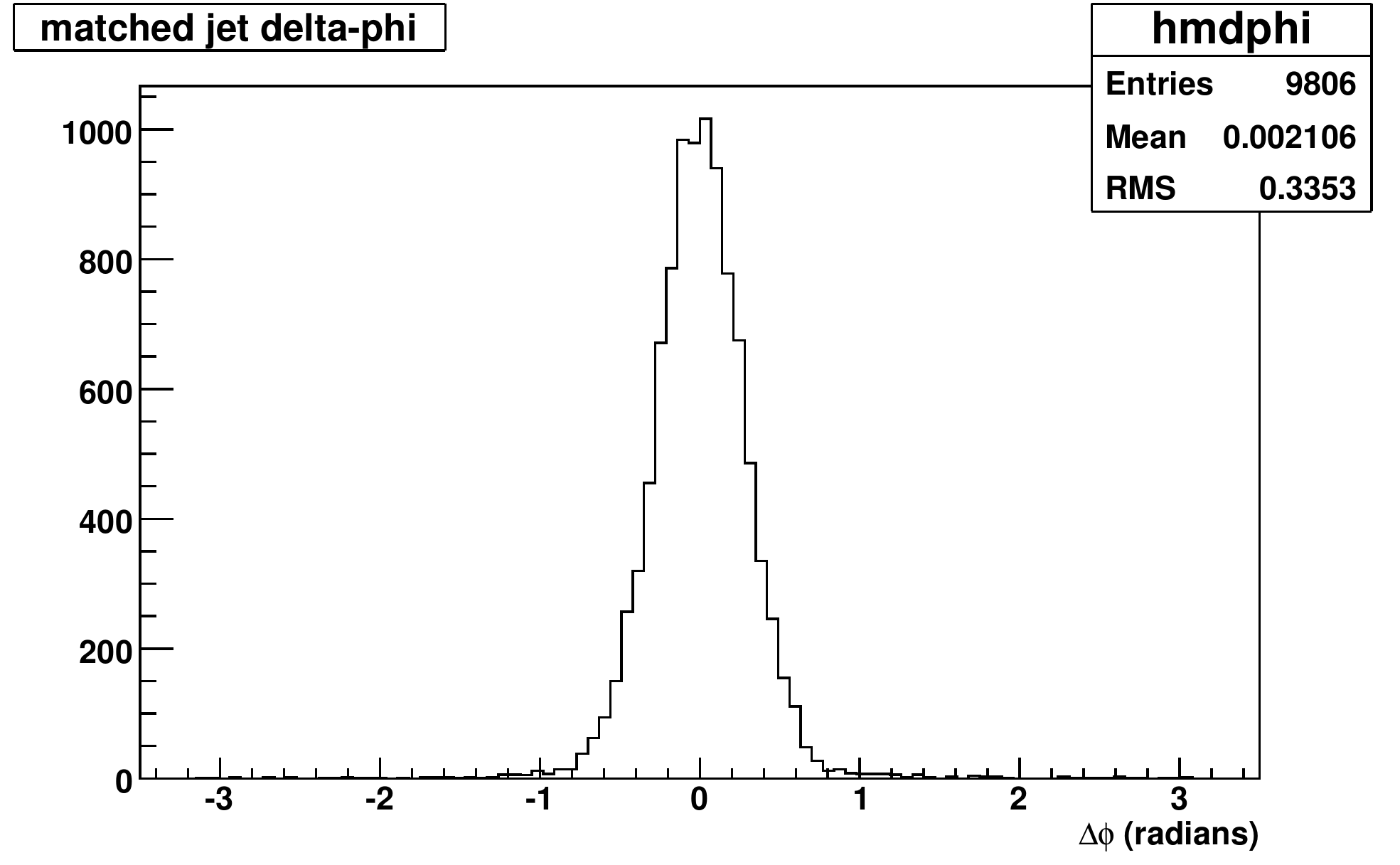}
\end{minipage}
\begin{minipage}[b]{0.5\linewidth}
\centering
\includegraphics[angle=90, width=0.95\linewidth]{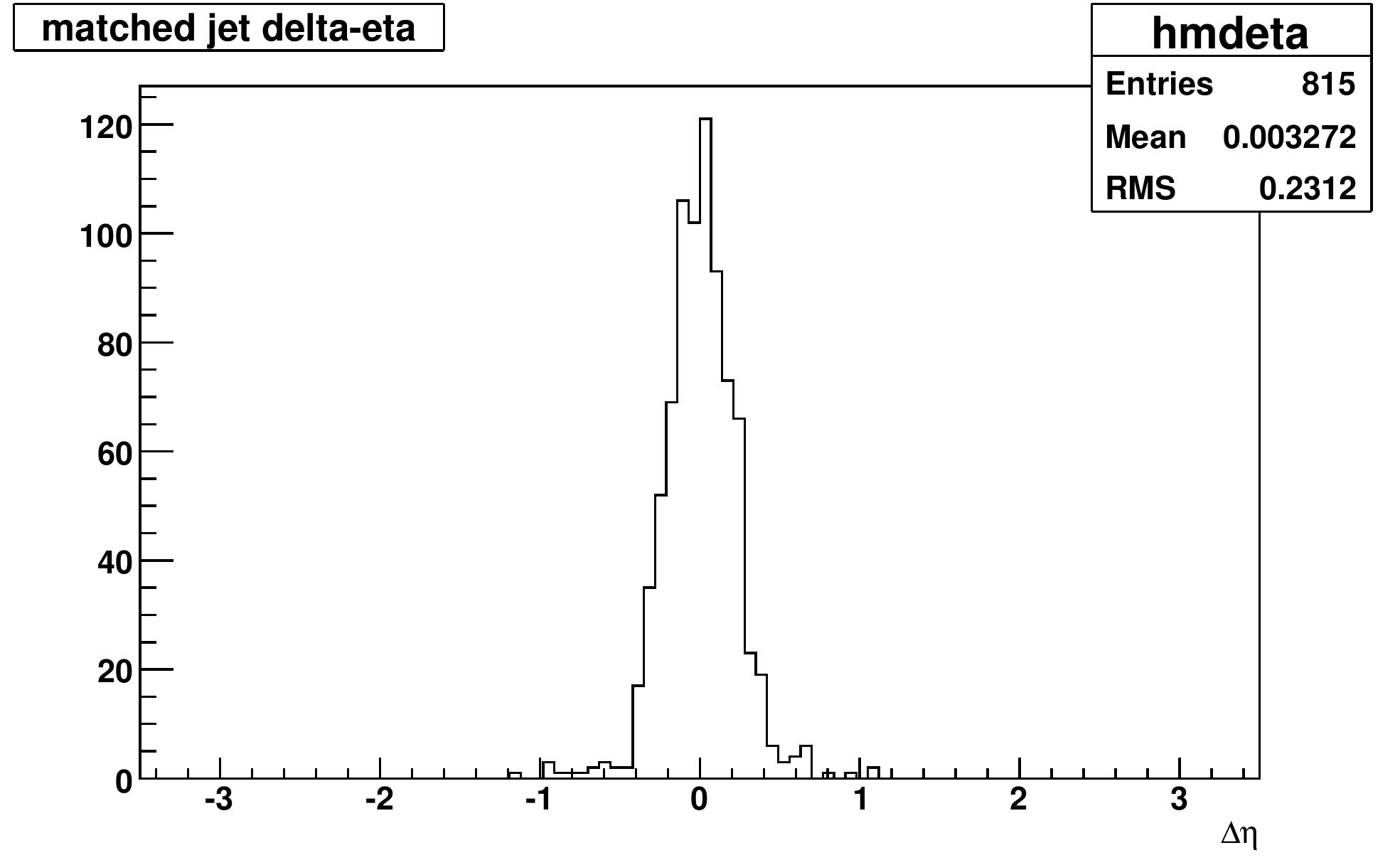}
\end{minipage}
\hspace{0.5cm}
\begin{minipage}[b]{0.5\linewidth}
\centering
\includegraphics[angle=90, width=0.95\linewidth]{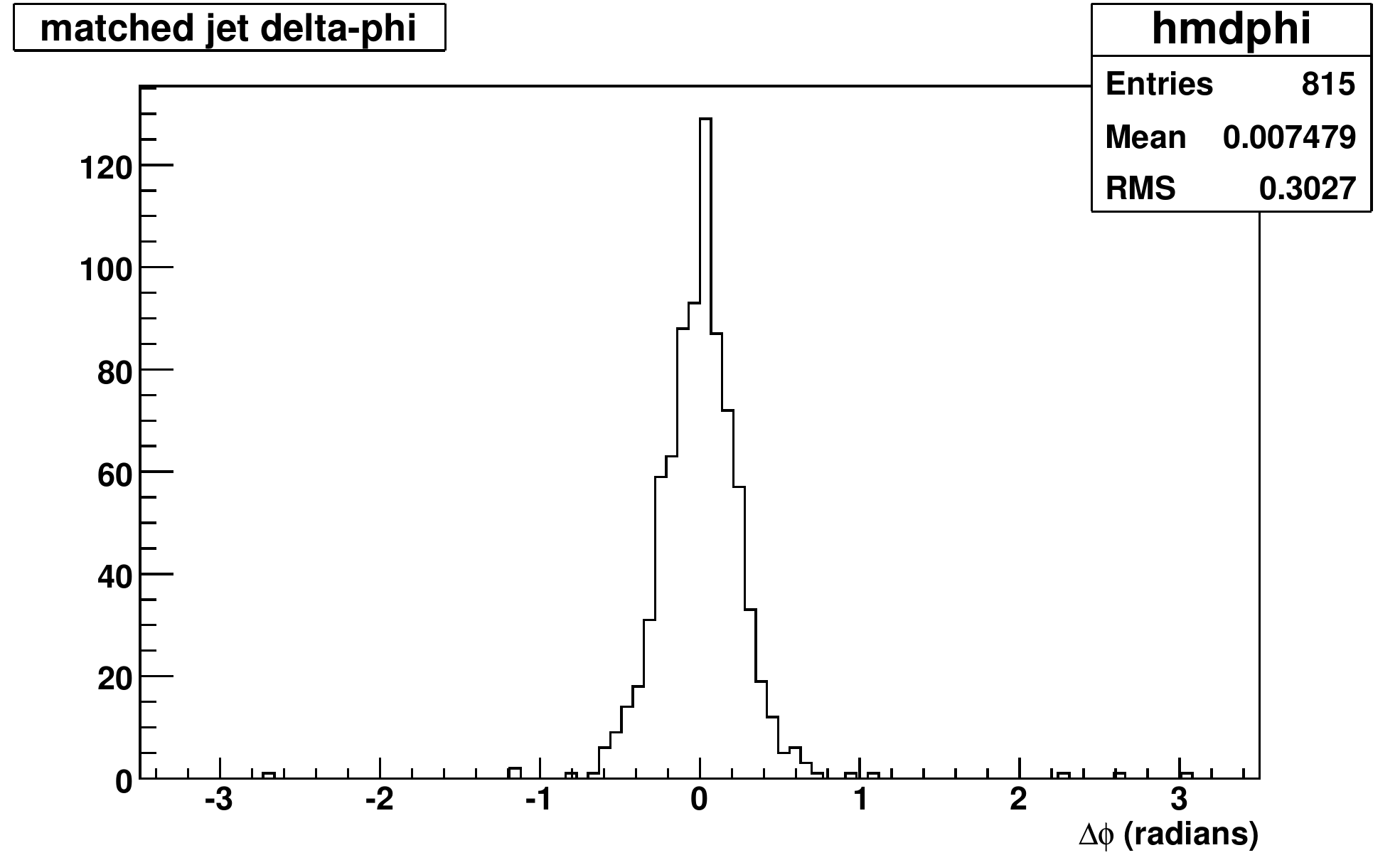}
\end{minipage}
\vspace*{-0.12in}
\caption{\label{fig:jets_no_pi0} Resolution in $\eta$ and $\phi$ for a ``jet'' axis determined from charged particles reconstructed
in the MPC-EX, for events with muons correlated with the ``jet'' axis without a $\pi^{0}$. The top row is for ``jet'' clusters determined
with two or more charged particles, the middle with three or more, and the bottom with four or more charged tracks.}
\end{figure}

Returning to our simple jet model consisting of four muons, 
in Figure~\ref{fig:jets_no_pi0} we show the resolution in $\eta$ and $\phi$ for charged track clusters with two, three and four charged particles for
``jet'' events with only charged tracks (the $\pi^{0}$ along the jet direction is not included in the simulation). The resolution
improves slowly with the number of charged tracks used to determine the ``jet'' direction, as expected.

\begin{figure}[hbt]
\hspace*{-0.12in}
\begin{minipage}[b]{0.5\linewidth}
\centering
\includegraphics[angle=90, width=0.95\linewidth]{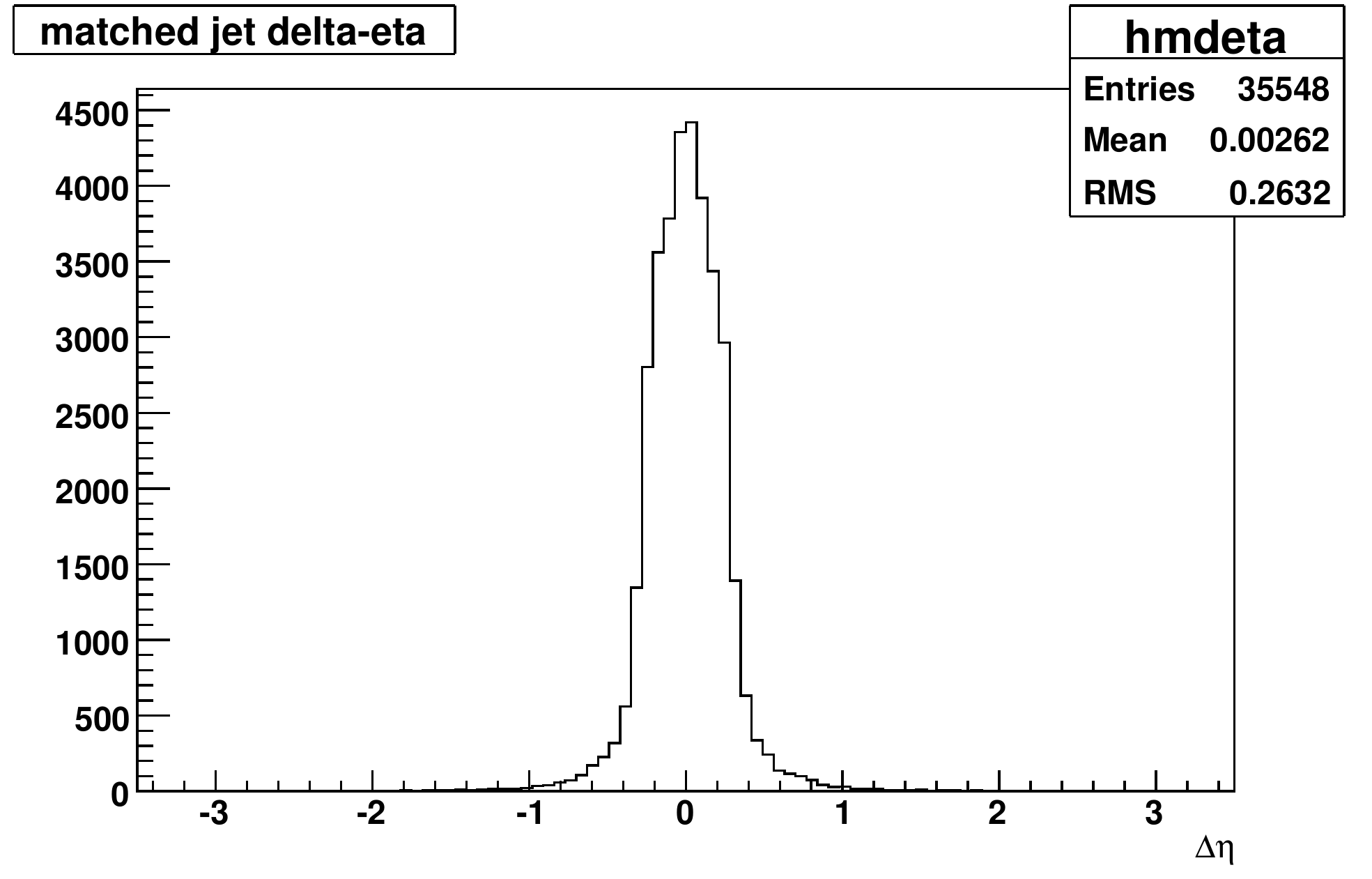}
\end{minipage}
\hspace{0.5cm}
\begin{minipage}[b]{0.5\linewidth}
\centering
\includegraphics[angle=90, width=0.95\linewidth]{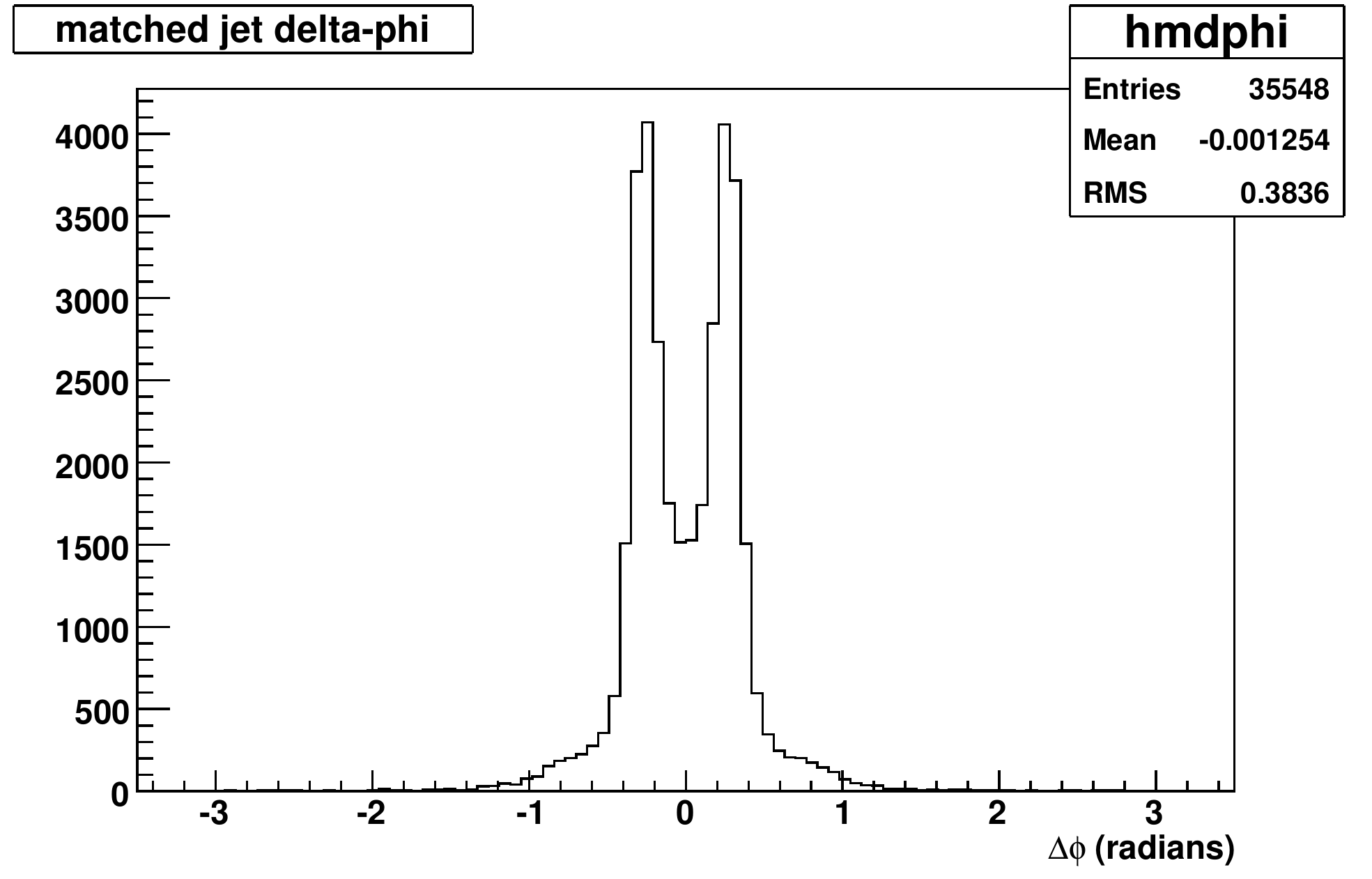}
\end{minipage}
\begin{minipage}[b]{0.5\linewidth}
\centering
\includegraphics[angle=90, width=0.95\linewidth]{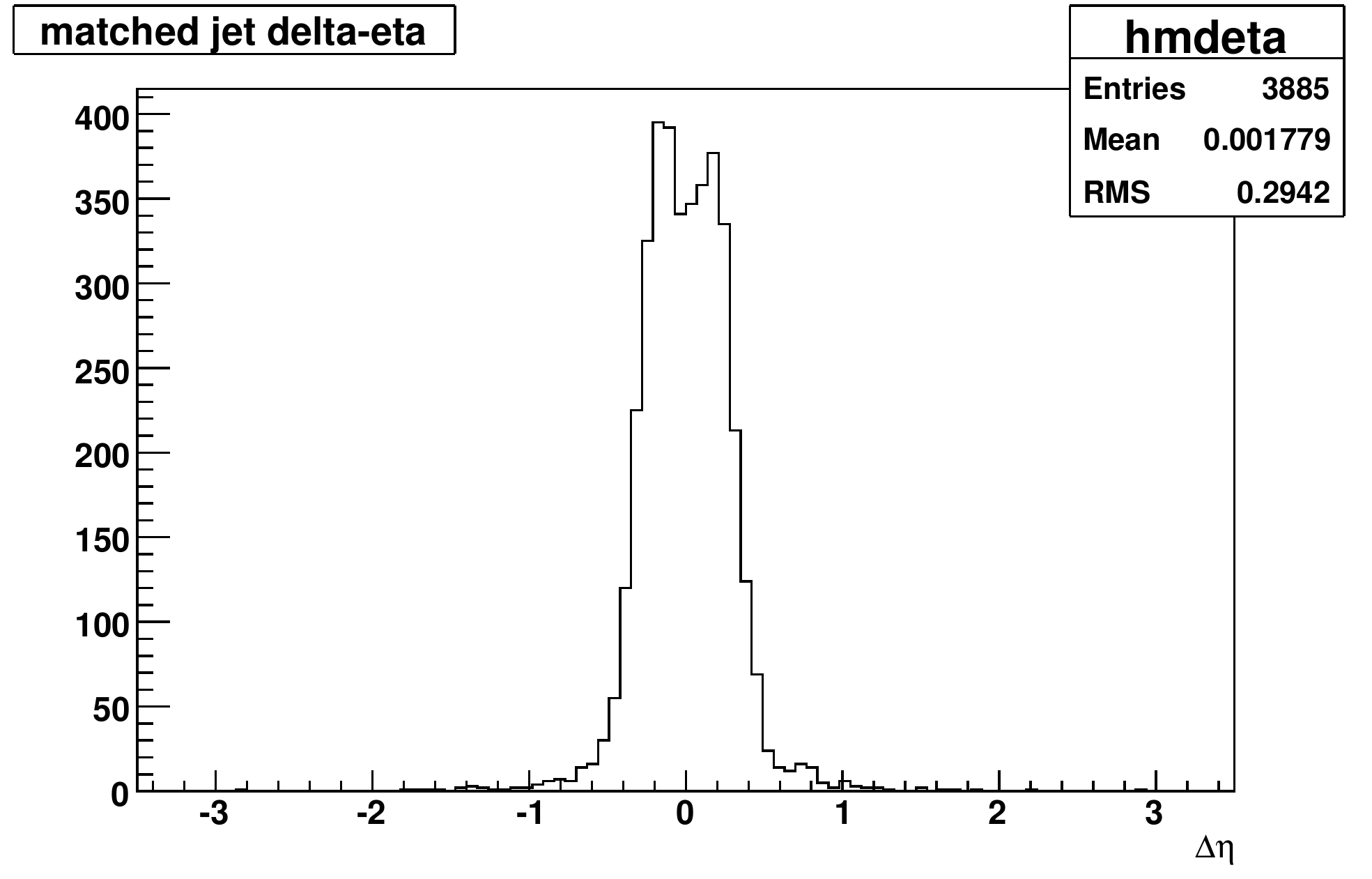}
\end{minipage}
\hspace{0.5cm}
\begin{minipage}[b]{0.5\linewidth}
\centering
\includegraphics[angle=90, width=0.95\linewidth]{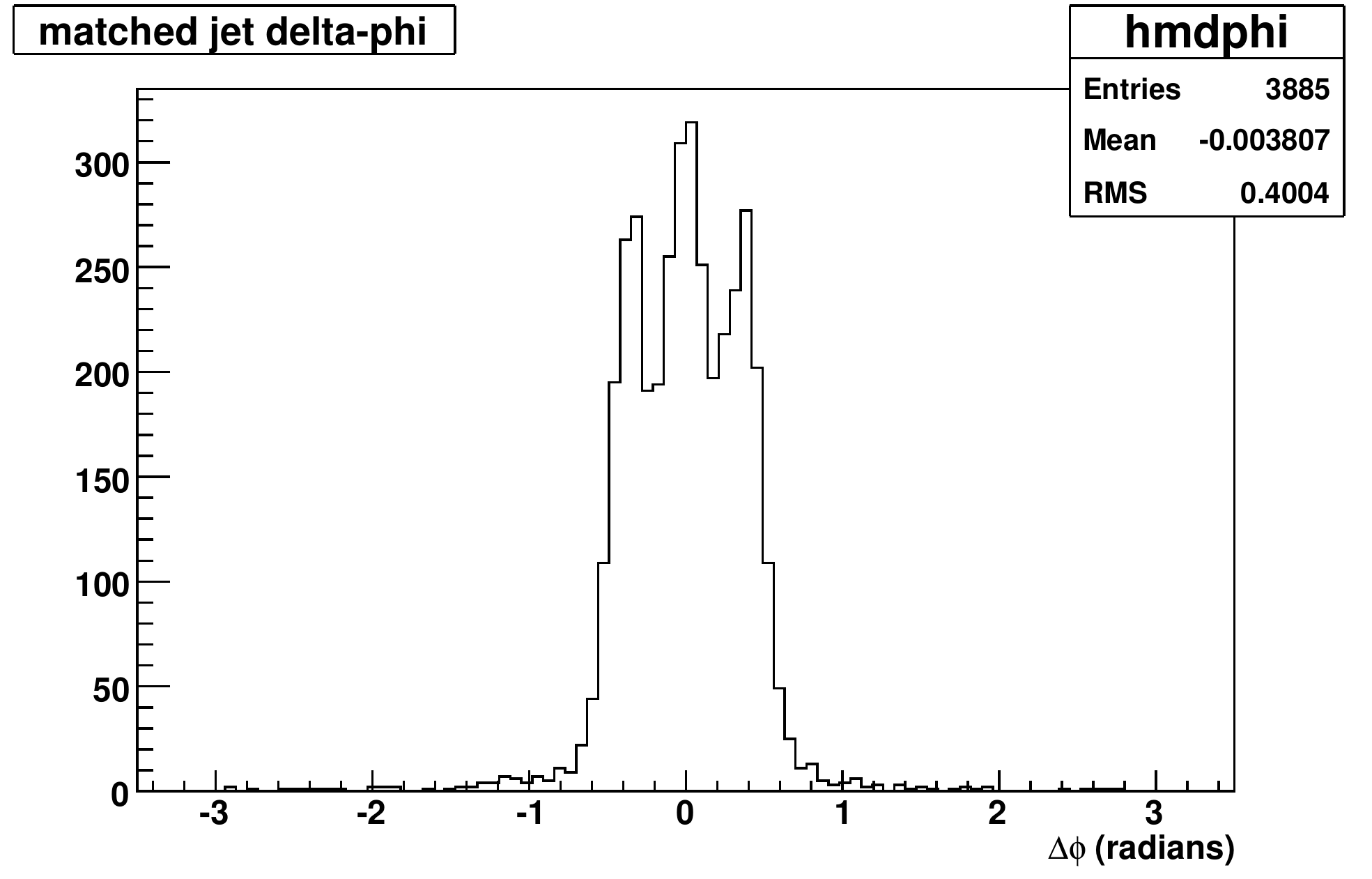}
\end{minipage}
\begin{minipage}[b]{0.5\linewidth}
\centering
\includegraphics[angle=90, width=0.95\linewidth]{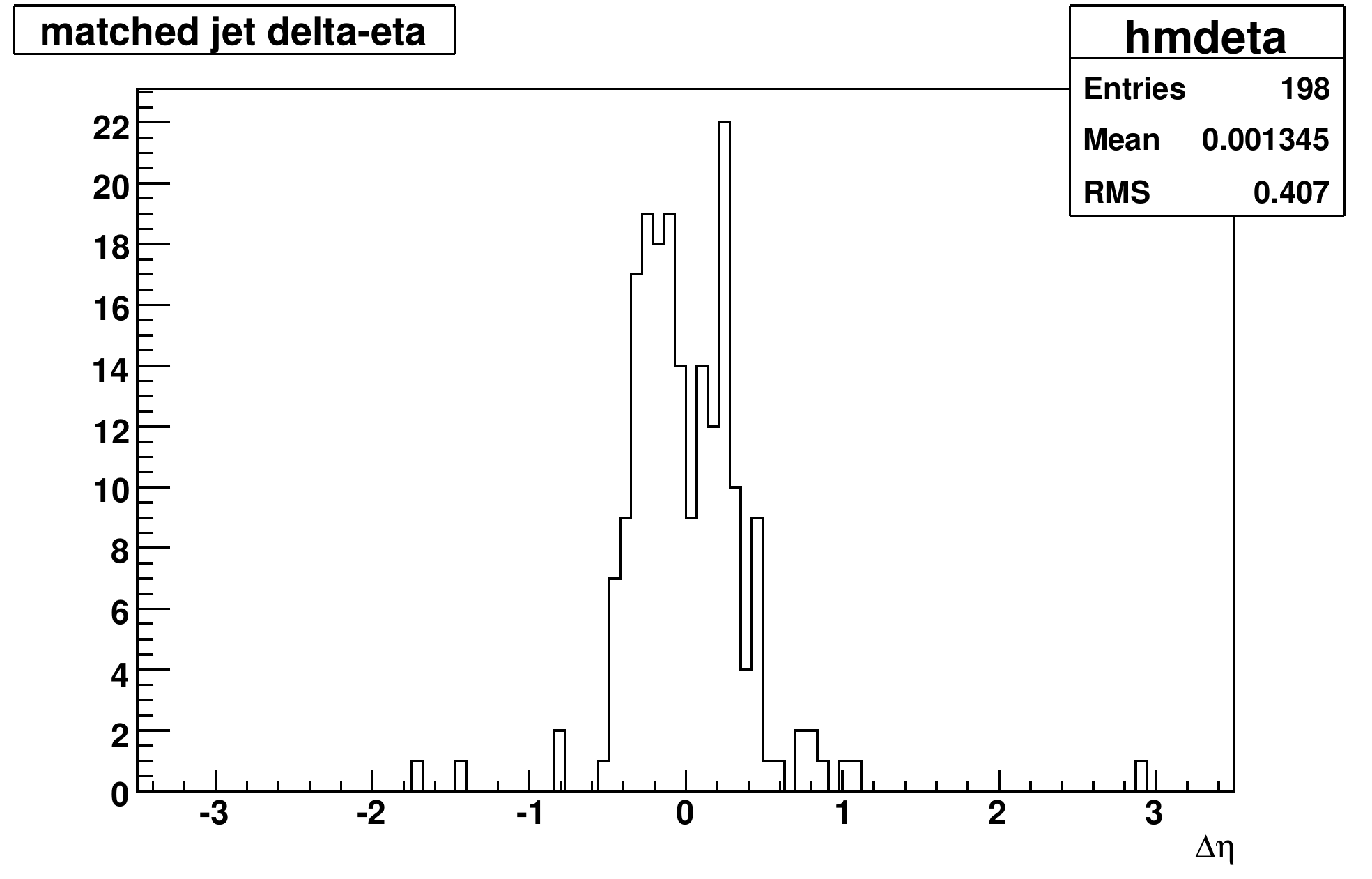}
\end{minipage}
\hspace{0.5cm}
\begin{minipage}[b]{0.5\linewidth}
\centering
\includegraphics[angle=90, width=0.95\linewidth]{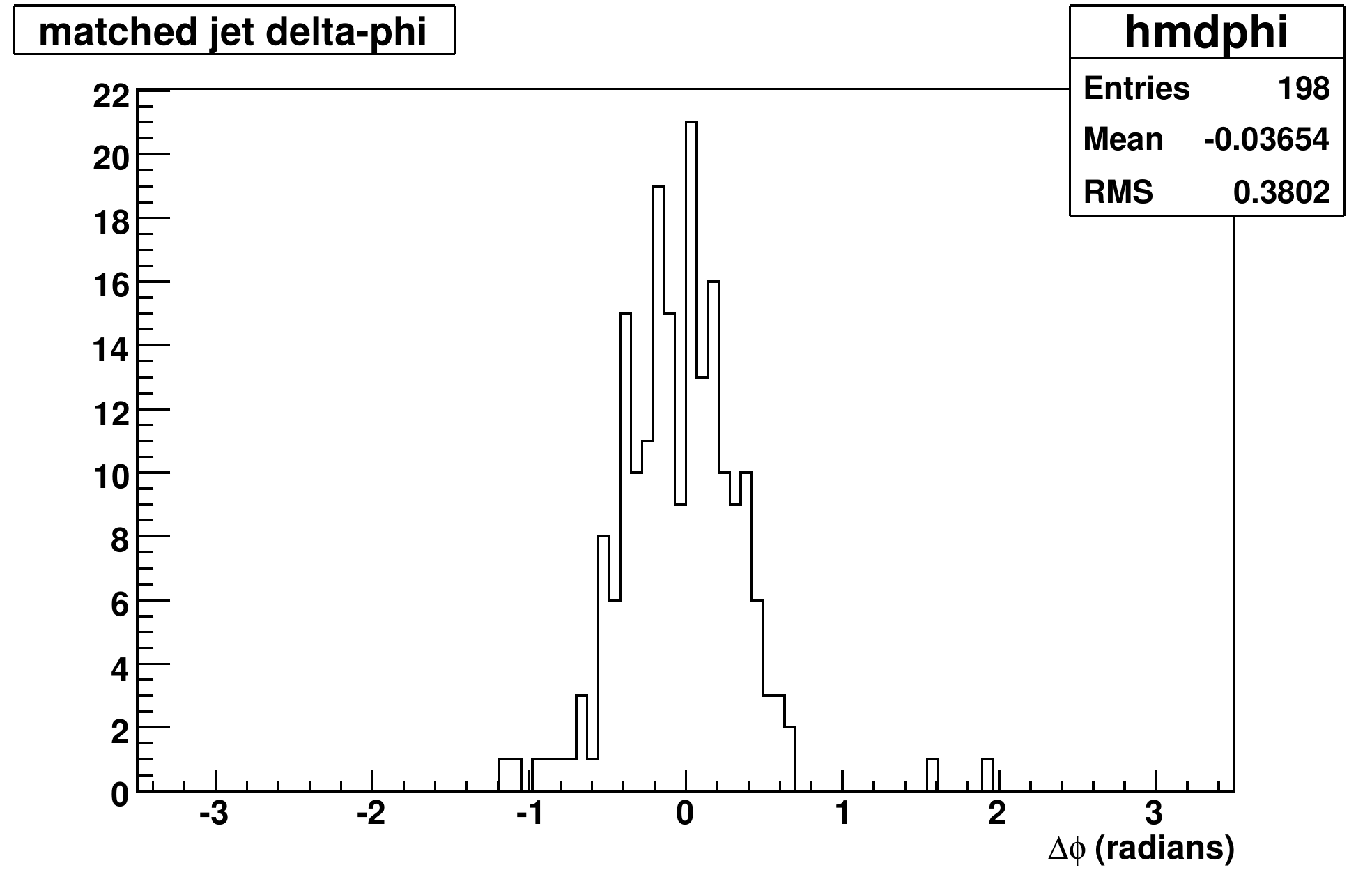}
\end{minipage}
\vspace*{-0.12in}
\caption{\label{fig:jets_with_pi0} Resolution in $\eta$ and $\phi$ for a ``jet'' proxy determined from charged particles reconstructed
in the MPC-EX, for events with muons correlated with the ``jet'' axis with a $\pi^{0}$ coincident with the jet axis. 
The top row is for ``jet'' clusters determined
with two or more charged particles, the middle with three or more, and the bottom with four or more charged tracks.} 
\end{figure}

In Figure~\ref{fig:jets_with_pi0} we show the resolution in $\eta$ and $\phi$ for charged track clusters with two, three and four charged particles for
``jet'' events with charged tracks and a $\pi^{0}$ along the jet direction. The charged tracks found under these circumstances are
clearly biased by the presence of the electomagnetic shower. In the case of only two tracks a clear splitting in the distribution of $\Delta\phi$ can 
be seen depending in whether both tracks are found on one side or another in $\phi$. The distribution in $\eta$ has also been affected, and is no longer
purely Gaussian. Because the cone algorithm uses a fixed radius in $\eta$-$\phi$ space the actual distribution at forward rapidities on the face of the 
MPC-EX is a distorted ellipse, physically wider in the $\phi$ dimension. This is the reason that the splitting appears largest in $\Delta\phi$.  
The resolution improves slowly with the number of charged tracks used to determine the ``jet'' direction, and the splitting observed in phi is diminished. 
Nevertheless, the resolution at a fixed number of charged particles in these events is worse by about a factor of two. 

Of course, this splitting effect is exaggerated in our toy model by placing the $\pi^{0}$ coincident with the jet axis. In real events there will be an 
event-by-event bias based on the number of charged tracks and their relative location with respect to any electromagnetic energy in the jet. If there is a 
spin dependence to this bias, the bias in the determination of the jet proxy axis will only exacerbate this effect. 

\subsection{Jet Axis Resolution in {\sc Pythia} Events}
\label{sim:pythia_jaxis_res}

In order to understand the resolution on the jet axis we can expect in real events, we simulated 
a set of {\sc Pythia} $2\rightarrow 2$ hard scattering events in the MPC-EX detector. For these events
the ``true'' jet axis was determined using the Fastjet \cite{Cacciari:2011ma} anti-$k_T$ jet-finding algorithm 
with a radius of 0.7 on the full {\sc Pythia} event (no detector acceptance). The ``true'' jets found in this manner were 
required to have at leats 10 constituent particles and a $p_T>2$~GeV. 

These events were then simulated in the full MPC-EX Monte Carlo, and the charged tracking and clustering 
algorithm described in Section~\ref{sim:jaxis_toy} was applied. In the sections that follow we will be interested 
in correlating the jet proxy axis with a high momentum $\pi^{0}$ that is within $|\Delta\phi|<\frac{\pi}{2}$ in azimuthal 
angle of the jet proxy axis, so this requirement is also placed on the {\sc Pythia} events. The jet proxy clusters
were required to have at least two charged particles. (This is effectively a three particle requirement when the $\pi^0$ is included, 
but the $\pi^0$ is not used in the determination of the jet proxy direction.)

\begin{figure}[hbt]
\hspace*{-0.12in}
\begin{minipage}[b]{0.5\linewidth}
\centering
\includegraphics[angle=90, width=0.95\linewidth]{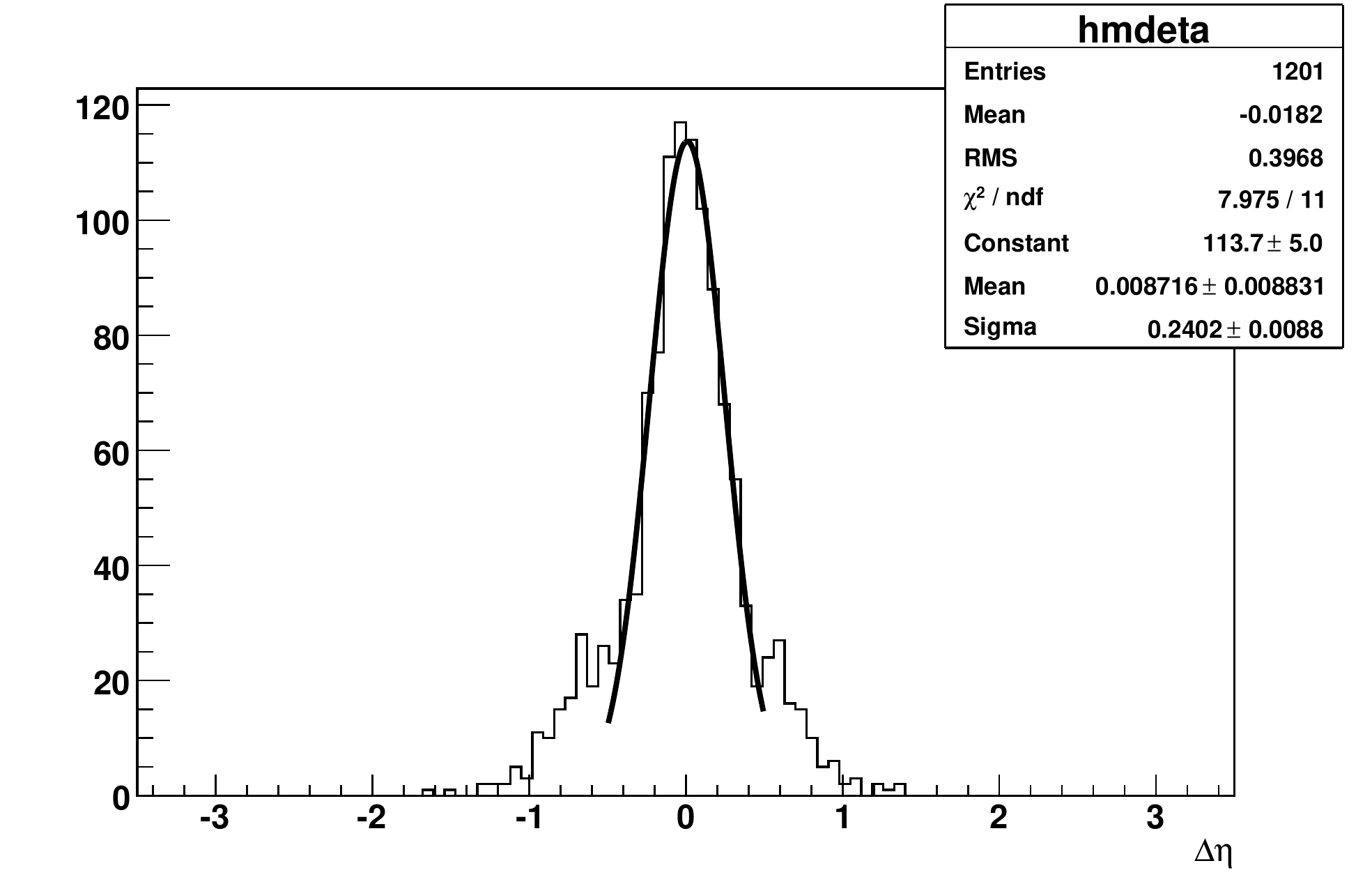}
\end{minipage}
\hspace{0.5cm}
\begin{minipage}[b]{0.5\linewidth}
\centering
\includegraphics[angle=90, width=0.95\linewidth]{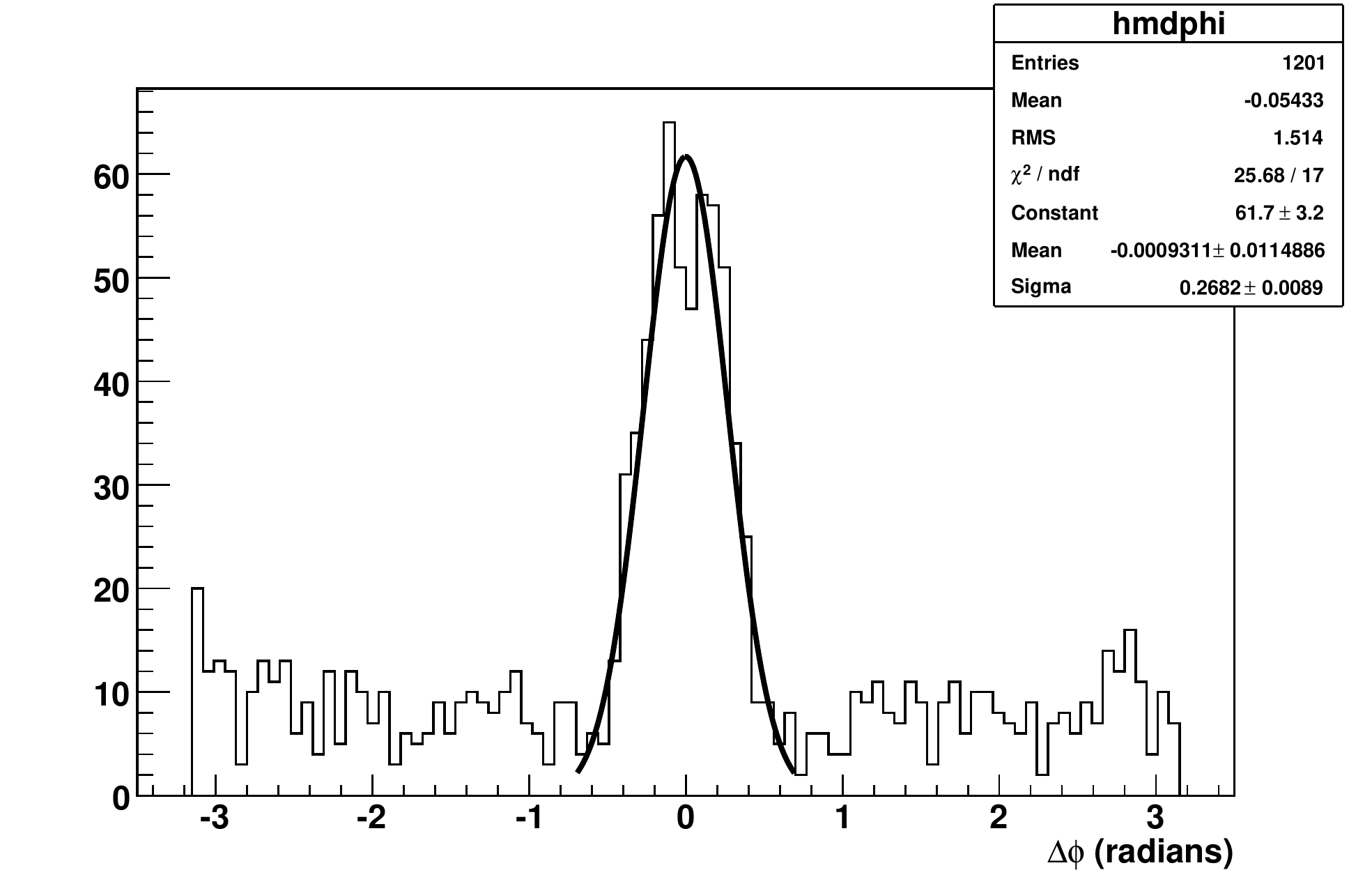}
\end{minipage}
\vspace*{-0.12in}
\caption{\label{fig:jaxis_pythia_resolution} Resolution in $\eta$ (left) and $\phi$ (right) for a jet proxy determined 
from charged particles reconstructed in the MPC-EX for {\sc Pythia} hard-scattering events. The resolution can be approximately 
described by a sigma of $\sim$0.3 units in both $\eta$ and $\phi$.} 
\end{figure}

We then compare the jet proxy axis determined as described above with the ``true'' {\sc Pythia} jet as determined 
by Fastjet. The jet proxy is associated with the closest ``true'' jet in $\Delta\eta$ - $\Delta\phi$ space, and the difference between 
the jet proxy and the ``true'' jet in $\Delta\eta$ and $\Delta\phi$ is determined. These distributions are shown in 
Figure~\ref{fig:jaxis_pythia_resolution}. 

\section[$\pi^0$ Correlations in Jets]{$\pi^0$ Correlations in Jets}
\label{sim:collins}

In this section we describe simulations to estimate the sensitivity of the MPC-EX detector
to an asymmetry of neutral pions around a fragmenting quark in transversely polarized \pp ~collisions. 
Our strategy is as follows. First, we describe a Monte Carlo model that was developed to include the
effects of finite transversity and Collins fragmentation in the final state distribution of 
hadrons from a fragmenting quark. Using this model we generate a sample of events with 
roughly the same single-particle asymmetry $A_N$ as that observed in neutral clusters in the 
PHENIX MPC, where the asymmetry is generated from the effects of transversity and Collins fragmentation. 
These events are 
put through a realistic simulation of the MPC-EX + MPC detectors, and are reconstructed 
and analyzed as physics data. The results of this exercise allows us to demonstrate 
the level of asymmetry due to transversity and Collins fragmentation 
that could be observed and make projections for a full event sample in 
real data. 

\subsection{The toyMC Monte Carlo Model}

Simulations of the Collins asymmetry in jets in 200~GeV \pp~collisions requires a model that can produce
a sample of particles from parton fragmentation with the asymmetry built-in to the kinematic distributions. 
At the start of the simulations for the MPC-EX no such model existed that implemented the best information 
available from SIDIS experiments in an event generator format. The toyMC model \cite{toymc} was developed to address 
the need for just such an event generator. 

The toyMC model starts with the event generator {\sc Pythia} (version 6.421)~\cite{Sjostrand:2006za}, configured only to 
perform the initial partonic event. All fragmentation of the partons is disabled. In addition, because we seek a leading-order 
model consistent with the SIDIS extractions we will use to implement transversity and the Collins fragmentation functions, 
all QED and QCD radiation from the parton legs is disabled. {\sc Pythia} is run with the standard QCD 2x2 hard scattering processes 
enabled (process ID's 11,12,13,28,53,68 and 96) and we use TuneA {\sc Pythia} parameter set, as this has been shown to better produce
the pion cross sections (albiet at lower rapidities). The cross section for all sampled processes is 22.3mb. Configured in this manner 
the toyMC model allows us to benchmark the performance of the MPC-EX against current SIDIS analysis and data. 

\begin{figure}
\centering
\includegraphics[width=0.55\linewidth]{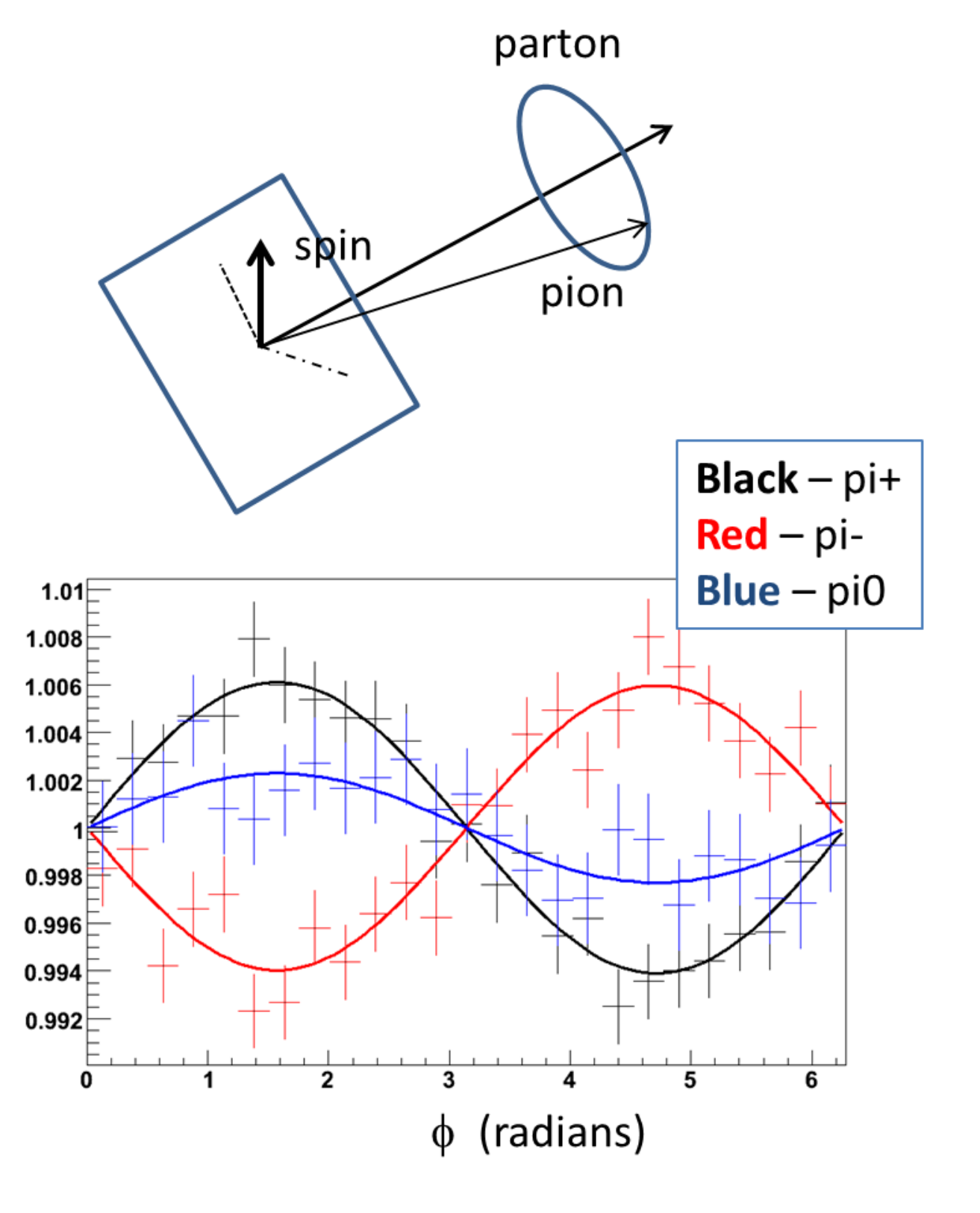}
\caption{\label{fig:Collins_pions} Sample toyMC distributions for pions, summed over all fragmenting quarks. The combination of transversity 
and the Collins FF's yields an asymmetry patern where the $\pi^+$ and $\pi^-$ asymmetries are equal and opposite, and the 
$\pi^0$ asymmetry is the same sign as the $\pi^+$ asymmtery, although somewhat reduced. Note that the actual value of the asymmetries depends
on the kinematic cuts and model parameters.}
\end{figure}

The spin of the incident protons are assigned randomly, and the
spin of the scattered partons is set from the transversity distribution as parametrized in ~\cite{Anselmino:2007fs} and 
~\cite{Anselmino:2008jk} at the scale of the hard interaction. Pions from quarks are generated according to the Collins fragmentations 
functions extracted from SIDIS and Belle data ~\cite{Anselmino:2008jk} and parametrized based on the DSS fragmentation functions (FF's), 
while pions from gluon fragmentation and all other particles are fragmented according to the spin-independent DSS FF's.  
The choice of favored/unfavored Collins FF's 
is made based on the fragmenting quark valence quark content of the hadron for charged pions, while it is assumed the the $\pi^0$ 
always fragments to pions according to the favored FF. The fragmentation functions are taken at the scale $\mu=p_T$ of the 
fragmenting parton. An example of the distribution of the Collins asymmetry as implemented in the toyMC model is shown in 
Figure~\ref{fig:Collins_pions}. 

The toyMC model also implements the Sivers distributions as a set of event weights which can also be used to generate
asymmetries for the final state particles. Because these are not relevant to the MPC-EX analysis they will not be discussed 
further here. 

\begin{figure}
\centering
\includegraphics[width=0.8\linewidth]{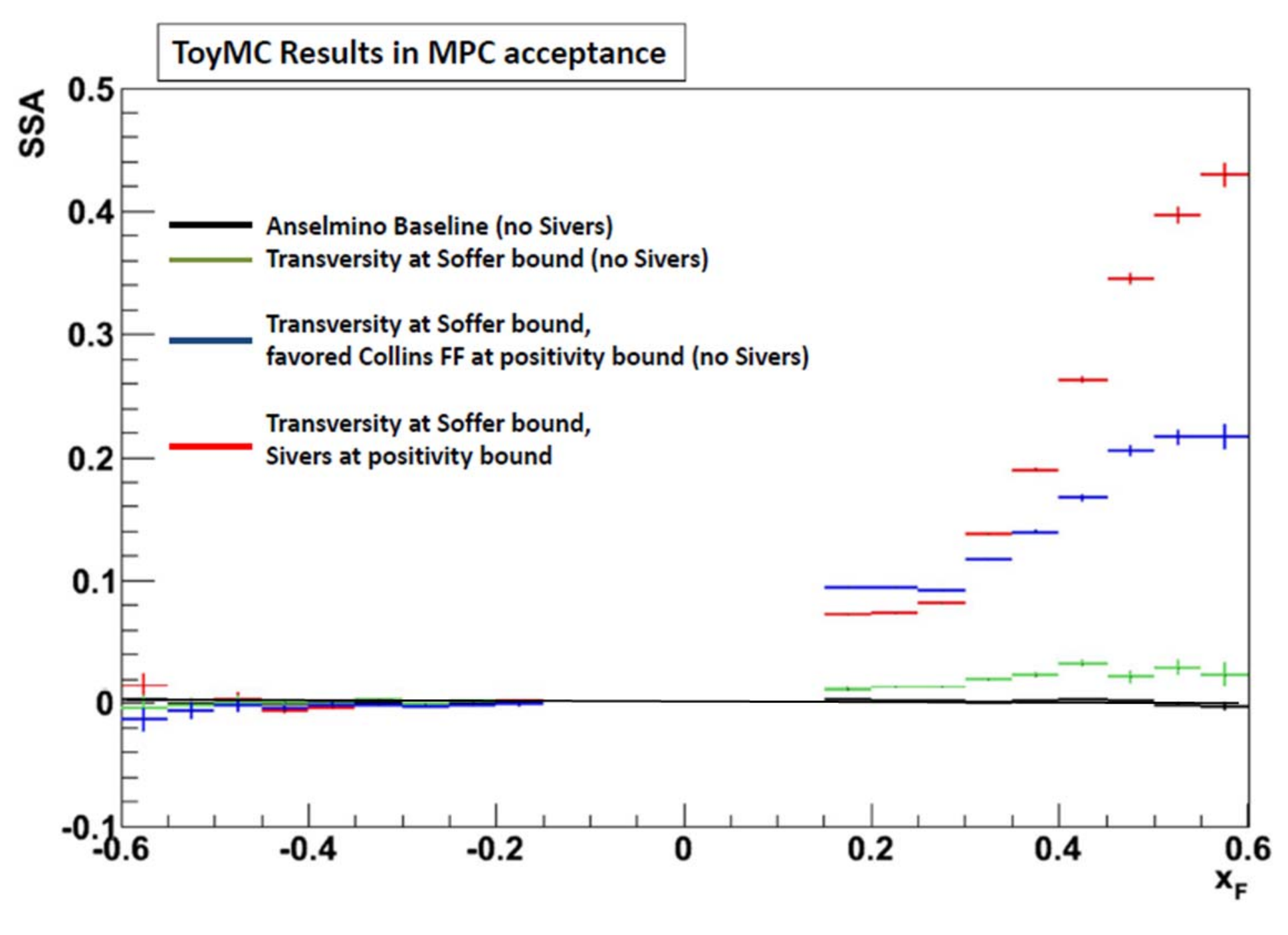}
\caption{\label{fig:toyMC_AN} $A_N$ single spin asymmetries (single particle) for $\pi^0$ in the acceptance of the PHENIX MPC under various tunings of the toyMC model.}
\end{figure}

\begin{figure}
\centering
\includegraphics[width=0.8\linewidth]{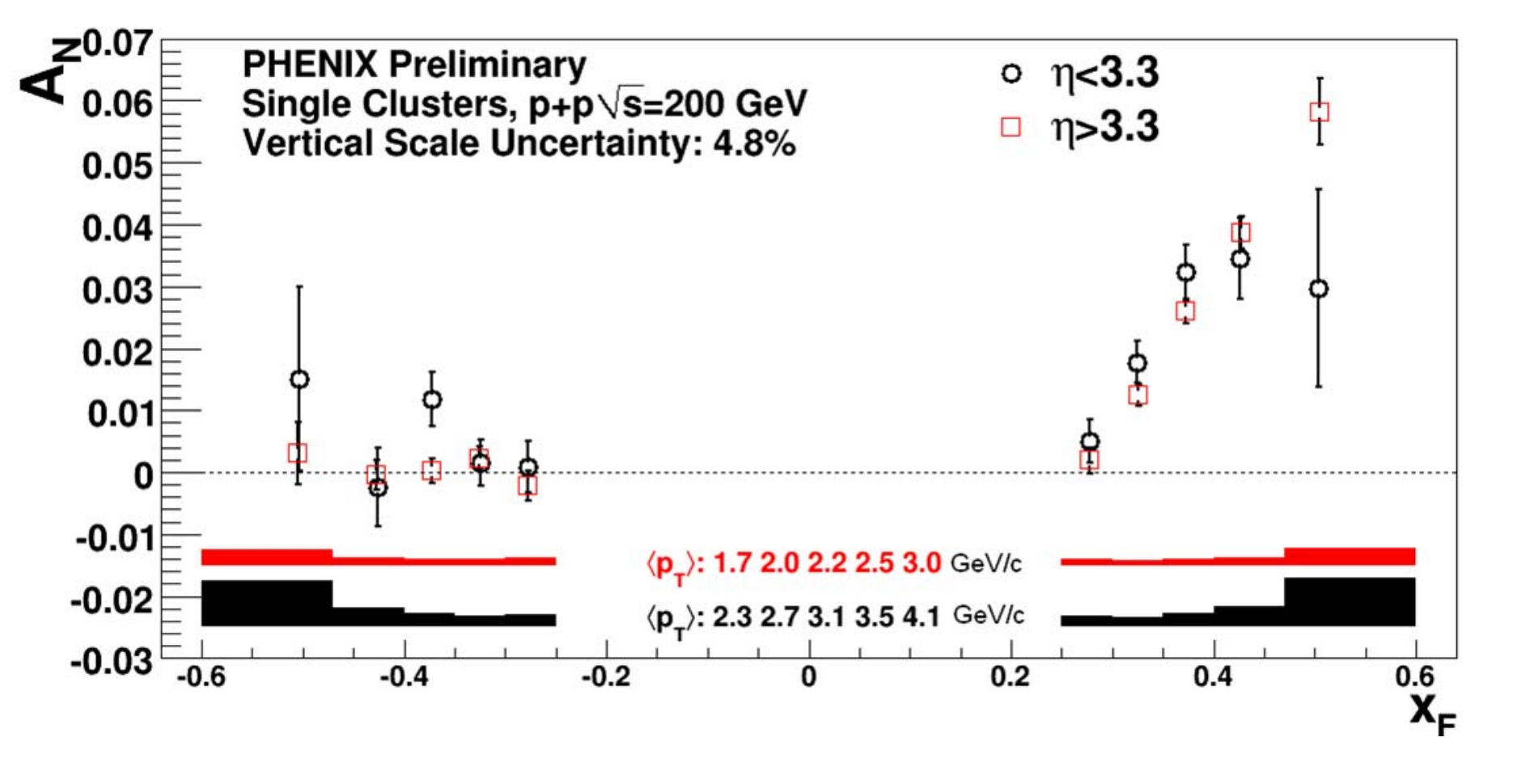}
\caption{\label{fig:MPC_cluster_AN} PHENIX preliminary results for the $A_N$ single spin asymmetries (single particle) for neutral clusters in the acceptance of the MPC.}
\end{figure}

In order to benchmark the model and tune the asymmtries it generated for MPC-EX studies, we compare the single-particle $A_N$ for 
$\pi^0$ mesons in the MPC under various conditions in Figure~\ref{fig:toyMC_AN}. It should be noted that using the standard parametrizations
for transversity and the Collins FF's yields a vanishingly small asymmtery at large $x_F$ in 200~GeV \pp~ collisions. This may in part 
be due to the fact that the transversity distribution above $x_{F}\sim0.3$ is an extrapolation of the functional fit form and is not constrained
by SIDIS data. Pushing transversity to the Soffer bound yields toyMC asymmetries that are similar to those observed in PHENIX 
(see Figure ~\ref{fig:MPC_cluster_AN}). 

\subsection{Event Generation and Statistics}

Two large samples of toyMC events were generated to simulate the extraction of the Collins asymmetry in the MPC-EX. The first sample of 4.9M 
events was generated with transversity set to the Soffer bound and the Collins FF's set to the positivity limit. 
This set of events was used to 
examine the asymmetry for small systematic effects, such as an angular correlation between the extracted jet axis and the axis of the 
fragmenting parton. A second, larger set of 19.5M events was generated with transversity at the Soffer bound in order to match the single 
particle $A_N$ observed in PHENIX. This sample is the main sample used to test the sensitivity of the MPC-EX analysis to the Collins asymmetry.
In both samples an event was written out from the Monte Carlo only if there was a $\pi^0$ and at least one charged particle in the acceptance
of the MPC-EX. The small event sample corresponds to a sample luminosity of 0.10~$pb^{-1}$, while the large sample corresponds to 0.41~$pb^{-1}$.

In both cases the event vertex along the beamline was chosen according to a Gaussian distribution matched to the real distrbution of events 
in 200~GeV \pp~collisions. In the final analysis only events with a vertex between $\pm70$ centimeters were used. 

The output of the toyMC event generator was then put through the GEANT3-based PHENIX Integrated Simulation Application (PISA) in order to 
simulate the response of the MPC-EX, MPC and BBC detectors to the event, and the PHENIX reconstruction and analysis framework was used 
to turn the simulated hits into raw data quantities in a simulated DST. These DSTs were then analyzed to produce physics quantities.

\subsection{$\pi^{0}$ Reconstruction}

In jet events, reconstruction of $\pi^{0}$ mesons is done by single-track reconstruction for tracks with a total energy $>20$~GeV, and by two-track combinations
for tracks with energies $<20$~GeV. The reconstruction method is described in section ~\ref{sim:photonrecomethod}. Electromagnetic tracks are only required to 
have an associated MPC cluster and be flagged as the ``closest'' track in hough space. In addition, an electromagnetic track must not pass the charged
particle cuts described above. This eliminates charged tracks that pass through the MPC-EX and shower in the MPC. 

\begin{figure}
\centering
\includegraphics[width=0.9\linewidth]{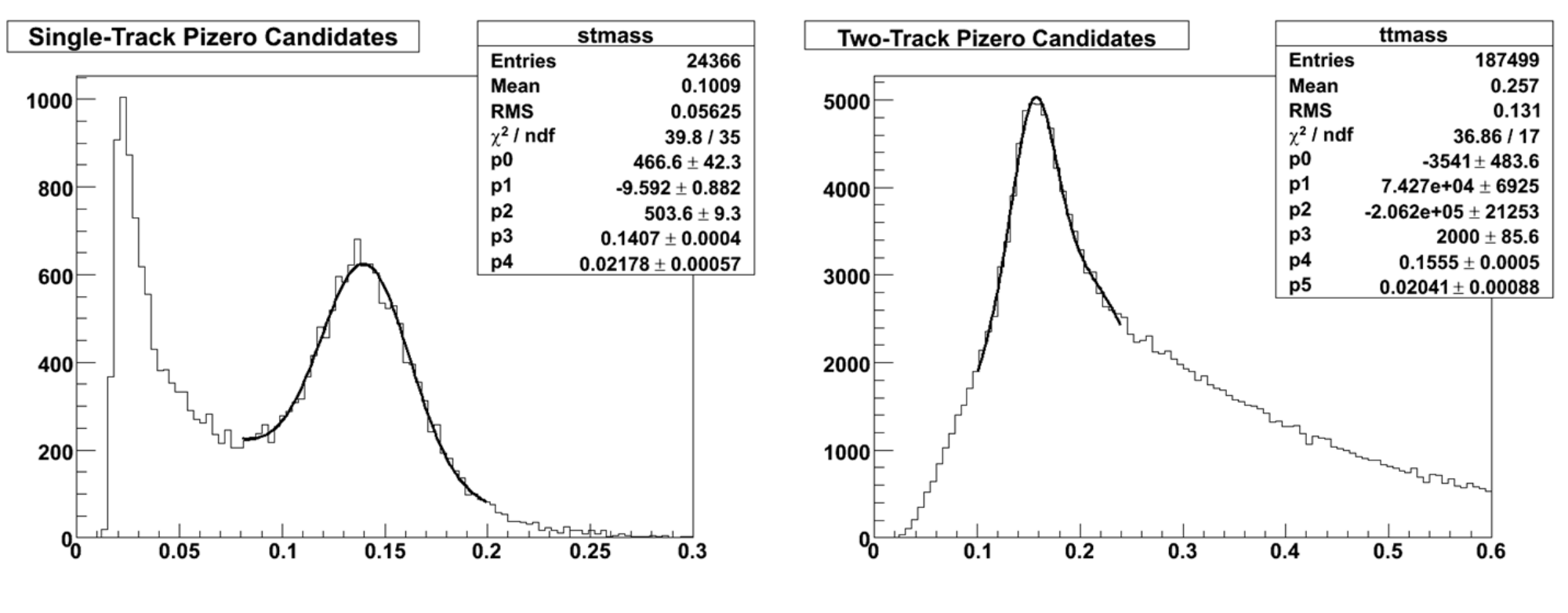}
\caption{\label{fig:Mass_jets} Invariant mass distributions for single-track (left) and two-track (right) $\pi^{0}$ reconstruction in toyMC jet events. 
The single-track $\pi^{0}$ distribution has less background, but lower statistics.}
\end{figure}

Figure ~\ref{fig:Mass_jets} shows the resulting invariant mass distributions for single-track and two-track $\pi^{0}$'s in toyMC jet events. In general the 
single-track $\pi^{0}$ distribution has less background, particularly under the $\pi^{0}$ mass peak. Candidate $\pi^{0}$ mesons for correlations are selected by
a $2.5\sigma$ mass cut in the reconstrcution, $0.105 < m < 0.205 GeV$ for two-track $\pi^{0}$'s and $0.085 < m < 0.195 GeV$ for the single-track reconstruction.

\begin{figure}
\centering
\includegraphics[width=0.9\linewidth]{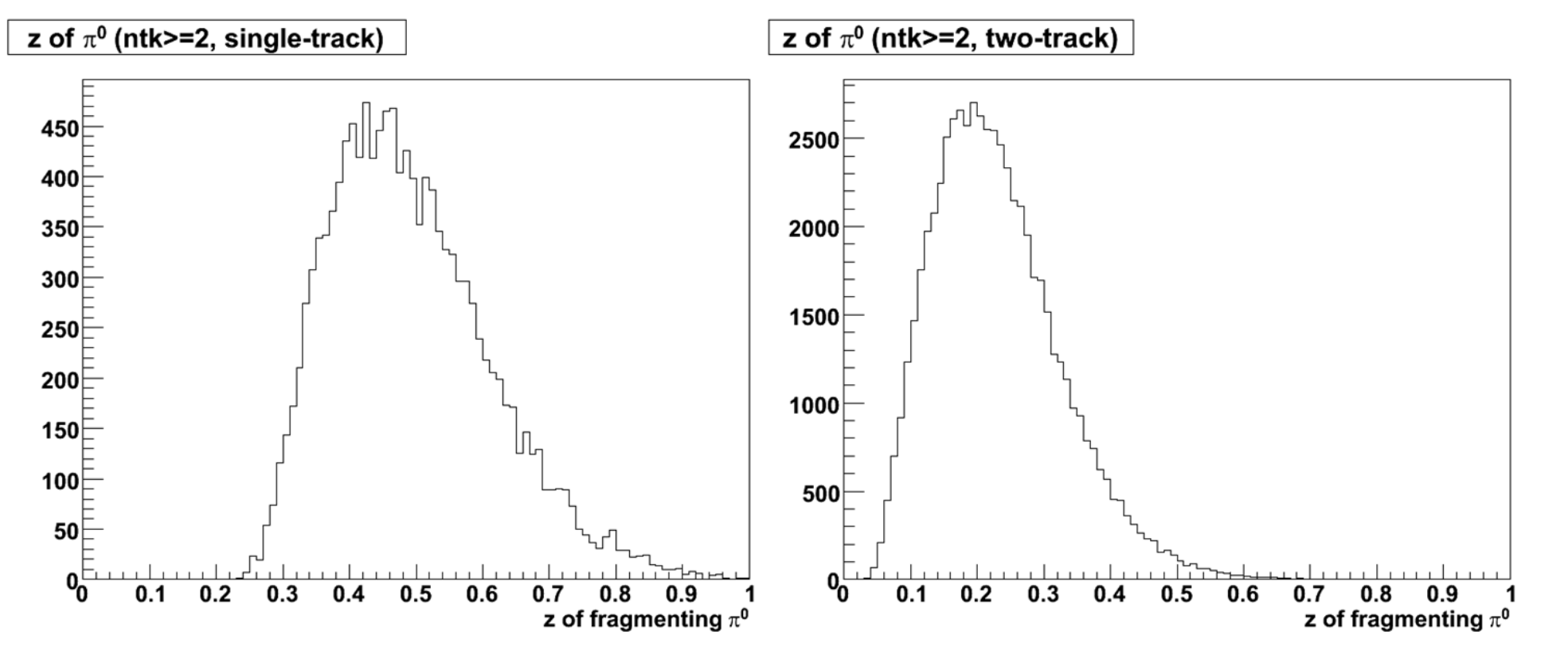}
\caption{\label{fig:frag_z} Distribution of fragmentation $z$ for single-track (left) and two-track (right) $\pi^{0}$ reconstruction. The sample of events
chosen were events with two tracks found making up a charged cluster, with the $\pi^{0}$ on the same side as the cluster in azimuthal angle.}
\end{figure}

Because the two-track and single-track $\pi^{0}$'s are reconstructed in different energy ranges they sample different kinematics. In particular, 
the fragmentation $z=\frac{p_{\pi^0}}{p_{jet}}$ is higher for the single-track $\pi^{0}$'s, as shown in Figure~\ref{fig:frag_z}. 
Because the Collins fragmentation function extracted from 
SIDIS measurements and used in toyMC is a strong function of the fragmentation $z$, the single-track pizero sample will have a 
larger asymmetry.  Conversely, measuring the two-track and single-track pizero samples would allow some experimental 
sensitivity to the $z$ dependence of the Collins fragmentation function and would provide a greater constraint to theoretical models. 

\subsection{Jet Cluster Reconstruction}\label{sec:jetreco}

Jet cluster reconstruction proceeds by sorting MPC-EX tracks into charged track and electromagnetic track lists. For electromagntic
tracks the total, calibrated energy deposited in the MPC-EX and MPC are used to define the track's momentum. However, we have no 
momentum measurement for charged tracks, so all charged tracks are arbitrarily assigned an equal momentum of 1~GeV. The same jet cluster reconstruction is
applied to both charged and EM tracks independently, but charged tracks and electromagnetic tracks are not combined in this analysis. 

The jet cluster algorithm is a seeded cone algorithm that uses every particle in the track list as seed for a cluster cone.
The cluster cone is take as a fixed radius in $\eta$ and $\phi$ space of 1.0 units for both electromagnetic and charged track 
clusters. For each selection of a seed track, the cone algorithm and cluster axis are iterated until further iteration produces 
no change in the cluster axis. This cluster is recorded and the next seed is analyzed. Finally, from the list of all found clusters 
the cluster that resulted in the highest transverse momentum is selected. It should be noted that in
principle we could be less sensitive to backgrounds from gluon jets by reducing the cone radius, but this has not been explored 
in this analysis. 

\begin{figure}
\centering
\includegraphics[width=0.9\linewidth]{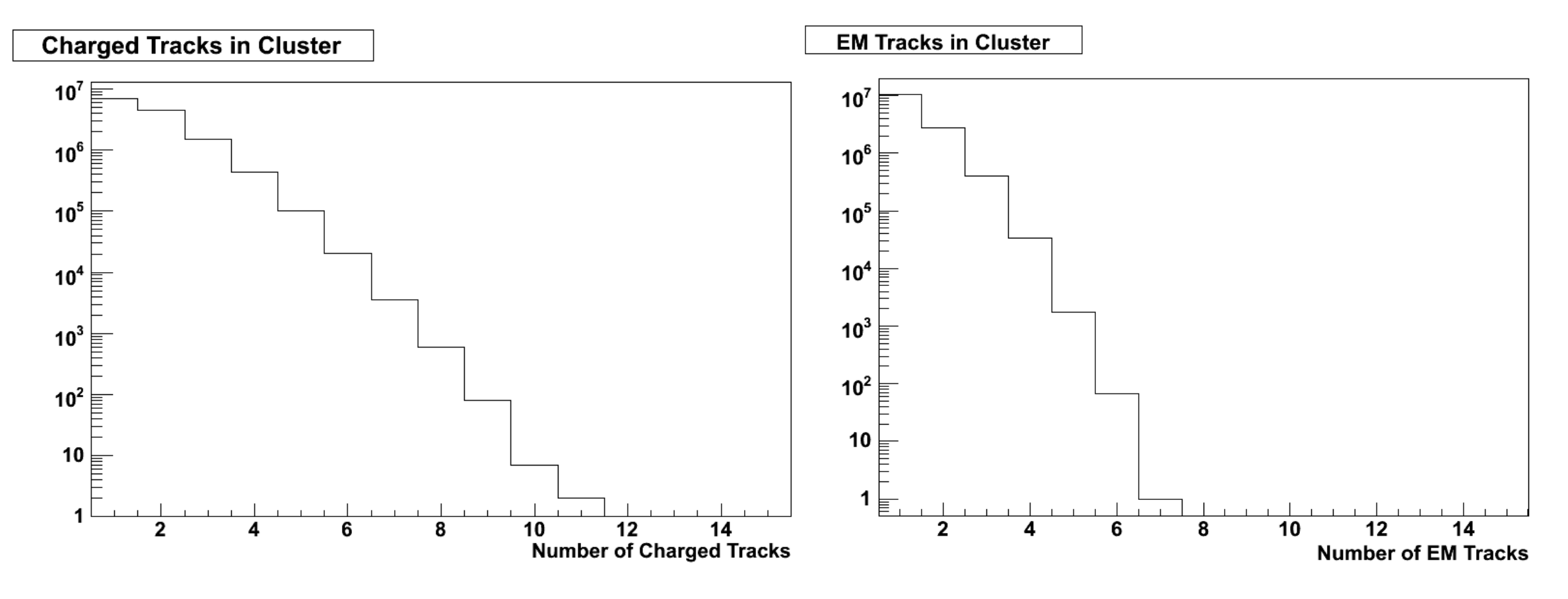}
\caption{\label{fig:npart} Distribution of number of tracks found in the highest $p_T$ cluster in a sample of toyMC jet events, for both charged jet
clusters (left) and electromagnetic jet clusters (right).}
\end{figure}

In the analysis of jet cluster + $\pi^{0}$ correlations that follows, we require the jet cluster be comprised of two or more, or three or more particles, 
effectively making a three-particle or four-particle requirement for the jet 
when the $\pi^{0}$ requirement is added. For a complete dataset of real collisions the asymmetries obtained
could be examined as a number of the particles required in the cluster in order to study systematic effects and improve signal-to-background, but that is
not possible with the limited statistics available in this simulation.

\subsection{Simulated Asymmetries}

In this section we describe the simulated asymmetries based on the large statistics dataset, with single-particle $A_N$ asymmetries that are
comparable to those observed in the MPC. The goal of this exercise is to benchmark the sensitivity of this analysis in terms of the 
minimum asymmetry that should be visible at a given integrated luminosity. 

\subsubsection{Charged Cluster - $\pi^0$ Correlations}

The low asymmetry toyMC event sample is used to gauge the size of the asymmetry expected as well as the total available statistics
given the simulated luminosity. Correlations are examined individually for both the single-track and two-track pizero samples. This statistical 
power in the Monte Carlo sample allows us to make meaningful correlations with charged particle jet clusters that have $>=2$ or $>=3$ charged 
partciles in the cluster. Charged cluster - $\pi^0$ correlations are obtained by selecting events with a charged cluster and a reconstructed 
$\pi^0$ within $\pm\pi/2$ in azimuthal angle. The Collins correlation angle $\phi$ is calculated and binned, and the result can be fit to extract
the spin asymmetry. 

\begin{figure}
\centering
\includegraphics[width=0.9\linewidth]{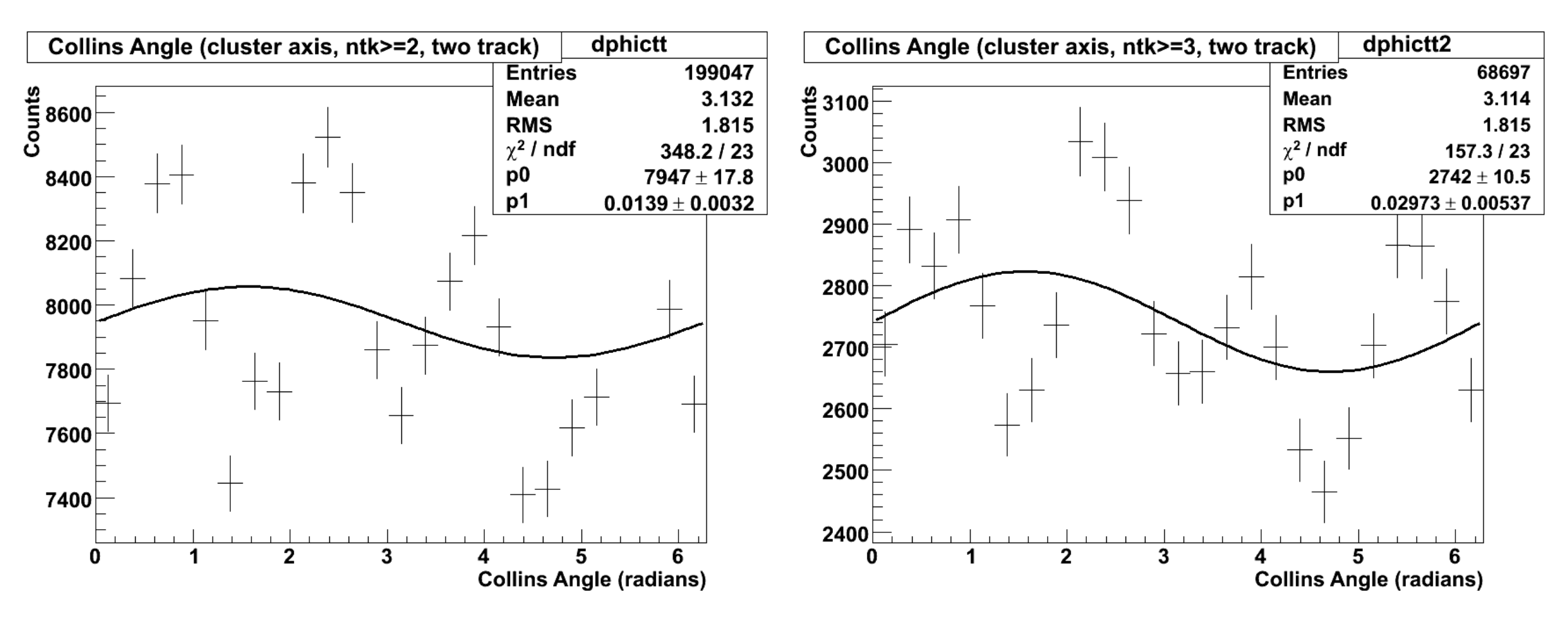}
\caption{\label{fig:tt_asymm} Collins angle asymmetries for two-track $\pi^0$'s correlated with charged clusters consisting of 
two or more particles in a charged cluster (left) and three or more particles (right). The shape of the distributions shows an acceptance effect that 
has not been accounted for.}
\end{figure}

\begin{figure}
\centering
\includegraphics[width=0.9\linewidth]{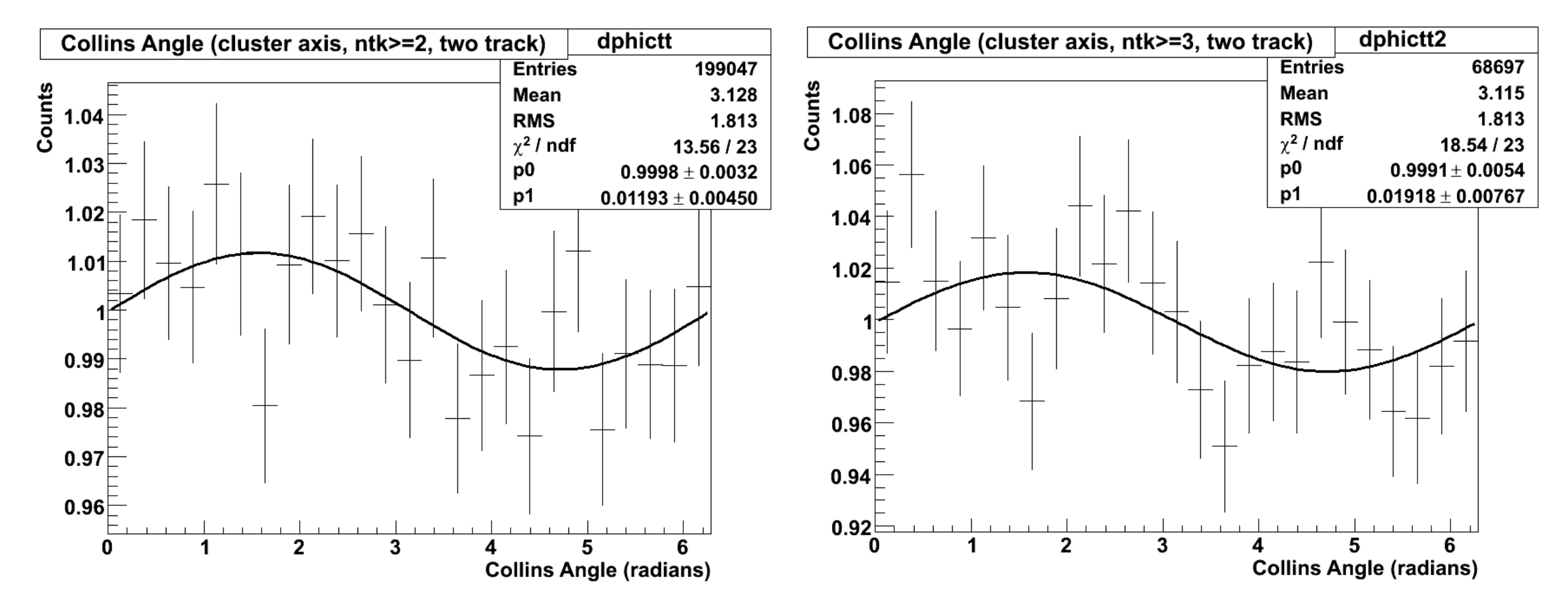}
\caption{\label{fig:tt_asymm_div} Correction of the acceptance effect in the two-track correlation sample after division by the spin-randomized
distribution. Note that this process artificially inflates the error bars in the distribution because the two samples are correlated. A small
Collins asymmtery is clearly visible after the correction. The correlation with charged track clusters containing two or more charged particles is shown
on the left, and with three or more charged particles on the right.}
\end{figure}

As anticipated, 
the asymmetries are small in the two track sample. In fact, they are small enough that acceptance corrections at the level of 1\% are
required to be able to reliably extract the asymmetries (see Figure ~\ref{fig:tt_asymm}). This can be demonstrated by dividing the asymmetry distributions
by a distribution generated by using the same events but with a random spin orientation, thus cancelling any spin-dependent effects and leaving only acceptance 
effects (see Figure ~\ref{fig:tt_asymm_div}). 
Figure ~\ref{fig:tt_asymm_div} shows the result of this exercise, and a small Collins asymmetry is visible after the correction is made. 
We note that dividing the spin-dependent by the spin-randomized distributions in this way does artificially inflate the error bars on each point 
because the two samples are not independent. However, a small but sigificant spin-dependent asymmetry is still visible. 

\begin{figure}
\centering
\includegraphics[width=0.9\linewidth]{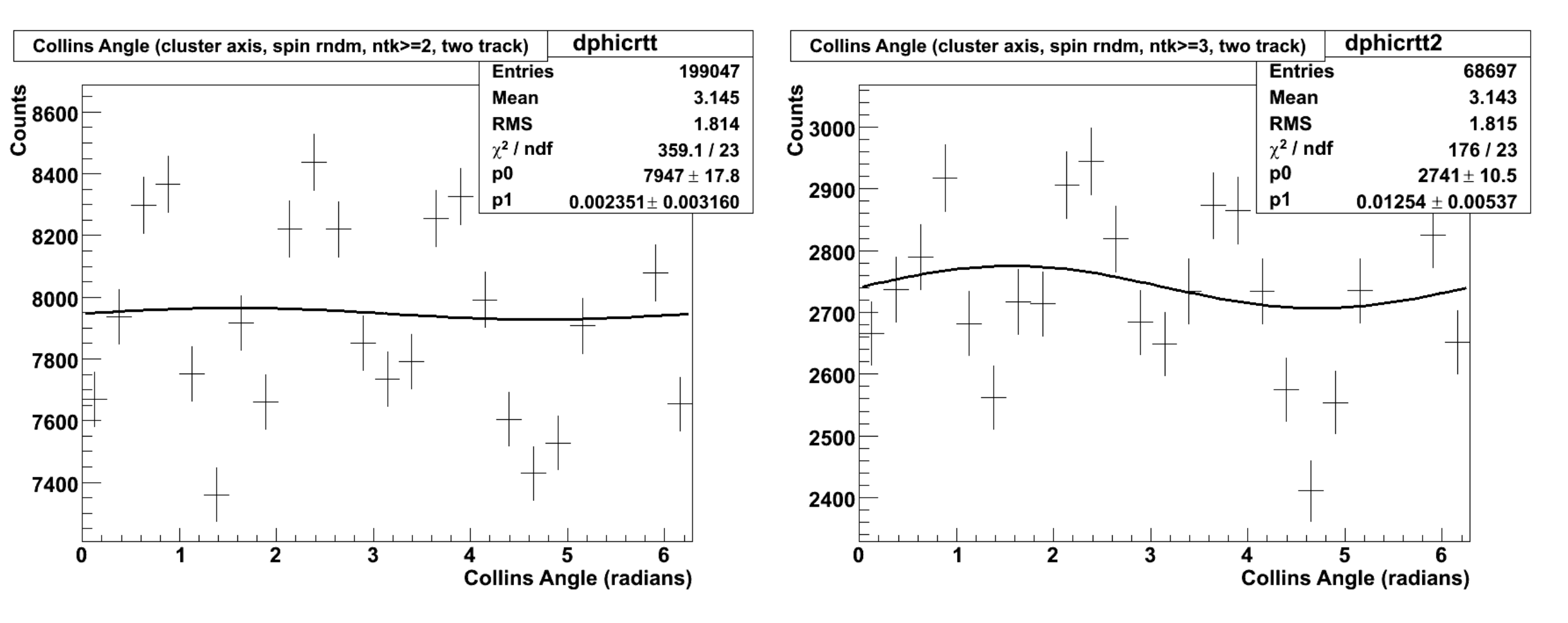}
\caption{\label{fig:tt_asymm_rndm} Collins angle asymmetries for two-track $\pi^0$'s correlated with charged clusters consisting of 
two or more particles in a charged cluster (left) and three or more particles (right). 
The spin asymmetry in these distributions is destroyed by randomizing the spins, 
leaving only acceptance effects.}
\end{figure}

\begin{figure}
\centering
\includegraphics[width=0.9\linewidth]{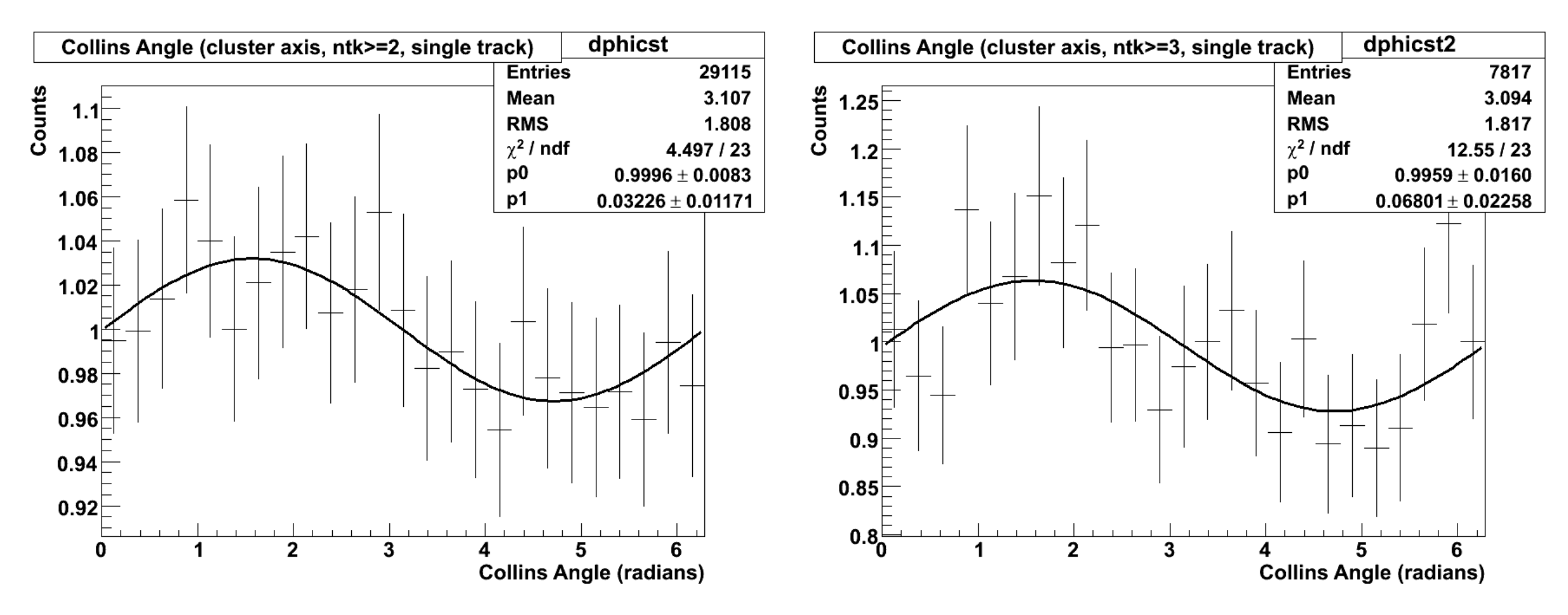}
\caption{\label{fig:st_asymm} Collins angle asymmetries for single-track $\pi^0$'s correlated with charged clusters consisting of 
two or more particles in a charged cluster (left) and three or more particles (right).}
\end{figure}

Figure ~\ref{fig:st_asymm} shows the single-track  $\pi^0$ asymmetries extracted from the large statistics toyMC sample for $\pi^0$'s correlated with different
charged cluster samples. As with the two-track correlations, the spin-dependent distributions are divided by the spin-randomized distributions
to eliminate acceptance effects. In the first sample two or more charged tracks are required to determine the charged cluster, while three or more are
requred in the second sample. As expected, the extracted asymmetries are larger for the single-track correlations.

The Monte Carlo samples allow us to estimate how many pairs will be obtained in a given sample of 
integrated luminosity as well as the magnitude of the anticipated asymmetry. The statistics in each bin in Collins correlation angle can be 
scaled to the total sampled luminosity of 49~$pb^{-1}$. These numbers then provide
the basis for the statistical error in each bin. We assume a 60\% beam polarization, and scale the anticipated asymmetry down by 
0.6 to account for partial beam polarization. In addition, we also impose the trigger requirement that total energy in the MPC
is greater than 35~GeV (see Section~\ref{sec:rates}). This requirement greatly reduces the statistical power of the two-track pizero sample by eliminating
low-energy jets. 
 
\begin{figure}
\centering
\includegraphics[width=0.8\linewidth]{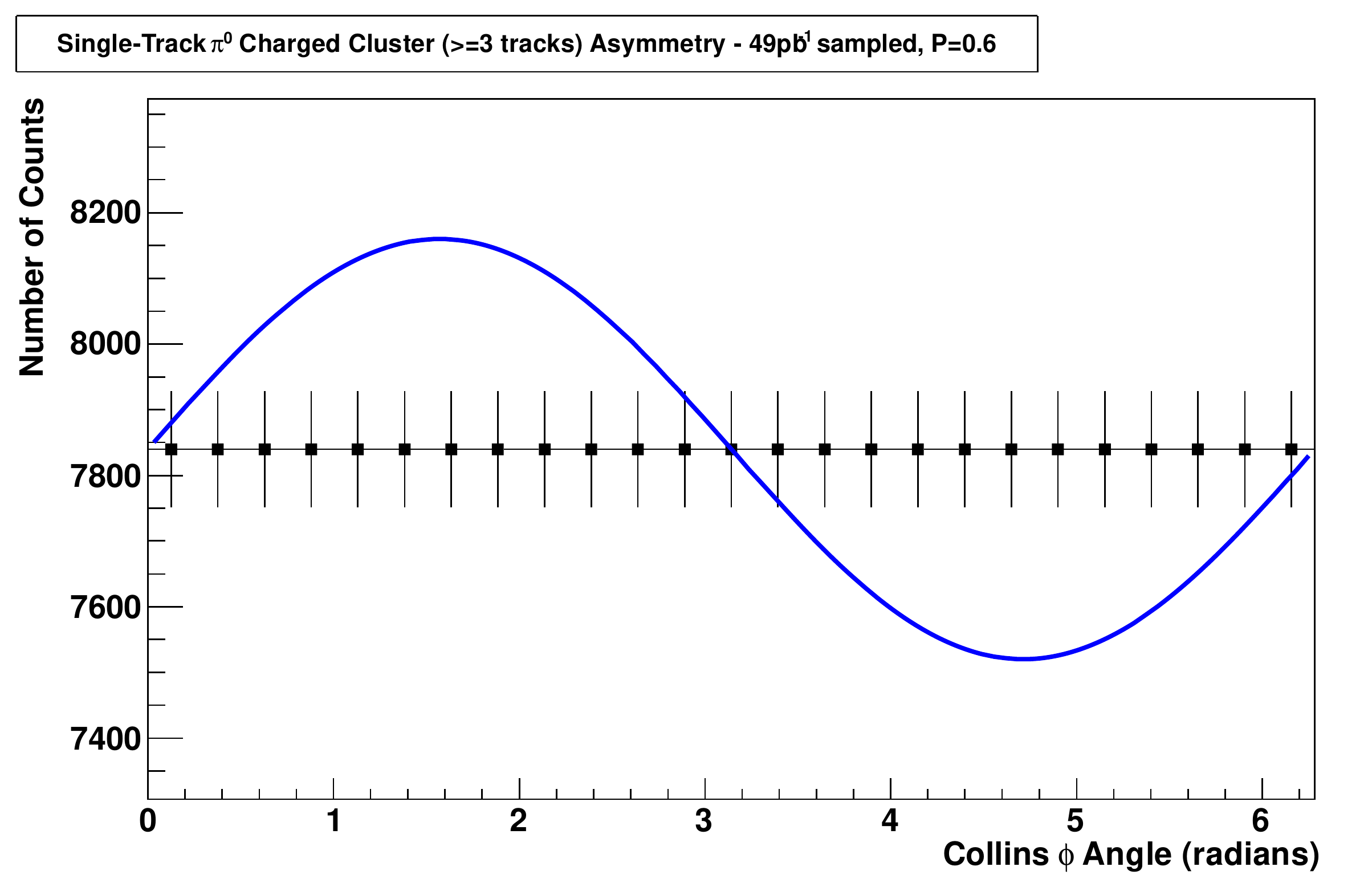}
\caption{\label{fig:st_proj} Anticipated statistics as a function of Collins angle for 49~$pb^{-1}$ sampled luminosity and 
average polarization of 60\% using single-track $\pi^0$'s correlated with a jet axis determined by three or more charged particles. 
The blue curve is the anticipated asymmetry for the data sample from the Monte Carlo, corrected for the beam polarization 
of 60\%.}
\end{figure}

In Figure ~\ref{fig:st_proj} we show the antipated asymmetry and statistical errors using single-track $\pi^0$'s 
correlated with a jet axis determined using three or more charged tracks. The anticipated statistical power is more 
than adequate to measure the expected asymmetries if the single particle $A_N$ is due to transversity and Collins fragmentation alone.  


The minumum asymmetry that can be observed is set by the ability to ascertain if more $\pi^0$'s go ``left'' with respect to the direction set by the 
spin axis and fragmenting parton, as opposed to ``right''. To estimate the smallest asymmetry that could be observed we make maximum
use of the statistical power of the data by dividing it into  
left and right, spin up and spin down samples, and calculating the asymmetry using the square root formula: 

\begin{equation}
A_{N}^{raw} = \frac{\sqrt{N_L^{\uparrow}N_R^{\downarrow}} - \sqrt{N_L^{\downarrow}N_R^{\uparrow}}}
{\sqrt{N_L^{\uparrow}N_R^{\downarrow}} + \sqrt{N_L^{\downarrow}N_R^{\uparrow}}}
\end{equation}

Based on the assumed statistics and error propagation of the above formula, we can estimate the statistical error on the 
raw asymmetry $A_{N}^{raw}$ for each event selection. For Table ~\ref{tab:3sigma_err} this analysis should be sensitive to 
a raw asymmetry down to one-seventh of the expected asymmetry, at the level of $3\sigma$. This means that this analysis will 
be capable of measuring a Collins asymmtery even if it is only responsible for as little as 27\% of the overall single
particle $\pi^0$ $A_{N}$ in the single-track correlations. 

\begin{table*}
\caption{\label{tab:3sigma_err} 
Three-sigma statistical errors assuming 49~$pb^{-1}$ and 60\% polarization for two events samples.   
}
\begin{center}
\begin{tabular}{||c||c|c||}
\hline \hline
                 & Raw Asymmetry  &  3-$\sigma$ statistical  \\
Event Selection  &  P = 0.6       &  error on $A_{N}^{raw}$   \\
\hline
two-track $\pi^0$, $>=3$ tracks in charged cluster             & 0.011   & 0.014  \\
single-track $\pi^0$, $>=3$ tracks in charged cluster          & 0.041   & 0.011  \\
\hline \hline
\end{tabular}
\end{center}
\end{table*}

\subsubsection{Electromagnetic Cluster - $\pi^0$ Correlations}

As a demonstration, we also reconstruct clusters of electromagnetic tracks in this analysis. While 
electromagnetic clusters have the added benefit of yielding energy information, they are not appropriate
for the determination of a Collins asymmetry through the correlation with $\pi^{0}$ mesons. The reason for this 
is simple - clusters determined with electromagnetic particles will contain the correlation that we are trying 
to measure, because most of the electromagnetic energy will come from $\pi^{0}$'s. This effect will become worse
for higher momentum $\pi^{0}$'s because the cluster axis determination will be dominated by the $\pi^{0}$. In this
case the correlation is increasingly like correlating the $\pi^{0}$ with itself. It is for this reason that the 
MPC alone cannot perform this type of analysis. Such an analysis has been attemped with the STAR FPD++ ~\cite{Poljak:2011pg}. 

\begin{figure}
\centering
\includegraphics[width=0.9\linewidth]{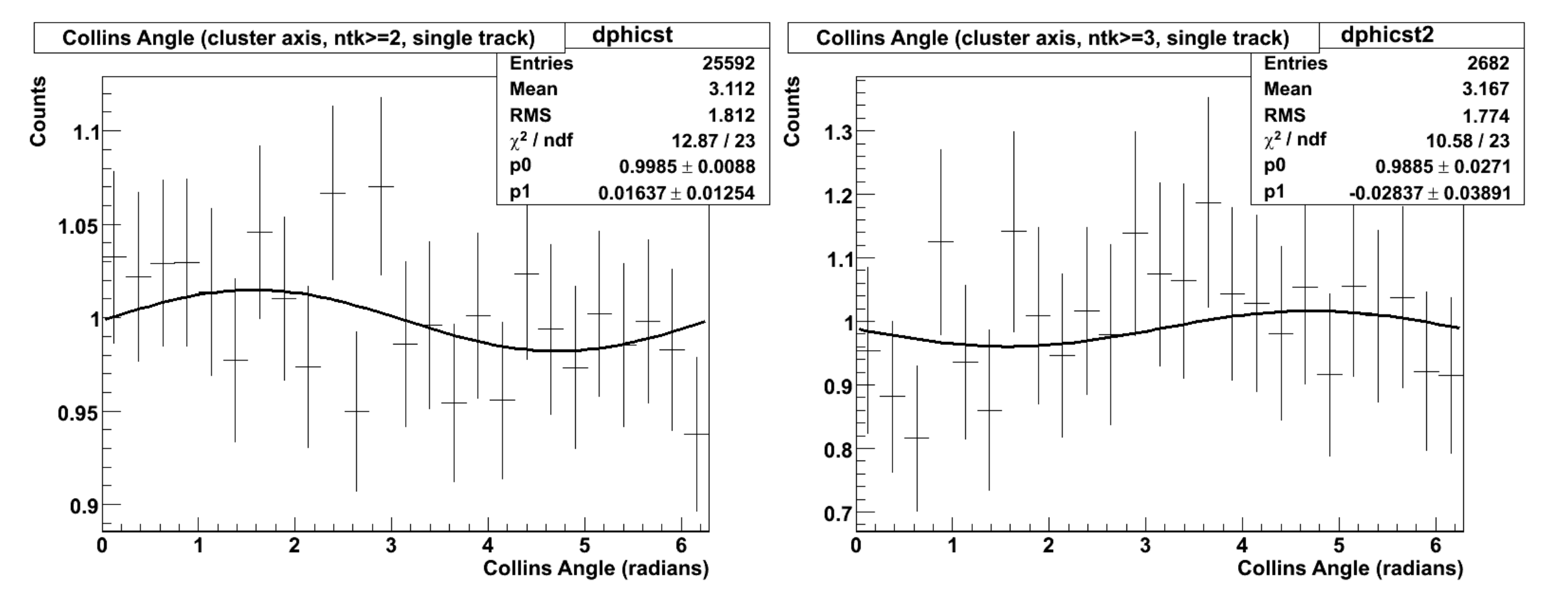}
\caption{\label{fig:st_asymm_em} Collins angle asymmetries for single-track $\pi^0$'s correlated with electromagnetic clusters consisting of 
two or more electromagnetic tracks (left) and three or more electromagnetic tracks (right). The asymmtery is dramatically reduced compared to 
the charged cluster correlations, see Figure ~\ref{fig:st_asymm}.}
\end{figure}

To demonstrate this effect, we show the correlations between $\pi^{0}$ mesons and an electromagnetic jet axis 
determined with three or more electromagnetic tracks in the MPC-EX in Figure ~\ref{fig:st_asymm_em}. The asymmetries when 
correlating with an electromagnetic axis are essentially zero within errors.  

\subsubsection{Jet Axis Resolution and Asymmetries}

\begin{figure}[hbt]
\hspace*{-0.12in}
\begin{minipage}[b]{0.5\linewidth}
\centering
\includegraphics[angle=90, width=0.95\linewidth]{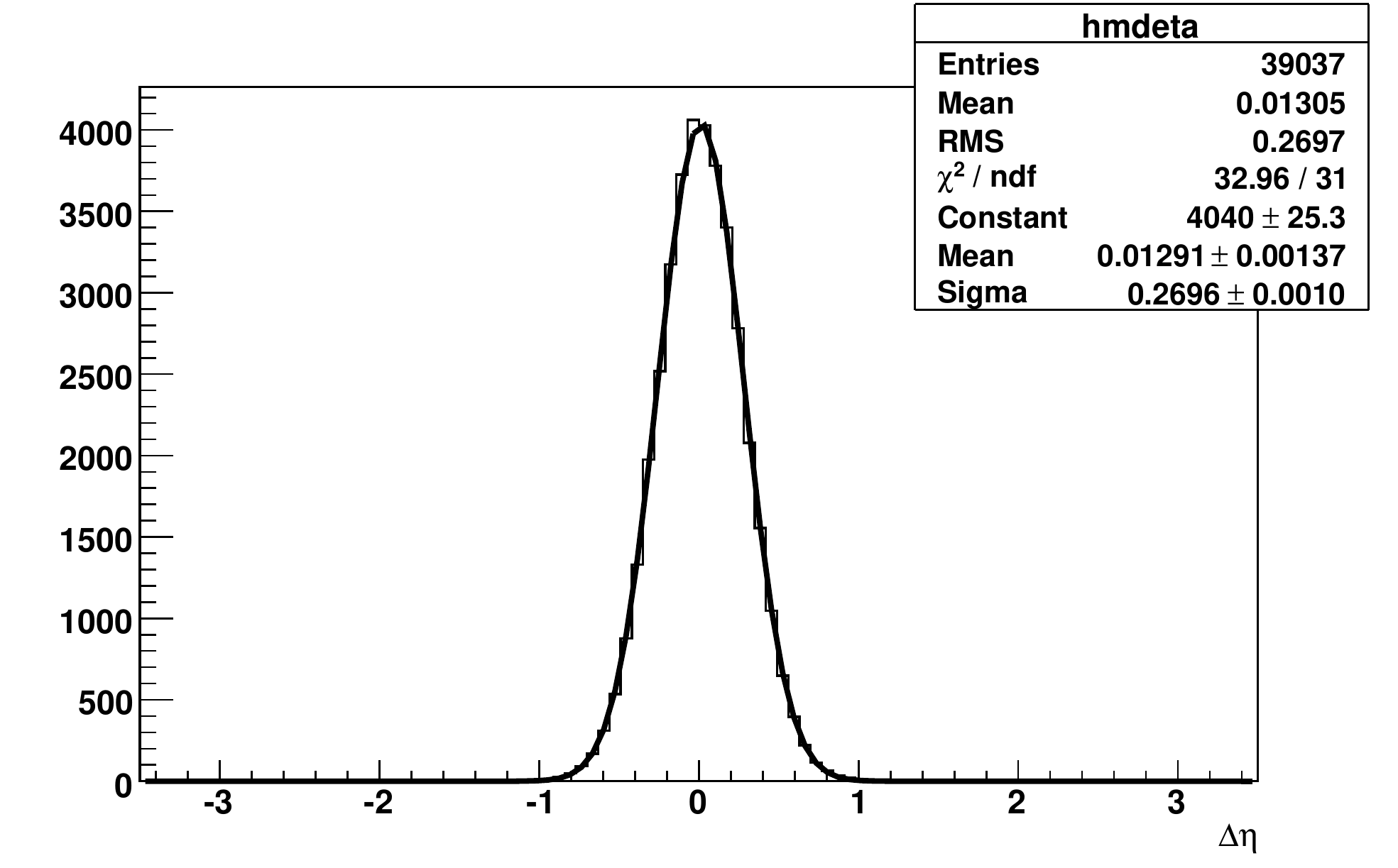}
\end{minipage}
\hspace{0.5cm}
\begin{minipage}[b]{0.5\linewidth}
\centering
\includegraphics[angle=90, width=0.95\linewidth]{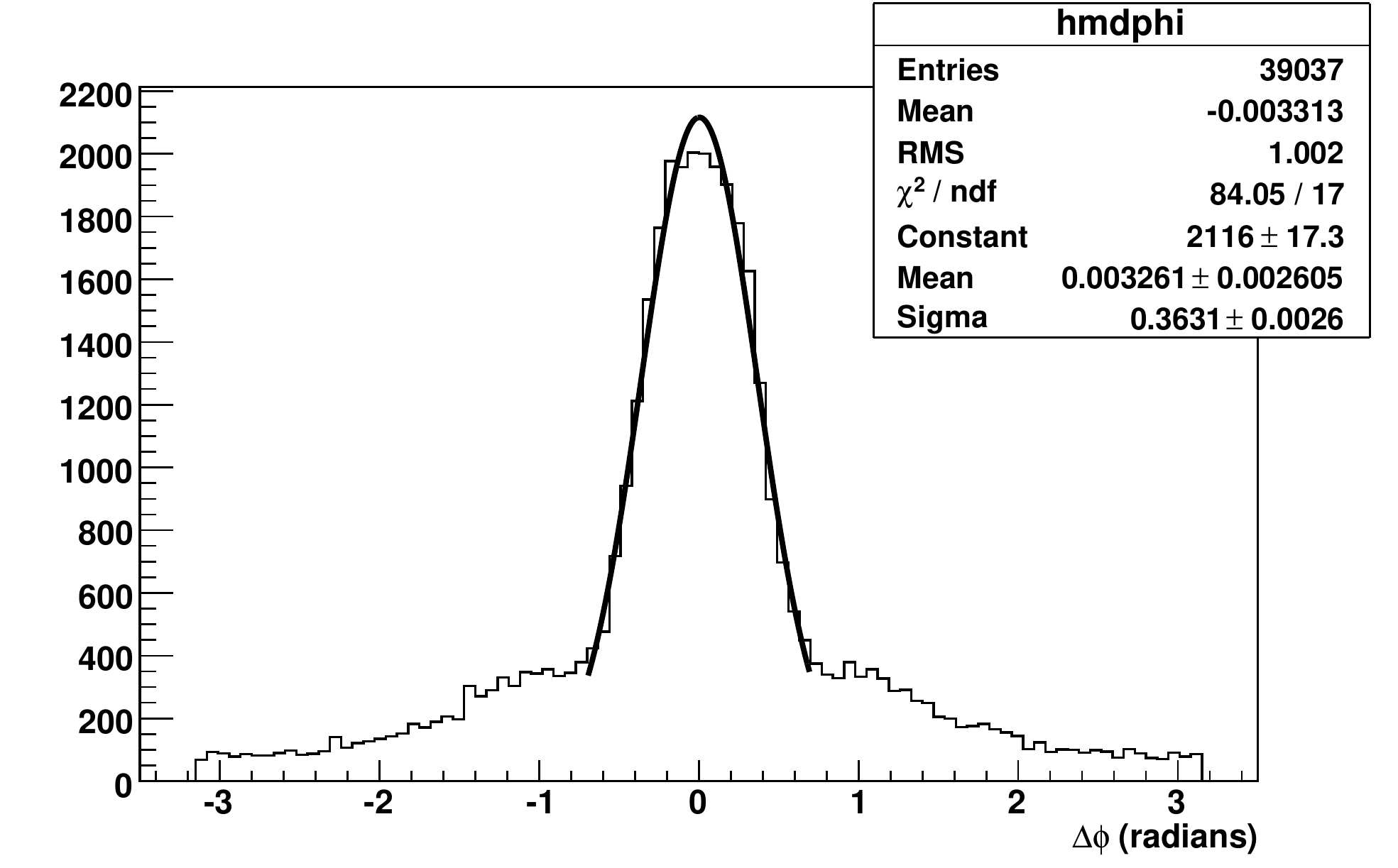}
\end{minipage}
\vspace*{-0.12in}
\caption{\label{fig:jaxis_toymc_resolution} Resolution in $\eta$ (left) and $\phi$ (right) for a jet proxy determined 
from charged particles reconstructed in the MPC-EX for toyMC events. The jet proxy is required to have three or more 
charged particles. The resolution can be approximately 
described by a sigma of $\sim$0.3 units in both $\eta$ and $\phi$ and is consistent with the resolution obtained 
from {\sc Pythia} events.} 
\end{figure}

The resolution of the jet axis critically determines how the asymmetry measurement is diluted due to an imperfect knowledge of 
the fragmenting parton direction. Because the toyMC model is based on a simplified fragmentation scheme it is possible that this 
could yield a better resolution than one would anticipate from {\sc Pythia} events. In Figure~\ref{fig:jaxis_toymc_resolution}
we show the jet proxy axis resolution in $\eta$ and $\phi$ for toyMC events (averaged over spin orientation). The jet proxy axis 
resolution is similar to {\sc Pythia} events, indicating that the simlified toyMC model has not resulted in an overly optimistic
set of assumptions.  

When charged particles (mostly $\pi^+$ and $\pi^-$) are used to reconstruct the jet direction it is possible that the 
asymmetries that these particles carry could bias the jet axis and either dilute or induce an asymmetry when correlated 
with $\pi^0$'s. However, any potential effect should be limited by the fact that the charged pions carry a roughly equal 
and opposite asymmetry, so that in the limit that a larger number of particles are used in the charged cluster any systematic
effect should vanish. 

\begin{figure}
\centering
\includegraphics[width=0.95\linewidth]{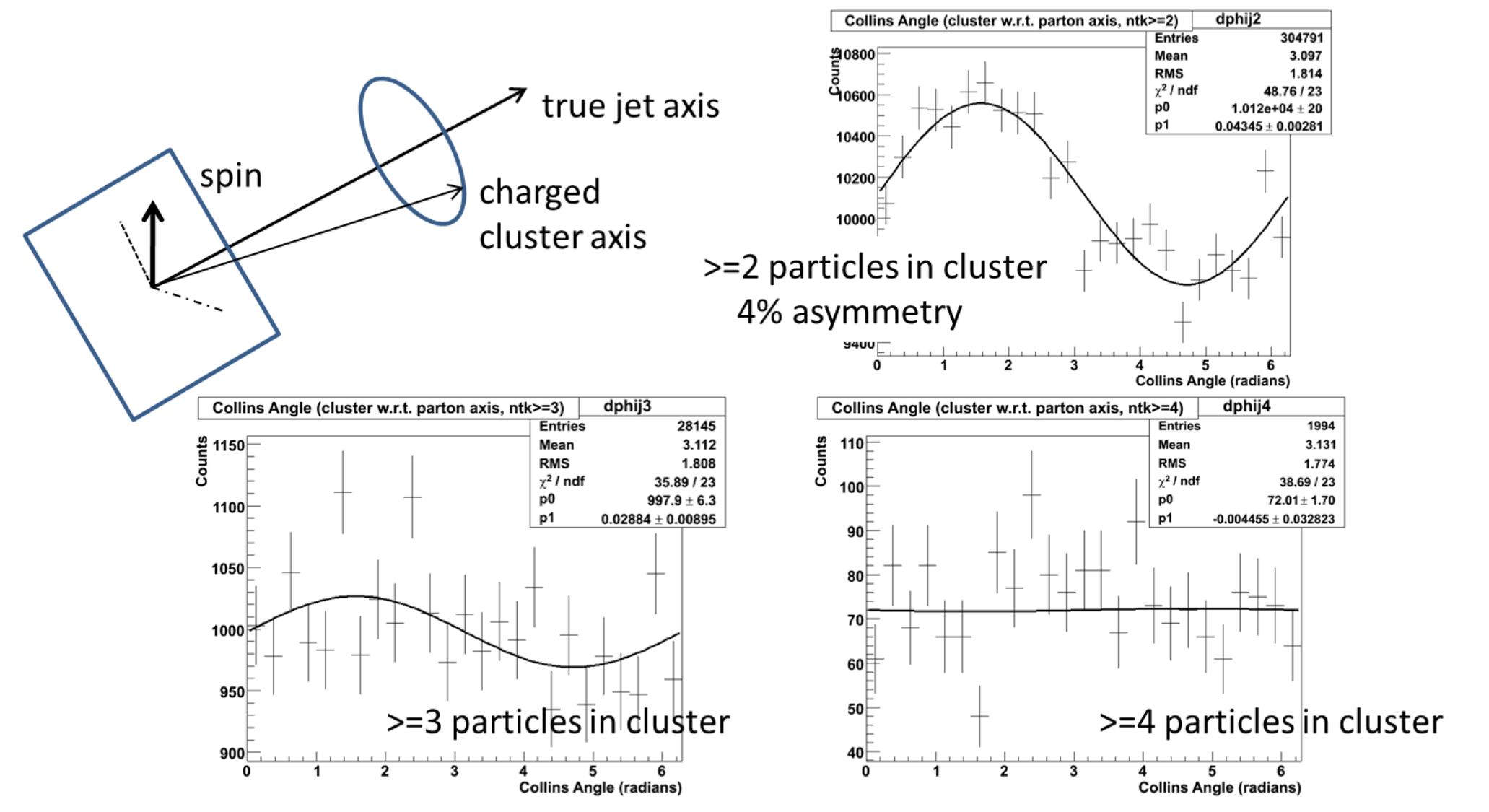}
\caption{\label{fig:jaxis_bigasymm} Asymmetry of the reconstructed charge cluster axis with respect to the parton axis for the
large asymmetry sample of events. As a comparison the single particle $A_N$ is ~20\% in these events, substantially larger than the 
observed $A_N$, which magnifies the asymmetry of the charged cluster axis. This effect essentially disappears when three or more 
charged particles are required in the jet determination.}
\end{figure}

This effect is too small to observe in the small asymmetry sample of events, 
but in Figure~\ref{fig:jaxis_bigasymm} we show the charged cluster asymmetry with respect to the jet axis for the large asymmetry sample of 
events. The large asymmetry sample enhances the effect, but it can be seen to diminish as a function of the number of 
charged particles required in the charged cluster. Studying the measured $\pi^0$ asymmetries as a function of the number of tracks
in the charged cluster will provide a way to constrain any potential systematic error. 

\subsection{Effect of the Underlying Event}

The previously described toyMC simulations included only the particles from jet fragmentation, 
and not additional particles from the underlying event (breakup of the target and projectile
protons). In order to estimate this affect, a small sample of approximately 300k toyMC events 
were generated where the final jet particles were merged with an independent {\sc Pythia} minimum bias event.  These
merged events were reconstructed following the same procedure as the jet events, and the 
resulting asymmetries compared. 

Because of the lower statistics, a detailed comparison between the jet + minbias events and
the low-asymmetry jet sample is not possible. However, we do note that even with the low statistics a significant
asymmetry is still visible in the correlations. This gives us confidence that the presence of the 
underlying event does not destroy the correlations observed in the pure jet events. 

\begin{figure}
\centering
\includegraphics[width=0.75\linewidth]{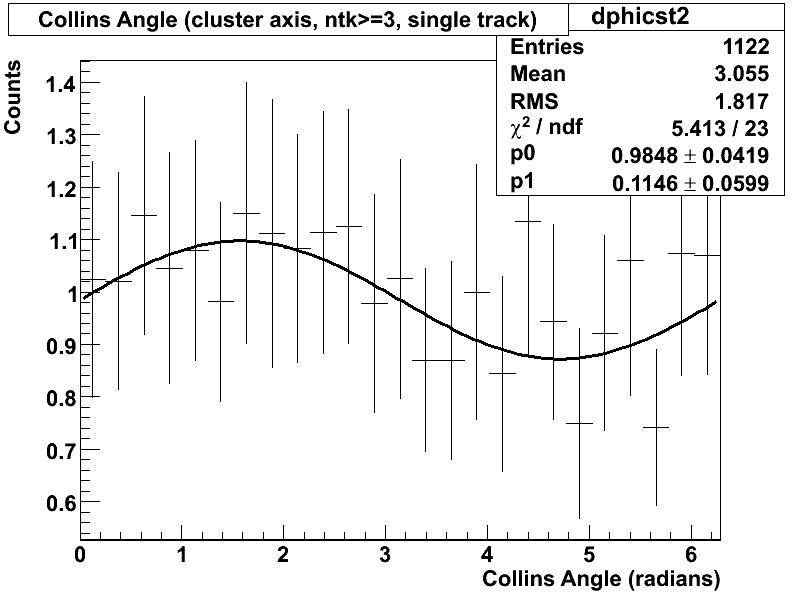}
\caption{\label{fig:j_mb_st2_div} Angular correlations of $\pi^{0}$ mesons for clusters with three or more charged tracks
in events where a toyMC jet event is merged with a {\sc Pythia} minbias event. While the statistics are lower due to the smaller
number of these events generated, a clear asymmetry is still visible.}
\end{figure}

\cleardoublepage

\resetlinenumber

\cleardoublepage





\appendix

\setcounter{chapter}{9}

\appendix

\cleardoublepage
\resetlinenumber

\chapter{Event Rates}
\label{sec:rates}
\label{sec:A}

\markboth{Appendix A}{Event Rates and Triggering}

\subsection{Event Rates}

In this section we estimate the event rates for selected processes in
p+p and d+Au collisions.  All rates are for an MPC-EX in both the south and north arms in p+p, and in the deuteron going direction only 
in the case of d+Au collisions. In calculating event rates we start with latest
guidance document from the RHIC Collider Accelerator Division (CAD)~\cite{CAD_Oct2011}, which lists
the store average luminosity for the various species and the delivered luminosity per week. We then assume a 60\% up time for the
PHENIX detector, a DAQ live time of 90\%, and minbias trigger efficiencies (see Table~\ref{tab:effs}) for a 12 week run 
to obtain the luminosity sampled by PHENIX (see Table~\ref{tab:lums}). Note that we do not include an event vertex cut efficiency, as the 
MPC-EX analysis will be able to use essentially the full vertex distribution delivered by RHIC.

\begin{table*}
\caption{\label{tab:effs} 
Efficiency factors used in the rate calculations.  For A+A collisions the minimum bias
trigger formed by the BBC is very close to 100\% efficient,
however in p+p and d+A collisions this is not the case.  
}
\begin{center}
\begin{tabular}{||c||c|c|c|c|c||}
\hline \hline
Species                & p+p 200 GeV  & p+p 500 GeV & d+Au   &Cu+Cu    & Au+Au  \\
\hline
min bias trigger eff   & 0.75         & 0.75        & 0.90    & 1.0    & 1.0   \\
\hline \hline
\end{tabular}
\end{center}
\end{table*}

\begin{table*}
\caption{\label{tab:lums} 
Luminosity guidance from CAD for RHIC. We then assume a 12 week run and a 60\% uptime for PHENIX, a DAQ live time of 90\% and a minbias
trigger efficiency to obtain a PHENIX sampled integrated luminosity.}
\begin{center}
\begin{tabular}{||c||c|c|c|c|c||}
\hline \hline
Species                    & p+p             & p+p            & d+Au               &Cu+Cu            & Au+Au \\
CM Energy                  & 200 GeV         & 500 GeV        & 200 GeV           &200 GeV         & 200 GeV \\
\hline
store average luminosity  &                  &                &                    &               &          \\
 ($s^{-1}cm^{-2}$)        & $3\times10^{31}$ &$1.2\times10^{32}$ &$1.8\times10^{29}$  &$5\times10^{28}$ &$3.5\times10^{27}$ \\
interaction rate (kHz)       & 1260            & 4200         & 328               & 124            & 17     \\
lum/wk (pb$^{-1}$wk$^{-1}$)  & 10              & 40            & 0.060            & 0.016          &0.0011      \\
int. sampled lum (pb$^{-1}$) & 49             & 190           & 0.350              & 0.100          &0.0071      \\
\hline \hline
\end{tabular}
\end{center}
\end{table*}

\begin{figure}[H]
  \begin{center}
  \includegraphics[width=0.8\linewidth]{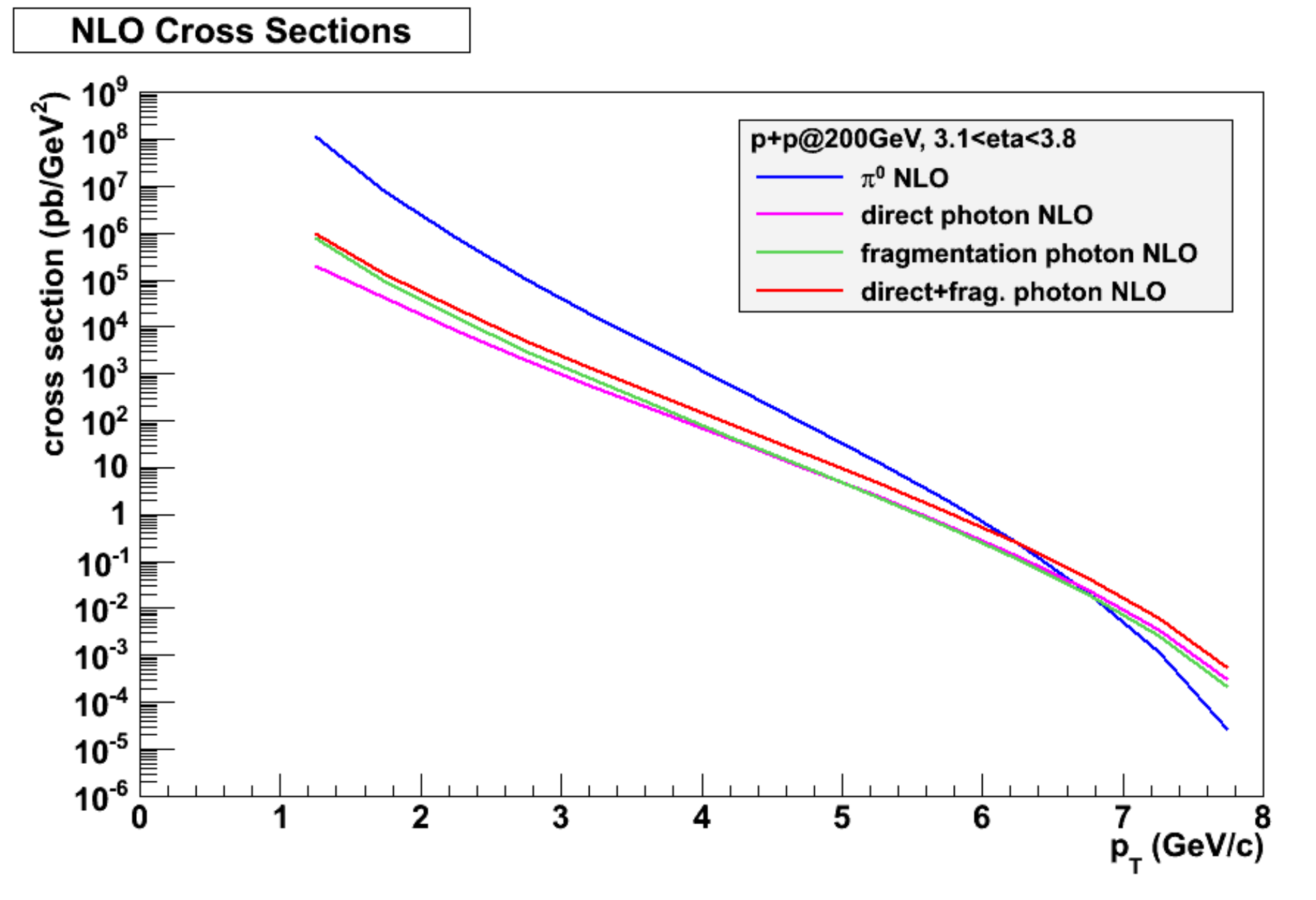}
  \end{center}
  \caption{\label{fig:Werner1} NLO cross sections for pizero, direct photons, and fragmentation photons
  in the MPC-EX acceptance at 200~GeV as a function of transverse momentum. }
\end{figure}

NLO cross sections for $\pi^{0}$, direct and fragmentatation photons at 200~GeV were obtained from Werner Vogelsang, 
for p+p collisions at 200~GeV~\cite{Werner_NLO}. 
These cross sections were calculated using the CTEQ6M5 pdf's, with the scale $\mu = p_T$, and no isolation cuts for the 
photons. The $\pi^{0}$ cross sections were calculated using the DSS fragmentation functions.These cross sections are plotted in Figure
~\ref{fig:Werner1}. As one approaches very high $p_T$ (near the luminosity limit) the direct photon production surpasses the pizero production due
to the fact that the pizeroes arise from fragmentation.

\begin{figure}[H]
  \begin{center}
  \includegraphics[width=0.8\linewidth]{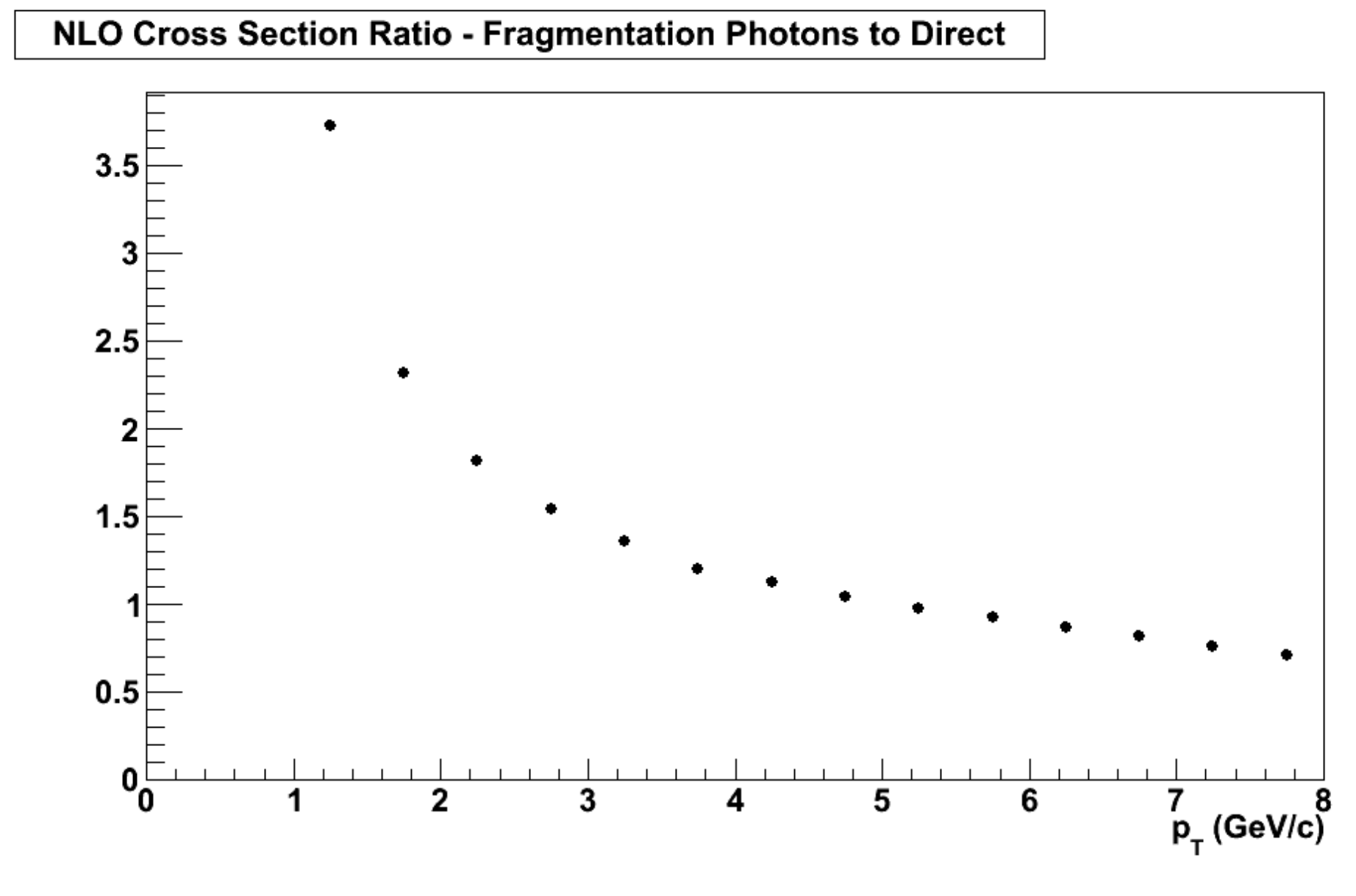}
  \end{center}
  \caption{\label{fig:Werner2} Ratio of the NLO cross sections for fragmentation photons to direct photons
  in the MPC-EX acceptance at 200~GeV as a function of transverse momentum. From Werner Vogelsang. }
\end{figure}

In order to estimate the available statistics in a d+Au run we start with the NLO cross sections and assume $<N_{coll}>$ scaling so that the 
cross sections are scaled up by a factor of 8.3, the mean number of binary collisions in a minbias d+Au collsions as determined by a Glauber 
simulation ~\cite{dAu_PHENIX_ANs}. The estimated statistics in a RHIC d+Au run with 350~$nb^{-1}$ PHENIX integrated sampled luminosity 
(consistent with Table~\ref{tab:lums}) is shown in Table~\ref{tab:dAu_stats} (without correction for suppression effects that will be present in 
d+Au collisions). The MPC-EX reconstruction efficiency is not included in these tables. These efficiencies are described for 
various cuts scenaries in Section~\ref{sim:dphotcuts}.

\begin{table*}
\caption{\label{tab:dAu_stats} 
Estimated statistics for dAu collisions, assuming PHENIX efficiencies (vertex and minbias trigger efficiency). MPC-EX reconstruction efficiencies 
are not included (see Section~\ref{sim:dphotcuts}). 
These numbers are not adjusted for any suppression or shadowing effects.}
\begin{center}
\begin{tabular}{||c|c|c|c||}
\hline \hline
$p_T$  (GeV/c)  & $\pi^{0}$ & photons (direct + frag.) & photon/$\pi^{0}$ \\
\hline
1.0-1.5      & $9.2 \times 10^{8}$  & $7.6 \times 10^{6}$  & 0.0084 \\
1.5-2.0      & $8.2 \times 10^{7}$  & $1.5 \times 10^{6}$  & 0.018 \\
2.0-2.5      & $1.1 \times 10^{7}$  & $3.5 \times 10^{5}$  & 0.031 \\
2.5-3.0      & $1.8 \times 10^{6}$  & $9.0 \times 10^{4}$  & 0.050 \\
3.0-3.5      & $3.3 \times 10^{5}$  & $2.5 \times 10^{4}$  & 0.075 \\
3.5-4.0      & $6.5 \times 10^{4}$  & $7.2 \times 10^{3}$  & 0.11 \\
4.0-4.5      & $1.3 \times 10^{4}$  & $2.0 \times 10^{3}$  & 0.16 \\
4.5-5.0      & $2.5 \times 10^{3}$  & 583                  & 0.24  \\
5.0-5.5      & 450                  & 173                  & 0.36 \\
5.5-6.0      & 72                   & 42		   & 0.58 \\
6.0-6.5      & 10                   & 10                   & 1.0 \\
\hline \hline
\end{tabular}
\end{center}
\end{table*}

The estimated statistics in a transversely polarized RHIC p+p run with 49~$pb^{-1}$ PHENIX integrated sampled luminosity 
(consistent with Table~\ref{tab:lums}) is shown in Table~\ref{tab:pp_stats}.  

\begin{table*}
\caption{\label{tab:pp_stats} 
Estimated statistics for p+p collisions, assuming PHENIX efficiencies (vertex and minbias trigger efficiency).
MPC-EX reconstruction efficiencies are not included (see Section~\ref{sim:dphotcuts}). }
\begin{center}
\begin{tabular}{||c|c|c||}
\hline \hline
$p_T$  (GeV/c)  & $\pi^{0}$ & photons (direct + frag.)  \\
\hline
1.0-1.5      & $3.2 \times 10^{10}$ & $2.7 \times 10^{8}$  \\
1.5-2.0      & $2.8 \times 10^{9}$  & $5.0 \times 10^{7}$  \\
2.0-2.5      & $3.7 \times 10^{8}$  & $1.2 \times 10^{7}$  \\
2.5-3.0      & $6.0 \times 10^{7}$  & $3.0 \times 10^{6}$  \\
3.0-3.5      & $1.1 \times 10^{7}$  & $8.3 \times 10^{5}$  \\
3.5-4.0      & $2.2 \times 10^{6}$  & $2.3 \times 10^{5}$  \\
4.0-4.5      & $4.3 \times 10^{5}$  & $7.0 \times 10^{4}$  \\
4.5-5.0      & $8.5 \times 10^{4}$  & $2.0 \times 10^{4}$  \\
5.0-5.5      & $1.5 \times 10^{4}$  & $5.2 \times 10^{3}$  \\
5.5-6.0      & $2.3 \times 10^{3}$  & 1433                 \\
6.0-6.5      & 317                  & 333                  \\
6.5-7.0      & 32                   & 63                   \\
7.0-7.5      & 2                    & 10                  \\
\hline \hline
\end{tabular}
\end{center}
\end{table*}

\begin{figure}[H]
  \begin{center}
  \includegraphics[width=0.8\linewidth]{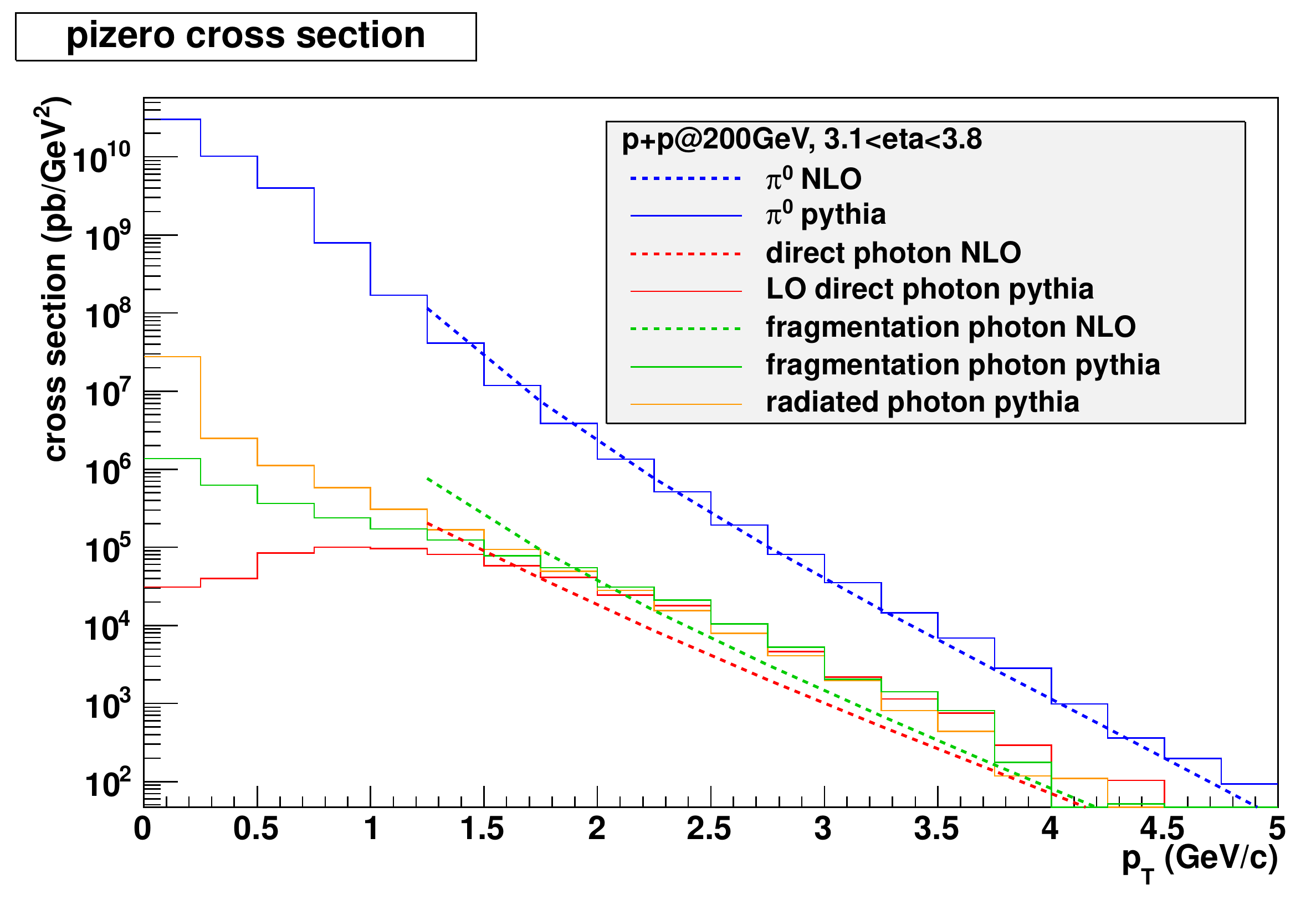}
  \end{center}
  \caption{\label{fig:xsect} NLO cross sections fom Werner Vogelsang compared the cross sections extracted from {\sc Pythia} for $\pi^{0}$'s and direct
photons.  Direct photons from {\sc Pythia} are selected only from processes that produce a photon at the hard scattering vertex, fragmentation photons and 
radiation from incoming quarks are shown as separate entries. The direct photon, fragmentation, and radiation cross sections are comparable in {\sc Pythia}. 
from the hard scattering vertex.}
\end{figure}

Finally, we compare the NLO cross sections obtained from Werner Vogelsang with the output of {\sc Pythia} (see Figure ~\ref{fig:xsect}). 
The {\sc Pythia} events were generated using the ``Tune A'' {\sc Pythia} tune, which is matched to CDF and D0 data. In general, there are three sources
of ``direct'' photons in {\sc Pythia}: photons generated at the hard scattering vertex, fragmentation photons, and photons from QED radiation off the incoming 
quarks. We find that the cross section for photons from the hard scattering vertex in {\sc Pythia} is a reasonable match to the NLO cross section, and fragmentation photons
are approximately equal, also in agreement with the NLO calculations.  

\subsection{Triggering}

Because the MPC-EX preshower has no direct trigger capabilities we will rely on triggering in the MPC to select our event 
sample. At the antipated interaction rate of 328~kHz a rejection of about 330 is required to keep the DAQ bandwidth allocated
to the MPC-EX physics at 1~kHz. As we are primarily interested in the deuteron-going direction for d+Au running we assume 
we will be triggering on only one MPC-EX. For polarized p+p collisions, we assume that the both north and south MPC's will be utilized for 
triggering, but that the DAQ bandwidth allocated to MPC-EX physics is correspondingly higher at 2~kHz. At the antipiated interaction 
rate of 1.3~Mhz the required event rejection is 1300.  

\begin{figure}
  \begin{center}
  \includegraphics[width=0.8\linewidth]{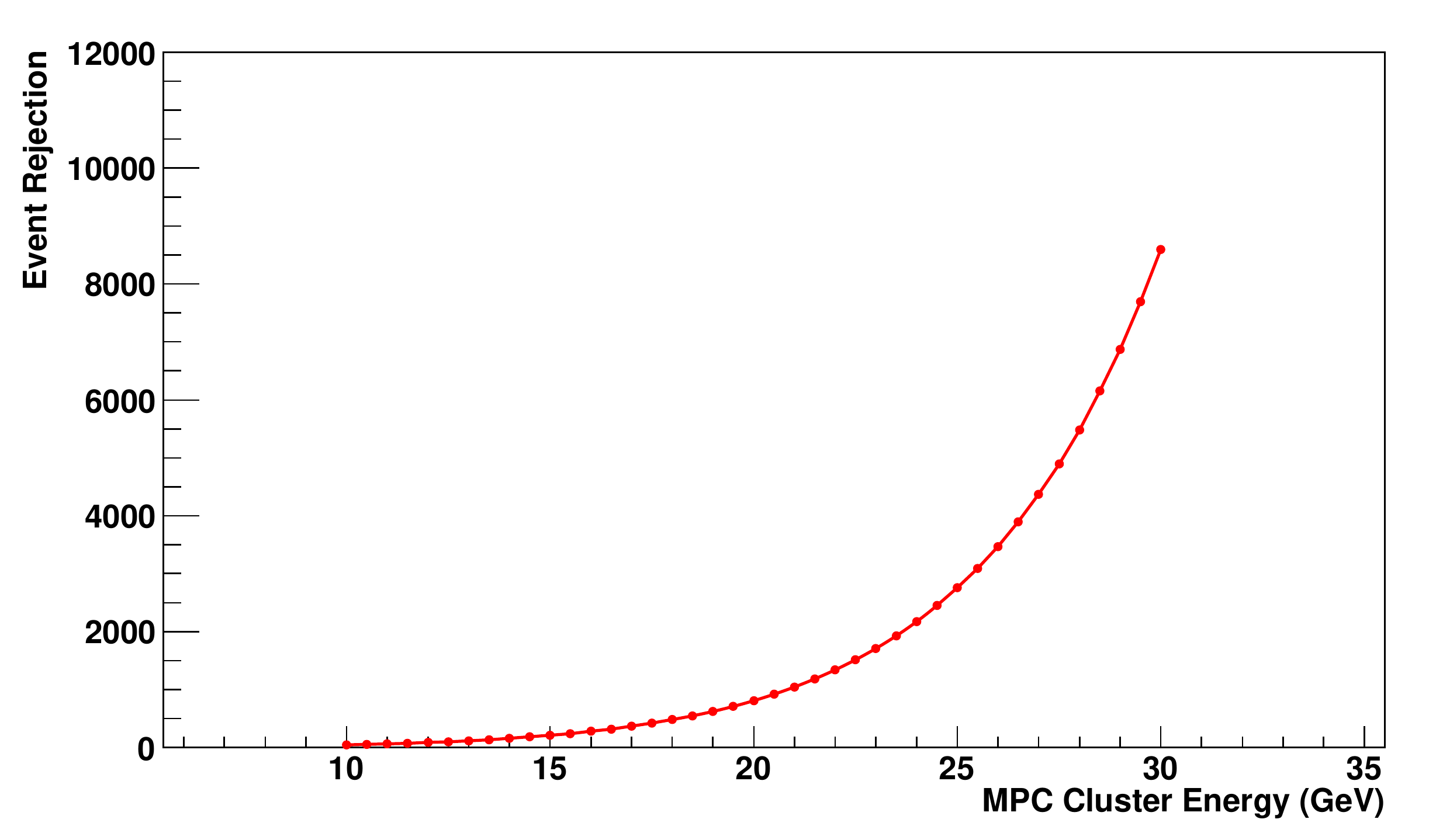}
  \end{center}
  \caption{\label{fig:TrigRej} Event rejection as a function of MPC cluster energy for the {\sc Pythia} p+p 200~GeV event sample. For d+Au
collisions the required rejection of approximately 330 is reached at a trigger cluster energy of 16.5~GeV. For p+p collisions the 
required event rejection of approximately 1300 is reached at a trigger energy of 22~GeV.}
\end{figure}

Figure~\ref{fig:TrigRej} shows the anticipated trigger rejection as a function of MPC cluster energy trigger threshold, based on the  
minbias {\sc Pythia} event sample used for the direct photon analysis with full simulation through the PHENIX GEANT-based package PISA. 
Sufficient rejection is achieved in d+Au collisons at a trigger threshold of 
16.5~GeV.

\begin{figure}
  \begin{center}
  \includegraphics[width=0.8\linewidth]{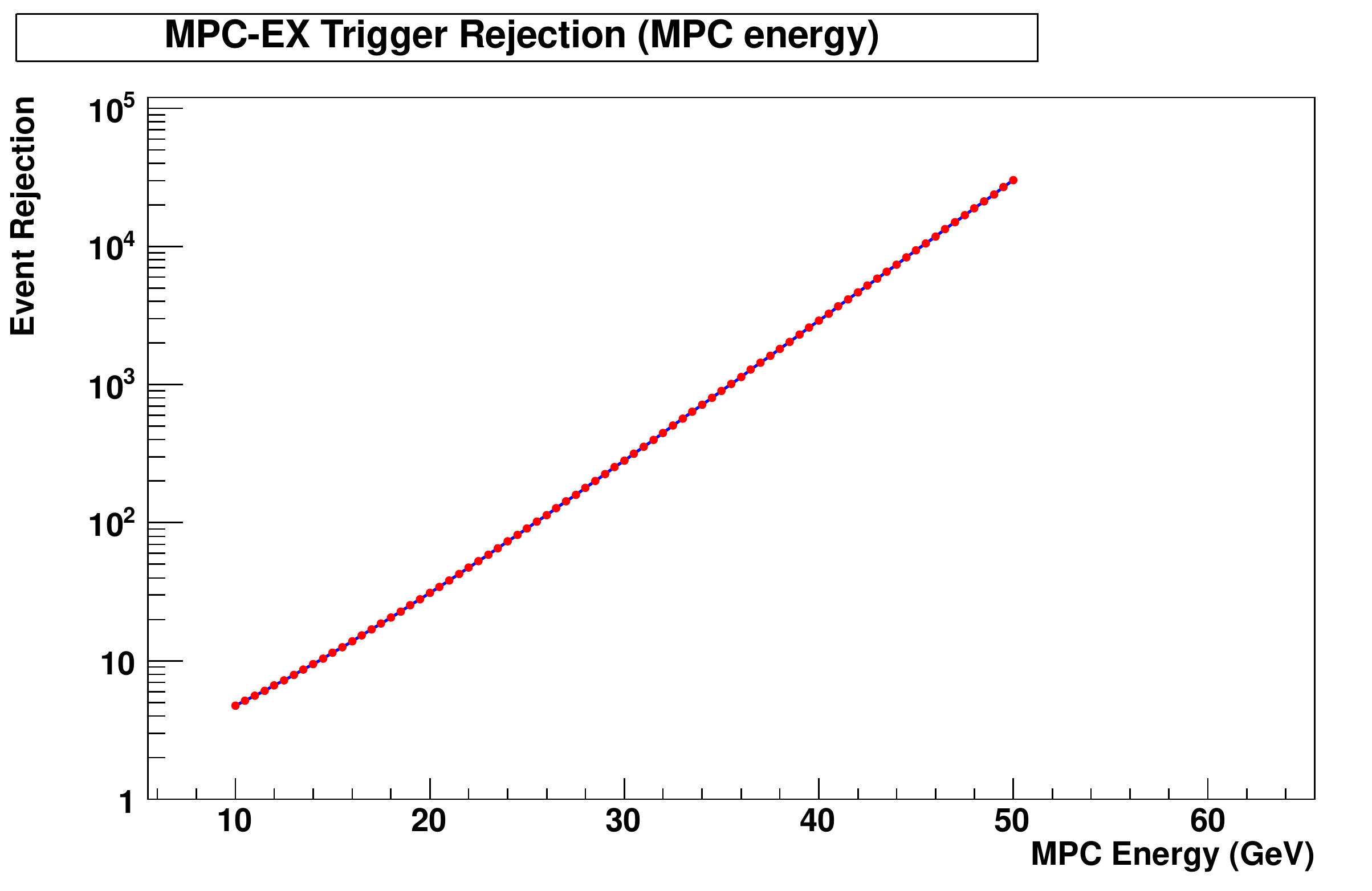}
  \end{center}
  \caption{\label{fig:TotTrigRej} Event rejection as a function of total MPC energy for the {\sc Pythia} p+p 200~GeV event sample. For p+p collisons the 
required event rejection of approximately 1300 is reached at a trigger energy of 35~GeV.}
\end{figure}

For jets in polarized p+p collsions we find that triggering on total energy in the MPC provides an improved trigger efficiency. 
Figure~\ref{fig:TotTrigRej} shows the anticipated trigger rejection as a function of total energy in the MPC, based on the  
minbias {\sc Pythia} event sample used for the direct photon analysis with full simulation through the PHENIX GEANT-based package PISA. 
Sufficient rejection is achieved in p+p collisions at a total energy trigger threshold of 35~GeV.

\cleardoublepage
\resetlinenumber


\chapter{MPC-EX Collaboration}
\label{sec:participants}
\label{sec:B}

\markboth{Appendix B}{MPC-EX Collaboration}

\begin{flushleft}
\begin{description}
 
\item[Brookhaven National Laboratory,] 
\ \\ {\em  Upton, NY 11973, USA} \\
	E.~Kistenev,
        A.~Sukhanov

\item[Chonbuk National University,] 
\ \\ {\em  Jeonju, Korea} \\
	Eun-Joo Kim

\item[Ewha Womens University,] 
\ \\ {\em  Seoul, Korea} \\
	K.I.~Hahn,
	D.H.~Kim,
	S.Y.~Han

\item[Hanyang University,] 
\ \\ {\em  Seoul, Korea} \\
	B.H.~Kang, 
	J.S.~Kang,
        Y.K.~Kim,
	J.S.~Park

\item[Iowa State University,] 
\ \\ {\em  Ames, Iowa 50011, USA} \\
	S.~Campbell,
	J.G.~Lajoie,
	R.~McKay,
	J.~Perry,
	A.~Timilsina

\item[Los Alamos National Laboratory,] 
\ \\ {\em  Los Alamos, New Mexico 87545, USA} \\
	J.~Huang,
        X.~Jiang,
	M.~Leitch,
	M.~Liu

\item[University of California - Riverside,] 
\ \\ {\em  Riverside, California 92521, USA} \\
	K.N.~Barish,
	D.~Black,
	L.~Garcia,
	R.S.~Hollis,
	A.~Iordanova,
	R.~Seto,
	W.~Usher
  
\item[Yonsei University, IPAP] 
\ \\ {\em  Seoul, Korea} \\
	J.H~Do,
	J.H.~Kang,
	H.J.~Kim,
	Y.~Kwon,
	S.H.~Lim,
	M.~Song
 
\end{description}
\end{flushleft}

\cleardoublepage

\backmatter

%

\cleardoublepage
\phantomsection
\addcontentsline{toc}{chapter}{References}


\bibliographystyle{plainurl}
\bibliography{0main}

\end{document}